\documentclass[10pt,
			a4paper,
			oneside,
			openright,
			titlepage,
               		headinclude,
			footinclude,
			BCOR5mm,
               		numbers=noenddot,
			cleardoublepage=empty,
               		tablecaptionabove
							]{scrreprt}
 
\usepackage[T1]{fontenc}
\usepackage[utf8]{inputenc}
\usepackage[english]{babel}
\usepackage{amssymb}
\usepackage{amsmath}
	\DeclareMathOperator{\arccosh}{arccosh}
	\DeclareMathOperator{\sech}{sech}
\usepackage{mathrsfs}
\usepackage{amsthm} 
	\theoremstyle{plain}
	
	\theoremstyle{plain}
	
	\theoremstyle{plain}
	
	\theoremstyle{plain}
	
\usepackage{indentfirst}
\usepackage[style=philosophy-modern,hyperref]{biblatex}
\usepackage{chngpage}
\usepackage{calc}
\usepackage{listings}
\usepackage{graphicx}
\usepackage{subfig}
\usepackage{lipsum}
\usepackage{shapepar}
\usepackage{pifont}
\usepackage{acronym}
 \usepackage{quoting}
	\quotingsetup{font=small}
\usepackage{chngcntr}

\usepackage[
		eulerchapternumbers,
		subfig,
		beramono,
		eulermath,
		pdfspacing,
		listings
				]{classicthesis}
\usepackage{arsclassica}

\newcommand{\myName}{Luca Bellino}
\newcommand{\myTitle}{Temperature and rate effects in damage and decohesion of biological materials}
\newcommand{\mySubTitle}{XXXIV Ph.D. program in Mechanical and Management Engineering\\\vspace{0.1cm}
MAT/07 - Mathematical Physics}

\graphicspath{{}}

\definecolor{lightergray}{gray}{0.99}

\lstnewenvironment{code}%
   {\setkeys{lst}{columns=fullflexible,keepspaces=true}%
   \lstset{basicstyle=\small\ttfamily}}{}

\bibliography{LucaBib}

\defbibheading{bibliography}{%
\cleardoublepage
\manualmark
\phantomsection
\addcontentsline{toc}{chapter}{\tocEntry{\bibname}}
\chapter*{\bibname\markboth{\spacedlowsmallcaps{\bibname}}
{\spacedlowsmallcaps{\bibname}}}}

\areaset[5mm]{400pt}{700pt}
\begin{document}
	\pagenumbering{roman}
	\pagestyle{plain}
	
%

\begin{titlepage}
\pdfbookmark{Titlepage}{Titlepage}
\changetext{}{}{}{((\paperwidth  - \textwidth) / 2) - \oddsidemargin - \hoffset - 1in}{}
    \begin{center}
        {\LARGE  

        \hfill

        \vfill

        {\spacedlowsmallcaps{\myName}} \\ \bigskip

        {\color{Maroon}\spacedallcaps{\myTitle}} 

        }

        \vfill

		{\spacedallcaps{\mySubTitle}}

        \vfill       
        
      {\color{Maroon} \textsc{Ph.D. Thesis}}\\\smallskip

    \end{center}   \vspace{3cm}   
    
    \vfill

        \noindent{\color{Maroon} \textsc{Thesis Supervisors:}}\\
        Prof. Giuseppe Maria Coclite - \textit{Polytechnic University of Bari}\\
        Prof. Giuseppe Florio - \textit{Polytechnic University of Bari}\\
        Prof. Alain Goriely - \textit{Oxford University, Mathematical Institute, OCIAM}\\
        Prof. Giuseppe Puglisi - \textit{Polytechnic University of Bari}\\
       
       \noindent{\color{Maroon} \textsc{Thesis Referees:}}\\
        Prof. Paolo Biscari - \textit{Polytechnic University of Milan}\\
        Prof. Sefi Givli - \textit{Technion, Israel Insitute of Technology}\\

\end{titlepage}

	\cleardoublepage
	\null
	\thispagestyle{empty}
	\cleardoublepage

%

\thispagestyle{empty}
 \null\vspace{\stretch{1}}

 \begin{flushright} 
 	\textit{to my sister Chiara}\\
	\bigskip
	\bigskip
 \end{flushright}
 
 \vspace{\stretch{2}}\null
 
\bigskip
\bigskip
\begin{quotation}
\footnotesize
\begin{flushright}
\textit{“Faithless is he that says farewell when the road darkens”}\\
(J. R. R. Tolkien)
\end{flushright}
\end{quotation}

	\cleardoublepage
	\null
	\thispagestyle{empty}
	\cleardoublepage

%
\pdfbookmark[1]{Acknowledgments}{acknowledgments}

\begin{flushright}{\slshape
    "We rise by lifting others"\\\vspace{0.3cm}
    I truly believe it.\\
    Luca} \\ \medskip
 \end{flushright}

\bigskip

\begingroup
\let\clearpage\relax
\let\cleardoublepage\relax
\let\cleardoublepage\relax

\chapter*{Acknowledgments}

\noindent When I started this journey, more than three years ago, I had no idea of how much effort and sacrifices it would have required, and of how great and beautiful satisfaction I gained. I say that alone I would never have made it, and I would like to express my deep and sincere gratitude to all those who have guided and joined me on this journey.

\vspace{0.15cm}
\noindent \textit{PS}: I do not want to lose the seriousness of these acknowledgements, therefore, since they are written at 4 a.m. the dawn of the final dissertation, and with a couple of pints of IPA, I consciously believe that they are an expression of my most sincere feelings.

\vspace{0.9cm}
\noindent Thanks to Professor Geppe Puglisi and Professor Giuseppe Florio. We often forget how much responsibility there is behind the task of leading someone, of teaching. And I don't think my thanks are enough to express how much gratitude I have towards you. You have always trusted me, assigned tasks that required maturity and responsibility and you have always treated me as an equal, both in our research work we are so passionate about and in the everyday life. For me, this has been the greatest motivation to always do my best. Working while having fun, smiling, surrounded by beautiful people is an added and rare value, which I feel lucky to have and share with you, with Salvo, who never fails to bring good humour and with Vincenzo, with whom I share, in addition to work, the friendship of a lifetime.

\vspace{0.2cm}
\noindent Thanks to Professor Giuseppe Coclite, for having always been present during my journey, for giving me valuable advice, opportunities for growth (wonderful Cetraro!), and for all the constructive and formative discussions we shared. 

\vspace{0.2cm}
\noindent I particularly want to express my gratitude to Professor Alain Goriely, because you have entrusted me as a stranger, and with extreme kindness and availability you have guided me in these last months of work, opening me to new possibilities and perspectives.

\vspace{0.9cm}
\noindent Thanks to my Mother and my Father. Your tireless support and unconditional love has allowed me to grow, and you have given me endless possibilities to do so. I have always been able to count on your support and you have my infinite gratitude for always sharing my choices, offering me criticisms and advice that have helped me to choose the right path to take.

\vspace{0.2cm}
\noindent Thanks to my sister Chiara, my strength and my light. You've never left me alone, even in the worst of times, and I'll never be grateful enough for that. I know that I can always count on you, and in spite of any distance, you are always closer than ever.

\vspace{0.2cm}
\noindent I don't have the right words to thank my whole family. My wonderful grandmothers, my uncles and aunts, my little and big cousins, who are my brothers. We have always been distant, but we have never been far, and this makes every occasion extremely precious and a reason for celebration and joy. Knowing that we can count on each other is a strong certainty. Our mutual support, inevitable in the beautiful moments, and even more so in the difficult ones, is hope, strength and love.

\vspace{0.9cm}
\noindent Thanks to my brothers, my best friends. Nicola, Vincenzo, Michele, Michele and Silvia. I don't know what I would do without you. You are always here, you have always been here, and I hope you will always be here. It is beautiful to know that we can count on such a strong and indissoluble bond. We have shared every emotion, joy, drama and adventure together so that I can no longer distinguish the difference between a life with you and one without. Thanks.

\vspace{0.9cm}
\noindent I would like to be able to express a special thanks to those who have faced with me these years, and have been part of this difficult journey with me, during which mutual support, friendships, laughter, outbursts, dramas, joys and evenings spent together made the weight of an uncertain future lighter and more bearable. I am more than certain that the passion that has guided us up to now, and that perhaps will lead us on different paths, will reserve for all of us beautiful surprises, because we deserved them because we put all our commitment and dedication into our work. I wish this to all of us. Thanks in particular to all my friends who have given life to our open space, a place for communication, discussion and fun. Thanks to Davide, Vito Ceglie, Alessandro Nitti, Vito Errico, Simone Carone, Michele Tricarico and all the others. Thanks to Donatella, because you are a beautiful person and one of the best friends I have, I know I can always count on you. Thanks to Simone, we have strengthened each other too many times, in work and in the friendship that binds us outside of it.

\vspace{3.0cm}
\noindent Finally, thanks to what happens by chance, \textit{and it's beautiful}. Too many times we dwell on how many things are wrong, on how many things could have gone differently and on how many we had the possibility to intervene to remedy. Too many times, however, we forget to appreciate all the beauty that we have around us, in its countless and changing faces.

\vspace{2.5cm}

\begin{flushright}
\textit{I wish to thank all of you who believed in me, especially when I saw nothing beyond the dark veil.}
\end{flushright}

\vspace{0.6cm}
\begin{flushright}
Deeply, Luca
\end{flushright}

\endgroup

	\cleardoublepage
	\null
	\thispagestyle{empty}
	\cleardoublepage

%
\pdfbookmark[1]{Abstract}{Abstract}

\chapter*{Abstract}

The incredible thermo-mechanical properties of biological materials arise from the microscopic scale due to a complex hierarchical mechanism, which is regulated by microinstabilities at the molecular level. The description of such complex structures is allowed by both the know-how introduced by the advent of single molecule force spectroscopy experiments, which gives the possibility of studying such systems in different thermal and mechanical conditions, and the possibility of correctly mimicking their behaviour at the lowest scale by introducing mathematical models based on non-convex energies. In this thesis, different classes of models are introduced to describe the important features of phase transition, decohesion and damage under different conditions of applied forces and displacement, thermal fields and rates of loading. By increasing the level of complexity of such models, different phenomena have been analyzed. For instance, by introducing a chain of bistable units to mimic the behaviour of a titin molecule undergoing unfolding, it has been described the effect of the handling device in single molecule experiments, which strongly affects the system's mechanical response, leading to large errors in the measure of the resulting force or displacement. Temperature effects are considered within a Statistical Mechanics framework, also in the case when non local interactions are introduced. Indeed, phenomena such as the presence of a stress peak in the force-extension diagram and corresponding to the generation and nucleation of a phase is experimentally observed in tensile tests on memory shape nanowires or polymer materials and can be described as a competition between interfacial energy terms and entropic effects. The cooperativity of weak interactions, such as hydrogen bonds, has been also studied to highlight phenomena such as decohesion and fracture in biological systems. Indeed, simple amino acids are arranged in a multiscale fashion resulting in high performing hierarchical materials and structures, with elevated mechanical properties. Specifically, considering elastic springs coupled with breakable units, in this thesis a micromechanical model of systems such as the double-stranded DNA helix or the bundles of microtubules (MT) and tau proteins arranged within the axons with thermal and rate effects has been deduced. The decohesion process is found to be highly regulated by the relative stiffness of the two pseudo-elastic units, and the type of fracture may range from an abrupt collapse (fragile behaviour) to a sequential detachment of the bonds (ductile behaviour). This effect is also enhanced when the loading rate is considered, where the ability to overcome energy barriers separating the metastable states becomes crucial. The results obtained in the thesis are compared to pieces of evidence from an extensive literature review and to the experimental behaviours of the systems described, and microscopic constitutive analytic laws are deduced illustrating the overall behaviour of such complex systems regulated by multiscale microinstabilities.

	\cleardoublepage
	\null
	\thispagestyle{empty}
	\cleardoublepage

%

\pdfbookmark[1]{Publications}{publications}
\chapter*{Publications}


\noindent List of publications and preprints of the author of the PhD thesis. 


\begin{refsection}[ownpubs]
    \small
    \nocite{*}
    \printbibliography[heading=none]
\end{refsection}



	\cleardoublepage
	\null
	\thispagestyle{empty}
	\cleardoublepage

%
\pagestyle{scrheadings}
\pdfbookmark[1]{\contentsname}{tableofcontents}
\setcounter{tocdepth}{2} 
\setcounter{secnumdepth}{3} 
\manualmark
\markboth{\spacedlowsmallcaps{\contentsname}}{\spacedlowsmallcaps{\contentsname}}
\tableofcontents
\automark[section]{chapter}
\renewcommand{\chaptermark}[1]{\markboth{\spacedlowsmallcaps{#1}}{\spacedlowsmallcaps{#1}}}
\renewcommand{\sectionmark}[1]{\markright{\textsc{\thesection}\enspace\spacedlowsmallcaps{#1}}}

%
\clearpage
\begingroup
    \let\clearpage\relax
    \let\cleardoublepage\relax
    \pdfbookmark[1]{\listfigurename}{lof}
    \listoffigures

    \vspace{8ex}

%

    \pdfbookmark[1]{\listtablename}{lot}
    \listoftables

    \vspace{8ex}

%

%
    \pdfbookmark[1]{Acronyms}{acronyms}
    \markboth{\spacedlowsmallcaps{Acronyms}}{\spacedlowsmallcaps{Acronyms}}
    \chapter*{Acronyms}
    \begin{acronym}[HS-AFM]
  	\acro{AFM}{atomic force microscopy}
	\acro{BPM}{biological protein materials}
	\acro{DAI}{diffuse axonal injury}
	\acro{dsDNA}{double-stranded DNA}
	\acro{FJC}{freely jointed chain}
	\acro{HS-AFM}{high-speed AFM}
	\acro{MD}{molecular dynamics}
	\acro{MEMS}{micro-electromechanical system}
	\acro{MT}{microtubules}
	\acro{NEMS}{nano-electromechanical system}
	\acro{NN}{nearest neighbor}
	\acro{NNN}{next to nearest neighbor}
	\acro{PSD}{position-sensitive detector}
	\acro{SMA}{shape memory alloys}
	\acro{SMFS}{single molecule force spectroscopy}
	\acro{ssDNA}{single-stranded DNA}
	\acro{TBI}{traumatic brain injury}
	\acro{TST}{transition-state theory}
	\acro{WLC}{worm like chain}    
    \end{acronym}

\endgroup

	\cleardoublepage
	\null
	\thispagestyle{empty}
	\cleardoublepage
\pagenumbering{arabic}
	
%


%
\renewcommand{\thefigure}
{\arabic{chapter}.\arabic{figure}}
\setcounter{figure}{0}

\renewcommand{\thetable}
{\arabic{chapter}.\arabic{table}}
\setcounter{table}{0}

\renewcommand{\theequation}
{1.\arabic{equation}}
\setcounter{equation}{0}

\chapter{Introduction}
\label{ch_1}

%
Nature has always been an enlightening source of inspiration for humankind across time and it represents an infinite vessel of ideas and smart solutions from which we may draw. We first learned how to use natural materials to meet primary needs such as building shelters with mud and straw or manipulating the fire to cook. Afterwards, with the necessity of coming up with man-made solutions for specific applications the development of new techniques was required to modify raw materials. Indeed, archaeologists distinguish different eras according to material use. Since then, the introduction of technologies and theories to predict the behaviour of newly discovered materials became essential. Meanwhile, scientists and philosophers started questioning the underlying physics behind certain phenomena and coming up with ideas, models and theoretical frameworks. 

More recently, materials such as steel, concrete, plastics, and composites were the base of both the Industrial Revolutions and fundamental new technologies in all fields of engineering and physics. Understanding new natural material behaviours have been at the source of scientists' and engineers' investigations and human technological revolutions. For instance, Aristotele was impressed by the ability of a gecko to ``run up and down a tree in any way, even with the head downward'' (\cite{aristotele}) and he tried to provide a physical interpretation based on the particular ``structure'' of the gecko pads even though he was limited by the knowledge of that age. It took two thousand years of experimental and theoretical development to understand this phenomenon based on a multiscale hierarchical fibrillar structure of the pads spreading down from millimetres to the nanoscale, where Van der Walls forces play a crucial role in the adhesion-decohesion process (\cite{autumn:2000}). 

\begin{figure}[b!]
\centering
  \includegraphics[width=0.95\textwidth]{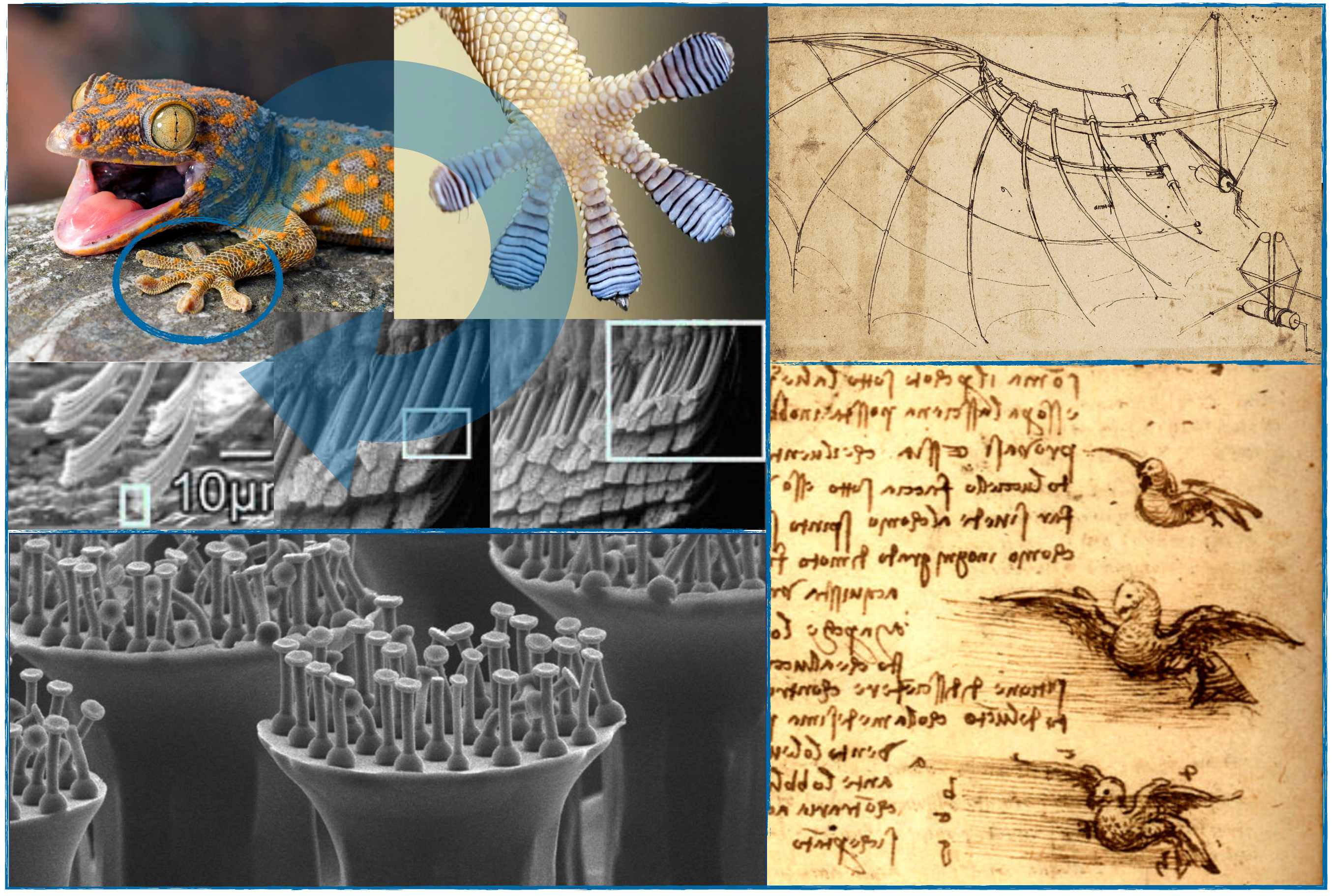}
  \caption[Biomimetics: some examples]{Example of biomimetics in history. In the left panel, the hierarchical structure of the gecko pads is shown, firstly sensed by Aristotele in $343$ B.C. at different scales ranging from the millimetres of the pads to the micrometres of the smallest \textit{lamellae}. In the right panel different sketches and drawings of Leonardo da Vinci represent his will to understand the flight of birds and design tools to fly, one of the ancient dreams of humankind.}
  \label{fig:ch1_biomimetics}
\end{figure}

Leonardo da Vinci was fascinated by nature and he based his life-work trying to dig deeper and deeper into biological secrets. He first studied the flight of birds and designed the first sketches of wings and aircrafts. He was inspired by the shells of some mollusks and turtles to design tanks and buildings with higher structural properties and he pioneered the anatomy and physiology of the human body (see Figure~\ref{fig:ch1_biomimetics}). Leonardo da Vinci can be considered the father of \textit{biomimetics}, which is the process of understanding nature, developing ideas and transferring them into technological solutions. Since then, the evolution in this field has slowed down by the complexity of natural material behaviour, requiring the development of experiments and new material theories.

Nowadays biomimetics is a source of innovative research and technologies, especially in the development of new experimental techniques. Similarly, the evolution of new technologies opened up the possibility of controlling material structure at the micro and nanoscale leading to the emergence of new fields such as \ac{NEMS}, \ac{MEMS} and new metamaterials. The impact has already been considerable in many different technological fields, ranging from aerospace, mechanical, electrical and civil engineering, to biotechnology, medicine and computer science.

%
%

Natural and biological phenomena are also fundamental for understanding human life in processes such as DNA and proteins chemo-thermo-mechanics,  growth and damage of cells and tissues and the related health pathologies associated with their misfunctioning. While the last century's evolution in this field was mainly related to genetics, in the last decades it has been realized that the knowledge of the response of biological systems to external or internal stimuli has to be undertaken also from a thermo-mechanical perspective.

One of the key features of biological systems is that the overall behaviour results from hierarchical structures arranged in a multiscale fashion. Consequently, an effective theory aimed at the deduction of bioinspired materials and devices requires a bottom-up approach that, starting from a detailed description at the lower involved scales, deduces macroscopic laws directly depending on the micro and nanostructure properties. Such complex multiscale organization is typically attained by repetitive units arranged in complex geometric domains with a self-similarly construction at different scales. The behaviour at each scale is regulated by local intra-domain (often strong) interactions and (often weak) inter-domain interactions. The macroscopic response is thus a homogenized result of complex interactions among different scales, with a continuous transition between the many local minimizers of the bumpy energy landscape (see Figure~\ref{fig:ch1_energy}). Therefore, complexity is an emerging property arising from interactions at small scales which reminds us of the words of Mr Feynman ``\textit{There is plenty of room at the bottom}'' (\cite{feynman:1959}). 

The challenge in the field of hierarchical \ac{BPM} and structures is the correct description of the structure at the molecular level. At this level, these systems are characterized by a multiplicity of configurations, with comparable energy. We need to find among the tools of statistical mechanics a correct framework whereas a multiscale approach to deduce macroscopic responses.

%
\section{Single molecule experiments}

%
\begin{figure}[b!]
\centering
  \includegraphics[width=0.95\textwidth]{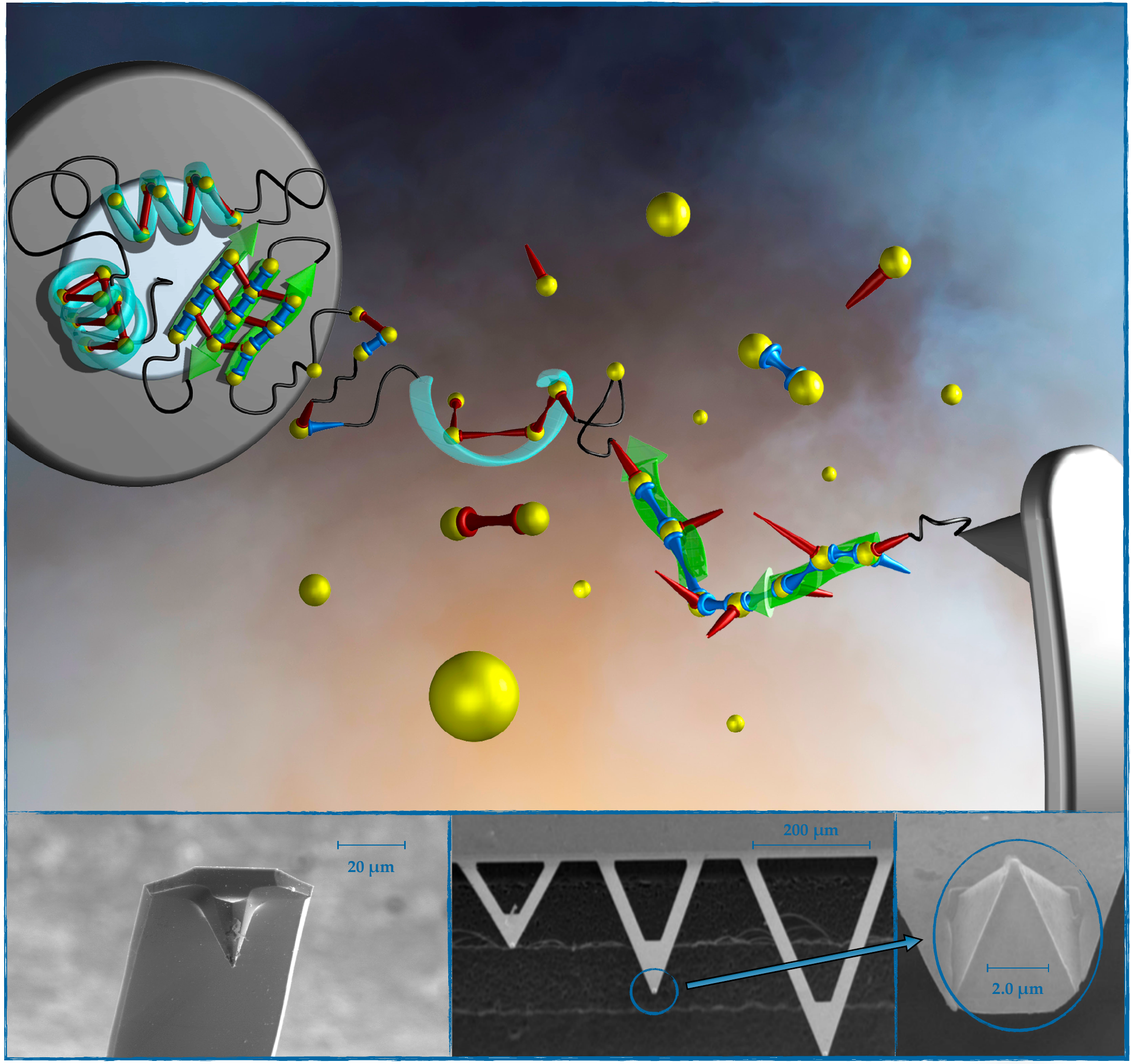}
  \caption[Atomic Force Microscopy and cantilever tips]{In the top panel, a graphical picture of an AFM is shown. The cantilever tip is pulling a molecule attached to the surface of the microscope causing the process of unfolding of the secondary structures of $\alpha$-helices and $\beta$-sheets. Created by the author. On the bottom line, different examples of cantilever tips are reported with particular focus on the dimension of such tools. Reproduced from (\cite{rounsevell:2004} and Southwest Center for Microsystems Education (SCME)).}
  \label{fig:ch1_SMFS}
\end{figure}
%

A fundamental example of interest for this thesis is given by materials and tissues, constituted at the molecular scales by proteins. The experimental investigation of the complex behaviour of these materials at the microscopic scale has been made possible in the last three decades by the advent of \acf{SMFS} techniques. Based on innovation in material science and nanotechnologies, these devices are based on complex tools such as small and stiff cantilevers as in the case of \acs{AFM}s or nanobeads to handle single protein molecules in optical tweezers, widely used to perform experiments on RNA or DNA hairpins, as shown in Figure~\ref{fig:ch1_SMFS}. Single molecule experiments allow the investigation of the inter- and intra-molecular interactions of biological structures and the measure of energy barriers separating energy wells, giving thus the possibility of elucidating the main properties of the multiwells energy landscape (\cite{bustamante:2000}). Different experimental techniques give the possibility to explore a vast operational range in terms of resolution, stiffness, probe size or boundary condition (\cite{neuman:2008}). 

We now discuss the most commonly adopted techniques of \acs{SMFS} related to the research presented in this thesis. 

Atomic force microscopy (\acs{AFM}) is typically used for pulling single molecule (\cite{dutta:2016}) or study protein unfolding (\cite{rounsevell:2004}) by fixing the total probe displacement. The first atomic force microscopy was invented in $1985$ by Berg Binnig and colleagues and adopted in the field of tribology for scanning and mapping rough surfaces with atomic resolution (\cite{binnig:1986}). The goal of such an experiment is to obtain the topography of the sample and to measure the roughness of the surface with great accuracy, but it was soon realized that it was also the perfect apparatus for single molecules pulling experiments to explore their mechanical response (\cite{francis:2010}), as shown in Figure~\ref{fig:ch1_SMFS}. The molecule attaches to the microcantilever tip, available in different shapes, materials and stiffnesses ranging approximately from $10$ to $10^{5}$ $pN/nm$, giving them a very high range of sensitivity (see Table~\ref{tab:stiffnesses}). A laser light hits the cantilever tip and it is deflected through a lens on a \ac{PSD} photodiode, that reads the position by converting the incident light into voltage. On the other side, the sample is fixed on a piezo-electric diode that may move in all directions when a scanning test is performed or it can be fixed to hold the molecule when pulled. A variable force is obtained, depending on the stiffness of the cantilever, by increasing the molecule's end-to-end length. This set-up in which the displacement is controlled is called a \textit{hard device}. The possibility of accurate control of the velocity of the photodiode is crucial because when a surface is scanned the slowest the movement the more accurate the surface topography. Similarly, when a molecule is pulled it may be of interest to study the molecule's rate-dependent mechanical response under different pulling velocities. With the introduction of \ac{HS-AFM}, this velocity may reach values ranging from few $nm/s$ to $4000$ $\mu m/s$. (\cite{garcia:2002}).  

%
\begin{figure}[b!]
\centering
  \includegraphics[width=0.95\textwidth]{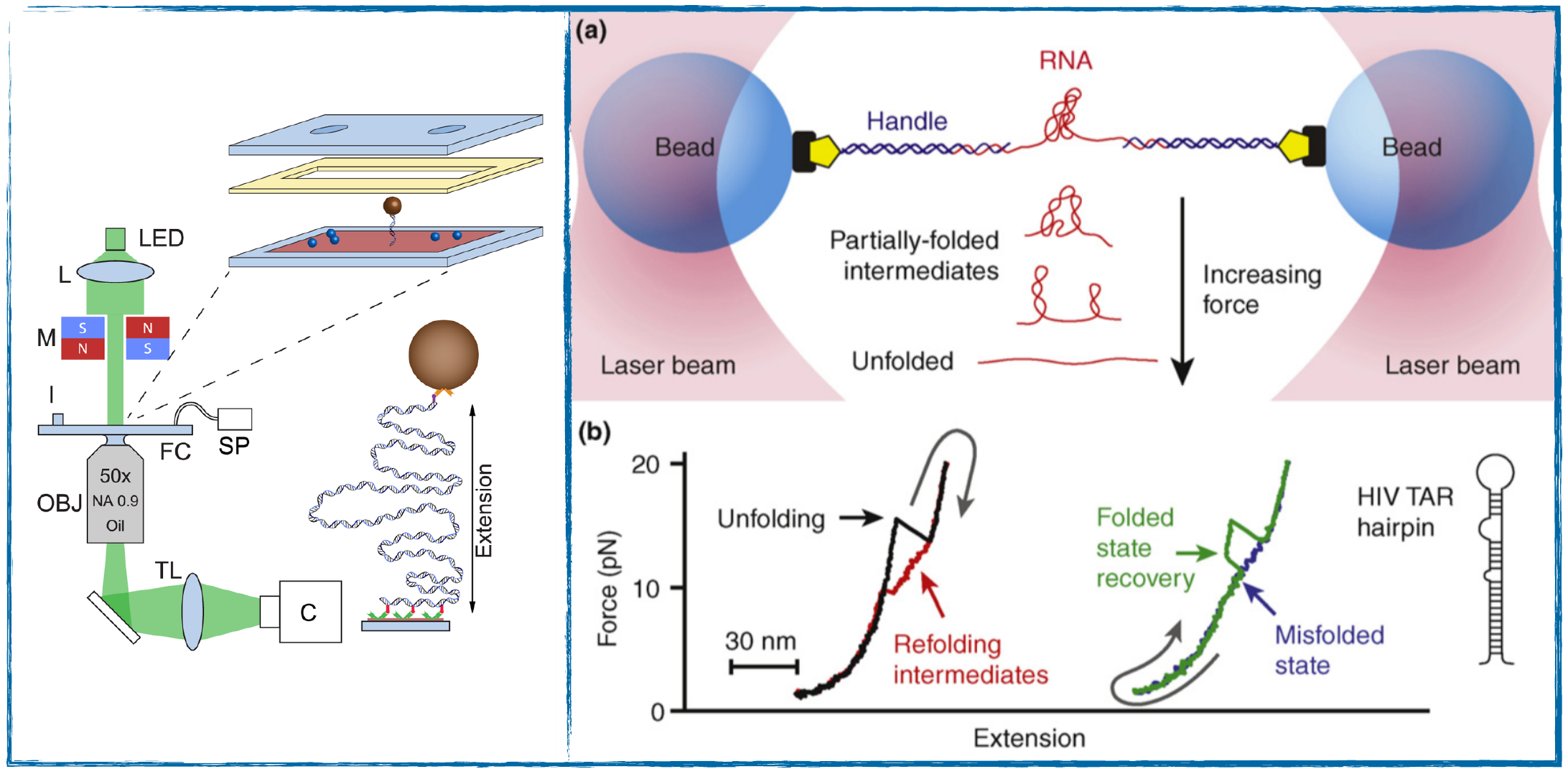}
  \caption[SMFS experimental apparatuses]{Schematic representation of  Magnetic (left) and Optical (right) Tweezers is shown and the resulting force-extension curves are represented when performing experiments on an RNA sequence. Reproduced from (\cite{woodside:2008,burnham:2014}).}
  \label{fig:ch1_SMFStw}
\end{figure}
%

Another experimental device for \ac{SMFS} is the \textit{optical tweezer}. In this case, it is possible to control the force applied to the probe and deduce the displacement, a set-up known as a \textit{soft device}. It is one of the most versatile techniques to manipulate single molecules and it allows the measurement of a three-dimensional displacement with nanometer resolution. It is used to study systems in which decohesion or denaturation processes happen, such as the breaking of the DNA double helix (\cite{woodside:2006,burnham:2014}), as shown in Figure~\ref{fig:ch1_SMFStw}. It was invented by the $2018$ Nobel Prize in Physics Arthur Asking and coworkers in $1986$ to demonstrate the possibility of generating the optical levitation of microspheres with two counter-propagating beams (\cite{ashkin:1986}). 

The mechanism relies on the forces generated by the emission and absorption of light. Specifically, the optical trap is generated by the radiation pressure exerted by a laser-focused beam that holds the probe in a fixed position. The narrowest point of the beam is known as the bea4 and it contains a very strong electric field gradient. Dielectric particles are attracted in the region where the electric field is strong, the centre of the beam. along the direction of the gradient. Moreover, the laser light tends to apply a force on particles in the beam along the direction of beam propagation due to momentum conservation. Indeed, photons that are absorbed or scattered by the dielectric particle impart momentum to this particle. The resulting force is known as \textit{scattering force} and it pulls the particle slightly downstream with respect to the beam waist, generating the load. Thanks to this particular layout, one of the main features of optical tweezers is the possibility of controlling the molecule without contact, ensuring a sterile environment during experiments. It is fundamental to highlight that the scattering force that pulls the particle is linear in the displacement, thus the optical trap can be compared to a spring-like system. Typical stiffness is much smaller than the cantilever's ones, as shown in Table~\ref{tab:stiffnesses}. A similar apparatus is the magnetic tweezer that, thanks to a magnetic field generated by permanent magnets is able to move and stretch particles constrained between the beads and a coated glass surface with an assigned force, that is proportional to the gradient of the field (see the left panel of Figure~\ref{fig:ch1_SMFStw}). It was invented in $1996$ by Strick, Bensimon and Croquette to study the elasticity of supercoiled DNA (\cite{strick:1996}). 

Although these techniques may be very different in terms of mechanical properties and specific applications, the common feature is that they concern study biological systems at the molecular scale with the possibility of deducing detailed key thermo-mechanical characteristics. For instance, one may analyse the relative stability of the multiple metastable configurations at varying pulling conditions (\cite{szabo:2005}). It is also possible to study the response of the probe when different directions of the applied load are considered (\cite{walder:2017}). A key topic that will be discussed further in this chapter is the possibility of varying the loading rates, thus enabling the analysis of the rate-dependent mechanical response (\cite{benichou:2016,cossio:2015}). 

%
\begin{table}[t!]
	\footnotesize
	\centering
    	\begin{tabular}{lllll}
		\toprule
		\textsc{Handler}	& \textsc{Typical Force ($N$)} 	& \textsc{Stiffness ($Nm^{-1}$)}  	& \textsc{Resolution ($m$)}\\
		\midrule
		Cantilevers 		&  $10^{-11}-10^{-7}$		& $0.001-100$ 					& $10^{-10}$		\\
		Photon Field		&  $10^{-13}-10^{-10}$		& $10^{-10}-10^{-3}$		 		& $10^{-9}$	  \\
		Magnetic Field		&  $10^{-14}-10^{-11}$		& n.a.						& $10^{-8}$	 \\
		Microneedles		&  $10^{-12}-10^{-10}$		& $10^{-6}-1$					& $10^{-9}$	\\
		Flow field			&  $10^{-13}-10^{-9}$			& n.a.						& $10^{-8}$	\\
		\bottomrule
		\end{tabular}
	\vspace{0.2cm}
	\caption[Overview of the typical \ac{SMFS} techniques stiffnesses]{Main mechanical properties of the handlers of the most used \ac{SMFS} experimental instruments. Reproduced from (\cite{bustamante:2000}).}  
\label{tab:stiffnesses}
\end{table}
%

However, one of the main drawbacks of \acs{SMFS}, which will be analyzed in this thesis, is that despite the technological advances in the field, the handling device unavoidably affects the experimental response of the tested molecules, and is often underestimated or even neglected in the literature (\cite{maitra:2010}). Indeed, as shown in Table~\ref{tab:stiffnesses}, depending on the specific apparatus the stiffness may change by orders of magnitude and it can be often comparable to the stiffness of the molecule under investigation. As an example, the finite value of the cantilever stiffness of an \ac{AFM} can produce a wrong estimate for both the dissipated energy and the unfolding force thresholds, leading to major discrepancies between the theoretical and experimental forces (\cite{biswas:2018}) as well as transition rates (\cite{dudko:2008,li:2014}). Therefore, it is crucial to take into account this feature when modelling the mechanical response of a biological system at the microscopic scale. In Chapter~\ref{ch_3} of this thesis, I tackle this problem by introducing a model describing a molecule undergoing a phase transition from the folded configuration to the unfolded one and its interaction with the pulling device considered as a whole thermodynamical system. In particular, following the work in (\cite{fp:2019}), the thermo-mechanical response of such a system is analyzed. I obtain fully analytical formulas in a statistical mechanics framework and consider the effect of the device stiffness. As I show, by varying such stiffness as compared with the tested molecule the response varies between the two limit cases of applied force (described by the Gibbs statistical ensemble), a soft device, and applied displacement, a hard device (Helmholtz ensemble). Such a behaviour can be extended also in the ``continuum'' and ``thermodynamical'' limits.

%
\section{Proteins and polymers structure}

\begin{figure}[b!]
\centering
  \includegraphics[width=0.95\textwidth]{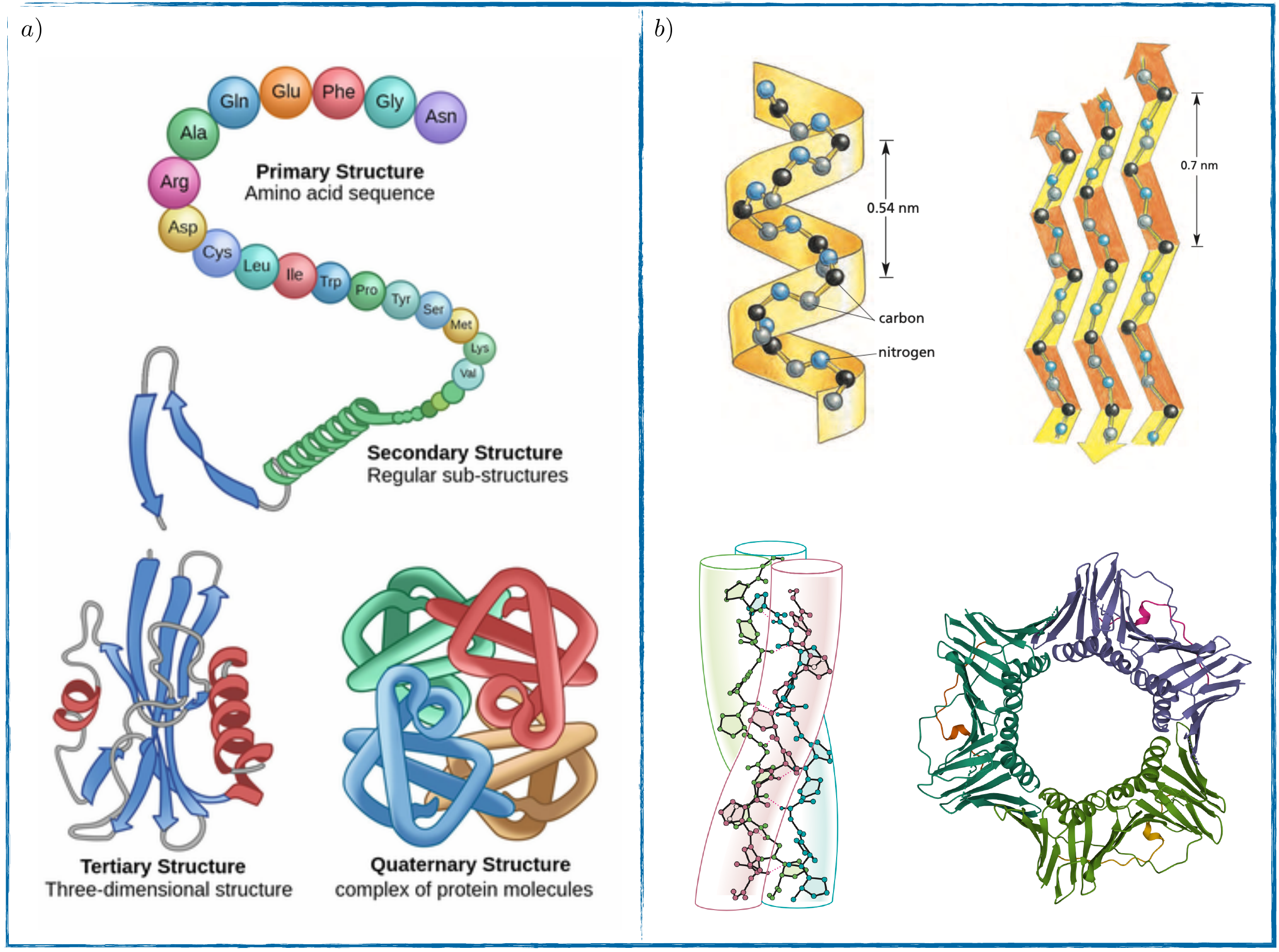}
  \caption[Structure of the Protein]{Primary, Secondary, Tertiary and Quaternary structures of proteins. In panel a) it is represented the evolution of the primary structure, composed of a chain of amino acids joined by covalent bonds that becomes arranged in regular substructures obtained when the polypeptide structures arrange themselves and some amino acids may bond with others on different sites of the chain through hydrogen bonds (secondary structure). Then, a three-dimensional tertiary structure is obtained and finally, complex proteins can be synthesized (quaternary structure). In panel b) different examples of structures are reported such as the elementary $\alpha$-helix and the typical $\beta$-sheet (top), a collagen-like tertiary structure made by triple helices (bottom left) and a quaternary, three-dimensional structure of a protein (bottom right). Reproduced from (\cite{roberts:2002}) and from the Protein Data Bank.}
  \label{fig:ch1_proteins}
\end{figure}

The correct description of biological materials from a modelling point of view requires detailed knowledge about the structures that can be found in different configurations. Although these materials are quite complex and may have an enormous spectrum of applications differing in the functional task they are made for, the basic structural organization is characterized by features common to almost all protein materials. At the very ground level, the $21$ amino acids that are found in nature are arranged in series forming a sequential chain of precise length measured in terms of number of amino acids, that are responsible for cell activities such as repair, survival, regeneration and growth (\cite{hughes:2016,goriely:2017}). This is the so-called \textit{primary structure} and it is schematized in Figure~\ref{fig:ch1_proteins}$_a$. From an energetic point of view, we may expect that the exact position of the amino acids along the chain characterizes how the polypeptide sequence folds based on the attracting-repulsive interactions between both amino acids and the surrounding environment (\cite{roberts:2002}). This represents a long-standing unsolved theoretical problem that is currently  tackled using various approaches such as machine learning (\cite{butler:2018}), statistical mechanics (\cite{luca2019,manca:2013,makarov:2009}) or molecular dynamics (\cite{gonzalez:2017,karplus:2002,hsu:2020}). Concerning the deduction of protein folding stability, the problem is particularly important due to the possibility of the onset of proteins mutations that may lead to malfunctioning and pathology. On the other hand, a mismatching in the protein sequence arrangement may cause different diseases (\cite{aliu:2018}) and in particular hereditary metabolic disorders, caused by the degradation of some particular amino acid. A well-known example is sickle cell anaemia, a blood disorder characterized by an anomaly in the transport of the red blood cells due to the rigid, sickle-like shape of haemoglobin (\cite{rees:2010}). 

The \textit{secondary structure} is a structure obtained when the polypeptide chain folds combine the amino acids' strong local covalent bonds with the weaker non-local hydrogen bonds. The two most common secondary configurations adopted by proteins that can be found in nature are the so-called $\beta$-sheet and $\alpha$-helix, which are reproduced in Figure~\ref{fig:ch1_proteins}$_b$ and b. Two conformational $\beta$-sheets configurations can be attained depending on the direction of the amino acids planes, which can be found in anti-parallel or parallel strands. Also, many types of helices can be found, depending on the pitch of the helix, as theoretically predicted by Linus Pauling in 1951, earning him the Nobel prize (\cite{pauling:1951}). He assumed that the peptide bonds are planar, that all the amino acid residues are equivalent with respect to the backbone conformation and that the bonds among amide protein to an oxygen atom of another residue are of the hydrogen type with a $N-O$ distance of $2.72$ $\AA$ (\cite{edison:2001}).

%
%

When the secondary structures --considered as an almost planar configuration-- fold in three dimensions in different geometric shapes the \textit{tertiary structures} are generated. Depending on the specific arrangement not only its shape is fixed, but also its function (see Figure~\ref{fig:ch1_proteins}$_a$). The three-dimensional shapes are mostly regulated by the so-called hydrophobic or hydrophilic interactions, with non-polar solutes linked in order to minimize their interaction with water (\cite{baldwin:2013}). For instance, tropoelastin, the most elastic protein in our body, is conformationally composed of a particular arrangement of the primary, secondary and tertiary structures, mediating hydrophobic and hydrophilic interactions through hydrogen bonds in order to obtain a particular shape capable of linking with other tropoelastin protein and undergo biological processes (\cite{baldock:2011,yeo:2016}).

The \textit{quaternary structure} refers to the coupling and interaction among several protein chains and subunits into a closely packed arrangement each one composed of its own primary, secondary and tertiary structures. It is regulated typically by non covalent bonds such as Van der Waals forces and hydrogen links and they were first discovered by Sir John Kendrew and Max Petrux with X-ray diffraction techniques in the $50$s (\cite{kendrew:1958}).

%
\subsection{Hierarchy in biological materials}

At the next scale, protein tissues and materials are typically arranged in a multiscale hierarchical fashion up to the macroscale (\cite{buehler:2009}). Thus, the behaviour of energy dissipation and fracture is strictly regulated by the properties and the functions resulting from the specific arrangement of the multiscale structure that depends both on the function of the tissue or cells and the surrounding environment (\cite{roberts:2002}).  The formation of such hierarchy is mediated by the process of assimilation that determines how the proteins structures are capable of attaching together and attaining a specific conformation. In general, during this evolution, other phenomena driven by specific functions of the components arise such as growth, fibres assembling or the generation of a network such as in the cases of lamins and spider silks (\cite{herrmann:2004,rammensee:2008,bini:2004}). These complex multiscale structured materials and their extraordinary properties are largely controlled by the weak interactions at the nano-scale and in particular by hydrogen bonds. The presence of hierarchies is responsible for a `smart' response with a contemporary involvement of structures at different scales (\cite{fratzl:2007}).

%
\begin{figure}[b!]
\centering
  \includegraphics[width=0.95\textwidth]{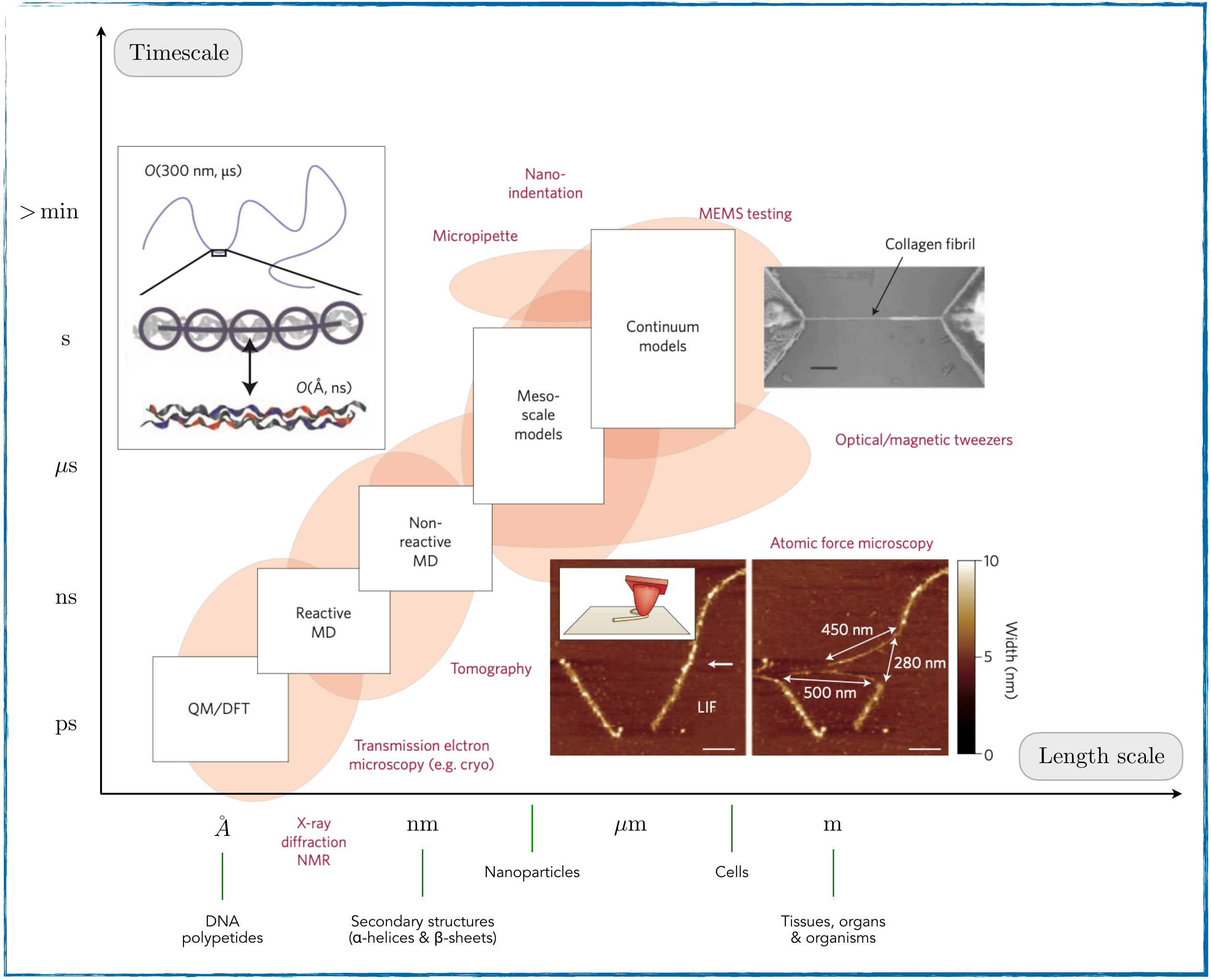}
  \caption[Scale dependent approach for biological materials]{Multiscale structure in biological materials in terms of length scale and time scale. Each area has its own domain of applicability in terms of the experimental method such as X-ray diffraction, transmission electron microscopy, atomic force microscopy (AFM), optical/magnetic tweezers, microelectromechanical systems (MEMS) testing and nanoindentation or theoretical models as quantum mechanics/density functional theory (QM/DFT), molecular dynamics (MD), coarse-grained models and mesoscale atomistically informed continuum theories, as well as continuum models. Adapted from (\cite{buehler:2009}).}
  \label{fig:ch1_hierarchy}
\end{figure}
%

The outstanding performance of biological materials depends on their intricate and ingenious hierarchical structure. Knowledge about these arrangements is crucial in deriving mathematical models to describe these materials and designing bioinspired materials. For instance, toughness is usually conferred by the presence of controlled interfacial features such as hydrogen bonds, friction or chain assembling straightening and stretching (\cite{meyers:2013}). Thus, the knowledge of the behaviour at the lowest scale coupled with the ability to control processes such as the recombination of DNA or the assembling of amino acids chains are the key ingredients for engineering structures at the lowest level (\cite{langer:2004,hest:2001,mandal:2014}). The scale hierarchy in time and length and the corresponding experimental techniques are schematized in Figure~\ref{fig:ch1_hierarchy}. 

The abundance of protein materials is accompanied by complexity and diversity generating an enormous number of different structures so it is often necessary to categorize them by introducing distinguishing features (\cite{alon:2007}). In this perspective, a crucial concept is the distinction between \textit{diffuse} and \textit{peculiar} properties. For instance, it is widely recognized that at lower scales protein secondary structures or synthesis and growth phenomena are characterized by diffuse (sometimes universal) properties, found in almost all biological materials (the described presence of primary, secondary and higher-order structures, the presence of sacrificial joints and weak links, the organization in hierarchic structures). On the other end, peculiar properties are observed in higher-order assembled textures such as the titin protein or the bone shells that are highly specialized depending on their function. Thus, a common denominator is that diversity increases with scales, suggesting that specific functions or specialized protein structures are associated with structural diversity. To this point, we remark that there is an extensive literature on the microscopic description as well as many phenomenological models describing macroscopic bulk properties. On the contrary, there is not adequate progress in theoretical, multiscale investigations that systematically connect mechanical properties at different length scales. A fundamental step in this direction consists in filling this gap, with insight at each hierarchical level highlighted with correct multiscale techniques (\cite{meyers:2008,bechtle:2010}).

%

\section{Objectives of the thesis}

%
In this thesis, I approach the study of biological multiscale systems from a bottom-up perspective by introducing microscale models to describe features such as conformational transitions and scission phenomena. In particular, I enter into the open challenge that nowadays tries to understand the mechanism of folding, failure and damaging across the involved vast length and time scales. A fundamental step in the description of biological systems and in the perspective of mimicking their behaviour is the introduction of models describing the effects of the number of microinstabilities observed at the molecular scales in the macroscopic `homogenized' response. Specifically, low scale properties are typically analyzed in the fields of biology, physics and chemistry. There are many problems derived from such features and scientists have dedicated their research to unveiling such interesting mechanisms in the case of DNA mechanics denaturation (\cite{busta:2000}) or for the conformational transition in proteins undergoing unfolding, as the Titin immunoglobulin domain studied from Matthias Rief and coworkers (\cite{rief:1997}). Also, focal adhesion phenomena in cell adhesion were first studied by Bell in $1978$ (\cite{bell:1978}) even though many questions are still open such as the relation between adhesion to substrate and stiffness or the effect of mechanical forces on binding dynamics (\cite{gao:2011,schwarz:2013}). Other examples are the behavior of muscle mechanics  and in particular the sarcomeres in skeletal muscles (\cite{huxley:1971,caruel:2013,darocha:2020}) or the unzipping of macromolecular hairpins (\cite{dudko:2007,liphardt:2001,mathe:2004}). 

Macroscopically, different behaviours at the molecular scale such as local and non-local interactions and cooperative or non-cooperative behaviour result in different history and rate-dependent behaviour at the macroscale, resulting in ductile or fragile behaviours, reversible or irreversible cyclic paths, localized or diffuse damage regions or material phases, typically phenomenologically studied in different fields of engineering and in particular based on continuum mechanics approaches (\cite{gurtin:2010,lemaitre:1994}). These have been successfully extended to consider such non-linear effects, history dependence and non-convexity of the energy, together with the possible importance growth phenomena (\cite{goriely:2017}). The common feature of all these examples is that at the microscale, it is always possible to identify elementary units with two or more (metastable) different configurations. 

The transition between multiple states regulated by inter- and intra-chains interactions often energetically equivalent is a very complex problem that has been tackled with different experimental, numerical and analytical techniques. A widely used approach in this field is \ac{MD} which opened up the description of important aspects of molecular-configuration transitions. However, the limitations in terms of computational costs and visualization are huge. Indeed, it is possible to study only single molecule or frames of systems in a very limited range of length and time scales even though crucial information can be drawn off. On the other hand, simplified analytical models, neglecting many molecule energy details, have been proposed to describe ideal molecular behaviours. For example, for polymeric and biopolymeric chains we recall the \ac{FJC} that considers a chain made by rigid segments (Kuhn segments) in which the end-to-end distance measures the elongation of the system. It is suitable for the description of single-stranded molecule such as polymer chains or \ac{ssDNA} and RNA sequences (\cite{weiner:1983,su:2009,benedito:2018}). Afterwards, the \ac{WLC} model was introduced and is better at describing more complex molecules such \ac{dsDNA}, in which the energy of the system depends on both the orientation of the segments and the actual configuration described by the so-called \textit{persistence length} measuring the chain `bending stiffness' (\cite{marko:1995,detom:2013,staple:2008}). 

When considering multiphase systems, which are very common in nature and in particular in the case of three-dimensional proteins and molecules undergoing a phase transition, such simple models are not effective in capturing all the topological properties and the variable natural configurations attainable. Thus, from a theoretical point of view, these features correspond to multiwells internal energies. The resulting mathematical problem is similar to the macroscale problem of describing the behaviour of a continuum characterized by a multiwells energy landscape. J. L. Ericksen proposed a variational energy approach in non-linear elasticity theory with non-convex energy densities to describe the microstructure of multiphase materials in his pioneering work (\cite{ericksen:1975}). Later, a fundamental inspection of such problems can be traced back to metallurgy and the description of solid-solid phase transition proposed by Khachaturyan and coworkers (\cite{khachaturyan:1983}). The resulting mathematical extensions of the variational theories in continuum mechanics (\cite{ball:1989,muller:1999}) were proposed to describe phase transition. Importantly, the role of non local energy terms to capture phenomena such as the nucleation and propagation of phase interfaces leading to periodic microstructures observed in solid-solid martensitic phase transitions proposed through continuum models (\cite{carr:1984}) and their discrete counterparts (\cite{truskinovsky:2000, puglisi:2006,puglisi:2007}) where the effect of boundary conditions was also clarified. In their pioneering work (\cite{villaggio:1977}), Villaggio and colleagues introduced discrete micro-mechanics models considering lattices of elements with non convex energy densities including intrinsic discreteness length-scale (\cite{zanzotto:1992}). Thus, it is possible to describe fundamental properties only visible at the microscopic level such as energy barriers separating phases, metastable states and quasi-plastic and pseudo-elastic behaviours (\cite{puglisi:2000,puglisi:2005}).  

\begin{figure}[b!]
\centering
  \includegraphics[width=0.95\textwidth]{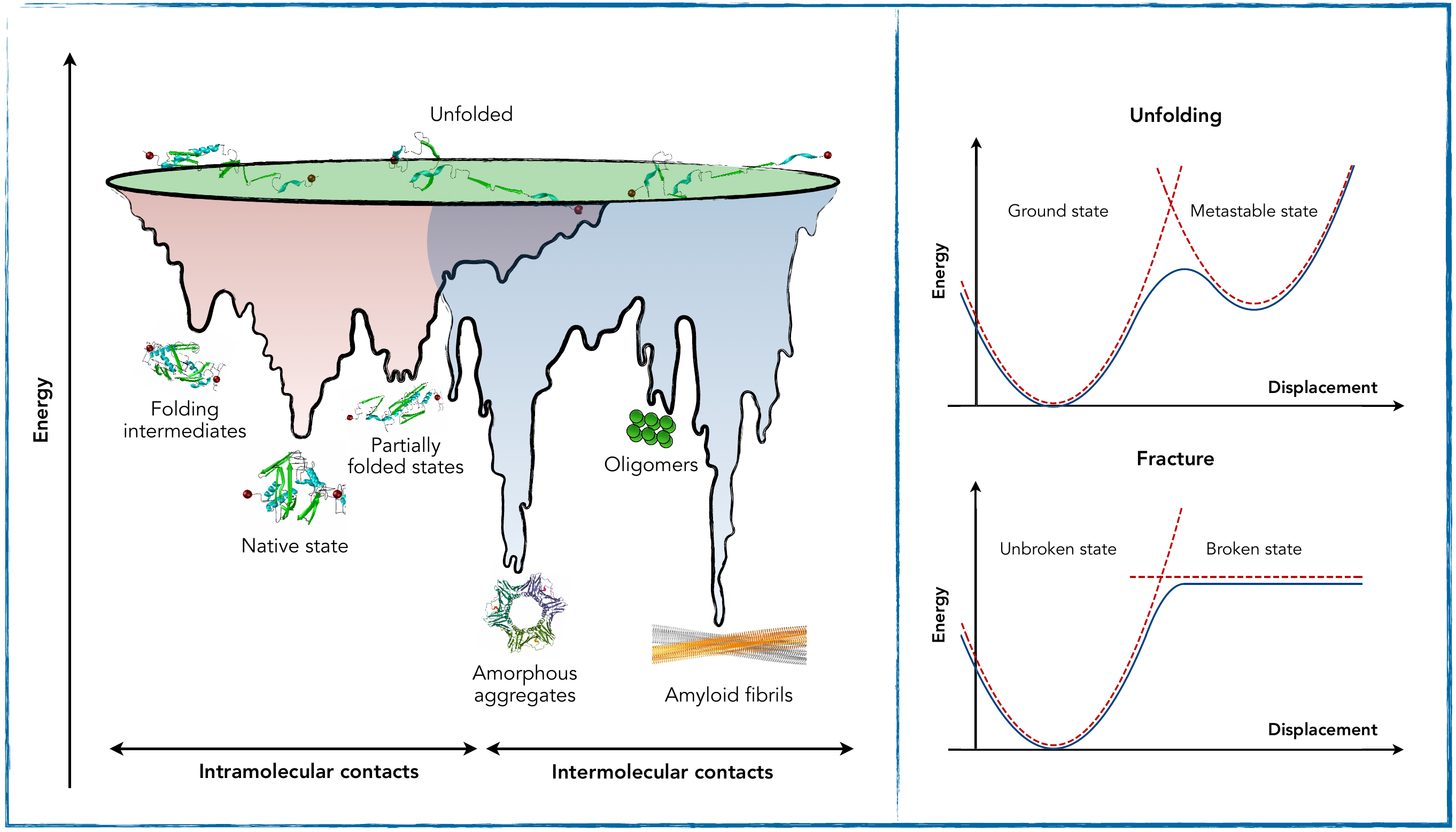}
  \caption[Bumpy energy landscape typical of system with microinstabilities]{In the left panel a typical bumpy energy landscape of system with microinstabilities is represented, and it is possible to observe the multiple local minima representing partially folded states or similar and the areas with higher energy in the case of unfolded molecule. In the right panel, two different types of potential energies used in this thesis to mimic the behaviour of such energy landscapes are represented. Created by the author.}
  \label{fig:ch1_energy}
\end{figure}

In this thesis, I focus on the behaviour of materials characterized by microinstabilities by introducing internal energies of single units assuming a specific form depending on if we study a phase transition or a decohesion --fracture-- process. Following (\cite{puglisi:2000}) I consider different responses representing the `folded' and `unfolded' (or broken) phases as a superimposition of two parabolas as shown in Figure~\ref{fig:ch1_energy}. Based on this simplifying, but effective, constitutive assumption, we are able to deduce analytical solutions for all equilibrium states, characterized by different phase states, both in the case of pure mechanics (zero temperature) and within the Statistical Mechanics' framework is considered. Moreover, depending on the choice of the type of energy one can study multiple phenomena. 

In particular, we consider two main energetic assumptions. In the first case, we consider two-wells energy as shown in Figure~\ref{fig:ch1_energy}$_a$, used to model the conformational transition in polymers or protein macromolecules described as the unfolding of secondary structure such $\alpha$ helices and $\beta$ sheets crystal domains. This framework can be considered also in the case when different crystal phases are observed in metallic materials such as in the case of solid-solid martensitic phase transitions in metallic alloys. 

In the other case, we jump from an attached (elastic) state to a broken one. Here the second state is characterized by constant (fracture/decohesion) energy corresponding to zero force, as schematized in Figure~\ref{fig:ch1_energy}$_b$. Such an assumption may mimic, among many other examples, force-induced denaturation phenomena such as for DNA, RNA hairpins or the focal adhesion. Similarly, by combining such types of elements to describe bond breaking effects and elastic phases to describe covalent bonds, with possible non local interactions, more complex systems can be described. As an example, I consider the interaction between \ac{MT} and tau proteins in axons or the complex denaturation process of a double-stranded DNA helix. 

It is worth remarking that in both cases the phase transition or bonds breaking can be reversible, partially reversible or irreversible. Moreover, in both situations, two special types of boundary conditions can be considered. When the end-to-end length is fixed and the force is measured we refer to the case of hard device and from a statistical point of view this corresponds to the system studied in the Helmholtz canonical ensemble (\cite{efendiev:2010}). On the other hand, when a force is applied such as in the case of magnetic tweezers, we are in the soft device regime, corresponding to the Gibbs ensemble. Moreover, a crucial point is that these two limiting regimes can be shown to be conjugate and in particular, the two ensembles can be recovered from the other one using a Laplace transform (\cite{giordano:2018,luca2019}). Many contributions have recently shown the effectiveness of such systems in describing thermal effects in multi-stable systems, and different problems regarding entropic effects and boundary conditions have been analyzed (\cite{makarov:2009,giordano:2017,benedito:2018,benedito:2018:2,benedito:2020,fp:2019,luca2019}). It is worth to remark Statistical Mechanics is an appropriate framework to introduce temperature effects in systems with microinstabilities when the energy barriers and the transition energies, regulated by weak interactions such as hydrogen bonds and hydrophilic/hydrophobic or Van der Walls forces, are comparable with thermal fluctuations (i.e. of the order of few $k_B T$ \cite{weiner:1983}). 

%
%

%
\subsection{Phase transitions}

Based on these general considerations I analyzed different possible specific applications. In the first part, I focus on phase transition phenomena and specifically I analyze the influence of devices on proteins unfolding as an example of biological materials. As anticipated, the stiffness of the measuring apparatus in single molecule force spectroscopy experiments is an important factor that has to be taken into consideration to interpret correctly experimental results. Even though the model we present and the results can be adapted to the general problem of molecular interactions (proteins-DNA, intermolecular, cell-molecular, etc.), as a paradigm we refer to the case of AFM induced unfolding experiments in bio-molecules such as titin (\cite{rief:1997,givli:2011}). In this case the molecules are characterized by domains undergoing a conformational (folded $\rightarrow$ unfolded) transition (\cite{rief:2002,chung:2013}). In particular, we consider the system made by device and molecule as a unique thermodynamical system and thus it is possible to obtain explicit formulas describing the thermo-mechanical response of the molecule for varying properties of the loading device. 

In Chapter~\ref{ch_4}, the model is extended to consider the possibility that the generation of interfaces can be described by non-local energy terms with cooperative effects. In the purely mechanical case, non local energy terms have been introduced in the works of A. Vainchtein and G. Puglisi (\cite{vainchtein:2004,puglisi:2006,puglisi:2007}), allowing a description of features such as the effect of the stress peak between the nucleation and propagation of phases, representing a well-known experimental effect in materials undergoing a phase transition. Even in this case, the mechanical response of the two different boundary conditions of hard and soft devices may be considered to affect the phase transition strategy (such as internal or boundary nucleation) with a variable number of phase fronts (\cite{puglisi:2006,puglisi:2007}). These models have been applied to study polymer necking localization phenomena, as well as they have been extended to the case of configurational transition in proteins and high-tech nanomaterials where the successive unfolding of phases has been studied (\cite{detom:2013,manca:2013}). As a paradigm, in the thesis, I focus on phase transition phenomena in shape memory nanowires. Indeed, I show in this case that the transition energy and the interfacial energy are comparable to entropic energy terms. Thus, I consider a statistical mechanics approach for systems with non-convex energies and non-local energy terms introducing temperature effects on both the Helmholtz and Gibbs statistical ensembles. In particular, I recover how crucial features such as the nucleation stress and the propagation stress depend on temperature. The comparison with molecular dynamics simulations shows the ability of the model to describe the temperature-dependent behaviour (\cite{luca2020}). Eventually, based on multiscale approaches (\cite{detommasi:2015,detommasi:2017}) and on the analytical results here obtained a future perspective is to deduce macroscale constitutive laws depending on the microscopic formulas I deduced in this thesis.

%
\subsection{Dechoesion and fracture}

Chapters~\ref{ch_5} and~\ref{ch_6} of the thesis are devoted to microscale models for fracture and decohesion phenomena in biological materials. In this case, we consider an energetic approach with the energy of the type sketched in Figure~\ref{fig:ch1_energy}$_b$. One of the major outcomes is the possibility of determining the mechanism of failure, which is governed by energy dissipation through the scales. As in the most general case of materials science, failure in biological materials can be related to the loss of the capacity of a component to completely provide the function it was designed for or the ability to do it safely.  

%
\begin{figure}[b!]
\centering
  \includegraphics[width=0.95\textwidth]{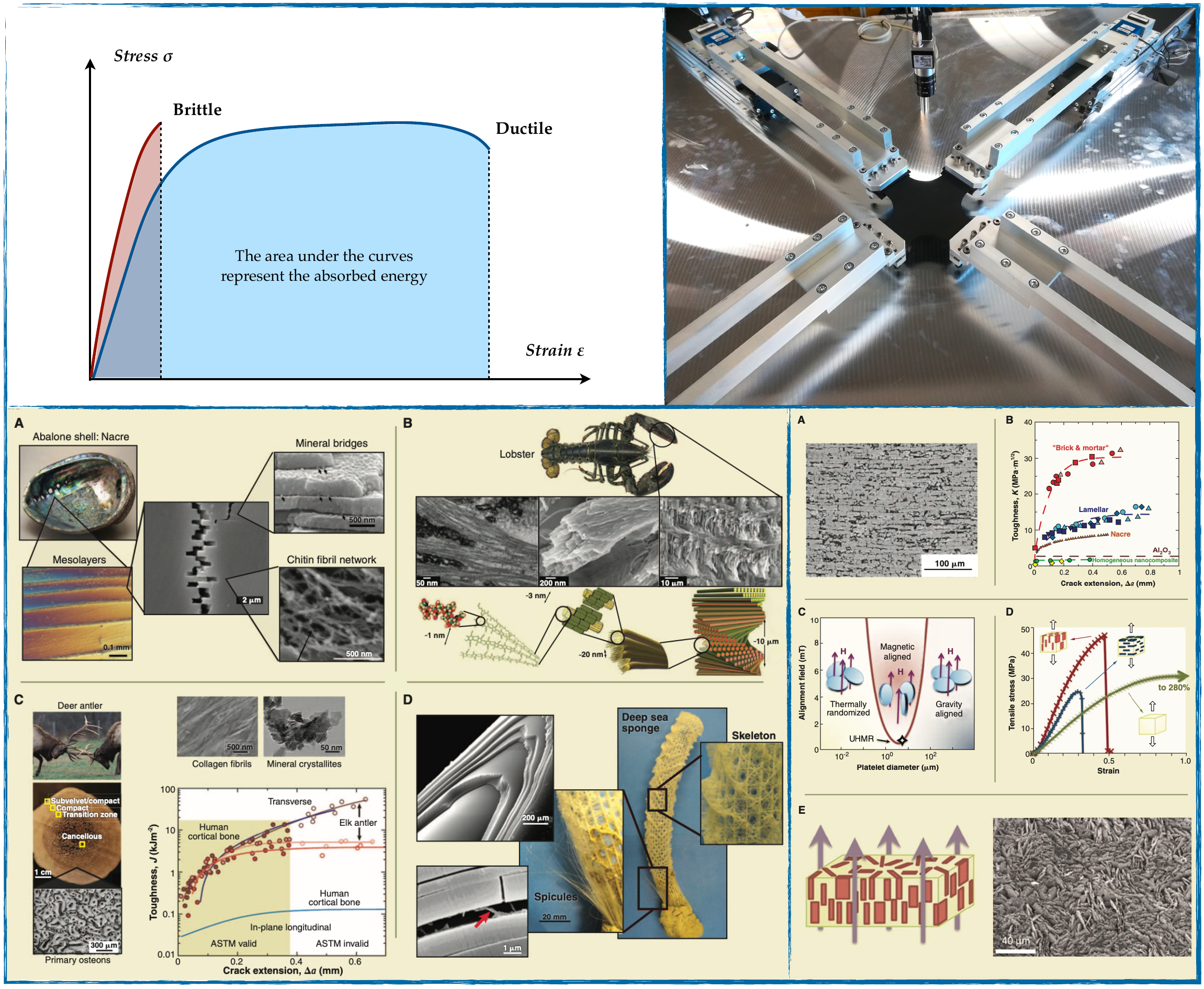}
  \caption[Fracture]{In the top left panel the two typical fracture behaviours are represented.  A fragile material (red curve) breaks at low strains and high stresses whereas a ductile specimen (blue curve) dissipates a higher amount of energy before rupture. On the top left panel, a typical biaxial machine used to test different materials (in this case soft polymers and structures) owned by our research group in Politecnico di Bari is represented. In the bottom row, different hierarchical structures of tough biological materials are represented demonstrating heterogeneous interfaces that provide crack deflection. Reproduced from (\cite{pineau:2016,meyers:2013}).}
  \label{fig:ch1_fracture_1}
\end{figure}
%

As in the case of metallic materials, when a biological material is deformed the typical experimental behaviour is characterized by different regimes. For small deformations the behaviour is reversible and it is known as the \textit{elastic regime}, whereas there exist load thresholds after which permanent deformations are observed. This regime corresponds to the \textit{plastic regime} widely studied for metallic materials (\cite{lubliner:2008,courtney:1990}). Then follows fracture, in which the material fails. It can happen abruptly, known as \textit{brittle behaviour}, typically observed in artificial materials such as ceramics or glasses where a fast propagation of cracks fronts leads to rupture (\cite{lemaitre:1994}). On the other hand, failure may occur after a high regime of permanent deformations with more dissipated energy, a regime known as \textit{ductile fracture} that can be observed in plastics and some metals such as copper, as shown in Figure~\ref{fig:ch1_fracture_1} (\cite{pineau:2016,timoshenko:1983}). 

\begin{figure}[b!]
\centering
  \includegraphics[width=0.95\textwidth]{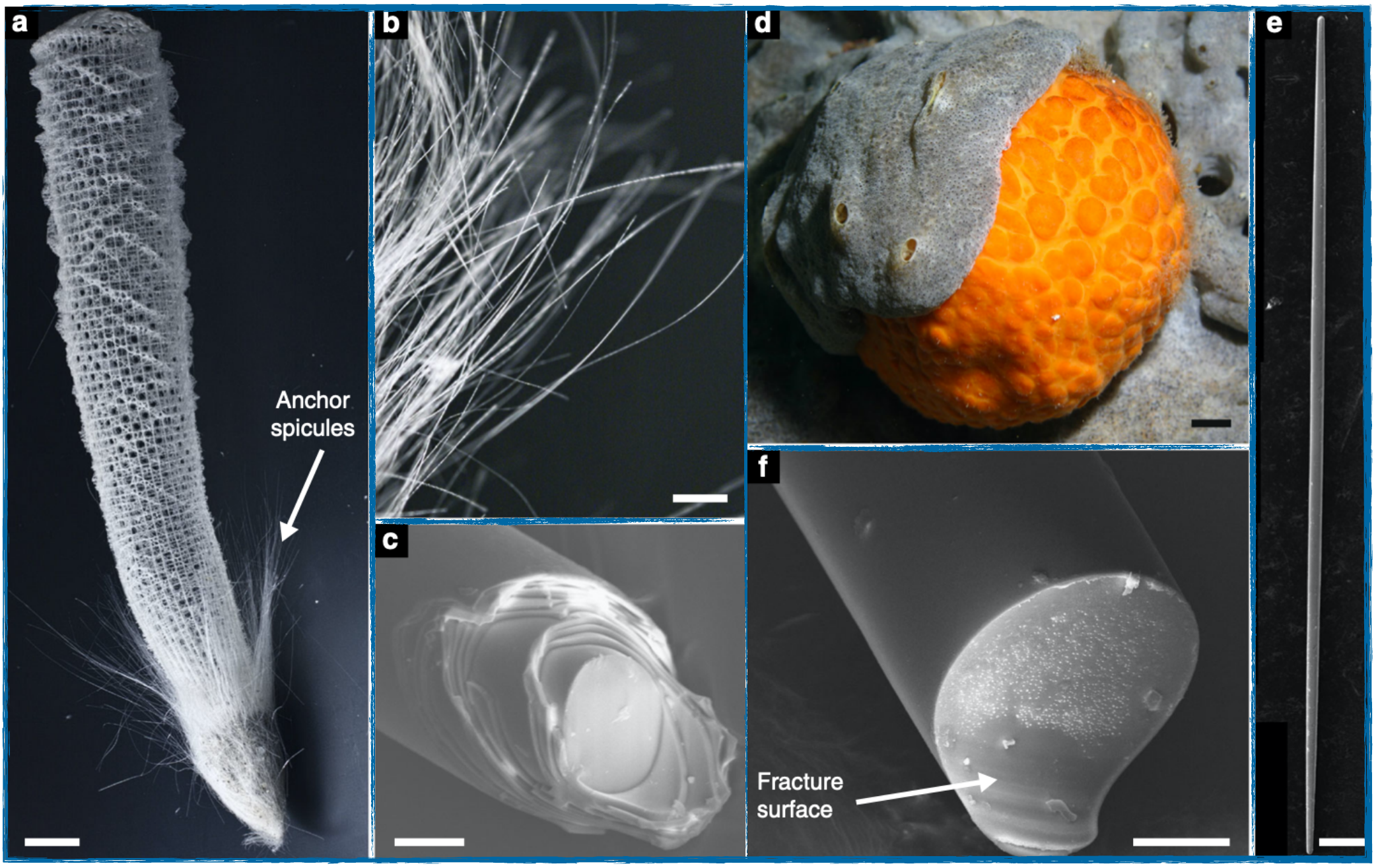}
  \caption[Fracture in biological materials]{Skeletons and spicules of the \textit{Euplectella aspergillum} (Ea) marine sponge. In particular in panel a) the skeleton of the sponge is shown. Panel b): the anchor spicules that fasten the sponge to the seafloor.  Panel c) shows the broken end of an Ea anchor spicule that was fractured in three-point bending showing its layered architecture. Panel f): A toothpick-like spicule found within the Ta sponge in panel d) that doesn't contain a layered architecture. Panel g): The exposed surface of a Ta. spicule that was fractured in the same way as the Ea. Reproduced from (\cite{monn:2020}).}
  \label{fig:ch1_fracture_2}
\end{figure}

In the case of failure or decohesion, a complete analysis requires the description of phenomena at different scales (see Figure~\ref{fig:ch1_fracture_2}). Specifically, in biological and polymeric materials, the macroscopic failure is the homogenized response of complex phenomena of bonds breaking and recrosslinking at the molecular network scales, with entropic energy terms that play a crucial role. In particular, the breaking of weak interactions such as hydrogen bonds (H-bonds) is at the base of protein unfolding (\cite{rief:1997}), interchains debonding phenomena, sliding of planar surfaces or molecules (\cite{fantner:2005}), misalignments in the polypeptide chain or denaturation of the DNA sequence (\cite{buehler:2008}). As described above, the availability of sophisticated experimental techniques allowed a much deeper understanding of the key rupture phenomena at the molecular scale (\cite{ferrer:2008}). An approach to study these features is again Molecular Dynamic, which is crucial for very small systems, but it has an unavoidable restriction to small length and time scales due to computational cost (\cite{ackbarow:2007}).

%
%

Based on the classical energetic approaches, the pioneering works concerning failure in materials are attributed to Griffith (\cite{griffith:1921}) and Irwin (\cite{irwin:1957}), who developed mathematical tools to study this phenomenon and describe this complex mechanism. First attempts were made also by Leonardo da Vinci, who guessed that there is an inverse proportionality of length and strength such that shorter wires are stronger for a given thickness (\cite{williams:1957}). I approach fracture by considering a microscopic description of the phenomena by introducing a lattice model with simple modifications that may mimic the behaviour of different systems such as double-stranded DNA helices, the bundles of microtubules and tau-proteins inside the axon (\cite{kuhl:2016,garcia:2017}), focal adhesion (\cite{cao:2015}) or polymeric networks. I apply our model mainly to the first two biological systems and we recover analytical expressions for their thermomechanical behaviour. In particular, I show that, as in the case of macroscopic decohesion and fracture, when considering shear loads applied to our system, we may define a `process zone' (in terms of base pairs) and we are able to reproduce a previous result of Pierre Gilles De Gennes (\cite{degennes:2001}), in which he described, with a simple lattice model, the law that links the length of a DNA sequence with the rupture force. Experimental comparison to our results (\cite{prakash:2011,hatch:2008}) supports the proposed model.

\subsection{Rate effects}

\begin{figure}[b!]
\centering
  \includegraphics[width=0.95\textwidth]{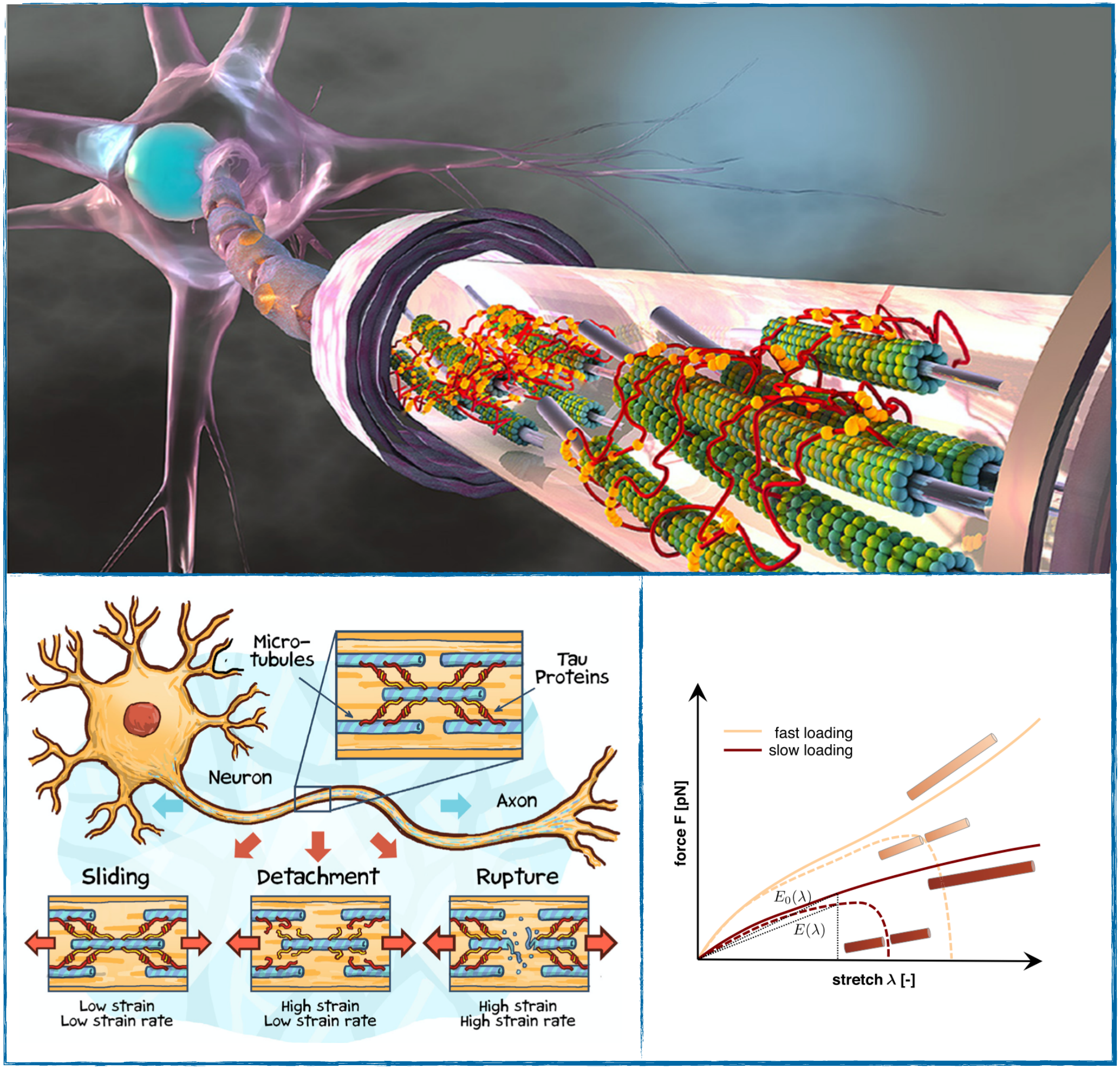}
  \caption[Microtubules and tau protein bundle inside the axon.]{In the top panel a graphical representation of the inner part of the axon is shown. In particular, it is possible to observe the bundle of microtubules (green rods) held together by the tau proteins (red filaments). In the bottom left panel the rate-dependent behaviour of such a system is schematized and in the right panel the typical stress-strain behaviour at varying rates is presented. Reproduced and adapted from (\cite{kuhl:2015,kuhl:2018}).}
  \label{fig:ch1_rate}
\end{figure}

With this simple shear model, we describe important effects related to neural damage, mimicking the behaviour at the microstructure of the interaction between microtubules and the tau-proteins, which are the main structural component of the axon, the longest part of the neuron cell, as shown in Figure~\ref{fig:ch1_rate}. In the last part of the thesis, I extend such a model to include important rate effects. Indeed, the microscopic rate-dependent behaviour results in damaged properties at the macroscopic scale and is fundamental in many health problems. It is shown that external forces acting on the brain, impacts and traumatic accidents, pathologically known as \textit{\ac{TBI}}, may cause severe brain damage not only at the moment of the hit, but also within days, months, and years (\cite{smith:2000,smith:2010,smith:2012}) and at the root of such behaviour there are phenomena of debonding and rupture at the scale of axons, which are part of the white matter of the brain (\cite{thompson:2020}). The neurological damage, due to the viscoelastic nature of the brain tissue, can spread through time from some months to years after the initial application of the force. One of the major related issues with this fact is the occurrence of neurodegenerative diseases such as Parkinson's and Alzheimer's. This phenomenon was firstly studied in boxers in the U.S.A. because a great percentage of them suffered from Alzheimer's diseases years after they retired. It has been shown that the mechanism of damage due to repetitive or impulsive external forces generates what is known as \textit{\ac{DAI}}, clinically recognized by scattered lesions at the level of axons. Axonal failure happens mainly in two ways: it can be abrupt in \textit{primary axotomy}, or with a progressive degradation leading to gradual failure, a process known as \textit{secondary axotomy}. The nature of this behaviour has to be found at the micro-scale level, where the complex energetic landscape and the transition between different configurations associated with bonds breaking and recrosslinking of the tau proteins are observed. The axon cytoskeleton is made of microtubules (\acs{MT}), neurofilaments and microfilaments, where the first ones represent the major structural components, arranged in bundles linked and kept together by tau proteins (\cite{rosenberg:2007}). The time-dependent behaviour of the white matter damage represents an overall emerging result from the rate dependency at the scale of microtubules and tau proteins. In particular, microtubules may slide with respect to each other in three main regimes. If both the strain and the strain rate are small, there is sliding within the axons and the microtubules roll by each other to recover their initial length. If the strain increases, the tau-protein connections start to break. The mechanism of this phenomenon is still poorly understood. If the strain rate is large, the force transmitted from the proteins to the MT is high and the microtubules can break. A complete understanding of this important physiological phenomenon remains elusive and is discussed in part of this thesis. As a matter of fact, complex phenomena of bond breaking and recrosslinking are observed with a continuous variation of the natural configuration, insurgence of self-stresses and possible healing (\cite{kuhl:2015,kuhl:2016}). 

Rate effects in hierarchical materials represent a very interesting theoretical subject, of interest in both previously described phenomena and, more generally, to describe viscoelastic effects with permanent deformations observed in many natural and artificial soft materials such as rubbers, polymers or protein materials (\cite{zhurkov:1965,lubliner:2008}). Indeed, the rate dependent analysis in single molecule force spectroscopy experiments at the molecular scale is often characterized by important effects depending on the pulling velocity and it may affect their behaviour. At this scale, rate effects can be described within statistical mechanics (in- and out-of-equilibrium), based on the possibility of overcoming energy barriers (\cite{hummer:2001}). Mainly three different time scales are involved and regulate the observed response: the time scale associated with the relaxation into the energy wells of the bumpy energy landscape $t_{well}$, the time scale associated with the overcome of energy barriers $t_{bar}$ and the time scale associated with the rate of loading $t_{load}$ (\cite{friddle:2012, friddle:2008, li:2014}). The response can be divided roughly into three regimes. At very small rates $t_{well}\ll t_{bar} \ll t_{load}$, the system operates in quasi-equilibrium conditions and it can ideally always relax to the global energy minimum. In this case, also known as \textit{Maxwell convention}, the behaviour is reversible with no hysteresis and dissipation. Under a moderate increase of loading $t_{well}\ll  t_{bar} \sim t_{load}$, another regime can be observed and the system can only relax to local energy minima so that the temperature-dependent capacity of overcoming energy barriers is crucial (\cite{dudko:2008,suzuki:2013}). This is the rate dependent regime I focus on in this thesis. In this regime, hysteresis can be observed even though it can be addressed to the transition from high energy to low energy configurations. When rates are very high, the system has no time to explore the energy landscape and the response is deterministic, and stochastic fluctuations are irrelevant. Moreover, in this regime hysteresis is actually due to the damping mechanism observed in materials (\cite{stillinger:2015, benichou:2016}).

In the considered time-scale ordering successful approaches (\cite{hanggi:1990}) are derived from the pioneering work of Bell in the $70$s (\cite{bell:1978}), based on Kramers' theory for Brownian motion (\cite{kramers:1940}). According to this approach, the evolution into the bumpy energy landscape is regulated by the energy barriers separating the wells. The probability of overcoming these barriers is a function of the intrinsic rate of the system, the thermal fluctuation due to temperature (Brownian motion) and the height of the energy barriers separating the different minima. The description of the energetic landscapes is still a crucial task, especially when three-dimensional evolution is involved (\cite{hummer:2001,hummer:2003}). I consider the classical minimum mountain pass problem, evaluating the height of successive energy barriers in the system evolution between different metastable states. As a result, it is possible to explicitly express the transition probability from one configuration to the other and obtain a rate-dependent model at the molecular scale. The original Bell theory has a different implementation that may range in a large framework of scales to include different effects (\cite{keten:2008,rief:1998}), and, in particular, for biological systems such as intermediate vimentin filaments, the extended theory proposed by Buehler and colleagues shows pretty good results (\cite{buehler:2007}). When considering biological structures one has to deal mainly with non-covalent bonds such as hydrogen bonds assembled in particular configurations attaining adhesion surfaces, secondary and tertiary structures and fibrillar linkages. Thus the dynamical behaviour of hydrogen bonds is crucial for many phenomena in biological materials (\cite{evans:1997}). Moreover, the relation between forces, energy landscape and bonds may offer clues and insights into the behaviour of this structure under the effects of external forces, as experimented with \acs{SMFS} (\cite{evans:2001,dudko:2006}). 

%
%

In summary, the aim of this thesis is to obtain analytical models at the molecular scales, describing a different kinds of instabilities, allowing for the analytical description of the system behaviour, possibly including rate effects and fundamental entropic effects. Such results, besides their intrinsic ability to describe important experimental features at the molecular scale as extensively demonstrated in this thesis, will be the starting point to derive macroscopic constitutive models whose material parameters are directly deduced from the microscopic ones (\cite{detommasi:2015, detommasi:2017}). This will be the base for explicit analyses of the evolution of multistable hierarchical biological materials, with damage and residual stretches directly descending from these laws, and for the design of new bioinspired materials, which is a task equally demanding undertaking.


	\clearpage


%
\renewcommand{\thefigure}
{\arabic{chapter}.\arabic{figure}}
\setcounter{figure}{0}

\renewcommand{\theequation}
{2.\arabic{equation}}
\setcounter{equation}{0}

\chapter{Thermo-mechanical methods for multistable materials: a brief introduction}
\label{ch_2}
\vspace{1.2cm}

In this chapter, we introduce the fundamental mathematical methods used within the thesis. To this end, we consider a simple typical example of a chain with three bistable elements under different thermo-mechanical external fields. After briefly introducing the energy models used to describe polymeric chains or biological materials, we focus our attention on the approximate approach used in this thesis.

In the first part of this chapter, I discuss the different possibilities allowed by our energetic approach used to model the behaviour of multiphase materials. In particular, I present a general model that can be adapted to reproduce a wide range of physical problems. Moreover, first we analyze the problem from a `purely' mechanical point of view, \textit{i.e} when thermal effects are neglected, \textit{i.e.} $T=0$, with $T$ the temperature. We refer to this case with the `zero temperature limit'. Doing this, it is possible to obtain from the energy model much crucial information concerning the microscopic description of the system, such as the possibility of having local and global energy minimizers of the system and describing reversible and hysteretic behaviours. We also show the possibility of deducing the minimum barrier paths for a given metastable state (local energy minimizers) fundamental in the analysis of thermal-induced rate effects. Also, for the global minimizers, we show that this case can be deduced as limit system behaviour when temperature and entropic energy terms decrease.  

In the second part, I introduce thermal effects through classical statistical mechanics method to analyze the effect of temperature on the system. Furthermore, we discuss Rate Theories, necessary to describe the possible rate-dependent behaviour of multistable systems.  In particular, based on the pioneering work of Bell (\cite{bell:1978}),  a rate-dependent microscopic model for multistable materials is obtained. In the final part of the chapter, the paradigm of a chain undergoing a conformational transition with three units is presented and is used to show how our theoretical tools can be applied to a simple case. Such theoretical approaches are then extended in the remaining part of the thesis to more complex systems.

\section{Constitutive hypotheses: different multiwells energies}

The complex, history and rate-dependent, macroscopic behaviours of biological and polymeric materials are homogenized effects of different micro-instabilities phenomena, leading to conformational transitions or breaking and recrosslinking mechanisms at the molecular network scale. In the development of this thesis, many examples captured my interest such as the topological transition observed in titin protein, which switches from a folded configuration (usually stiffer) to an unfolded one (usually softer) (\cite{rief:1997,givli:2011,rounsevell:2004}). Another interesting conformational transition is found in skeletal muscle sarcomeres, where two different phases at the level of the actin-myosin mechanism result in two `stable' muscle positions (\cite{huxley:1971,caruel:2013,caruel:2018}). The second type of micro-instabilities is related to link-breaking effects. One important example is observed in cell adhesion and decohesion where fibrils can be in different states such as attached and detached due to the action of mechanical forces or chemical and thermal fields (\cite{erdmann:2007,gao:2011,schwarz:2013}), whereas a similar phenomenon is the unzipping of macromolecular hairpins that can be described with a switch from an on/off state (\cite{liphardt:2001,dudko:2007}). In our variational approach, these features and phenomena result in the complex evolution of the system throughout the whole \textit{multiwells energy landscape} (see Figure~\ref{fig:ch1_energy}). 

Discrete systems with non-convex energies were introduced in the pioneering work of Villaggio and M{\"u}ller (\cite{villaggio:1977}) to describe the elastic-plastic behaviour of metallic materials. Later, this approach was exploited in the framework of shape memory alloys phase transitions, to mimic different transition strategies based on the ability of the system to overcome energy barriers (\cite{puglisi:2000}). Following these approaches, I have used these models to describe the properties of biological and bioinspired systems, both within discrete and continuum approaches, as shown in (\cite{luca2019,fp:2019,luca2020,florio:2020}).

\subsection{Unfolding and fracture}

From this perspective, to mimic the behaviour of a multiphase system the problem reduces to correctly interpreting the potential energy of single units that establish the building block of the overall structure, which examples are given in Figure~\ref{fig:ch1_energy}. In a very general configuration one may consider the situation in which different `phases' can be attained, \textit{i.e.} a folded configuration, an unfolded phase or a broken state. In general, it could be possible that a material under external loads from a generic initial stiff phase (first well) becomes softer (second lower well) under the action or the competition of both mechanical forces and entropic therms and finally may break (straight energy line) attaining zero force transmitted. A typical example that may be described with this general configuration is the high-temperature behaviour of strength in sapphire whiskers and some high-entropic and medium-entropy alloys. Indeed, in the first situation (\cite{brenner:1962}) the phase transition happens due to the evolution of the damage from brittle fracture propagation at low temperature to the nucleation of dislocations before crack spreading at a higher temperature. On the other hand, for the case of medium-entropy and high-entropy alloys (\cite{miracle:2017,gali:2013}), the conformational transition is identified by different structures and dislocations mobility methods at different strains and temperatures. These two situations, however, are not covered in this thesis.

Here, I focus my attention on two main cases. In the first one, I consider only two elastic wells neglecting fracture, represented in Figure~\ref{fig:ch1_energy} top right, corresponding to two equilibrium branches one softer than the other. Indeed, the main idea is to model phenomena such as the unfolding of proteins where the second (unfolded) phase is much softer than the folded configuration (\cite{rief:1997}). This behaviour is also observed in the case of some particular engineered material such as the memory shape nanowires, in which the distinction between the two phases is regulated by the presence of phase interfaces, modelled by introducing non-local terms, as discussed in Chapter~\ref{ch_4} (\cite{li:2010,seo:2013,ma:2013}). 

In the second case, I tackle the problem of decohesion and fracture by considering the unit energy of the form elastic-broken, thus by considering only one well representing the metastable elastic phase followed by a constant energy term representing rupture (see Figure~\ref{fig:ch1_energy} bottom right). This situation is used to mimic a very wide range of phenomena, and I will use this simple model to mimic the denaturation process in DNA, where the base pairs may detach to undergo processes such as transcription and recombination (\cite{degennes:2001,busta:2000}). Similarly, I apply the same assumption to study the intrinsic mechanism of detachment and rupture between microtubules and tau-proteins, the main structural constituent of the axon, the longest part of the neuron cell (\cite{kuhl:2015,rosenberg:2007}).

\section{Notes of Statistical Mechanics}
\label{sec:ch2_SM}

Consider the properties of a gas in a certain space, for instance, a room or a climatic chamber where you are performing experiments. In theory, to obtain mechanical properties such as the pressure or the velocity you should write down the equations of motion of all the particles inside the gas volume but this approach is untractable. Indeed, the number of particles is of the order of the Avogadro constant, $N_{A}=6.02214076\times10^{23}$ mol$^{-1}$. Statistical Mechanics has been introduced to fulfil this task and its main purpose is to avoid counting all the equally probable ways in which a system may divide up its energy. It is a topic touched by many scientists, but, quoting Goodstein, we should approach the subject with caution (\cite{goodstein:2014}): 
\begin{quoting}
Ludwig Boltzmann, who spent much of his life studying Statistical Mechanics, died in 1906, by his own hand. Paul Ehrenfest, carrying on the work, died similarly in 1933. Now it is our turn to study Statistical Mechanics. Perhaps it will be wise to approach the subject cautiously.\\\vspace{-0.4cm}
\flushright{(David L. Goodstein)}
\end{quoting}

I am willing to take this chance, and in this section, I introduce basic concepts of Statistical Mechanics. In particular, it is crucial to define the \textit{partition function} $\mathcal{Z}$, which will help us to include temperature effects in our microscopic models. When systems with microinstabilities are considered, the energy regulating bond linking and transitions states are comparable with thermal fluctuations (\cite{weiner:1983}) so that such an approach is quite convenient and suitable to explain experimental behaviours (\cite{efendiev:2010,makarov:2009,giordano:2017,luca2019}).

\subsection{Thermodynamics and canonical ensembles}
\label{sec:ensembles}

\begin{figure}[t!]
\centering
  \includegraphics[width=0.95\textwidth]{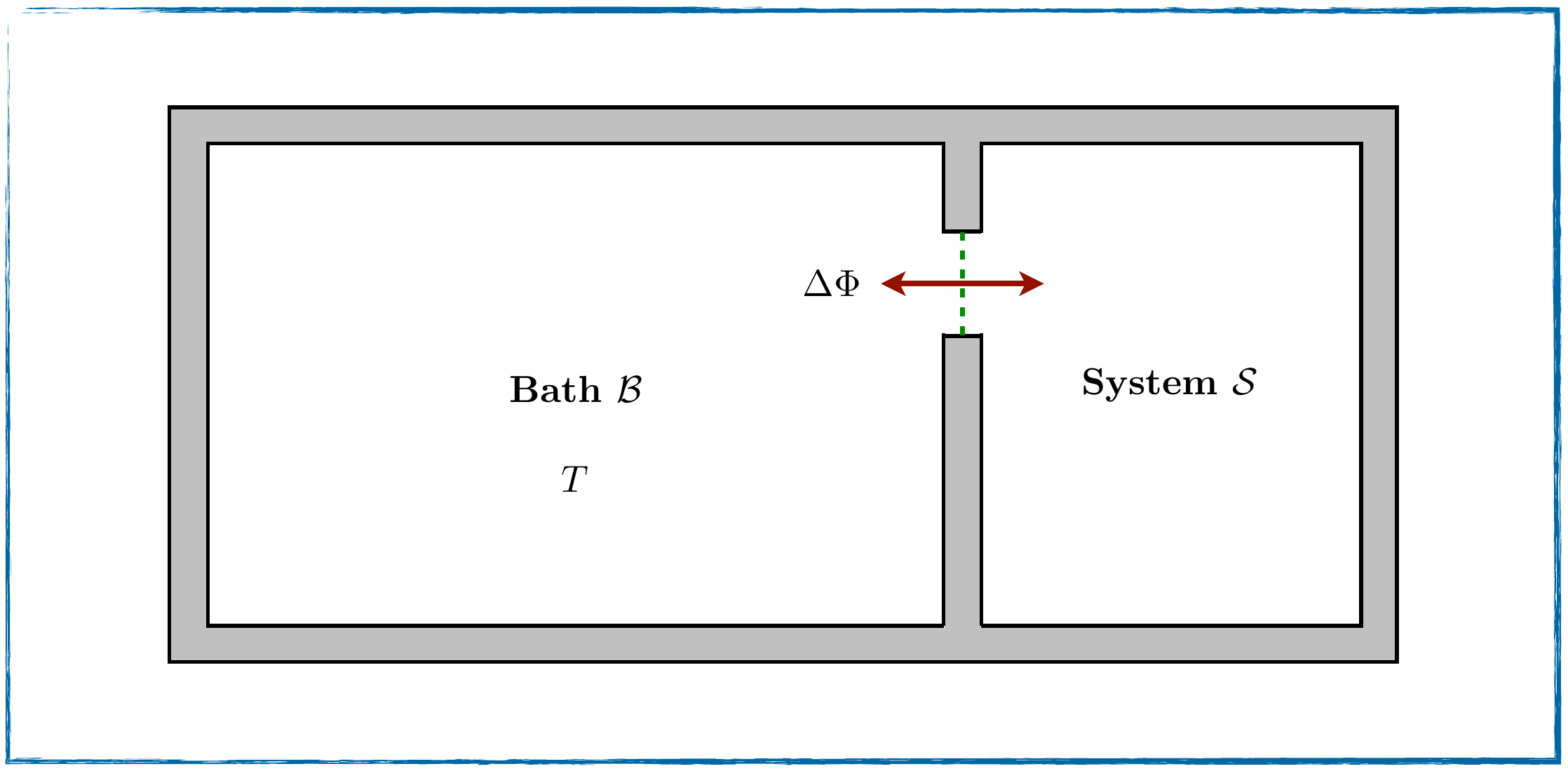}
  \caption[The canonical ensemble]{The figure shows a representation of the canonical ensemble that describes equilibrium systems that can exchange energy with a thermal bath at temperature $T$. The probability of a state with energy $\varphi_i$ is given by the Boltzmann distribution. Adapted from (\cite{sethna:2021}).}
  \label{fig:ch2_thermodynamics}
\end{figure}

We consider a system with fixed volume $V$ and total energy $\Phi$, with energy fluctuations $\Delta\ll\Phi$. To define an observable, such as temperature or pressure, one considers different `replicas' of the system, keeping volume $V$, energy $\Phi$ and number of particles $N$ fixed. The ensemble of copies of the system obtained in this way is the \textit{microcanonical ensemble}. If we consider a thermal bath $\mathcal{B}$ and a system $\mathcal{S}$ with total energy fixed, but where exchanges of energy are allowed among the two, still maintaining both volume $V$ and the number of particles $N$ fixed, we obtain a different collection of replica called the \textit{canonical ensemble}, shown in Figure~\ref{fig:ch2_thermodynamics}. 

%
%

For a given system $\mathcal{S}(\Phi, V, N)$ consider two subsystems $\mathcal{S}_1$ and $\mathcal{S}_2$, with $\mathcal{S}_1(\Phi_1, V_1, N_1)$ and $\mathcal{S}_2(\Phi_2, V_2, N_2)$, and total energy
\begin{equation}
\Phi=\Phi_{1}+\Phi_{2}.
\end{equation}
Suppose that the two subsystems are weakly interacting, \textit{i.e.} with almost zero energy flux between them and with volume and number of particles still fixed. To search for the relation among the possible microscopic states attainable by the two systems we suppose that $\Phi_2$ is assigned. The probability density of the microscopic configuration of the whole system is related to the configurations of the two subsystems by
\begin{equation}
\Omega(\Phi_1,\Phi_2)=\Omega_{1}(\Phi_{1})\times\Omega_{2}(\Phi-\Phi_{1}),
\end{equation}
wherewith $\Omega(\cdot)$ we denote the number of `equiprobable' microscopic states. We then obtain
\begin{equation}
\text{ln}\,\Omega(\Phi_{1},\Phi-\Phi_{1})=\text{ln}\,\Omega_{1}(\Phi_{1})+\text{ln}\,\Omega_{2}(\Phi-\Phi_{1}).
\end{equation}

Suppose now that we want to maximize the possible state attainable by system $\mathcal{S}_1$ by searching for the stationary point of
\begin{equation}
\left.\frac{\partial}{\partial \Phi_{1}}\Omega(\Phi_{1},\Phi-\Phi_{1})\right|_{N,V,\Phi}=0\rightarrow\left.\frac{\partial}{\partial \Phi_{1}} \text{ln}\,\Omega_{1}(\Phi_{1})\right|_{V_{1},N_{1}}=\left.\frac{\partial}{\partial \Phi_{2}} \text{ln}\,\Omega_{2}(\Phi_{2})\right|_{V_{2},N_{2}},
\label{eq:ch2_condition}
\end{equation}
which is the condition that assures thermal equilibrium between the two subsystems. Indeed, we recall that in a system in equilibrium condition the \textit{entropy} has its maximum value because all the possible spontaneous transformations have already occurred. Thus, in this case, the entropy is maximum and is defined as
\begin{equation}
S(\Phi,V,N)=k_B \text{ln}\,\Omega(\Phi,V,N)
\label{eq:ch2_entropy}
\end{equation}
where $k_B=1.38\times10^{-23}$ J/K is the Boltzmann constant. The formula in~\eqref{eq:ch2_entropy} is the Boltzmann's entropy formula\footnote{A curiosity: the formula represents such an important results in thermodynamics that it is carved on the grave of Ludwig Boltzmann, in Vienna's cemetery.} and it represents the connection between thermodynamics and Statistical Mechanics. With this in mind, we interpret the previous microscopic experiment of searching for a maximum number of states as a correspondent macroscopic test in which we search for thermal equilibrium, with the system attaining at a given temperature the maximum number of states corresponding to the maximum value of entropy. Accordingly, by using
\begin{equation}
\left.\frac{1}{T}=\frac{\partial S}{\partial \Phi}\right|_{V,N},
\label{eq:ch2_entropy2}
\end{equation}
where $T$ is the absolute temperature, we obtain that the stationarity condition~\eqref{eq:ch2_condition} ensures thermal equilibrium, thus 
\begin{equation}
T_1=T_2.
\end{equation}
%


%
%

Following, consider again a system $\mathcal{S}$ immersed in a thermal bath $\mathcal{B}$ at thermal equilibrium and suppose that the energy, the number of particles and the volume of  $\mathcal{S}$ are much smaller than the ones of the bath. We assume also that the total energy $\Phi=\Phi_{\mathcal{S}}+\Phi_{\mathcal{B}}$ is fixed with possible energy exchanges between the bath and the system, whereas the number of particles and volume of $\mathcal{S}$ are fixed. We recall that in this case the ensemble of replicas (with possible energy exchanges with the bath) is called \textit{canonical ensemble}, as shown in Figure~\ref{fig:ch2_thermodynamics}. The energy of the system depends on the specific $i$-th configuration of the system itself, thus the energy of the bath can be expressed as
\begin{equation}
\Phi_{\mathcal{B}}=\Phi-\varphi_{i},
\label{eq:ch2_energybath}
\end{equation}
where we have defined $\Phi_{\mathcal{S}}:=\varphi_i$ to indicate the dependence on the microscopic configuration. Accordingly, in this situation, all the microscopic possible states of the bath also depend on the ones of the system. If we introduce $\rho(\varphi_i)$, the probability density of being in a certain state in the phase space for the system at a given energy $\varphi_i$, we have (\cite{sethna:2021}) 
\begin{equation}
\rho(\varphi_{i})\propto\Omega_{\mathcal{B}}(\Phi-\varphi_{i}).
\label{eq:ch2_probability}
\end{equation}
By recalling that $\Phi_{\mathcal{B}}\gg\varphi_{i}$, it is reasonable to assume $\Phi_{\mathcal{B}}\simeq \Phi$, which means $\varphi_{i}/\Phi\ll 1$. The entropy can be thus evaluated as
 \begin{equation}
 S_{\mathcal{B}}=\text{ln}\,\Omega_{\mathcal{B}}(\Phi-\varphi_{i}).
 \end{equation}
Let us perform a Taylor expansion around $\varphi_{i}/\Phi=0$ up to the first order, and using the definition of entropy in~\eqref{eq:ch2_entropy}, and~\eqref{eq:ch2_entropy2} and the expression in~\eqref{eq:ch2_energybath} we obtain
\begin{equation} 
	\begin{split}
\text{ln}\,\Omega_{\mathcal{B}}(\Phi-\varphi_{i})	&\simeq\text{ln}\,\Omega_{\mathcal{B}}(\Phi)+\left.\frac{\partial \text{ln}\,\Omega_{\mathcal{B}}}{\partial \Phi_{\mathcal{B}}}\right|_{\varphi_{i}/\Phi=0}(\Phi_{\mathcal{B}}-\Phi)\vspace{0.2 cm}\\
	&=\text{ln}\,\Omega_{\mathcal{B}}(\Phi)-\frac{\varphi_{i}}{k_B T}\\
	&=\text{ln}\,\Omega_{\mathcal{B}}(\Phi)-\beta \varphi_{i},
	\end{split}
\label{eq:ch2_expression1}
\end{equation}
where 
\begin{equation}
\beta=\frac{1}{k_B T}.
\label{eq:ch2_beta}
\end{equation}
Accordingly we rewrite~\eqref{eq:ch2_expression1} as
\begin{equation}
\text{ln}\,\Omega_{\mathcal{B}}(\Phi-\varphi_{i})-\text{ln}\,\Omega_{\mathcal{B}}(\Phi)=-\beta\varphi_{i}\rightarrow \Omega_{\mathcal{B}}(\Phi-\varphi_{i})=\Omega_{\mathcal{B}}(\Phi)e^{-\beta\varphi_{i}},
\end{equation}
so that the probability in~\eqref{eq:ch2_probability} becomes 
\begin{equation}
\rho(\varphi_{i})\propto e^{-\beta\varphi_{i}},
\label{eq:ch2_boltzdistr}
\end{equation}
the \textit{Boltzmann distribution}. 

Knowing the distribution probability, one may compute the expectation value of the energy of the system $\langle\Phi_{\mathcal{S}}\rangle$ which, from a statistical point of view, is the average concerning all the possible microscopic states of the total energy $\varphi_{\mathcal{S}}$. By definition, including the normalization condition, the expectation value of the energy reads 
\begin{equation}
\Phi_{\mathcal{S}}=\langle\varphi_i\rangle=\frac{\sum_{i}\varphi_i\,e^{-\beta\varphi_{i}}}{\sum_{i}e^{-\beta\varphi_{i}}}=-\frac{\partial}{\partial \beta}\left(\text{ln}\,\sum_{i}e^{-\beta\varphi_{i}}\right)=-\frac{\partial}{\partial \beta}\text{ln}\,\mathcal{Z},
\label{eq:ch2_avgen}
\end{equation}
where 
\begin{equation}
\mathcal{Z}=\sum_{i}e^{-\beta\varphi_{i}}
\label{eq:ch2_partitionfunc}
\end{equation}
is called the \textit{Partition Function} and it is denoted by the letter $\mathcal{Z}$ from the German word `\textit{Zustandssumme}', which means "sum over states". This quantity is of crucial importance in Statistical Mechanics because it allows the evaluation of the average of all the possible microscopic states and is used as a statistical weight to compute all the other relevant macroscopic quantities. In particular, it links Thermodynamics to Statistical Mechanics by the \textit{Free Energy}, defined as
\begin{equation}
\mathcal{F}=-k_B T \,\text{ln}\, \mathcal{Z}.
\label{eq:ch2_freeen}
\end{equation}
%

\subsection{Statistical Ensemble}

In this thesis, I will mainly focus on one-dimensional systems with two external loading conditions. Indeed, from a mechanical point of view, we consider the two different regimes of external applied force and displacement, and this distinction holds also from a statistical point of view where two different ensembles have to be considered. 

Specifically, when the controlled parameter is the generalized displacement $\delta$ we refer to the Helmholtz statistical ensemble (indicated with the subscript $\mathscr{H}$) and we define its partition function $\mathcal{Z}_{\mathscr{H}}$  (corresponding to the states at assigned $\delta$) and free energy $\mathcal{F}$. With the knowledge of these two quantities, it is then possible to derive all the expected values of the mechanical and statistical parameters. For instance, we may compute the conjugate force to the displacement that, by definition reads 
\begin{equation}
\langle f \rangle=\frac{\partial\mathcal{F}}{\partial \delta}=-\frac{1}{\beta\mathcal{Z}_{\mathscr{H}}}\frac{\partial\mathcal{Z}_{\mathscr{H}}}{\partial \delta},
\label{eq:ch2_fedef}
\end{equation}
I also consider the case in which the force $f$ is the control parameter, a condition well described by the Gibbs ensemble (subscript $\mathscr{G}$), with partition function $\mathcal{Z}_{\mathscr{G}}$ and Gibbs free energy $\mathcal{G}$. As an example, it is possible to compute the conjugate displacement to the force as 
\begin{equation}
\langle\delta \rangle=-\frac{\partial\mathcal{G}}{\partial f}=\frac{1}{\beta\mathcal{Z}_{\mathscr{G}}}\frac{\partial\mathcal{Z}_{\mathscr{G}}}{\partial f},
\label{eq:ch2_dedef}
\end{equation}
as well as the expected values of other observables. Finally, we highlight the fundamental properties of the two ensembles which consist in the fact that they are not independent. Indeed, it can be proved that the two energy functions are related by a Legendre transform (\cite{callen:1960,arnold:2013}) and the partition function of the Gibbs ensemble can be obtained from the Helhmotz one by a Laplace transform, and vice versa with the inverse transform (\cite{zinn:2004,weiner:1983}), as shown in Appendix~\ref{appA}.

\section{Rate theories}
\label{sec:ch2_rate}

In this section I briefly discuss some historical notes and mathematical concepts about rate theories, that will be used in the following chapters to describe the rate-dependent behaviour of multi-stable biological and bioinspired materials, accordingly to the topics presented in this thesis. In particular, a more detailed literature on the topic can be found in (\cite{weiner:1983,hanggi:1990,evans:1997}) and references therein. 

The dynamics evolution of multi-phase systems depends on the possibility of overcoming energy barriers separating different metastable configurations. Historically, reaction-rate theories have been developed in the field of chemical kinetics to study the rate of dissociation in chemical reactions but the theory has been extended and applied to many fields such as diffusion processes in solids and crystals or electrical transport, to name but a few. The first theoretical approach based on experimental evidence was obtained by Svante Arrhenius at the end of the $19$th century, when he found that the experimental data on reaction rates and the inverse of the absolute temperature $T$ are linked by a logarithmic scale. This pioneering results is described in Subsection~\ref{sec:arrhenius} and in Figure~\ref{fig:ch2_diffusion}$_b$, and it provides the first rate law that goes under the name of \textit{Van't Hoff-Arrhenius law} for the rate coefficient $\nu$ (\cite{vanthoff:1884,arrhenius:1889}) 
\begin{equation}
\nu=\nu_0 e^{-\beta \Phi_b},
\label{eq:ch2_arrhenius_1}
\end{equation}
where $\Phi_b$ is the activation energy, \textit{i.e} the energy necessary to overcome the barrier, $\beta=1/(k_b T)$ introduced in~\eqref{eq:ch2_beta} and $\nu_0$ is a constant factor whose meaning will be clarified later. 

It was soon realized that spontaneous escape from a configuration of local stability may happen only with noise-assisted hopping events. Thus, it was necessary to introduce a theory of fluctuations, that started with the work of Lord Rayleigh (\cite{rayleigh:1891}) Einstein (\cite{einstein:1905}) and Lindemann (\cite{lindemann:1922}) and developed by many others (\cite{fokker:1914,planck:1917,ornstein:917,pontryagin:1933}). A cornerstone idea underlying any rate calculation was described by Farkas (\cite{farkas:1927}), where he states that the rate of escape from a metastable configuration is characterized by the flux of particles that pass through the `bottleneck' separating the two `phases', or, in his nomenclature, the products from the reactants in chemical reactions. It is worth noticing that this concept is analogous to the idea that the minimum energy path is the one that passes through the saddle point of a bumpy energy landscape. Then Eyring presented the seminal rate formula deduced in a quantum statistical mechanics framework by using the definition of Partition Function (\cite{eyring:1935})
\begin{equation}
\nu=\nu_0\left(\frac{1}{\beta h}\right)\frac{\mathcal{Z}^{\mathscr{U}}}{\mathcal{Z}^{\mathscr{S}}}e^{-\beta \Phi_b},
\label{eq:ch2_eyring}
\end{equation}
where $\mathcal{Z}^{\mathscr{S}}$ and $\mathcal{Z}^{\mathscr{U}}$ are the partition functions of the stable state ($\mathscr{S}$) and the saddle point unstable state ($\mathscr{U}$), respectively, $\nu_0$ is a correction factor analogous to the one introduced in~\eqref{eq:ch2_arrhenius_1} and $h$ is Planck's constant. Most importantly, if $\nu_0=1$ this result is commonly known as \textit{\acf{TST}} of rate, indicated with $\nu_{TS}$. 

In $1940$ Kramers published the celebrated paper "\textit{Brownian motion in a Field of Force and the Diffusion Model of Chemical Reactions}" (\cite{kramers:1940}), in which he realized that escape processes are regulated by thermal noise reactions. Thus, he studied the evolution from a metastable configuration to an adjacent one as a process governed by Brownian motion dynamics, obtained by the Fokker-Planck equation and driven by thermal forces, that are connected by means of the fluctuation-dissipation theorem to the temperature $T$ and the friction coefficient $\eta$. Thus, he derived his seminal result
\begin{equation}
\nu=\nu_0 f(\eta) e^{-\beta \Phi_b}
\label{eq:ch2_kramers_1}
\end{equation}
where $f(\eta)$ is a function of the friction coefficient (see Subsection~\ref{sec:kramers}) and $\nu_0$ is a factor and $\Phi_b$ the energy barrier. 

From these classical results, since the late $20$th century to today's works (\cite{buehler:2007}) many scientists have deduced rate formulas to describe specific problems such as in the case of Bell, who in 1978 obtained a law to describe the adhesion among cells by mimicking the dissociation process of hydrogen bonds (\cite{bell:1978}).

\subsection{Atom diffusion in crystals}
\label{sec:arrhenius}

\begin{figure}[t!]
\centering
  \includegraphics[width=0.95\textwidth]{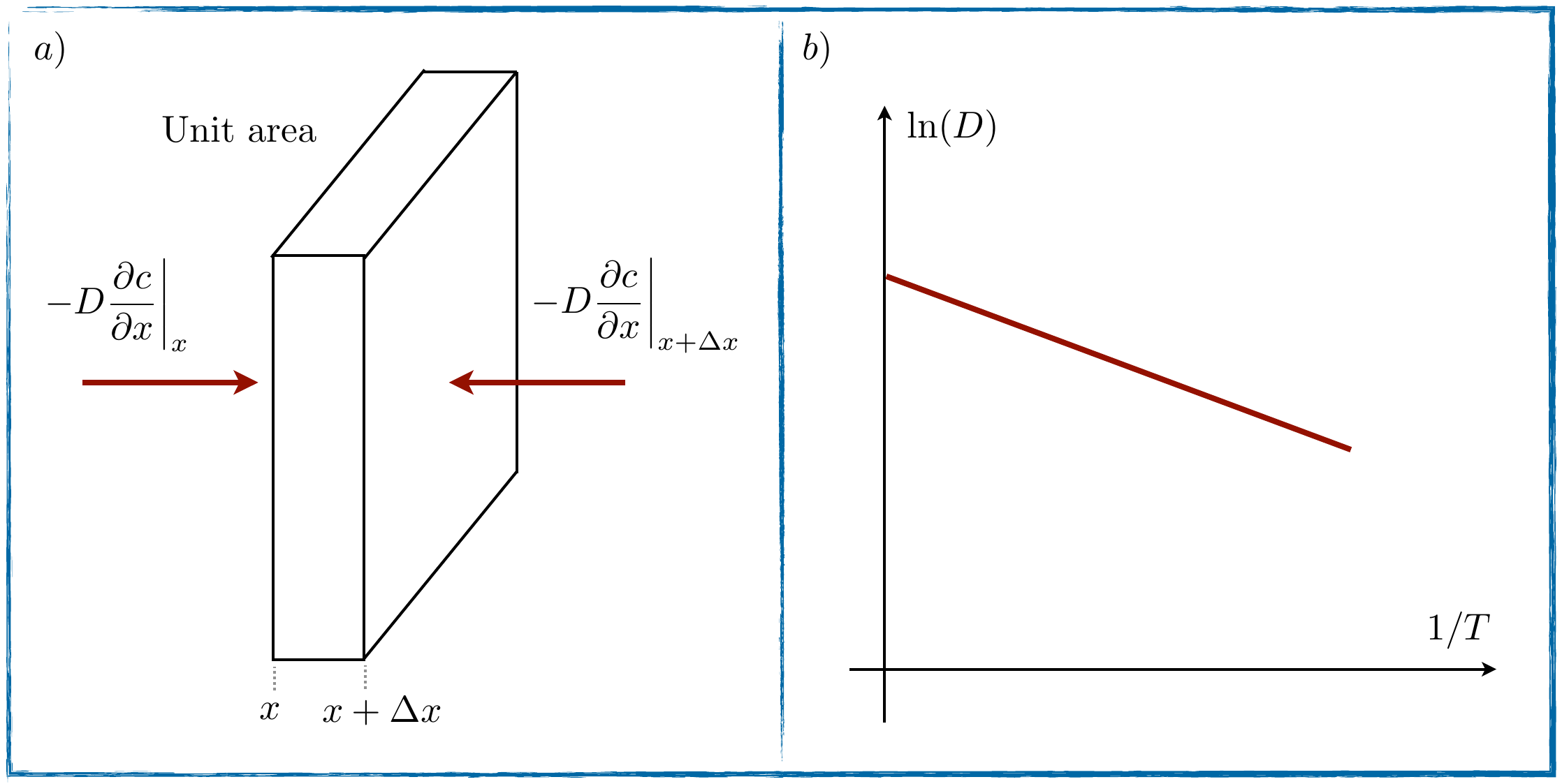}
  \caption[Impurity atom diffusion in crystals]{Panel a): the diffusion process of impurity atoms in a portion of crystal is schematized from a macroscopic point of view. Panel b): the Arrhenius rate law is shown. Adapted from (\cite{weiner:1983}).}
  \label{fig:ch2_diffusion}
\end{figure}

To begin with, we consider a macroscopic manifestation of diffusion phenomena that depend on the rate effects acting at the microscopic scale, to show how it is possible to link the macroscopic experimentally observed effects described by Arrhenius to the microscopic phenomenon. We consider the typical process of atom diffusion in crystals, that can be experimentally reproduced. We consider a sample of crystal with cubic symmetry and we deposit a uniform surface layer of interstitial impurity atoms on a plane perpendicular to one of its principal axes at a time $t=0$, as shown in Figure~\ref{fig:ch2_diffusion}$_a$. We assume that the chosen axes coincide with the $x$ direction, and assuming that this dimension of the unit area is much smaller than the other two it is possible to neglect boundary conditions and the process can be considered as one dimensional with a certain concentration on the plane $x=0$. Let $c(x,t)$ be the concentration of atom impurity. By Ficks' law, the flux $J$ of impurity atoms crossing the plane area in the positive direction is
\begin{equation}
J(x,t)=-D\frac{\partial c(x,t)}{\partial x},
\end{equation}
where $D$ is called \textit{diffusion constant}. Then, by the conservation of the mass of impurity atoms, one may write
\begin{equation}
D\left(\left.\frac{\partial c}{\partial x}\right|_{x+\Delta x}-\left.\frac{\partial c}{\partial x}\right|_{x}\right)=\frac{\partial c}{\partial t}\cdot\Delta x,
\end{equation}
which becomes, in the limit of $\Delta x\to0$, 
\begin{equation}
D\frac{\partial^{2} c}{\partial x^{2}}=\frac{\partial c}{\partial t},
\label{eq:ch2_pde}
\end{equation}
with $-\infty<x<+\infty$ and initial conditions
\begin{equation}
c(x,0)=c_{0}\delta(x),
\label{eq:ch2_bc}
\end{equation}
where $c_0$ is the initial deposited concentration and $\delta(x)$ is the Dirac delta function. The solution of the PDE in~\eqref{eq:ch2_pde} with boundary condition~\eqref{eq:ch2_bc} is
\begin{equation}
c(x,t)=\frac{c_{0}}{\left(4\pi D t\right)^{\frac{1}{2}}}e^{-\frac{x^2}{4 D t}}.
\label{eq:ch2_sol1}
\end{equation}
%

%
%

In general, the value of the diffusion constant $D$ can be obtained by properly tuning the analytical solution in~\eqref{eq:ch2_sol1} compared with the observed experimental data. Indeed, by performing experiments at different temperatures it is found that the relation between the diffusion coefficient and the inverse of the temperature is given by the \textit{Arrhenius relation} on a logarithmic scale, as shown in Figure~\ref{fig:ch2_diffusion}$_b$. The law is 
\begin{equation}
D=Ae^{-\beta C},
\label{eq:ch2_diffusionconstant}
\end{equation}
where $A$ and $C$ are generic constants obtained from experimental data. We observe that from an atomistic point of view if we assume that the concentration is sufficiently dilute such that the atoms do not interact with each other, we may focus our attention on a single atom. If it executes a random walk from an interstitial position to another at an average rate of $\nu$ jumps per unit time in the $x$ direction, after a certain time $t$ it has done $n=\nu t$ jumps of interatomic length $a$, where $a$ is the lattice parameter of the crystal. Thus, the probability distribution $p(x,t)$ of finding the atom at the position $x$ after time $t$ performing a random walk is (\cite{weiner:1983})
\begin{equation}
p(x,t)=\frac{1}{\sqrt{2\pi\nu t a^2}}\,e^{-\frac{x^2}{2 \nu  t a^2}}.
\end{equation}
Then, having initially $c_0$ atoms per unit area in the plane $x=0$, the expected concentration at time $t$ is 
\begin{equation}
c(x,t)=c_0 p(x,t)=\frac{c_0}{\sqrt{2\pi \nu  t a^2}}\,e^{-\frac{x^2}{2 \nu  t a^2}}.
\label{eq:ch2_sol2}
\end{equation}
Finally, by comparing~\eqref{eq:ch2_sol1} and~\eqref{eq:ch2_sol2} one obtains 
\begin{equation}
D=\frac{1}{2}\nu a^2,
\end{equation}
that represents a clear relation among the macroscopic diffusion coefficient $D$ and microscopic rate of jump $\nu$ by means of the interatomic distance $a$. Accordingly, it can be concluded that the rate $\nu$ has to obey an Arrhenius relation because the diffusion constant does, that is exactly the Van't Hoff-Arrhenius rate law in~\eqref{eq:ch2_arrhenius_1}.

\subsection{A simple one-dimensional rate theory}
\label{sec:ch2_1D}

\begin{figure}[t!]
\centering
  \includegraphics[width=0.95\textwidth]{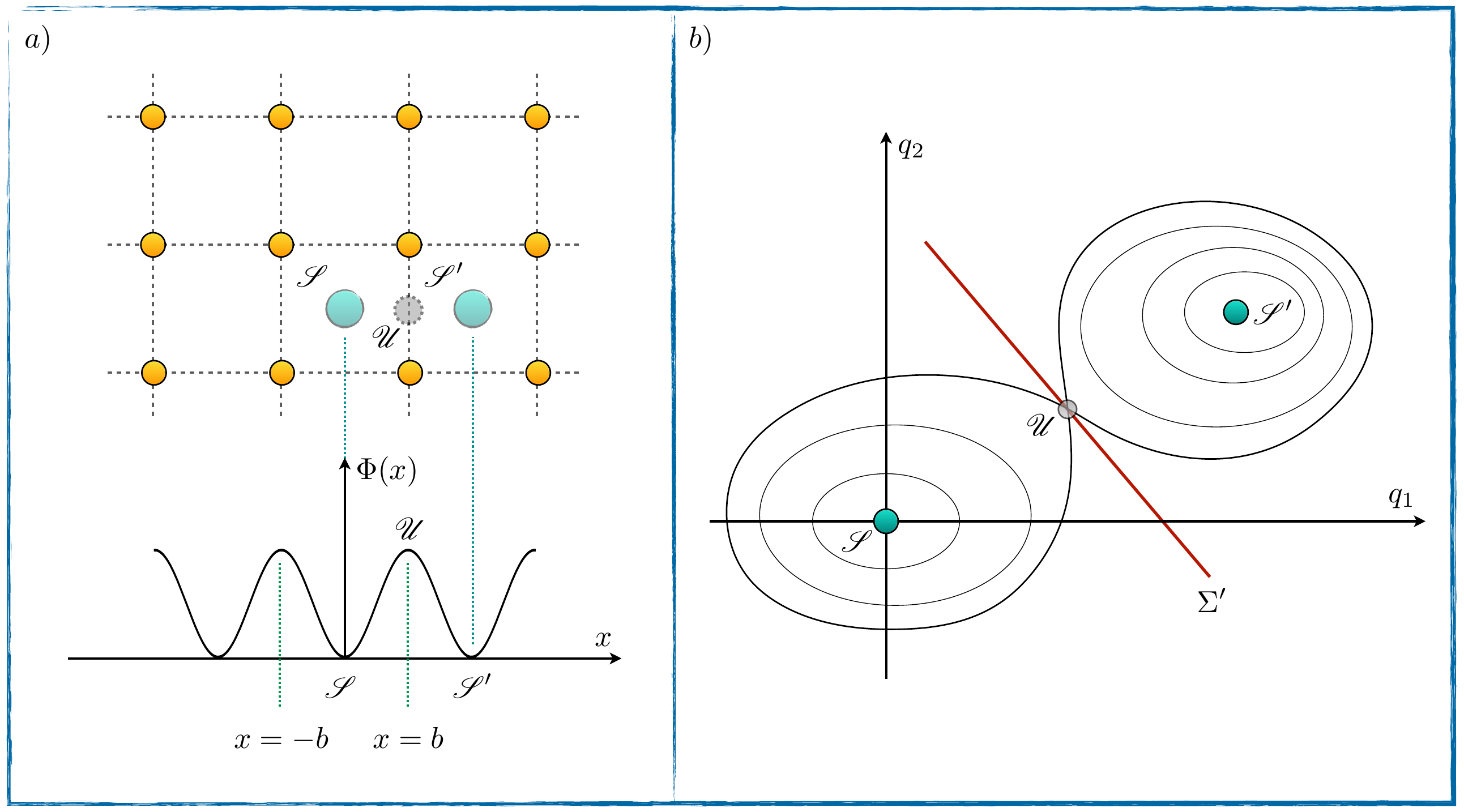}
  \caption[Impurity atom diffusion in crystals]{Panel a): the impurity atoms dispersed in the crystal lattice are shown with light green dots within the atoms of the crystal (yellow dots). In particular, two different stable positions $\mathscr{S}$ and $\mathscr{S}'$ of the impurity atoms are represented separated by an unstable configuration $\mathscr{U}$, and the associated potential energy $\Phi(x)$ is shown below. Panel b): the energy landscape of the potential energy in the configurational space is shown with respect to the configuration in panel a). Adapted from (\cite{weiner:1983}).}
  \label{fig:ch2_lattice}
\end{figure}

Following the example, in the previous section, we introduce a simple one-dimensional rate theory to describe the process from a microscopic point of view. Consider the crystal lattice in Figure~\ref{fig:ch2_lattice}$_a$, where the atoms in each node are represented, whereas in the interstitial sites there are the impurity atoms of mass $m$. We focus on the motion in one direction and consider the impurity atom in the equilibrium position $\mathscr{S}$, that, due to thermal motion may move equivalently to the adjacent interstitial position $\mathscr{S}'$. 

%
%

The interaction of the atoms with the crystal is characterized by the periodic potential energy function $\Phi(x)$, because both $\mathscr{S}$ and $\mathscr{S}'$ are stable equilibrium positions.  From a classical mechanics point of view, it is sufficient to know the position and momentum of the atoms to describe their dynamical state, thus in the phase space, we have a point with coordinates $(x,p)$, where $-b<x<b$ defined among two equilibrium position (see Figure~\ref{fig:ch2_lattice}$_a$). To describe the entire collection of impurity atoms we use the distribution function $\rho(x,p)$ so that the integral
\begin{equation}
\int_{\mathcal{D}}\rho(x,p)\text{d}x\,\text{d}p
\end{equation}
is equals to the fraction of all the impurity atoms in an arbitrary region $\mathcal{D}$ of the phase space and must satisfy the normalization condition 
\begin{equation}
\int_{-b}^{+b}\text{d}x\int_{-\infty}^{+\infty}\rho(x,p)\text{d}p=1.
\label{eq:ch2_norm}
\end{equation}
We remark that $\rho(x,p)$ does not depend explicitly on time so we are in steady-state conditions. This means that even though there is a continuous motion in phase space, the fraction in a certain region remains constant. It is, therefore, possible to compute the rate at which atoms leave one stable equilibrium configuration and move to an adjacent one by crossing the line $x=b$ at time $\text{d}t$ and with momentum from $p$ to $p+\text{d}p$ that is fixed in the phase space. The atoms are the fraction
\begin{equation}
\frac{1}{m}\rho(b,p)(p\text{d}t)\text{d}p
\end{equation}
of the total impurity atoms in the collection. Thus the average fractional rate $\nu$ at which a certain number of impurity atoms cross the potential barrier at $x=b$ is 
\begin{equation}
\nu=\frac{1}{m}\int_0^{\infty}p\rho(b,p)\text{d}p.
\label{eq:ch2_nu1}
\end{equation}

The distribution function $\rho(x,p)$ can be derived explicitly by considering that the atom, being diluted in the crystal, do not interact with each other so that they are in thermal equilibrium with the rest of the hosting space at a common temperature $T$. This is exactly the physical representation of the concept of a canonical ensemble of replicas of a single impurity atom, introduced in Section~\ref{sec:ensembles}. Then we get
\begin{equation}
\rho(x,p)=Ce^{-\beta \mathcal{H}(x,p)},
\end{equation}
where
\begin{equation}
\mathcal{H}(x,p)=\frac{p^2}{2m}+\Phi(x),
\end{equation}
is the Hamiltonian of the system and $C$ is the normalization constant that can be obtained by the~\eqref{eq:ch2_norm}. To compute explicitly the rate in~\eqref{eq:ch2_nu1} it is necessary to assume the form of the potential energy. In the case of periodic wells it can be approximated by elastic energy:
\begin{equation}
\Phi(x)=\frac{1}{2}k x^2.
\end{equation}
Moreover, it is important to highlight that the contributions due to temperature in the integral are much less probable away from the minimum of the well and they are exponentially suppressed. Thus, the range of integration can be extended to $-\infty<x<\infty$ and $C$ reads
\begin{equation}
C=\nu_0(T),
\end{equation}
where $\nu_0(T)$ represents the natural frequency of oscillation in the harmonic region of the well depending on the absolute temperature, and, at low temperature it reads
\begin{equation}
\nu_0=\frac{1}{2\pi}\sqrt{\frac{k}{m}}.
\end{equation}

The physical meaning of this approximation is very important: in a general experiment at a certain time almost all impurity atoms are oscillating in the neighbourhood of the well's bottom, and few of them are going up the slope to cross the energy barrier with an exponential probability. With $\rho(x,p)$ explicit, the integration of~\eqref{eq:ch2_nu1} can be performed leading to 
\begin{equation}
\nu=\nu_0(T) e^{-\beta \Phi_b},
\label{eq:ch2_rate1d}
\end{equation}
where $\Phi_b$ represents the height of the energy barrier. This expression is of the Arrhenius type, \eqref{eq:ch2_arrhenius_1}. By compare~\eqref{eq:ch2_diffusionconstant} with~\eqref{eq:ch2_rate1d} we recognize that the factor $A$ coincides with $\nu_0(T)$ whereas the coefficient $C$ in the exponent is the energy barrier $\Phi_b$. It is thus possible to take experimental data for $D$, link them to the jump frequency $\nu$ and extrapolate crucial data such as the activation energy and $\nu_0(T)$.

\subsection{Extension to multi-degrees of freedom in a thermodynamic framework}

We may now consider a more general framework introducing the role of thermal motion of the host atoms that prevent or favour the transition among different stable configurations and the cooperative motion of the impurity atoms. We consider again the crystal portion with a certain quantity of impurity atoms uniformly distributed and we take a certain number of this portion. Then, by introducing the mass reduced coordinates $p$ and $q$, the new Hamiltonian reads 
\begin{equation}
\mathcal{H}(q,p)=\frac{1}{2}\sum_{i=1}^{n}p_{i}^{2}+\Phi(q_1,\dots,q_n).
\end{equation}

As shown in Figure~\ref{fig:ch2_lattice}$_a$, we consider two stable configurations $\mathscr{S}$ and $\mathscr{S}'$ and an unstable configuration $\mathscr{U}$. In the configuration space $\Gamma_q$ the constant energy contours of $\Phi(q_1,\dots,q_n)$ appear as in Figure~\ref{fig:ch2_lattice}$_b$. $\mathscr{S}$ and $\mathscr{S}'$ are potential minima while $\mathscr{U}$ corresponds to a saddle --unstable-- point. 

As before, the distribution function $\rho(q_1,\dots,q_n,p_1,\dots,p_n)=\rho(\boldsymbol{q},\boldsymbol{p})$ is equal to the fraction of crystallites in a certain region $\mathcal{D}$ and in analogy with~\eqref{eq:ch2_norm}, must satisfy the normalization condition 
\begin{equation}
\int_{\mathcal{D}_\mathcal{S}}\text{d}q\int_{\Gamma_p}\rho(\boldsymbol{q},\boldsymbol{p})\text{d}p=1,
\label{eq:ch2_norm2}
\end{equation}
where $\mathcal{D}_\mathscr{S}$ corresponds to the maximal region in configuration space in which $\mathscr{S}$ is the only stable equilibrium configuration for the crystallite and $\mathscr{U}$ is the only other equilibrium configuration. From a physical point of view, this is the representation of the concept of canonical ensemble of replicas at thermal equilibrium and hence the distribution function can be expressed as 
\begin{equation}
\rho(\boldsymbol{q},\boldsymbol{p})=C  e^{-\beta \mathcal{H}(q,p)},
\end{equation}
with $C$ to be determined by the normalization condition. Similarly to what was introduced in Section~\ref{sec:ch2_1D} one may define the harmonic approximation of $\Phi(\boldsymbol{q})$ that reads
\begin{equation}
\Phi(q_1,\dots,q_n)=\frac{1}{2}A_{i,j}q_i q_j,
\end{equation}
where $A_{i,j}$ is a positive defined matrix. Then, by considering again that the major contribution of the integral is the region in which $\Phi$ is harmonic, one may consider the whole $\Gamma_q$ as the integration domain for~\eqref{eq:ch2_norm2}, thus obtaining  
\begin{equation}
C^{-1}=\int_{\Gamma_q}e^{-\frac{\beta}{2}A_{i,j}q_i q_j}\text{d}q\int_{\Gamma_p}e^{-\frac{\beta}{2}p_i p_j}\text{d}p.
\end{equation}
By transforming to normal coordinates one obtains
\begin{equation}
A_{i,j}q_iq_j=\sum_{i=1}^{n}\lambda_i^{A}\left(Q_i^{A}\right)^2
\end{equation}
where, since $A_{i,j}$ is positive definite, its eigenvalues $\lambda_i^{A}>0$ $\forall i=1,\dots,n$ we get 
\begin{equation}
C^{-1}=\left(\frac{2\pi}{\beta}\right)^{n}\prod_{i=1}^{n}\left(\lambda_i^{A}\right)^{-\frac{1}{2}}=\frac{1}{\sqrt{\text{det}(A)}}\left(\frac{2\pi}{\beta}\right)^{n}.
\end{equation}

Consider now the flux of replicas of the adjacent potential well. Let $q_i^{\mathscr{U}}$ be the coordinates of the unstable state $\mathscr{U}$ and consider the harmonic approximation of $\Phi$ in the neighbourhood of $\mathscr{U}$:
\begin{equation}
\Phi=\Phi_{\mathscr{U}}+\frac{1}{2}B_{i,j}\left(q_i-q_i^{\mathscr{U}}\right)\left(q_j-q_j^{\mathscr{U}}\right).
\end{equation}
Since $\mathscr{U}$ is an unstable equilibrium configuration, the matrix $B_{i,j}$ must have at least one negative eigenvalue $\lambda_1^{B}<0$, and we assume that it is the only one. Let $Q_{i}^{B}$ be the normal coordinates of $B_{i,j}$, with origin at $\mathscr{U}$, with corresponding momenta $P_{i}^{B}$. Then, $Q_{i}^{B}$ is in the direction of the maximum negative curvature of the saddle point in the potential energy surface at $\mathscr{U}$, with its sign taken so that it points in the direction away from $\mathscr{S}$ and towards $\mathscr{S}'$. Define now the hypersurface passing through $\mathscr{U}$ with the property that any trajectory that begins in $\mathcal{D}_{\mathscr{S}}$ and crosses $\Sigma$ goes on to the neighbourhood of $\mathscr{S}'$, as shown in Figure~\ref{fig:ch2_lattice}$_b$.  If the temperature level is sufficiently low, then most of the transitions among the two stable states take place along trajectories that will pass through the neighbourhood of the saddle point at $\mathscr{U}$ and, consequently, we may consider as the critical surface the hyperplane $\Sigma'$ perpendicular to with coordinates $Q_i^{B}$, with  $i = 2,\dots, n$ lying in $\Sigma'$. Then, by the same reasoning used in the one-dimensional case, the fractional rate $\nu$ of crossing the critical hyperplane $\Sigma'$ is
\begin{equation}
f=C\int_0^{\infty}P_{1}^{B}e^{-\frac{\beta}{2}\left(P_1^B\right)^2}\int_{\Gamma'}e^{-\beta \mathcal{H}'(Q_i^{B},P_{i}^{B})}\prod_{i=2}^{n}\text{d}Q_{i}^{B}\text{d}P_{i}^{B},
\end{equation}
where $\mathcal{H}'(Q_i^{B}, P_{i}^{B})$ is the Hamiltonian of the system contained to move in the hyperplane $\Sigma'$ so that it has only $n-1$ degrees of freedom. In this case, the normalization constant coincides with the canonical partition function of the system in the vicinity of the equilibrium configuration so that 
\begin{equation}
C=\frac{1}{\mathcal{Z}}
\end{equation}
The integration with respect to $P_1^{B}$ leads to 
\begin{equation}
\nu=\frac{1}{\beta}\frac{\mathcal{Z}_{\Sigma'}}{\mathcal{Z}},
\end{equation}
where $\mathcal{Z}_{\Sigma'}$ is the partition function corresponding to $\mathcal{H}'(Q_i^{B},P_{i}^{B})$. Let then $\Sigma_0$ be a hyperplane passing through the equilibrium configuration at $\mathscr{S}$ parallel to $\Sigma'$, and let $\mathcal{Z}_{\Sigma_0}$ be the partition function of the system when constrained to move in $\Sigma_0$. Then we get 
\begin{equation}
\nu=\frac{1}{\beta}\frac{\mathcal{Z}_{\Sigma_0}}{\mathcal{Z}}\frac{\mathcal{Z}_{\Sigma'}}{\mathcal{Z}_{\Sigma_0}}.
\end{equation}
We may consider two different types of boundary conditions of applied stress or strain. As an example, consider the case of applied strains $\varepsilon$. The free energies for the equilibrium configuration and the saddle points are, respectively
\begin{equation}
\mathcal{F}_0=-\frac{1}{\beta}\text{ln}\mathcal{Z}_{\Sigma_0}(\varepsilon),
\end{equation}
and
\begin{equation}
\mathcal{F}'=-\frac{1}{\beta}\text{ln}\mathcal{Z}_{\Sigma'}(\varepsilon),
\end{equation}
so that the rate becomes
\begin{equation}
\nu=\nu_0(T)\,e^{-\beta \Delta\mathcal{F}},
\end{equation}
where
\begin{equation}
\Delta\mathcal{F}=\mathcal{F}'(\varepsilon)-\mathcal{F}_0(\varepsilon),
\end{equation}
is the free energy of activation that coincides with the height of the energy barrier $\Phi_b$ and the frequency factor is 
\begin{equation}
\nu_0(T)=\frac{1}{\beta}\frac{\mathcal{Z}_{\Sigma_0}}{\mathcal{Z}},
\end{equation}
that formally coincides with the Eyring formula in~\eqref{eq:ch2_eyring} derived in a non-quantum framework.

\subsection{Kramers rate formula}
\label{sec:kramers}

The transition state theory introduced in Section~\ref{sec:ch2_1D} can be extended to include the effect of the interaction of the system with a heat bath. The resulting rate expression is originally formulated to deal with chemical reactions rates (\cite{kramers:1940}). We first generalize the framework of~\eqref{eq:ch2_rate1d}, where it is assumed that the rate corresponds to the transition at a barrier located at $x=b$. Indeed, we may compute the rate at which the atoms cross the plane at $x=a$, with $0<a<b$, and with momentum such that it is sufficient to carry them over the barrier. The `critical' momentum $p_a$ is 
\begin{equation}
p_a=\sqrt{2m\left(\Phi(b)-\Phi(a)\right)}.
\end{equation}
With the same reasoning that leads to~\eqref{eq:ch2_nu1}, the rate can be then expressed as 
\begin{equation}
\nu=\frac{1}{m}\int_{p_a}^{\infty}p\rho(a,p)\text{d}p
\end{equation}
which gives 
\begin{equation}
\nu_{TS}=\frac{C}{m}e^{-\beta \Phi(a)}\int_{p_a}^{\infty}p e^{-\frac{\beta p^2}{2m}}\text{d}p=\frac{C}{\beta}e^{-\beta\left(\Phi(a)+\frac{p_a^2}{2m}\right)},
\end{equation}
which becomes 
\begin{equation}
\nu_{TS}=\nu_0e^{-\beta \Phi_b},
\end{equation}
as previously derived. 

%
\begin{figure}[t!]
\centering
  \includegraphics[width=0.75\textwidth]{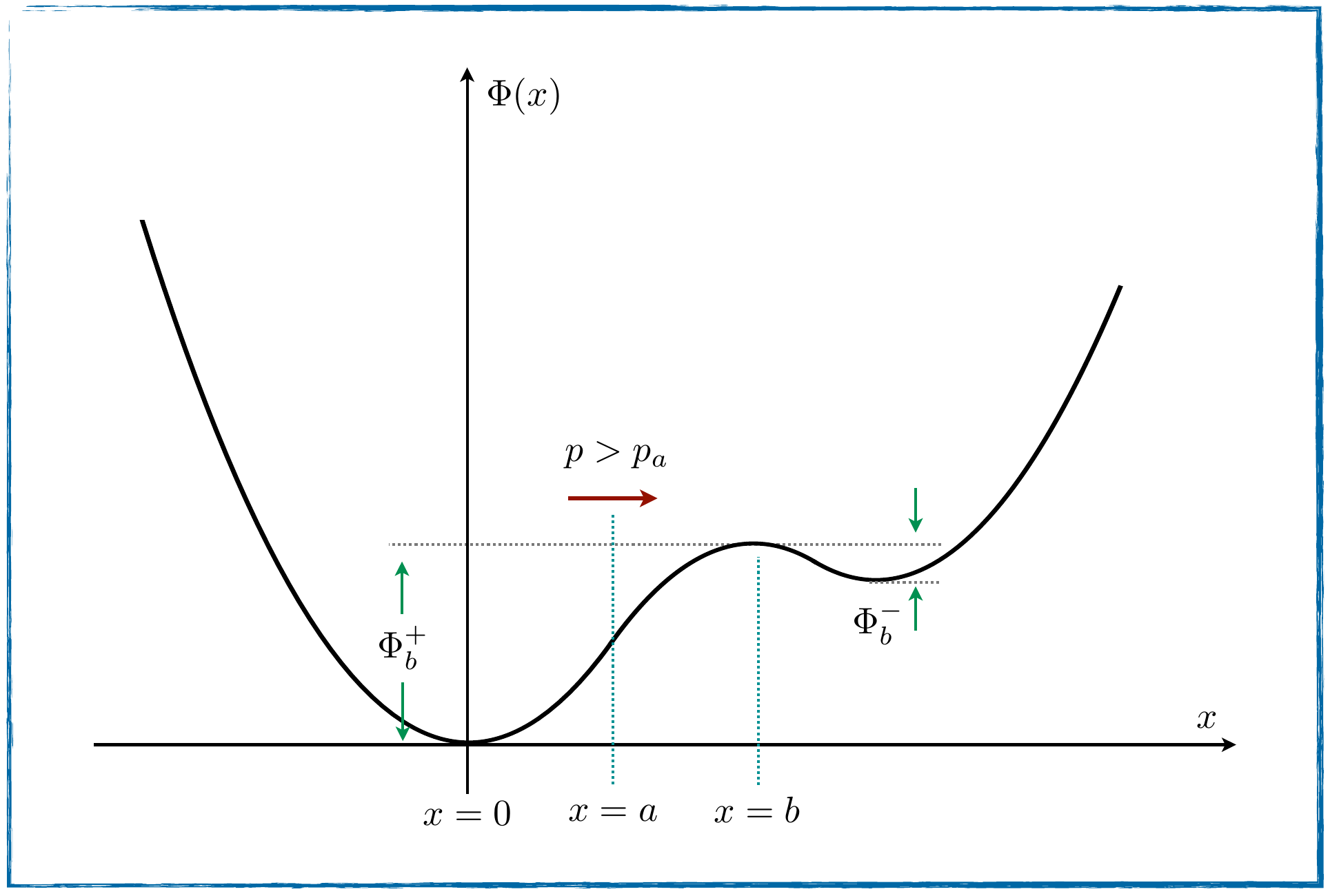}
  \caption[General potential energy]{Potential energy of a typical multistable material with two stable phases and an unstable configuration at $x=b$. The energy barriers to go from the first (\textit{e.g.} folded) phase to the second one (\textit{e.g.} unfolded) $\Phi_b^+$ and viceversa ($\Phi_b^-$) are represented. Adapted from (\cite{weiner:1983}).}
  \label{fig:ch2_kramers}
\end{figure}
%

We have denoted the rate here as $\nu_{TS}$ to empathize that it is derived based on the transition-state assumption. It is this assumption, in which the effect of interactions with the heat bath on the particle trajectory is neglected, which permitted us to state that all particles crossing $x=a$ with  $p>p_a$  will surmount the barrier peak and that none can cross if  $p<p_a$.  To keep on going with our discussion, we assume that the atom interactions with the thermal bath are characterized by a viscosity parameter $\eta$. Thus, it may happen that a particle with $p<p_a$ in $x=a$ may still cross the barrier with the aid of the thermal interactions and, on the other hand, a particle with $p>p_a$ may be turned back. Moreover, the two stable minima configurations may have different ground energy so that the energy to cross in one direction may be different from the other, as shown in Figure~\ref{fig:ch2_kramers}. For this reason, we assume a quadratic --elastic-- function for the potential energy
\begin{equation}
\Phi(x)=\Phi_b-\frac{1}{2}k_b\left(x-b\right)^2
\end{equation}
By performing the same passages developed in the previous section it is possible to obtain the Kramers rate formula introduced in~\eqref{eq:ch2_kramers_1}: 
\begin{equation}
\nu=\nu_0\left(\sqrt{\xi^2+1}-\xi\right)e^{-\beta \Phi_b},
\label{eq:ch2_kramers}
\end{equation}
where 
\begin{equation}
\nu_0=\frac{1}{2\pi}\sqrt{\frac{k_{\mathscr{S}}}{m}},
\end{equation}
is the natural frequency of oscillation of the particle in the harmonic region well and $\xi$ is a function of the viscosity parameter such that
\begin{equation}
\xi=\frac{\eta}{2\sqrt{k_b m}}.
\end{equation}

The results obtained in~\eqref{eq:ch2_kramers} depend directly on the activation energy $\Phi_b$ that is the same as the transition state theory whereas the frequency factor is reduced by the factor $\sqrt{\xi^2+1}-\xi$. Moreover, one may observe that as the interaction strength goes to zero, \textit{i.e.} when $\eta\to0$, Kramers rate formula in~\eqref{eq:ch2_kramers} reduces just to the transition state formula~\eqref{eq:ch2_rate1d}. On the other hand, the effect of increasing $\eta$ is to decrease both the frequency factor and the rate $\nu$ and in the limit of large $\xi$,~\eqref{eq:ch2_kramers} becomes 
\begin{equation}
f\simeq\frac{\nu_0}{2\xi}e^{-\beta E_b}=\nu_0\frac{\sqrt{k_b m}}{\eta}e^{-\beta E_b}.
\label{eq:ch2_kramers2}
\end{equation}

Similarly, a decrease in the factor $\sqrt{k_b m}$ with $\eta$ fixed also causes a decrease in the effective rate of transition. This is understandable on physical grounds since a decrease in $k_b$ corresponds to a flatter barrier peak, and therefore the thermal random force can more easily return the particle after it has crossed the peak. A smaller particle mass also facilitates returns. The fact that the Kramers rate formula gives a more detailed description of the process than the one-dimensional transition-state rate formula of \eqref{eq:ch2_rate1d} is indicated by its dependence on two additional parameters: $\eta$, a measure of the interaction strength with the heat bath and $k_b$, the curvature of the barrier peak. On the other hand, the value of $x=a$ (see Figure~\ref{fig:ch2_kramers}), which may be regarded as the point of transition between the regions of positive and negative curvature of the potential function $\Phi(x)$, drops out in the integration as it does in the transition-state analysis and does not appear in the final formula.

\section{Prototypical example}
\label{sec:ch2_pt1}

\begin{figure}[t!]
\centering
  \includegraphics[width=0.95\textwidth]{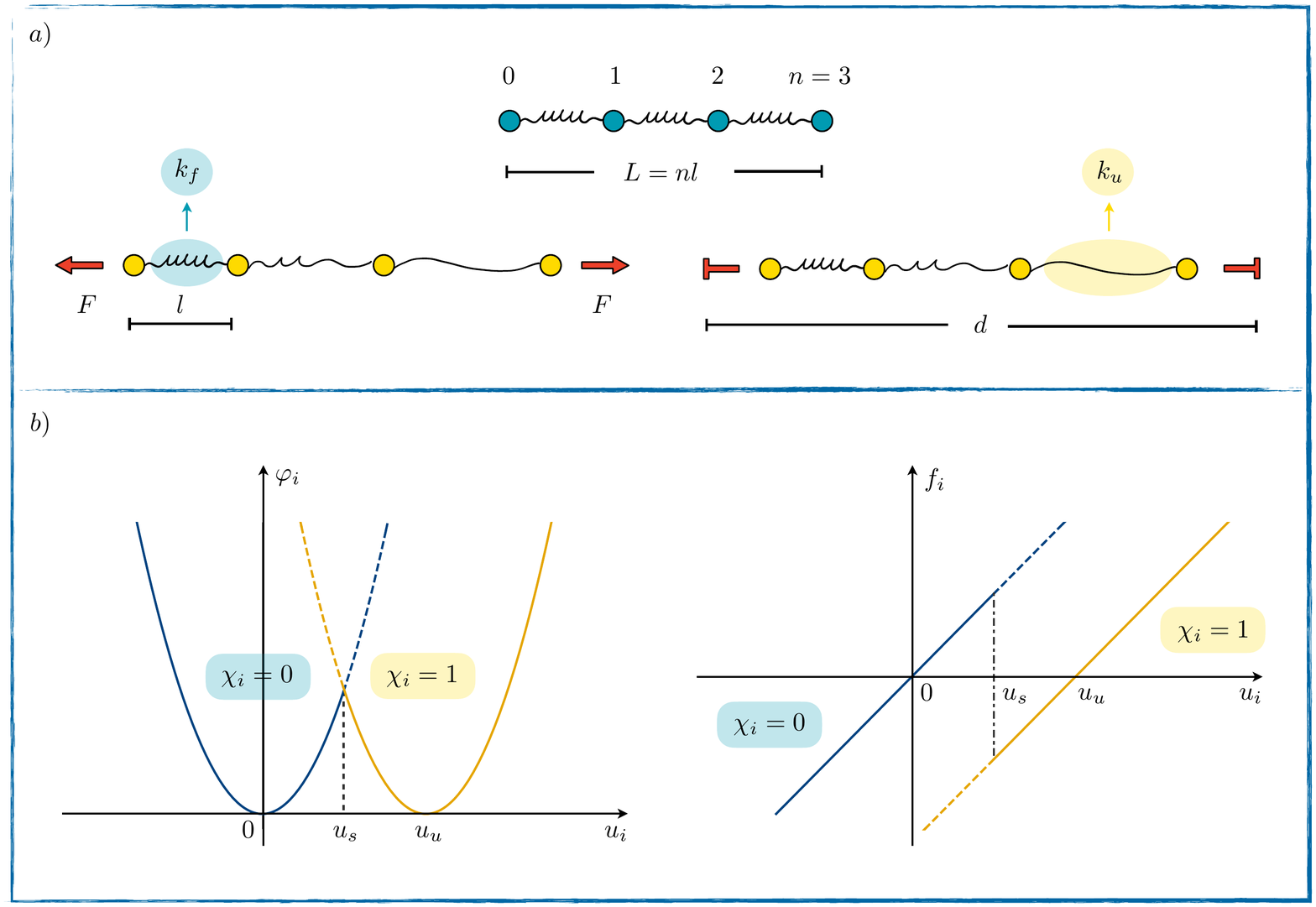}
  \caption[Model of the chain composed by three unit element]{The prototypical model is made by three units and the relative energy and force of the single elements are shown. In panel a) the chain is represented with some of the mechanical properties displayed. In particular, in the bottom left figure, the case of soft device is shown whereas in the bottom right the applied displacement (hard device) is represented. In panel b), left, the energy of a single unit with the folded (phase 1) and unfolded (phase 2) wells are shown and the corresponding force of the single units is on the right side of the panel.}
  \label{fig:ch2_modelloesempio}
\end{figure}

In this section we introduce a simple model to present and apply the main theoretical frameworks discussed above that will also be adopted in the following chapters. We consider a chain of only three identical elements, each characterized by the simplest form of non-convex energy, \textit{a biparabolic energy}, as shown in Figure~\ref{fig:ch2_modelloesempio}$_b$. To fix the ideas, we consider the case of topological transition, as observed in protein molecules, where one energy well corresponds to the folded state whereas the second well corresponds to the unfolded configuration. 

%
%

Consider the chain constituted by $n=3$ units in series linked by massless points, as shown in the scheme in Figure~\ref{fig:ch2_modelloesempio}$_a$.  Let then $L=nl=3l$ be the overall unstressed length of the chain, where $l$ is the unit reference length. In this simple example, we assume that the folded and unfolded configurations have the same stiffness and there is no transition energy, thus we define $\kappa:=k_f=k_u$. Accordingly, the energy of each unit can be expressed as 
\begin{equation}
\Phi_i=\frac{1}{2}\kappa l \left(\frac{u_i-\chi_i u_u}{l}\right)^2,
\end{equation}
where $u_i$ is the displacement from the equilibrium position of the $i-th$ element and $u_u$ is the displacement at which the phase transition to the unfolded configuration occurs. Moreover, we have introduced the `spin' variable $\chi_i$ that assumes value $\chi_i=0$ when the element is in the folded state and $\chi_i=1$ in the case in which the spring is in the unfolded configuration. After introducing the spring mechanical \textit{strain}, defined as 
\begin{equation}
\varepsilon_i=\frac{u_i}{l},
\end{equation}
we may express the total energy of the chain as
\begin{equation}
\Phi=\frac{1}{2}\kappa l \sum_{i=1}^{n}\left(\varepsilon_i-\chi_i\varepsilon_u\right)^2,
\label{eq:ch2_bip_en}
\end{equation}
where $\varepsilon_u=u_{u}/l$ is the unloaded strain of the second well. 

From a mechanical point of view, two different boundary conditions are typically applied. In one case we may think to fix the total length of the chain by assigning the displacement $d$, given by
\begin{equation}
d=\sum_{i=1}^{n}u_i=l\sum_{i=1}^{n}\varepsilon_i.
\label{eq:ch2_displacement}
\end{equation}
This situation corresponds, for instance, to atomic force microscopy (\acs{AFM}) pulling experiments and it is known as \textit{isomeric} condition or, more frequently, as the \textit{hard device} hypothesis. The other situation is when the force $F$ acting on the system is assigned such as in the case of optical tweezers tests. We refer to this case as \textit{isotensional} or the \textit{soft device} hypothesis. As we will discuss in detail in Chapter~\ref{ch_3} of this thesis, this simplification of dividing the whole range of possible applications into only two opposite regimes is a condition that may play a crucial theoretical and experimental role. Indeed, typical experiments are in none of these two theoretical regimes but may range continuously between them. Moreover, when the important case of the \textit{thermodynamical limit} is considered, \textit{i.e.} under the assumption that the number of elements $n\rightarrow \infty$, the equivalence of the response of the system in the two boundary conditions described above represents an open problem that must be analyzed depending on the specific system considered. 

A convenient approach that we follow in this thesis is to use non-dimensional quantities. Thus the total energy --adimensionalized with respect to $\kappa l$-- reads 
\begin{equation}
\varphi=\frac{\Phi}{\kappa l}=\frac{1}{2}\sum_{i=1}^{n}(\varepsilon_i-\chi_i\varepsilon_u)^2,
\label{eq:ch2_energynondimensionale}
\end{equation}

With the same reasoning, based on the introduction of the internal variables $\chi_i$ defining the non-convex aspect of the energy, it is useful to introduce the strain and internal variable vectors
\begin{equation}	\qquad
	\boldsymbol{\varepsilon}=
	\begin{pmatrix}
	\varepsilon_1	\\
	\varepsilon_2 	\\
	\varepsilon_3	\\ 
	\end{pmatrix}, \qquad \qquad
	\boldsymbol{\chi}=
	\begin{pmatrix}
	\chi_1	\\
	\chi_2	\\
	\chi_3	\\
	\end{pmatrix}, 
	\label{eq:ch2_vect}
\end{equation}
so that we may rewrite the energy as 
\begin{equation}
\varphi=\frac{1}{2}\boldsymbol{\varepsilon}\cdot\boldsymbol{\varepsilon}-\varepsilon_u \,\boldsymbol{\varepsilon}\cdot\boldsymbol{\chi}+\frac{1}{2}\varepsilon_u^{2}\,\boldsymbol{\chi}\cdot\boldsymbol{\chi}.
\label{eq:ch2_matrixenergy}
\end{equation}
%

\subsection{Mechanics}
\label{sec:ch2_pt2}

We first consider the case of `pure' mechanics, in which we neglect temperature effects and only take into account the mechanical response of the system under the two conditions of soft and hard devices, represented in Figure~\ref{fig:ch2_modelloesempio}$_{a}$. In the first case, we fix the total displacement of the whole chain $d$, defined in~\eqref{eq:ch2_displacement} that in dimensionless form reads 
\begin{equation}
\delta=\frac{d}{l},
\label{eq:ch2_displacementnd}
\end{equation}
to explore the \textit{hard device} condition. In the second case, we study the system in the \textit{soft device} regime, by introducing the non-dimensional force
\begin{equation}
f=\frac{F}{\kappa}.
\label{eq:ch2_force}
\end{equation}

To begin with, we fix the overall displacement $\delta$ defined in~\eqref{eq:ch2_displacementnd}:
\begin{equation}
\sum_{i=1}^{n=3}\varepsilon_i=\boldsymbol{\varepsilon}\cdot\boldsymbol{1}=\delta,
\end{equation}
where $\boldsymbol{1}=\{1,1,1\}$. We search for the minima of
\begin{equation}
h=\varphi-f(\boldsymbol{\varepsilon}\cdot\boldsymbol{1}-\delta).
\end{equation}
where $f$ is a Lagrange multiplier and represents the force conjugated to the imposed displacement. Thus, we study the variational problem
\begin{equation}
\min_{\boldsymbol{\varepsilon},\,\,\,\boldsymbol{\varepsilon}\cdot\boldsymbol{1}=\delta} \,h,
\end{equation}
At equilibrium, to obtain the Euler equations, \textit{i.e.} the equations of motion, we study the first variation with respect to the variables $\boldsymbol{\varepsilon}$ obtaining 
\begin{equation}
\frac{\partial h}{\partial \boldsymbol{\varepsilon}} = \boldsymbol{\varepsilon}-\varepsilon_u \boldsymbol{\chi}-f \boldsymbol{1}= \boldsymbol{0}.
	\label{eq:ch2_1var}
\end{equation}

From~\eqref{eq:ch2_1var} we obtain the equilibrium strains depending on both the resulting equilibrium force $f_{eq}$ and the phase state $\boldsymbol{\chi}$, 
\begin{equation}
\boldsymbol{\varepsilon}_{eq}=f_{eq}\boldsymbol{1}+\varepsilon_u\boldsymbol{\chi}.
\label{eq:ch2_eqstrains}
\end{equation}
By definition, the average strain of the chain is simply
\begin{equation}
\bar{\varepsilon}:=\frac{1}{n}\sum_{i=1}^{n=3}\varepsilon_i=\frac{1}{n}\left(\boldsymbol{\varepsilon}_{eq}\cdot\boldsymbol{1}\right),
\end{equation}
which, using~\eqref{eq:ch2_eqstrains}, gives
\begin{equation}
\bar{\varepsilon}=\frac{1}{n}\left(f_{eq}\boldsymbol{1}\cdot\boldsymbol{1}+\varepsilon_u\boldsymbol{1}\cdot\boldsymbol{\chi}\right)=f_{eq}+\frac{p}{n}\varepsilon_u=f_{eq}+\bar{\chi}\varepsilon_u,
\label{eq:ch2_avgstrain}
\end{equation}
where $p$ is the number of elements in the second phase, \textit{i.e.} the number of unfolded units, and
\begin{equation}
\bar{\chi}:=\frac{1}{n}\sum_{i=1}^{n=3}\chi_i=\frac{1}{n}\left(\boldsymbol{\chi}\cdot\boldsymbol{1}\right)=\frac{p}{n}\in[0,1]
\label{eq:ch2_chi}
\end{equation}
is the unfolded fraction. From a physical point of view, this fraction measures the size of the conformational transition in terms of units that undergoes unfolding. In particular when $\bar{\chi}=0$ all the elements are in the first (folded) state. On the other hand, when $\bar{\chi}=1$ the whole system has undergone a conformational folded$\rightarrow$unfolded transition and the system is in a new homogeneous state in the second phase. 

Eventually, the non dimensional equilibrium force and energy in the hard device hypothesis obtained by using~\eqref{eq:ch2_matrixenergy}, and~\eqref{eq:ch2_avgstrain} read 
\begin{equation}
	\begin{split}
	\frac{\varphi_{eq}}{n}	&	=\frac{1}{2}\bar{\varepsilon}^{2}-\bar{\chi}\varepsilon_u\bar{\varepsilon}+\frac{1}{2}\bar{\chi}^2\varepsilon_u^2,\\
	&\\
	f_{eq}				&	=\bar{\varepsilon}- \bar{\chi}\varepsilon_u.
	\end{split}
\label{eq:ch2_hardforceenergy}
\end{equation}

For the soft device case, we apply a constant force at the free ends of the chain (see Figure~\ref{fig:ch2_modelloesempio}$_{a}$). We introduce the Gibbs energy  
\begin{equation}
g=\varphi-f\left(\boldsymbol{\varepsilon}\cdot\boldsymbol{1}\right),
\label{eq:ch2_g}
\end{equation}
and the variational problem is now 
\begin{equation}
\min_{\boldsymbol{\varepsilon}} g.
\end{equation}
Equilibrium requires 
\begin{equation}
\frac{\partial g}{\partial \boldsymbol{\varepsilon}}=\boldsymbol{\varepsilon}-\varepsilon_u \boldsymbol{\chi}-f \boldsymbol{1}= \boldsymbol{0}\rightarrow \boldsymbol{\varepsilon}_{eq}=f\boldsymbol{1}+\varepsilon_u\boldsymbol{\chi},
\end{equation}
that gives the equilibrium solution already obtained in~\eqref{eq:ch2_eqstrains}, as expected. In this case the force is fixed, and we search for the conjugate equilibrium average strain of the chain, obtained from~\eqref{eq:ch2_eqstrains} and~\eqref{eq:ch2_avgstrain} and the equilibrium energy derived by substituting~\eqref{eq:ch2_avgstrain} in~\eqref{eq:ch2_g}:
\begin{equation}
	\begin{split}
	\frac{g_{eq}}{n}			&=-\frac{1}{2}f^{2}-\bar{\chi}\varepsilon_uf,\\
	&\\
	\bar{\varepsilon}_{eq}	&=f+\bar{\chi}\varepsilon_u.
	\end{split}
\label{eq:ch2_softforceenergy}
\end{equation}
%

%
\begin{figure}[t!]
\centering
  \includegraphics[width=0.95\textwidth]{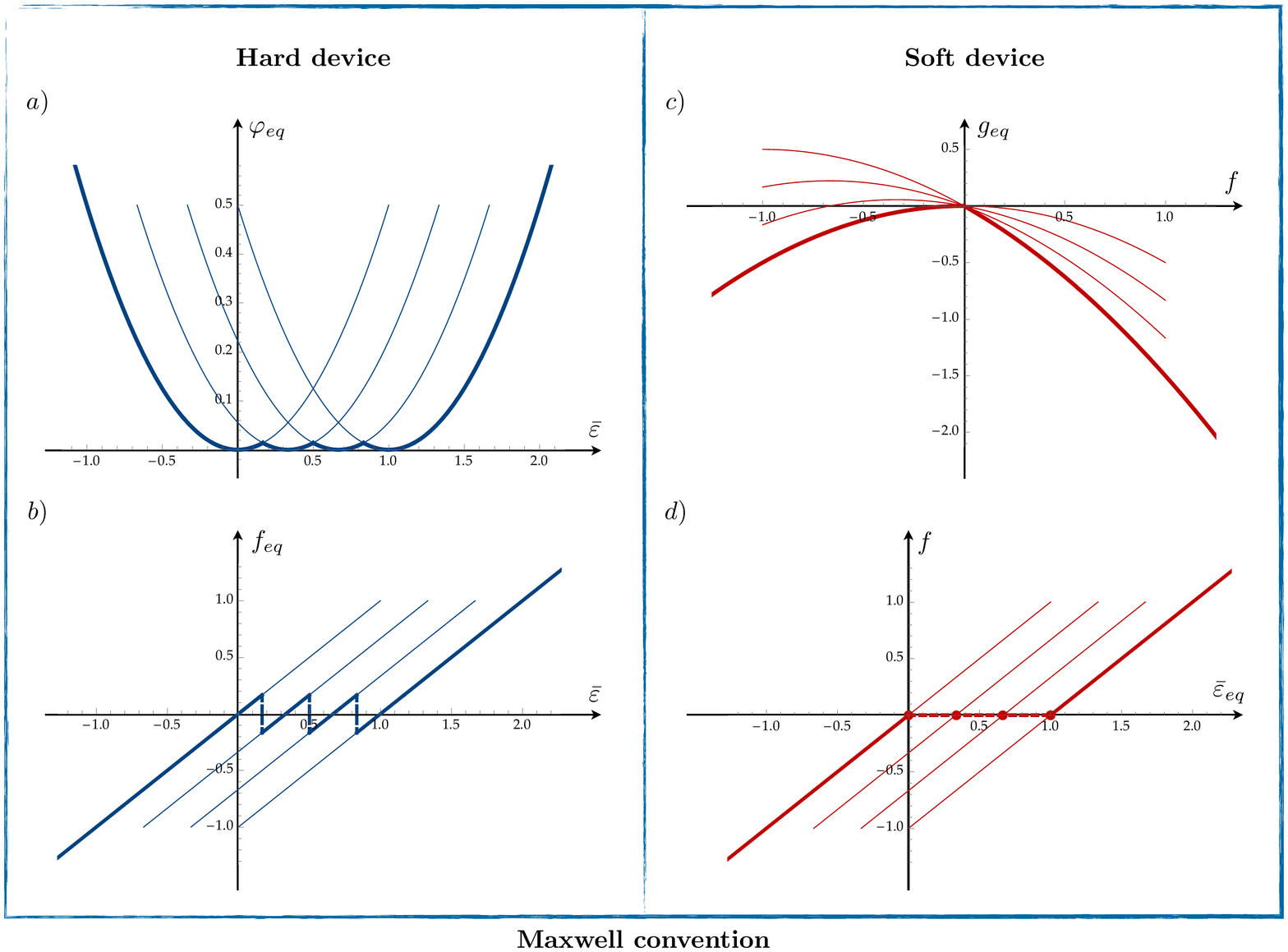}
  \caption[Hard and Soft Device in the Maxwell convention]{The figure shows the two different cases of hard (left) and soft (right) hypotheses developed in the section, corresponding to the case of applied displacement (or total strain) and force, respectively, in the Maxwell hypothesis. In panels a) and b) the energy and the force of the hard device are shown whereas in panel c) and d) there are the energy and the force in the soft device case. In all four graphs, the bold lines represent the global minima of the system whereas the thin lines are the local minima. Parameters: $\varepsilon_u=1$, $\bar{\chi}=p/n$, with $p=1,2,3$ and $n=3$.}
  \label{fig:ch2_hardsoftesempio}
\end{figure}
%

The two-phases equilibrium branches, identified by the average phase fraction $\bar{\chi}\in\,\,]0,1[$, are defined in a specific range of strains representing the relative stability of such solutions. In particular, by observing the expressions for the equilibrium force and displacement in~\eqref{eq:ch2_softforceenergy} and~\eqref{eq:ch2_hardforceenergy} it is clear that the branches exists only for $|f|\le\varepsilon_u$.  On the other hand, the fully attached configuration (the first equilibrium branch) and the fully detached phase ($n$-th equilibrium branch) exist in a larger --one way unbounded-- domain. Correspondingly in terms of strain, we have the following existence domains: 
\begin{equation}
\frac{\bar{\varepsilon}}{\varepsilon_u}\in
	\begin{cases}\vspace{0.2cm}
	\displaystyle\left(-\infty,1\right) & \text{if} \quad \bar{\chi}=0,\\\vspace{0.2cm}
	\displaystyle\left(\bar{\chi}-1,\bar{\chi}+1,\right) & \text{if} \quad \bar{\chi}\in\,\,]0,1[,\\
	\displaystyle\left(0,+\infty\right) & \text{if} \quad \bar{\chi}=1.
	\end{cases}
\end{equation}
Observe that in these ranges the equilibrium solutions are \textit{metastable}, \textit{i.e.}, due to the local convexity of the energy wells, they correspond to local energy minima both in the case of applied force and displacement.   
 
Conversely, when investigating the \textit{global stability}, the system `jumps' from a locally stable configuration to the next one at the moment its energy becomes greater than the one of the other solutions. In this case, by minimizing the equilibrium energy with respect to the internal variable $\bar \chi$ it is easy to verify that the transition strain corresponds to the strain attained at the intersection of two successive parabolas corresponding to a generic $p$ and $p\pm1$ configurations. Thus, in the case of a hard device one obtains the strain domain 
\begin{equation}
\frac{\bar{\varepsilon}}{\varepsilon_u}\in
	\begin{cases}\vspace{0.3cm}
	\displaystyle\left(-\infty,\frac{1}{2n}\right) & \text{if} \quad \bar{\chi}=0,\\\vspace{0.3cm}
	\displaystyle\left(\bar{\chi}-\frac{1}{2n},\bar{\chi}+\frac{1}{2n},\right) & \text{if} \quad \bar{\chi}\in\,\,]0,1[,\\
	\displaystyle\left(1-\frac{1}{2n},+\infty\right) & \text{if} \quad \bar{\chi}=1,
	\end{cases}
\end{equation}
whereas in the case of soft devices we get 
\begin{equation}
	f\in
	\begin{cases}\vspace{0.3cm}
	\displaystyle\left(-\infty,0\right) & \text{if} \quad \bar{\chi}=0,\\
		\displaystyle\left(0,+\infty\right) & \text{if} \quad \bar{\chi}=1. 
	\end{cases}
\end{equation}
%

%
\begin{figure}[t!]
\centering
  \includegraphics[width=0.95\textwidth]{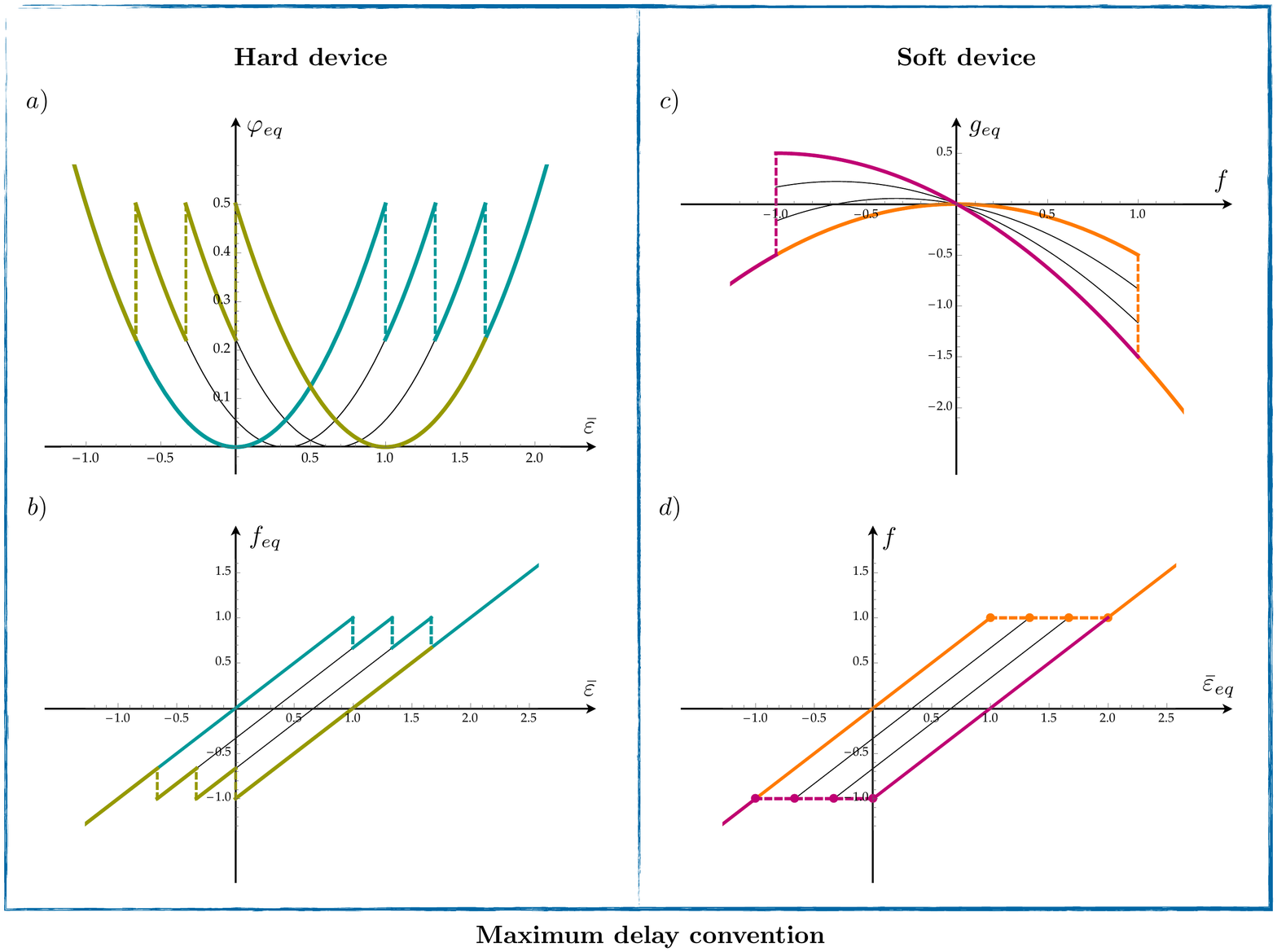}
  \caption[Hard and Soft Device in the maximum delay convention]{The figure shows the two different cases of hard (left) and soft (right) hypotheses developed in the section, corresponding to the case of applied displacement (or total strain) and force, respectively, in the maximum delay convention. In panels a) and b) the energy and the force of the hard device are shown whereas in panel c) and d) there are the energy and the force in the soft device case. In all the four graphs the bold lines represent the path attained when the maximum delay strategy is followed. Parameters: $\varepsilon_u=1$, $\bar{\chi}=p/n$, with $p=1,2,3$ and $n=3$.}
  \label{fig:ch2_maximumdealy}
\end{figure}
%

According to previous results, we may consider two different transition strategies of the system. In the \textit{Maxwell convention}, the systems always stay in the global minima of the energy and the resulting behaviour is shown in Figure~\ref{fig:ch2_hardsoftesempio}. To clarify this strategy, let us first observe the energy of the hard device in Figure~\ref{fig:ch2_hardsoftesempio}$_a$. When the strain has been applied the system follows the equilibrium branch of the fully folded configuration. Then, even though this phase is locally stable beyond the intersection point with the second branch, which represents the configuration with one of the three units switched to the unfolded lines, the system change phase. This process is repeated for each phase fraction, thus following the path representing the global minima pictured with thick lines in the figure. In this case, the elements change phase one at a time, following a sawtooth equilibrium path in terms of force-displacement behaviour as shown in Figure~\ref{fig:ch2_hardsoftesempio}$_b$. On the other hand, when the system is loaded as a soft device, we observe the energy evolution represented in Figure~\ref{fig:ch2_hardsoftesempio}$_c$ where, from the fully folded equilibrium branch the system switches with a cooperative transition to the fully unfolded phase, corresponding to the lower energy branch. In terms of displacement conjugate to the assigned force we observe a force plateau that links the two phases of $\bar{\chi}=0$ and $\bar{\chi}=1$ (see Figure~\ref{fig:ch2_hardsoftesempio}$_d$). Observe that due to the absence of non-local interactions, that are introduced in Chapter~\ref{ch_4}, the analysis cannot distinguish the specific spring that changes phase. 

A different behaviour is attained in the so-called \textit{maximum delay convention}, when the system is allowed to stay in local minima until it is locally stable, before switching to another equilibrium configuration. In this hypothesis, when the displacement is fixed, we observe in Figure~\ref{fig:ch2_maximumdealy}$_a$ how the first equilibrium configuration is fully explored (thick line) and only when the local solution doesn't exist anymore the system switches to the second locally stable configuration in which one of the units has changed phase. The detailed proof that the system can reach this new branch with continuously descending energy can be found in (\cite{puglisi:2000}). The system then follows this new locally stable configuration for increasing overall strain until a second element switches to the unfolded state. In terms of force-displacement relation, the sawtooth path is always observed, but for different values of strain and forces as represented in Figure~\ref{fig:ch2_maximumdealy}$_b$. In the soft device regime, the transition is again cooperative (see Figure~\ref{fig:ch2_maximumdealy}$_c$) but for different values of the force, it is attained when all the fully folded configuration becomes unstable and only the fully unfolded state is locally stable (lower energy), as shown in Figure~\ref{fig:ch2_maximumdealy}$_d$. It is worth highlighting that both the Maxwell convention and the maximum delay hypothesis may be reversible but only in the second case, it is possible to observe hysteresis effects, as shown in Figure~\ref{fig:ch2_maximumdealy} where the paths in the two ways, \textit{i.e} from the fully folded to the fully unfolded and vice-versa, are represented with different colours. Indeed, intermediate hysteresis cycles can be attained, as shown in  (\cite{puglisi:2000,puglisi:2005}) when a limited ability to overcome energy barriers is assigned.

\subsection{Thermal effects}
\label{sec:ch2_pt3}

\begin{figure}[t!]
\centering
  \includegraphics[width=0.95\textwidth]{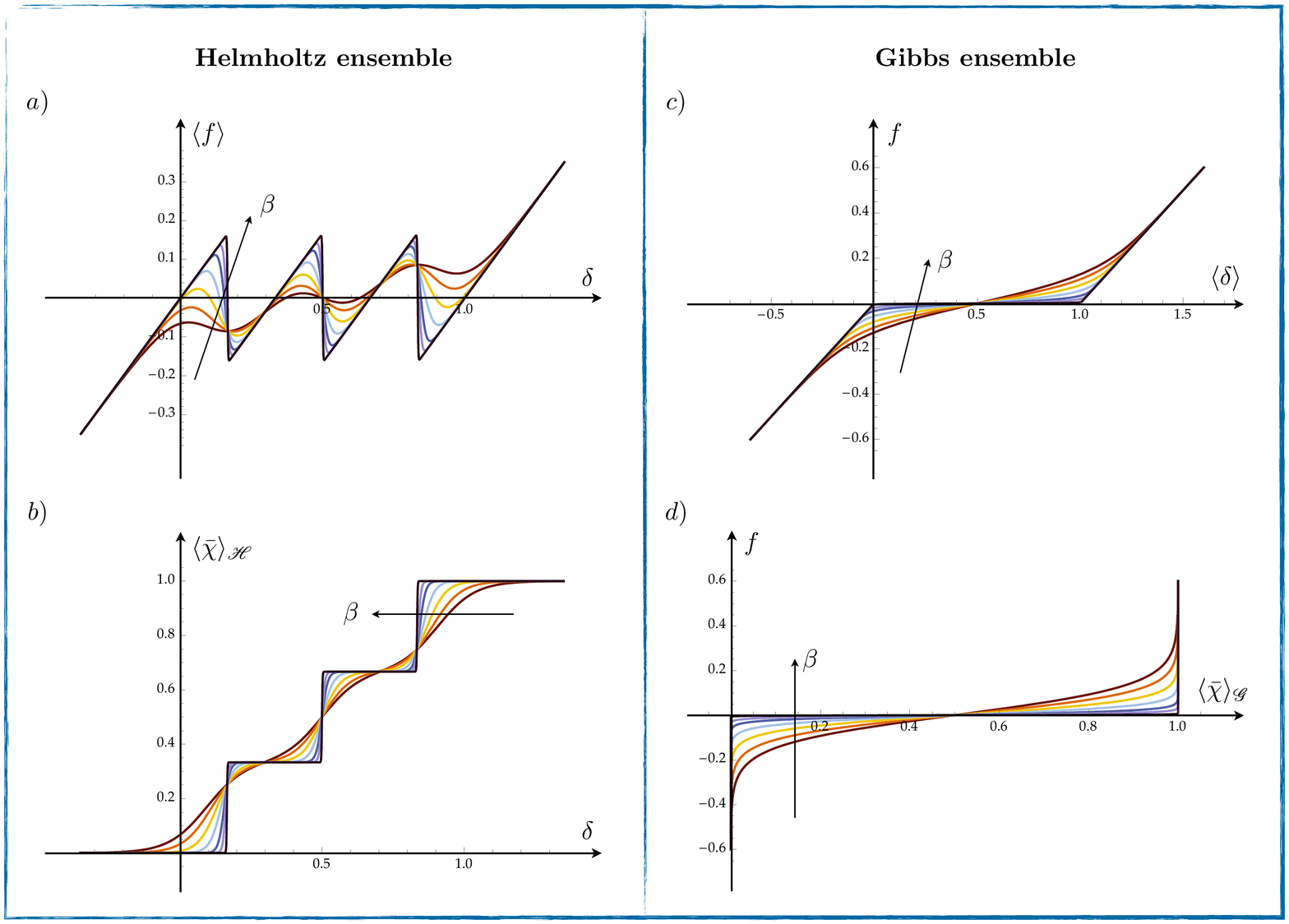}
  \caption[Temperature effects for the Helmholtz and Gibbs ensembles]{Temperature effects in the cases of applied displacement (Helmholtz) and applied force (Gibbs). In panels a) and b) the force-displacement curve and the expectation value of the phase fraction are represented for different values of temperature, respectively, in the case of fixed displacement. Similarly, in panel c) and d) the two graphs are shown in the case of the Gibbs ensemble. Parameters: $\varepsilon_u=1$, $\beta=15, 20, 30, 50, 100, 200, 1000$ and $n=3$. }
  \label{fig:ch2_temperature}
\end{figure}

We use statistical mechanics to introduce temperature effects in our prototypical example. First, we consider the case of applied force $f$, defined in \eqref{eq:ch2_force}, \textit{i.e.} the soft device hypothesis. In this situation, the Gibbs ensemble correctly describes the system with applied force. Thus, by definition, the Gibbs partition function reads 
\begin{equation}
\mathcal{Z}_{\mathscr{G}}(f)=\sum_{\boldsymbol{\chi}}\int_{\mathbb{R}^{3}}e^{\,-\beta g}\text{d}\boldsymbol{\varepsilon}=\sum_{\boldsymbol{\chi}}\int_{\mathbb{R}^{3}}e^{\,-\beta\left(\varphi-f\delta\right)}\text{d}\boldsymbol{\varepsilon},
\end{equation}
where
\begin{equation}
\beta:=\frac{\kappa l}{k_B T},
\end{equation}
correctly rescaled to take into account the adimensionalization of the energy, $\varphi$ is the energy in~\eqref{eq:ch2_matrixenergy} and $f$ is the assigned force. By recalling that the total displacement is $\delta=\boldsymbol{\varepsilon}\cdot\boldsymbol{1}$, and performing a direct Gaussian integration (see Appendix~\ref{appB} and Appendix~\ref{appC} for the detailed calculation), we obtain
\begin{equation}
\mathcal{Z}_{\mathscr{G}}(f)=\left(2\pi\right)^{\frac{3}{2}}\sum_{p=0}^{n}\binom{n}{p}e^{\,\frac{n\beta}{2}f^{2}+\beta\varepsilon_u p f},
\label{eq:ch2_zgtemp}
\end{equation}
where the binomial is required to take into account all the possible configurations attainable due to the phase fraction $\bar{\chi}=p/n$ defined in~\eqref{eq:ch2_chi}. In this particular case the sum can be expressed explicitly, giving the canonical partition function in the Gibbs ensemble:
\begin{equation}
\mathcal{Z}_{\mathscr{G}}(f)=\left(2\pi\right)^{\frac{3}{2}}\left(1+e^{\,\beta\varepsilon_u f}\right)^{n}e^{\,\frac{n\beta}{2}f^{2}}.
\label{eq:ch2_zg}
\end{equation}
Therefore, the Gibbs free energy reads 
\begin{equation}
\mathcal{G}=-k_B T \,\text{ln}\, \mathcal{Z}_{\mathscr{G}},
\label{eq:ch2_fe}
\end{equation} 
and the expectation value of the displacement conjugated to the assigned force, using~\eqref{eq:ch2_dedef}, is 
\begin{equation}
\langle \delta \rangle=f+\varepsilon_u\langle\bar{\chi}\rangle_{\mathscr{G}},
\label{eq:ch2_dg}
\end{equation}
with 
\begin{equation}
\langle\bar{\chi}\rangle_{\mathscr{G}}=\frac{e^{\frac{\beta\varepsilon_u}{n}f}}{1+e^{\frac{\beta\varepsilon_u}{n}f}}
\end{equation}
the expectation value of the unfolded fraction in the Gibbs ensemble. It is worth noticing that we recognize the same expression of the displacement in~\eqref{eq:ch2_dg} obtained for the case with $T=0$, in equation~\eqref{eq:ch2_softforceenergy}. These two expressions differ only for the expectation value of the unfolded fraction $\langle\bar{\chi}\rangle$, which defines the dependence on temperature and regulates the transition phenomena, that is `cooperative'. The effect of temperature leads to a hardening phenomenon resulting in a steeper force-strain diagram as temperature increases (see Figure~\ref{fig:ch2_temperature}$_c$). Similarly, we observe that the transition probability is different from zero before the transition Maxwell stress $f=0$ is attained as the expectation value of the phase fraction in Figure~\ref{fig:ch2_temperature}$_d$ shows.  

Now we consider the case of assigned displacement $\delta$ given by~\eqref{eq:ch2_displacementnd}, \textit{i.e.} the hard device hypothesis, when the Helmholtz ensemble must be considered. The partition function in this ensemble can be obtained from the Gibbs one through an inverse Laplace transform. In particular, by performing the change of variable $f\to i\omega$, it is possible to compute a Fourier transform in the complex plane instead of the Laplace one. Thus, we may write
\begin{equation}
\mathcal{Z}_{\mathscr{H}}(\delta)=\int_{-\infty}^{+\infty}\mathcal{Z}_{\mathscr{G}}(f)\,e^{\,-\beta i \omega\delta}\text{d}\omega,
\end{equation}
where $\mathcal{Z}_{\mathscr{G}}(f)$ is given by~\eqref{eq:ch2_zgtemp}. After calculation, we obtain the canonical partition function for the Helmholtz ensemble (see Appendix~\ref{appC} for all the detailed calculations) that reads 
\begin{equation}
\mathcal{Z}_{\mathscr{H}}(\delta)=\frac{2\pi^2}{n\beta}\sum_{p=0}^{n=3}\binom{n}{p}\,e^{\,-\frac{\beta n}{2}\left(\delta-\varepsilon_u\frac{p}{n}\right)^2}.
\label{eq:ch2_hz}
\end{equation}

Again, the binomial takes into account all the possible attainable configurations at varying phase fractions, \textit{i.e.} all the possible arrangements of unfolded elements with $p=0, 1, 2, 3$. It is important to point out that, as we will discuss in the following chapters, the fundamental assumption to obtain an analytical expression of the partition function as in \eqref{eq:ch2_hz}, is that the two energy functions of the wells extend over all the strains domain. This assumption is acceptable because all the configurations with higher energy are exponentially suppressed so that their weight is negligible. Under this hypothesis, by using~\eqref{eq:ch2_hz} and the definition in~\eqref{eq:ch2_freeen}, we may compute the free energy 
\begin{equation}
\mathcal{F}=-k_B T \,\text{ln}\, \mathcal{Z}_{\mathscr{H}},
\label{eq:ch2_feh}
\end{equation} 
and, consequently, using~\eqref{eq:ch2_fedef} we compute the expectation value of the force 
\begin{equation}
\langle f \rangle=\delta-\varepsilon_u\langle\bar{\chi}\rangle_{\mathscr{H}},
\label{eq:ch2_fh}
\end{equation}
where 
\begin{equation}
\langle\bar{\chi}\rangle_{\mathscr{H}}=\frac{\displaystyle\sum _{p=0}^{n=3} \binom{n}{p} \,\frac{p}{n}\,e^{\,-\frac{\beta n}{2}\left(\delta-\varepsilon_u\frac{p}{n}\right)^2}}{\displaystyle\sum _{p=0}^{n=3} \binom{n}{p} \,e^{\,-\frac{\beta n}{2}\left(\delta-\varepsilon_u\frac{p}{n}\right)^2}}
\end{equation}
is the expectation value of the fraction of unfolded domains. As one may observe, as in the Gibbs ensemble case the expression of the force in~\eqref{eq:ch2_fh} is formally the same as the force in the zero temperature case derived in~\eqref{eq:ch2_hardforceenergy}, except for the expectation value of the phase fraction, which depends directly on temperature and regulates the transition among the different microscopic configurations. Indeed, as shown in Figure~\ref{fig:ch2_temperature}$_a$, the sawtooth path of the force attained in the zero temperature limit is modified for decreasing values of  $\beta$, \textit{i.e.} when temperature increases. In particular, as temperature increases, force peaks are smoothed. Loosely speaking, this effect is due to an entropic energy effect such that the probability of phase transition from the homogeneous phase configuration is non zero before they lose their global stability. On the other hand, a hardening phenomenon due to temperature is also observed, which is an effect typically encountered in experiments that we will compare with our model in Chapter~\ref{ch_3}. These features are well understood also by the analysis of the expectation value of the unfolded fraction, reported in Figure~\ref{fig:ch2_temperature}$_b$, that, for varying beta, defines the typical  `non-cooperativity' of the transition behaviour under isomeric (fixed displacement) conditions. 

%
%


\subsection{Energy Barriers}
\label{sec:ch2_pt4}

The aim of this section is to illustrate the properties of the energy landscape of the bi-parabolic approximation introduced in this thesis to model biological and bio-inspired materials and in particular here I use the prototypical example to describe the non-equilibrium paths among different metastable configurations governed by the possibility of overcoming energy barriers. Consider again the `pure' mechanical case and the analysis derived in Section~\ref{sec:ch2_pt3}, where temperature effects are neglected, \textit{i.e.} $T=0$. We have observed that the transition between the fully folded state to the fully unfolded state depends on both the boundary conditions (hard or soft device) and the evolution strategy (Maxwell or maximum delay convention). Thus, the different metastable configurations represented by the different equilibrium branches can be explored in a range that depends on the local stability of the system. In particular, following (\cite{puglisi:2002}) where a similar analysis has been performed for a trilinear energy approximation, we deduced that an equilibrium configuration is a minimum of the potential energy if and only if all the bi-stable units are centred within the convex wells of the energy (see Figure~\ref{fig:ch2_modelloesempio} and Equation~\eqref{eq:ch2_bip_en}), and none of them is the spinodal (intersection of the parabolas) region. Indeed, when the bi-parabolic approximation is considered the saddle point that represents a metastable unstable configuration that has to be crossed by a unit in order to change the phase does not exist so that there are only two possibilities in terms of phases instead of three: the folded, here referred as phase $1$ and the unfolded one (phase $2$). Consequently, the concave region in the trilinear approximation, (see Figure~\ref{fig:ch2_kramers} as an example), is substituted by the spinodal point that is the intersection among the two wells and represents the analogous of the saddle point with the concave region reduced to a cusp. Moreover, according to the Mountain pass theorem, the energy difference between the first well and the spinodal point is the energy barrier to overcome in order to switch from phase $1$ to phase $2$ as well as the barrier necessary to attain the inverse transition is represented by the difference between the spinodal point and the second well. This means that in the simple bi-parabolic approximation the passage among different equilibrium branches is `instantaneous' and is characterized by a sudden jump from a metastable configuration to the other, hence all the equilibrium branches are characterized only by the fraction of folded $n-p$ and unfolded $p$ elements. It is worth remarking that when non-local terms are introduced, the generations and propagation of phase interfaces regulate the phenomenon and in this case, the fraction of elements alone is not sufficient to describe the transition (see Chapter~\ref{ch_4} for details). This fundamental property opens the possibility of studying different scenarios of quasi-static evolution of the system ranging among different metastable configurations and that depends on the ability of overcoming energy barriers separating the different states. 

To clarify these features, here I study the non-equilibrium paths in the soft device hypothesis for $n=2$ and in the hard device case for $n=3$, applying the analysis to the prototypical example proposed in this section. In this way it is possible to visualize the transition pathways in the energy landscape and in the configuration space identifying both local minima configuration ($m$) and the spinodal --transient-- points, that, for the sake of simplicity, will be indicated with the letter $s$. Finally, following (\cite{puglisi:2002}), it will be demonstrated that for the generic case of a chain made by $n$ units and with the specific case of bi-parabolic energy the \textit{minimum energy barrier path} is the one across the spinodal points.

\vspace{0.5cm}
\noindent\textsc{Soft device for n=2}\vspace{0.2cm}

Let us begin with the case with only two bi-stable units, \textit{i.e.} $n=2$ when the force $f$ is applied. Following (\cite{puglisi:2002}), the total energy of the system can be then rewritten in terms of average strain $\bar{\varepsilon}$ given by~\eqref{eq:ch2_avgstrain} and $\zeta$, the order parameter that measures the deviation from the equilibrium symmetric configuration such that
\begin{equation}
\varepsilon_1=\bar{\varepsilon}-\zeta, \qquad \varepsilon_2=\bar{\varepsilon}+\zeta. 
\end{equation}
Thus, by introducing the vector 
\begin{equation}	\qquad
	\boldsymbol{\varepsilon}_{\zeta}=
	\begin{pmatrix}
	\bar{\varepsilon}-\zeta	\\
	\bar{\varepsilon}+\zeta 	\\
	\end{pmatrix}
\end{equation}
the Gibbs energy in~\eqref{eq:ch2_g} can be rewritten as 
\begin{equation}
\frac{g(\boldsymbol{\varepsilon}_{\zeta})}{n}=\frac{\varphi(\boldsymbol{\varepsilon}_{\zeta})}{n}-f\,\bar{\varepsilon}.
\label{eq:ch2_enbarrier_1}
\end{equation}

It is worth highlighting that also the energy in~\eqref{eq:ch2_enbarrier_1} depends on the specific configuration of the system prescribed by the value of $\boldsymbol{\chi}$ so that we may have different situations, depending on the value of the strains of the two units, in which the two springs can be found in the folded ($1$) or in the unfolded ($2$) phase. Specifically, considering the strain at the spinodal point representing the transition between the phases that reads 
\begin{equation}
\varepsilon_{s}=\frac{\varepsilon_u}{2},
\label{eq:ch2_strainspinodal}
\end{equation}
we may express the energy as
\begin{equation}
\frac{g(\boldsymbol{\varepsilon}_{\zeta})}{n}=
		\begin{cases}
		\varepsilon_1 < \varepsilon_{s}\,\,(\chi_1=0)\quad \&\quad\varepsilon_2 < \varepsilon_{s}\,\,(\chi_2=0)  \quad \rightarrow \quad g_{1,1}\\
		\varepsilon_1 < \varepsilon_{s} \,\,(\chi_1=0)\quad\&\quad \varepsilon_2 > \varepsilon_{s}\,\,(\chi_2=1)  \quad \rightarrow \quad g_{1,2}\\
		\varepsilon_1 > \varepsilon_{s}\,\,(\chi_1=1)\quad \&\quad\varepsilon_2 < \varepsilon_{s}\,\,(\chi_2=0)  \quad \rightarrow \quad g_{2,1}\\
		\varepsilon_1 > \varepsilon_{s} \,\,(\chi_1=1)\quad\&\quad \varepsilon_2 > \varepsilon_{s}\,\,(\chi_2=1)  \quad \rightarrow \quad g_{2,2}\\
		\end{cases},
		\label{eq:ch2_softenergycases}
\end{equation}
%
where $g_{i,j}$ with $i,j=1,2$ we refer to the phases folded ($1$) and unfolded ($2$) of the first and the second spring, respectively. 

%
\begin{figure}[t!]
\centering
  \includegraphics[width=0.95\textwidth]{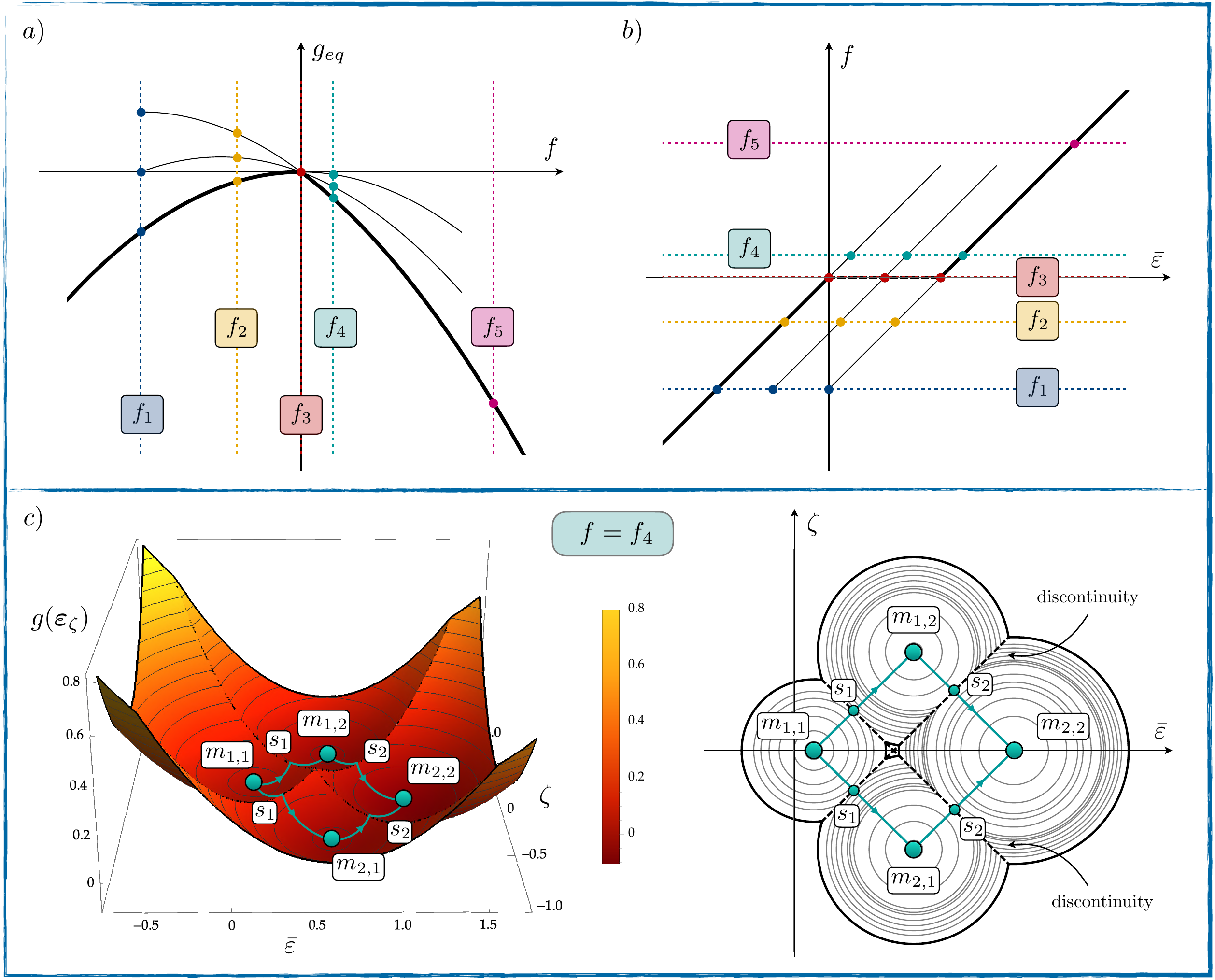}
  \caption[Energy landscape for $n=2$ in soft device]{Energy landscape and configuration space for the case of soft device with $n=2$. In panel a) the gibs energy is represented and different value of the force are assigned highlighting the local minima. In panel b) the corresponding force is shown. In panel c) the the energy landscape in the configuration space $(\bar{\varepsilon},\zeta)$ is represented and the local minima are shown with dots. The minimum energy path is represented with oriented lines both in the energy landscape (left) and in the configurational space (right). Here $f_1=-1$, $f_2=-0.4$, $f_3=0$, $f_4=0.2$, $f_5=1.2$ and $\varepsilon_u=1$.}
  \label{fig:ch2_landscape_1}
\end{figure}
%

The configurational space identified by the average strain and the order parameter $\zeta$ is two dimensional and hence it is possible to directly visualize the energy landscape and the minimum energy barrier paths. On this matter, it is important to observe that not always the system exhibits multiple local configurations depending on whether or not the assigned force falls into the interval of local stability $|f|\le \varepsilon_u$ (see $f_5$ in Figure~\ref{fig:ch2_landscape_1}$_a$). To explain this features, consider the value of the force $f=f_4>0$, represented on the Gibbs energy and on the average strain diagram in Figures~\ref{fig:ch2_landscape_1}$_{a,b}$. Here the global minimum is attained for the equilibrium branch corresponding to the configuration in which both the bi-stable units are in the unfolded configuration, $m_{2,2}$ (lower green dots on the energy). There exist other two metastable equilibrium branches (see Figure~\ref{fig:ch2_landscape_1}$_{a}$, dots on the thin lines) corresponding to two local minima solutions from which, starting from the position $m_{1,1}$ the minimum can be reached following the minimum energy barrier path. To highlight the physical meaning of these paths, let us suppose that we are moving from $m_{1,1}$ to $m_{1,2}$ following the straight line 
\begin{equation}
\zeta=\bar{\varepsilon}-f,
\end{equation}
along which $\varepsilon_1$ is equilibrated because $\bar{\varepsilon}-\zeta=f\rightarrow\varepsilon_1=f$ that corresponds to the equilibrium equation in~\eqref{eq:ch2_softforceenergy} for $\chi_1=0$. Simultaneously, the other unit is not in equilibrium along this line and it switches from phase $1$ to phase $2$. The transition happens when the spinodal point $s_1$ is reached and the discontinuity line where it is located is crossed, as shown in Figure~\ref{fig:ch2_landscape_1}$_{c}$. Notice that $s_1$ is also the point on the discontinuity line with the minimum energy and corresponds to the minimum barrier separating the metastable states. Then, the remaining path connecting the spinodal point to the metastable configuration $m_{1,2}$ is the path of steepest descent having the maximum rate of dissipation. To this matter, let us recall that when non-local interactions are considered the system only recognizes the number of elements in a certain phase and not the position. Thus the two configurations $m_{1,2}$  and $m_{2,1}$ are energetically equivalent and symmetric in the configurational space ($\bar{\varepsilon},\zeta$). With the same reasoning, the global minimum $m_{2,2}$ corresponding to the case of both the springs in the unfolded phase can be reached by crossing the second spinodal point $s_2$  along the line 
\begin{equation}
\zeta=-\bar{\varepsilon}+f+\varepsilon_u. 
\end{equation}

In general, for $n=2$ it is possible to deduce all the four minimum energy barrier paths pictured in Figure~\ref{fig:ch2_landscape_1}$_{d}$ that are identified by the straight lines 
\begin{equation}
\zeta=\pm\bar{\varepsilon}\mp\left(f+\varepsilon_u\chi\right),
\end{equation}
with $\chi=0$ for the transition between $m_{1,1}$ and $m_{1,2}$ ($m_{2,1}$) and $\chi=1$ when $m_{2,2}$ is reached. Moreover, the discontinuity lines representing the barriers on which the spinodal point are located can are described by the equations  
\begin{equation}
\zeta=\bar{\varepsilon}-\frac{\varepsilon_u}{2},\qquad \zeta=-\bar{\varepsilon}+\frac{\varepsilon_u}{2}
\label{eq:ch2_intersezione_1}
\end{equation}
Eventually, all the points representing the local minima in the configurational space ($\bar{\varepsilon},\zeta$) can be identified by 
\begin{equation}
\begin{split}
m_{1,1} & =\Big(f,0\Big)\\
m_{1,2} & =\Big(f+\frac{\varepsilon_u}{2},\frac{\varepsilon_u}{2}\Big)\\
m_{2,1} & =\Big(f+\frac{\varepsilon_u}{2},-\frac{\varepsilon_u}{2}\Big)\\
m_{2,2} & =\Big(f+\varepsilon_u,0\Big)
\end{split}.
\end{equation}
%

%
\begin{figure}[]
	\centering
	 \includegraphics[width=0.94 \textwidth]{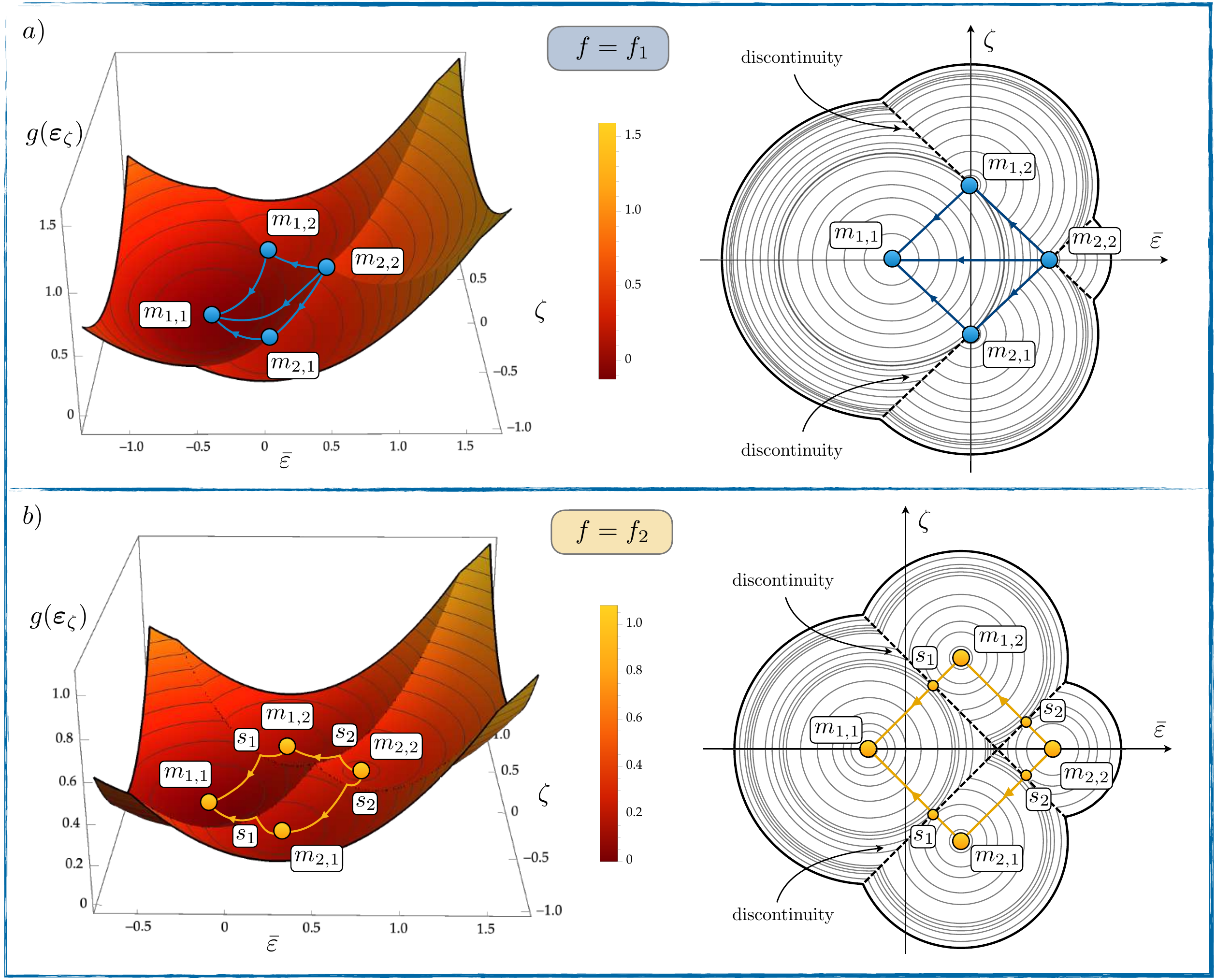}\\\vspace{-0.15cm}
	  \includegraphics[width=0.942 \textwidth]{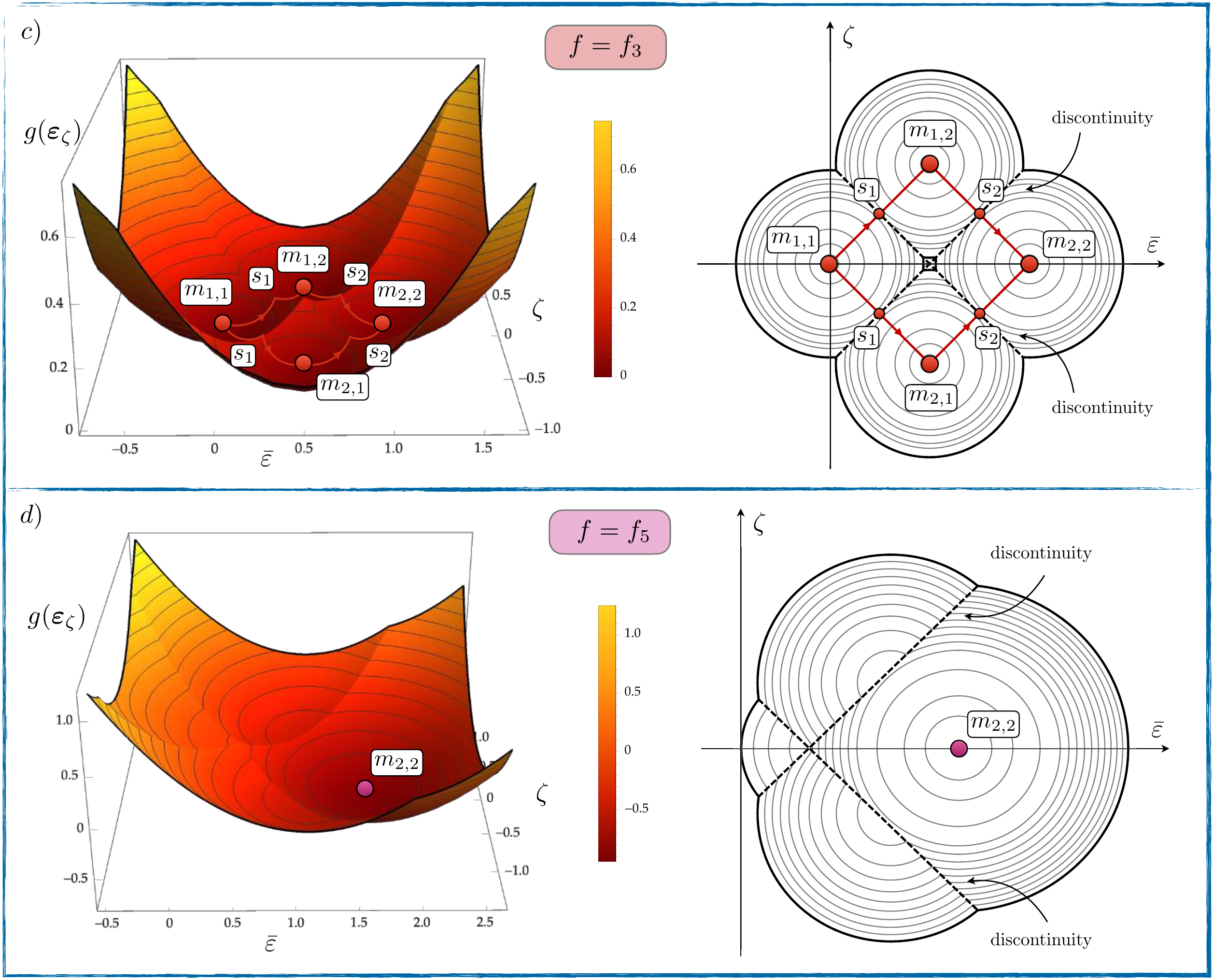}
	  \caption[Minimum energy paths for different values of the force]{Energy landscapes and configurational spaces for different values of the assigned force corresponding to the values in Figure~\ref{fig:ch2_landscape_1}. In particular in panel a) the force is equal to the limit force $f=-\varepsilon_u$ and the path corresponds to the intersection lines, in panel b) $f_3=0$ and the three minima have the same energy and coexist. In panel c) $f_4>0$ and in panel d) $f_5>\varepsilon_u$ so that only the global minimum exists. Here $\varepsilon_u=1$.}
	\label{fig:ch2_landscape_2}
\end{figure}
%

The assigned value of the force may change drastically the behaviour of the system in terms of phase transition equilibrium paths. In Figure~\ref{fig:ch2_landscape_2} different energy landscapes and corresponding configurational spaces are represented for different values of the assigned parameter force $f$ and they have to be analyzed depending on whether or not the transition is folded$\to$unfolded or vice-versa. For example, consider the case in Figure~\ref{fig:ch2_landscape_2}$_a$ where the value of the force corresponds to the limiting value fixed by the constraint $|f|\le\varepsilon_u$, thus $f_1=-\varepsilon_u$. This situation is the representation of the maximum delay convention from the fully unfolded state to the fully folded one (see Figure~\ref{fig:ch2_maximumdealy}). In this case, the global solution is attained at $m_{1,1}$ that is stable whereas both the equivalent metastable configuration in which one spring is changing phase corresponds to the saddle points $s_1$ and are located on the discontinuity line. Similarly, also the configuration $m_{2,2}\equiv s_2$ and consequently neither $m_{1,2}$ ($m_{2,1}$) nor $m_{2,2}$ are stable. Then, when considering the evolution from the minimum with higher energy to the one with the lower one two possible strategies can ideally be followed. The first one leads from $m_{2,2}$ to $m_{1,1}$ passing through $m_{1,2}$ or $m_{2,1}$ whereas the second one goes directly to the global solution. This last path is the one that will be followed because it corresponds to the path of steepest descent with the maximum rate of dissipation. When $f=f_2<0$ (see Figure~\ref{fig:ch2_landscape_2}$_b$) the situation is the opposite of the one describe in the previous paragraph ($f=f_4>0$) but here the global minimum is for both the units in the first folded phase at $m_{1,1}$. Then the path of minimum energy barriers, with the same reasoning of the opposite case goes from the higher energy solution $m_{2,2}$ to the minimum following $m_{2,2}\rightarrow s_{2}\rightarrow m_{1,2}\rightarrow s_{1}\rightarrow m_{1,1}$, or equivalently passing through $m_{2,1}$. Another interesting situation is found when $f_3=0$ (see Figure~\ref{fig:ch2_landscape_2}$_{c}$) in which the three equilibrium branches intersect in a single point. This means that all the local solutions have the same energy and coexist at a single point. Being the equilibrium path equivalent this is also the case reproducing the evolution of the system within the Maxwell convention, where the transition happens simultaneously with the cooperative behaviour of the units reaching the unfolded phase altogether. Finally, when a value of the force such as $f=f_5$ represented in Figure~\ref{fig:ch2_landscape_2}$_d$ is chosen, there exists only one minimum and all the possible paths come back on $m_{2,2}$.

The evolution of the system among three different metastable configurations along the `minimal energy barrier path' in the configurational space allows us to draw some general conclusions. The transition proceeds as a sequence of events with one bi-stable unit changing phase at a time when the other a fixed, and this behaviour is general for a system with $n$ units, as it will be shown in the following section. In some cases, as the one in Figure~\ref{fig:ch2_landscape_2}$_d$, there exists only one configuration and all the possible paths moving away from the minima return exactly in $m_{2,2}$, without allowing the possibility of switching to a different phase. Moreover, we remark again that the only thing that matters is the number of elements in each phase. Also, in the configuration space, the metastable symmetric units can be observed as a variation from the equilibrium condition measured by the order parameter $\zeta$. Indeed, when $\zeta=0$, the horizontal path connecting directly $m_{1,1}$ to $m_2,2$ is the so-called \textit{Cauchy-Born} path and represents the simultaneous cooperative transition of the whole units of the chain, reproducing the typical behaviour of the soft device in the Maxwell convention.

\vspace{0.5cm}
\noindent\textsc{Hard device for n=3}\vspace{0.2cm}

%
\begin{figure}[t!]
\centering
  \includegraphics[width=0.95\textwidth]{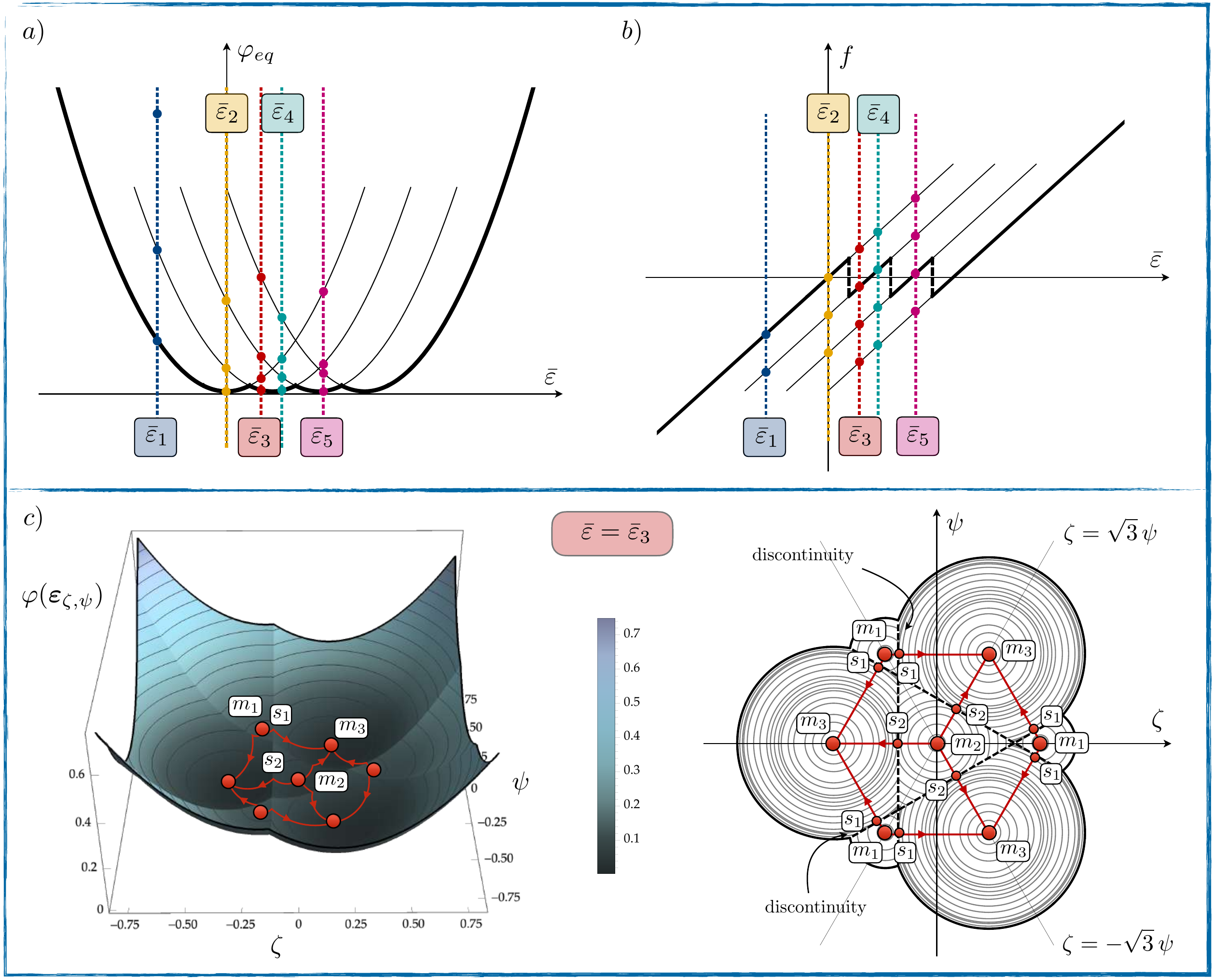}
  \caption[Energy landscape for $n=3$ in hard device]{Energy landscape and configuration space for the case of hard device with $n=3$. In panel a) the energy is represented and different value of the average strain are assigned highlighting the local and global minima. In panel b) the corresponding force is shown. In panel c) the the energy landscape in the configuration space $(\zeta,\psi)$ is represented and the local minima are shown with dots. The minimum energy path is represented with oriented lines both in the energy landscape (left) and in the configurational space (right). Here $\bar{\varepsilon}_1=-0.5$, $\bar{\varepsilon}_2=0$, $\bar{\varepsilon}_3=0.25$, $\bar{\varepsilon}_4=0.4$, $\bar{\varepsilon}_5=0.7$ and $\varepsilon_u=1$.}
  \label{fig:ch2_landscape_hard_1}
\end{figure}
%

\begin{figure}[!]
	\centering
	 \includegraphics[width=0.94 \textwidth]{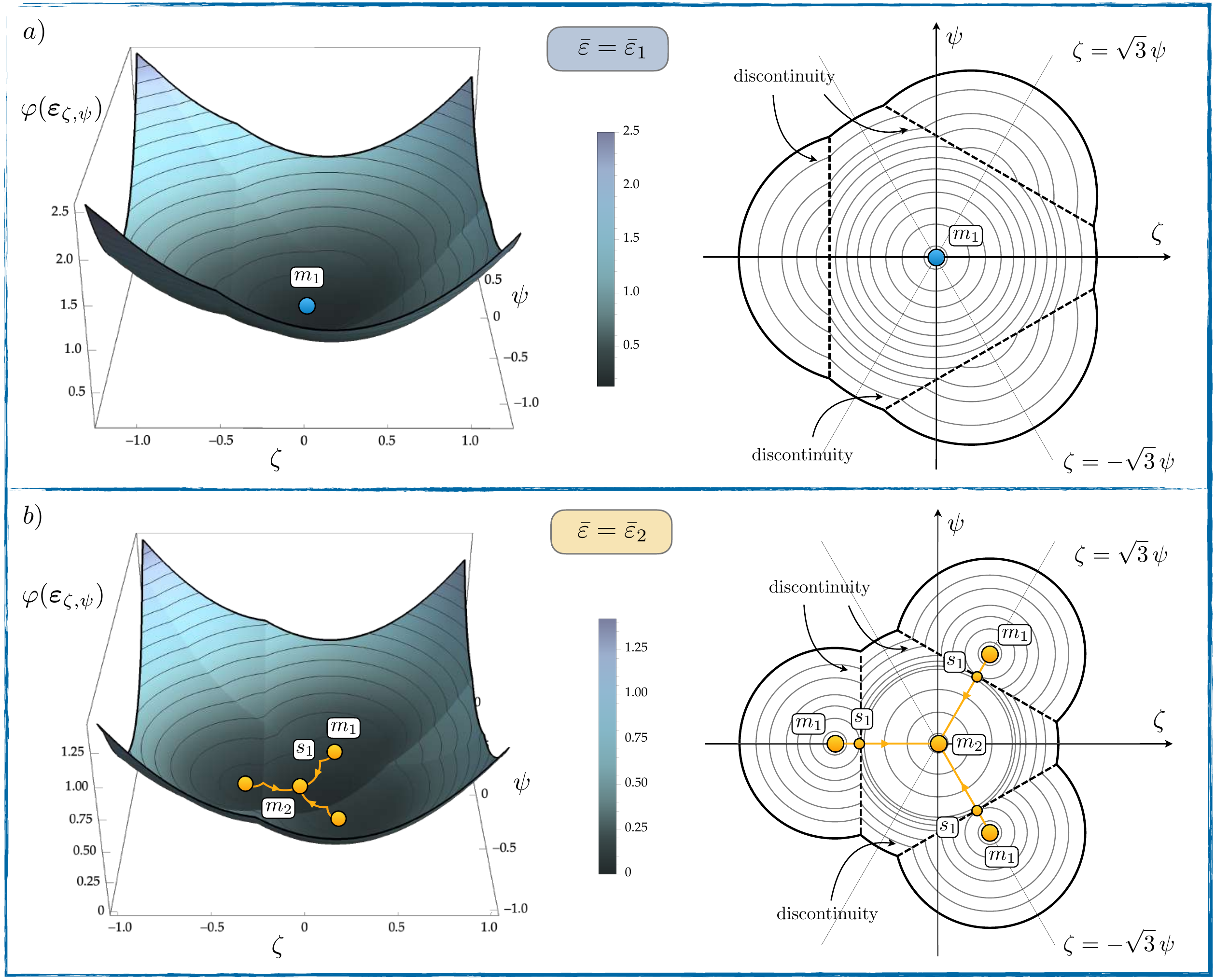}\\\vspace{-0.15cm}
	  \includegraphics[width=0.942 \textwidth]{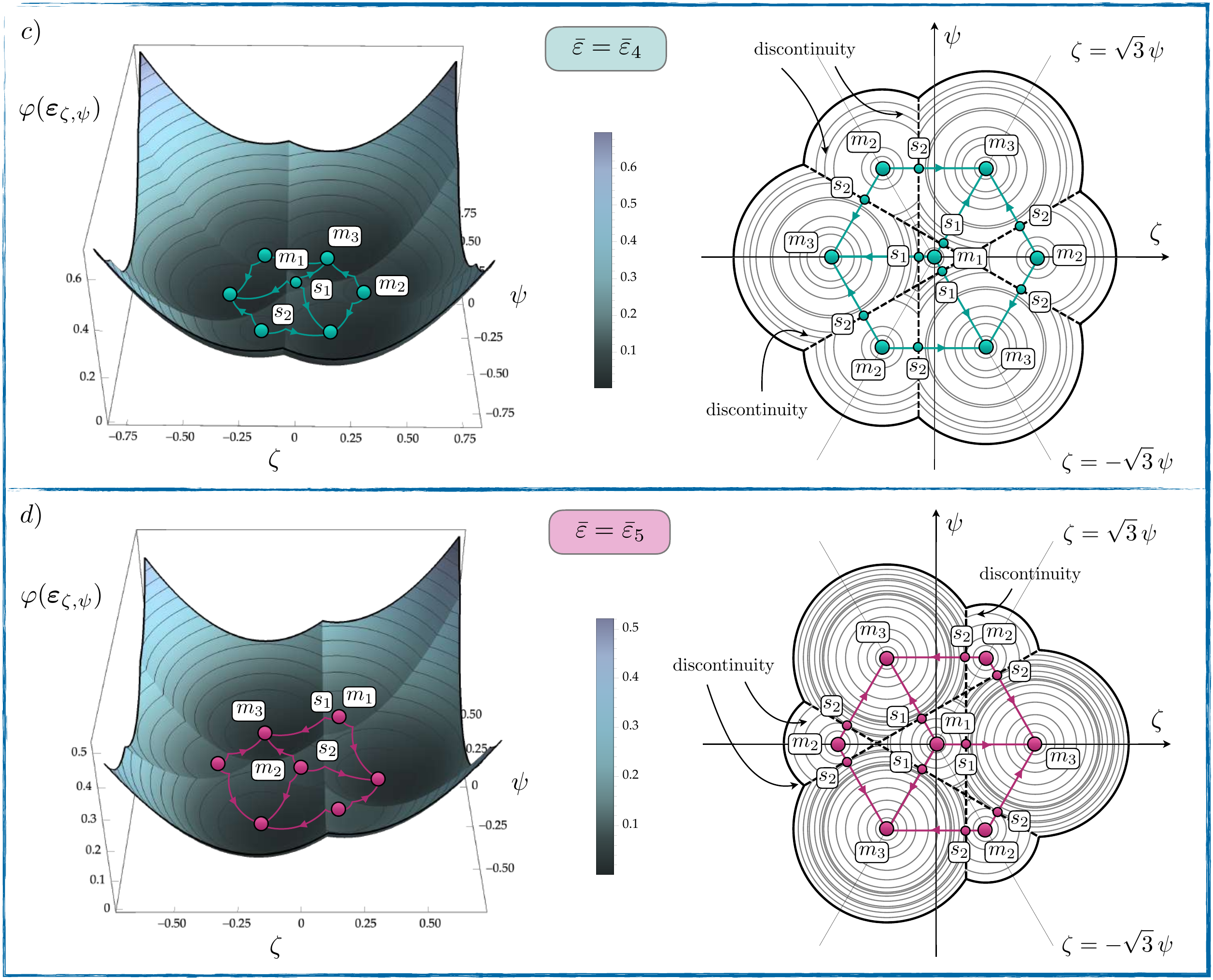}
	  \caption[Minimum energy paths for different values of the strain]{Energy landscapes and configurational spaces for different values of the average strain corresponding to the values in Figure~\ref{fig:ch2_landscape_hard_1}. In particular in panel a) the strain corresponds to the first equilibrium branch and only the global minimum exists. In panel b) $\bar{\varepsilon}_2=0$ and two possibile configuration can be attained. In panel c) $\bar{\varepsilon}_4>0$ representing the possible evolution from the fully folded configuration to the one with two units changing phase. In panel d) $\bar{\varepsilon}_5$ represents the evolution among three metastable configurations. Here $\varepsilon_u=1$.}
	\label{fig:ch2_landscape_hard_2}
\end{figure}

Consider now the prototypical example with $n=3$ and fixed total displacement (or average strain), so that the hard device hypothesis is applied, as shown in Figure~\ref{fig:ch2_landscape_hard_1}$_{a,b}$. Here, due to the size of the system, one more order parameter has to be introduced. Following (\cite{puglisi:2000,puglisi:2005}), to describe the evolution of the system in the hard device we introduce 
\begin{equation}
\varepsilon_1=\bar{\varepsilon}+\zeta+\sqrt{3}\,\psi,\qquad\varepsilon_2=\bar{\varepsilon}+\zeta-\sqrt{3}\,\psi,\qquad\varepsilon_3=\bar{\varepsilon}-2\zeta.
\label{eq:ch2_strains_hd_b}
\end{equation}
In particular, the parameter $\zeta$ describes the case in which two of the three springs have the same strain whereas the other one is changing phase. Consequently, it can be said that $\zeta$ represents the degree of `non Cauchy-Born' behaviour. On the other hand, $\psi$ is introduced in~\eqref{eq:ch2_strains_hd_b} in such a way that the configurational space is divided into six half-planes, one each $60$ degree, and it measures a further reduction of symmetry of the system characterizing the state with all three springs having different strains, as shown in Figure~\ref{fig:ch2_landscape_hard_1}$_c$. By introducing the vector 
\begin{equation}	\qquad
	\boldsymbol{\varepsilon}_{\zeta,\psi}=
	\begin{pmatrix}
	\bar{\varepsilon}+\zeta-\sqrt{3}\,\psi	\\
	\bar{\varepsilon}+\zeta+\sqrt{3}\,\psi	\\
	\bar{\varepsilon}-2\zeta	\\
	\end{pmatrix},
\end{equation}
the energy in~\eqref{eq:ch2_hardforceenergy} reads 
\begin{equation}
\frac{\varphi_{eq}}{n}:=\frac{\varphi(\boldsymbol{\varepsilon}_{\zeta,\psi})}{n}.
\label{eq:ch2_enbarrier_2}
\end{equation}
It is important to remark that also in this case~\eqref{eq:ch2_enbarrier_2} still depends on the specific configuration of the system prescribed by $\boldsymbol{\chi}$ (or, equivalently, $\bar{\chi}$). Thus~\eqref{eq:ch2_enbarrier_2} has to be expressed as the soft device energy in~\eqref{eq:ch2_softenergycases} depending on the different cases and I will not report here the full expression to not be redundant. Moreover, to be consistent with the previous analysis, I should indicate the energy according to the phase of the three units. For instance, if the first two are in phase one and the last in phase two I should use $\varphi_{1,1,2}$ and the associated local minimum is $m_{1,1,2}$ but in this case, this would result in an overload of the notation especially in the Figures~\ref{fig:ch2_landscape_hard_1} and~\ref{fig:ch2_landscape_hard_2} that will result unreadable. Thus, here I use $m_1$, $m_2$ or $m_3$ to indicate the minima from the higher to the lower and similarly for the spinodal points $s_1$ and $s_2$ separating the minima.

%
%

Thus, by varying $(\zeta,\psi)$ it is possible to describe all the equilibrium configurations where symmetries can be founded. In particular, when considering the constraint assigned by fixing the average strain $\bar{\varepsilon}$, the straight horizontal line attained at $\zeta=0$ and $\psi=0$ represents the Cauchy-Born path, thus the two equilibrium branches with all the three units in the same phase corresponding to the first and last straight lines in Figure~\ref{fig:ch2_landscape_hard_1}$_b$. On the other hand, the `rotate' planes with $\psi=0$, $\psi=-\sqrt{3}\zeta$ and $\psi=\sqrt{3}\zeta$ represent the solutions with two units in one phase and the third in the other one. Finally, a generic point of the space with $\zeta\neq0$ and $\psi\neq0$ is a state with all the three units having different phases. 

These paths are represented in Figure~\ref{fig:ch2_landscape_hard_2} for different values of the average strain. For instance, let us consider increasing values of $\bar{\varepsilon}$ and in particular let us start from $\bar{\varepsilon}=\bar{\varepsilon}_1$ (see Figure~\ref{fig:ch2_landscape_hard_2}$_a$). Here there is only one equilibrium configuration that represents the trivial state in which all the units are in the folded phase and there is still no possibility of transition. By increasing the strain we encounter $\bar{\varepsilon}=\bar{\varepsilon}_2$ where the higher minima represent the situation with one unfolded unit attained before the complete refolding of the system (unfolded$\to$folded transition). As shown in Figure~\ref{fig:ch2_landscape_hard_2}$_b$ this path starts at the minimum $m_1$ and crosses the barrier $s_1$ along the lines $\zeta=\pm\sqrt{3}\,\psi$ and $\zeta=0$, that are the solution with only one units in a different phase, finally reaching $m_2$. Indeed, the oriented lines represented in Figure~\ref{fig:ch2_landscape_hard_2} are the lines in which two units are mechanically equilibrated while the other is changing. When $\bar{\varepsilon}=\bar{\varepsilon}_3$, there can be observed two different situations. The path from $m_1$ to $m_3$, is, as in the previous case, the refolding case with two units in the unfolded phase and the other one changing whereas when the strain is increased from the fully folded solution one unit may switch on the unfolded configuration at a higher metastable state $m_2$ and undergo a transition overcoming the spinodal point $s_2$. The situation described by $\bar{\varepsilon}=\bar{\varepsilon}_4$ corresponds to the unfolding process that leads from a fully folded configuration $m_1$ to the case with one unit unfolded across $s_1$ and to $m_3$ as shown in Figure~\ref{fig:ch2_landscape_hard_2}$_c$ Similarly, the last case $\bar{\varepsilon}=\bar{\varepsilon}_5$ (see Figure~\ref{fig:ch2_landscape_hard_2}$_d$) is an intermediate transition that leads to the case with two units still folded to the equilibrium branch representing the situation in which two units up to three have changed phase.

\vspace{0.5cm}
\noindent\textsc{General case}\vspace{0.2cm}

\begin{figure}[!]
	\centering
	 \includegraphics[width=0.94 \textwidth]{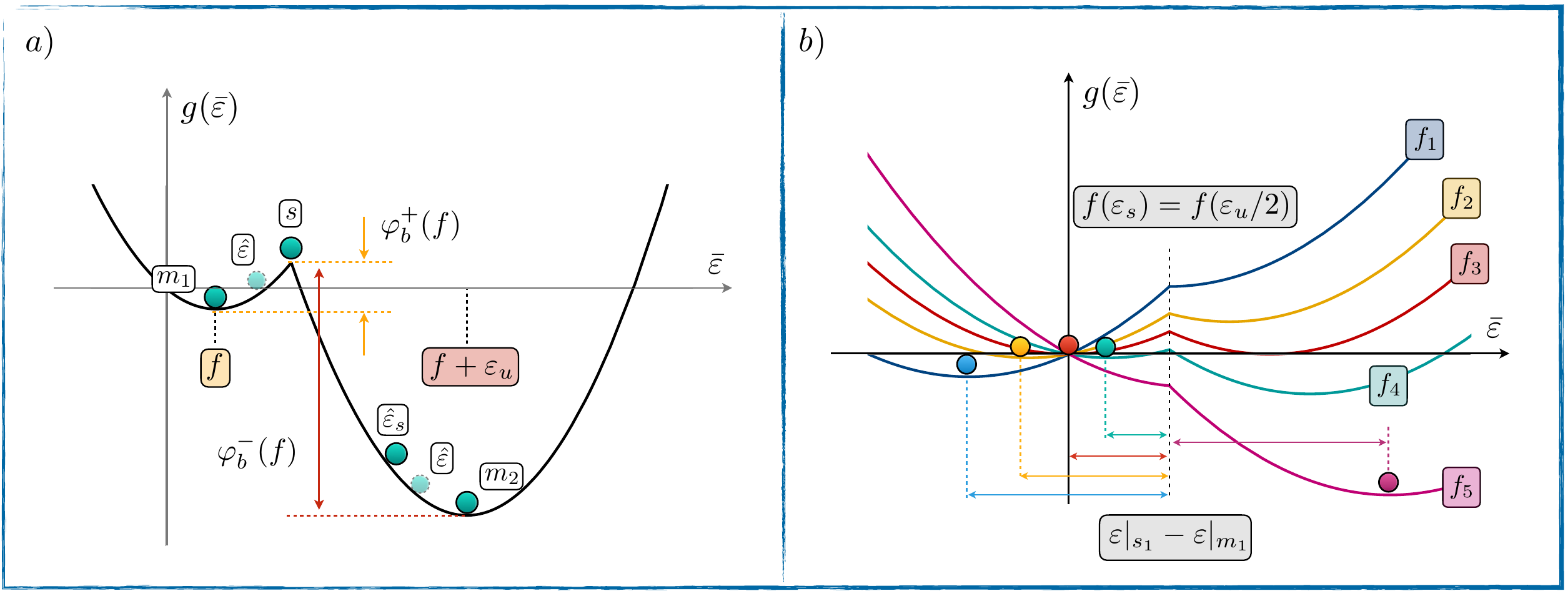}
	  \caption[Demonstration of the minimum energy barrier path]{Demonstration of the minimum barrier energy path. Panel a) shows the energy $g(\bar{\varepsilon})$ for the value $f=f_4$ of Figure~\ref{fig:ch2_landscape_1}$_c$ and it is possible to observe that when a single units is allowed to change phase while the other are fixed, the minimum energy barrier $\varphi_b^+$ going from the first phase $m_1$ to the second $m_2$ is the one through the spinodal point. Panel b) shows different energies with varying forces showing that the spinodal point doesn't depends on $f$.}
	\label{fig:ch2_barrierpath}
\end{figure}

The analysis performed in the previous sections for a system of $n=2$ and $n=3$ elements can be generalized for a generic $n$ both for the soft and hard device hypothesis. Indeed, the evolution of a multistable system can be always be represented as a combination of successive switching events of a single unit while the others are equilibrated. This phase transition occurs along the minimum energy barrier path and it depends only on the height of the energy barriers $\varphi_b$. For simplicity and to not become redundant, in this section I show that the minimum barrier is always achieved at the spinodal point $s$ for the case of the soft device, even though it is also possible to perform a similar analysis in the hard device hypothesis. In our reasoning, it is crucial to remark that, as shown by Eq.~\eqref{eq:ch2_strainspinodal}, the strain at the spinodal point is constant at varying forces. The height of the energy barrier attained at the spinodal point separating phase $1$ to phase $2$ reads 
\begin{equation}
\varphi_b^+(f):=h_{SD}(f)=g(\varepsilon_{s})-g(\varepsilon_1)=\frac{f^2}{2}-\frac{\varepsilon_u}{2}f+\frac{\varepsilon_u^2}{8}. 
\label{eq:ch2_softbarrier}
\end{equation}
where $h_{SD}(f)$ is the height of the barrier in the soft device and $g(\cdot)$ are given by~\eqref{eq:ch2_softforceenergy}. The barrier is zero when the force is equal to $f=\varepsilon_u/2$ and, being fixed the position of the spinodal point, \eqref{eq:ch2_softbarrier} tells that the minimum $m_1$ become closer to the spinodal point as the force increases until the barrier disappears, as shown in Figure~\ref{fig:ch2_barrierpath}$_b$.

%
%

Consider the general situation represented in Figure~\ref{fig:ch2_barrierpath}$_a$. The global minimum of the energy at a fixed size of the system $n$ is located at the trivial equilibrium branches corresponding to the fully folded solution $m_1$ and the fully unfolded configuration $m_2$, whereas the metastable states are attained for the partially evolved system, where a certain number $n-p$ of bi-stable units are in phase $1$ and the remaining $p$ are unfolded. Let us choose one of these metastable configurations, that, according to the equilibrium condition in~\eqref{eq:ch2_softforceenergy}, has the strain of the $i-$th unit such that
\begin{equation}
\hat{\varepsilon}_i=
	\begin{cases}
	f & i=1,\dots, n-p,\\
	f+\varepsilon_u & i=n-p+1,\dots,n
	\end{cases}.
\end{equation}
Then, let us keep all the other units fixed while only this configuration is allowed to move from its equilibrium position $m_1$, with varying energy $g(\hat{\varepsilon})$. A similar situation can be considered for a configuration in the other equilibrium position, $m_2$, as shown in Figure~\ref{fig:ch2_barrierpath}$_a$. Consider also the fact that both in the regions $\bar{\varepsilon}<\varepsilon_s$ and $\bar{\varepsilon}>\varepsilon_s$ the energy is convex so that at a certain strain difference corresponds to the same energy difference between the two minima and $g(\hat{\varepsilon})$. The transition point is then reached when $\hat{\varepsilon}$ is at the spinodal point $s$ and this corresponds also to the minimum path required to attain this configuration. Specifically, in the case represented in Figure~\ref{fig:ch2_barrierpath}$_a$, the minimum path going from phase $1$ to phase $2$ is exactly the one from $m_1$ to $s$ and the energy difference correspond to the barrier $\varphi_b^+(f)$. The same reasoning can be repeated to study the opposite transition leading from the fully unfolded state to the folded configuration. Indeed, in this case, the barrier is $\varphi_b^-(f)$ but this does not correspond to the minimum energy path. It is worth remarking that this transition from phase $2$ to phase $1$ which is not the one favoured, can still occur if its probability is appropriately considered in a rate theory framework, as we will show in Chapter~\ref{ch_6}.

\subsection{Rate effects}
\label{sec:ch2_pt5}

In this section, I propose a simple approach to study rate effects on the thermo-mechanical response of the multistable system considered in the example. In particular, the method is based on Kramers' rate formula, as described in Section~\ref{sec:ch2_rate}. The dynamical behaviour of systems with micro instabilities is an open theoretical challenge, as will be widely discussed in Chapter~\ref{ch_6}, and several theories destining the microscopic evolution of biological materials are mainly based on the pioneering works of Kramers (\cite{kramers:1940}) and Bell (\cite{bell:1978}). 
aa
Then, let us consider the chain with three units in a soft device hypothesis. I assume that the force is applied with a certain velocity $v_f$ so that
\begin{equation}
f(t)=v_f t,
\end{equation}
where $t$ is the time. Moreover, we have shown that when a multistable system is considered, the probability of phase transition is given by the possibility of overcoming energy barriers that vary depending on the assigned force. Although it is possible to consider also the probability of jumping back from the second to the first phase or the probability that more units change phase at the same time, in this simple example we consider only the folded$\to$unfolded transition and the corresponding energy barrier obtained for the soft device in~\eqref{eq:ch2_softbarrier}. Thus we have 
\begin{equation}
\varphi_b^+:=\varphi_b^+(f(t))=\frac{f(t)^2}{2}-\frac{\varepsilon_u}{2}f(t)+\frac{\varepsilon_u^2}{8}.
\end{equation}

Accordingly, to capture the main features of this method and avoid unnecessary computations, I use a simplified version of the Kramers rate formula derived in~\eqref{eq:ch2_kramers}, where the viscosity parameter is neglected, so that the transition rate between metastable configurations reads 
\begin{equation}
\nu(t)=\nu_0 e^{-\beta \varphi_b^+(f(t))},
\end{equation}
where $\beta=1/(k_B T)$, $\nu_0$ is the natural frequency of oscillation of the bond and $\varphi_b^+(f(t))$ is the height of the energy barriers to cross from phase $1$ to phase $2$ at a given force $f$. It is important to remark that Bell's formula as the one presented in the work of Buehler (\cite{buehler:2008}), considers at the exponent of the rate a fixed energy barrier at zero force $E_b$ minus the variation of the energy due to the applied force with respect to the transition strain $-f \,x_b$, where $x_b\equiv\varepsilon_u$. On the other hand, in this thesis, I explicitly evaluate the height of the energy barriers due to a given force so that the two approaches are equivalent. 

Following (\cite{hummer:2003}), we may define the probability that the transition is not yet occurred, \textit{i.e.} the probability that a certain unit is still in the folded configuration at a certain instant of time $\mathcal{P}_f(t)$, that satisfy the first-order rate equation
\begin{equation}
\frac{\text{d}\mathcal{P}_f(t)}{\text{d}t}=\dot{\mathcal{P}}_f(t)=-\nu(t)\mathcal{P}_f(t),
\end{equation}
that is
\begin{equation}
\mathcal{P}_f(t)=e^{-\int_{0}^{t}\nu(\tau)\text{d}\tau}.
\label{eq:ch2_probabilityfold}
\end{equation}
%

%
\begin{figure}[t!]
	\centering
	 \includegraphics[width=0.7 \textwidth]{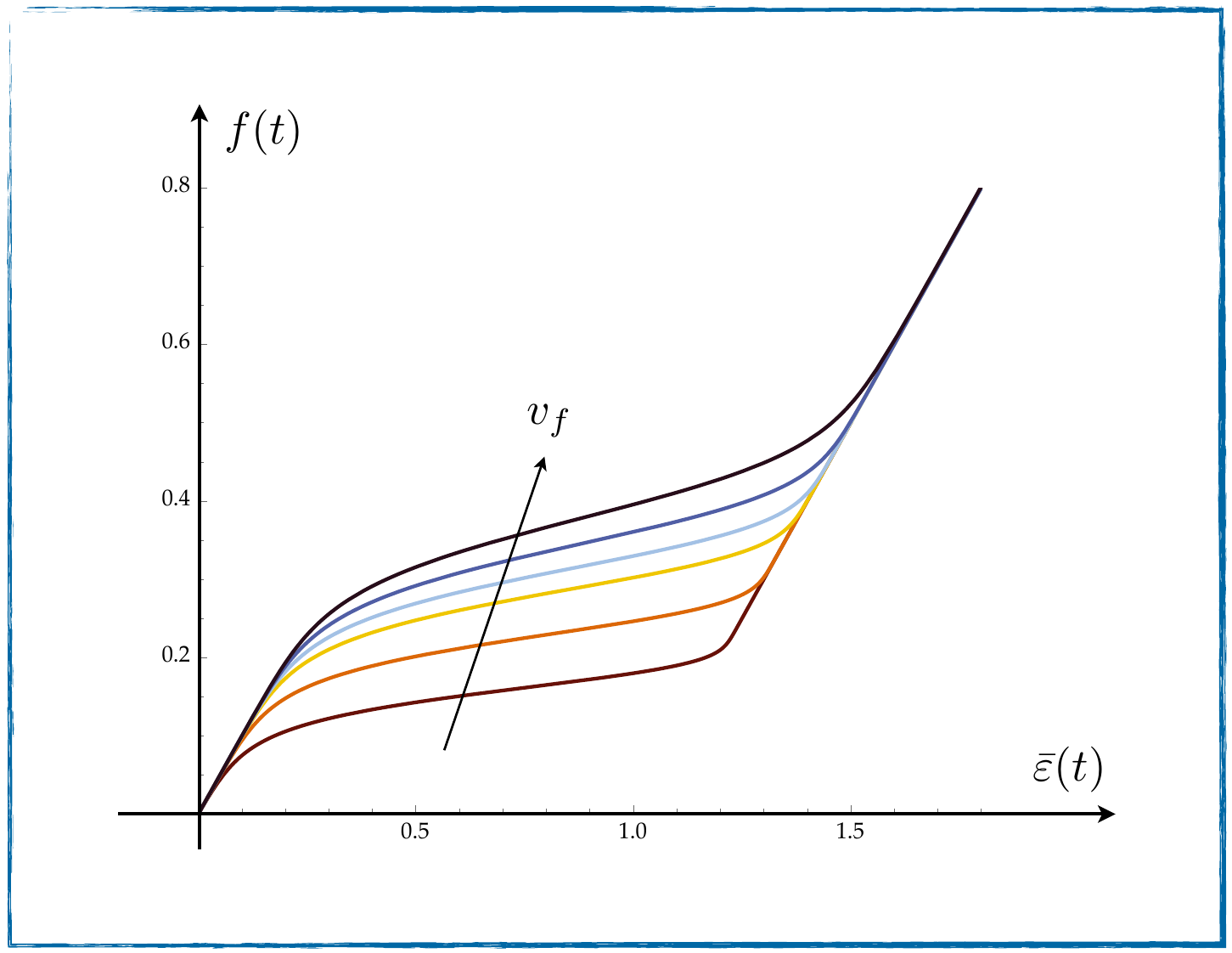}
	  \caption[Stress-strain diagram at different rates]{Stress-strain diagram at different pulling velocity. It is possible to observe that as the velocity increases the Maxwell stress increases and the hardening phenomena is observed. Here parameters are chosen to highlight the main phenomenological features of the approach: $\varepsilon_u=1$, $\nu_0=10^3$, $\beta=100$, $v_f=0.1,1,5,10,20,40$.}
	\label{fig:ch2_rate}
\end{figure}
%

The~\eqref{eq:ch2_probabilityfold} tells that as the rate increases, the probability to jump the energy barrier due to the applied force increases. Indeed, this probability depends on the number of elements in each phase. At the beginning of the transition, all the $n$ units are in the folded configuration and when the force is applied there is a certain probability $\mathcal{P}_f(t)$ that one of those units changes phase. Then, the next metastable equilibrium branch will have $n-1$ elements still folded and the probability increases so that transition may occur easily. This reasoning can be repeated for each $n-p$ folded unit and $p$ unfolded elements that have their own probability to change phase. Thus, the time-dependent average strain can be expressed in terms of how many units are in the first configuration times and the probability to be there similarly for the unfolded one. Moreover, one may obtain the unfolded probability $\mathcal{P}_u(t)$, \textit{i.e.} the probability of being in the unfolded configuration at a certain time $t$ as 
\begin{equation}
\mathcal{P}_u(t)=1-\mathcal{P}_f(t).
\label{eq:ch2_probabilityfold}
\end{equation}

Finally, we obtain 
\begin{equation}
\bar{\varepsilon}(t)=\mathcal{P}_f(t)f(t)+\left(1-\mathcal{P}_f(t)\right)\left(f(t)+\varepsilon_u\right)=\mathcal{P}_f(t)f(t)+\mathcal{P}_u(t)\left(f(t)+\varepsilon_u\right).
\label{eq:ch2_probabilityfold}
\end{equation}

In figure~\ref{fig:ch2_rate} the rate-dependent response of the chain is represented for different rates of the applied force $v_f$. We may observe that the main effect of the increasing velocity is that the transition --Maxwell-- stress increases and a hardening phenomenon occurs. Indeed, when the velocity grows, the probability of overcoming the energy barrier also increases and more units are found in the unfolded configuration earlier leading to higher stresses. The aim of this section is to introduce the rate theory applied to a simple case to highlight the main phenomenological features. The effects of the parameters such as the frequency of oscillation of the bonds, the temperature or the time scale will be better discussed in Chapter~\ref{ch_6}, where rate effects are considered in a wider framework.




	\clearpage


%
\renewcommand{\thefigure}
{\arabic{chapter}.\arabic{figure}}
\setcounter{figure}{0}

\renewcommand{\thetable}
{\arabic{chapter}.\arabic{table}}
\setcounter{table}{0}

\renewcommand{\theequation}
{3.\arabic{equation}}
\setcounter{equation}{0}

\chapter{Influence of the handling device in single molecule experiments}
\label{ch_3}
\vspace{1cm}

The possibility of experimentally testing the mechanical response of molecules and biological systems represents a crucial topic in many fields such as biomechanics, biology, medicine and material science (\cite{goriely:2017}). Many examples that are nowadays under investigation by the scientific community captured my attention, as discussed in the introduction, and to mention some of them I refer to the morphogenesis of the neuronal network (\cite{recho:2016}), the influence of contour instabilities in tumour growth (\cite{benamar:2011}), the phenomena of misfolding of $\alpha$-synuclein protein in Parkinson's disease (\cite{krasnoslobodtsev:2013}),  or the different processes happening at the genetic level as gene transcription in DNA (\cite{gonzalez:2017,woodside:2008}) and folding/unfolding processes in RNA (\cite{li:2008}). 

The main breakthrough that opened up the possibility of investigating the force response of these systems at the molecular level can be addressed to Single Molecule Force Spectroscopy (SMFS) techniques. An outstanding innovation in the last decades delivered high-precision instruments such as Atomic Force Microscopes (AFM), optical tweezers, magnetic tweezers, and micro-needles (\cite{bustamante:2000}), letting the possibility of unreached precision during micro and nanoscale mechanical experiments. These set-ups are characterized by very different operational ranges in terms of spatial resolution, stiffness, displacement and probe size, and the choice of the instrument depends on the specific application (\cite{neuman:2008}). For instance, AFM experiments are adopted for molecule pulling and interaction tests (\cite{dutta:2016,hughes:2016}) whereas optical tweezers are better suited for DNA synthesis analysis (\cite{woodside:2006}). Moreover, \acs{SMFS} experiments allowed the opportunity of deducing the detailed properties of the complex energy landscape of multistable molecules due to the possibility of considering different loading directions and rates, enabling the analysis of the relative stability of the multiple metastable configurations (\cite{szabo:2005}). Nevertheless, to attain effective results, several delicate aspects need to be analyzed, ranging from a detailed definition of the force direction (\cite{walder:2017}) to the analysis of loading rate effects (\cite{benichou:2016,cossio:2015}).

However, the most important issue which is the main focus of this chapter concerns the strong influence of the handling device stiffness on the experimental response (\cite{maitra:2010}), an aspect often underestimated or even neglected. This influence depends on the specific apparatus but is a common feature in SMFS experiments. For instance, the finite value of the cantilever stiffness of an AFM can induce a wrong estimation of both the dissipated energy and the unfolding force thresholds, leading to important discrepancies between the theoretical \textit{vs} experimental forces (\cite{biswas:2018}) and transition rates (\cite{dudko:2008,li:2014}). Moreover, when the rate is considered the energy barrier separating the metastable states is highly sensitive to the overall stiffness, which depends on factors such as the number of domains in the chain, the ones that have already changed phase and the stiffness of the experimental apparatus, that have to be considered to correctly interpret the experimental results (\cite{shoham:2020}). Indeed, since the adopted SMFS devices are characterized by a device stiffness differing by several orders of magnitude (\cite{bustamante:2000}), the analysis of this effect cannot be neglected.

To address this feature, in this chapter I introduce a model capable of describing the experimental device's influence in the whole range of its stiffnesses on a molecule undergoing a two-states transition, within the framework of equilibrium Statistical Mechanics (\cite{weiner:1983}) and aiming at obtaining analytical results. While our approach and analytical results are general, as a paradigmatic example I refer to the fundamental case of AFM induced unfolding experiments in bio-molecules such as titin (\cite{rief:1997,givli:2011}). In this case, the molecules are characterized by domains undergoing a conformational from a folded co figuration to an unfolded one (\cite{rief:2002}). In the model presented in this chapter, I start by assuming that rate effects are neglected, so it is restricted to low rates of loading corresponding to a rate-independent transition behaviour (\cite{chung:2013}). Indeed, in a typical experiment three different time scales are involved in the evolution of the system in its multiwells energetic landscape: the time scale of the external loading, the time scale of relaxation to the local minimizer of the energy and the time scale characterizing the overcoming of the energy barriers. As a result, two different regimes have been experimentally and theoretically determined~(\cite{friddle:2012, friddle:2008, li:2014}).  At a high rate of loading, a `far from equilibrium regime' is observed with the force regulated by the escaping rate from the energy wells. In this case, a rate-dependent behaviour is observed and, as can be shown based on Kramer's type relations, the unfolding force logarithmically grows with the rate of loading~(\cite{dudko:2008,suzuki:2013}). On the other hand, at a low rate of loading a `quasi-equilibrium regime' is attained. Moreover, it is observed that under a certain rate threshold the transition force are rate-independent and they are influenced only by the device stiffness $k_d$ (\cite{friddle:2008}).

Following the approach of (\cite{detom:2013,manca:2013,benichou:2013,makarov:2009,bustamante:2003}), in this chapter I apply the multi wells energy-based description introduced in Chapter~\ref{ch_2} to mimic the conformational transition of a molecule from a folded state to an unfolded one. It is worth remarking that this approximation is acceptable as far as the extension of the molecule is much shorter than the contour length both in the folded and unfolded configuration. The main difference with respect to other classical models that can be used to describe molecules, such as the Freely Jointed Chain (FJC) (\cite{su:2009,benedito:2018}), or the Worm Like Chain (WLC)  (\cite{detom:2013,staple:2008}), is that with our method analytical results, both in the case of zero temperature and in the Statistical Mechanical framework, can be obtained. Thus, the focus of this chapter is to extend the prototypical model presented in Chapter~\ref{ch_2} to a more complex case, studying the thermo-mechanical response of the molecule under different loading conditions by including also the interaction with the handling system. Also, I remark that this model can be extended to the important field of the analysis of the elastic interaction between different molecules (\cite{rosa:2004}). 

I consider the two limit regimes of the hard and soft devices, \textit{i.e.}, fixed displacement or applied force, respectively, and I show that these two are ideal conditions never attained during real experiments. Indeed, the true response of a biological component is affected by the loading device and it lives in between the `\textit{ideal hard device}' and `\textit{ideal soft device}' cases, as I refer to these two limit conditions, and it was firstly elucidated by (\cite{kreuzer:2001}), where the interaction of a molecule with convex energy loaded by a device with variable stiffness has been studied. As introduced in Chapter~\ref{ch_2}, in the ideal hard device hypothesis, the molecule elongation is assigned and the force is a fluctuating conjugated variable. In this case, as the extension is increased, a sequence of localized transitions occurs with a typical sawtooth force-elongation diagram. In the ideal soft device, isotensional experiments are considered. A force is applied at the last element of the chain and the overall elongation represents the conjugate unknown variable. In this case, the transition path is monotonic with a more cooperative transition behaviour. As remarked above, such behaviour is observed in stretching experiments of other multistable systems such as deformation localizations and sawtooth transition paths in metal plasticity, known as the Portevin-LeChatelier effect (\cite{froli:2000}), in shape-memory materials (\cite{puglisi:2000}) and in muscle contraction mechanics (\cite{hudson:2019}). 

In the framework of such ideal boundary conditions, for systems characterized by convex energies, the analysis of the hard and soft device cases represents a simple problem in the field of Statistical Mechanics (\cite{weiner:1977}), where the Helhmotz ensemble suits the description of the hard device whereas the Gibbs ensemble describes the case of the soft device. Conversely, the problem of a multistable material is more subtle and only a partly solved subject. For instance, in (\cite{efendiev:2010}) the authors model the thermalization of a Fermi Pasta Ulam system and semi-analytical results are obtained in both soft and hard devices, yet neglecting the device stiffness effect. In (\cite{mancagiordano:2014}) a chain of links with three-parabolic energy wells has been considered and the authors analyze the equivalence --in terms of mechanical response-- of the Gibbs and Helmholtz ensembles in the thermodynamic limit. Moreover, the case of non-convex energy was recently analyzed in (\cite{fp:2019}), where the influence of the device stiffness in the case of assigned displacement acting on the system has been deduced. The analytical results are in very good agreement with the observations in (\cite{zhanget:2008}) where the authors experimentally show the strong influence of the pulling device on the P-Selectin molecule stretching response. 

Thus, by considering the molecule and the device as a single thermodynamical system, the aim of this chapter is to extend the approach in (\cite{fp:2019}) and describe, in a fully analytical and self-consistent model, all the possible experimental boundary conditions. Specifically, I consider the main two realizations of \ac{SMFS} experiments. For instance, in AFM pulling tests one of the free ends of the molecule is fixed whereas the other one is attached to an elastic handle that is subjected to a fixed displacement (\cite{rief:1997}), representing the hard device case, as shown in Figure~\ref{fig:ch3_model}. On the other hand, in cases such as magnetic and optical tweezers, the generated field is used to apply a force to a handle (\cite{kim:2009}), whereas the handle and macromolecule displacements are experimentally determined, as described in (\cite{capitano:2013,borgia:2008}). Moreover, as I show, the Gibbs ensemble for the isotensional case can be obtained from the Helmholtz one by a Laplace transform (\cite{weiner:1983}), also in the case when the device influence and non-convex energies are considered. Furthermore, I obtain that the ideal hard (soft) device can be deduced as limit regimes when the device stiffness is much larger (smaller) than the molecule stiffness. Finally, I show the equivalence of the two ensembles in the thermodynamic limit. This result was analytically shown in (\cite{mancagiordano:2014}) for the convex Freely Jointed Chain energy whereas it was numerically deduced for the non-convex case. Here I extend this result by proving this equivalence even in the non-convex energy model,  both when the device is considered or not. Eventually, the main outcome of the analysis performed in this chapter is that it is not possible to define uniquely the response of a multistable molecule, because both the relative stability of the multiple minima and the unfolding thresholds depend on the stretching system. The proposed approach lets us analytically predict this (temperature-dependent) variable behaviour depending on the relative stiffness of the stretched and stretching system.

\section{Model}
\label{sec:ch3_model}

%
\begin{figure}[t!]
\centering
  \includegraphics[width=0.95\textwidth]{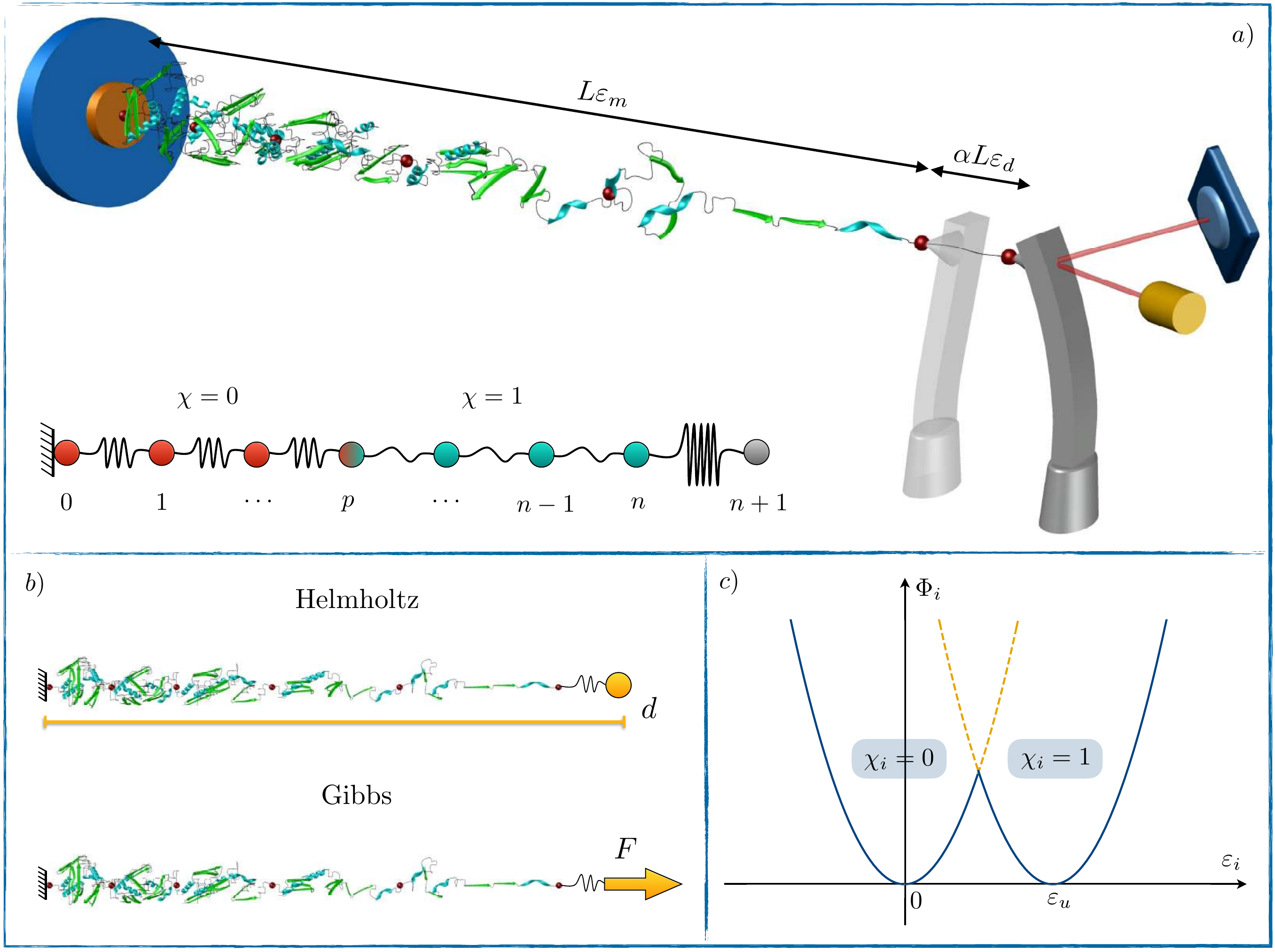}
  \caption[Mechanical model of a typical AFM experiment]{Model. Panel a): Scheme of an SMFS unfolding experiment with an AFM of a molecule with domains undergoing folded$\rightarrow$unfolded transition and the mechanical schematization of the system. Panel b) Boundary conditions: applied displacement (Helmholtz) or applied force (Gibbs). Panel c): Two-wells elastic energy of the bi-stable elements.}
  \label{fig:ch3_model}
\end{figure}
%

To describe the (typically all or none) folded $\rightarrow$ unfolded element conformational transition, we model the molecule as a chain of $n$ bistable springs with reference length $l$ and total reference length $L = n l$.  Each spring has a bi-parabolic energy of the type in Figure~\ref{fig:ch3_model}$_c$, and we introduce the `spin' variable $\chi_i$ for each bi-stable element such that  $\chi_i=0$ in the folded state and $\chi_i=1$ in the unfolded one. Moreover, with the aim of keeping analytical compactness, we consider identical wells for the two phases with stiffness $k_m$, but analytical results can be easily deduced also in the case of different stiffnesses and non zero transition energy, as we show in Section~\ref{sec:ch3_aimeta}. The total elastic energy of the molecule can be written as
\begin{equation}
\Phi_m=\sum_{i=1}^{n}\frac{1}{2}\,k_m l \bigl(\varepsilon_i-\varepsilon_u\chi_i\bigr)^2,
\label{eq:ch3_eq:1}
\end{equation}
where $\varepsilon_i$ is the strain of the $i$-th element and $\varepsilon_u$ is the unloaded strain of the second well (see Figure~\ref{fig:ch3_model}$_c$). 

As anticipated, the key feature of the proposed approach is that effective analysis of the influence of the loading device on the molecular behaviour requires considering the stretched molecule and the stretching device as a whole thermodynamical system. Following (\cite{fp:2019}), the device influence is described by an auxiliary spring with variable stiffness $k_d$, reference length $\alpha L$, strain $\varepsilon_d$ and energy
\begin{equation}
\Phi_d = \frac{1}{2}\alpha L \, k_d \,\varepsilon_{d}^{2}.
\label{eq:ch3_eq:2}
\end{equation}

According to the previous discussion we need to introduce the {\it total elongation} (molecule plus handle) 
\begin{equation}
d=\sum_{i=1}^{n}\frac{L}{n}\,\varepsilon_i+\alpha L \varepsilon_d = L\bigl(\varepsilon_m+\alpha \, \varepsilon_d\bigr),
\label{eq:ch3_eq:3}
\end{equation}
where
\begin{equation}
\varepsilon_m = \frac{1}{n}\sum_{i=1}^{n}\varepsilon_i
\label{eq:ch3_eq:4}
\end{equation}
is the molecule's average strain. Similarly, by using Equations \eqref{eq:ch3_eq:3} and \eqref{eq:ch3_eq:4}, we introduce the {\it total  averaged strain}
\begin{equation}
\varepsilon_t =\frac{d}{L(1+\alpha)} \Rightarrow (1+\alpha)\,\varepsilon_t = \varepsilon_m + \alpha \varepsilon_d.
\label{eq:ch3_eq:5}
\end{equation}
Here and in the following, we use the pedex $m$ to denote the molecule, $d$ to denote the device and $t$ to denote the total (device plus molecule) system quantities. Finally, we need to introduce the \textit{total elastic energy} of the system 
\begin{equation}
\Phi_t = \Phi_m + \Phi_d = \sum_{i=1}^{n}\frac{1}{2}\,k_m l \bigl(\varepsilon_i -\varepsilon_u\chi_i\bigr)^2 + \frac{1}{2}\alpha L \, k_d \varepsilon_{d}^{2}.
\label{eq:ch3_eq:6}
\end{equation}
%

\section{Mechanical limit}
\label{sec:ch3_mechanics}

Following the approach of Chapter~\ref{ch_2}, and with the aim of getting physical insight into the introduced model, we first consider the case when entropic energy terms are neglected, namely the mechanical limit, and then we extend the results to the general case measuring temperature effects. We consider the two different boundary conditions described in Figure~\ref{fig:ch3_model}$_b$. In one case, denoted as \textit{hard device}, we suppose that a fixed total displacement $d$ is applied and we solve the constrained problem
\begin{equation}
\min_{\vspace{-0.1 cm}\scriptsize \begin{array}{c}{\varepsilon_{1},\dots,\varepsilon_n,\varepsilon_d}\vspace{+0.0 cm}\\ \sum_{i}\varepsilon_i/n+ \alpha \varepsilon_d =\varepsilon_t(1+\alpha)\end{array}} \Phi_t\bigl(\varepsilon_{1},\dots,\varepsilon_n,\varepsilon_d\bigr).
\label{eq:ch3_eq:7}
\end{equation}
In the other case, known as \textit{soft device}, a fixed constant force is applied to the chain and we search for the minima of the total potential energy:
\begin{equation}
\min_{\varepsilon_{1},\dots,\varepsilon_n,\varepsilon_d}\,\, G_t(\varepsilon_{i},\varepsilon_d) = \Phi_t(\varepsilon_{i},\varepsilon_d) - F \left(l\sum_{i=1}^n \varepsilon_i+ \alpha L \varepsilon_d\right).
\label{eq:ch3_eq:8}
\end{equation}
In both cases, by deriving with respect to the variables $\varepsilon_i$, with $i=1,\dots,n$ and $\varepsilon_d$, equilibrium requires a constant force $F$ such that
\begin{equation}
\varepsilon_i = \frac{F}{k_m}+ \varepsilon_u\chi_i \quad \text{with} \quad i = 1, \dots, n \quad \text{and} \quad \varepsilon_d = \frac{F}{k_d}.
\label{eq:ch3_eq:9}
\end{equation}
Due to the absence of non-local interactions, the equilibrium force and energy only depend on the number $p$ of unfolded elements, here assigned by the unfolded fraction 
\begin{equation}
\bar{\chi}=\sum_i=1^{n} \frac{\chi_{i}}{n}=\frac{p}{n}\in [0,1].
\label{eq:ch3_eq:pn}
\end{equation}
In particular, $\bar \chi=0$ and $\bar \chi=1$ correspond to the initial fully folded state and to the fully unfolded state, respectively.  Thus, by using~\eqref{eq:ch3_eq:4},~\eqref{eq:ch3_eq:5} and~\eqref{eq:ch3_eq:9} we obtain a compact expression for the equilibrium force 
\begin{equation}
F=k_m\gamma\Bigl[(1+\alpha)\,\varepsilon_t-\varepsilon_u\bar{\chi}\Bigr],
\label{eq:ch3_fepsilont}
\end{equation}
and for the equilibrium strain of the molecule
\begin{equation}
\varepsilon_m = \varepsilon_u\bar{\chi}+ \gamma\Bigl[(1+\alpha)\,\varepsilon_t-\varepsilon_u\bar{\chi}\Bigr],
\label{eq:ch3_epsilontm}
\end{equation}
where we introduced the main non-dimensional parameter of the model  
\begin{equation}
\gamma=\frac{k_d}{k_d+\alpha k_m}\quad\text{with}\quad \gamma \in \,\,]0,1[ .
\label{eq:ch3_gamma}
\end{equation}
Finally, by using \eqref{eq:ch3_fepsilont} and \eqref{eq:ch3_epsilontm}, we obtain the force-strain relations of the molecule for the equilibrium branches with different unfolded elements $p$:
\begin{equation}
F=k_m \bigl(\varepsilon_m - \varepsilon_u\bar{\chi}\bigr).
\label{eq:ch3_Fo}
\end{equation}

\begin{figure}[t!]
	\centering
	 \includegraphics[width=0.95\textwidth]{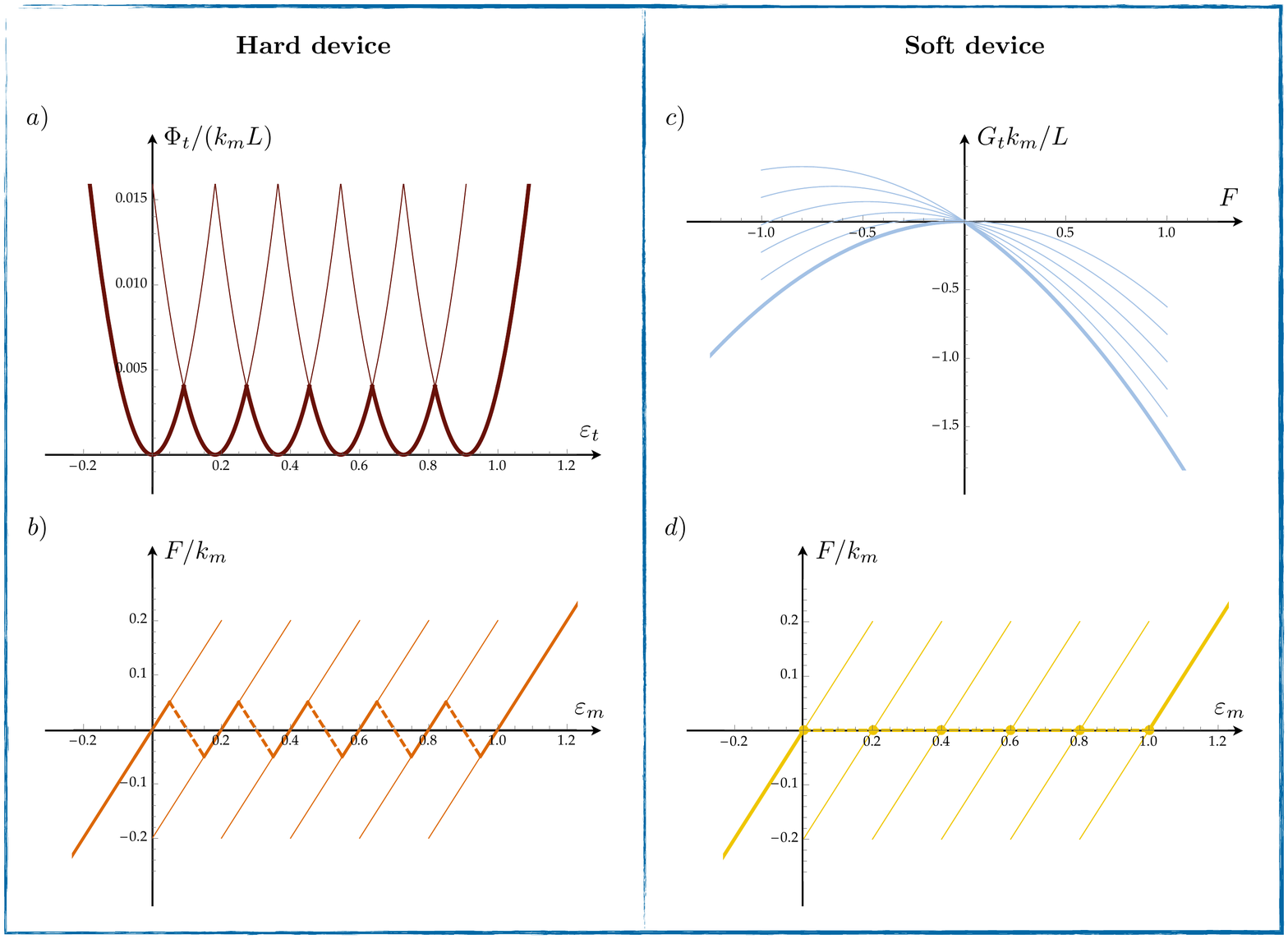}
	  \caption[Mechanical limit of chain with $n=5$.]{Equilibrium branches (thin lines) and global energy minima (thick lines) in the mechanical limit. Total elastic energy in the hard device in panel a) and potential energy in the soft device in panel b).  Force-extension diagrams under the Maxwell convention in the hard soft device, in panels c) and d) respectively: dashed lines represent strain discontinuities, \textit{i.e.} the partially unfolded configurations. Here  $\alpha=0.1$, $n=5$ and $\gamma = 0.5$, $L=1$, $k_m=1$.}
	\label{fig:ch3_energy}
\end{figure}

Observe that by using \eqref{eq:ch3_fepsilont} and \eqref{eq:ch3_epsilontm}  the  two-phases metastable equilibrium branches ($\bar{\chi}\in]0,1[$) are defined only for $|F|/k_m\le \varepsilon_u$ corresponding to a strain domain
\begin{equation}
\frac{(1+\alpha)}{\varepsilon_u} \varepsilon_t\in 
\left \{ \begin{array}{ll} 
\displaystyle\left( -\infty , \,\,\frac{1}{\gamma}\right )  &  \mbox{ if } \quad \bar \chi=0  \vspace{0.2 cm}\\
\displaystyle\left(\bar{\chi}-\frac{1}{\gamma}, \,\, \bar{\chi}+\frac{1}{\gamma}\right) & \mbox{ if } \quad \bar \chi\in \,\, ]0,1[ \vspace{0.2 cm}\\
\displaystyle\left( 1-\frac{1}{\gamma}, \,\,+\infty \right)&  \mbox{ if } \quad \bar \chi=1
\end{array} \right..
\label{eq:ch3_brancheseq}
\end{equation}
Moreover, due to the convexity of the wells, these solutions are locally stable in the case of both assigned force and displacement. To obtain the global minima of the energy we need to distinguish the two cases of hard and soft devices and minimize the energy with respect to the remaining variable $\bar \chi$. For the equilibrium solution of the case of assigned displacement, the total elastic energy \eqref{eq:ch3_eq:7} can be rewritten as
\begin{equation}
\Phi_t(\varepsilon_t)=k_m L\frac{\gamma}{2}\biggl[(1+\alpha)\varepsilon_t-\varepsilon_u\bar{\chi}\biggr]^2.
\label{eq:ch3_pot}
\end{equation}
As represented in Figure~\ref{fig:ch3_energy}$_a$, one can easily show that the branch  $\bar{\chi}$ corresponds to the global minimum for 
\begin{equation}
\frac{(1+\alpha)}{\varepsilon_u}\varepsilon_t\in 
\left \{ \begin{array}{ll} 
\displaystyle\left( -\infty , \,\,\frac{1}{2n}\right)  &  \mbox{ if } \quad \bar \chi=0  \vspace{0.2 cm}\\
\displaystyle\left(\bar{\chi}-\frac{1}{2n}, \,\, \bar{\chi}+\frac{1}{2n}\right) & \mbox{ if } \quad \bar \chi\in \,\, ]0,1[ \vspace{0.2 cm}\\
\displaystyle\left( 1-\frac{1}{2n}, \,\,+\infty \right)&  \mbox{ if } \quad \bar \chi=1
\end{array} \right..
\label{eq:ch3_globaleq}
\end{equation}
%

%
%

Consequently, the force-extension behaviour is assigned by \eqref{eq:ch3_Fo} with the phase fraction depending on the total assigned strain as follows:
\begin{equation}
\bar \chi=\bar \chi(\varepsilon_t) =
\left \{ \begin{array}{ll}
\displaystyle0 &\varepsilon_t<\varepsilon_t^f \vspace{0.2 cm}\\
\displaystyle \sum_{p=0}^n\frac{p}{n}\,\, {\bf 1}_{\Omega_p} (\varepsilon_t) &\varepsilon_t^f<\varepsilon_t<\varepsilon_t^u  \vspace{0.2 cm}\\
\displaystyle1 &\varepsilon_t> \varepsilon_t^u 
\end{array} \right . 
\end{equation}
where we indicated with ${\bf 1}_{\Omega_p}$ the characteristic function of the set
\begin{equation}
\Omega_p=\Biggl
(\frac{\varepsilon_u}{1+\alpha}\, \left (\frac{p}{n}-\frac{1}{2n}\right ), \frac{\varepsilon_u}{1+\alpha}\, \left(\frac{p}{n}+\frac{1}{2n}\right ) \Biggr),
\end{equation}
 and
\begin{equation}
\varepsilon_t^f=\frac{\varepsilon_u}{1+\alpha}\, \frac{1}{2n}, \quad \quad
\varepsilon_t^u=\frac{\varepsilon_u}{1+\alpha}\left(1-\frac{1}{2n}\right ).
\end{equation}

Differently, in the case of assigned force, we have to minimize the potential energy with respect to the phase fraction $\bar \chi$; in particular, for the equilibrium solutions~\eqref{eq:ch3_eq:8} the energy can be written as
\begin{equation}
G_t(F)=-\frac{L}{2 k_m \gamma}F^2-\varepsilon_u L \,\bar{\chi}\, F.
\label{eq:ch3_gibbsen}
\end{equation}

Thus, the global energy minimum corresponds to the fully folded state $\bar \chi=0$ for $F<0$ and to the fully unfolded state $\bar \chi=1$ for $F>0$, as shown in Figure~\ref{fig:ch3_energy}$_b$. Finally, the force-displacement relation is again given by \eqref{eq:ch3_Fo} with %
\begin{equation}
\bar \chi=\bar \chi(F)={\bf 1}_{]0,\infty[}(F).
\label{eq:ch3_sdml}
\end{equation}

The behaviour of the system under the hypothesis that its configurations correspond to the global minima of the energy (Maxwell convention) is represented in Figure~\ref{fig:ch3_energy}$_c$ with thick lines for the case of the hard device. As the figure shows the transition corresponds to a sawtooth path with  the elements unfolding one at a time at a constant transition force $F=F_{un}$ that using \eqref{eq:ch3_fepsilont} and \eqref{eq:ch3_globaleq}  is given by 
\begin{equation}
F_{un}=\frac{k_m\,\varepsilon_u\gamma}{2n}.
\end{equation}

This behaviour reflects the experimental results of the behaviour of AFM unfolding experiments (\cite{rief:1997}) with a periodic sawtooth path corresponding to the successive transition of the single domain. The case of assigned force is represented in Figure~\ref{fig:ch3_energy}$_d$. Observe that under these boundary conditions the transition is cooperative, with a single value force threshold independent of the relative stiffness parameter $\gamma$. It is important to remark that the experiments show both in the case of hard and soft devices a hardening behaviour with the unfolding force increasing with the unfolded fraction (\cite{rief:1997}). Interestingly, in the following, we show that this hardening behaviour can be associated with an entropic effect. 

The main point that we can already observe in the mechanical limit is the strong dependence of the stability domains and the unfolding force of the different unfolded configurations on the device stiffness, with a linear dependence of the force on both $\gamma$ and the discreteness parameter $n$. While we can already deduce that the behaviour of the system in the hard device reproduces the behaviour of the soft device in both cases of $\gamma \rightarrow 0$ and $n\rightarrow \infty$, we postpone this discussion to Section~\ref{sec:ch3_tl}. There, we obtain analytically this new result even in the case when we do not neglect entropic energy terms.

\section[Temperature effects]{Temperature effects within Helmholtz and Gibbs statistical ensembles}
\label{sec:ch3_temp}

In this section, I analyze the temperature effects in the case of hard and soft devices, corresponding, respectively, to the Helmholtz and Gibbs ensembles in the framework of  Statistical Mechanics. As in the mechanical limit, we consider the system and the measuring device as a whole and we study separately the cases of assigned displacement and assigned force acting on the handle of the experimental device. To not weigh down the notation, full calculations for this case are reported in Appendix~\ref{appE}.

\subsection{Hard device: Helmholtz ensemble}

To describe the system in thermal equilibrium in the case of assigned displacement, we consider the canonical partition function in the Helmholtz statistical ensemble $\mathscr{H}$. Due to the absence of non nearest neighbourhood interactions, the chain's energy depends only on the number $p$ of unfolded domains and not on the specific phase configuration $\chi$. As a result  (see Appendix~\ref{appE} and (\cite{fp:2019})) the partition function assumes the simple form
 \begin{equation}
 \mathcal{Z}_{\mathscr{H}} = \mathcal{K}_{\mathscr{H}}\sum_{p=0}^{n} \binom{n}{p} \, e^{-\frac{\beta k_m l \gamma n}{2}\left(\varepsilon_u \frac{p}{n}-(1+\alpha)\,\varepsilon_t\right)^2},
 \label{eq:ch3_eq:19}
 \end{equation}
where $\mathcal{K}_{\mathscr{H}}$ is a constant, taking into account also the kinetic energy, $\beta = 1/k_B T$, with $k_B$ the Boltzmann constant and $T$ the absolute temperature. Also, the binomial coefficient gives the number of configurations of the chain with $p$ unfolded domains among the $n$ bistable elements. \vspace{0.3 cm}

\noindent \textbf{Remark --} We point out that in order to obtain this analytical expression we assume that the two wells are extended beyond the spinodal point (\cite{efendiev:2010, fp:2019}) and for fixed phase configuration $\chi_i$ we integrate each $\varepsilon_i$ in $\mathbb{R}$. In (\cite{fp:2019}) the authors numerically showed that this approximation does not influence the energy minimization in the temperature regimes of interest for real experiments. In Appendix~\ref{appD}, the approximation is verified also in the Gibbs ensemble studied in the following section.\vspace{0.3 cm}. 

\begin{figure}[t!]
\centering
\includegraphics[width=0.95\columnwidth]{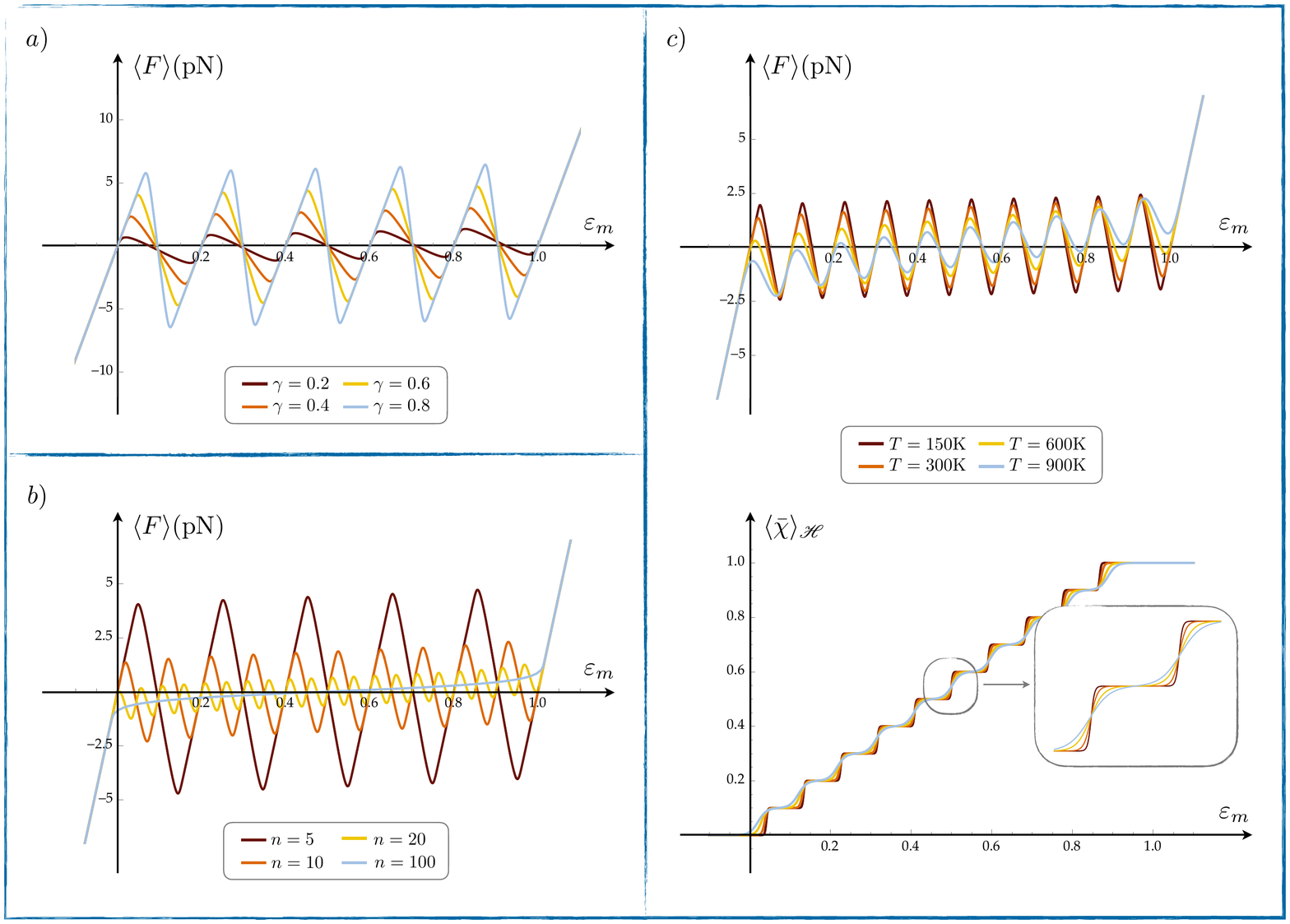}
\caption[Temperature effects in the hard device hypothesis]{Helmholtz ensemble force-elongation diagrams (hard device). Panel a): Effect of $\gamma$, of temperature. Panel b): Effect of the discreteness parameter $n$ on the strain of the molecule. In panel c) the temeperature effects are represente in terms of force-displacement diagram (top) and expectation value of the unfolded fraction (bottom). Parameters:  $k_m=90\;\mbox{pN}$, $\alpha = 0.1$, $l=20\;\mbox{nm}$, $\varepsilon_u=1$, $n=5$ in a) $n=10$ in b).}
  \label{fig:ch3_helmholtz}
\end{figure}

The Helmholtz free energy ($\mathcal{F}$) is, by definition,
\begin{equation}
\mathcal{F} = -\frac{1}{\beta} \ln \mathcal{Z}_{\mathscr{H}}.
\label{eq:ch3_henergy}
\end{equation}
Consequently, following the explicit evaluation in Appendix~\ref{appE}, we can obtain the expectation value of the force conjugated to the applied displacement $d$
\begin{equation}
\langle F \rangle = \frac{1}{L(1+\alpha)}\frac{\partial \mathcal{F}}{\partial \varepsilon_t}= k_m \gamma \Bigl[(1+\alpha)\,\varepsilon_t-\varepsilon_u\langle \bar{\chi} \rangle \Bigr],
\label{eq:ch3_hforce}
\end{equation}
where 
\begin{equation}
\langle \bar{\chi} \rangle = \langle \bar{\chi} \rangle_{\mathscr{H}}(\beta, \varepsilon_t) = \frac{\displaystyle \sum_{p=0}^n\binom{n}{p}\frac{p}{n}\,e^{-\frac{\beta k_m l n \gamma}{2}\left(\varepsilon_u\frac{p}{n}-(1+\alpha)\,\varepsilon_t\right)^2}}{\displaystyle \sum_{p=0}^n\binom{n}{p}\,e^{-\frac{\beta k_m l n \gamma}{2}\left(\varepsilon_u\frac{p}{n}-(1+\alpha)\,\varepsilon_t\right)^2}},
\label{eq:ch3_hchi}
\end{equation}
is the expectation value of the fraction $\bar{\chi}$ of unfolded domains. After some manipulation, we obtain the expectation value of the molecule strain 
\begin{equation}
\langle \varepsilon_m \rangle = \varepsilon_u\langle \bar{\chi} \rangle  + \gamma \Bigl[(1+\alpha)\varepsilon_t-\varepsilon_u\langle \bar{\chi} \rangle\Bigr].
\label{eq:ch3_hemet}
\end{equation}
Finally, by using \eqref{eq:ch3_hforce} and \eqref{eq:ch3_hemet} we obtain 
\begin{equation}
\langle F \rangle = k_m\Bigl[\langle \varepsilon_m \rangle - \varepsilon_u \langle \bar{\chi} \rangle\Bigr],
\label{eq:ch3_hfem}
\end{equation}
by which we can study the effect of temperature and device stiffness (through the parameter $\gamma$) on the molecule response. Observe that \eqref{eq:ch3_hforce}, \eqref{eq:ch3_hemet}, \eqref{eq:ch3_hfem} are formally identical to the  equations \eqref{eq:ch3_fepsilont}, \eqref{eq:ch3_epsilontm}, and \eqref{eq:ch3_Fo} obtained in the mechanical limit case,  with the only difference in the expression of the fraction, which is temperature dependent, consistently with \eqref{eq:ch3_hchi}. In Figure~\ref{fig:ch3_helmholtz} we show the influence of temperature,  device stiffness and number of elements $n$ of the chain on the unfolding behaviour of the molecule.  A detailed interpretation of these results is provided in Section~\ref{sec:ch3_discussion}, where all the cases are compared.

%
%

%
\begin{figure}[t!]
\centering
\includegraphics[width=0.95\columnwidth]{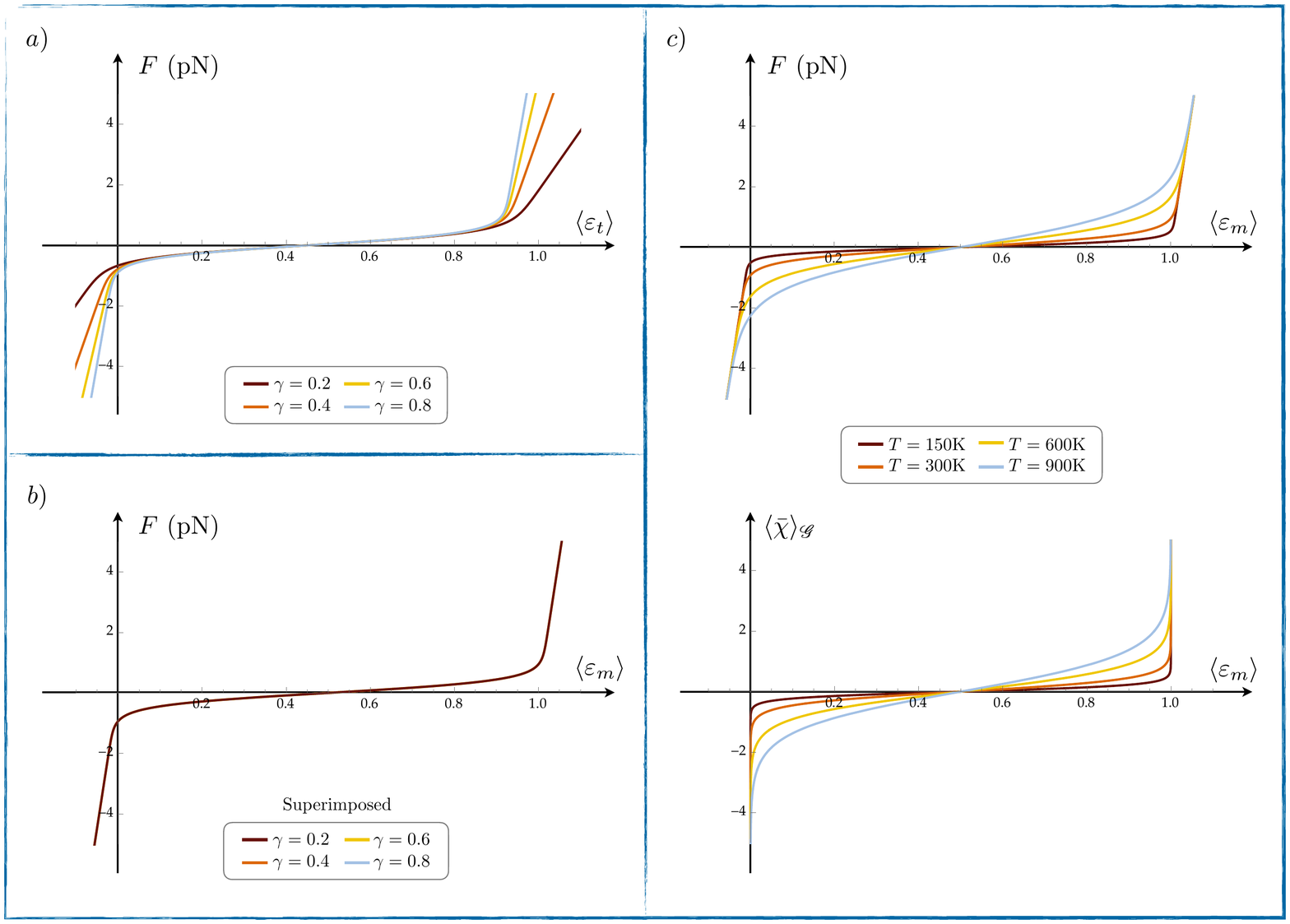}
  \caption[Temperature effects in hard and soft device hypothesis]{Stress-strain curves of the Gibbs ensemble (soft device). Panel a): Effect of $\gamma$. As $\gamma$ increases, the molecule becomes stiffer. Panel b): Effect of $\gamma$. As $\gamma$ increases, the force acting on the molecule is not affected. Panel: c): Effect of the temperature $T$. As the temperature increases, the curves become steeper. The values used for the molecule properties are $n=5$, $k_m=90\;\mbox{pN}$, $\alpha = 0.1$, $l=20\;\mbox{nm}$, $\varepsilon_u=1$.}
  \label{fig:ch3_gibbs}
\end{figure}
%

\subsection{Soft device: Gibbs ensemble}

The partition functions of Gibbs ($\mathscr{G}$) and Helmholtz ($\mathscr{H}$) ensembles are related by a Laplace transform with force $F$ and displacement $d$ as conjugate variables (\cite{weiner:1983}). Thus, using \eqref{eq:ch3_eq:5}, we have
\begin{equation}
\mathcal{Z}_{\mathscr{G}}= \int \mathcal{Z}_{\mathscr{H}}\,e^{\,\,\beta\, Fd}d d= (1+\alpha)L\int \mathcal{Z}_{\mathscr{H}}\,e^{\beta \bigl[L(1+\alpha)\,\varepsilon_t\,F \bigr]}d\varepsilon_t.
\label{eq:ch3_eq:25}
\end{equation}

A detailed calculation leads to a Gaussian integral whose solution is the partition function in the Gibbs canonical ensemble (see Appendix~\ref{appE}), which can be written as 
\begin{equation}
\mathcal{Z}_{\mathscr{G}}= \mathcal{K}_{\mathscr{G}}\sum_{p=0}^{n}\binom{n}{p}\,e^{\frac{\beta l n}{2 k_m \gamma}\left(F^2 + 2 k_m \gamma \varepsilon_u \frac{p}{n} F\right)},
\label{eq:ch3_pfg2}
\end{equation}
where $\mathcal{K}$ is a constant. Again we used the simplifying result that the energy depends only on the number of unfolded elements. In this case, it is possible to evaluate explicitly this summation in order to obtain
\begin{equation}
\mathcal{Z}_{\mathscr{G}}= \mathcal{K}_{\mathscr{G}}\,\left(1+e^{\beta l \varepsilon_u F}\right)^ne^{\frac{\beta l n}{2 k_m \gamma}F^2}.
\label{eq:ch3_pfg}
\end{equation}

The Gibbs free energy ($\mathcal{G}$) is 
\begin{equation}
\mathcal{G}=-\frac{1}{\beta}\text{ln}\, \mathcal{Z}_{\mathscr{G}},
\label{eq:ch3_genergy}
\end{equation}
and, therefore, we can evaluate the expectation value of the total strain
\begin{equation}
\langle \varepsilon_t \rangle = \frac{1}{\beta L(1+\alpha)}\frac{1}{\mathcal{Z}_{\mathscr{G}}}\frac{\partial \mathcal{Z}_{\mathscr{G}}}{\partial F} = \frac{1}{(1+\alpha)}\left(\frac{F}{k_m\gamma}+\varepsilon_u\langle \bar{\chi} \rangle \right),
\label{eq:ch3_gfet}
\end{equation}
where
\begin{equation}
\langle \bar{\chi} \rangle =\langle \bar{\chi} \rangle_{\mathscr{G}} (\beta, F)= \frac{e^{\,\beta l \varepsilon_u F}}{1+e^{\,\beta l \varepsilon_u F}}.
\label{eq:ch3_gchi}
\end{equation}
Also in the case of the soft device (see Appendix~\ref{appE}) it is possible to show that  the total and the molecule strain are related by \eqref{eq:ch3_hemet} so that by using \eqref{eq:ch3_gfet}  we obtain 
\begin{equation}
\langle \varepsilon_m \rangle = \frac{F}{k_m} +\varepsilon_u \langle \bar{\chi} \rangle
\label{eq:ch3_gfem}.
\end{equation}

Once again, we point out the analogy among~\eqref{eq:ch3_gfet}, \eqref{eq:ch3_gfem} and~\eqref{eq:ch3_hemet} which are consistent with the analogous relations in the mechanical limit~\eqref{eq:ch3_fepsilont},~\eqref{eq:ch3_epsilontm} and ~\eqref{eq:ch3_Fo} and in the Helmholtz ensemble ~\eqref{eq:ch3_hforce},~\eqref{eq:ch3_hfem} and ~\eqref{eq:ch3_hemet} with the only difference given by \eqref{eq:ch3_gchi}. In Figure~\ref{fig:ch3_gibbs} we show the effects of temperature. It is important to remark that in this case, differently from the case of assigned displacement, the molecular response \eqref{eq:ch3_gfem}, \eqref{eq:ch3_gchi} is independent from the number  $n$ of elements of the chain and from the relative stiffness $\gamma$. In Section \ref{sec:ch3_discussion} a detailed discussion of these results will be provided.  As anticipated above, in Appendix~\ref{appE} we numerically verify the effectiveness of our approximation on the extension of the biparabolic energies beyond the spinodal point also in the case of the Gibbs ensemble. The comparison of the results obtained from analytical formulas with those obtained numerically show an excellent agreement for a wide range of temperatures.

\section{Thermodynamic limit}
\label{sec:ch3_tl}

Many important biological molecules  undergoing conformational transitions, \textit{i.e.} titin (\cite{rief:1997}) or DNA (\cite{busta:2000}), are constituted by a very large number of domains. Therefore, it is interesting to explore the thermodynamic limit, which is the limit where $n \to \infty$ whereas $l=const$ so that $L\to\infty$. Firstly, let us consider the case of \textit{hard device}. In order to perform the thermodynamic limit, we use the saddle point method as described in (\cite{zinn:1996,fp:2019} and Appendix~\ref{appE}, and we obtain the expectation value of the unfolded fraction 
\begin{equation}
\langle \bar{\chi} \rangle\simeq \chi_c(\varepsilon_t),
\end{equation}
where $\chi_c$ is the solution of 
\begin{equation}
\text{ln}\left (\frac{x}{1-x}\right )+\varepsilon_u \beta k_m l \gamma \Bigl[x \, \varepsilon_u-(1+\alpha)\,\varepsilon_t\Bigr]= 0. 
\label{eq:ch3_limitH}
\end{equation}

It is straightforward to see that, consistently with the previous results of this chapter, we obtain the same form of the mechanical response of the molecules as in \eqref{eq:ch3_hforce}, \eqref{eq:ch3_hemet}, \eqref{eq:ch3_hfem}.  The same equations can be extended also to the case of the \textit{soft device} thermodynamic limit, after observing that the phase fraction in \eqref{eq:ch3_gchi} does not depend on $n$. 

We want now to prove the equivalence of the molecule response under the hard and soft device in the thermodynamic limit for systems with non-convex energies. This result has been analytically shown in (\cite{winkler:2010, manca:2012}) for flexible polymers in the case of convex energy. In the same papers, this result has been numerically shown also in the case of two-wells energy. Notice that, in the mechanical limit, the observed equivalence can be deduced following the approach in (\cite{puglisi:2000}) and in (\cite{puglisi:2002}) where the authors consider also metastable configurations and hysteresis. On the other hand, this equivalence is not true when non-local interactions are considered, as shown in (\cite{vainchtein:2004, puglisi:2007}) and in the following Chapter~\ref{ch_4}. In this section, we extend the results to the case of non-convex energy when the handle device is considered and of course, the result can be extended also to the ideal case when we can neglect its effect.

\begin{figure}[t!]
	\centering
	 \includegraphics[width=0.7\textwidth]{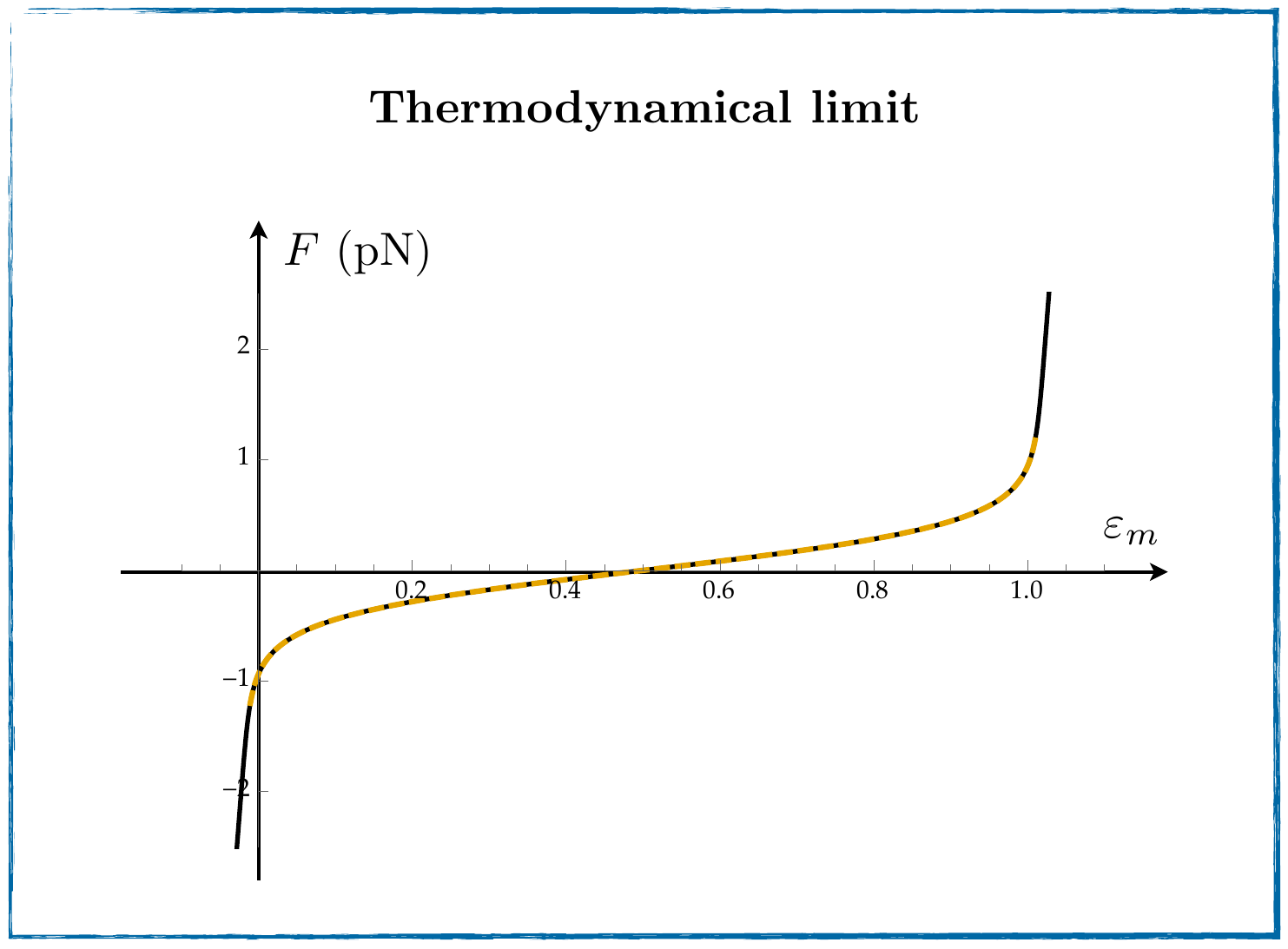}
	  \caption[Equivalence of the thermodynamic limit in the two ensemble]{The figure shows the thermodynamic limit in the Helmholtz (yellow line) and Gibbs (black line) ensemble. It is evident how the two limits coincides. Here  $k_m=90\;\mbox{pN}$, $\alpha = 0.1$, $\gamma = 0.6$, $l=20\;\mbox{nm}$, $\varepsilon_u=1$.}
	\label{fig:ch3_tlim}
\end{figure}

Since the force-elongation relation has the same expression in the two ensembles (see \eqref{eq:ch3_hfem} and \eqref{eq:ch3_gfem}), to prove the analytical equivalence in the thermodynamic limit in terms of molecular response we need to show that the expressions of the unfolding fraction in \eqref{eq:ch3_hchi} and \eqref{eq:ch3_gchi} at given equilibrium forces coincide. We can rewrite the expectation value of the total strain in the Gibbs ensemble~\eqref{eq:ch3_gfet} as
\begin{equation}
\langle \varepsilon_t \rangle = \rho(F).
\label{eq:ch3_eq:etF2}
\end{equation}
On the other hand, we want to show that the expectation value $\langle F \rangle$ in the Helmholtz ensemble converges to $F$ in the thermodynamical limit. Since in both cases we found the same strain-force relations with the only difference in the expectation value of the phase fraction, we only need to show that the two expressions~\eqref{eq:ch3_hchi} and~\eqref{eq:ch3_gchi} attain the same limit as $n$ diverges. This can be done by  simply verifying that $\langle\bar{\chi}\rangle_{\mathscr{G}}$ is the only solution of \eqref{eq:ch3_limitH}: 
\begin{equation}
\text{ln}\left (\frac{\langle\bar{\chi}\rangle_{\mathscr{G}}}{1-\langle\bar{\chi}\rangle_{\mathscr{G}}}\right )+\varepsilon_u \beta k_m l \gamma \Bigl[\varepsilon_u\,\langle\bar{\chi}\rangle_{\mathscr{G}}-(1+\alpha)\langle \varepsilon_t \rangle \Bigr]=F l \beta \varepsilon_u-F l \beta \varepsilon_u= 0. 
\label{eq:ch3_hlimitt2}
\end{equation} 
As stated, this proves the equivalence of the two ensembles in the thermodynamic limit and the two curves are represented in Figure~\ref{fig:ch3_tlim}. 

%
%

\section{Transition with softening}
\label{sec:ch3_aimeta}

%
\begin{figure}[t!]
	\centering
	 \includegraphics[width=0.95\textwidth]{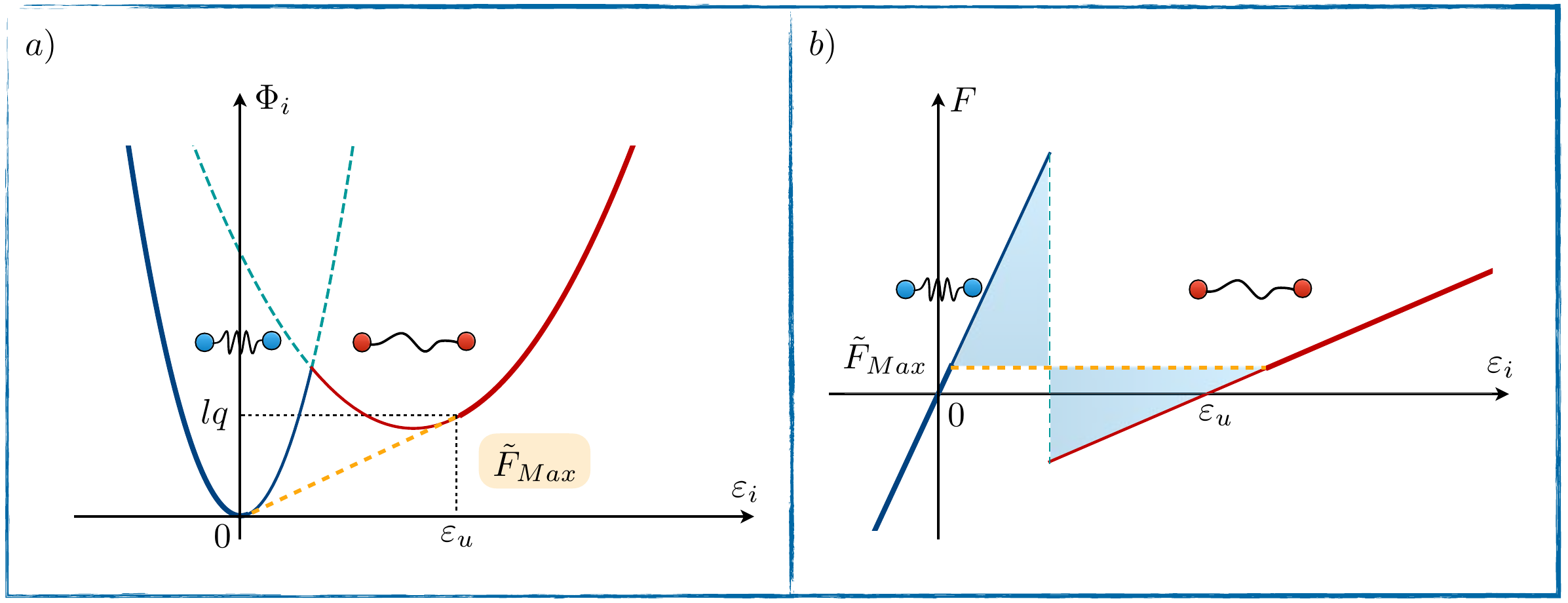}
	  \caption[Energy model of an unfolding protein with softening]{The figure shows the energy a) and the associated force b) for the generalized case introduced in the chapter, where different stiffnesses are considered for the two wells and non zero transition energy. In particular, the unfolded phase is softer than the folded one. The Maxwell stress is also represented with a dashed yellow line.}
	\label{fig:ch3_modelsoftening}
\end{figure}
%

In this section we generalize the model introduced in Figure~\ref{fig:ch3_model} of Section~\ref{sec:ch3_model} by considering different stiffnesses for the folded and the unfolded phases and a non-zero transition energy (\cite{luca2020}). In particular, here we analyze only the soft device hypothesis, both in the pure mechanical case and when the temperature is considered. As demonstrated in the previous section, also in this case the hard device (Helmholtz ensemble) can be obtained by performing a Laplace transform (see Appendix~\ref{appE}), but here we focus on the soft device highlighting the main features emerging when considering softening. We consider the energy in Figure~\ref{fig:ch3_modelsoftening}$_a$ for each bi-stable unit, and consequently, the potential energy of the molecule can be expressed as 
\begin{equation}
\Phi_m=l \sum_{i=1}^n \left(\frac{1}{2}\,k_i(\varepsilon_i-\varepsilon_u \chi_i)^2+q\,\chi_i\right),
\label{eq:ch3_softpotential}
\end{equation}
where, in addition to the same quantities introduced in~\ref{eq:ch3_eq:1}, $q$ represents the unfolding energy density per unit length. We consider different stiffnesses $k_i$ (having the dimension of a force) in the two wells such that
\begin{equation}
\begin{array}{l}
\chi_i = 0 \rightarrow k_i = k_f \quad \text{folded well},\vspace{0.2 cm}\\
\chi_i = 1 \rightarrow k_i = k_u \quad\text{unfolded well}.
\end{array}
\label{eq:ch3_chisoft}
\end{equation}
In particular, in the presence of the (enthalpic) transition energy term $q$, we obtain a non-null Maxwell stress, that is given by
\begin{equation}
\tilde{F}_{Max}=\frac{{F}_{Max}}{k_f}=\frac{\varepsilon_u}{\zeta}\left(\sqrt{1+\frac{2\,q\,\zeta}{k_f\,\varepsilon_u^2}}-1\right).
\label{eq:ch3_fMaxwell}
\end{equation}
This force corresponds to the classical equal-area condition in the force-extension diagram, represented by the blue regions and yellow line in Figure~\ref{fig:ch3_modelsoftening}$_b$. In~\eqref{eq:ch3_fMaxwell} we introduced the non-dimensional parameter $\zeta$, which measures the ratio between the stiffnesses of the two wells and reads
\begin{equation}
\zeta=\frac{ k_f-k_u}{k_u} \in  (0,+\infty).
\label{eq:ch3_zetasoft}
\end{equation}
Observe that for $\zeta \to 0$ the value of $k_f$ tends to $k_u$, and the case analyzed in the previous sections is obtained. On the other hand, if $\zeta \to +\infty$ we have that $k_u \to 0$. In this particular case, the well of the second phase becomes almost equal to a straight line, \textit{i.e.} the position of the unloaded strain of the second well $\varepsilon_u\to\infty$ and the phenomena of rupture can be described, as we show in the following chapters. Also in this case we consider the pulling device, which potential energy is again given by~\eqref{eq:ch3_eq:2} and hence the total energy is Equation~\eqref{eq:ch3_eq:6} where now $\Phi_m$ is given by~\eqref{eq:ch3_softpotential}.

\subsection{Mechanics with softening}
\label{sec:ch3_aimeta_2}

%
\begin{figure}[t!]
	\centering
	 \includegraphics[width=0.95\textwidth]{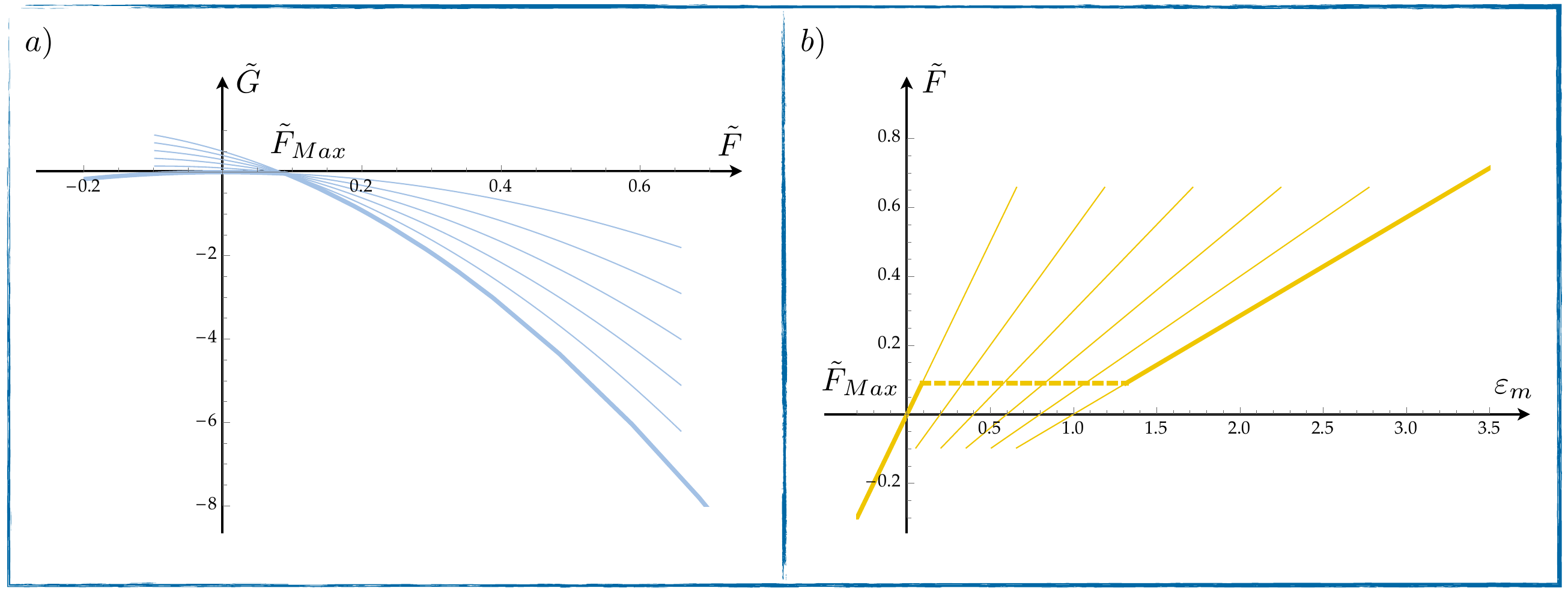}
	  \caption[Mechanical response of the system with softening]{Mechanical response of the system. Equilibrium energy in panel a) and force branches in panel b). Thin lines represent metastable configurations (local minima). Thick lines represent global energy minima, corresponding to the~\textit{Maxwell convention}, with a cooperative transition from the fully folded to the fully unfolded state at $\tilde{F}=\tilde{F}_{Max}$. Parameters: $k_f=1$ pN, $l=1$ nm, $\varepsilon_u=1$, $q=0.2$ pN, $\zeta=2.5$, $\gamma=0.6$ and $n=5$.}
	\label{fig:ch3_mecsoft}
\end{figure}
%

Following the analysis in Section~\ref{sec:ch3_mechanics}, we consider the mechanical limit in which temperature effects are neglected. The variational problem for the soft device is again
\begin{equation}
\min_{\varepsilon_{1},\dots,\varepsilon_n,\varepsilon_d}\,\, G_t(\varepsilon_{i},\varepsilon_d) = \Phi_t(\varepsilon_{i},\varepsilon_d) - F \left(l\sum_{i=1}^n \varepsilon_i+ \alpha L \varepsilon_d\right).
\label{eq:ch3_eq:8soft}
\end{equation}
but for the energy of the molecule, given by~\eqref{eq:ch3_softpotential}. Equilibrium gives 
\begin{equation}
\varepsilon_i = \frac{F}{k_i}+\varepsilon_u\chi_i \quad \text{with} \quad i=1,\dots,n \quad \text{and} \quad \varepsilon_d = \frac{F}{k_d},
\label{eq:ch3_eqsoft}
\end{equation}
and by using the definition of $G_t$ in~\eqref{eq:ch3_eq:8soft} and~\eqref{eq:ch3_eqsoft} the Gibbs (soft device) energy can be expressed as 
\begin{equation}
\tilde{G}_t=\frac{G_t}{L\,k_f}=-\frac{1}{2}\left(\frac{1}{\gamma}+\zeta \,\bar{\chi}\right)\tilde{F}^2-\varepsilon_u\,\bar{\chi}\left(\tilde{F}-\frac{q}{k_f\,\varepsilon_u}\right),
\label{eq:ch3_gsoft}
\end{equation}
where 
\begin{equation}
\tilde{F}=\frac{F}{k_f},
\end{equation}
is the (adimensionalized) assigned force and
\begin{equation}
\gamma=\frac{k_d}{k_d+\alpha k_f}\quad\text{with}\quad \gamma \in \,\,]0,1[ .
\label{eq:ch3_gammasoft}
\end{equation}
We highlight the difference of~\eqref{eq:ch3_gammasoft} with~\eqref{eq:ch3_gamma}, that is given by the stiffness $k_f$ instead of $k_m$. As previously described, by varying $\gamma$ it is possible to evaluate the effect of different values of the device stiffness with respect to the macromolecule stiffness. Notice also that, typically, in a biological macromolecule undergoing conformational transition the unfolded configuration is softer than the folded one and to reproduce this feature here we then that $k_f\ge k_u$ ($\zeta>0$) even though the extension to the opposite hypothesis is straightforward. 

The relation between force and average strain of the macromolecule can be evaluated by using~\eqref{eq:ch3_eqsoft}, obtaining 
\begin{equation}
\varepsilon_m = \bigl(1+\zeta\bar{\chi}\bigr)\tilde{F}+\varepsilon_u\bar{\chi}.
\label{eq:ch3_emsoft}
\end{equation}
From Equation~\eqref{eq:ch3_emsoft}, we notice again that the force response of the macromolecule is independent of the stiffness of the device when the force is applied. On the other hand, its effect can be observed in the total strain, when we consider macromolecule and device as a whole system. Indeed, by using~\eqref{eq:ch3_eq:5} and~\eqref{eq:ch3_emsoft} we get 
\begin{equation}
(1+\alpha)\,\varepsilon_t = \left(\frac{1}{\gamma}+\zeta\bar{\chi}\right)\tilde{F}+\varepsilon_u \bar{\chi}.
\label{eq:ch3_etsoft}
\end{equation}
Moreover, the minimization of the free energy with respect to the phase fraction $\bar \chi$ for the equilibrium solutions (Maxwell convention) leads to
\begin{equation}
\tilde{G}_{Max}=-\frac{\tilde{F}^2}{2\gamma}-\frac{\chi}{2}\,\left(\tilde{F}-\tilde{F}_{Max}\right)\left(\tilde{F}-\tilde{F}_{Max}+\frac{2\,\varepsilon_u}{\zeta}\,\sqrt{1+\frac{2\,q\,\zeta}{k_f\,\varepsilon_u^2}}\right)
\label{eq:ch3_Gfmax}
\end{equation}
where $\tilde{F}_{Max}$ is defined in \eqref{eq:ch3_fMaxwell}. 

Based on \eqref{eq:ch3_Gfmax}, we observe that the global energy minima correspond to the fully folded state with $\bar \chi=0$ for $\tilde F<\tilde F_{Max}$ and to the fully unfolded state with $\bar \chi=1$ for $\tilde F>\tilde F_{Max}$ as represented in Figure~\ref{fig:ch3_mecsoft} with bold lines. As the figure shows, the molecule behaviour under the \textit{Maxwell hypothesis} is cooperative with the molecule undergoing a fully conformational folded$\rightarrow$unfolded transition.

\subsection{Effects of temperature on a softening molecule}
\label{sec:ch3_aimeta_3}

Following the same procedure and calculation introduced in Section~\ref{sec:ch3_temp} and in Appendix~\ref{appE}, to consider temperature effects we may express the canonical partition function for the Gibbs ensemble as 
\begin{equation}
\mathcal{Z}_{\mathscr{G}} = \mathcal{K}_{\mathscr{G}}\sum_{p=0}^n\binom{n}{p}e^{\frac{\beta  l n}{2} \left[\left(\frac{1}{\gamma }+\zeta\frac{p}{n}\right)\frac{F^2}{k_f} +2 \varepsilon_u \frac{p}{n} \left(F-\frac{q}{\varepsilon_u}\right)\right]},
\end{equation}
that is simplified in
\begin{equation}
\mathcal{Z}_{\mathscr{G}}=\mathcal{K}_{\mathscr{G}} \,e^{\frac{\beta l n}{2 k_f \gamma} F^2} \biggl[1+e^{\frac{\beta l}{2 k_f }\bigl( \zeta F^2 +2k_f \varepsilon_u \left(F -\frac{q}{\varepsilon_u}\right)\bigr)}\biggr]^n,
\label{eq:ch3_zgsoft}
\end{equation}
where $\mathcal{K}_{\mathscr{G}}$ is the constant term
\begin{equation}
 \mathcal{K}_{\mathscr{G}}= l^n\, \alpha L (2 \pi)^{\frac{n+1}{2}}\left(\frac{\beta}{m}\right)^{\frac{n}{2}}\left(\frac{\beta}{M}\right)^{\frac{1}{2}}\left(\frac{2\pi}{\beta \alpha L k_d}\right)^{\frac{1}{2}}\left(\frac{2\pi}{\beta l}\right)^{\frac{n}{2}}\left(\frac{1}{k_u}\right)^\frac{p}{2}\left(\frac{1}{k_f}\right)^\frac{n-p}{2}.
\end{equation}

By definition, the Gibbs free energy is 
\begin{equation}
\mathcal{G}=-\frac{1}{\beta}\ln \mathcal{Z}_{\mathscr{G}}.
\end{equation}
Thus, we can evaluate the expectation value of the total strain as
\begin{equation}
\langle \varepsilon_t \rangle = \frac{1}{\beta L(1+\alpha)}\frac{1}{\mathcal{Z}_{\mathscr{G}}}\frac{\partial}{\partial F}\mathcal{Z}_{\mathscr{G}} = \frac{1}{(1+\alpha)}\Biggl[\left(\frac{1}{\gamma}+\zeta\,\langle \bar{\chi} \rangle\right)\frac{F}{k_f} +\varepsilon_u\langle \bar{\chi} \rangle \Biggr],
\label{EPST}
\end{equation}
where
\begin{equation}
\langle \bar{\chi} \rangle = \frac{e^{\frac{\beta l}{2 k_f }\bigl[\zeta F^2 +2k_f \varepsilon_u \left(F -\frac{q}{\varepsilon_u}\right)\bigr]}}{1+e^{\frac{\beta l}{2 k_f }\bigl[ \zeta F^2 +2k_f \varepsilon_u \left(F -\frac{q}{\varepsilon_u}\right)\bigr]}}
\end{equation}
 is the expectation value of the unfolded fraction. Eventually, the expectation value of the average strain of the molecule is 
\begin{equation}
\langle \varepsilon_m \rangle = \Bigl(1+\zeta\langle \bar{\chi} \rangle\Bigr)\frac{F}{k_f}+\varepsilon_u \langle \bar{\chi} \rangle.
\end{equation}
%

%
\begin{figure}[t!]
	\centering
	 \includegraphics[width=0.95\textwidth]{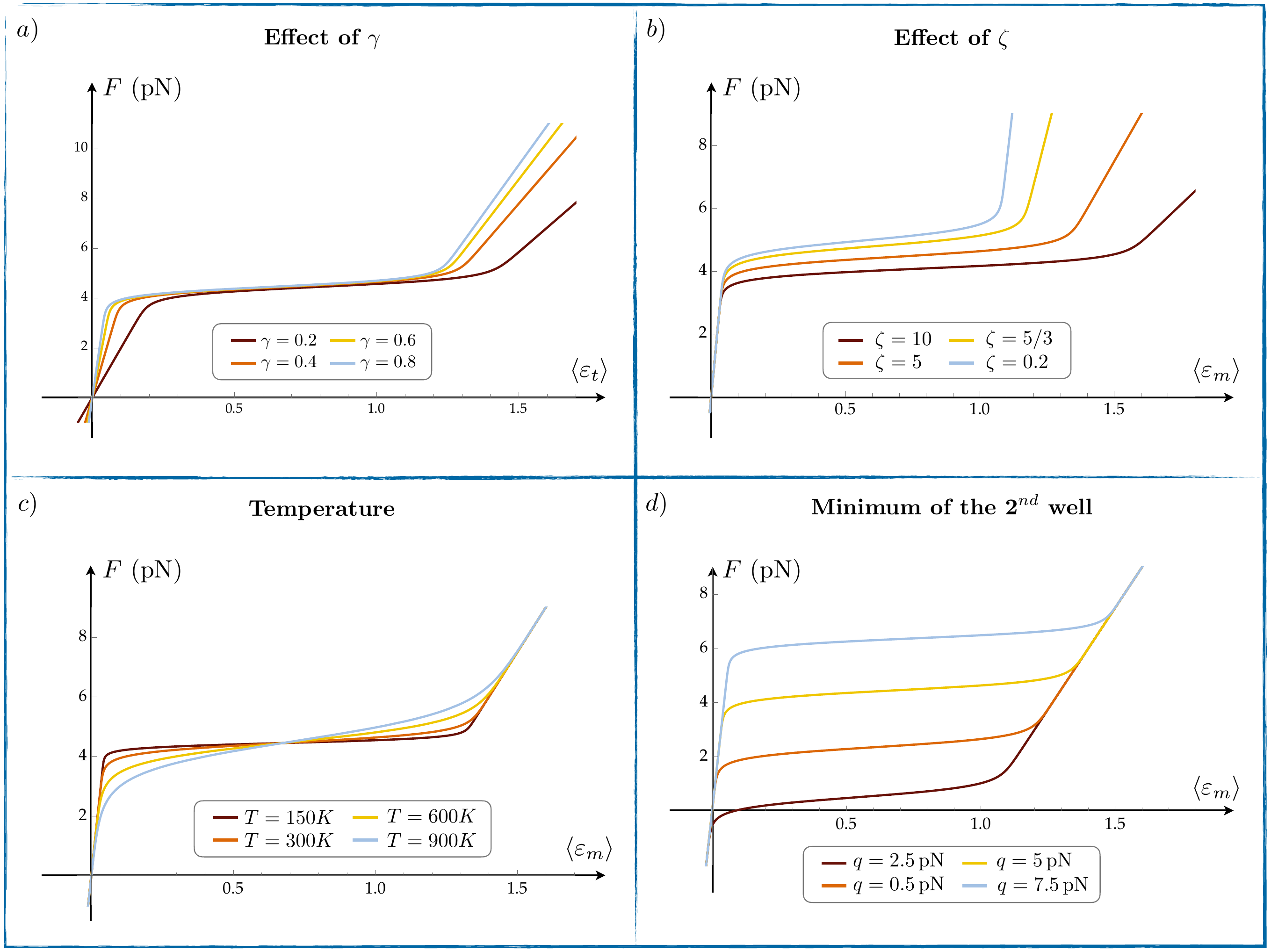}
	  \caption[Temperature effects on the molecule with softening]{Molecular response in the Gibbs ensemble. In panel a) the effect of the device is shown whereas in panel b) the effect of the stiffness ratio between the two phases is represented. In panel c) the hardening effect due to temperature is observed. In panel d) is shown the effect of the transition energy. The parameter varying in each figure are displayed in the legends while the unspecified ones are given by the following value:  $n=5$, $T=300$ K, $l=20$ nm, $\varepsilon_u=1$, $q=0.5$ pN, $k_f=90$ pN, $\alpha=0.1$, $\zeta=5$, $\gamma=0.6$.}
	\label{fig:ch3_tempsoft}
\end{figure}
%

In Figure~\ref{fig:ch3_tempsoft} we show the behaviour of the system in terms of molecular response under the hypothesis of applied force when different stiffnesses for the two wells and non-zero transition energy are considered. The pulling device is also part of the system, and it is interesting to observe that the limit case for $\gamma = 1$ corresponds to an extremely rigid device while in the opposite limit, $\gamma = 0$, the force goes to zero because the extremely soft device doesn't transmit any effect on the molecule. The effect of the relative stiffnesses of the phases, measured by $\zeta$ is represented in Figure~\ref{fig:ch3_tempsoft}$_b$, where, as the ratio increases the second well becomes softer. In Figure~\ref{fig:ch3_tempsoft}$_d$ we show the effect of different values of the energy density $q$ leading to different values of the Maxwell stress which increases at increasing value of $q$. Lastly, due to temperature, the system shows a hardening behaviour with the unfolding force that grows as the unfolded fraction $\bar \chi$ grows.

\section{Discussion}
\label{sec:ch3_discussion}

In this chapter, I developed an exact \textit{unified} mathematical model, in the framework of equilibrium Statistical Mechanics, quantifying the effect of the handling device stiffness in stretching experiments on a chain of bi-stable elements. Among the many important examples, one can think of SMFS tests on biomolecules. To fix the idea, here we referred to AFM experiments on molecules constituted by a chain of domains (\textit{e.g.} $\alpha$-helix and $\beta$-sheets) undergoing conformational (folded $\rightarrow$ unfolded) transition, but the model is quite general and it is important to remark that the proposed framework can be extended to the case of mechanical molecular interactions (\cite{rosa:2004}).

\begin{figure}[h!]
	\centering
	 \includegraphics[width=0.94 \textwidth]{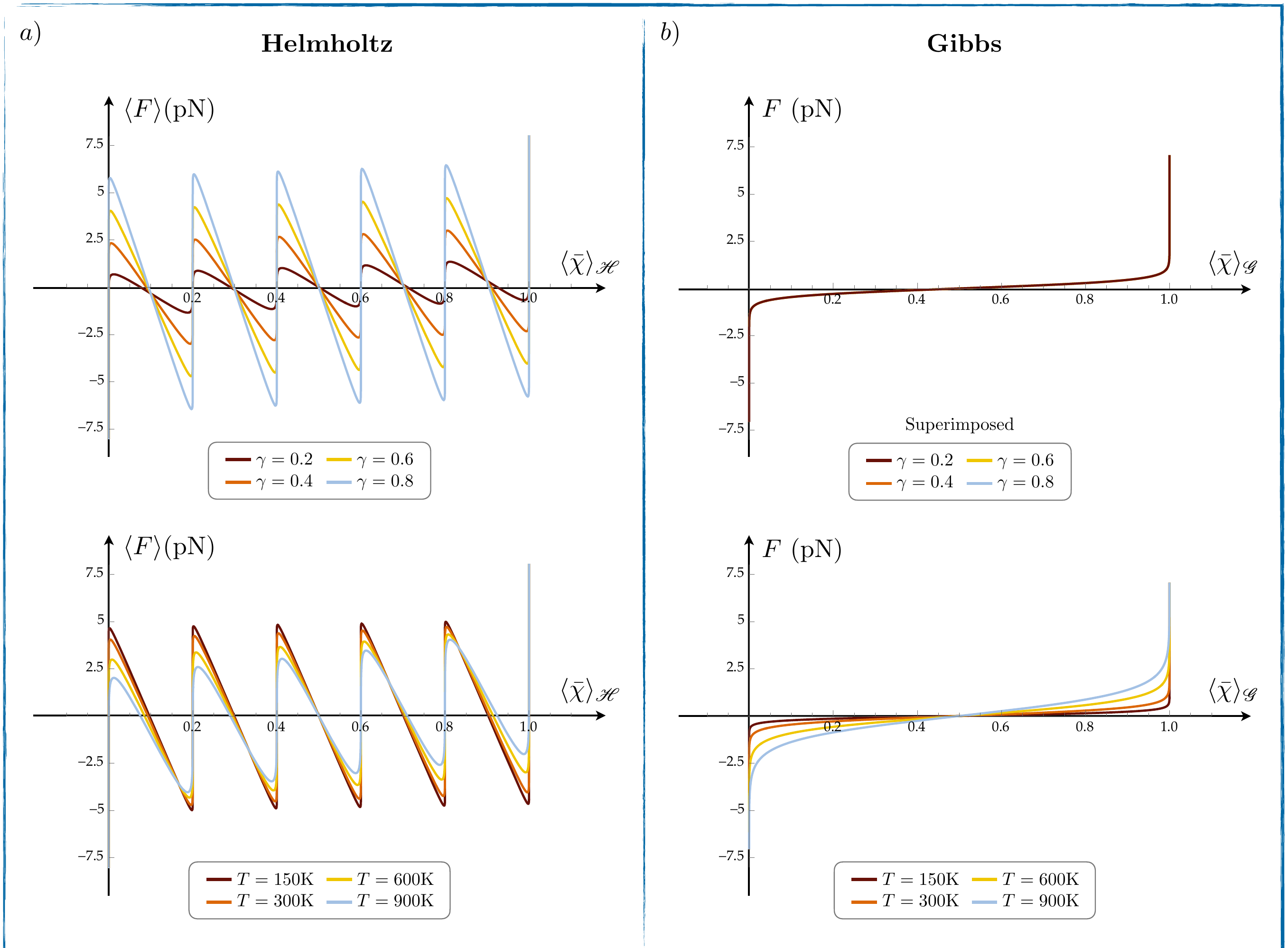}\\\vspace{-0.15cm}
	  \includegraphics[width=0.938 \textwidth]{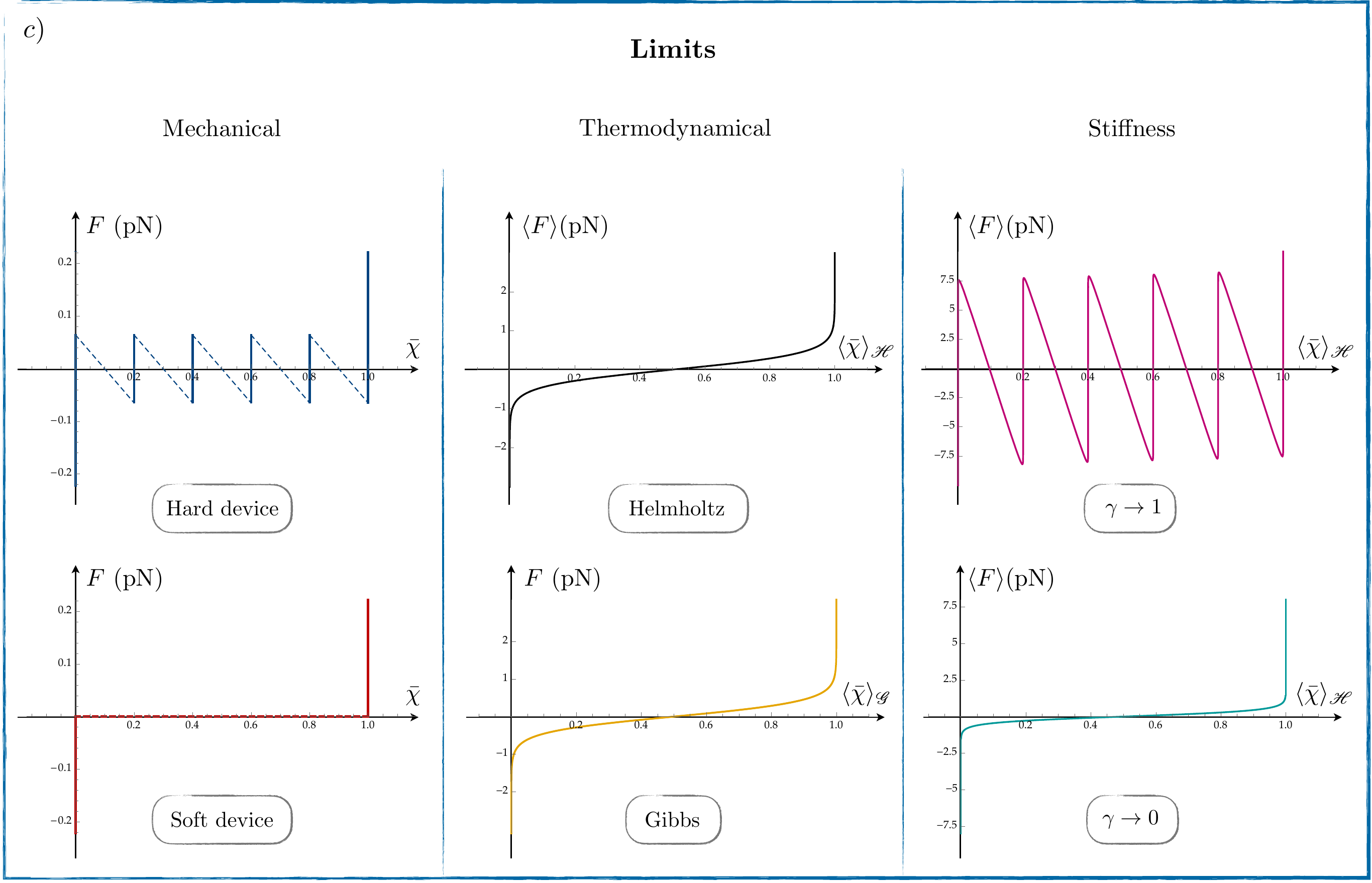}
	  \caption[Summary of the molecule behavior under different limit regimes and boundary conditions]{Summary of the molecule behavior in terms of phase fraction evolution under different boundary conditions and limit regimes.}
	\label{fig:ch3_chis}
\end{figure}

Specifically, following the approach in (\cite{fp:2019}), I considered the chain and the device microcantilever (or ancilla molecule) as a unique thermodynamical system. I analyzed both the assumptions of assigned displacement and applied force acting on the cantilever.  The former case is described by the so-called Helmholtz ensemble, whereas the latter is described by the so-called Gibbs ensemble, linked by an integral Laplace transform. As we show, several limit cases of theoretical interest can be deduced by our general approach: the ideal hard and soft device (neglecting the device stiffness), the thermodynamic limit and the mechanical limit (neglecting entropy effects). In particular, I deduced that in both statistical ensembles and all considered limits, the molecular response can be formally described by the following relations:
\begin{equation}
\begin{array}{l}
 F  =  k_m \gamma \Bigl[(1+\alpha)\,\varepsilon_t-\varepsilon_u \bar{\chi}\Bigr],\vspace{0.2 cm}\\
 \varepsilon_m = \varepsilon_u \bar{\chi}   + \gamma \Bigl[(1+\alpha)\,\varepsilon_t-\varepsilon_u\bar{\chi} \Bigr],
 \vspace{0.2 cm}\\
  F = k_m( \varepsilon_m  -  \varepsilon_u  \bar{\chi}),  
  \end{array}
  \label{eq:ch3_vfa}
\end{equation}
where, with a slight abuse of notation, we identified the value of the variables with their expectation values depending on the specific case. Indeed, the \textit{only} difference among all considered possibilities lies in the expectation value of the unfolded fraction $\bar \chi$. Accordingly, to summarize all different behaviours analyzed in previous sections in Figure~\ref{fig:ch3_chis} we represent the phase fraction evolutions during the molecule unfolding. 

%
%

The important influence of the device on the mechanical response in the hard device is shown in Figure~\ref{fig:ch3_helmholtz} and Figure~\ref{fig:ch3_chis}$_a$.  As the device stiffness decreases, the unfolding force decreases and the transition becomes more cooperative. This behaviour is coherent with the experimental results in (\cite{bustamante:2010}) where the authors use an optical tweezer to study topological transitions in proteins. It is important to remark that this can lead to huge overestimation or underestimation of the force thresholds of the molecule. In this perspective, we point out that extended literature in the field neglects the stiffness effect and considers the ideal cases when the total molecule displacement is assigned (\textit{ideal hard device}) or the force acting on the molecule is assigned (\textit{ideal soft device}). 

In passing, I observe that these limit behaviours can be deduced in the model by considering the limit of rigid device-molecule connection: $\gamma \to 1$. Indeed, as I show in Appendix~\ref{appE}, in these ideal cases I obtain the same formal expressions in \eqref{eq:ch3_vfa} with phase fractions
\begin{equation}
\begin{array}{l}\displaystyle
 \langle \bar{\chi}^{ideal} \rangle_{\mathscr{H}} (\beta, \varepsilon_t)=\frac{\displaystyle\sum_{p=0}^{n}\binom{n}{p} \frac{p}{n}\,e^{-\frac{\beta k_m l n}{2}\bigl(\varepsilon_u\frac{p}{n}-\varepsilon_t\bigr)^2}}{\displaystyle\sum_{p=0}^{n}\binom{n}{p} e^{-\frac{\beta k_m l n}{2}\bigl(\varepsilon_u\frac{p}{n}-\varepsilon_t\bigr)^2}},
\vspace{0.4 cm}\\ \displaystyle
\langle \bar{\chi}^{ideal} \rangle_{\mathscr{G}} (\beta, F) = \frac{\displaystyle e^{\beta l  \varepsilon_u F}}{\displaystyle 1+e^{\beta l  \varepsilon_u F}}.
\end{array}
\end{equation}
The same expressions are obtained by the general model in the $\gamma\rightarrow1$ limit. Moreover, it is interesting to observe that if I consider the hard device, when $\gamma \rightarrow 0$, the molecule response approaches the behaviour of the ideal soft device (see Figure~\ref{fig:ch3_chis}$_c$ down-right). Conversely, when $\gamma\rightarrow 1$ the molecular response coincides with the one of the ideal hard device (see Figure~\ref{fig:ch3_chis}$_c$ up-right). Thus, by varying the stiffness ratio $\gamma$ all the ranges of behaviour between the two limit cases can be attained. The important result is that I \textit{cannot} define a unique material response of the molecule. To this matter, it is important to highlight that the stiffness ratio $\gamma$ can be estimated to predict the experimental response based on two considerations (\cite{fp:2019}). The cantilever stiffness, or more generally the stiffness of the device is usually a known parameter provided by the owner of the instrument whereas the stiffness of the molecule can be deduced by the unfolding energy (see (\cite{detom:2013,linke:2002})), as in the case of the titin PEVK described in (\cite{fp:2019}). 

The soft device boundary conditions are described in Figure~\ref{fig:ch3_gibbs} and Figure~\ref{fig:ch3_chis}$_b$. According to the experimental behaviour, I observe a monotonic transition force-elongation path. Interestingly, I obtain that in this case, the molecule behaviour is independent of $\gamma$  and $n$, as shown in Figure~\ref{fig:ch3_gibbs}$_b$. In this specific case of the soft device, I have also considered the case of softening (see Section~\ref{sec:ch3_aimeta}), where the second unfolded configuration is softer than the folded one, a typical effect attained in real molecules. 

The temperature effect is shown in Figure~\ref{fig:ch3_helmholtz}$_c$, Figure~\ref{fig:ch3_gibbs}$_c$ and in Figure~\ref{fig:ch3_chis}$_{a,b}$ for the two ensembles. This should be compared with the limit regime when temperature effects are neglected, studied in Section~\ref{sec:ch3_mechanics} and represented in Figure~\ref{fig:ch3_energy} and Figure~\ref{fig:ch3_chis}$_c$ left. Interestingly, while when I neglect temperature effects the unfolding transition corresponds to constant forces thresholds when temperature effects are considered I may observe a hardening behaviour with the unfolding force growing with the unfolded percentage. This effect is in accordance with the experiments in (\cite{rief:1997}) showing such a hardening behaviour of a recombinant titin molecule with identical $\beta$-sheets domains. Observe that the hardening grows as the temperature grows.

The influence of the discreteness size is shown in Figure~\ref{fig:ch3_helmholtz}$_b$. The force threshold decreases as the number of elements $n$ increases. This result agrees with the experimental observations on native and reconstructed titin molecules in (\cite{rief:1997}). The thermodynamic limit ($n\rightarrow\infty$) has been studied in Section~\ref{sec:ch3_tl} and it is represented in terms of phase fraction $\langle \bar \chi\rangle$ in Figure~\ref{fig:ch3_chis}$_c$ center. Specifically, by extending the results in (\cite{manca:2012}), where the authors study generalizations of the Freely Jointed Chain and of the Worm Like Chain models with extensible bonds, we demonstrated the equivalence in the thermodynamic limit of the molecule response under the hard and soft device. This result was numerically described in (\cite{mancagiordano:2014}). Moreover, in (\cite{fp:2019}) a numerical comparison with the assumption of  WLC energies is performed, and the obvious curvature effects on the force-displacement curves that are numerically obtained in the WLC model, the force thresholds and the locations of the jumps are shown to be consistent in both cases.

In conclusion, In this Chapter I deduced a general framework, able to analytically describe in the rate-independent regime the importance of the handle stiffness, considering also temperature effects. As I show, all other cases analyzed in the literature on SMFS experiments and models, such as thermodynamic limit (\cite{mancagiordano:2014,efendiev:2010}), mechanical limit (\cite{puglisi:2002}) and ideal hard and soft devices (\cite{puglisi:2000,manca:2012,makarov:2009}) can be obtained as limit cases of this general model.

	\clearpage


%
\renewcommand{\thefigure}
{\arabic{chapter}.\arabic{figure}}
\setcounter{figure}{0}

\renewcommand{\thetable}
{\arabic{chapter}.\arabic{table}}
\setcounter{table}{0}

\renewcommand{\theequation}
{4.\arabic{equation}}
\setcounter{equation}{0}

\chapter{Nucleation and propagation phenomena mediated by competing entropic and interface energy terms}
\chaptermark{Introduction}
\label{ch_4}
\vspace{1.8cm}

The description of finite microstructure domains evolution for multistable materials represents an important topic in material science that has been longly at the centre of research activities for both its theoretical and technological importance (\cite{allen:1979}). Indeed, multi-states systems are of interest in several fields such as material science, biology, medicine, and engineering, both for natural and artificial materials. More recently, a large effort in this field has been devoted to the design of new high-performing metamaterials. In this case, a successful new inverse approach is adopted, with a multistable microstructure designed to obtain the desired macroscopic behaviour (\cite{bilal:2017,shang:2018}). 

In the thermomechanical three-dimensional case, the problem is quite complex and it asks for a description not only of the local structure of an interface but also of its curvature and topological properties, necessary for the deduction of its excess energy (\cite{gurtin:1989}). Ericksen's pioneering work (\cite{ericksen:1975}) proposed a variational energetic approach in non-linear elasticity theory with non-convex energy densities whereas the hypothesis of describing the microstructure evolution on the base of energy minimization was already proposed in the field of linear elasticity by Khachaturyan (\cite{khachaturyan:1983}) and coworkers. Grounded on these works, important theoretical advances in the field of modern continuum mechanics (\cite{ball:1989}) and calculus of variation (\cite{muller:1999}) successfully described basic aspects of phase transitions and associated effects at the microstructure level. In particular, the analysis of the complex problem of minimization of the non (quasi) convex energy allowed important results regarding the phase fraction, the orientation of the interfaces and the stress corresponding to the phase evolution. Nonetheless, continuum variational approaches minimizing the elastic energy density are not able to capture important aspects unless they introduce interfacial energy effects and non-local interactions. As a matter of fact, these energetic contributions are crucial for a detailed description of the observed microstructure evolution (\cite{shaw:1997}) and they are fundamental not only from a theoretical point of view but also for a correct portrayal of the material behaviour and for the design of new materials (\cite{truskinovsky:2000}). In particular, the minimization of the (non-convex, bulk) elastic energy only cannot describe the existence of internal length scales regulating the effectively observed phenomena leading to discrete processes of transition with nucleation and propagation of finite domains, as it can be found in the works of (\cite{abeyaratne:1996,benichou:2013}) and references therein. Based on these observations, different `augmented' energetic models have been proposed. Schematically, on one side higher gradient (\cite{carr:1984,triantafyllidis:1993}) antiferromagnetic energy terms, representing surface energy contributions, have been added to the energy density, possibly deduced as a result of compatibility effects as in the case of polymer necking in the classical work presented in Reference (\cite{coleman:1983}). On the other side, non-local, both ferromagnetic and antiferromagnetic energy terms have been considered, thus introducing the desired length scale needed to describe the insurgence of periodic microstructures typically observed in solid-solid martensitic phase transitions as in the works of (\cite{truskinovsky:2000,muller:1999}) and references therein. 

Starting from the pioneering work of Villaggio (\cite{villaggio:1977}), a parallel framework in the field of discrete mechanics was proposed. In the simpler setting, one can consider a lattice of elements with non-convex energy densities and the presence of an intrinsic discreteness length-scale (\cite{zanzotto:1992}). As a result, it is possible to describe fundamental properties such as energy barriers, metastable states, quasi-plastic and pseudo-elastic behaviours (\cite{puglisi:2000,puglisi:2005}). However, also this approach was not able to distinguish between different microstructures with the same phase fraction and, therefore, important details of the nucleation and phase fronts evolution phenomena could not be described. The model has been generalized by considering non local energy terms in Refs. (\cite{vainchtein:2004,puglisi:2006,puglisi:2007}) where the authors were able to describe the important effect of a stress peak to distinguish the different phenomena of nucleation and propagation. In analogy with higher gradient approaches in elasticity, only single wall solutions with a fully cooperative phase transition could be described. On the other hand, in Refs. (\cite{puglisi:2006,puglisi:2007}) a more detailed analysis of the influence of boundary conditions lets the author energetically distinguish between internal and boundary nucleation, with possible two-phase fronts. These features are often observed for example in polymer necking localization phenomena. Moreover, the model has been extended to the case of configurational transitions in protein macromolecules to study the successive unfolding events observed in protein stretched under single molecule force spectroscopy experiments (\cite{detom:2013,manca:2013}). In this case, the two energy wells correspond to two different configurations of the protein domains, with a folded $\rightarrow$ unfolded transition. Finally, based on multiscale approaches this class of models has been adopted to deduce the macroscopic features of protein materials in Refs. (\cite{detommasi:2015,detommasi:2017}), showing a significant predictivity of the experimental behaviour.

In this chapter, I am interested in a description of the possibility of entropic effects delivering a temperature dependence of the stress peak, the size of the nucleation domain and, in general, the cooperative nature of the transition. Schematically, temperature effects can have crucial effects on the behaviour of multistable materials for three different reasons. First, increasing or decreasing temperature may result in a change in the potential energy of the different possible configurations such as in the case of \acf{SMA} (\cite{tanaka:1986}). Second, temperature effects regulate the capability of overcoming energy barriers, thus influencing the key role of metastability, hysteresis and rate effects, as can be seen in References (\cite{tanaka:1986,keller:1998,puglisi:2002}). Third, when rate effects are excluded and the system can relax to the global minimum of the free energy, the entropic term can favour solutions with multiple interfaces (\cite{muller:2001}). This is particularly relevant when the transition is controlled by low enthalpies such as in the case of conformational transitions of units with weak bonds in biological materials or when the scale of the system is very small. In this case, we have a competition between interfacial energy (favouring solutions with low numbers of interfaces) and entropic energy terms (favouring solutions with a high number of interfaces). As a matter of fact, under these hypotheses, the behaviour of the system is described by the expectation values of relevant observables with respect to the thermal fluctuations and, therefore, the correct framework for the analysis is given by a Statistical Mechanics approach. 

Specifically, by extending the results in (\cite{puglisi:2006,puglisi:2007}) and including temperature effects, in this chapter, I consider the model previously introduced in Chapter~\ref{ch_3} of the chain with bi-stable elements undergoing a two-phase transition, such as the folded$\to$unfolded configurations in a macromolecule or austenitic and martensitic phases in SMA. In analogy with the continuous model with higher-order gradient energy terms (\cite{triantafyllidis:1993}), here we mimic the important role of interfacial energy by considering next to nearest neighbour (\acs{NNN}) interactions. In a recent paper (\cite{florio:2020}), these models have been used to consider the important temperature effects to study a peeling behaviour such as in geckoes pad, thus extending the approach in Ref. (\cite{puglisi:2013}). In this case, the bistable elements are breakable so that the second state (detached state) is characterized by zero force and stiffness. Such type of model has been widely adopted also in the literature on DNA denaturation in analogy with the Peyrard-Bishop model (\cite{peyrard:2004}). As before, here I first reproduce the zero-temperature limit behaviour, that has been studied in (\cite{vainchtein:2004,puglisi:2006,puglisi:2007}). Then, due to the effectiveness of the Statistical Mechanics approach in describing thermal effects observed in multistable systems (\cite{efendiev:2010,makarov:2009,giordano:2018,giordano:2017,benedito:2018,benedito:2018:2,benedito:2020,fp:2019,luca2019}), here I study the effect of temperature on the model introduced, and I apply the theoretical analytical results obtained to the specific case of Memory Shape Nanowires, reproducing the experimental behaviour.

\begin{figure}[h!]
\centering
  {\includegraphics[width=0.95\textwidth]{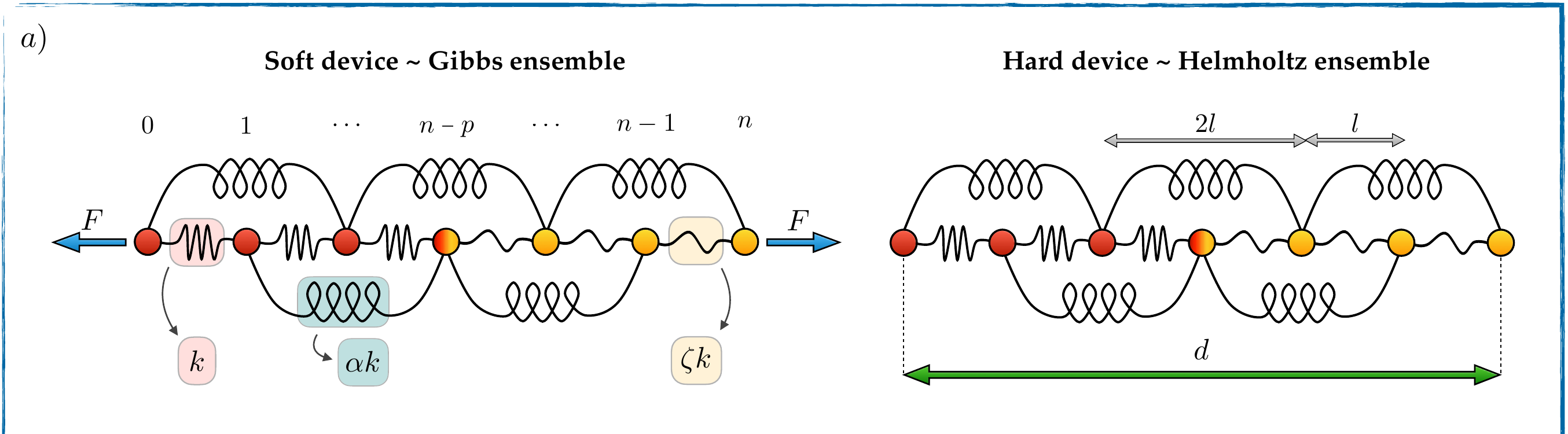}}\\\vspace{-0.08cm}
  {\includegraphics[width=0.952\textwidth]{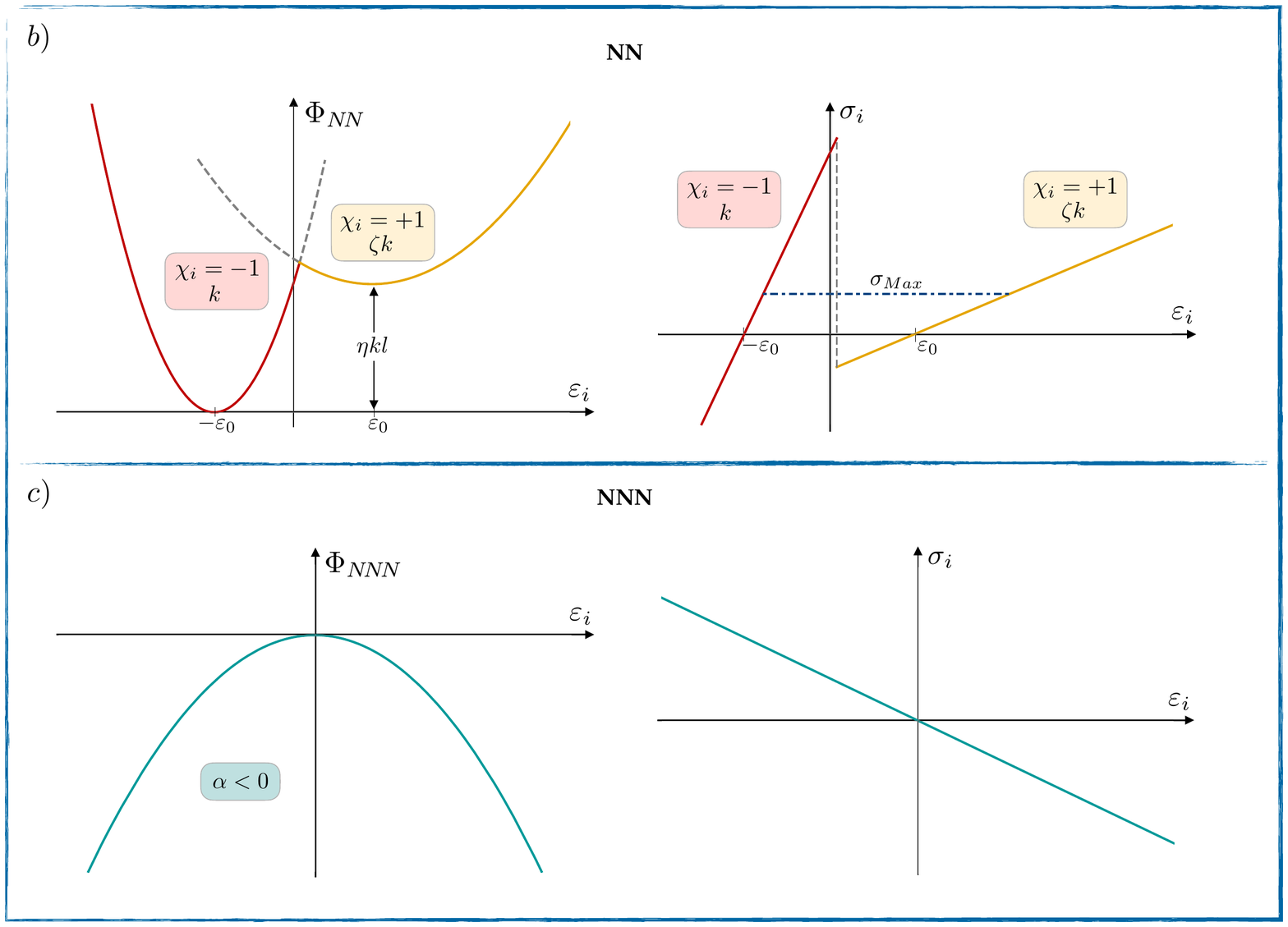}}
  \caption[Mechanical scheme of the bistable chain with NNN interactions]{In panel a) the scheme of the mechanical model in the two considered cases of assigned force $F$ and assigned total displacement $d$ is shown. The energy and force for the NN units is represented in panel b) whereas the harming energy and force of the NNN is shown in panel c).}
  \label{fig:ch4_model}
\end{figure}

\section{Model}

To introduce non local interaction in the unfolding chain model of Chapter~\ref{ch_3}, following the approaches of (\cite{vainchtein:2004,puglisi:2006,puglisi:2007}), I consider a chain of $n+1$ units, schematized in Figure~\ref{fig:ch4_model}$_a$, linked by \ac{NN} bistable units with reference length $l$ and next-to-nearest-neighbor (\acs{NNN}) harmonic springs with natural length $2l$.  Each NN link is characterized by two wells' energy, corresponding to two different material states. They can represent two different phases (\cite{puglisi:2000}), as in solid-solid phase transition, or two different conformational (folded and unfolded) states as in the case of unfolding phenomena in protein molecules (\cite{detom:2013}).

In the same spirit of the previous chapter and keeping the same notation for the common quantities, we introduce the ``spin" variable $\chi_j$, which defines the phase of each unit and in this case takes the values 
\begin{equation}
\chi_j=\begin{cases}  	-1  	& \text{phase one} \\
				+1	& \text{phase two}
\end{cases}.
\end{equation}
The potential energy of the NN springs reproduced in Figure~\ref{fig:ch4_model}$_b$, can be expressed as follows:
\begin{equation}
\Phi_{N\!N}(\chi_{j},\varepsilon_{j})=\frac{1}{2}k\, l\Bigg[\left(\varepsilon_j+\varepsilon_0\right)^2\frac{\left(1-\chi_j\right)}{2}+\bigg(\zeta \left(\varepsilon_j-\varepsilon_0\right)^2+2\eta\bigg)\frac{\left(1+\chi_j\right)}{2}\Bigg],
\end{equation}
where $j=1,...,n$, $-\varepsilon_0$ and $\varepsilon_0$ are the reference strains of the first and second well, respectively, $k/l$  is the stiffness of the first well ($k$ has the dimension of a force), $\zeta k/l$ is the stiffness of the second phase and $\eta$ measures the transition energy with respect to the ground state. 

Next, we consider the simplest case of non local interactions by introducing  NNN harmonic oscillators of length $2l$, with energy
\begin{equation}
\Phi_{N\!N\!N}(\varepsilon_{j},\varepsilon_{j+1})=\frac{1}{2}\alpha\,k\,l \,\big(\varepsilon_j+\varepsilon_{j+1}\big)^2, \qquad j=1,...,n-1,
\label{eq:ch4_nnn}
\end{equation}
where $\alpha$ measures the relative stiffness of the non local vs local springs. Observe that while in this chapter we are interested in the (concave) case with $\alpha <0$, when non local energy terms penalize interfaces formation (ferromagnetic hypothesis), the following calculations hold also in the (convex) case with $\alpha>0$, when the formation of the interface is favoured (antiferromagnetic hypothesis). Moreover, while we deduce our theoretical relations for general values of $\alpha$, with the aim of getting analytic results we then focus on the case $|\alpha|\ll1$, so that the NNN energy represents a perturbative term that let us distinguish between solutions with the same phase fraction, but with a different number of interfaces.

The total (adimensionalized with respect to $lk$) energy of the system is then
\begin{multline}
\varphi(\{\chi_{j}\},\{\varepsilon_{j}\})=\frac{1}{2} \left\{\sum_{j=1}^{n}\Bigg[\left(\varepsilon_j+\varepsilon_0\right)^2\frac{\left(1-\chi_j\right)}{2}\right.\\
\left.+\bigg(\zeta \left(\varepsilon_j-\varepsilon_0\right)^2+2\eta\bigg)\frac{\left(1+\chi_j\right)}{2}\Bigg]+\sum_{j=1}^{n-1}\alpha \Big(\varepsilon_j+\varepsilon_{j+1}\Big)^2\right\}.
\label{eq:ch4_toten}
\end{multline}
With the aim of simplifying the notation, we may introduce the following functions of the spin variable:
\begin{equation}
\begin{split}
L(\chi_j)=&\begin{cases}	1 \quad &\text{if} \quad \chi_j=-1\\
				\zeta \quad &\text{if} \quad \chi_j=+1
	\end{cases},\\
	&\\
m(\chi_j)=&\begin{cases}	-\varepsilon_0           & \text{if} \quad \chi_j=-1\\
					\zeta\varepsilon_0   &\text{if} \quad \chi_j=+1
		\end{cases},\\
		&\\
q(\chi_j)=&\begin{cases}	\varepsilon_0^2/2  &\text{if} \quad \chi_j=-1\\
					\zeta\varepsilon_0^2/2+\eta \quad &\text{if} \quad \chi_j=+1
		\end{cases}.\\
		\label{eq:ch4_func}
\end{split}
\end{equation}
Accordingly, we may define the vectors
\begin{equation}	\quad
	\boldsymbol{\varepsilon}=
	\begin{pmatrix}
	\varepsilon_1	\\
	\dots 		\\
	\varepsilon_j 	\\
	\dots		\\
	\varepsilon_n	\\ 
	\end{pmatrix}, \quad
	\boldsymbol{\chi}=
	\begin{pmatrix}
	\chi_1	\\
	\dots 		\\
	\chi_j 	\\
	\dots		\\
	\chi_n	\\ 
	\end{pmatrix}, \quad
	\boldsymbol{m}=
	\begin{pmatrix}
	m_1	\\
	\dots 		\\
	m_j 	\\
	\dots		\\
	m_n	\\
	\end{pmatrix},
	\quad
	\boldsymbol{q}=
	\begin{pmatrix}
	q_1	\\
	\dots 		\\
	q_j 	\\
	\dots		\\
	q_n	\\
	\end{pmatrix},
	\quad
	\boldsymbol{1}=
	\begin{pmatrix}
	1		\\
	\dots 	\\
	\dots	\\
	\dots	\\
	1		\\
	\end{pmatrix},
	\label{eq:ch4_vect1}
\end{equation}
where $m_j=m(\chi_j)$, $q_j=q(\chi_j)$ and each element of the vector $\boldsymbol{\chi}$ is the ``spin'' of the $j$-th NN unit. In addition, we may introduce the matrices
\begin{equation}
\boldsymbol{L}=
	\begin{pmatrix}
	L_1						& 		& 		& 		& \boldsymbol{0} \\
	 				 	 	& \ddots 	& 		& 		& \\
	 				  		&  		& L_j		& 		& \\
	 				  		&  		& 		& \ddots	& \\
	\boldsymbol{0} 	 	 	&  		& 		& 		& L_n\\
	\end{pmatrix}, 
	\quad
	\boldsymbol{A}=
	\begin{pmatrix}
	1				& 1	& 		& 			& \boldsymbol{0} \\
	1				& 2	&  1		& 	 		& \\
	 			  	& \ddots 	& \ddots	& \ddots 	& \\
	 				&  		& 1		& 2		& 1\\
	\boldsymbol{0} 	&  		& 		& 1 		& 1\\
	\end{pmatrix},	
	\label{eq:ch4_mat}
\end{equation}
where $L_j=L(\chi_j)$. Finally, by using~\eqref{eq:ch4_func},~\eqref{eq:ch4_vect1} and~\eqref{eq:ch4_mat} we can compactly rewrite  the energy of the system as
\begin{equation}
\varphi(\boldsymbol{\chi},\boldsymbol{\varepsilon})=\frac{1}{2}\boldsymbol{J}\boldsymbol{\varepsilon}\cdot\boldsymbol{\varepsilon}-\boldsymbol{m}\cdot\boldsymbol{\varepsilon}+q,
\label{eq:ch4_en}
\end{equation}
where $\boldsymbol{J} = \boldsymbol{J}(\boldsymbol{\chi}) = \boldsymbol{L}(\boldsymbol{\chi})+\alpha \boldsymbol{A}$, $\boldsymbol{m}=\boldsymbol{m}(\boldsymbol{\chi})$ and $q=q(\boldsymbol{\chi})=\boldsymbol{q}(\boldsymbol{\chi})\cdot\boldsymbol{1}$.

\vspace{0.4cm} \noindent \textbf{Remark --} While the previous description shows that the following analysis can be extended to the fully general case, to focus on the physical results and get simply interpretable analytical formulas, in the following we study the case of $\zeta=1$ and $\eta=0$, \textit{i.e.} I assume identical wells with no energy gap. Thus, with the aforementioned hypothesis we have 
\begin{equation}
\boldsymbol{L}=\boldsymbol{I}, \qquad q=n\,\varepsilon_0^2/2, \qquad\boldsymbol{m}=\varepsilon_0\boldsymbol{\chi}.
\label{eq:ch4_9}
\end{equation}
Observe that in the following we first obtain all the formulas without this assumption in terms of $\boldsymbol{J}$ and $\boldsymbol{m}$ and then we specialize them to the simpler case of \eqref{eq:ch4_9}.

\section{Equilibrium mechanics}
\label{sec:ch4_eq}

In the same spirit of the analysis performed in the previous chapter and with the aim of getting physical insight into the problem of non local interactions and highlighting their role, in this section, we summarize the equilibrium configurations of the system when temperature effects are neglected. Here we study both the case of soft and hard devices, which, as demonstrated in references (\cite{luca2019}), can be considered as limit regimes of a general problem in which the stiffness of the pulling device regulates the effective boundary conditions. We refer to (\cite{puglisi:2006,puglisi:2007}) for a more detailed analysis when also the interaction with the loading experimental apparatus is considered.

\subsection{Soft device}

First, let us consider the case of the soft device with applied force $F$  (Figure~\ref{fig:ch4_model}$_a$). The potential energy of the systems is
\begin{equation}
g(\boldsymbol{\chi},\boldsymbol{\varepsilon},\sigma)=\varphi(\boldsymbol{\chi},\boldsymbol{\varepsilon})-\sigma\sum_{j=1}^{n}\varepsilon_j=\frac{1}{2}\boldsymbol{J}\boldsymbol{\varepsilon}\cdot\boldsymbol{\varepsilon}-\left(\boldsymbol{m}+\sigma \boldsymbol{1}\right)\cdot\boldsymbol{\varepsilon}+q,
\label{eq:ch4_gen}
\end{equation}
where $\sigma=F/k$ is the adimensional load. By minimizing the energy~\eqref{eq:ch4_gen} equilbrium at assigned phase configuration $\boldsymbol{\chi}$ requires
\begin{equation}
\frac{\partial g(\boldsymbol{\chi},\boldsymbol{\varepsilon},\sigma)}{\partial \boldsymbol{\varepsilon}}=\boldsymbol{J}\boldsymbol{\varepsilon}-(\boldsymbol{m}+\sigma \boldsymbol{1})=\boldsymbol{0}.
\end{equation}
Therefore, the equilibrium strains are
\begin{equation}
\boldsymbol{\varepsilon}_{eq}=\boldsymbol{J}^{-1}(\boldsymbol{m}+\sigma\boldsymbol{1}).
\label{eq:ch4_eqstrain}
\end{equation}
By definition, the average strain is 
\begin{equation}
\bar{\varepsilon}:=\frac{1}{n}\sum_{j=1}^{n}\varepsilon_j,
\end{equation}
so that we obtain equilibrium strain and energy 
\begin{equation}
\begin{split}
\bar{\varepsilon}_{eq}&=\frac{1}{n}\big(\boldsymbol{\varepsilon}_{eq}\cdot\boldsymbol{1}\big)=\frac{1}{n}\Big[\boldsymbol{J}^{-1}(\boldsymbol{m}+\sigma\boldsymbol{1})\cdot\boldsymbol{1}\Big],\\
&\\
g_{eq}&=-\frac{1}{2}\boldsymbol{J}^{-1}(\boldsymbol{m}+\sigma\boldsymbol{1})\cdot(\boldsymbol{m}+\sigma\boldsymbol{1})+q.
\end{split}
\label{eq:ch4_eq}
\end{equation}
Following (\cite{puglisi:2006}) and the evaluation in Appendix~\ref{appF}, the inverse of the tridiagonal Hessian matrix $\boldsymbol{J}$ that is non-singular in the hypothesis of small $\alpha$, can be expressed as 
\begin{equation}
\boldsymbol{J}^{-1}=\sum_{j=0}^{\infty}(-\alpha)^{j}\boldsymbol{L}^{-1}(\boldsymbol{A}\boldsymbol{L}^{-1})^{j}
\label{eq:ch4_JJ}.
\end{equation}
In the case here analyzed, \textit{i.e.} for small $\alpha$ and considering the hypotheses in~\eqref{eq:ch4_9}, we have $\boldsymbol{L}=\boldsymbol{I}$ so that~\eqref{eq:ch4_JJ} reads 
\begin{equation}
\boldsymbol{J}^{-1}\simeq\boldsymbol{I}-\alpha \boldsymbol{A}.
\label{eq:ch4_J}
\end{equation}
A remark is in order. In the case analyzed in this chapter the non-local energy term in~\eqref{eq:ch4_nnn} is chosen with $\alpha<0$ in order to prevent the formation of too many phase interfaces, thus a ferromagnetic-type interaction is considered. For a generic $\alpha$, it may be then possible that the total energy in~\eqref{eq:ch4_toten} is not positive-defined and may assume negative values. In this case, to describe the physical phenomenon illustrated in the introduction of this chapter, we choose small values of $\alpha$ and we consider the expansion of the inverse matrix $\boldsymbol{J}^{-1}$ developed in Appendix~\ref{appF} up to the first order. From a direct inspection of~\eqref{eq:ch4_J} it is easy to show that the condition that ensures the positivity of the energy and that the equilibrated reference configurations are stable is $0<\alpha<1/4$ and all the values used both in the theoretical analysis and in the experimental comparison (see Section~\ref{sec:ch4_exp}) are chosen within this range. 

Then, let us introduce the  identities 
\begin{equation}
\begin{split}
	\boldsymbol{\chi}\cdot\boldsymbol{\chi}&=n,\\
	\boldsymbol{\chi}\cdot\boldsymbol{1}&=2\,p-n,\\
	\boldsymbol{1}\cdot\boldsymbol{1}&=n,\\
	\end{split} \qquad \qquad
	\begin{split}
	\boldsymbol{A}\boldsymbol{\chi}\cdot\boldsymbol{\chi}&=4(n-i-1),\\
	\boldsymbol{A}\boldsymbol{\chi}\cdot\boldsymbol{1}&=4(2\,p-n)-2(\chi_1+\chi_n),\\
	\boldsymbol{A}\boldsymbol{1}\cdot\boldsymbol{1}&=4(n-1),
\end{split}
\label{eq:ch4_id}
\end{equation}
where $p$  the  {\it is number of elements in the second phase} and $i$ is {\it   the number of interfaces}, \textit{i.e.} the number of times NN adjacent links have different phases, $\chi_1$ and $\chi_n$ assign the phase of the boundary elements. Finally, by using~\eqref{eq:ch4_9},~\eqref{eq:ch4_J} and~\eqref{eq:ch4_id}, Equation~\eqref{eq:ch4_eq} can be rewritten as  
\begin{equation}
\begin{split}
 \bar{\varepsilon}_{eq}&=\left(1-4\alpha\right)\bigg[\sigma+\big(2 \bar{\chi}-1\big)\varepsilon_0\bigg]+\frac{2\alpha}{n}\bigg[2\sigma+\big(\chi_1+\chi_n\big)\varepsilon_0\bigg],\\
&\\
\frac{g_{eq}}{n}&=-\left(1-4\alpha\right)\Bigg[\frac{\sigma^2}{2}+\big(2 \bar{\chi}-1\big)\varepsilon_0\sigma\Bigg]-\frac{2\alpha}{n}\Bigg[\sigma^2+\big(\chi_1+\chi_n\big)\varepsilon_0\sigma +i\,\varepsilon_0^2+\big(n-1\big)\varepsilon_0^2\Bigg],
\end{split}
\label{eq:ch4_eqgeppe}
\end{equation}
where  $\bar{\chi}:=p/n$ is the {\it phase fraction}.

\begin{figure}[h!]
\centering
  \includegraphics[width=0.95\textwidth]{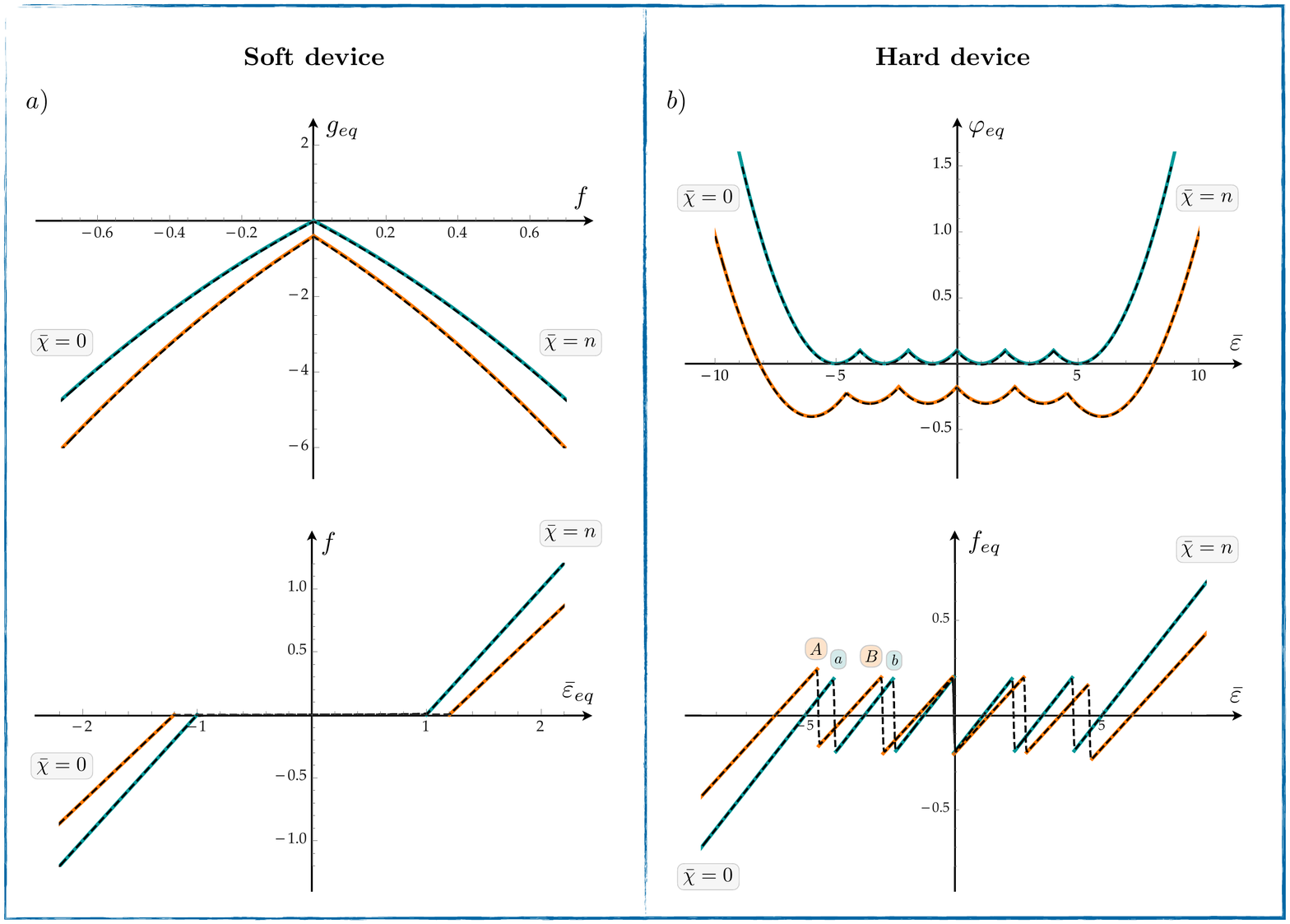}
  \caption[Mechanical limit of the soft and hard device with NNN interactions]{Mechanical equilibrium. Energies and stress-strain diagrams in the soft a) and hard b) device. Cyan curves represent the cases when only NN interactions are considered, orange curves show the behavior with the presence of non local terms. Dashed black lines are calculated with the Statistical Mechanics approach when the temperature goes to zero (see Sections~\ref{sec:ch4_gibbs}, \ref{sec:ch4_helmholtz}). Parameter: $n=5$, $\alpha=0$ in cyan curves and $\alpha=-0.05$ in the orange ones, $\tilde\beta\rightarrow\infty$ in the black dashed lines, $\varepsilon_0=1$.}
  \label{fig:ch4_mechanics}
\end{figure}

Observe that the equilibrium configurations in~\eqref{eq:ch4_eqgeppe}, when NNN interactions are not considered ($\alpha=0$), only depend on the phase fraction $\bar{\chi}$. On the contrary, when non local terms are introduced, the number of interfaces distinguishes configurations with the same phase fraction. In particular, solutions with a larger number of interfaces are energetically penalized. It is easy to verify that the \textit{global minimum} of the energy is attained when all bistable units are in the first phase for $\sigma<0$ and in the second phase when $\sigma>0$, as shown in figure~\ref{fig:ch4_mechanics}$_a$. Consequently, under the Maxwell hypothesis, when the configurations of the system correspond always to the global energy minimum, we observe that as the force is increased the chain undergoes an `instantaneous' transition from the homogeneous state in the first phase to the homogeneous state in the second one with a fully cooperative phase transition and no interfaces.

\subsection{Hard device}

Consider now the case of assigned total displacement $d$, as schematized on the right side of Figure~\ref{fig:ch4_model}$_a$, and introduce its dimensionless measure
\begin{equation}
\delta = \frac{d}{l} = \sum_{j=1}^{n}\varepsilon_j=\boldsymbol{\varepsilon}\cdot\boldsymbol{1}.
\label{eq:ch4_d}
\end{equation}

The equilibrium strain equations at fixed configuration $\boldsymbol{\chi}$ are again given by~\eqref{eq:ch4_eqstrain}, where $\sigma$ is the Lagrange multiplier measuring the force conjugate to the fixed total displacement in~\eqref{eq:ch4_d}. By using Equation~\eqref{eq:ch4_9}, in the case of small $\alpha$, the epression for the average strain~\eqref{eq:ch4_d} can be rewritten as
\begin{equation}
\bar \varepsilon=\frac{\delta}{n}=\frac{1}{n}\Bigl[\boldsymbol{J}^{-1}\left(\boldsymbol{m}+\sigma_{eq}\boldsymbol{1}\right)\cdot\boldsymbol{1}\Bigr]\simeq\frac{1}{n}\biggl[\left(\boldsymbol{1}\cdot\boldsymbol{1}-\alpha\boldsymbol{A}\boldsymbol{1}\cdot\boldsymbol{1}\right)\sigma_{eq}+(\boldsymbol{\chi}\cdot\boldsymbol{1}-\alpha\,  \boldsymbol{A}\boldsymbol{\chi}\cdot\boldsymbol{1})  \varepsilon_0 \biggr],
\label{eq:ch4_delta}
\end{equation}
where we indicate by $\sigma_{eq}$ the equilibrium stress. Accordingly, we obtain the equilibrium energy
\begin{equation}
\begin{split}
\varphi_{eq}&=-\frac{1}{2}\boldsymbol{J}^{-1}\big(\boldsymbol{m}+\sigma_{eq}\boldsymbol{1}\big)\cdot\big(\boldsymbol{m}+\sigma_{eq}\boldsymbol{1}\big)+q+\delta\,\sigma_{eq}\simeq\\
&\simeq-\frac{1}{2}\big(\boldsymbol{1}\cdot\boldsymbol{1}-\alpha\boldsymbol{A}\boldsymbol{1}\cdot\boldsymbol{1}\big)\sigma_{eq}^2-\big(\boldsymbol{\chi}\cdot\boldsymbol{1}-\alpha\,  \boldsymbol{A}\boldsymbol{\chi}\cdot\boldsymbol{1}-\delta\big)\,  \varepsilon_0 \, \sigma_{eq}+\frac{1}{2}\alpha\, \varepsilon_0^2 \,\boldsymbol{A}\boldsymbol{\chi}\cdot\boldsymbol{\chi}.
\label{eq:ch4_heneq}
\end{split}
\end{equation}
Eventually, by using ~\eqref{eq:ch4_id} we obtain
\begin{equation}
\sigma_{eq}=\frac{\displaystyle\bar\varepsilon-\bigg[\big(1-4\alpha\big)\big(2 \bar{\chi}-1\big)+\frac{2\alpha}{n}\big(\chi_1+\chi_n\big)\bigg]\varepsilon_0}{\displaystyle1-\frac{4\alpha}{n}(n-1)},
\label{eq:ch4_eqstress}
\end{equation}
and the equilibrium energy per element is given by 
\begin{equation}
\frac{\varphi_{eq}}{n}=\frac{1}{2}\Biggl\{ \frac{\Big[\big(\left(1-4\alpha\right)\left(2\,\bar{\chi}-1\right)+\frac{2\alpha}{n}(\chi_1+\chi_n)\big)
\varepsilon_0-\bar \varepsilon\Big] ^2}{1-\frac{4\alpha}{n}\left(n-1\right)}+\frac{4\alpha}{n}\big(n-i-1\big)\varepsilon_0^2\Biggr \}.
\label{eq:ch4_harden}
\end{equation}

In this case, the \textit{global minima} of the energy with respect to the assigned rescaled displacement $\delta$ are given by solutions either homogeneous or with one single interface, as shown in Figure~\ref{fig:ch4_mechanics}$_b$. More in detail, in agreement with well known experimental observations, under the hypothesis of hard device the system changes state between the two homogeneous phase branches following a `sawtooth' equilibrium path. Interestingly, the stress corresponding to the nucleation of the new phase, represented by the point $A$ in the force-displacement diagram of Figure~\ref{fig:ch4_mechanics}$_b$, is higher than the stress corresponding to the propagation of the interface (\textit{i.e.} point $B$ in the Figure). It is important to notice that this effect is not observed in absence of non local interactions (points $a$ and $b$). Moreover, one may show that by increasing the value of $n$ or of the non-local energy $\alpha$ the nucleation can correspond to the cooperative transition of more elements. Such behaviour is discussed in detail in (\cite{puglisi:2007}), whereas here we are interested in the energetic competition of surface energy terms (measured by the parameter $\alpha$) and entropic energy terms.

\section{Applied stress: Gibbs ensemble}
\label{sec:ch4_gibbs}

To describe the important effect of temperature, consider now our prototypical model with non local interactions in the framework of equilibrium Statistical Mechanics. Under isotensional conditions (\textit{i.e} assigned stress $\sigma$, soft device) we derive the canonical partition function in the Gibbs ensemble (denoted by the subscript $\mathscr{G}$), in order to study the system in thermal equilibrium (\cite{weiner:1983}). By definition and by using the  energy in~\eqref{eq:ch4_gen}, we have
\begin{equation}
\mathcal{Z}_{\mathscr{G}}(\sigma)=\sum_{\{\chi_{j}\}}\int_{\mathbb{R}^n}e^{-\tilde{\beta} g}\mbox{d}\boldsymbol{\varepsilon}=\sum_{\{\chi_{j}\}}\int_{\mathbb{R}^n}e^{-\tilde{\beta}\bigl[\frac{1}{2}\boldsymbol{J}\boldsymbol{\varepsilon}\cdot\boldsymbol{\varepsilon}-(\boldsymbol{m}+\sigma\boldsymbol{1})\cdot\boldsymbol{\varepsilon}+q\bigr]}\mbox{d}\boldsymbol{\varepsilon},
\label{eq:ch4_z1}
\end{equation}
where  $\tilde{\beta}=\beta l k$ and $\beta=1/(k_BT)$, with $k_B$ the Boltzmann constant and $T$ the absolute temperature. Here, we have summed over the discrete spin variables and we integrated the continuous strains.\vspace{0.2 cm}

\noindent{\bf Remark --} It is again important to point out that by following (\cite{efendiev:2010,florio:2020}) to obtain the analytical result we assume that the two wells are extended beyond the spinodal point so that at the given configuration $\boldsymbol{\chi}$ we may integrate all the strain all over $\mathbb{R}$ and avoid error functions. In (\cite{florio:2020}) the authors numerically showed that in the temperature regimes of interest for real experiments this approximation does not influence the energy minimization.\vspace{0.2 cm}

 A straightforward evaluation of the Gaussian integral leads to
\begin{equation}
\mathcal{Z}_{\mathscr{G}}(\sigma)=\sum_{\{\chi_{j}\}}\sqrt{\frac{(2\pi)^n}{\mbox{det}(\boldsymbol{J})}}e^{\,-\tilde{\beta}\bigl[-\frac{1}{2}\boldsymbol{J}^{-1}(\boldsymbol{m}+\sigma\boldsymbol{1})\cdot(\boldsymbol{m}+\sigma\boldsymbol{1})+q\bigr]},
\label{eq:ch4_gformal}
\end{equation}
where, as expected, at the exponent, we find the equilibrium energy in~\eqref{eq:ch4_eq}. Accordingly, preserving the same first-order approximation in~\eqref{eq:ch4_J} with $\alpha$ kept small and by using~\eqref{eq:ch4_9}, we obtain
\begin{equation}
\mathcal{Z}_{\mathscr{G}}(\sigma)\simeq\sqrt{\frac{(2\pi)^n}{\mbox{det}(\boldsymbol{I}-\alpha \boldsymbol{A})}}\sum_{\{\chi_{j}\}}e^{\,\tilde{\beta}\bigl\{\frac{1}{2}\bigl[\left(\varepsilon_0\boldsymbol{\chi}+\sigma\boldsymbol{1}\right)\cdot\left(\varepsilon_0\boldsymbol{\chi}+\sigma\boldsymbol{1}\right)-\alpha\boldsymbol{A}\left(\varepsilon_0\boldsymbol{\chi}+\sigma\boldsymbol{1}\right)\cdot\left(\varepsilon_0\boldsymbol{\chi}+\sigma\boldsymbol{1}\right)\bigr]-q\bigr\}}.
\label{eq:ch4_z2}
\end{equation}

Here we want to remark that while~\eqref{eq:ch4_z2} gives numerically the possibility of evaluating the partition function, we may observe that by using~\eqref{eq:ch4_id}, in the approximation of small $\alpha$ the exponent in~\eqref{eq:ch4_z2} reduces to the equilibrium energy in \eqref{eq:ch4_eqgeppe}. This expression depends on the configuration ${\chi_j}$ only by the number of unfolded elements $p$, the number of interfaces $i$ and by the boundary phases $\chi_1$ and $\chi_n$. In order to obtain explicit formulas of the partition function and optimize the numerical calculation, we can neglect the contribution of the boundary energy terms in~\eqref{eq:ch4_eqgeppe} given by $-\frac{2\alpha}{n}(\chi_1+\chi_n)$, while keeping the contribution of $\chi_1$ and $\chi_n$ in the evaluation of the phase fraction $\bar{\chi}=p/n $ and of the number of interfaces $i$. This is justified by the observation that this energy term is proportional to $\alpha/n$ and can be shown to be small for large $n$ and small enough $\alpha$ as compared with the other energy terms. We refer to Refs. (\cite{puglisi:2006,puglisi:2007}) for a more detailed discussion about boundary effects, that instead can be particularly important when temperature effects are neglected.

Under the described assumption we need to evaluate the combinatorial coefficient $\mathcal{W}_{p,i}$ counting the number of configurations corresponding to the values of $p$ and $i$. Following the approach in References (\cite{ising:1925,denisov:2005,seth:2016}), it is possible to obtain that 
\begin{equation}
\mathcal{W}_{p,i}=\begin{cases}  \mathcal{W}_{odd}(p,i) &\text{if} \quad i \quad \text{odd},\\
\mathcal{W}_{even}(p,i) &\text{if} \quad i \quad \text{even},\\
1  &\text{if} \quad i=0 ,  p=0 \text{ or } p=n,\\
0 &\text{if} \quad i=0 ,  0<p<n,\\
\end{cases}
\label{eq:ch4_comb}
\end{equation}
with
\begin{equation}
\mathcal{W}_{odd}(p,i)=2\binom{p-1}{\frac{i-1}{2}}\binom{n-p-1}{\frac{i-1}{2}}\Theta\left(p-\frac{i+1}{2}\right)\Theta\left(n-p-\frac{i+1}{2}\right),
\end{equation}
and
\begin{equation}
\begin{split}
\mathcal{W}_{even}(p,i)&=\binom{p-1}{\frac{i}{2}}\binom{n-p-1}{\frac{i}{2}-1}\Theta\left(p-\frac{i+2}{2}\right)\Theta\left(n-p-\frac{i}{2}\right)+\\
&\\
&+\binom{n-p-1}{\frac{i}{2}}\binom{p-1}{\frac{i}{2}-1}\Theta\left(p-\frac{i}{2}\right)\Theta\left(n-p-\frac{i+2}{2}\right),
\end{split}
\end{equation}
where $\Theta$ is the Heaviside step function ($\Theta(x)=0$ if $x<0$ and $\Theta(x)=1$ if $x\geq 0$). 

Finally, using the combinatorial coefficient in~\eqref{eq:ch4_comb}, we obtain the canonical partition function in the Gibbs ensemble
\begin{equation}
\mathcal{Z}_{\mathscr{G}}(\sigma)=\mathcal{K}_{\mathscr{G}}\sum_{p=0}^{n}\sum_{i=0}^{n-1}\mathcal{W}_{p,i}\,e^{\displaystyle\,\Gamma_{p,i}(\sigma)},
\end{equation}
where
\begin{equation}
\mathcal{K}_{\mathscr{G}}=\sqrt{\frac{(2\pi)^n}{\mbox{det}(\boldsymbol{I}-\alpha \boldsymbol{A})}}
\label{eq:ch4_kkk}
\end{equation}
is a noninfluential constant and the exponent, by using~\eqref{eq:ch4_id}, can be written as
\begin{equation}
\Gamma_{p,i}(\sigma)=\frac{n\tilde{\beta}}{2}\Bigg\{\big(1-4\alpha\big)\bigg[4\,\frac{p}{n}\,\varepsilon_0\,\sigma+\big(\sigma-\varepsilon_0\big)^2\bigg]+\frac{4\alpha}{n}\, i \,\varepsilon_0^2-\varepsilon_0^2+\frac{4\alpha}{n}\big(\varepsilon_0^2+\sigma^2\big)\Bigg\}.
\end{equation}

By definition, the Gibbs free energy is (\cite{weiner:1983,manca:2012})
\begin{equation}
\mathcal{G}(\sigma)=-\frac{1}{\tilde{\beta}}\mbox{ln}\,\mathcal{Z}_{\mathscr{G}}(\sigma),
\end{equation}
and one may now evaluate the expectation value of the strain, which is the conjugate variable (up to $n$ since $\delta=n\bar \varepsilon$) to the applied force $\sigma$, as
\begin{equation}
\langle \bar \varepsilon\rangle =-\frac{1}{n}\frac{\partial}{\partial \sigma}\mathcal{G}(\sigma)=\frac{1}{n\tilde{\beta} \mathcal{Z}_{\mathscr{G}}(\sigma)}\frac{\partial}{\partial \sigma}\mathcal{Z}_{\mathscr{G}}(\sigma),
\end{equation}
that takes the form
\begin{equation}
\begin{split}
\langle \bar \varepsilon\rangle &=\displaystyle \left(1-4\alpha\right)\Big[2\varepsilon_0\langle \bar{\chi}\rangle_{\mathscr{G}}+(\sigma-\varepsilon_0)\Big]+\frac{4\alpha}{n}\sigma,\\
&\\
\langle \bar{\chi}\rangle_{\mathscr{G}} &=\frac{\displaystyle\sum_{p=0}^{n}\sum_{i=0}^{n-1}\mathcal{W}_{p,i}\,\bar{\chi} \,e^{\displaystyle\,\Gamma_{p,i}(\sigma)}}{\displaystyle \sum_{p=0}^{n}\sum_{i=0}^{n-1}\mathcal{W}_{p,i}\,e^{\displaystyle\,\Gamma_{p,i}(\sigma)}},
\end{split}
\end{equation}
where $\langle \bar{\chi}\rangle_{\mathscr{G}}$ is the expectation value of the phase fraction $\bar{\chi}=p/n$.

Interestingly we formally obtain (up to the neglected boundary term) the same expression of the zero-temperature limit in~\eqref{eq:ch4_eqgeppe} with the value of $\bar{\chi}$ substituted by its expectation value. Moreover, we can also compute the expectation value of the number of interfaces as
\begin{equation}
\langle i\rangle_{\mathscr{G}}=\frac{\displaystyle\sum_{p=0}^{n}\displaystyle\sum_{i=0}^{n-1}\mathcal{W}_{p,i}\,i\, e^{\displaystyle\,\Gamma_{p,i}(\sigma)}}{\displaystyle \sum_{p=0}^{n}\sum_{i=0}^{n-1}\mathcal{W}_{p,i}\,e^{\displaystyle\,\Gamma_{p,i}(\sigma)}}.
\end{equation}
%

%
\begin{figure}[t!]
\centering
  {\includegraphics[width=0.95\textwidth]{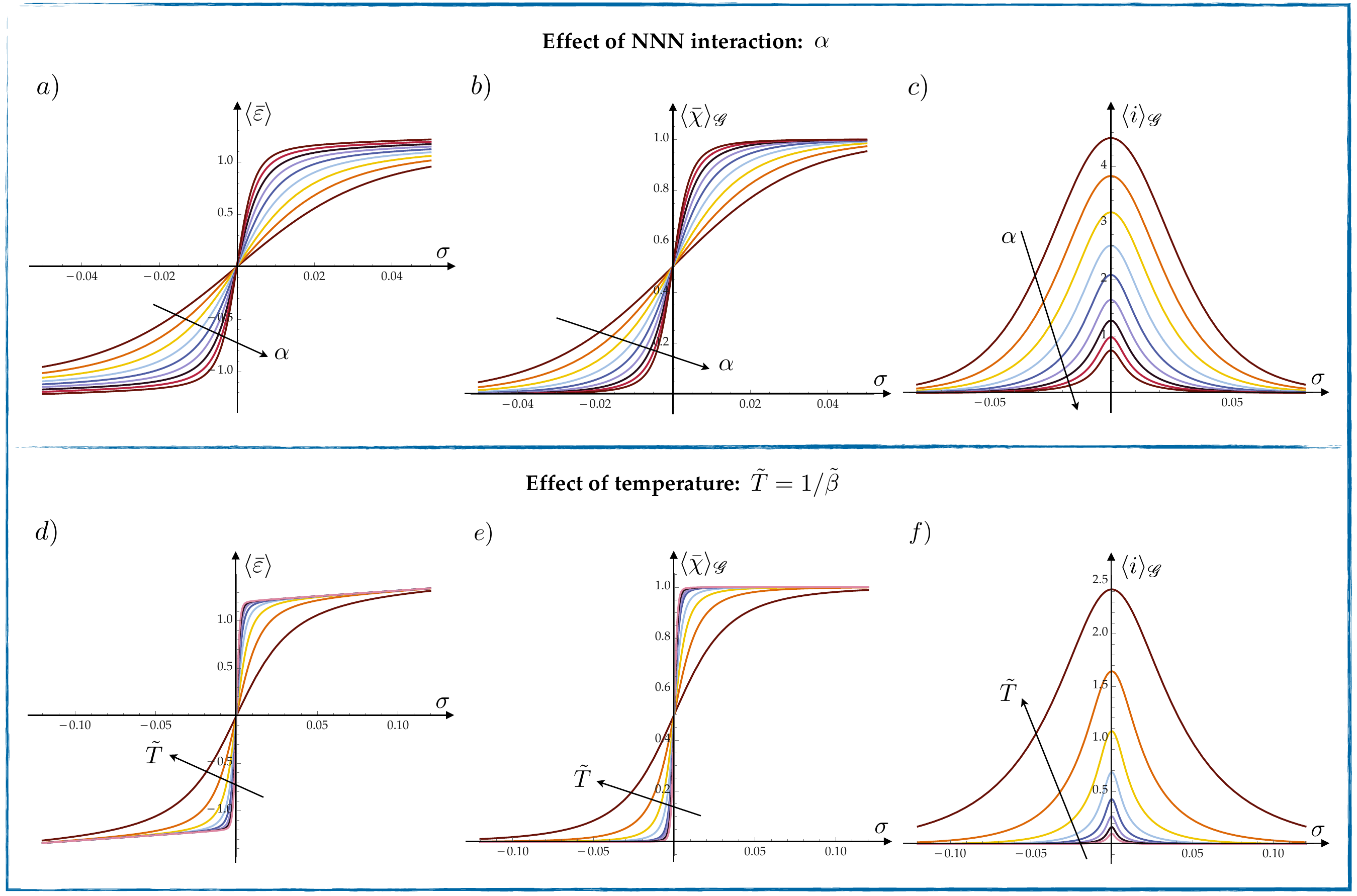}}
  \caption[Temperature effects in the Gibbs ensemble]{Gibbs ensemble. Effects of interfacial energy (measured by $\alpha$) and temperature ($\tilde T=1/\tilde \beta$) on the transition behavior.  We show the effects of $\alpha$ and $\tilde T$ on the stress-strain diagrams in a) and d), on the expectation value of the phase fraction $\chi$ in b) and e) and on the expectation value of the number of interfaces $i$ in c) and f), respectively. Parameters: $n=10$, $\varepsilon_0=1$. In the first row $\tilde{\beta}=30$, $\alpha=0.0\rightarrow-0.04$ with a step of $0.005$. In the second row $\alpha=-0.05$, $\tilde{\beta}=10\rightarrow45$ with a step of $5$.}
  \label{fig:ch4_gibbs}
\end{figure}
%

In Figure~\ref{fig:ch4_gibbs} we show the effect of temperature and interfacial energy on the phase transition strategy of the system. Observe that as the surface energy (measured by $\alpha$) grows (at fixed $\tilde T$) the cooperativity increases, whereas the expectation value of the number of interfaces  $\langle i \rangle_{\mathscr{G}}$ decreases and the transition is localized more and more around the Maxwell stress $\sigma=0$. In other words, the behaviour is a remnant of the one observed when temperature effects are neglected, as can be seen from Section~\ref{sec:ch4_eq} and Figure~\ref{fig:ch4_gibbs}$_d$ at low values of $\tilde T$. On the other hand, an opposite regime is obtained when the interfacial energy decreases (smaller values of $\alpha$). In this case, a non-cooperative transition may be observed, with a sloped transition plateau and a large number of interfaces. The same effect is attained by increasing the value of the temperature due to the growing importance of the entropic energetic component. Indeed, as we show in the discussion section, effective behaviour can be influenced by the competition between the entropic and interfacial energy terms in several fundamental physical problems.

\section{Applied strain: Helmholtz ensemble}
\label{sec:ch4_helmholtz}

To study the system under isometric conditions, that correspond to assigned displacement or, equivalently, hard device, and evaluate the canonical partition function in the Helmholtz ensemble $\mathscr{H}$, we may observe that it is related to the Gibbs one through a Laplace transform, as shown in Appendix~\ref{appA} and (\cite{weiner:1983,manca:2012}). In particular, as well known, to switch from Gibbs to Helmholtz ensemble, we can evaluate the inverse Laplace transform as a Fourier transform through the change of variable $\sigma\rightarrow \iota\omega$. Using the expression of $\mathcal{Z}_{\mathscr{G}}$ in~\eqref{eq:ch4_gformal} we get
\begin{equation}
\begin{split}
\mathcal{Z}_{\mathscr{H}}(\delta)&=\displaystyle \int_{-\infty}^{+\infty}\mathcal{Z}_{\mathscr{G}}(\iota\omega)\,e^{\,-\tilde{\beta} \iota \omega \delta}\mbox{d}\omega\\
&=\sum_{\{\chi_{j}\}}\mathcal{K}_{\mathscr{G}} \int_{-\infty}^{+\infty}e^{\,\tilde{\beta}\Bigl[\frac{1}{2}\boldsymbol{J}^{-1}\big(\boldsymbol{m}+(\iota\omega)\boldsymbol{1}\big)\cdot\big(\boldsymbol{m}+(\iota\omega)\boldsymbol{1}\big)-q - \iota\omega \delta \Bigr]}\mbox{d}\omega\\
&=\displaystyle \sum_{\{\chi_{j}\}}\mathcal{K}_{\mathscr{G}}\, e^{\,\tilde{\beta}\Bigl(\frac{1}{2}\boldsymbol{J}^{-1}\boldsymbol{m}\cdot\boldsymbol{m}-q\Bigr)}\int_{-\infty}^{+\infty}e^{\,\frac{\tilde{\beta}}{2}\Bigl[-\boldsymbol{J}^{-1}\boldsymbol{1}\cdot\boldsymbol{1}\,\omega^2+2\left(\boldsymbol{J}^{-1}\boldsymbol{m}\cdot\boldsymbol{1}-\delta\right)\,\iota\omega\Bigr]}\mbox{d}\omega\\
&=\displaystyle\sum_{\{\chi_{j}\}} \mathcal{K}_{\mathscr{G}}\, e^{\frac{\tilde{\beta}}{2}\bigg[\boldsymbol{J}^{-1}\boldsymbol{m}\cdot\boldsymbol{m}-2q-\frac{\left(\boldsymbol{J}^{-1}\boldsymbol{m}\cdot\boldsymbol{1}-\delta\right)^2}{ \boldsymbol{J}^{-1}\boldsymbol{1}\cdot\boldsymbol{1}}\biggr]}\int_{-\infty}^{+\infty}e^{-\frac{\tilde{\beta}}{2}\boldsymbol{J}^{-1}\boldsymbol{1}\cdot\boldsymbol{1}\bigg(\omega-\iota\frac{\boldsymbol{J}^{-1}\boldsymbol{m}\cdot\boldsymbol{1}-\delta}{\boldsymbol{J}^{-1}\boldsymbol{1}\cdot\boldsymbol{1}}\bigg)^2}\mbox{d}\omega,
\end{split}
\end{equation}
that leads to the expression 
\begin{equation}
\mathcal{Z}_{\mathscr{H}}(\delta)=\displaystyle \sum_{\{\chi_{j}\}}\mathcal{K}_{\mathscr{G}}\sqrt{\frac{2\pi}{\tilde{\beta}\left(\boldsymbol{J}^{-1}\boldsymbol{1}\cdot\boldsymbol{1}\right)}}\, e^{\frac{\tilde{\beta}}{2}\bigg[\boldsymbol{J}^{-1}\boldsymbol{m}\cdot\boldsymbol{m}-2q-\frac{\left(\boldsymbol{J}^{-1}\boldsymbol{m}\cdot\boldsymbol{1}-\delta\right)^2}{ \boldsymbol{J}^{-1}\boldsymbol{1}\cdot\boldsymbol{1}}\bigg]}.
\end{equation}
In the perturbative regime of small $\alpha$ given by~\eqref{eq:ch4_J} and based on the hypothesis in~\eqref{eq:ch4_9} we get
\begin{equation}
\mathcal{Z}_{\mathscr{H}}(\delta)\simeq\mathcal{K}_{\mathscr{G}}\sqrt{\frac{2\pi}{\tilde{\beta}\left(\boldsymbol{1}\cdot\boldsymbol{1}-\alpha \boldsymbol{A}\boldsymbol{1}\cdot\boldsymbol{1}\right)}}\sum_{\{\chi_{j}\}}e^{\frac{\tilde{\beta}}{2}\bigg[\varepsilon_0^2(\boldsymbol{\chi}\cdot\boldsymbol{\chi}-\alpha \boldsymbol{A}\boldsymbol{\chi}\cdot\boldsymbol{\chi})-\frac{\left(\varepsilon_0\boldsymbol{\chi}\cdot\boldsymbol{1}-\alpha\varepsilon_0 \boldsymbol{A}\boldsymbol{\chi}\cdot\boldsymbol{1}-\delta\right)^2}{\boldsymbol{1}\cdot\boldsymbol{1}-\alpha \boldsymbol{A}\boldsymbol{1}\cdot\boldsymbol{1}} -n \varepsilon_0^2 \bigg]}.
\label{eq:ch4_h2}
\end{equation}
Once more, the exponent of~\eqref{eq:ch4_h2} reduces to the equilibrium energy \eqref{eq:ch4_harden} by applying the identities in~\eqref{eq:ch4_id}. Following the same reasoning of the Gibbs ensemble, we may observe that if we neglect the boundary energy term depending on $\chi_1+\chi_n$ we may express the equilibrium energy as a function of the phase fraction $\bar{\chi}=p/n$, of the number of interfaces $i$ and of the averaged strain $\bar \varepsilon=\delta/n$, so that we obtain
\begin{equation}
\Omega_{p,i}(\bar \varepsilon)=-\frac{n\tilde{\beta}}{2}\Biggl\{\frac{\big[\left(1-4\alpha\right)\left(2\,p/n-1\right)\varepsilon_0-\bar \varepsilon\big]^2}{1-\frac{4\alpha}{n}\left(n-1\right)}+\frac{4\alpha}{n}\big(n-i-1\big)\varepsilon_0^2\Biggr\}.
\end{equation}
Accordingly, the canonical partition function is
\begin{equation}
\mathcal{Z}_{\mathscr{H}}(\bar\varepsilon)=\mathcal{K}_{\mathscr{H}}\sum_{p=0}^{n}\sum_{i=0}^{n-1}\mathcal{W}_{p,i}\,e^{\displaystyle\,\Omega_{p,i}(\varepsilon)},
\label{eq:ch4_zh}
\end{equation}
where
\begin{equation}
\mathcal{K}_{\mathscr{H}}=\mathcal{K}_{\mathscr{G}}\sqrt{\frac{2\pi}{\tilde{\beta}\left[n-4\alpha\left(n-1\right)\right]}}
\end{equation}
is a noninfluential constant. 

%
\begin{figure}[t!]
\centering
  {\includegraphics[width=0.95\textwidth]{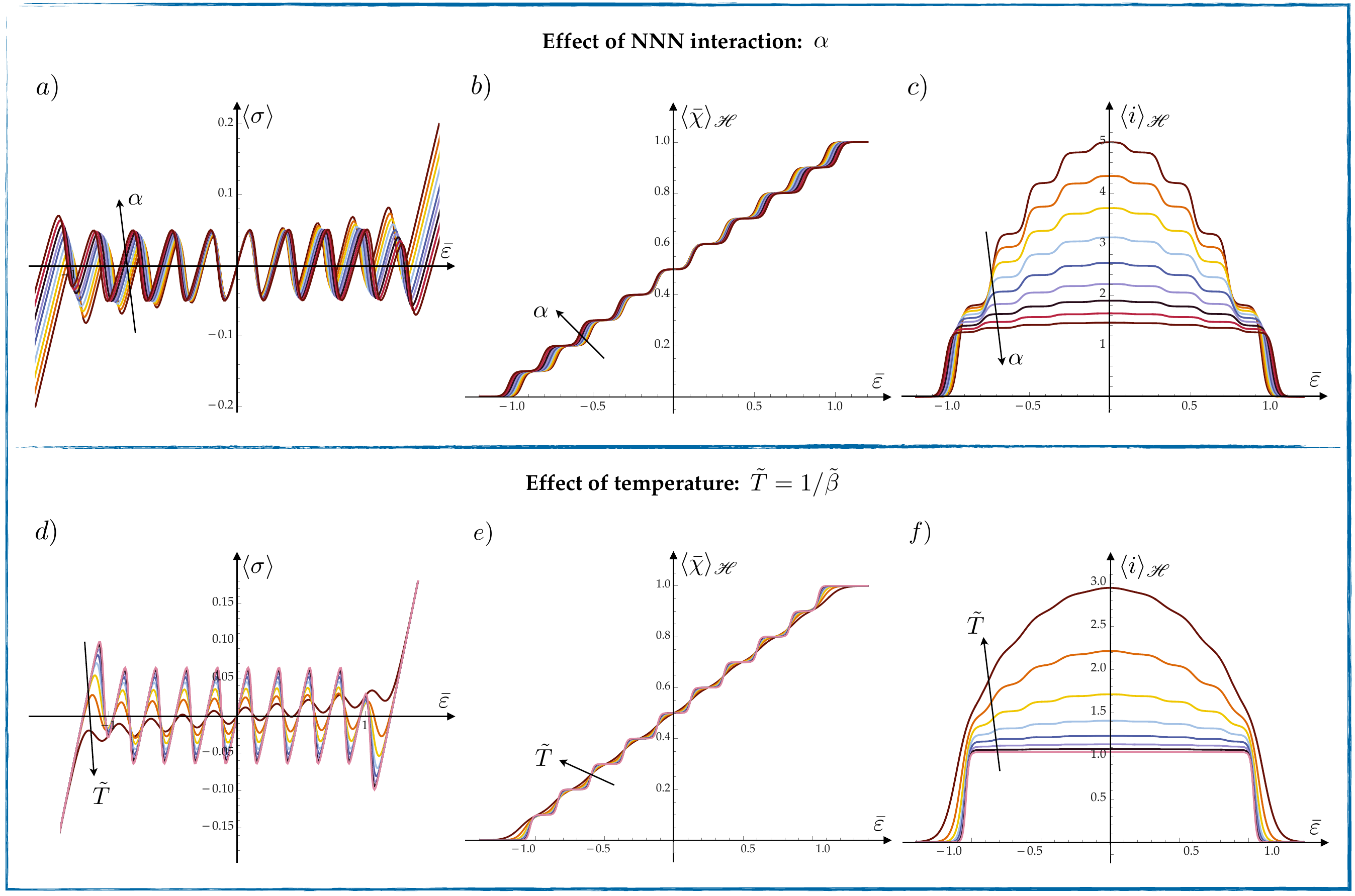}}
  \caption[Temperature effects in the Helmholtz ensemble]{Helmholtz ensemble.  In the first row we show the effect of $\alpha$ and in the second row the effect of $\tilde T$. In a) and c) we the stress-strain curves are represented, in b) and e) the expectation value of $\langle\bar{\chi}\rangle_{\mathscr{H}}$ and in c) and f) the expectation value of $\langle i\rangle_{\mathscr{H}}$. Parameters: $n=10$, $\varepsilon_0=1$. In the first row $\tilde{\beta}=30$ and $\alpha=0.0\rightarrow-0.05$ with a step of $0.005$. In the second row $\alpha=-0.05$ and $\tilde{\beta}=10\rightarrow45$ with a step of $5$.}
  \label{fig:ch4_helmholtz}
\end{figure}

By definition, the Helmholtz free energy is (\cite{weiner:1983,manca:2012})
\begin{equation}
\mathcal{F}(\bar \varepsilon)=-\frac{1}{\tilde{\beta}}\mbox{ln}\,\mathcal{Z}_{\mathscr{H}}(\bar \varepsilon),
\end{equation}
so that the expectation value of the stress, conjugate variable to the displacement $\delta=n\bar \varepsilon$, is (\cite{weiner:1983,manca:2012})
\begin{equation}
\langle \sigma \rangle=\frac{1}{n}\frac{\partial}{\partial \bar \varepsilon}\mathcal{F}(\bar \varepsilon)=-\frac{1}{n \tilde{\beta}}\frac{1}{\mathcal{Z}_{\mathscr{H}}(\bar\varepsilon)}\frac{\partial}{\partial\bar \varepsilon}\mathcal{Z}_{\mathscr{H}}(\bar \varepsilon),
\end{equation}
that leads to
\begin{equation}
\langle \sigma \rangle=\frac{1}{1-\frac{4\alpha}{n}(n-1)}\bigg[\bar\varepsilon-\left(1-4\alpha\right)\Big(2\langle \bar{\chi} \rangle_{\mathscr{H}}-1\Big)\varepsilon_0\bigg],
\label{eq:ch4_hstress}
\end{equation}
where we have introduced the expectation value of the unfolded fraction in the Helmholtz ensemble $\langle \chi\rangle_{\mathscr{H}}$, defined as
\begin{equation}
\langle  \bar{\chi} \rangle_{\mathscr{H}}\displaystyle  =\frac{\displaystyle \sum_{p=0}^{n}\sum_{i=0}^{n-1}\mathcal{W}_{p,i}\,\bar{\chi} \,e^{\displaystyle\,\Omega_{p,i}(\bar\varepsilon)}}{\displaystyle \sum_{p=0}^{n}\sum_{i=0}^{n-1}\mathcal{W}_{p,i}\,e^{\displaystyle\,\Omega_{p,i}(\bar\varepsilon)}},
\label{eq:ch4_chid}
\end{equation}
with $\chi=p/n$. Thus, also in the case of assigned displacement, we obtain formally, up to the neglected boundary term in $\chi_1+\chi_n$, the same expression of the mechanical limit in \eqref{eq:ch4_eqstress}, where we have to consider the expectation value of the phase fraction in \eqref{eq:ch4_chid}. 
The expectation value of the number of interfaces, in the Helmholtz ensemble, using~\eqref{eq:ch4_h2}, is given by
\begin{equation}
\langle i\rangle_{\mathscr{H}} =\frac{\displaystyle \sum_{p=0}^{n}\sum_{i=0}^{n-1}\mathcal{W}_{p,i}\,i\, e^{\displaystyle\,\Omega_{p,i}(\bar\varepsilon)}}{\displaystyle \sum_{p=0}^{n}\sum_{i=0}^{n-1}\mathcal{W}_{p,i}\,e^{\displaystyle\,\Omega_{p,i}(\bar\varepsilon)}}.
\end{equation}
%

%
\begin{figure}[t!]
\centering
  {\includegraphics[width=0.95\textwidth]{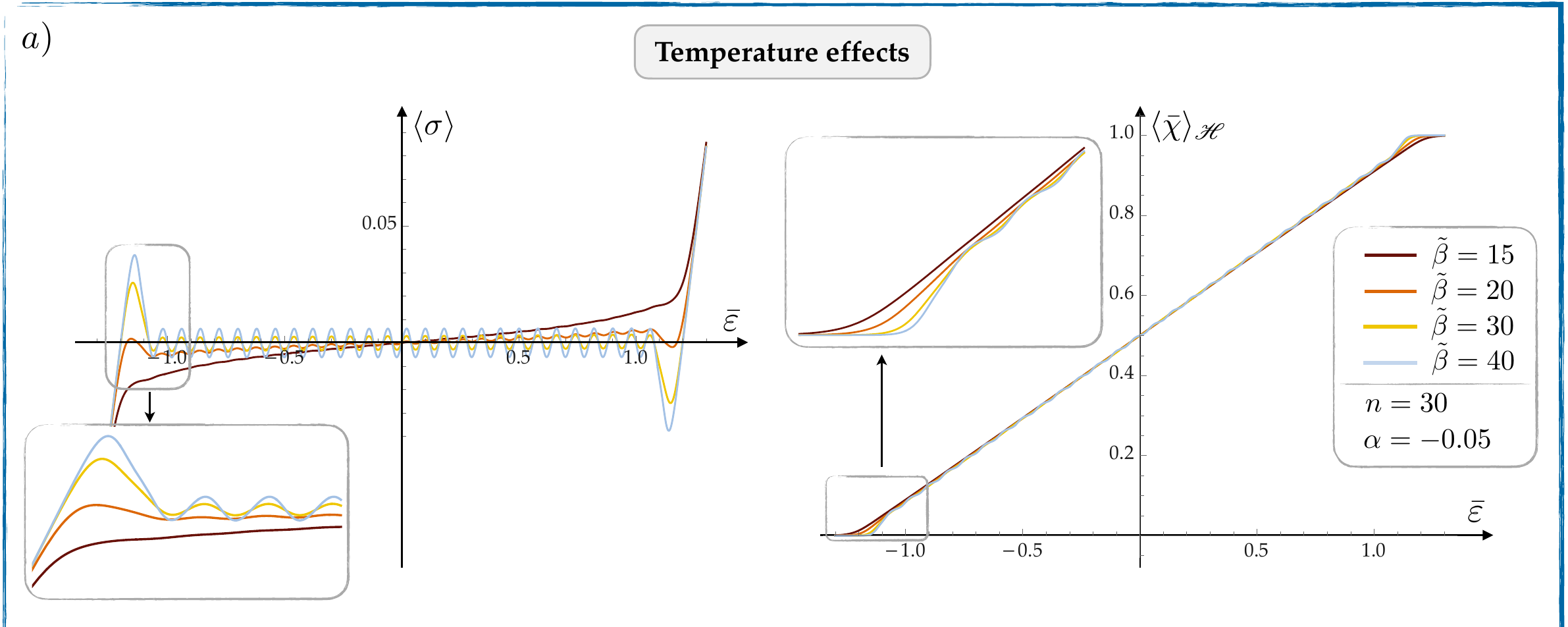}}\vspace{-0.1cm}
    {\includegraphics[width=0.95\textwidth]{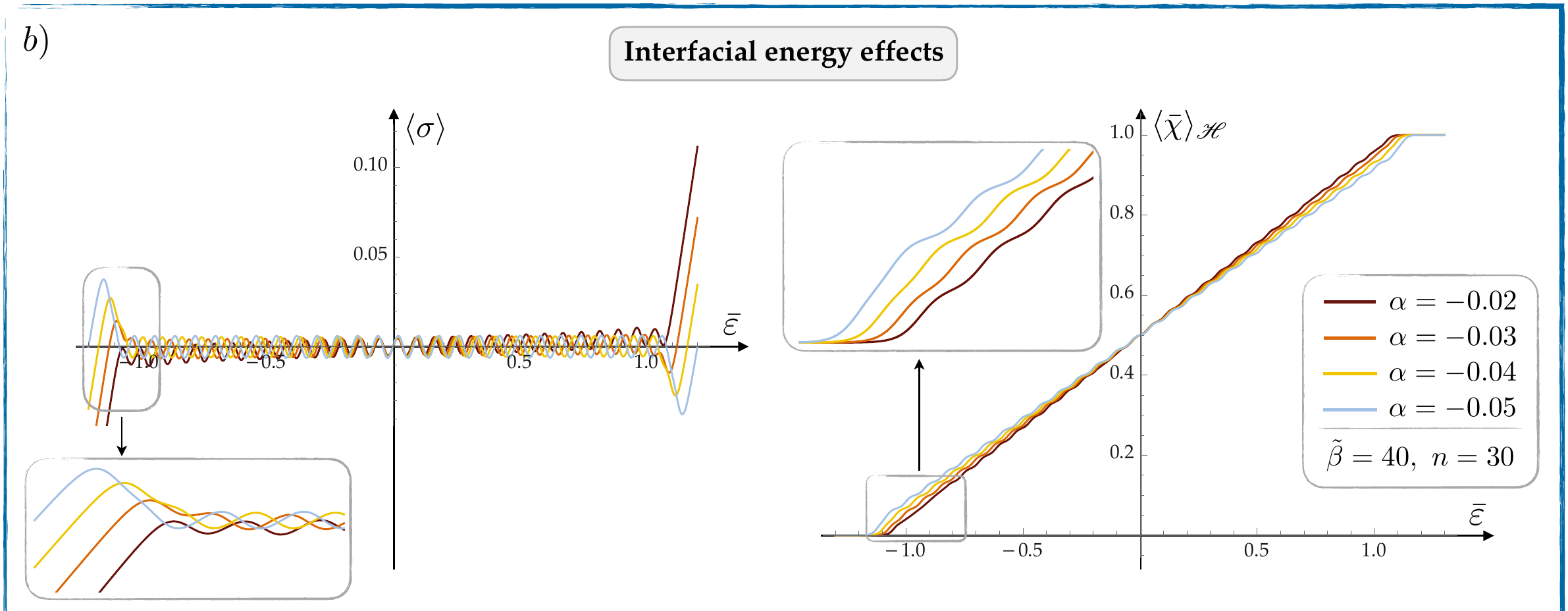}}\vspace{-0.01cm}
  {\includegraphics[width=0.95\textwidth]{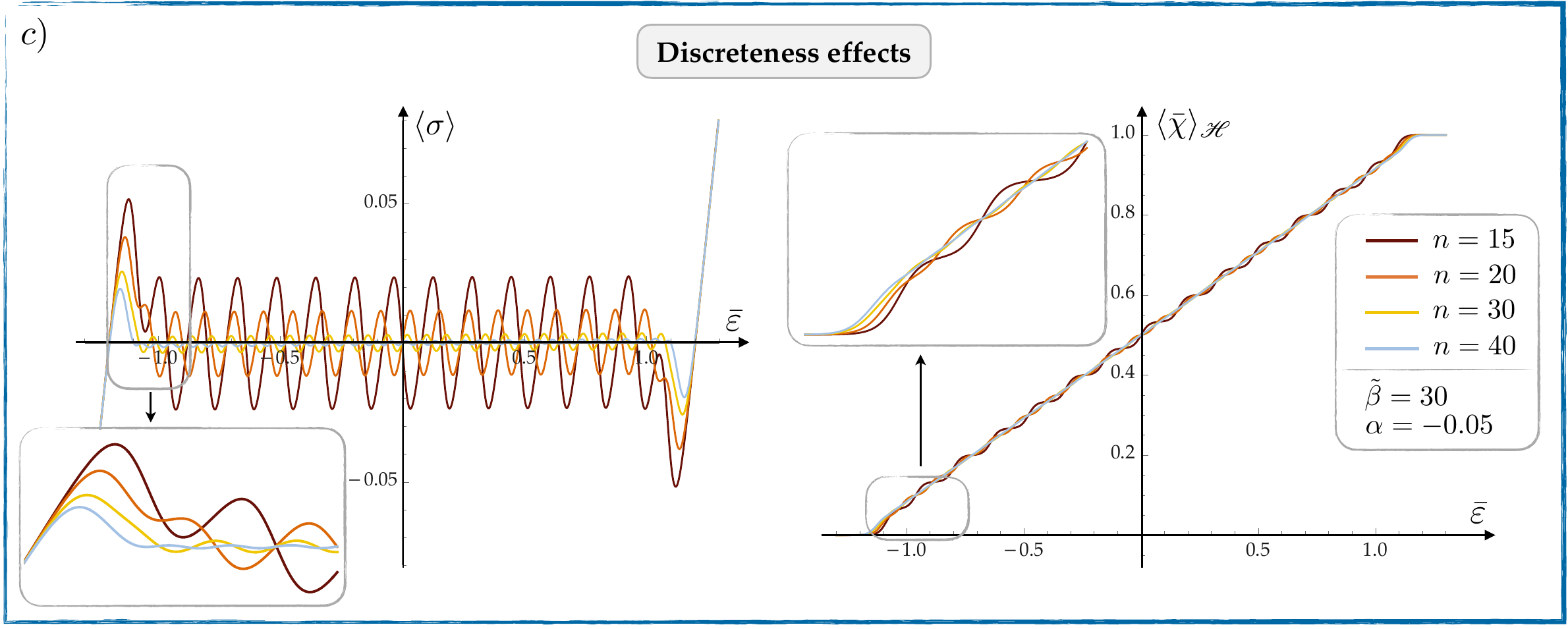}}
  \caption[Nucleation and propagation of the first phase]{Nucleation of the first phase in the Helmholtz ensemble. In panel a) the effect of the increasing $n$ is shown. In panel b) the effect of the strength of the non local interactions is displayed. In panel c) the effect of the temperature is represented. Parameters displayed on the legends.}
    \label{fig:ch4_phasen}
\end{figure}

The effect of temperature and interface energy in the case of assigned displacement is described in Figure~\ref{fig:ch4_helmholtz}. As in the isotensional setting, the influence of temperature and interfacial energy terms affect the cooperativity, which increases as $\alpha$ grows and $T$ decreases. The main differences between isotensional and isometric boundary conditions reflect the discussion anticipated in the mechanical case in Section~\ref{sec:ch4_eq}. Indeed, in the case of assigned displacement, the minimum expected value of interfaces, attained for larger values of $\alpha$ and low values of $\tilde T$ during phase nucleation is $\langle i\rangle_{\mathscr{H}}=1$. Moreover, the transition corresponds to sawtooth stress-strain diagrams. As  figure~\ref{fig:ch4_helmholtz}$_{a,c}$  shows, due to the presence of non local energy terms, the nucleation and propagation stresses may differ considerably. Observe that also the size of the first nucleated domain can be, when entropic effects are considered, significantly regulated by both parameters.

More in detail, in Figure~\ref{fig:ch4_phasen} we show how fundamental aspects such as the different nucleation and propagation stresses and the size of the initial nucleation domain are affected by variable interfacial energy, temperature, and discreteness size. In particular, the nucleation peak is strongly influenced by all these parameters and we observe that larger values are obtained for high values of $\alpha$ and low values of temperature. On the other hand, the size of the first nucleated domain grows as $\alpha$ increases and $\tilde T$ decreases. Another important effect is that by increasing temperature the stress-strain diagrams attain a `wiggly' shape instead of a `sawtooth', as shown in Figure~\ref{fig:ch4_phasen}$_a$. It is important to remark that these diagrams represent expectation values and are a counterpart of the process of switching, due to temperature effects, between solutions with similar energy. Moreover, as the temperature grows we may observe the transition from a constant average of the transition stress plateaux to a sloped one. This behaviour is limited by the presence of non-local terms, as shown in Figure~\ref{fig:ch4_phasen}$_b$. The effect of discreteness is described in Figure~\ref{fig:ch4_phasen}$_c$ that shows decreasing values of the peak and an increasing number of nucleated elements as $n$ increases. In this case, we also observe that as the discreteness increases, also the other peaks decrease and the ratio between the first peak and the following increases, as we will show in the following section. This particular behaviour is also obtained in the continuum limit, where only the first peak is observed with a finite value of stress and the transition is represented by a force plateau (\cite{puglisi:2007}).

\section{Comparison with molecular dynamics tests}
\label{sec:ch4_exp}

To test the effectiveness of the proposed model in deducing the transition behaviour taking into account the role of interfacial energy and temperature effects, we compare our theoretical results with numerical experiments based on Molecular Dynamics (MD) of shape memory nanowires (NWs).  These systems acquired large attention due to their incredible properties in terms of energy storage (no hysteresis), large actuation strain ($>30\%$ as compared to $\sim$~$5\%$ of the bulk ones) and stress ($>3$ GPa as compared to $\sim$~$0.5$GPa of the bulk ones) (\cite{li:2010}). Several MD simulations, also supported by experimental tests (see \textit{e.g.} references (\cite{seo:2013,ma:2013})),  showed that such superior behaviour results from a different mechanism of microstructure evolution during the phase transition. Indeed, in SMA nanowires the phase transition process is accompanied by a shear dominant diffusionless transformation, leading to the formation and the migration of defect-free twins connecting the two (kinematically incompatible) phases. The typical applications are based on face-centred cubic (FCC) metallic (Cu, Ni, Au and Ag) nanowires whose transition is simply accompanied by a crystal reorientation enabled by the notion of coherent twin boundaries. This leads to a different type of pseudoelasticity, as compared to bulk materials, that is not regulated by martensitic phase transition but by the propagation of twins leading to crystal reorientations (\cite{liang:2007}). As a result, the same bulk materials do not show pseudoelastic behaviour. The energetic competition between dislocation and twinning nucleation at a crack tip has also been described in bulk materials (\cite{tadmor:2003}). In the case of nanowires, coherent twin interfaces are formed between the two phases that then propagate along the wire axis when the average strain is increased. More detailed metallographic analysis on the underlying mechanism of twin regulated deformation in nanowires can be found in (\cite{park:2006}). This process is reversible and is responsible for the incredible pseudoelastic behaviour observed in these materials.

%
\begin{figure}[t!]
	\centering
	{\includegraphics[width=0.95\textwidth]{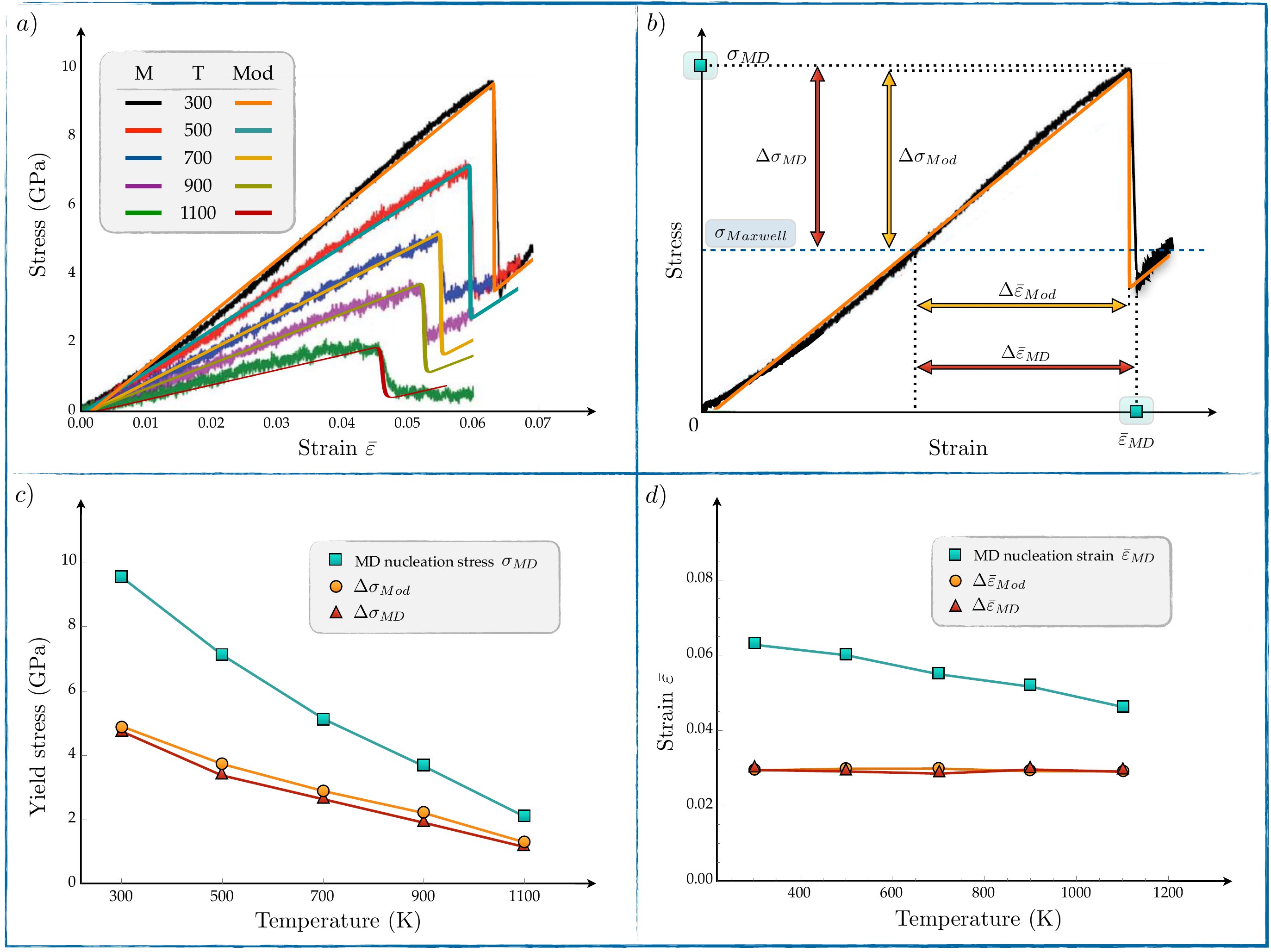}}
	\caption[Comparison with molecular dynamic experiments at varying temperature]{Comparison of results obtained from the model presented in this Chapter with those from Molecular Dynamics (MD) simulations. Simulation adapted from (\cite{lu:2019}), published by the Royal Society of Chemistry. Details about the values of the parameters can be found in the main text. Panel a): Stress-strain curves. Panel b): Relevant quantities used for the comparison in panels c) and d), see main text for details. Panel c): Dependence of the yield stress on temperature. Panel d): Dependence on the temperature of the strain corresponding to the drop of the yield stress.}
	\label{fig:ch4_exp}
\end{figure}

We begin by testing our model to predict the behaviour obtained in the simulations performed on Ni-Co alloy nanowires (\cite{lu:2019}). Some remarks are in order. As previously stated, our model is one-dimensional. On the other hand, as we deduce qualitatively in the following, the role of the lateral size of the system can be related to the strength of non local interactions included in our model, but also in the presence of non local interactions with the loading device. Moreover, it is important to stress that physical constitutive parameters of the system such as the Young modulus depend on both temperature and lateral size. This dependence is here used to fit the numerical results in MD.  Among many effects such as the presence of Co in the alloy and the role of surface/internal defects, in Reference (\cite{lu:2019}) the authors study the temperature dependence of the stress behaviour. They simulate a nanowire with dimensions $2.465\mbox{nm}\times2.464\mbox{nm}\times19.712\mbox{nm}$. In order to adapt this system to our one-dimensional model, we consider a unit reference length $l$ equal to the lattice constant $l=a_{Ni}=0.3499$ nm (for Nickel) and, therefore, $n=55$ oscillators in our formulas. It is important to observe that \textit{in order to calibrate the system and fit the experimental data we have only two parameters to be fixed: $\alpha$ and $\varepsilon_0$}. Here we set the value of the reference strain of the second phase $\varepsilon_0=0.18$ and $\alpha=-0.2$. I remark that even though this value for the parameter $\alpha$ is not small compared with the assumption in~\eqref{eq:ch4_J}, still it is within the range of admissible values ($\alpha<1/4$) and the error associated with this choice (tested numerically of about $5\%$) does not affect the obtained results. The values of the modulus $E$ are the ones proposed in (\cite{lu:2019}) and accordingly we deduce $k=EA$ (with $A$ the section considered in (\cite{lu:2019})). Similarly, given the temperatures, we may deduce the corresponding values of $\tilde\beta=\beta k l$. 

The comparison of the theoretical and numerical results is exhibited in Figure~\ref{fig:ch4_exp}$_a$, where the stress-strain curves for different values of the temperature are represented. In particular, using Equation~\eqref{eq:ch4_hstress} we obtain that $E(1-4\alpha (n-1)/n)\langle \sigma\rangle$ corresponds to the stress measured in the MD simulations. We may observe that, despite the oversimplification of the model, the theoretical response is in very good agreement with the MD simulations. We remark that although we may observe that for $T=500, 700, 900$ K the description of the size of the first stress drop is not accurate, the authors show in the paper that effects such as internal defects and the percentage of diluted Cobalt can be responsible for this effect. On the other hand, as shown in Figure~\ref{fig:ch4_phasen}$_c$, this value differs from the physically meaningful difference between the nucleation and propagation stress (not reported by the authors), because it strictly depends on the precise position of the first drop. 

In order to obtain a more clear comparison with the MD results, in Figure~\ref{fig:ch4_exp}$_{c,d}$ we extracted the values of the nucleation stresses and strains depending on temperature. It is important to remark that we assumed for simplicity of notation a zero Maxwell stress in our model. To reconcile the theoretical results and the MD experiments, in Figure~\ref{fig:ch4_exp}$_b$ we show how we deduce the value of the propagation (Maxwell) stress and the corresponding strain. The agreement between the MD results (red triangles) and those obtained from the model (orange circles) is excellent in both cases. 

%
\begin{figure}[t!]
	\centering
	\includegraphics[width=.95\textwidth]{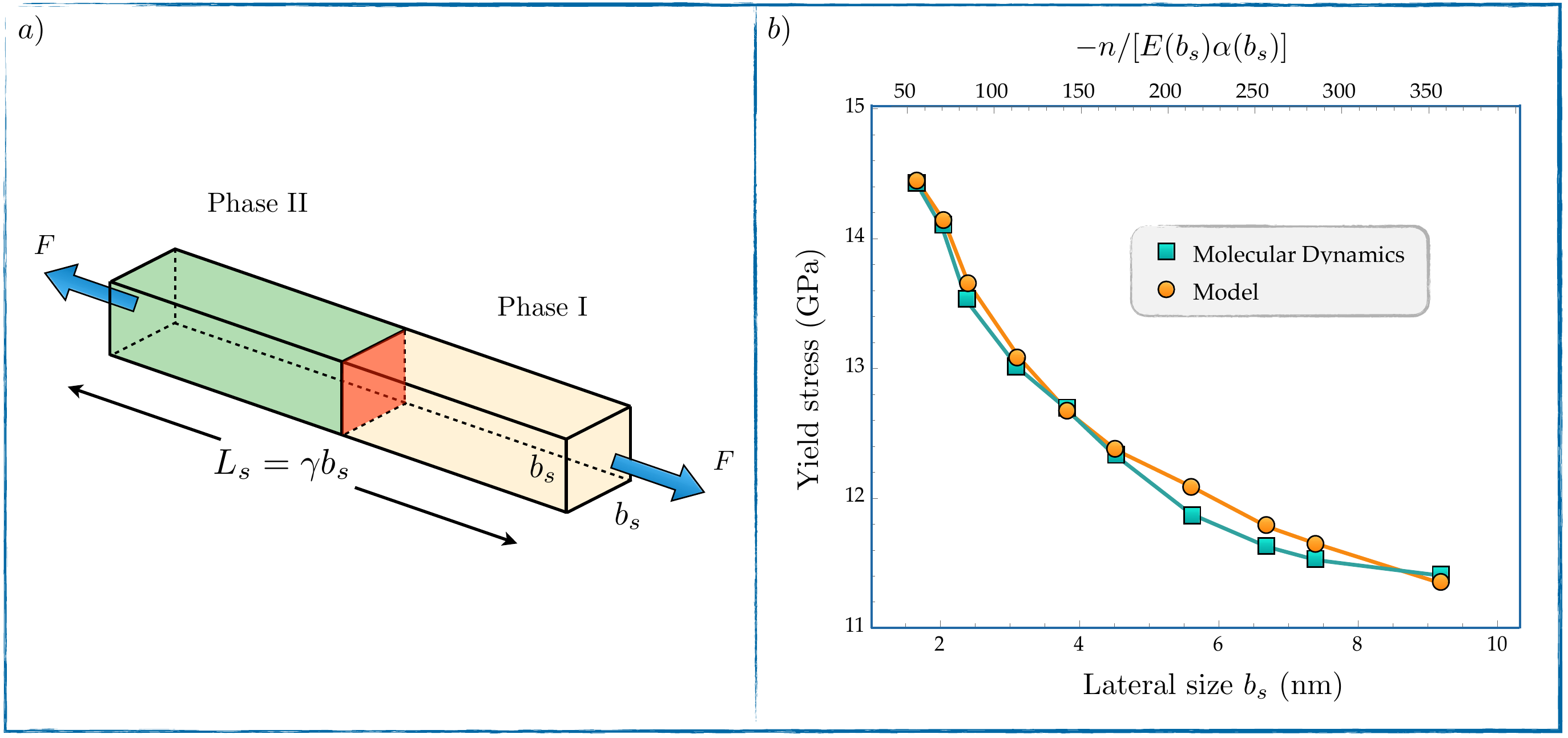}
	\caption[Comparison with molecular dynamic experiments at varying interface energy]{Panel a): One dimensional scheme of the shape memory wires. Panel b): Dependence of the nucleation (yield) stress on the lateral size $b_s$ (MD simulations reproduced from Wu and collegues (\cite{wu:2006})) and theoretical model (see the text for the deduction of the parameters from the MD experiments).}
	\label{fig:ch4_exp2}
\end{figure}

We have also tested our model to verify the possibility of describing the fundamental size effects observed in shape memory nanowires, showing an increasing role of the interface energy as the size of the specimen decreases. In particular, we analyzed the size effect on the nucleation (yield) stress studied by Wu and colleagues (\cite{wu:2006}). There, the author performed MD simulations considering Cu nanowires with fixed aspect ratio 1:1:3 ($\gamma=L_s/b_s=3$ in Figure~\ref{fig:ch4_exp2}$_a$) and variable lateral size. Following the same procedure described above, with the aim of fitting the MD results, we have fixed the reference length $l$ with the value of Copper lattice constant $l=a_{Cu}=0.3609$ nm, temperature $T=1$K (as indicated in the paper) and $\varepsilon_0=0.23$. Thus, we need to consider a variable number of elements $n\simeq L_s/l$ ranging from 14 to 77 where $L_s$ is the length of the sample considered in Ref. (\cite{wu:2006}). Another important feature to point out is that the (effective) elastic modulus depends on the lateral size $E=E(b_s)$. This dependence is reported in the paper (\cite{wu:2006}) and is used in our computation to evaluate the stiffness $k=k(b_s)=E(b_s)b_s^2$. 

To evaluate how the change of length influences the parameter $\alpha$ we may, in a first approximation, consider the following energy rescaling reasoning. A rough estimate of the total energy density of the MD system can be obtained, separating the contribution of bulk and surface energy, as follows:
\begin{equation}
\Phi^{MD}_{tot}=\Phi^{MD}_{b}+\Phi^{MD}_{s}=\frac{E(b_s) \gamma b_s^3+ C b_s^2}{\gamma b_s^3}=E(b_s)+\frac{C}{\gamma b_s},
\end{equation} 
where $C$ is the surface energy constant. In particular, we have that the relative role of the surface energy is
\begin{equation}
\xi^{MD}:=\frac{\Phi^{MD}_{s}}{\Phi^{MD}_{b}}=\frac{C}{E(b_s) \gamma b_s}. 
\label{eq:ch4_xia}
\end{equation}
On the other hand, in the proposed model we have that the analogous quantities are given by
\begin{equation}
\Phi^{Mod}_{tot}=\Phi^{Mod}_{b}+\Phi^{Mod}_{s}=\frac{k(b_s)n l+\alpha k(b_s) l}{n l b_s^2}=E(b_s)+E(b_s)\frac{\alpha}{n},
\end{equation}
so that
\begin{equation}
\xi^{Mod}:=\frac{\Phi^{Mod}_{s}}{\Phi^{Mod}_{b}}=\frac{\alpha(b_s)}{n}. 
\label{eq:ch4_xib} 
\end{equation}
We obtain then the main assumption for the comparison of molecular dynamic and model rescaling effects by comparing \eqref{eq:ch4_xia} and \eqref{eq:ch4_xib} that gives 
\begin{equation}
\frac{\alpha(b_s)}{n}\sim \frac{1}{E(b_s)b_s},
\end{equation}
and since
\begin{equation}
b_s=\gamma L_s\propto n,
\end{equation}
we get
\begin{equation}
\alpha(b_s)\sim \frac{1}{E(b_s)}. 
\label{eq:ch4_mmm} 
\end{equation}
The behavior resulting from this rescaling is reproduced in Figure~\ref{fig:ch4_exp2}$_b$ where according with the assumption~\eqref{eq:ch4_mmm} the parameter $\alpha$ ranges in the interval $\alpha\in(-0.25,-0.215)$. Observe that again we have two only parameters to fit the experimental data, \textit{i.e.} the value of $\varepsilon_0$ and the value of the constant of proportionality in~\eqref{eq:ch4_mmm}. Again, despite the simplification of the model, we obtain a very satisfying agreement.

\section{Analysis of the results}
\label{sec:ch4_disc}

In the chapter, I derived a fully analytical approach to describe the competition of entropic and interfacial energy terms in the transition strategy of a bistable material. More specifically, I considered a prototypical system constituted by a lattice of bistable elements and non local NNN interactions with a concave energy density reproducing the presence of an interfacial energy term penalizing the formation of new interfaces. The extension to the opposite case of NNN terms with convex energy, playing the (antiferromagnetic) role of favouring the presence of interfaces is straightforward in our framework, but not explicitly discussed and applied in this work. The presence of this energetic term plays an important role leading in the case of isometric loading to the presence of an initial stress-peak distinguishing the nucleation and propagation stresses. Roughly speaking, due to the presence of interfacial energy terms, the transition from a homogeneous reference phase to a two-phase configuration is delayed because of the energetically unfavourable event of nucleation of a new phase and the formation of interfaces. All these features are observed in phase transition for different materials, some of which are explicitly discussed in this section. On the other hand, the presence of interfacial energy terms, in the absence of temperature effect or other non local interactions terms would lead to the possibility of one single interface. The experimental behaviour contradicts this result and this effect may be ascribed to several different phenomena such as interaction with the loading device (\cite{truskinovsky:2000,puglisi:2006,puglisi:2007}), compatibility and dynamical effects (\cite{abeyaratne:1996,muller:1999}), inhomogeneities and rate effects (\cite{shaw:1997}) and temperature effects as considered in this chapter.

\begin{figure}[t!]
\centering
  {\includegraphics[width=0.95\textwidth]{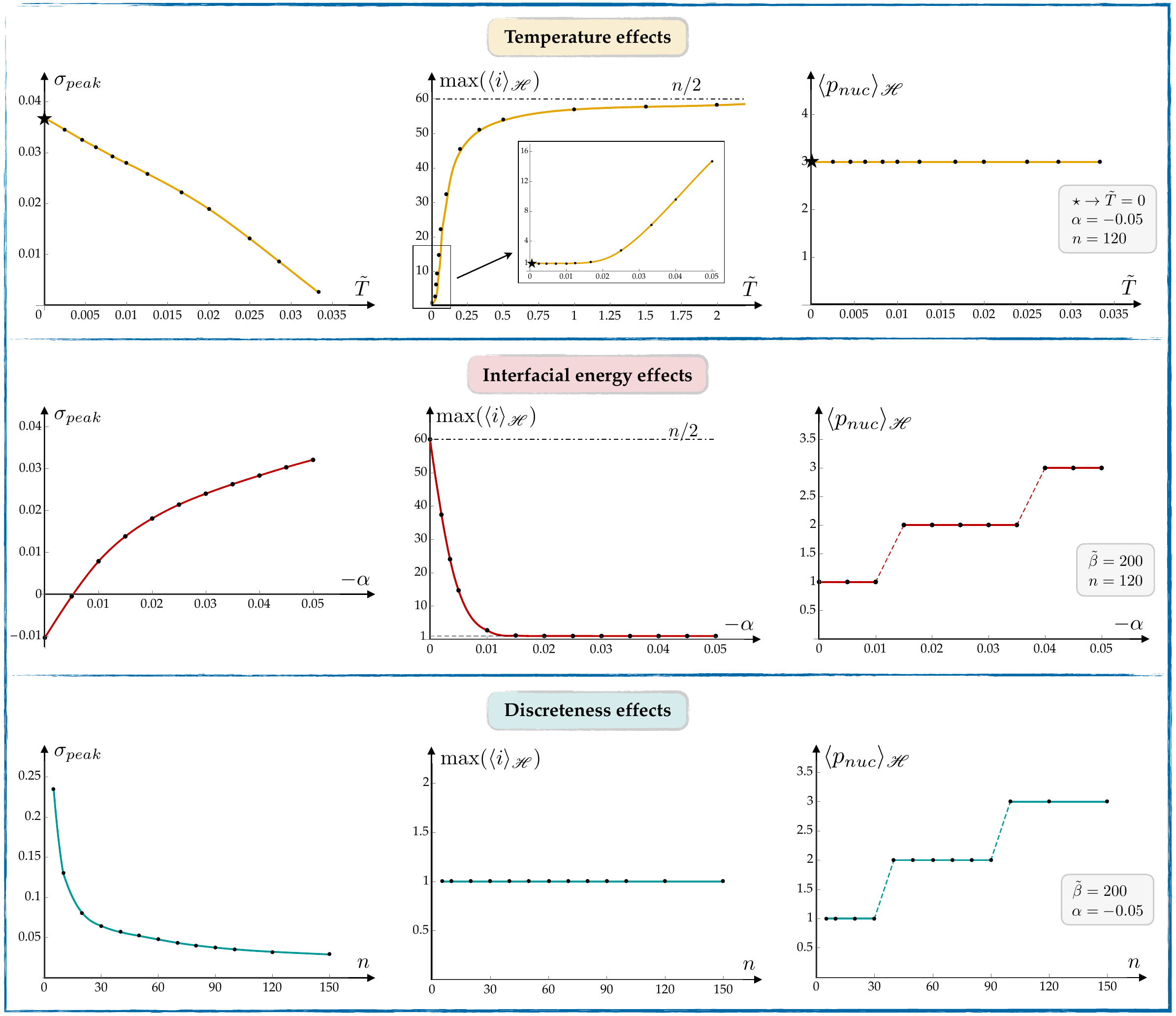}}
  \caption[Summary of the analytical results]{Synthetic description of the analytical results. We show the influence of the temperature $T$ (top), of the interfacial energy parameter $\alpha$ (middle) and of the discrete parameter $n$ (bottom) on the transition behavior in terms of nucleation peak (left), maximum number of interfaces (center) and size of the nucleated domain (right). In the first row we indicate with $\star$ the behavior of the system in Section~\ref{sec:ch4_eq} coinciding with the limit of $\tilde T\rightarrow 0$ when entropic effects can be neglected.}
  \label{fig:ch4_figfin}
\end{figure}

Specifically, I have introduced a model whose behaviour is of interest in the physical cases where there are competing energy terms, namely in situations where non-local interaction effects are of the same order of magnitude as temperature-driven entropic effects, such as in nanowires or several biological materials discussed in the following. In these cases, as already introduced in the previous chapter, the system behaviour has to be described in the framework of Equilibrium Statistical Mechanics and hence we are able to analytically describe the effect of temperature.

The results of the model have been summarized in Figure~\ref{fig:ch4_figfin}. Observe that the behaviour of the system when entropic energy terms can be neglected, coinciding with the system studied in Section~\ref{sec:ch4_eq} in the limit of $\tilde T=1/\tilde \beta\rightarrow 0$, is represented in the first row of Figure~\ref{fig:ch4_figfin} with the $\star$ symbol. The figure shows the fundamental role that may be played by entropic effects. In particular,  the size of the first nucleated domain is essentially temperature independent but, on the other hand, it strongly depends on the interfacial energy. Indeed, as shown in the central row of Figure~\ref{fig:ch4_figfin}, we observe that the size of the stress peak increases with increasing interfacial energy. It is important to remark that the negative values attained for low $\alpha$ depend on how the $\sigma_{peak}$ is measured and on the competition with entropic effects. Indeed, the hardening phenomenon attained at varying temperatures may lead to negative values of the Maxwell stress, as shown in Figure~\ref{fig:ch4_phasen}$_a$, and if the stress peak is evaluated as the difference between the peak and the Maxwell line at nucleation strain negative values can be found for low $\alpha$. Moreover, as already discussed, the presence of the stress peak in this specific case is generated by the competition between interface and entropic terms whereas in other physical phenomena it can be possible to observe a similar behaviour due to other effects such as non-local terms or qthe influence of the handling device. On the other hand, both the stress peak and the number of interfaces may strongly depend on temperature. In particular, we may observe that the stress peak decreases as temperature grows whereas the cooperativity decreases, with an increasing number of a maximum number of interfaces. This is an expected result because by increasing the temperature the solution is characterized by high values of the entropy --here measured by the number \eqref{eq:ch4_comb} of permutations corresponding to the same values of $p$ and $i$-- are energetically favoured. In these cases, we may interpret the experimental observation of less cooperative transition as induced by temperature entropic effects.

As discussed in Section \ref{sec:ch4_exp}, our model can be considered a prototype for the analysis of the temperature and size-dependent transition behaviour of SMA nanowires. The main reason for the different behaviour of SMA nanowires as compared with bulk materials is due to the well known important role of surface energy terms as compared with the bulk one as the scale of the specimens decreased. Specifically, two different scale effects have to be considered (\cite{seo:2013}). The first one corresponds to the energy of the wire surface that regulates, together with the bulk energy, the (diameter dependent) effective energy of the phases. As a result, the energy wells and, in particular, the Maxwell stress corresponding to the propagation threshold, are scale-dependent parameters. Of course, since our model considers global energy minimizers only, we neglect hysteretic effects, but this is in agreement with the typical behaviour of nanowires that often are characterized by a reversible (\cite{lao:2013}) transition. On the other hand, extensions to consider local energy minimizers can be considered (\cite{puglisi:2005,benichou:2013}) (see also the low scale interpretation in (\cite{liang:2007}) of possible dissipation associated with the discrete process of twin propagation). The second surface energy term, important at these scales, is the interfacial energy associated with the formation of a twin, here measured by non-local energy terms through the parameter $\alpha$. While the size dependence of this energy can be theoretically estimated depending on the specific twin connecting the two phases (\cite{zhang:2008}), in Section~\ref{sec:ch4_exp} we deduced this dependence in our one-dimensional setting by simple scaling arguments. When the size of the wire is fixed and the temperature is changed, we expect that the bulk energy of the wires does not change whereas the nucleation stress can be influenced by temperature effects as measured by our model. This is reflected by a variation of both the nucleation and propagation stress with the size of the wire (both $\alpha$ and the energy wells are modified), whereas temperature only influences the nucleation stress (see Figure~\ref{fig:ch4_exp}). Such behaviour reproduced from our model, is reflected in different papers based on MD simulations (\cite{seo:2011,lao:2013}). Moreover, as suggested by our model, the possibility of more twin formations has been described in several contributions (\cite{rezaei:2017,guo:2009}). Interestingly, as predicted from our results, the experimental behaviour (\cite{guo:2009}) shows that also the number of interfaces is influenced by temperature with a `Bell type' growth as described in Figure~\ref{fig:ch4_helmholtz}$_{c,f}$. We then considered explicitly the possibility of describing the dependence of the nucleation and propagation stresses from the wire diameter. In this case, as described above also the bulk energy depends on the size and can be deduced from MD experiments. Once the corresponding rescaling quantities are adopted in our model, we showed the possibility of well describing the size-dependent transition behaviour (see Figure~\ref{fig:ch4_exp2}).

\begin{figure}[b!]
\centering
  {\includegraphics[width=0.95\textwidth]{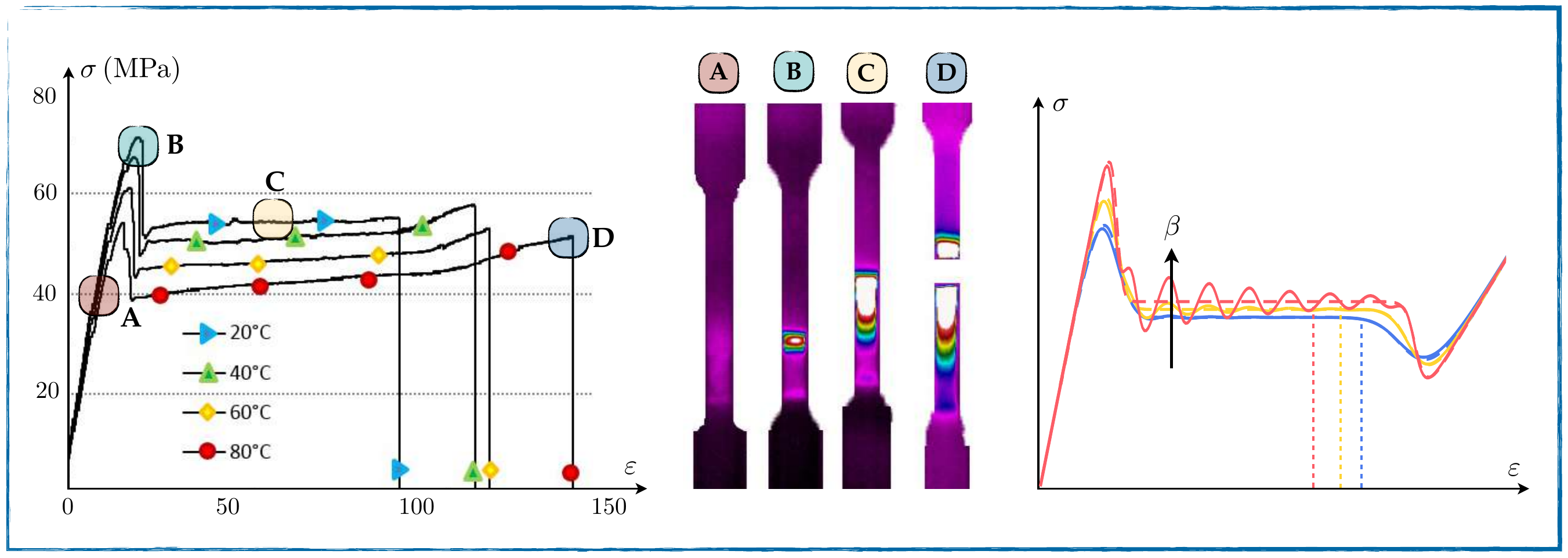}}
  \caption[Depedence of the Maxwell stress on temperature]{Experiments performed on specimens of polycarbonate Makrolon UV Climate Control blue at different temperatures, done by our research group in collaboration with (\cite{fuina:2016,fuina:2019}). It is possible to observe the described stress peak associated with the nucleation of a phase interface (see central picture) and the dependence on the temperature of the Maxwell stress, which decreases with increasing temperature. The nucleated phase can be observed with the thermo camera on the central panel propagating into the specimen until rupture occurs. On the left side the theoretical curves derived in our recent work (\cite{luca2022:1}) are shown and in particular the curves are represented until a complete transition to a softer phase is attained whereas rupture is indicated with dashed lines (force dropping to zero).}
  \label{fig:ch4_maxT}
\end{figure}

A second important field of application for the results of our model that we would like just to sketch in this chapter and that will be the subject of our future work is related to the important role that interdomains interactions can play in the stability and cooperativity transition of multidomain proteins, undergoing a conformational folded$\rightarrow$unfolded transition, as compared with the stability of isolated domains (\cite{law:2003,bhaskara:2011,xu:2013}). This subject has achieved in recent years increasing attention after a long period when the research was focused only on single domains behaviour both experimentally and theoretically, as reported in (\cite{law:2003,bhaskara:2011,xu:2013}) and references therein. In this second field, the possibility of describing important experimental effects in the framework of Statistical Mechanics with non-convex energies has been shown in recent works (\cite{detom:2013,manca:2013,giordano:2017,fp:2019,luca2019}). In these papers, the two wells correspond to the folded and unfolded states. On the other hand, in these models, no NNN interactions were considered. Experimental and theoretical literature showed the important role of interdomain interactions in the unfolding behaviour of proteins and their possible roles in the insurgence of diseases (\cite{bhaskara:2011}). A typical well-known example is the stabilization effect in spectrin repeat unfolding. Thermal stability analysis clearly shows the important stabilization effects (\cite{macdonald:2001}) due to non-local interactions between different two-states domains. Single molecule force spectroscopy shows that this effect can induce a contemporary transition of two domains at the first stress peak (\cite{rief:1999}). Interestingly, as theoretically deduced with our model, cooperativity is lost if the temperature is increased (\cite{batey:2005}). Moreover,  a similar behaviour has been observed in Filamin A proteins, where domain-domain interactions lead to a hierarchy of unfolding forces that may be properly studied by nonlocal models (\cite{xu:2013}). 

A recent approach to model these macromolecular behaviours is based on an Ising scheme suitably coupled with the chain of bistable units representing the protein domains (\cite{benedito:2018:2}). The results presented in this chapter show the possibility of describing such behaviour in the framework of Statistical Mechanics based on the consideration of non-local interactions but it is also possible to use the Ising model to describe the same system. Specifically, I report in Appendix~\ref{appG} how it is possible to obtain the classical Ising model starting from the Hamiltonian energy with non-local NNN interactions introduced within this chapter. This important result is used in a very recent paper in which I am one of the co-authors, though it is not included in this thesis, to describe the effect of temperature when more general energy with different stiffnesses for the wells and non-zero transition energy is considered (\cite{luca2022:1}). As shown in Figure~\ref{fig:ch4_maxT}, it is experimentally observed that the nucleation and generation of a phase interface (see the pictures on the centre of Figure~\ref{fig:ch4_maxT} captured with thermo cameras) results in the stress peak observed in the stress-strain diagram (left side). Moreover, it is also observed that the transition --Maxwell-- stress decreases with increasing temperatures, and in this recent work, we found this dependence also within the assumptions of the zipper model, \textit{i.e.} when only one interface is considered, and the stationary phase (the thermodynamic limit), and the stress-strain curves obtained at varying temperature are shown in the right side of Figure~\ref{fig:ch4_maxT}. 

Eventually, beyond the possibility of the contemporary transition of domains, the model shows that cooperativity can be increased by the effect of NNN interactions whereas it can be hidden by entropic effects when temperature increases. It is important anyway to remark that other important effects, such as allosteric effects of binding sites and fully non-local interactions in the tertiary and quaternary structures cannot be described in the simple one-dimensional setting here considered (\cite{daily:2009}).

	\clearpage


\renewcommand{\thefigure}
{\arabic{chapter}.\arabic{figure}}
\setcounter{figure}{0}

\renewcommand{\theequation}
{5.\arabic{equation}}
\setcounter{equation}{0}

\chapter{Decohesion and fracture of biological systems: application to DNA denaturation and Axonal damage}
\chaptermark{Decohesion and fracture of biological systems}
\label{ch_5}
\vspace{1.6cm}

Polypeptide molecules are life-key components in the majority of living systems and their correct functioning is of paramount importance to avoiding diseases and genetic malformations. Indeed, not only chemical reactions and electrical signalling are involved in the proper effectiveness of such systems, but also mechanical forces and thermal effects play a crucial role (\cite{rief:1999:2,smith:1996,bustamante:2000:2}). To this regard, it is nowadays possible to study these biosystems thanks to Single Molecule Force Spectroscopy (SMFS) experiments (\cite{lavery:2002}). Conformational transitions in biomolecules are very common to perform their operational tasks and both the relative stability of the different metastable states and the reversibility of the transition between them represent a crucial aspect for the associated physiological functions. In particular, among many examples here I refer to the folding and misfolding effects in proteins (\cite{dobson:2003})  taking care of the peptide sequences and possible mutations (\cite{pugno:2021}), to the DNA replication (\cite{watson:1953}) or denaturation (\cite{bustamante:2000}) and to the axonal damage in neurons (\cite{kuhl:2015}) or the mechanosensing cell behavior in focal adhesion (\cite{cao:2015, buehler:2007,salvo:2022}).  

In this framework, here we focus on the cooperativity phenomenon. Indeed, it is well known that a system can behave cooperatively or not depending, among many effects, on the external field that drives the transition. While the cooperativity is typically studied when the driving field is of thermal (\cite{lumry:1966}) or chemical (\cite{jha:2014}) nature, it is well known that the case of mechanical loading can lead to very different paradigms. In particular entropic effects can strongly decrease the cooperativity of the transition (\cite{hyeon:2005,florio:2020}). On the other hand, the stability and cooperativity have been recognized as concepts strictly related (\cite{abkevich:1995}), for instance inferring the known Anfinsen thermodynamic hypothesis in protein folding which suggests that the native state corresponds to the global energy minimum of the protein determined by the amino-acid sequence. In this chapter, we propose a model able to show in a purely mechanical setting, the cooperativity and stability effects induced by local interactions in double-stranded molecules. The obtained results can be conceptually extended to the analysis of DNA and RNA hairpins (\cite{antao:1991}) or to support the hypothesis of a zipper cooperativity model for protein folding (\cite{dill:1993}) often considered a way to answer the Levinthal paradox, or to the analysis of repeat-protein folding (\cite{kloss:2008}).

It is important to highlight that other phenomena such as non-local interactions (\cite{abkevich:1995}), or entropic cooperativity effects (\cite{searle:1992}) are also fundamental in the cooperativity behaviour with different mechanisms, especially in the case of macromolecules, but it is also known that secondary structure largely depends on local interactions with most of the small protein undergoing an all-or-none two-state transition (\cite{baldwin:1999}). Even though the model is quite general, to fix notation and to provide a specific example here we describe the cooperative behaviour of a double-stranded DNA helix, whereas in the following Chapter the same model is used to mimic the behaviour of the Microtubules and tau proteins inside the neuronal axon. The main non-dimensional parameter $\mu$ here introduced in the continuum limit (see Section~\ref{sec:ch5_cl}) regulates the behaviour: if $\mu\ll 1$ the behaviour is non cooperative and low denaturation forces are required, whereas if $\mu\gg 1$ the transition is an all-or-none phenomenon characterized by higher stability. Interestingly the cooperative effect is a purely `local' effect, that has instead been obtained in other models by introducing phenomenological non-local interactions (\cite{dauxois:1995,luca2020}).

This observed effect can have a crucial role in the experimental observation of the cooperativity measured, for example, in dsDNA that shows a narrow range of forces over which the transition occurs (\cite{bustamante:2000:2}). This behaviour is theoretically supported in this paper based on purely energetic considerations with only local interactions. We also remark that cooperativity and stability depend on the direction of applied force. Here we study shear-type loading, but similar results can be deduced when orthogonal forces are applied. The case of orthogonal loading was studied for example in (\cite{puglisi:2005}), whereas temperature effects have been recently analyzed in (\cite{puglisi:2022}). Interestingly, the purely mechanical effect obtained in this Chapter for lattice systems has been extensively discussed in the field of force transmitted by single fibres of composite materials (see (\cite{pugno:2003,salvo:2022}) and references therein), at least in the fully attached configuration, indicating the existence of a saturation force and an internal length scale leading to the so-called `shear-lag theory'. Finally, we remark that thermal and rate effects can have a fundamental role in the evolution of the system within its bumpy energy landscape with the possibility of spontaneous transition up to a certain critical temperature (\cite{schumakovitch:2002,puglisi:2022}). Specifically, if the rate is low, there can be time for thermal fluctuations to drive the system over the energy barrier with a small unbinding force whereas for higher regimes the mechanical potential quickly overwhelms the most prominent energy barrier along the reaction pathway, leaving all states free to unbind (\cite{evans:1997,pope:2001}). These effects will be analysed in the following chapter, where the rate is considered on the same prototypical model here introduced, whereas here, to focus the attention on the main physical effects described above, we focus on the case when entropic energy terms can be neglected as compared with the enthalpic ones in the case of low rate of loading. 

In this chapter, as already discussed, we apply our model to the specific case of a double-stranded DNA helix. The DNA is a life-key component in biological systems and it undergoes processes such as transcription, replication and recombination. From a structural point of view, a double-stranded DNA (dsDNA) helix is made by two polynucleotide strands held together by the base pairs (bp) through hydrogen bonds. The fundamental constraint required to obtain the helix structure is that two base sequences on opposite strands must be complementary and in particular adenine (A) binds with thymine (T) and guanine (G) is linked to cytosine (C) (\cite{saenger:2013}).  To this matter, the possibility of studying and understanding the role of intramolecular forces in dsDNA is crucial to determine how these processes are activated or in which situation their functionality can be compromised. 

In particular, different experimental techniques can be used to study the denaturation process of the DNA helix. For instance, Atomic Force Microscopy (\acs{AFM}) tests are conducted by fixing the displacement by pulling the molecule from a free end with the cantilever tip. An important feature in such experiments is the possibility of deducing the unbinding force at varying pulling rates (\cite{strunz:1999}). Similarly, both the interaction forces among complementary strands and the rupture force of interchain base pairs have been measured with high accuracy by the work of Lee and colleagues (\cite{lee:1994}). Here two kinds of forces have been identified: interchain forces associated with Watson-Crick base pairing between complementary strands of DNA (\cite{watson:1953}) and intrachain forces associated with the elasticity of single strands of DNA, where also thermal fluctuations play a crucial role (\cite{schumakovitch:2002}). Specifically, if the rate is low there can be time for thermal fluctuations to drive the system over the energy barrier with a small unbinding force, whereas for higher rate regimes the mechanical potential quickly overwhelms the most prominent energy barrier along the reaction pathway, leaving all states free to unbind (\cite{evans:1997}). These features were also studied in (\cite{pope:2001}), where anomalies in the rupture force by mapping the energy landscape of a DNA separating into a strand are analyzed. 

On the other hand, when a shear force is applied to the DNA chain, it can be observed that for strands with small lengths in the range of $10$ to $30$ base pairs, the force required to break the bonds increases linearly with increasing length. When the number of base pairs increases, the force saturates at a critical value approaching a plateau. While developing our model we found that a very similar approach has been pursued by Pierre Gilles De Gennes, in his work of $2001$ (\cite{degennes:2001}), where he tried to understand the intrinsic mechanism hidden under this behaviour by analysing a system with (formally) the same energy I introduced in this chapter. He derived, with a simple ladder model, a continuum equation correlating the rupture force to the total length of the DNA chain and he obtained the fundamental results that the force saturates for long chains (in terms of bp length). In his wake, Hatch and coworkers performed pulling experiments on dsDNA helices with magnetic tweezers and they fitted the data with the De Gennes' equation without any reference to the physical meaning of the quantities in the formulas (\cite{hatch:2008}). Indeed, the underlying reason for this behaviour is that the two backbones, which are able to freely move in the space, may slide in the direction of the load, depending on both the force and the length of the chain. If the backbone is elastic and stiffer than the base pairs, the shear decreases from the endpoints towards the centre of the system and consequently, the effect of the shear force is distributed only in a few pairs at the ends of the DNA. Consequently, if the chain is short all the length is affected by the shear force whereas if the number of the base pairs increases the central region will become gradually unaffected by the force. In this Chapter, we prove this feature by introducing a simple lattice model to describe such families of systems. In particular, I consider two elastic backbones plus breakable links representing the base pairs of the DNA. I study the system under applied shear by imposing the forces at the free upper and lower and of the chain (see Figure~\ref{fig:ch5_modenergy}) or by fixing the total displacement of the system and I obtain the equilibrium mechanical response under different hypotheses. Moreover, to compare our theoretical analysis with the results presented in (\cite{degennes:2001}), I study the system in the so-called continuum limit, where the discreteness of breaking events becomes a fraction of the broken portion of the system. In this regime, our analytical results are identical to the De Gennes's formula, and it is possible to recover the physical meaning of quantities such as the `persistence length' or the relative stiffness of the chain, in terms of the stiffnesses of the elements and internal lengths.





\section{Mechanical model}
\label{sec:ch5_model}

\begin{figure}[t!]
\centering
  \includegraphics[width=0.95\textwidth]{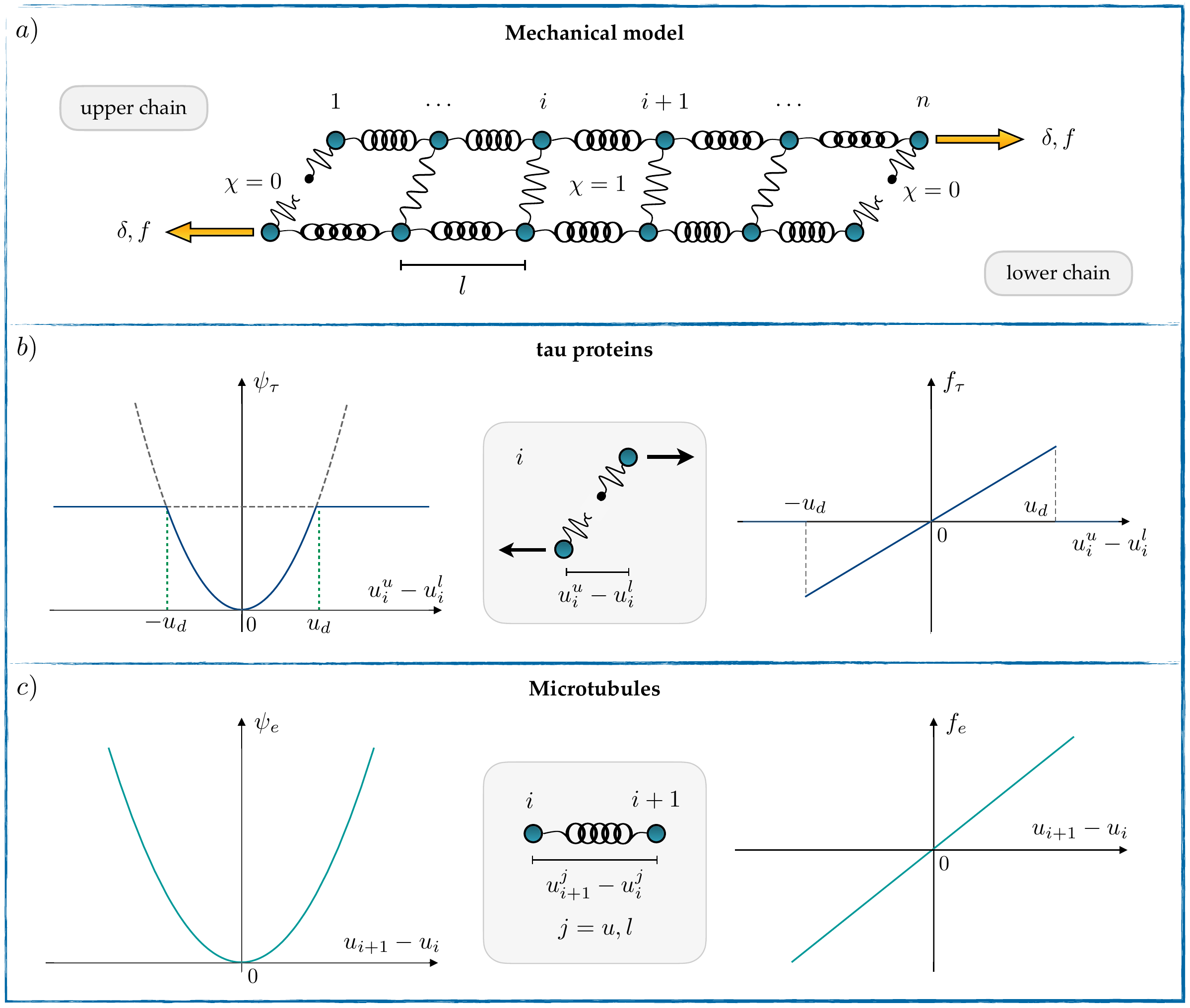}
  \caption[Mechanical model of the shear lattice]{Mechanical model. Panel a): Scheme of the lattice describing the microtubules (horizontal springs) and the $\tau$-proteins (shear links) made by $n$ elements. Panel b): Energy of a single $\tau$ protein where the behavior is elastic since the detachment displacement $u_i<u_d$ and broken for $u_i>u_d$. Panel c): Energy of the elastic units representing the MTs.}
  \label{fig:ch5_modenergy}
\end{figure}

Let us consider a bundle of two bi-stable chains connected by breakable units, as shown in Figure~\ref{fig:ch5_modenergy}$_c$. This mechanical model can be used to describe different physical microscopic systems. As two paradigmatic examples, I consider the viscoelastic rate-dependent behaviour of the axonal damage, mimicking the structure of the microtubules connected with $\tau$-proteins inside the axons, and the debonding effects under shear load in the denaturation process of the DNA, where our model describes the two strands joined by the base pairs. To focus our attention and fix the notation I refer in this draft to the first example, nevertheless, I analyze both debonding phenomena in the two cases at the end of the theoretical treatment. 

%

To mimic a portion of the axon I consider a section of a microtubules (MTs) bundle where I model two microtubules as elastic chains connected by shear breakable links describing the tau proteins. Specifically, I consider both the upper (denoted with the apex $u$)  and the lower (denoted with the apex $l$) chains made by $n$ identical linear elastic springs with stiffness $k_e$ and length $l$. The energy (see Figure~\ref{fig:ch5_modenergy}$_b$) is then
\begin{equation}
\psi_e=\frac{1}{2}k_{e}l\sum_{j=u,l}\sum_{i=1}^{n-1}\left(\frac{u_{i+1}^j-u_{i}^j}{l}\right)^2,
\end{equation}
%
where $u_i$ is the displacement of the $i-$th spring both in the upper and in the lower chain. On the other hand, following (\cite{puglisi:2013}), aiming at analytical clearness, I describe the energy of the breakable links as two states elements (see Figure~\ref{fig:ch5_modenergy}$_a$). For $u_i<u_d$ the behavior is linearly elastic with stiffness $k_{\tau}$, while for $u_i>u_d$ the link is detached and the force is zero. Thus, following  (\cite{fp:2019,luca2019}), after  introducing a ``spin'' variable $\chi_i$ that assumes value $\chi_i=1$ if the link is attached and $\chi_i=0$ if the link is broken, the energy of the breakable units is written as 
\begin{equation}
\psi_{\tau}=\frac{1}{2}k_{\tau}l\sum_{i=1}^{n}\left[\chi_i\left(\frac{u_i^u-u_{i}^l}{u_{d}}\right)^2+\left(1-\chi_i\right)\right],
\label{eq:ch5_tauenergy}
\end{equation}
where $l$ is the undeformed length of the spring.  In order to obtain non-dimensional quantities let us introduce the rescaled displacement $w_i=u_i/u_d$ and the main dimensionless parameter $\nu^2$, representing the ratio between the stiffness of the tau-protein with respect to the microtubules rigidity
\begin{equation}
\nu^2=\frac{k_{\tau}}{k_e}\frac{l^2}{u_{d}^{2}}.
\label{eq:ch5_nu}
\end{equation}
Eventually, we obtain the total non-dimensional energy of the system 
\begin{multline}
n\varphi=\frac{L}{k_e u_d^2}\left(\psi_e+\psi_{\tau}\right)
=\frac{1}{2}\nu^2\sum_{i=1}^{n}\left[\chi_i\left(w_{i}^{u}-w_{i}^{l}\right)^2+\left(1-\chi_i\right)\right]\\
+\frac{1}{2}\sum_{j=u,l}\sum_{i=1}^{n-1}\left(w_{i+1}^{j}-w_{i}^{j}\right)^2,
\label{eq:ch5_en}
\end{multline}
where $L=nl$ is the total natural length of both upper and lower chains.

To further simplify the manipulation of the mechanical quantities let us introduce the matrices
\begin{equation}
\boldsymbol{L}=
	\begin{pmatrix}
	2			&-1 		& 		& 		&\boldsymbol{0} 	\\
	-1 			&2		& -1		& 		&				\\
				&\ddots	&\ddots	&\ddots 	&				\\
				&  		& -1		& 2		&-1				\\
	\boldsymbol{0}	& 		&		& -1		& 2				\\
	\end{pmatrix}_{n\times n}, \qquad
	\boldsymbol{D}=
	\begin{pmatrix}
	\chi_1	&		& 		& 		&\boldsymbol{0} 		\\
			&\ddots	&		&		&		\\
			&		& \chi_i	& 		&		\\																&	 	& 		& \ddots	&		\\
	\boldsymbol{0} 		&		&		& 		&\chi_{n}	\\
	\end{pmatrix}_{n\times n},
\end{equation}
where $\boldsymbol{L}$ takes into account the elastic interaction in the upper and lower chains whereas $\boldsymbol{D}$ describes the overall state of the spins describing the attached/detached configuration of the breakable links. Moreover,  we introduce the displacement and the phase vectors
\begin{equation}
\boldsymbol{w}^{u}=
	\begin{pmatrix}
	w_1^{u}	\\
	w_2^{u}		\\
	\dots	\\
	w_i^{u} 		\\
	\dots	\\
	w_n^{u}		\\ 
	\end{pmatrix}_{n}, \qquad
	\boldsymbol{w}^{l}=
	\begin{pmatrix}
	w_1^{l}		\\
	w_2^{l}		\\
	\dots 	\\
	w_i^{l} 		\\
	\dots	\\
	w_n^{l}		\\ 
	\end{pmatrix}_{n}, \qquad
	\boldsymbol{\chi}=
	\begin{pmatrix}
	\chi_1	\\
	\chi_2	\\
	\dots	\\
	\chi_i	\\
	\dots	\\ 
	\chi_{n}	\\
	\end{pmatrix}_{n}.
\end{equation}
and the first and last bases of the $n$-th dimensional space
\begin{equation}
	\boldsymbol{i}_{n}=
	\begin{pmatrix}
	0	\\
	0	\\
	\dots\\
	\dots\\
	0 	\\
	1	\\ 
	\end{pmatrix}_{n}, \qquad
	\boldsymbol{i}_{1}=
	\begin{pmatrix}
	1	\\
	0	\\
	\dots\\
	\dots\\
	0 	\\
	0	\\ 
	\end{pmatrix}_{n}.
\end{equation}
Accordingly, we may rewrite the energy as 
\begin{multline}
n\varphi=\frac{1}{2}\sum_{j=u,l}\left[\left(\boldsymbol{L}+\nu^2\boldsymbol{D}\right)\boldsymbol{w}^{j}\cdot\boldsymbol{w}^{j}-\left(\boldsymbol{w}^{j}\cdot\boldsymbol{i}_{1}\right)^2-\left(\boldsymbol{w}^{j}\cdot\boldsymbol{i}_{n}\right)^2\right]\\
-\nu^2\boldsymbol{D}\boldsymbol{w}^{u}\cdot\boldsymbol{w}^{l}+\frac{1}{2}\nu^2(n-\boldsymbol{\chi}\cdot\boldsymbol{\chi}).
\label{eq:ch5_phi1}
\end{multline}

In order to solve the equilibrium problem, we introduce the vectors $\boldsymbol{v}$ and $\boldsymbol{z}$ 
\begin{equation}
\boldsymbol{v}=\boldsymbol{w}^{u}-\boldsymbol{w}^{l},\qquad\qquad\boldsymbol{z}=\boldsymbol{w}^{u}+\boldsymbol{w}^{l},
\end{equation}
where $v_i$ and $z_i$ represent the shear and the displacement of the $i$-th shear spring centre of mass,  respectively. We introduce also the matrix 
\begin{equation}
\boldsymbol{J}=\boldsymbol{L}+2\nu^2\boldsymbol{D}.
\end{equation}
Thus, we obtain
\begin{multline}
n\varphi=\frac{1}{4}\left[\boldsymbol{J}\boldsymbol{v}\cdot\boldsymbol{v}-\left(\boldsymbol{v}\cdot\boldsymbol{i}_{1}\right)^2-\left(\boldsymbol{v}\cdot\boldsymbol{i}_{n}\right)^2\right]+\frac{1}{4}\left[\boldsymbol{L}\boldsymbol{z}\cdot\boldsymbol{z}-\left(\boldsymbol{z}\cdot\boldsymbol{i}_{1}\right)^2-\left(\boldsymbol{z}\cdot\boldsymbol{i}_{n}\right)^2\right]\\
+\frac{1}{2}\nu^2(n-\boldsymbol{\chi}\cdot\boldsymbol{\chi}).
\label{eq:ch5_phi1}
\end{multline}
Notice that with this notation we have
\begin{equation}
\boldsymbol{w}^{u}=\frac{\boldsymbol{z}+\boldsymbol{v}}{2},\qquad\qquad\boldsymbol{w}^{l}=\frac{\boldsymbol{z}-\boldsymbol{v}}{2}.
\label{ww}
\end{equation}
%

\section{Equilibrium}

Let us study our system under assigned displacement, \textit{i.e.} in the \textit{hard device} hypothesis. We impose (see Figure~\ref{fig:ch5_modenergy}$_c$) the displacement of the upper right and lower left ends $\delta=w_{n}^u=-w_1^l$, whereas the two other end-pont masses are unconstrained.  We then minimize the new function
\begin{equation}
h=\varphi-\frac{f}{n}\Bigl(w_{n}^{u}-\delta\Bigr)+\frac{f}{n}\left(w_{1}^{l}+\delta\right),
\end{equation}
namely 
\begin{equation}
h=\varphi-\frac{f}{n}\left(\frac{\boldsymbol{z}+\boldsymbol{v}}{2}\cdot\boldsymbol{i}_{n}-\delta\right)+\frac{f}{n}\left(\frac{\boldsymbol{z}-\boldsymbol{v}}{2}\cdot\boldsymbol{i}_{1}+\delta\right),
\end{equation}
where $f$ is the Lagrange multiplier
\begin{equation}
f:=\frac{FL}{k_e u_d}
\label{eq:ch5_forceoriginal}
\end{equation}
having the meaning of (non-dimensional) force conjugated to $\delta$. Thus one has to study the variational problem
\begin{equation}
\min_{w_{1}^{l}=-\delta, w_{n}^{u}=\delta} h.
\end{equation}
Following (\cite{puglisi:2000}) we begin by minimizing the energy at fixed configuration $\boldsymbol{\chi}$. Equilibrium requires  
\begin{equation}
\begin{split}
	&	\frac{\partial h}{\partial \boldsymbol{v}}	= \boldsymbol{J}\boldsymbol{v}-v_{1}\,\boldsymbol{i}_{1}-v_{n}\,\boldsymbol{i}_{n}-\frac{f}{n}\boldsymbol{i}_{n}-\frac{f}{n}\boldsymbol{i}_{1}=\boldsymbol{0},\\
	&	\frac{\partial h}{\partial \boldsymbol{z}}	= \boldsymbol{L}\boldsymbol{z}-z_{1}\,\boldsymbol{i}_{1}-z_{n}\,\boldsymbol{i}_{n}-\frac{f}{n}\boldsymbol{i}_{n}+\frac{f}{n}\boldsymbol{i}_{1}=\boldsymbol{0},
\end{split}
\end{equation}
that give
\begin{equation}
\begin{split}
	& \boldsymbol{v} = v_{1}\,\boldsymbol{J}^{-1}\boldsymbol{i}_{1}+v_{n}\,\boldsymbol{J}^{-1}\boldsymbol{i}_{n}+\frac{f}{n}\boldsymbol{J}^{-1}\boldsymbol{i}_{n}+\frac{f}{n}\boldsymbol{J}^{-1}\boldsymbol{i}_{1},\\
	&\boldsymbol{z} = z_{1}\,\boldsymbol{L}^{-1}\boldsymbol{i}_{1}+z_{n}\,\boldsymbol{L}^{-1}\boldsymbol{i}_{n}+\frac{f}{n}\boldsymbol{L}^{-1}\boldsymbol{i}_{n}-\frac{f}{n}\boldsymbol{L}^{-1}\boldsymbol{i}_{1}.
	\label{eq:ch5_vec}
\end{split}
\end{equation}

Observe that $\boldsymbol{L}$ and $\boldsymbol{J}$ (that depends on $\boldsymbol{\chi}$) are tridiagonal symmetric matrices. They are always non-singular except in the special case when all $\tau$ proteins are broken ($\chi_i=0, i=1,...,n$) that will be considered separately. A direct inspection of $\boldsymbol{L}^{-1}$ shows that 
\begin{equation}
\boldsymbol{L}^{-1}_{1,n}=1-\boldsymbol{L}^{-1}_{n,n}=1-\boldsymbol{L}^{-1}_{1,1},
\end{equation}
hence 
\begin{equation}
z_{1}=-z_{n}=\frac{2\boldsymbol{L}^{-1}_{1,n}-1}{2\boldsymbol{L}^{-1}_{1,n}}\frac{f}{n}.
\label{eq:ch5_z1n}
\end{equation}
Thus we obtain
\begin{equation}
\boldsymbol{z}=\frac{\boldsymbol{L}^{-1}\boldsymbol{i}_{n}-\boldsymbol{L}^{-1}\boldsymbol{i}_{1}}{2\boldsymbol{L}^{-1}_{1,n}}\frac{f}{n}.
\label{eq:ch5_zz}
\end{equation}

Similarly, since $\boldsymbol{J}^{-1}_{1,n}=\boldsymbol{J}^{-1}_{n,1}$ we obtain 
\begin{equation}
v_{1}=\frac{\left(\boldsymbol{J}^{-1}_{1,n}\right)^2+\boldsymbol{J}^{-1}_{1,n}+\boldsymbol{J}^{-1}_{1,1}\left(1-\boldsymbol{J}^{-1}_{n,n}\right)}{\left(1-\boldsymbol{J}^{-1}_{1,1}\right)\left(1-\boldsymbol{J}^{-1}_{n,n}\right)-\left(\boldsymbol{J}^{-1}_{1,n}\right)^2}\frac{f}{n}
\end{equation}
and
\begin{equation}
v_{n}=\frac{\left(\boldsymbol{J}^{-1}_{1,n}\right)^2+\boldsymbol{J}^{-1}_{1,n}+\boldsymbol{J}^{-1}_{n,n}\left(1-\boldsymbol{J}^{-1}_{1,1}\right)}{\left(1-\boldsymbol{J}^{-1}_{1,1}\right)\left(1-\boldsymbol{J}^{-1}_{n,n}\right)-\left(\boldsymbol{J}^{-1}_{1,n}\right)^2}\frac{f}{n}.
\end{equation}
The shear vector can then be rewritten as
\begin{equation}
\boldsymbol{v}=\frac{\left(1+\boldsymbol{J}^{-1}_{1,n}-\boldsymbol{J}^{-1}_{n,n}\right)\boldsymbol{J}^{-1}\boldsymbol{i}_{1}+\left(1+\boldsymbol{J}^{-1}_{1,n}-\boldsymbol{J}^{-1}_{1,1}\right)\boldsymbol{J}^{-1}\boldsymbol{i}_{n}}{\left(1-\boldsymbol{J}^{-1}_{1,1}\right)\left(1-\boldsymbol{J}^{-1}_{n,n}\right)-\left(\boldsymbol{J}^{-1}_{1,n}\right)^2}\frac{f}{n}.
\label{eq:ch5_vv}
\end{equation}

By reimposing the kinematic constraints it is possible to evaluate the overall stiffness of the system for a generic $\boldsymbol{\chi}$. Using~\eqref{ww} we obtain 
\begin{equation}
	\delta =w_{n}^{u}=\frac{z_n+v_n}{2}, \quad 
		-\delta= w_{1}^{l}=\frac{z_1-v_1}{2}
		\label{eq:ch5_wd}
\end{equation}
that gives
\begin{equation}
	\delta=\frac{1}{4}\left(z_n-z_1+v_n+v_1\right).
	\label{eq:ch5_deltadelta}
\end{equation}
If we introduce the (configuration dependent) stiffness 
\begin{equation}
	\kappa(\boldsymbol{\chi})=\frac{f}{\delta}=4\bigl(z_n-z_1+v_n+v_1\bigr)^{-1}f
\end{equation}
we have
\begin{equation}
	\kappa(\boldsymbol{\chi})=4n\left(\frac{1-2\boldsymbol{L}^{-1}_{1,n}}{\boldsymbol{L}^{-1}_{1,n}}+\frac{2\left(\boldsymbol{J}^{-1}_{1,n}\right)^2+2\boldsymbol{J}^{-1}_{1,n}+\boldsymbol{J}^{-1}_{1,1}\left(1-\boldsymbol{J}^{-1}_{n,n}\right)+\boldsymbol{J}^{-1}_{n,n}\left(1-\boldsymbol{J}^{-1}_{1,1}\right)}{\left(1-\boldsymbol{J}^{-1}_{1,1}\right)\left(1-\boldsymbol{J}^{-1}_{n,n}\right)-\left(\boldsymbol{J}^{-1}_{1,n}\right)^2}\right)^{-1}.
	\label{eq:ch5_kchi}
\end{equation}

The inverse of the tridiagonal matrix $\boldsymbol{J}$  can be evaluated through recursive expressions hard to handle that can hide the underlying physics. On the other hand, using the results in (\cite{hu:1996}), $\boldsymbol{J}^{-1}$ can be explicitly evaluated in the special case when $\boldsymbol{D}=\boldsymbol{I}$, \text{i.e.} when we consider a system in which all the $\tau$-proteins are attached. In what follows we show that all the relative minimizers of the energy are characterized by a single block of connected elements --whose stiffness can be evaluated using previous considerations on $\boldsymbol{J}$-- with the remaining part of the system in the detached state (see Figure~\ref{fig:ch5_raccordo}). Moreover, it is easy to see that all solutions with the same number $p$ of connected elements, corresponding to different positions of the connected block in the system, are energetically equivalent. 

To show this  result, we start by demonstrating that for a generic $\delta$ and $p$ the largest component of the shear vector $v_i$ is always attained at the boundaries of the connected block. Indeed, let us consider the inverse of a matrix $\boldsymbol{J}$ with dimension $p\in (0,n)$ corresponding to the block of $p$ attached elements. This matrix is always positive definite and $\boldsymbol{J}^{-1}_{i,j}>\boldsymbol{J}^{-1}_{i,j+1}, j\ge i$ (\cite{nabben:1999,nabben:1999:2}), {\it i.e.} the elements decrease as the distance from the diagonal increases. In this hypothesis $\boldsymbol{J}$ is double symmetric, so that using \eqref{eq:ch5_vv} the required monotonicity of the components of $\boldsymbol{v}$ coincide with that of the vector $\boldsymbol{J}^{-1}\boldsymbol{i}_{1}+ \boldsymbol{J}^{-1}\boldsymbol{i}_{n}$. 
We need to verify that 
\begin{equation}
\boldsymbol{J}^{-1}_{1,i}+\boldsymbol{J}^{-1}_{n,i}>\boldsymbol{J}^{-1}_{1,i+1}+\boldsymbol{J}^{-1}_{n,i+1} \Leftrightarrow \boldsymbol{J}^{-1}_{1,i}-\boldsymbol{J}^{-1}_{1,i+1}>\boldsymbol{J}^{-1}_{n,i+1}-\boldsymbol{J}^{-1}_{n,i}.
\end{equation}
Thus we need to show that  $\boldsymbol{J}^{-1}_{1,i}-\boldsymbol{J}^{-1}_{1,i+1}$ decreases as $i$ grows. To get this result, we may use the expression of the inverse matrix in (\cite{hu:1996}) and by considering $i$ as a continuous variable observe that
\begin{equation}
\frac{\partial\bigl(\boldsymbol{J}^{-1}_{1,i}-\boldsymbol{J}^{-1}_{1,i+1}\bigr)}{\partial i}=\frac{\cosh\left[(p-i)\lambda\right]-\cosh\left[(p+1-i)\lambda\right]}{\sinh\left[(p+1)\lambda\right]} <0,
\label{eq:ch5_mono}
\end{equation}
with 
\begin{equation}
\quad i<p, \quad \nu^2 \ge 0,
\end{equation}
where 
\begin{equation}\label{lambda}
\lambda=\arccosh(1+\nu^2).
\end{equation}
This proves the monotonicity of the shears in the attached block. The resulting configuration evolution of the system is discussed in the following section.

\subsection{Di-Block solution}
\label{sec:ch5_diblock}

\begin{figure}[t!]
\centering
  \includegraphics[width=0.95\textwidth]{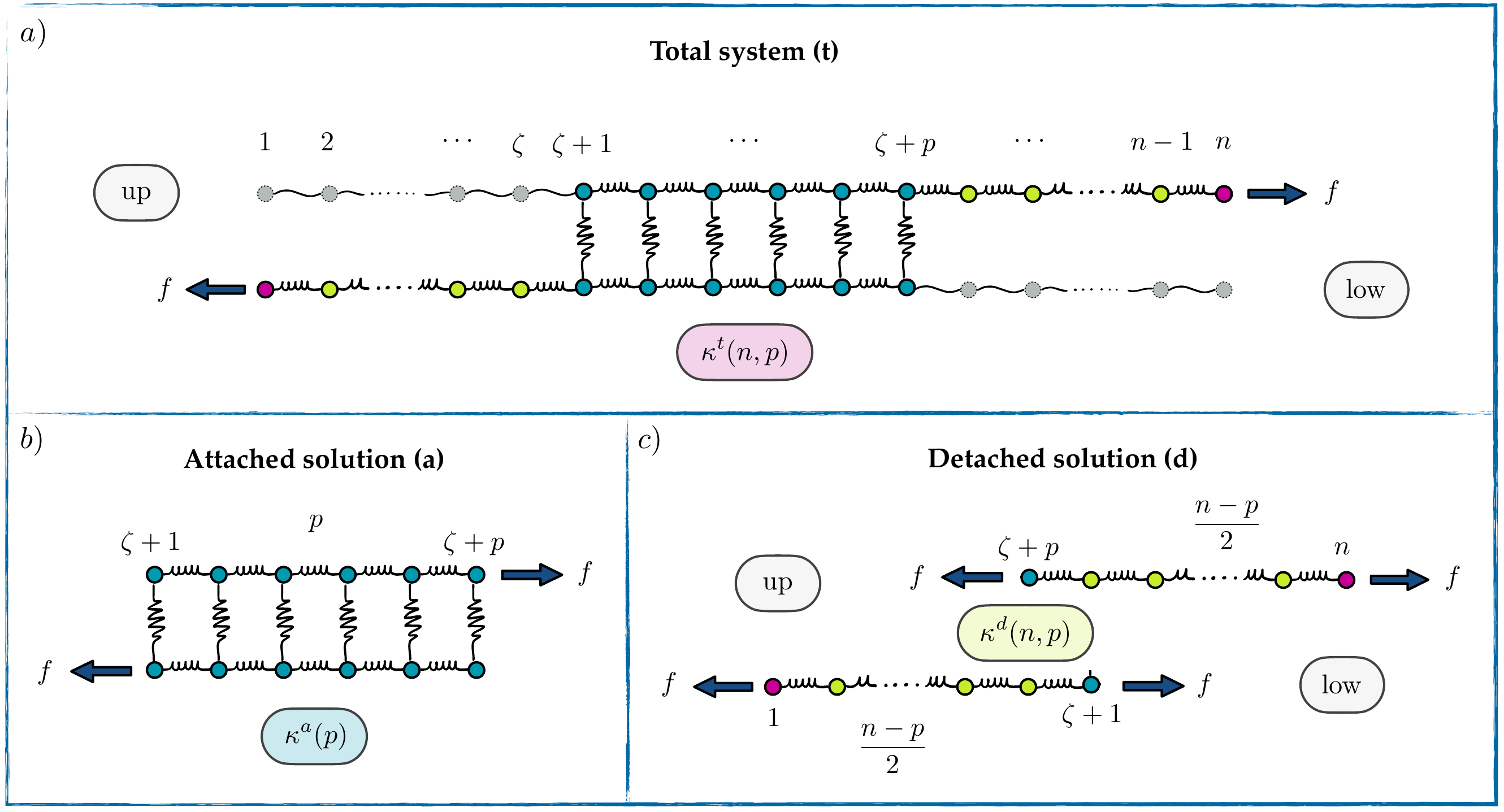}
  \caption[Di-Block solution]{Di-Block solution. Panel a): The total (apex t) system is shown and a shear load is applied by imposing two forces (or displacements). The variable $\zeta=(n-p)/2$ is a counter to locate the attached part with respect to the detached part. Panel b): The attached solutions and relative stiffness is represented. Panel c): The detached loaded parts are shown.}
  \label{fig:ch5_raccordo}
\end{figure}

Consider then a path of increasing displacement $\delta$, starting from the virgin configuration $\chi_i=1$, $i=1,...\,,n$ and $\boldsymbol{J}$ with dimension $n$. Following the discussion in the previous section, $v_1=v_n$ are the largest shears and the first ones that reach the limit condition $v_i=1$.  Once the first link is broken, we have a new system with a connected block with $n-1$ elements and one broken link. If we further increase the parameter $\delta$ we can repeat the monotonicity argument, by restricting our monotonicity considerations to the remaining  $n-i$ unbroken elements and so on. Accordingly, in the generic configuration the mechanical system is in a double-blocks configuration as the one represented in Figure~\ref{fig:ch5_raccordo},  composed by an attached part of dimension $p$ and a detached part of elastic units of dimension $(n-p)$ where only axial stretches are present. As the Figure shows the remaining links (indicated with grey colour in the Figure) are unloaded and unstretched. Observe that our system does not distinguish energetically the real position of the connected block.  Observe once again that in our model, neglecting non-local interactions (\cite{luca2020}), the position of the attached block is energetically indifferent. Here and in the following, we use the apexes $t, a, d$ to indicate the functions of the total, attached and detached parts.

After previous considerations, let us first determine the displacement of the attached block. Using~\eqref{eq:ch5_z1n} we obtain 
\begin{equation}
\boldsymbol{L}^{-1}_{1,1}=\boldsymbol{L}^{-1}_{p,p}=\frac{p}{p+1},
\qquad\qquad 
\boldsymbol{L}^{-1}_{1,p}=1-\boldsymbol{L}^{-1}_{1,1}=\frac{1}{p+1}.
\label{eq:ch5_l2}
\end{equation}
Moreover, for the connected block, we have a $p \times p$ matrix
\begin{equation}
\boldsymbol{J}=-
	\begin{pmatrix}
	D			&1 		& 		& 		&\boldsymbol{0} 	\\
	1 			&D		& 1		& 		&				\\
				&\ddots	&\ddots	&\ddots 	&				\\
				&  		& 1		& D		&1				\\
	\boldsymbol{0}	& 		&		& 1		& D				\\
	\end{pmatrix},
\end{equation}
where $D=-2(1+\nu^2)$. In this case, the inverse of $\boldsymbol{J}$ can be analytically computed as shown in (\cite{hu:1996}), where the generic element is 
\begin{equation}
\boldsymbol{J}^{-1}_{i,j}=\frac{\cosh\left[(p+1-|j-i|)\lambda\right]-\cosh\left[(p+1-i-j)\lambda\right]}{2\sinh\lambda \sinh(p+1)\lambda}.
\label{eq:ch5_j1}
\end{equation}
where $\lambda$ is given in~\eqref{lambda}. 

Let us simplify the expression of the overall stiffness in~\eqref{eq:ch5_kchi} by considering that $\boldsymbol{J}^{-1}_{1,1}=\boldsymbol{J}^{-1}_{p,p}$, obtaining 
\begin{equation}
	\kappa^a(p)=p\left(\frac{1-2\boldsymbol{L}^{-1}_{1,p}}{4\boldsymbol{L}^{-1}_{1,p}}+\frac{\boldsymbol{J}^{-1}_{1,1}+\boldsymbol{J}^{-1}_{1,p}}{2\left(1-\boldsymbol{J}^{-1}_{1,p}-\boldsymbol{J}^{-1}_{p,p}\right)}\right)^{-1},
	\label{eq:ch5_kchib}
\end{equation}
so that by using~\eqref{eq:ch5_l2} and \eqref{eq:ch5_j1} we obtain the stiffness of the attached part 
\begin{equation}
\kappa^a(p)=\frac{4p}{p-1+4\gamma(p)},
\end{equation}
where
\begin{equation}
\gamma(p)=\frac{\sinh(\lambda)+\sinh(p\lambda)}{2\left\{\sinh[\left(p+1\right)\lambda]-\sinh(\lambda)-\sinh(p\lambda)\right\}}.
\end{equation}
It's easy to verify that $\kappa^a(p)$ decreases giving a measure of the increasing damage effect.

We can compute explicitly also the displacement vector~\eqref{eq:ch5_zz} by computing firstly 
\begin{equation}
\boldsymbol{L}^{-1}\boldsymbol{i}_p=\frac{i}{p+1}, \quad 
\boldsymbol{L}^{-1}\boldsymbol{i}_p=\frac{p+1-i}{p+1},
\end{equation}
that give
\begin{equation}
z_i=\frac{2\left(2 i-p-1\right)}{p-1+4 \gamma (p)} \, \delta, \quad \quad i=1,\dots,p.
\label{eq:ch5_va}
\end{equation}
Moreover, by~\eqref{eq:ch5_vv} we get 
\begin{equation}
\boldsymbol{v}=\frac{1}{1-\boldsymbol{J}^{-1}_{p,p}-\boldsymbol{J}^{-1}_{1,p}}\left(\boldsymbol{J}^{-1}\boldsymbol{i}_{p}+\boldsymbol{J}^{-1}\boldsymbol{i}_{1}\right)\, \kappa^a(p) \, \delta,
\label{eq:ch5_v}
\end{equation}
that, by using~\eqref{eq:ch5_j1}, gives
\begin{equation}
\centering
	v_i  =\frac{2\left\{\cosh\left[(p-i+2)\lambda\right]-\cosh\left[(p-i)\lambda\right]+\cosh\left[(i+1)\lambda\right]-\cosh\left[(i-1)\lambda\right]\right\}}{\sinh(\lambda)\left\{\left(p-1\right)\sinh\left[(p+1)\lambda\right]-\left(p-3\right)\left[\sinh(\lambda)+\sinh(p\lambda)\right]\right\}}\, \delta,
	\label{eq:ch5_vi}
\end{equation}
with $i=1,\dots,p$. Here, based on the previous invariance of the position of the connected attached block, we relabeled the indexes of the corresponding displacements. 

%

Let us then consider the detached elastic part, where  the displacement respects the trivial  equilibrium condition 
\begin{equation}
w^j_{i+1}-w_{i}^{j}=\frac{f}{n} , \qquad\qquad i=p+1,\dots,n, \qquad j=l,r,
\end{equation}
where again we relabelled the indexes to take care of the arbitrariness of the position of the detached elements. To evaluate the stiffness of the whole (attached plus detached) system (see Figure~\ref{fig:ch5_raccordo}) we may just obtain the total displacement $\delta$, as the sum of the elongation of the attached block plus one of the detached segments
\begin{equation}
\delta=\frac{p}{\kappa^a(p)}\frac{f}{n}+\frac{n-p}{2}\frac{f}{n}.
\end{equation}
The total system stiffness  at given configuration $p$ is then
\begin{equation}
\kappa^{t}(n,p)=n\left(\frac{p}{\kappa^a(p)}+\frac{n-p}{2}\right)^{-1}=\frac{4n}{2n-p-1+4\gamma(p)}.
\label{eq:ch5_ktot}
\end{equation}

Eventually, the total energy  of the system is 
\begin{equation}
\varphi_{eq}(n,p,\delta)=\kappa^t(n,p)\delta^2+\frac{1}{2}n\nu^2\left(n-p\right),
\label{eq:ch5_phitot}
\end{equation}
and the force is 
\begin{equation}
f(n,p,\delta)=\kappa^t(n,p)\delta.
\label{eq:ch5_f}
\end{equation}

Finally, the existing domain of each equilibrium branch  can be computed by considering the limit condition when, by using~\eqref{eq:ch5_vi}, and the shown monotonicity, the external links reach the threshold $v_p=v_1=1$:
\begin{equation}
v_p=\frac{1}{\kappa_b(p)}\frac{f}{n}=\frac{\kappa^{t}(n,p)}{\kappa_b(p)}\frac{\delta_{max}}{n}=1,
\end{equation}
where we introduced the stiffness factor $\kappa_b(p)$ relating the displacement of the endpoint element of the attached block to the force, which can be evaluated by the simple relation
\begin{equation}
\kappa_b(p)=\frac{\gamma(p)}{2}.
\end{equation}

Thus for the $p$ branch, we have
\begin{equation}
\delta \in(0,\delta_{max}), \quad \delta_{max}:=\frac{1}{4} \left\{\frac{(2 n-p-1) \sinh \left[(p+1) \lambda\right]}{\sinh(\lambda)+\sinh(p \lambda)}-2 n+p+3\right\},
\label{eq:ch5_dmax}
\end{equation}
with $p=1,...,n$.

\subsection{Decohesion strategies}
\label{sec:ch5_strategie}

\begin{figure}[t!]
\centering
  \includegraphics[width=0.95\textwidth]{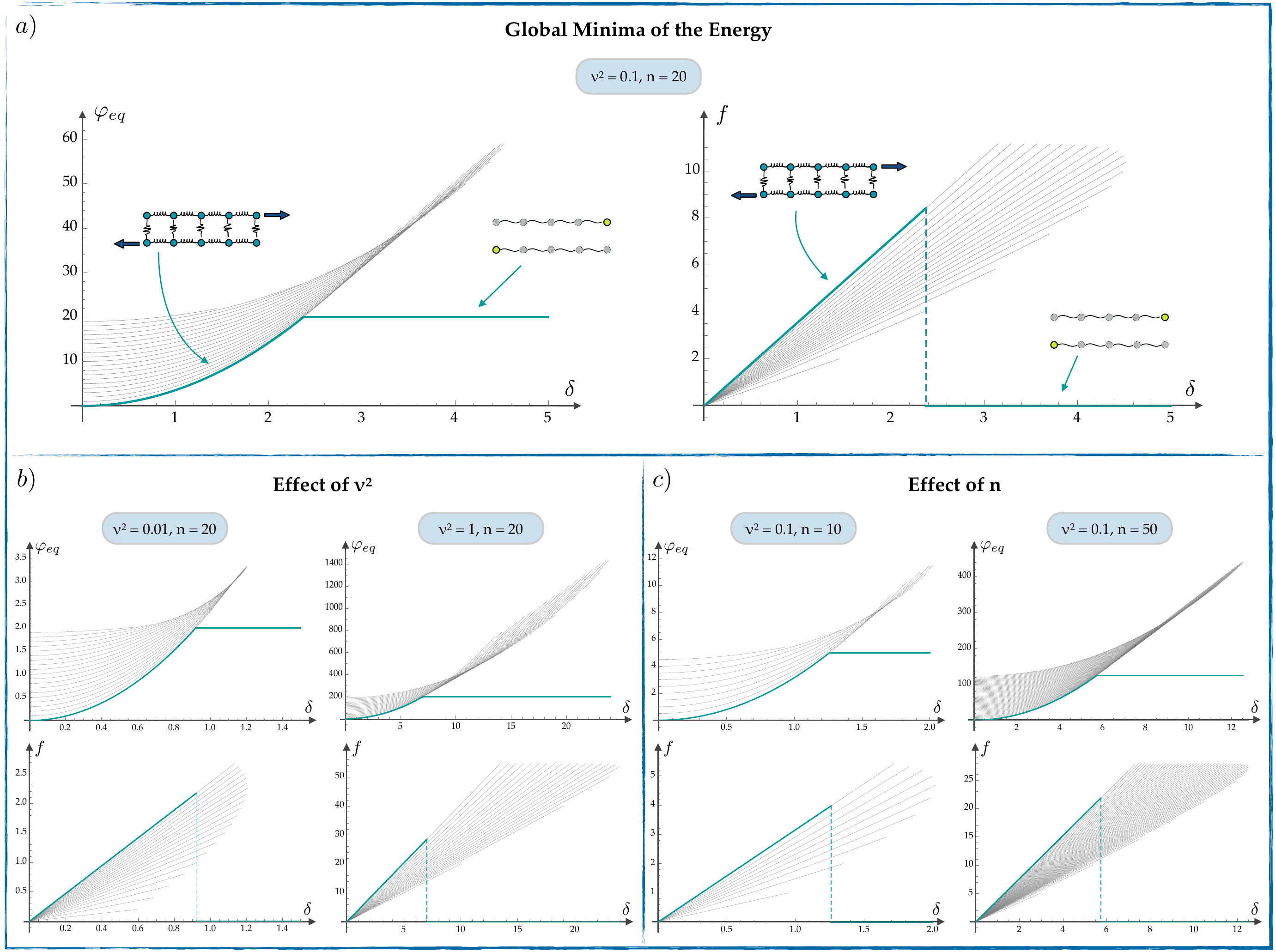}
  \caption[Maxwell convention]{Maxwell convention. Panel a): The behaviour of the system in the global minima is always characterized by a transition from the fully attached state to the fully detached configuration, i.e. it is always a fragile rupture. Panel b): Effect of different values of the parameter $\nu^2$. Panel c): Effect of different values of the number of objects $n$. In both b) and c) there is always a fragile rupture.}
  \label{fig:ch5_global}
\end{figure}

In this section, following (\cite{puglisi:2005}), we study the evolution of the system at varying displacement under two extreme hypotheses concerning the possibility of overcoming energy barriers. First, we may consider the \textit{Maxwell convention}, in which the system stays always in the global minimum of the energy. We observe that under this hypothesis the system shows a `fragile' behaviour with a sudden transition from the fully attached state to the fully detached one, independently both from the stiffness parameter $\nu^2$ and the number of objects $n$, as pictured in Figure~\ref{fig:ch5_global}.

%

To demonstrate this result we may prove that the energy branch representing the fully attached configuration (\textit{i.e.} $\varphi_{eq}(n,1,\delta)$) intersects the fracture energy line (fully detached, \textit{i.e.} $\varphi_{eq}(n,0,\delta)$) before intersecting the energy diagrams of all other partially detached configurations. Accordingly, at fixed $n$ and $\nu^2$ we evaluate the displacement $\delta_p$ of each intersection  by solving the following equality: 
\begin{equation}
\varphi_{eq}(p)=\varphi_{eq}(p+1) \Leftrightarrow \delta_p=\sqrt{\frac{n}{2}}\frac{\nu }{\sqrt{\kappa^t(p+1)-\kappa^t(p)}}.
\label{eq:ch5_deltap}
\end{equation}
We observe that if~\eqref{eq:ch5_deltap} decreases as $p$ decreases the thesis is demonstrated. Thus we need to demonstrate that 
\begin{equation}
\frac{\partial \delta_p}{\partial p}=\frac{\nu}{2}\sqrt{\frac{n}{2}}\frac{\frac{\partial \kappa^t(p)}{\partial p}-\frac{\partial \kappa^t(p+1)}{\partial p}}{\Big[\kappa(p+1)-\kappa(p)\Big]^{\frac{3}{2}}}>0\quad \forall \,p <n
\end{equation}
that reduces to 
\begin{equation}
\frac{\partial \kappa^t(p)}{\partial p}-\frac{\partial \kappa^t(p+1)}{\partial p}>0\quad \forall \,p <n,
\end{equation}
By a direct inspection of the derivative of~\eqref{eq:ch5_ktot} one observes that the condition is fulfilled, and this result proves our thesis. 

We remark that the obtained result is coherent with the experimental observation (\cite{smith:2000}) that in a slow rate regime, \textit{i.e.} when the system can relax to the global energy minimum and equilibrium condition is approached, all the $\tau$-proteins may detach together and eventually reattach to other sites along the microtubules (sliding regime) (\cite{kuhl:2015}). Here, as recalled in the introduction, in a first hypothesis we neglect re-crosslinking effects. Moreover, according to the mechanical quantities evaluated for the discrete system under investigation the exact analytical values of the force and displacement require the solution of a transcendental equation that can be easily obtained numerically at assigned parameters. On the other hand, explicit solutions will be obtained in the continuum approximation (\textit{i.e.} $n\to \infty$, see section~\ref{sec:ch5_cl}).

%
\begin{figure}[t!]
\centering
  \includegraphics[width=0.95\textwidth]{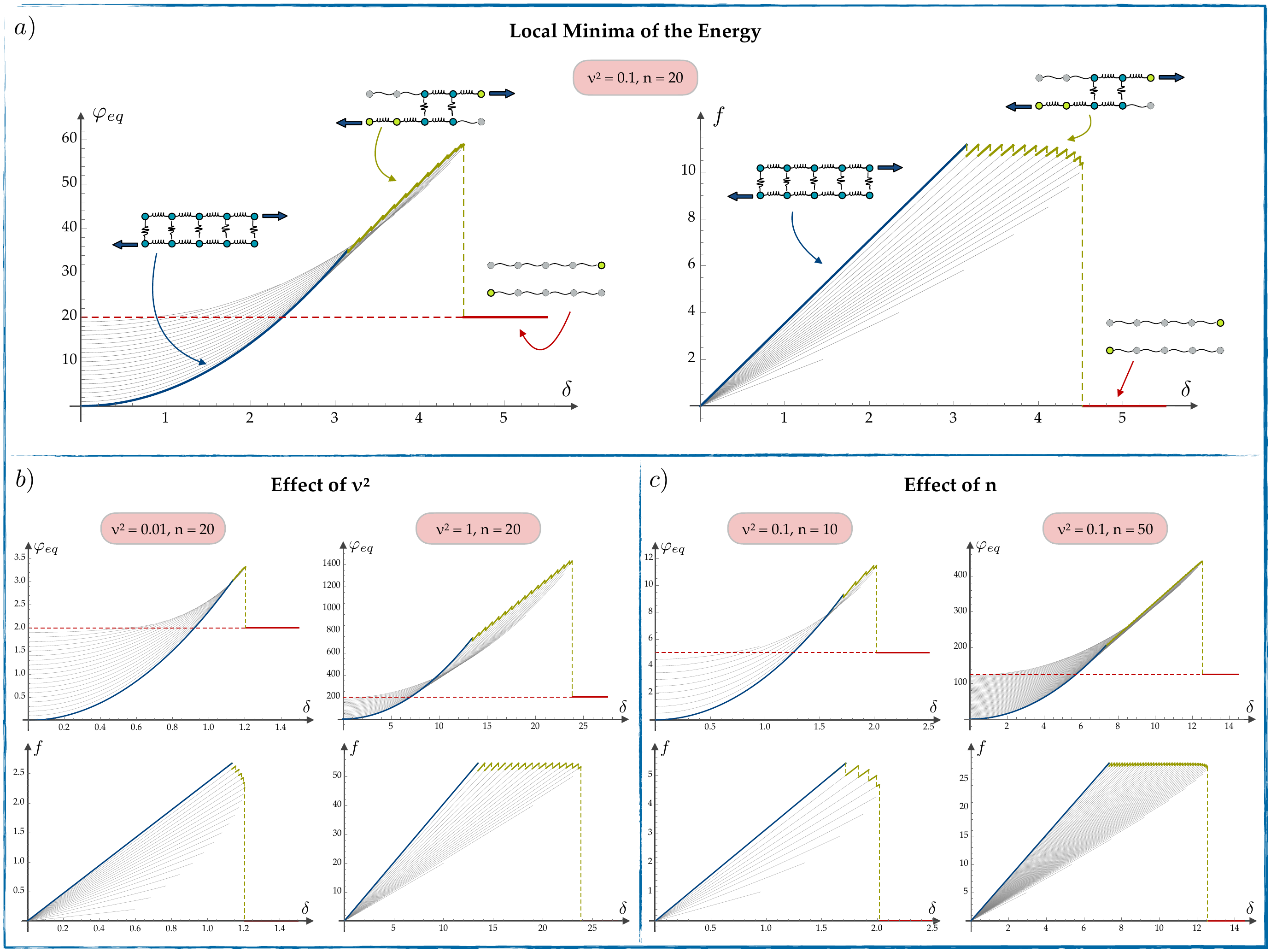}
  \caption[Maximum delay convention]{Maximum delay convention. Panel a): The behaviour of the system when it can explore the equilibrium branches until they exist is characterized by a locally stable phase in which the $\tau$-proteins break-in sequence until the maximum value is reached. In this case, a ductile fracture is observed, depending on the different parameters of the system. Panel b): Effect of $\nu^2$. As this parameter increases the decohesion becomes more and more ductile. Panel c): Effect of the discreteness. As $n$ grows the continuum limit is approached (see the following section) and the system becomes more ductile.}
  \label{fig:ch5_local}
\end{figure}

Consider now the opposite hypothesis, known as \textit{maximum delay convention}, where the system stays in the $p$-th equilibrium branch until it is locally stable. In this case, the system follows a given branch $p$ until the condition \eqref{eq:ch5_dmax} holds and one of the outermost $\tau$ proteins of the bidden block breaks. In this case, we observe an initial process of decohesion with a sequential phenomenon of a single unbinding effect, as demonstrated in~\eqref{eq:ch5_mono}. Following, when a critical displacement threshold is attained, a sudden decohesion of the remaining bonded domain is observed, with the force jumping to zero. Globally, in this case, a `ductile' type rupture characterized by a sawtooth decohesion plateau ending with a sudden jump to the fully detached state, as shown in Figure~\ref{fig:ch5_local}$_a$. Also in this case the evaluation of the dimension of the plateau and the values of the displacements at the beginning and the end of this metastable phase as well as the propagation and rupture stress requires the solution of transcendental equations and analytic formulas will be obtained in the continuum approximation. We may point out that, as shown in Figure~\ref{fig:ch5_local}$_b$ and~\ref{fig:ch5_local}$_c$, the values of the stiffness parameter $\nu^2$ and the discreetness of the system plays a crucial role in the evolution of the system affecting the above recalled mechanical quantities. In particular, it is important to observe that as $\nu^2$ increases --{\it i.e.} when the stiffness of the $\tau$ protein grows as compared with the microtubule one-- the system exhibits a more and more ductile behaviour (\cite{ahmadzadeh:2015}) (see Figure~\ref{fig:ch5_local}$_b$). On the other hand, the inspection of Figure~\ref{fig:ch5_local}$_c$ shows that the ductility grows as we increase the number of $\tau$ proteins.

\subsection{Continuum approximation}
\label{sec:ch5_cl}

To obtain analytical results with a more clear physical description of the axon damage, here we consider the case of a large number of $\tau$ proteins. Specifically, following  (\cite{puglisi:2000}), we consider the \textit{continuum approximation} obtained by fixing the total length of the chain $L=ln=const$ with $n\to\infty$ and $l\to 0$. We define the fraction of unbidden elements, described in this limit as a continuum variable and assuming values in the interval $(0,1)$, as
\begin{equation}
\pi:=\frac{p}{n}.
\end{equation}
Coherently, we consider the rescaled parameter
\begin{equation}
\mu^2:=\frac{\nu^2 n^2}{2}
\label{eq:ch5_mumu}
\end{equation}
so that the correct scaling of the energy in~\eqref{eq:ch5_en} is obtained for growing $n$ at fixed $\mu$.

In the continuum limit, where here and in the following we indicate the relevant quantities by an overbar, we find that the total rescaled stiffness $\kappa^{t}(n,p)$ of the chain in~\eqref{eq:ch5_ktot} reads 
\begin{equation}
\bar \kappa^{t}(\pi)=\lim_{n\,\to+\infty}\kappa^{t}(n,p)=\frac{4\mu}{\mu(2-\pi)+\coth\left(\mu\pi\right)}.
\end{equation}
Accordingly, the force-displacement relation is given by 
\begin{equation}
\bar f(\pi,\delta)=\bar \kappa^{t}(\pi)\delta
\label{eq:ch5_flc}
\end{equation}
and the energy reads
\begin{equation}
\bar \varphi(\pi,\delta)=\bar \kappa^t(\pi)\delta^2+\mu^2\left(1-\pi\right).
\label{eq:ch5_enlc}
\end{equation}
Similarly, by using~\eqref{eq:ch5_dmax}, we obtain the expression in the continuum approximation of the maximum displacement at given $\pi$, \textit{i.e.} 
\begin{equation}
\bar \delta_{max}(\pi)=\lim_{n\,\to+\infty}\delta_{max}(n,p)=\frac{1}{2} \Bigl[1+\mu (2-\pi) \tanh \left(\mu\pi\right)\Bigr].
\label{eq:ch5_deltamlc}
\end{equation}
%

%
\begin{figure}[t!]
\centering
  \includegraphics[width=0.95\textwidth]{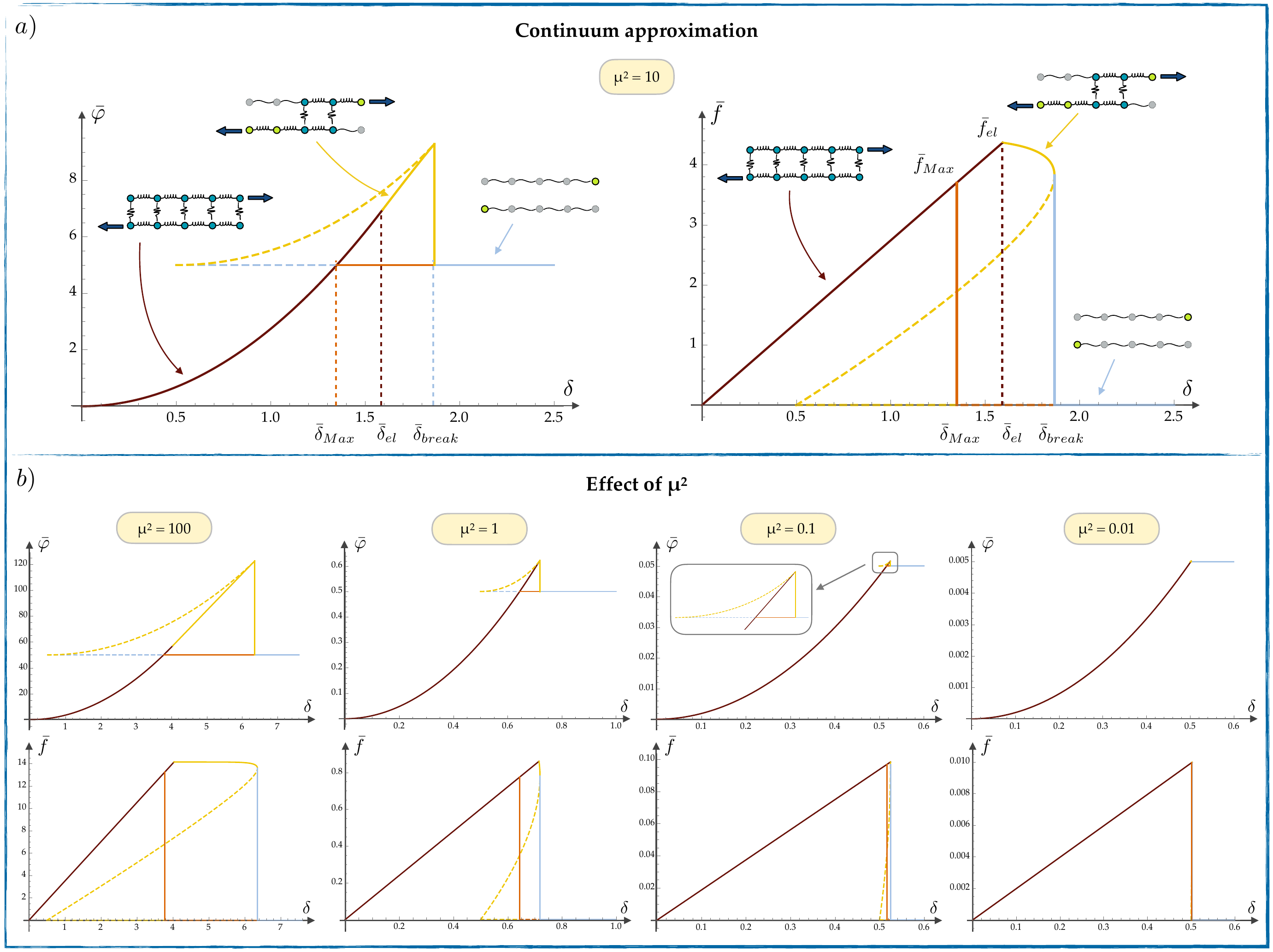}
  \caption[Continuum approximation]{Continuum approximation. Panel a): Energy and force with respect to the displacement in the continuum limit. The rupture in the Maxwell convention is represented by the orange line, whereas the fracture under the Maximum delay convention is represented by the cyan line. In the first case, we have always a fragile behaviour, while in the second one the decohesion is gradual along the path represented by the yellow curves, and it exhibits a ductile rupture. Panel b): The ductility increases as $\mu$ grows.}
  \label{fig:ch5_lc}
\end{figure}

The mechanical response of the system in the continuum approximation can now be explored in the two hypotheses of the Maxwell and maximum delay convention recalled above. \vspace{0.3 cm}

\noindent \textbf{\textsc{Maxwell convention --}} Consider first the case when the configurations correspond always to the global minima of the energy. Previously we demonstrated that, for a discrete system, the transition is always from the fully attached state to the fully detached one~\eqref{eq:ch5_dim1}. As a matter of fact, also in this approximation, we may determine the intersection between the two energy branches of the energy in~\eqref{eq:ch5_enlc} corresponding to the cases of $\pi=1$ and $\pi=0$ and prove that it is always the first intersection attained, confirming the `fragile' behaviour of the system under the Maxwell convention. With this in mind, we distinguish the energy and the force for the two extreme cases of fully attached system (apex \textit{fa}) and fully detached (apex \textit{fd}) configuration.  In the former case, we impose the condition $\pi=1$ in~\eqref{eq:ch5_enlc} and~\eqref{eq:ch5_flc}, obtaining
\begin{equation}
\bar \varphi_{fa}(\delta)=\frac{4\mu}{\mu+\coth\left(\mu\right)}\delta ^2 , \quad\quad \bar f_{fa}(\delta)=\frac{4\mu}{\mu+\coth\left(\mu\right)}\delta.
\label{eq:ch5_fa}
\end{equation}
With the same reasoning, when $\pi=0$, we obtain the energy and force:
\begin{equation}
\bar \varphi_{fd}=\mu^2, \quad\quad \bar f_{fd}=0.
\end{equation}
Afterwards, we compute the value of the fracture threshold $\delta_{Max}$ in the Maxwell hypothesis, that is obtained by imposing $\bar \varphi_{fa}(\delta_{Max})=\bar \varphi_{fd}$:
\begin{equation}
\bar \delta_{Max}=\frac{1}{2}\sqrt{\mu^2+\mu\coth \left(\mu\right)}.
\label{eq:ch5_deltaglobal}
\end{equation}
This displacement corresponds to the fracture (Maxwell) force
\begin{equation}
\bar f_{Max}=\frac{2\mu ^2}{\sqrt{\mu^2+\mu\coth \left(\mu\right)}}.
\end{equation}
In Figure~\ref{fig:ch5_lc}$_a$ we observe that the behaviour under the Maxwell hypothesis is always fragile, and the breaking path is represented by the orange line in the curves of the energy and of the force with respect to the displacement.

 \vspace{0.3 cm}

\noindent \textbf{\textsc{Maximum delay convention --}} Under this hypothesis the system stays in a local minimum until that configuration is locally stable. In particular, the fully attached solutions holds until the first $\tau$ protein detaches at 
\begin{equation}
\bar \delta_{el}:=\bar \delta_{max}(1)=\frac{1}{2} \Bigl[1+\mu\tanh \left(\mu\right)\Bigr],
\label{eq:ch5_del}
\end{equation}
where the subscript $el$ indicates the `elastic' --reversible-- path attained until this value of the displacement. The corresponding force is then 
\begin{equation}
\bar f_{el}:=\bar f_{fa}(\bar \delta_{el})=2\mu\tanh\left(\mu\right).
\label{eq:ch5_fel}
\end{equation}
%

%
\begin{figure}[t!]
\centering
  \includegraphics[width=0.95\textwidth]{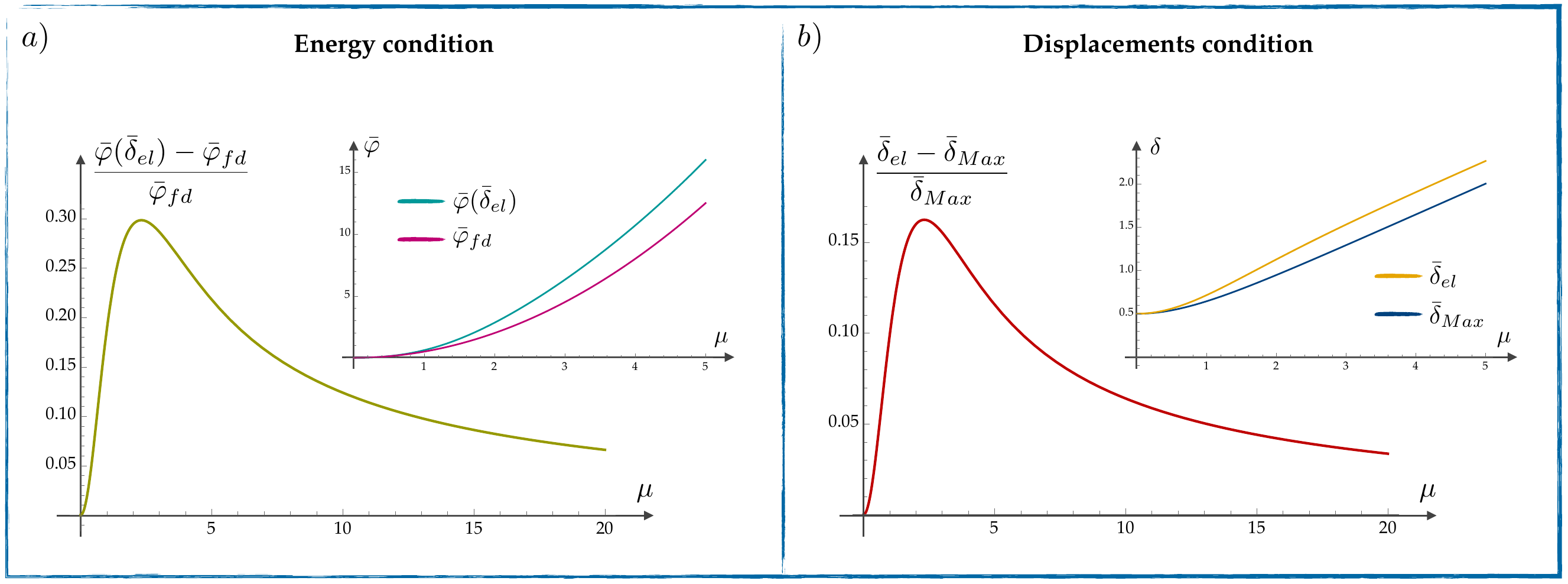}
  \caption[Condition on the global stability]{Global minima solution in the continuum limit. Panel a): The energy associated with the $\delta_{el}$ is always greater than the energy associated at the intersection displacement (fracture energy). Panel b): The breaking displacement of the fully attached branch is always greater than its intersection displacement with the fracture energy.}
  \label{fig:ch5_deltaenergy}
\end{figure}

In passing, as further proof that in the continuum limit the behaviour of the system in the Maxwell convention is always fragile, we observe that the conditions
\begin{equation}
\bar \varphi_c(\bar \delta_{el})>\bar \varphi_{fd} 
\label{eq:ch5_energycondition}
\end{equation}
and 
\begin{equation}
\bar \delta_{el}>\bar \delta_{Max},
\label{eq:ch5_dcond}
\end{equation}
are fulfilled. Indeed we have
\begin{equation}
\bar \varphi(\bar \delta_{el})-\bar \varphi_{fd}=\mu^2\tanh^2\left(\mu\right)+\mu\tanh\left(\mu\right)-\mu^2>0,\qquad\text{for}\quad \mu>0
\end{equation}
and
\begin{equation}
\bar \delta_{el}-\bar \delta_{Max}=
\frac{1}{2} \Bigl[1+\mu\tanh \left(\mu\right)\Bigr]-\frac{1}{2}\sqrt{\mu^2+\mu\coth \left(\mu\right)}>0.
\end{equation}
In Figure~\ref{fig:ch5_deltaenergy}$_a$ we show the inequality of the energy at varying $\mu$ and we observe that the condition is fulfilled and the two energy are equal only when $\mu=0$. Similarly, in Figure~\ref{fig:ch5_deltaenergy}$_b$ the condition concerning the displacements is represented. Also in this case the two values are the same when the stiffness parameter is zero and they are equal to $1/2$. These two limit conditions in $\mu$ will be discussed in detail in the following.

%
\begin{figure}[t!]
\centering
  \includegraphics[width=0.95\textwidth]{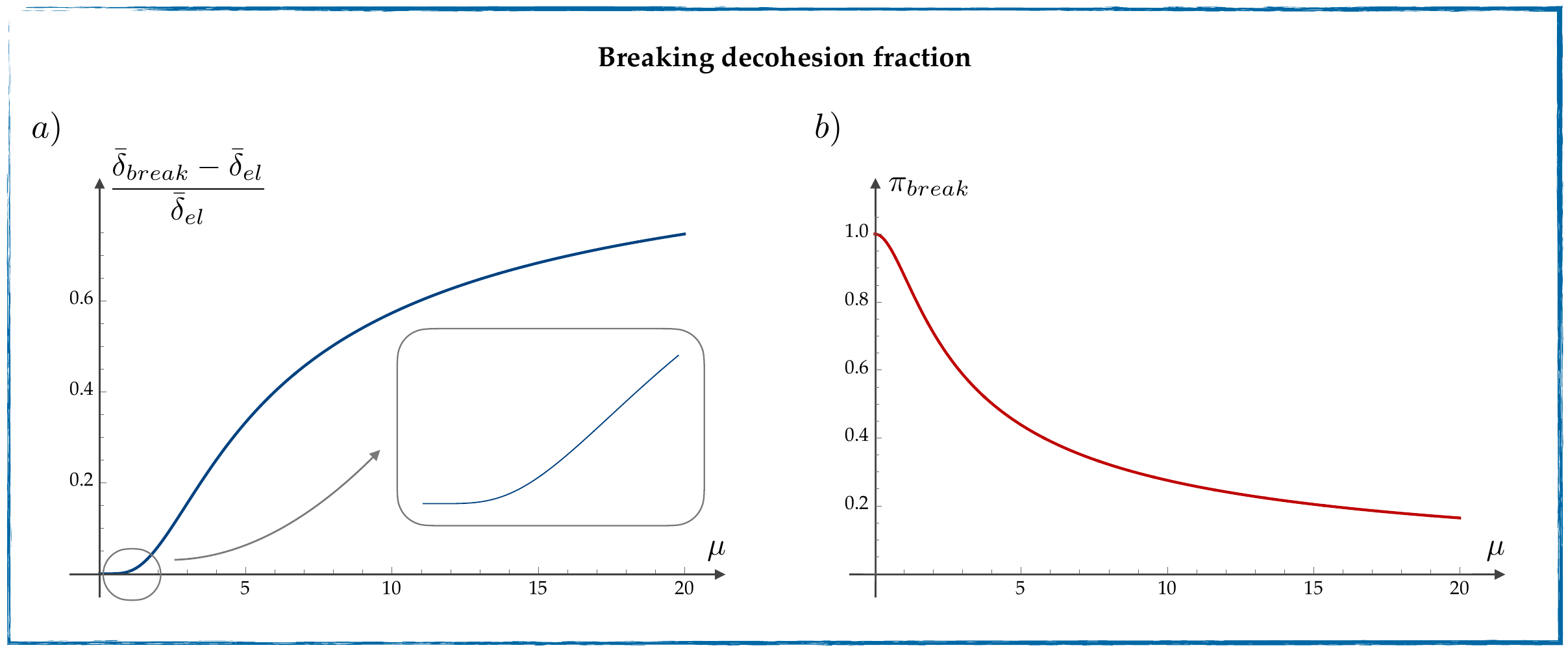}
  \caption[Behavior of the breaking threshold at varying stiffness]{Breaking decohesion fraction with respect to $\mu$. Panel a): The length of the decohesion plateau is represented with respect to varying values of $\mu$. Panel b): The numerical solution of~\eqref{eq:ch5_pib} is shown for different values of $\mu$.}
  \label{fig:ch5_deltapi}
\end{figure}

As we have observed for the discrete case, in the maximum delay convention the system is allowed to follow an equilibrium branch until it is locally stable. When we perform the continuum approximation, the displacements representing the stability condition for each branch of the energy (or force) in~\eqref{eq:ch5_dmax} shrinks to the continuous values given by~\eqref{eq:ch5_deltamlc} in the interval $[0,1]$. Indeed, once decohesion begins, this curve represents the unbinding path of our system by assigning at each displacement the corresponding phase fraction $\pi$. By substituting \eqref{eq:ch5_deltamlc} in~\eqref{eq:ch5_enlc} and~\eqref{eq:ch5_flc} we obtain the corresponding values of energy, indicated with the pedex \textit{md} (maximum delay):
\begin{equation}
\bar \varphi_{md}(\pi)=\frac{\mu}{2}\sech^2(\mu\pi)\Bigl[\sinh (2\mu\pi )+\mu(3-2 \pi )\cosh (2 \mu\pi)-\mu\Bigr],
\label{eq:ch5_encm}
\end{equation}
and force 
\begin{equation}
\bar f_{md}(\pi)=2\mu\tanh\left(\mu\pi\right).
\label{eq:ch5_fcm}
\end{equation}

The decohesion curve is followed until the breaking threshold of the maximum delay convention is reached and one can show that this condition is attained when the maximum of the curve in~\eqref{eq:ch5_deltamlc} is reached. This maximum value corresponds, at given $\mu$, at the fraction at which the system breaks, $\pi_{break}$, obtained as solution of
\begin{equation}
\frac{\partial \bar \delta_{max}(\pi)}{\partial \pi}:=\frac{1}{2}\Bigl[\mu ^2 (2-\pi) \sech^2\left(\mu \pi\right)-\mu\tanh\left(\mu\pi\right)\Bigr]=0.
\label{eq:ch5_pib}
\end{equation}
Unfortunately,~\eqref{eq:ch5_pib} is a transcendental equation and can be solved only numerically. From the analysis of this derivative, we find that the value of $\pi_{break}$ decreases as $\mu$ increases, as shown in Figure~\ref{fig:ch5_deltapi}$_b$. It is easy to show that the solution is unique and we indicate it by introducing $\bar\delta_{break}:=\bar\delta_{max}(\pi_{break})$. Indeed, an always greater portion of our continuous system breaks before reaching the threshold corresponding to the collapse of the system. Moreover, from a stress-strain point of view, we observe that the force path between the initial rupture of the system and the breaking value of displacement attained when it collapses increases as $\mu$ grows and can be observed in~\ref{fig:ch5_deltapi}$_a$, exhibiting a more and more ductile behaviour. In other words, when the system is allowed to follow the maximum delay convention the ductility of our `material' increases as $\mu$ grows, whereas in the Maxwell convention only a sudden, fragile breaking is observed. In Figure~\ref{fig:ch5_lc}$_a$ we picture both the two recalled behaviour. 

%
\begin{figure}[t!]
\centering
  \includegraphics[width=0.95\textwidth]{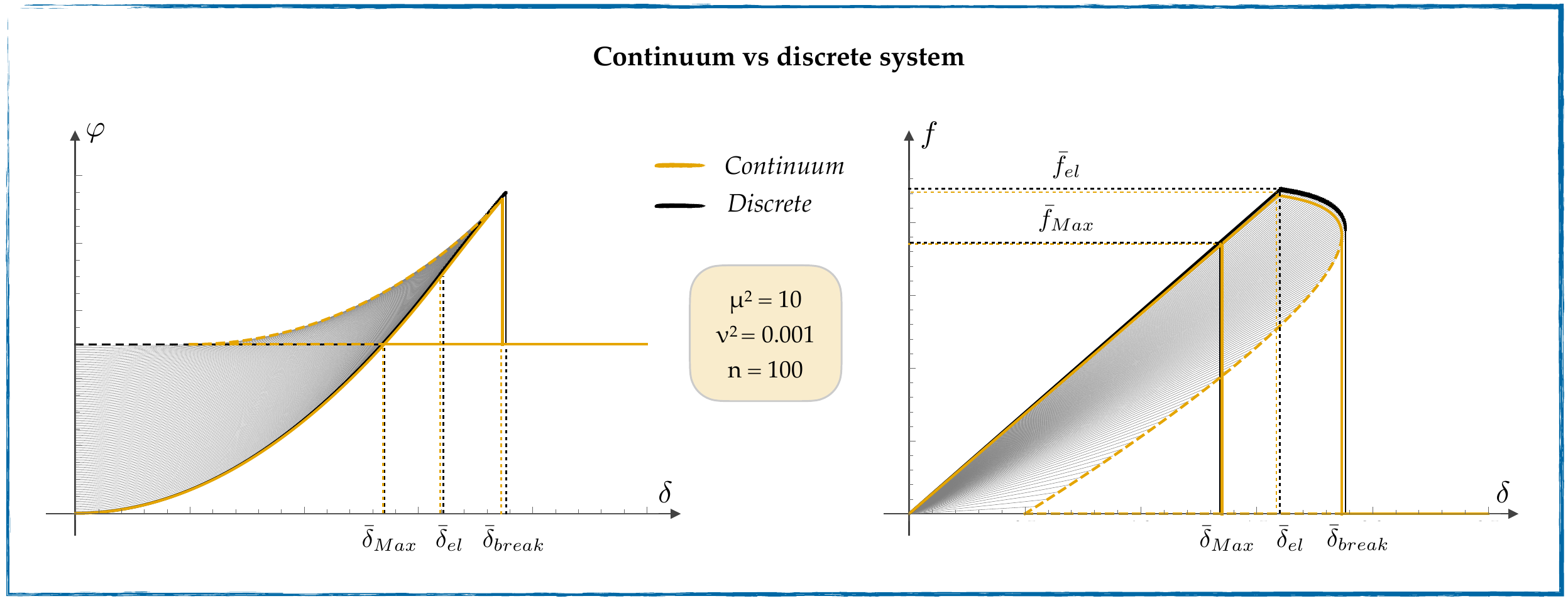}
  \caption[Comparison between continuum approximation and discrete system]{Comparison between a continuous system and the corresponding discrete one. For a system with $n=100$ objects, \textit{i.e.} `close' to infinity, we use $\mu^2=10$, $\nu^2=0.001$ to ensure the correct scaling of the force and energy. We observe that the match is almost perfect.}
  \label{fig:ch5_cvsd}
\end{figure}

In particular, in the maximum delay hypothesis, we may observe that when the initial rupture starts (endpoint of the brown curve) there is the yellow decohesion path that can be observed both from an energetic point of view or from a force-displacement mechanical response. When the threshold value of $\bar\delta_{max}(\pi_{break})$ is reached, the metastable path (dotted yellow line) is no longer allowed and the system collapses following the cyan curve that represents the breaking in the maximum delay convention. Moreover, in Figure~\ref{fig:ch5_lc}$_b$ we observe that the ductility is affected merely from the values of $\mu$. Indeed, as $\mu$ increases the metastable path is longer and the breaking thresholds in the two conventions are far from each other. Conversely, as $\mu$ decreases, the system becomes more and more ductile until the two breaking displacements are almost the same. 

For the sake of completeness, in Figure~\ref{fig:ch5_cvsd} we compare the continuum case to the discrete case with a high number of objects $n$, \textit{i.e.} when it is reasonable to assume that we are close to the limit $n\to\infty$. To perform the comparison the parameter of the discrete system $\nu^2$ is scaled by means of~\eqref{eq:ch5_mumu} to obtain the correct continuous $\mu^2$. The comparison shows very good agreement between the two curves.

\subsection{Small and large $\mu$ limits.}
\label{sec:ch5_mulimit}

\begin{figure}[t!]
\centering
  \includegraphics[width=0.95\textwidth]{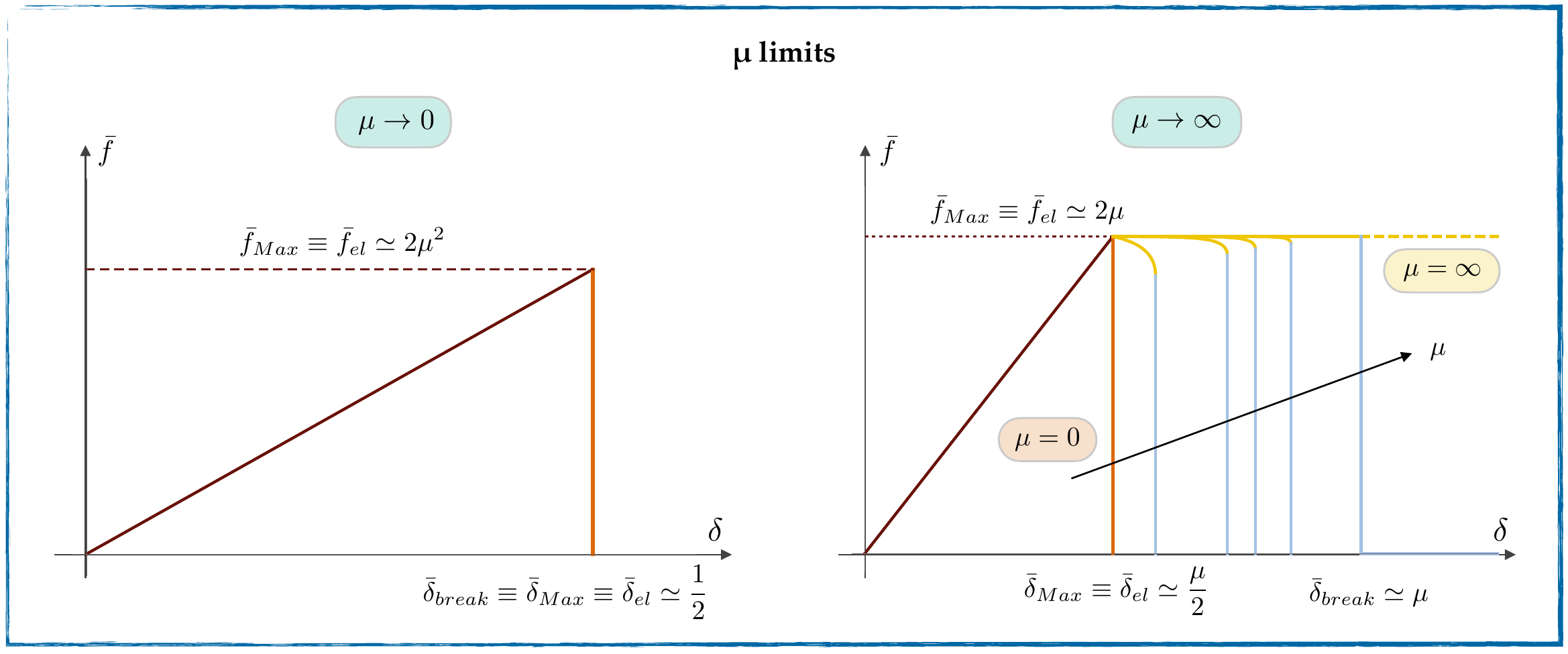}
  \caption[$\mu$ limits]{Limits of the stiffness parameter $\mu$. For small values, a fragile behavior is always attained. When $\mu$ grows a ductile type of fracture is shown and the decohesion process is completed in the interval of $\pi\in(0,1)$.}
  \label{fig:ch5_mulimit}
\end{figure}

Here we consider two limit regimes of the stiffness parameter $\mu$, representing the behaviour of the system when the $\tau$-proteins are extremely soft with respect to the microtubules ($\mu\to 0$) or when the links are much stiff than the elastic part ($\mu\to\infty$). To not overload the notation, we will use the same symbols of the continuum approximation even though the quantities in the two limits are different. Let us consider the first case.  When $\mu$ goes to zero, we distinguish the two recalled decohesion strategies analyzed in the previous subsections. In the Maxwell hypothesis, we have that the system follows the fully attached solution
\begin{equation}
\bar \varphi_{fa}(\delta)\simeq 4\mu^2 \delta^2+\mathcal{O}(\mu^{4}) , \quad\quad \bar f_{fa}(\delta)\simeq 4\mu^2 \delta+\mathcal{O}(\mu^{4}).
\end{equation}
The Maxwell displacement where the system breaks when the minimum energy path is followed reads
\begin{equation}
\bar \delta_{Max}\simeq\frac{1}{2}+\frac{\mu^2}{3}+\mathcal{O}(\mu^{4}),
\label{eq:ch5_deltamaxwell}
\end{equation}
with a corresponding force of 
\begin{equation}
\bar f_{Max}\simeq 2\mu^2+\mathcal{O}(\mu^{4}).
\label{eq:ch5_fmaxwell}
\end{equation}
With the same reasoning, we may compute the values of the displacement and the force in the maximum delay hypothesis in the limit of $\mu$ small and we get
\begin{equation}
\bar \delta_{el}\simeq\frac{1}{2}+\frac{\mu^2}{2}+\mathcal{O}(\mu^{4}),
\label{eq:ch5_deltaelmupiccolo}
\end{equation}
and 
\begin{equation}
\bar f_{el}\simeq2\mu^2+\mathcal{O}(\mu^{4}).
\label{eq:ch5_felmupiccolo}
\end{equation}
Furthermore, the maximum displacement at which the whole chain breaks, as we have observed, can be found looking for the numeric solution of equation~\eqref{eq:ch5_deltamlc}. Thus, to obtain $\bar\delta_{break}$ in this limit, we perform the limit~\eqref{eq:ch5_deltamlc} and we obtain the corresponding displacement, that for this case is  
\begin{equation}
\bar \delta_{break}\simeq\frac{1}{2}+\mathcal{O}(\mu^{4}).
\label{eq:ch5_dbreakmupiccolo}
\end{equation}
In Figure~\ref{fig:ch5_mulimit} we may observe the behaviour of our system in the limit of $\mu\to 0$. We notice that the type of fracture is always fragile attained at the value of $1/2$, where the displacements~\eqref{eq:ch5_deltamaxwell},~\eqref{eq:ch5_deltaelmupiccolo} and~\eqref{eq:ch5_dbreakmupiccolo} coincides. Moreover, also the forces in~\eqref{eq:ch5_fmaxwell} and~\eqref{eq:ch5_felmupiccolo} are the same. This fundamental result proves that when the $\tau$ proteins are soft with respect to the microtubules the breaking threshold is the one corresponding at the Maxwell convention, resulting in a fragile behaviour of the mechanical system under loads.

%

On the other hand, when $\mu$ grows, we have an initial elastic behaviour with associated energy and force 
\begin{equation}
\bar \varphi_{fa}(\delta)\simeq 4\delta ^2+\mathcal{O}(\mu^{4})  , \quad\quad \bar f_{fa}(\delta)\simeq 4\delta+\mathcal{O}(\mu^{4}) .
\end{equation}
When the global minima of the system are attained we have that the Maxwell displacement reads 
\begin{equation}
\bar \delta_{Max}\simeq\frac{\mu}{2}+\mathcal{O}(\mu^{4}),
\label{eq:ch5_deltmaxmdmugrande}
\end{equation}
with a corresponding force threshold 
\begin{equation}
\bar f_{Max}=\simeq 2\mu+\mathcal{O}(\mu^{4}).
\label{eq:ch5_fMaxmugrande}
\end{equation}
Conversely, when the reversible path is followed until the system starts to break, the elastic associated displacement reads
%
\begin{equation}
\bar \delta_{el} \simeq \frac{\mu}{2}+\mathcal{O}(\mu^{4}),
\label{eq:ch5_deltaelastic}
\end{equation}
with an elastic force of 
\begin{equation}
\bar f_{el}\simeq 2\mu+\mathcal{O}(\mu^{4}).
\label{eq:ch5_felasticmugrande}
\end{equation}
Also in this case we may ask which is the value of the plateau in this limit, obtained from~\eqref{eq:ch5_deltamlc}. As remarked before, we perform the limit of~\eqref{eq:ch5_deltamlc} and then we found the solution. Indeed, it exists only for $\mu$ large, as demonstrated in the previous sections. We have
\begin{equation}
\bar \delta_{max}(\pi)\simeq\frac{\mu}{2} (2-\pi)+\mathcal{O}(\mu^{4}).
\end{equation}
In this limit, one observes that the decohesion fraction at which the system collapses is $\pi=0$, \textit{i.e.} all the links (material in the continuous approximation) are allowed to break sequentially. The corresponding displacement is
\begin{equation}
\bar \delta_{break}=\mu.
\label{eq:ch5_deltabreakmugrande}
\end{equation}
As demonstrated, in the limit of large $\mu$ the mechanical response of the system under shear loads is comparable with the one of a ductile material, and the ductility depends on the value of $\mu$, as shown in Figure~\ref{fig:ch5_mulimit}. In particular, in this case the plateau of the force is constant when $\mu\to\infty$, and it is attained at the displacement at which the decohesion process starts,~\eqref{eq:ch5_deltaelastic} that corresponds to the Maxwell displacement~\eqref{eq:ch5_deltmaxmdmugrande}, where also~\eqref{eq:ch5_felasticmugrande} and~\eqref{eq:ch5_fMaxmugrande} coincides. Even though the plateau is constant and goes to infinite as the displacement grows, the system breaks when the $\bar\delta_{break}$ in~\eqref{eq:ch5_deltabreakmugrande} is reached, that in this limit corresponds to a phase fraction $\pi_{break}=0$. 

Eventually, we can rewrite the analytical expressions of the decohesion force and displacement in the two studied limits expressed in terms of real physical parameters. In particular, for the case of $\mu$ small we have that the `denaturation' (Maxwell) force is equal to the elastic force and reads 
\begin{equation}
\bar F_{Max}\equiv\bar F_{el}\simeq \frac{\kappa_{\tau}L}{u_d},
\label{eq:ch5_forcesmallmu}
\end{equation}
while the displacements~\eqref{eq:ch5_deltamaxwell},~\eqref{eq:ch5_deltaelmupiccolo} and~\eqref{eq:ch5_dbreakmupiccolo} correspond to the dimensional value 
\begin{equation}
\bar d_{Max}\equiv\bar d_{el}\equiv \bar d_{break}\simeq \frac{u_d}{2}.
\end{equation}

With the same reasoning, for the opposite limit attained when $\mu$ is large, we have the forces 
\begin{equation}
\bar F_{Max}\equiv\bar F_{el}\simeq \sqrt{2\kappa_{\tau}\kappa_e},
\label{eq:ch5_forcelargemu}
\end{equation}
and the corresponding displacements
\begin{equation}
\bar d_{Max}\equiv\bar d_{el}\simeq \frac{L}{2u_d} \sqrt{\frac{\kappa_{\tau}}{2\kappa_e}}.
\end{equation}

In this case, the breaking threshold is given by 
\begin{equation}
\bar d_{break}\simeq \frac{L}{u_d}\sqrt{\frac{\kappa_{\tau}}{2\kappa_e}}.
\end{equation}
%

\section{Experimental Comparison}

Although in the recent decades SMFS experiments opened up the possibility of testing at the microscopic scale even single molecules, the task of obtaining a clear answer concerning the mechanical or biological features of such systems is not so obvious. Indeed, these kinds of tests, especially on living biomolecules, are performed in vitro, thus they are always diluted in a certain solution and some techniques have been developed to isolate, for instance, the DNA chain. Another drawback is that is usually necessary to apply ligands or other molecules to the probe to correctly assign boundary conditions such as applied force or displacement when experimenting. All these effects combined may be responsible for a misinterpretation of the experimental results (\cite{lavery:2002}). Concerning double-stranded DNA experiments, there are two main possibilities: the force or the displacement can be applied in the transversal direction of the helix, transmitting normal stresses at the base pairs (\cite{bustamante:2000:2,rief:1999:2}), or longitudinal to the DNA axis, thus applying shear forces.

While we were developing and studying the model introduced in this Chapter, we found that a similar model was introduced by Pierre Gilles De Gennes, in his work of $2001$ (\cite{degennes:2001}). Here, starting from an Hamiltonian energy formally equal to the one in~\eqref{eq:ch5_en}, he described a double-stranded hybridized portion of DNA with a simple ladder model and, by directly using a continuum approach, he obtained the dependence of the rupture force $F_c$ of the chain with respect to the length, in terms of base pairs (called $\ell$ in the paper), of the DNA sequence. The fundamental result is that the critical force, representing the force at which the DNA chain starts to unbidden, is linear when the length is short, whereas it saturates to a plateau when the number of base pairs increases. As one may see, this result is crucial to comprehending the role of the forces in biological processes such as DNA transcription, denaturation and replication. While he focused his work only on this aspect, by directly computing the continuum limit, we highlight the fundamental discrete behaviour of such a system, analyzing different evolution strategies and the resulting fragile-ductile rupture transition at varying consitutitent parameters such as the relative stiffness and the size of the DNA. Moreover, we study the decohesion process in the continuum limit, analyzing the force plateau, the rupture displacements and the limit regimes attained when the relative stiffness goes to $0$ or to $\infty$. To compare the two results and evaluate the resulting quantities in terms of the mechanical properties of the system, we report the formula obtained by De Gennes: 
\begin{equation}
F_c=2f_c\chi^{-1}\tanh\left(\frac{\chi\ell}{2}\right),
\label{eq:ch5_degennesform}
\end{equation}
where using the notation of (\cite{degennes:2001}), $f_c$ is the rupture force of a single binding and $\chi=\sqrt{2R/Q}$ is the ratio of the stiffness of the polyphosphate backbone ($Q$) and the base pairs ($R$). 

After some years, in 2008, Hatch and coworkers performed a pulling experiment with magnetic tweezers on ds-DNA sequences pulled by either $3'3'$ and $5'5'$ ends (\cite{hatch:2008}). In particular, they applied constant force on probes of lengths ranging from $12$ to $50$ base pairs and used biological linkers to attach the molecule (see (\cite{hatch:2008}) for the detailed information concerning the experimental set-up). The experiment is performed at zero loading rate and at least $50$ molecules are tested. Also, they used the formula in~\eqref{eq:ch5_degennesform} to fit the data considering also the force of the last elastic units before the helix starts to open and a total length depending on the total base pairs with respect to the open ones. This last assumption is due to the fact that short chains are unstable at room temperature and thus some of them may be in an open state at the moment of the experiment. In particular, it is necessary that for rich sequences almost $8-12$ base pairs are necessary for the molecule to be stable at room temperature. The formulas of (\cite{hatch:2008}) reads
\begin{equation}
F_c=2f_c\left[\chi^{-1}\tanh\left(\frac{\chi L_{bound}}{2}\right)+1\right],
\label{eq:ch5_hatchform}
\end{equation}
where $L_{bound}=L_{sequence}-L_{open}$ is the length, in terms of base pairs of the attached chain, $L_{sequence}=\ell=n$ is the total length and $L_{open}$ is the number of base pairs that are already detached when the experiment is performed.

Following, Prakash and colleagues (\cite{prakash:2011}), in the wake of the work of Chakrabarti \textit{et al.} (\cite{chakrabarti:2009}) obtained numerically the expectation values of the force and displacement in a Statical Mechanics framework by assuming a hard-core repulsion potential for the base pairs and they show that at room temperature the agreement with the experiment of (\cite{hatch:2008}) is quite good despite the comparison with the quantities $\chi$ and $f_c$ of~\eqref{eq:ch5_degennesform} depends strictly on the potential used instead of the mechanical properties of the system. It is important to highlight that both (\cite{hatch:2008}) and (\cite{prakash:2011}) only fitted the experimental results without any reference to the physical meaning of the quantities in the formula such as $\chi$ and $f_c$. 

To do this, and to compare our model, consider the continuum approximation, where the `critical' force is the force at which the system starts to break defined in~\eqref{eq:ch5_fel}. In this section, to compare the results I denote with a bar the quantities referred to our model, and to use the same notation of De Gennes I have $\bar F_c:=\bar f_{el}$. To catch the physical meaning of this expression let us now substitute the true mechanical quantities assigned in~\eqref{eq:ch5_forceoriginal},~\eqref{eq:ch5_mumu} and~\eqref{eq:ch5_nu} in~\eqref{eq:ch5_fel}, obtaining 
\begin{equation}
\bar F_c=\sqrt{2 \kappa_{\tau}\kappa_e}\tanh\left(\sqrt{\frac{\kappa_{\tau}}{2 \kappa_e}}\frac{l}{u_d}n\right),
\label{eq:ch5_fcnostra}
\end{equation}
that represent the critical force with respect to the increasing length of the DNA. By comparing this result with~\eqref{eq:ch5_degennesform} one gets the internal length
\begin{equation}
\bar\chi=\sqrt{\frac{2\kappa_{\tau}}{\kappa_e}}\frac{l}{u_d},
\label{eq:ch5_chinostra}
\end{equation}
and the force threshold of a single base pair 
\begin{equation}
\bar f_c= \frac{\kappa_{\tau}l}{u_d},
\label{eq:ch5_fcnostra}
\end{equation}
that, as one may observe from~\eqref{eq:ch5_tauenergy}, it is exactly the derivative of the energy of a single base pair with respect to the displacement, thus the force required to break it. It is worth remarking that for low $n$ the trend of $\bar F_c$ is linear while at increasing length there is a force plateau. De Gennes studied these two limit regimes, thus when $\chi\ell<1$, \textit{i.e} short strands are considered, the force is linear
\begin{equation}
F_c=f_c \ell,
\label{eq:ch5_flineardeg}
\end{equation}
and by using~\eqref{eq:ch5_chinostra} and~\eqref{eq:ch5_fcnostra} we get that the~\eqref{eq:ch5_flineardeg}, in our model called $\bar F_c^{el}$, reads
\begin{equation}
\bar F_c^{el}= \frac{\kappa_{\tau} l}{u_d}n,
\label{eq:ch5_flinearnostra}
\end{equation}
and it exactly corresponds to the force obtained in the limit of small $\mu$ (that depends on $n$ in the continuum limit) in~\eqref{eq:ch5_forcesmallmu} representing the breaking threshold both in the Maxwell and maximum delay convention. With the same reasoning, we compare the value of the force plateau obtained in (\cite{degennes:2001}) for infinitely long strands
\begin{equation}
F_c \rightarrow F_m=2 f_c \chi^{-1}
\label{eq:ch5_fmdeg}
\end{equation}
with our results. Substituting~\eqref{eq:ch5_chinostra} and~\eqref{eq:ch5_fcnostra} in~\eqref{eq:ch5_fmdeg} we get  
\begin{equation}
\bar F_m = \sqrt{2\kappa_{\tau}\kappa_e},
\label{eq:ch5_fmnostra}
\end{equation}
that again corresponds to the force threshold in the two decohesion strategies analyzed in the previous section when $\mu$ is large~\eqref{eq:ch5_forcelargemu}. 

%
\begin{figure}[t!]
\centering
  \includegraphics[width=0.95\textwidth]{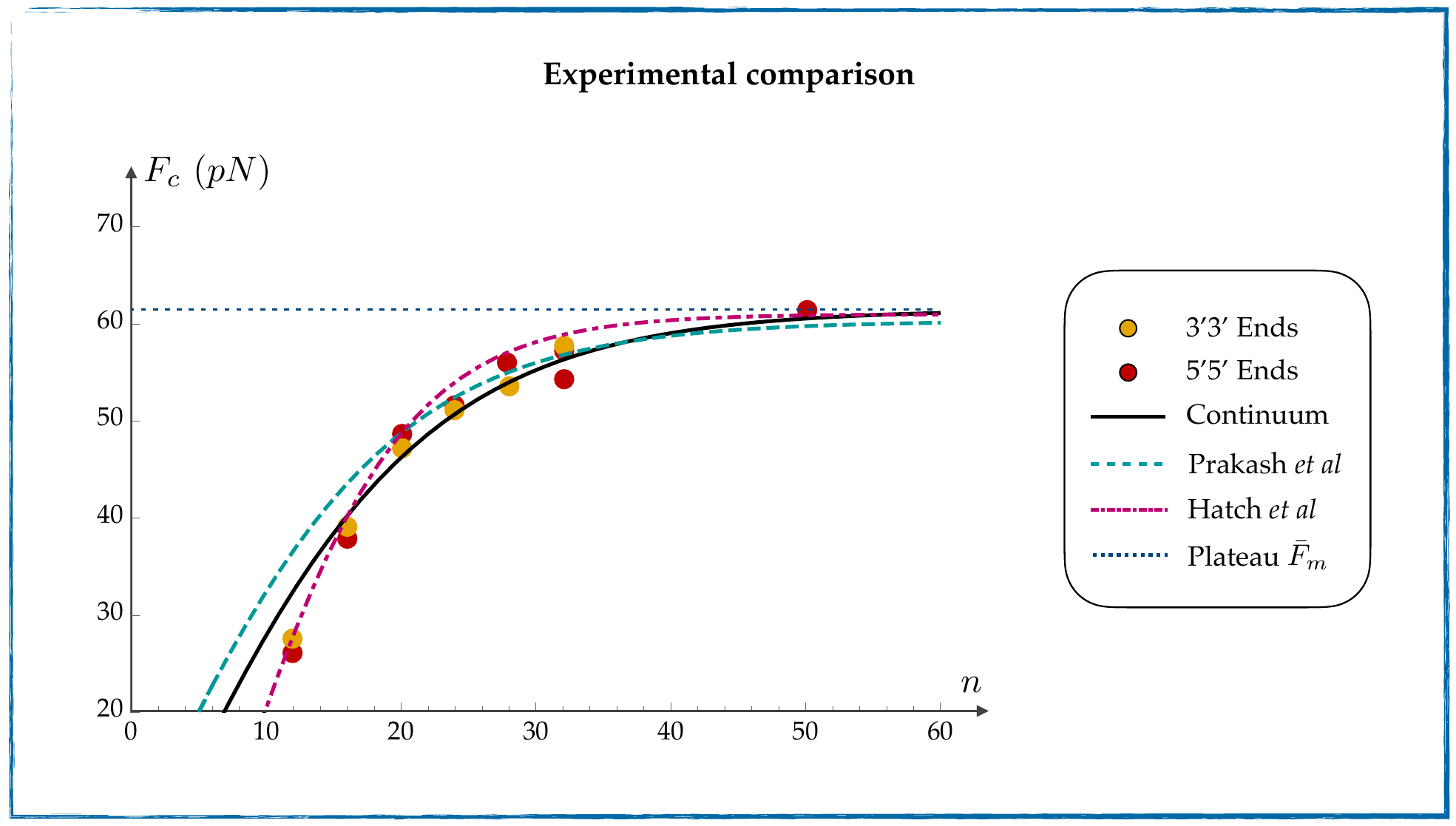}
  \caption[Experimental comparison with denaturation force at varying size of the system]{Comparison with the experiment performed by Hatch \textit{et al.} (\cite{hatch:2008}) with the theoretical curves. The dots represent the $3'3'$ ends DNA type molecule and the $5'5'$ ends type, in yellow and red, respectively. The black line is the continuum model derived in this work with the following parameters: $\kappa_{\tau}\kappa_e=1.89\times10^{-21}$ (pN)$^{2}$, $l=0.34$ nm and $u_d=0.034$ nm. The dashed sky-blue dashed line is the discrete one of Prakash \textit{et al.}, with parameters $\chi^{-1}=9.4$ and $f_c=4.1$. The dot-dashed purple line is the one fitted by Hatch and colleagues with $\chi^{-1}=6.4$ and $f_c=4.1$ and $L_{open}=7$.}
  \label{fig:ch5_experiment}
\end{figure}

Now, I use this result and the data in (\cite{hatch:2008,prakash:2011}) to fit the experiments. The distance between the base pairs of a DNA helix is a known parameter, and it is equal to $3.4$ \AA, thus $l=0.34$ nm. To what concerns $u_d$, the detachment distance of the base pairs, reasonable values are around $u_d\simeq 0.02-0.05$ nm depending on the fact the single base can be loaded longitudinally or at shear. By considering the experimental value of the force plateau provided by (\cite{hatch:2008}) and using~\eqref{eq:ch5_fmnostra} we get the product of the two stiffness that is $\kappa_{\tau}\kappa_e\simeq 1.89\times10^{-21}$ pN. In Figure~\ref{fig:ch5_experiment} we show the comparison between the experimental values with our model as well as with the previous results of Hatch and colleagues and Prakash \textit{et al.} The agreement between our theoretical results and the experimental values is excellent, and it is represented by the black line in Figure~\ref{fig:ch5_experiment}. With a dotted blue line is also represented the force plateau given by~\eqref{eq:ch5_fmnostra} fixed at the experimental value of $61.4$ pN attained with a sequence of $50$ base pairs. 

There are some important remarks to be outlined. Concerning the experimental data, Hatch and coworkers do not provide any additional information about the stiffnesses of the DNA sequences and the displacement at which rupture occurs. As a matter of fact, they only perform the `best fit' of~\eqref{eq:ch5_hatchform} obtaining $f_c=3.9$ pN, $L_{open}=7$ bp and $\chi^{-1}=6.8$ bp, that corresponds to a ratio of $Q/R=92.5$, and their curve is represented with a dot-dashed purple line in Figure~\ref{fig:ch5_experiment}. Prakash and colleagues also fitted the data without any comparison with the physical quantities. They address the fact that for a few base pairs there is mismatching to the fitting parameter used instead of the temperature, and their curve is shown in Figure~\ref{fig:ch5_experiment} with the dashed sky-blue line. Thus, the main difference between previous results and our approach is that we are able to provide the mechanical quantities such as the internal length and the rupture forces in terms of physical parameters, \textit{i.e.} stiffnesses and lengths. Moreover, we use the experimental data and known values found in the literature to describe the experiments, obtaining an excellent agreement.

\section{Discussion}

In this chapter I proposed a discrete lattice model to describe the mechanical behaviour of a double-stranded DNA helix and I derived analytical formulas. As discussed in the introduction, this model is very general and may be used to describe also the interaction between the microtubules and the tau-proteins that can be found in bundles inside the neuron's axons. I remark that while in the case of the DNA we studied the mechanical response at equilibrium conditions when the model is applied to describe the damage in the axon rate effects are fundamentals and are the topic of the following chapter. 

To mimic the intramolecular bonds holding together the base pairs of a DNA helix, I choose non-convex energy characterized by an elastic regime that results in a broken phase if a certain threshold $u_d$ is reached. The pairs are aligned between polyphosphate stiffer backbones that are described with two elastic chains, as shown in Figure~\ref{fig:ch5_modenergy}. With the idea of describing an SMFS experiment performed on a DNA sequence and understanding the debonding mechanism, I study the system under shear loads. In particular, I apply a force or a displacement at the $n$-th upper element of the chain and the $1$st lower spring in opposite directions, thus assigning the shear displacement to the whole system. 

With a variational approach already introduced in this thesis, I minimized the energy under the applied displacement $\delta$ by introducing force $f$, which is the conjugate variable of $\delta$ and represents the Lagrange multiplier of the minimization and I obtained the mechanical quantities describing the total stiffness of the system, the force with respect to the assigned displacement and the total energy of the system. To obtain analytical formulas it is necessary to invert the matrix $\boldsymbol{J}$ in~\eqref{eq:ch5_kchi}, which depends on the specific configuration of the system. However, one discovers that the shear displacements in the chain are monotonic, and they decrease from the onwards base pairs moving towards the centre of the lattice and it can be analytically proved, as shown in~\eqref{eq:ch5_mono}. Thus one may study a di-block system fully attached and an elastic chain representing the elastic backbones already detached and join the two solutions. In this way, all the mechanical quantities are analytically derived in terms of the size of the system $n$ and the number of attached base pairs $p$. 

I also studied the system under two different hypotheses. When the Maxwell convention is considered, \textit{i.e.} the system is always in the global minima of the energy, a fragile type of rupture is observed for both the value of the stiffness parameter $\nu$ and the size of the system $n$, as shown in Figure~\ref{fig:ch5_global}. On the other hand, when the maximum delay convention is followed, the behaviour of the debonding process of the DNA sequences may vary widely depending on both the two aforementioned parameters, ranging from a fragile kind of rupture to a more ductile decohesion process, characterized by a sawtooth force plateau representing the sequential detachment of the base pairs, as represented in Figure~\ref{fig:ch5_local}. Next, following (\cite{puglisi:2000}), I studied the so-called continuum approximation obtained when the total length of the chain $L$ is fixed and $n\to\infty$ while $l\to 0$. With this limit, a more sound description of a continuum-like material is proposed and I proved that the decohesion strategies studied in the discrete case also hold in this limit with the difference in the fraction of detached base pairs, which now is the continuum variable $\pi \in (0,1)$ and it represents a measure of the damage of the system. These results, widely demonstrated in Section~\ref{sec:ch5_cl}, are the key to deriving macroscopic quantities with a multiscale approach. It is also interesting to study two specific limit regimes, attained when the continuum stiffness parameter $\mu$ goes to zero or infinity. Indeed, in the first limit, the base pairs are extremely soft with the respect to the backbones, and always a fragile-type rupture occurs, as shown in Figure~\ref{fig:ch5_mulimit}. Conversely, when $\mu\to\infty$, the base pairs are stiffer than the backbone, the detachment is increasingly ductile and a force plateau given by~\eqref{eq:ch5_felasticmugrande} is reached. 

%
\begin{figure}[t!]
\centering
  \includegraphics[width=0.95\textwidth]{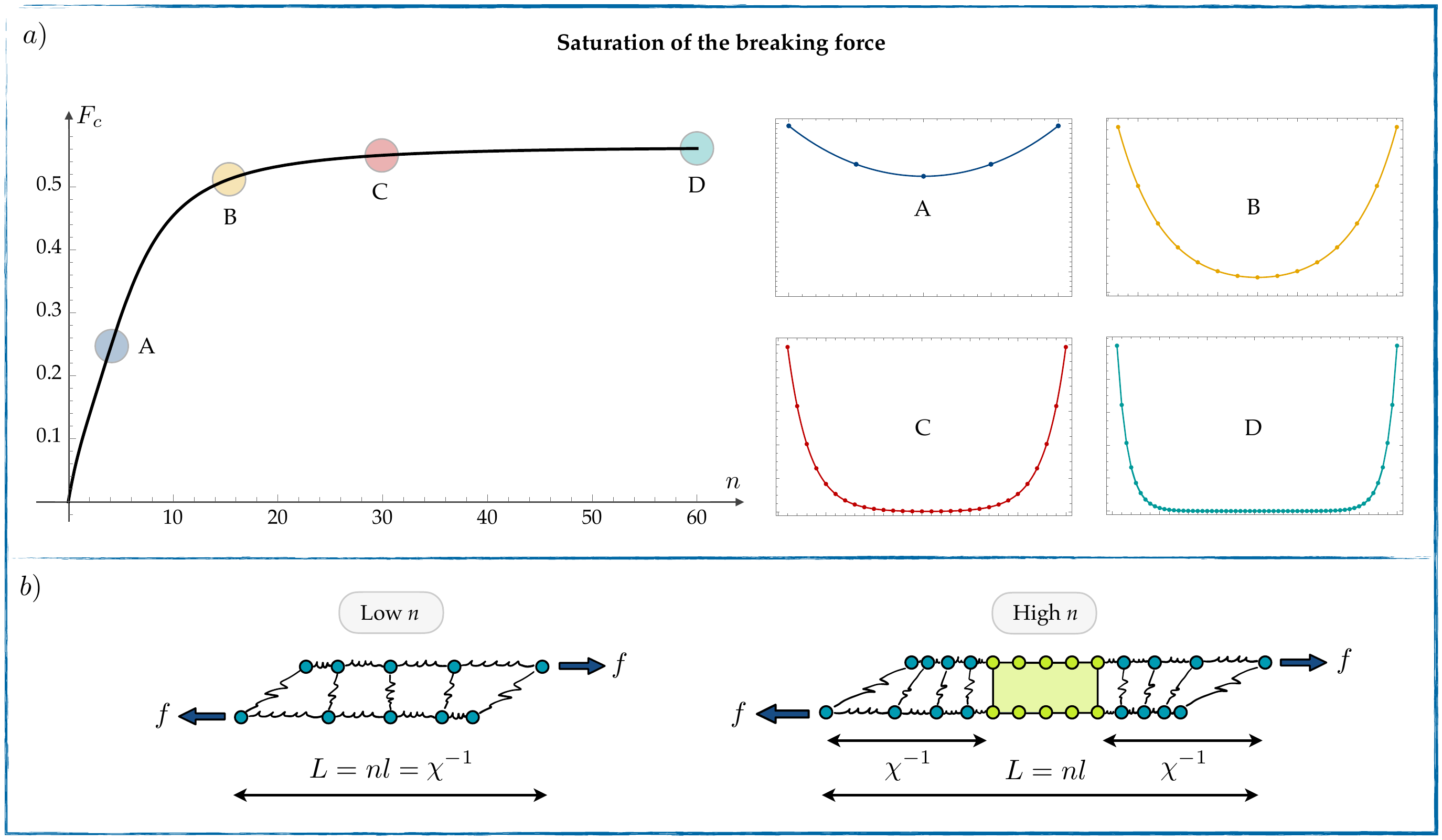}
  \caption[Saturation of the breaking force]{Saturation of the threshold critical force with respect to the number of objects, \textit{i.e} the length of the chain. In panel a) we observe the critical force $F_c$ for different $n$ and the corresponding shears along the chain. If the number is low (see $A$ and $B$) all the elements are affected by shear loads whereas when $n$ increases (see $C$ and $D$) only the external parts are subjected to the effect of the applied force. In panel b) we represent two schematic representations of the regime above described.}
  \label{fig:ch5_saturation}
\end{figure}

This result is fundamental when a system such as the double-stranded DNA sequence or the microtubule and the tau-proteins bundle is loaded under shear forces, depending also on the specific mechanical properties of the system, which may lead to a fragile or ductile type of fracture. Indeed, the two limit regimes attained at varying $\mu$ given by~\eqref{eq:ch5_mumu} in the continuum limit, define the shape of the debonding force with respect to the size of the system. Indeed, according to the experimental comparison and (\cite{degennes:2001}), where first was tried to answer the question about the rupture force of the DNA sequence, we observe that the saturation force in Figure~\ref{fig:ch5_saturation}$_a$ (see Eq.~\eqref{eq:ch5_fcnostra}) with respect to the number of base pairs $n$ saturates to plateau for increasing $n$. 

Eventually, observe in Figure~\ref{fig:ch5_saturation} the distribution of the shear displacements in the sequence at different values of $n$, indicated by the capital letters $A$, $B$, $C$ and $D$ and represented in the boxes on the right side of the panel. When $n$ is small, \textit{i.e.} boxes $A$ and $B$, the shear force acts on the elements of the chain leading to a non-zero displacement that decreases from the external regions towards the internal ones. When the size increases, (see panels $C$ and $D$), the effect of the shears can be observed only on the borders whereas it is zero in the centre part of the system. Thus, we identify a `persistence length', also known as \textit{shear lag} in fracture mechanics (\cite{lubliner:2008,courtney:1990}), that it is given by $\bar \chi^{-1}$. This length is a constant value that depends on the properties of the system (see Eq.~\ref{eq:ch5_chinostra}) and consequently measures the part of the materials affected by the shear force. From a physical perspective, when the regime of the rupture force is linear, all the system is influenced by the applied external loads and all the base pairs are loaded because the persistence length covers all the dimensions of the system, as shown on the left side of Figure~\ref{fig:ch5_saturation}$_b$. Conversely, when $n$ is large the force acts only on the external regions of the material along the length $\bar \chi^{-1}$ that defines the `process zone' in which the decohesion process occurs whereas the remaining inner part of the system is unloaded.


	\clearpage


\renewcommand{\thefigure}
{\arabic{chapter}.\arabic{figure}}
\setcounter{figure}{0}

\renewcommand{\theequation}
{6.\arabic{equation}}
\setcounter{equation}{0}

\chapter{Rate effects on microtubules and tau proteins under shear loads}
\label{ch_6}
\vspace{1.2cm}

The human brain is the most complex organism that can be found in nature and it is made by billions of neurons, the fundamental brain cell. The neuron, an electrically excitable cell, is made by a cell body containing the nucleus, a great number of dendrites, that are extensions of the nerve cell with the function of propagating electrochemical signals. One of these dendrites is longer than the other ones and it is called the axon, which transmits information through an impulse known as \textit{action potential} to other cells, muscles, tissues and glands. Despite the electrochemical nature of these cells, there is evidence that the mechanical properties of such a complex system also play a crucial role at the micro-scales and changes in these properties can lead to malfunctioning and damages, often resulting in neurodegenerative diseases.  

As a matter of fact, external forces acting on the brain, impacts and traumatic accidents, are pathologically knowns as traumatic brain injury (\acs{TBI}), and they may cause severe brain damage not only at the moment of the hit but also within the years (\cite{smith:2000,smith:2010,smith:2012}). The neurological damage caused by the external force may spread within the brain tissue due to its viscoelastic nature in a time ranging from days to years. The resulting effect of this phenomenon is the occurrence of neurodegenerative diseases such as Parkinson's and Alzheimer's. First shreds of evidence were found in boxers in the U.S.A. because a great percentage of them suffered Alzheimer's years after they retired (\cite{ballatore:2007,asken:2017}). It has been shown that the mechanism of damage due to repetitive or impulsive external forces generates what is known as diffuse axonal injury (\acs{DAI}), clinically recognized by scattered lesions at the level of axons, which is part of the white matter of the brain (\cite{cowan:2013,shi:2021}). Indeed, axonal failure can happen abruptly, and it is called primary axotomy, or with a progressive degradation that results in a gradual failure, the secondary axotomy (\cite{hawkins:2010}). The nature of this observed behaviour has to be searched at the micro-scale level, where the complex energetic landscape and the resulting damages due to mechanical effects are of crucial importance. 

The axon cytoskeleton is made by microtubules (\acs{MT}), neurofilaments and microfilaments. The first ones are the major structural components, composed by heterodimers of $\alpha$ and $\beta$ tubulin arranged in $13$ circularly joined protofilaments, making the microtubules a hollow cylinder with an external diameter of $d^{e}_{\tiny{ MT}}=24$ nm and an internal diameter of $d^i_{\footnotesize{MT}}=14$ nm (\cite{ouyang:2013,dubey:2020}). Their length does not cover the whole dimension of the axon but microtubules are arranged in bundles with a unit length varying from $2$ to $100$ $\mu$m. Typically, the Young Modulus is about $1.9$ GPa and in axon sections, the radial distance among the MTs is about $23 - 38$ nm whereas the spatial distribution of the microtubules within the axons can be considered almost constant with about $20$ MTs per section. Microtubules are linked and kept in bundles by the tau proteins (\cite{elbaum:2012,eisenberg:2017,li:2015}). The tau-proteins main function is to stabilize microtubules and keep the correct arrangements of the bundles inside along the axon length. These proteins are made by $352$ to $441$ amino-acids in length divided into three parts: the positive charged C-terminus, the ligand part with the microtubules and a negative charged N-terminus that is connected to another tau-proteins forming an electrostatic zipper between two adjacent MTs (\cite{rosenberg:2007,morris:2011}). Axons and neurons made the white matter of the brain, which viscoelastic behaviour is far to be fully understood. One hypothesis widely accepted is that its behaviour can be generated by a multiscale mechanism that starts from the mechanics of axons and microtubules and spread up across the scales.  At the level of the axons, three main regimes are highlighted. If both the strain and the strain rate are low, there is sliding within the axons and the microtubules roll by each other and can come back to their initial length. If the strain increases, the $\tau$-proteins connections start to break. If the strain rate is also high, the force transmitted from the proteins to the MT is high and the microtubules can break (\cite{kuhl:2015,kuhl:2016}). During dynamic injury, there is evidence of interruption of axonal transport, and the material transported accumulates in axonal swellings that appear in a periodic arrangement in the form of beads or strings along connected axons, forming pathological phenotype called `axonal varicosities'. Although this problem may arise after injuries and transported materials are lost, the transport can still proceed via other intact microtubules (\cite{smith:2010,smith:2012}).

From a theoretical point of view, the challenge of describing the thermo-mechanical rate-dependent behaviour of multiphase systems is widely tackled, due to the many open problems still to be unveiled (\cite{evans:1997,evans:1999}). Indeed, if on one side both the direct modelling of the intertial dynamics  (\cite{balk:2001,cherkaev:2005,vainchtein:2010,efendiev:2010,cohen:2014}) and the description of the evolution of the system with the use of kinetic theories (\cite{puglisi:2002,givli:2009,raj:2011}) are well-established methods, the possibility of obtaining analytical results based on the microscopic characterization of the energy landscape including phaenomena such as hysteresis, residual stretches and damage under different loading conditions has yet to be thoroughly investigated. 

As a matter of fact, when systems with microinstabilities are considered the dynamic evolution between different metastable configurations is regulated by the possibility of overcoming energy barriers, which are usually of the order of a few hundreds of $k_B T$, thus comparable with the energy associated to the external loads. Specifically, we consider as limit regimes of load the two cases of hard and soft devices. In the first hypothesis, the energy barrier depends on both the applied strain or displacement and the system's actual configuration in terms of units that have already changed phase. Conversely, when the soft device is considered, only the applied force regulates the evolution among metastable states (\cite{puglisi:2000,puglisi:2002}). When also the temperature is introduced and the experimental behaviour of molecules or polymeric chains is described within the Helmholtz or Gibbs ensembles, the boundary conditions play a key role in determining the overall response of the system and different behaviours are observed (\cite{bustamante:2003,manca:2014,luca2019}) On the other hand, in the thermodynamic limit the persistence length is of the same order as the contour length of the chain and the equivalence or inequivalence of the systems depends on the specific case considered (\cite{sinha:2005,giordano:2018,winkler:2010,luca2020}). The literature describing such systems when the rate is considered is divided roughly into three areas. We found continuum approaches based on a Fokker-Planck formulation (\cite{herrmann:2012}), numerical based simulations where FEM methods (\cite{kuhl:2016}), Langevin dynamics (\cite{efendiev:2010,darocha:2020}) or intensive molecular dynamics (\cite{buehler:2008}) are used and the direct modelization based on the two-state theory (\cite{kramers:1940,bell:1978,evans:1997,dudko:2006}). 

On this last approach, the pioneering work of Evans and colleagues (\cite{evans:1991}) first described such systems from a phenomenological point of view, and, based on the hypothesis that the applied external force accelerates exponentially the rate of rupture, single-molecule experiments where described. Indeed, in this family of models, it is important both the effect of the load and the position of the transition state, \textit{i.e.} the evolution of the energy barrier under the applied boundary condition. Specifically, the Bell relation describing the adhesion of molecular bonds is based on the idea that as a force $f$ is applied, the barrier height decreases of a quantity $f\delta$, where $\delta$ is the conjugate displacement. This law is valid only for low forces and consider the transition state as fixed (\cite{dembo:1988,bell:1978}). However, as the force increases the height of the barrier decreases, thus weakening the dependence of the rupture rate on the force, a phenomenon that has been called \textit{Hammond effect}, by Thirumalai and coworkers (\cite{hyeon:2006,koculi:2006}). Kramers-based theories improved these simple methods, also allowing the possibility of obtaining analytical expressions for the rate of rupture at a given force (\cite{evans:1997,evans:1999,izrailev:1997,shapiro:1997}). 

Concerning the specific theme of the viscoelastic response of the brain tissue, the topic has been approached both from a macroscopic --continuum-- point of view and with a microscopic based perspective. Due to the difficulty in obtaining clear pieces of information in terms of cause-effect relations from direct experiments on such delicate cells and tissues, computational simulations and models may help to elucidate the underlying mechanisms dictated by the interplay of forces and thermal effects (\cite{goriely:2015}). The axons can be modelled at different levels of complexity. When looking at the viscoelastic behaviour under relaxation and creep tests, continuum models considering the one-dimensional structure as s solid for short timescale and as a liquid for longer ones well represent the phenomena (\cite{ahmed:2013,otoole:2008}). In recent years, due to the innovation in experimental and diagnostic techniques, it has been recognized that both the structural function of the internal component of the axon (\cite{howard:2002}) and the `active' mechanisms that generate forces (\cite{otoole:2015}) is of key importance. Indeed, the interplay between the tension and compression generated in the actin cortex and the one resulting from the interaction between microtubules and tau proteins may generate phenomena such as stall, collapse or growth (\cite{recho:2016,garcia:2017}). Consequently, it is very important to provide a correct description of the axon from the perspective of the microtubules and tau proteins bundles, both from a static and dynamic point of view (\cite{suter:2011,soheilypour:2015,peter:2012,kasas:2004}). Moreover, the mechanism of attaching and detaching (or breaking) of the cross-links (tau proteins) at varying loading rates from the passive and active point of view is one of the main challenges (\cite{ahmadzadeh:2015,ahmadzadeh:2014,jakobs:2015}). Recently, Kuhl and colleagues developed a computational model based on the FEM method to address these features, considering the dynamics of microtubule polymerization and depolymerization, the biology of crosslink attachment and detachment, and the physics of stretching (\cite{kuhl:2016,kuhl:2018,kuhl:2018:2}).

In Chapter~\ref{ch_5} we have introduced a microscopic based model describing a lattice system composed of elastic chains and breakable cross-links, and we have that both the DNA mechanism and the microtubules and tau proteins system are correctly described within our framework. In particular, we have introduced non-convex energy and we have provided a mechanical description of such physical systems under different loading conditions and in different situations, such as the Maxwell and the maximum delay hypothesis obtaining always analytical results. In this chapter, we aim at introducing rate effects using a Kramers based theory based on the possibility of computing the height of the energy barriers at varying applied force or displacement. Indeed, the idea is to derive a model capable of interpreting the dynamic evolution of the microtubule bundles at different regimes of loading and loading rates. 

%
\begin{figure}[t!]
\centering
  \includegraphics[width=0.95\textwidth]{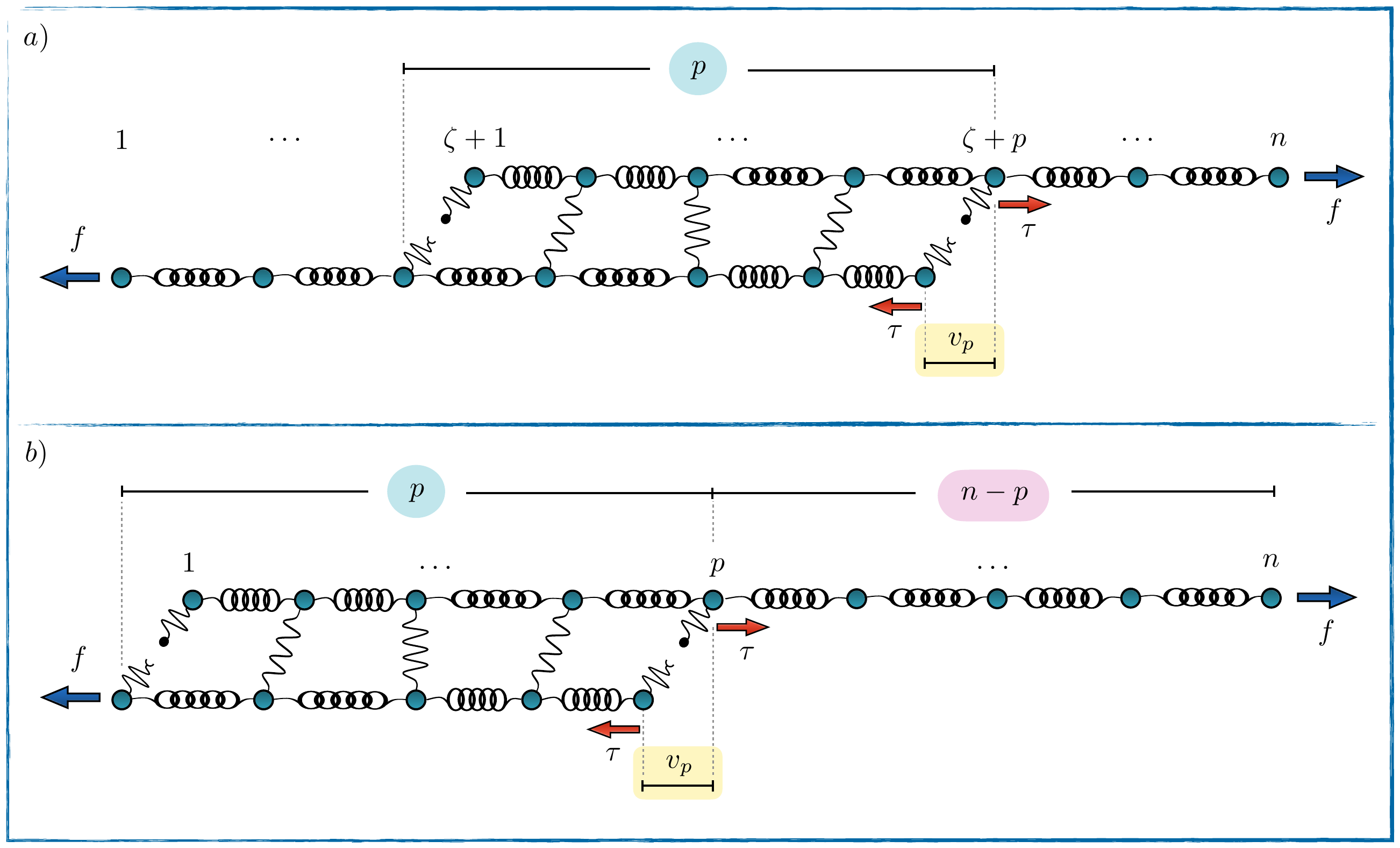}
  \caption[Mechanical model for the rate dependent system]{The mechanical model with the application of the shear force $\tau$ on the $p$-th element to evaluate the energy barrier is represented in panel a). This system, as demonstrated in Chapter~\ref{ch_5}, is mechanically equivalent to a system with $p$ attached element and $n-p$ detached units, independently from the position.}
  \label{fig:ch6_modelrate}
\end{figure}

\section{Mathematical framework}

The mechanical model we use to describe the rate-dependent behaviour of the system composed of microtubules and tau proteins is the same introduced in Chapter~\ref{ch_5}, where the mechanical response under assigned force and displacement has been studied in the zero-temperature limit. Thus, all the results obtained for the equilibrium case are still valid here. In particular, we remark that when temperature and non-local effects are neglected, the rupture propagates from the most external units through the internal ones, along the direction of the load. Moreover, we have demonstrated, through Equation~\eqref{eq:ch5_mono}, that the first element to break is always the most external so that it is possible to introduce the di-block solution (see Section~\ref{sec:ch5_diblock}). Accordingly, as shown in Figure~\ref{fig:ch6_modelrate}, we consider a system with a block of $p$ attached elements and a chain of $n-p$ unfolded units still carrying the load, whereas the broken springs on the side without the load are unloaded and hence negligible. Thus, the resulting mechanics is independent of the position of the breakable springs.

\subsection{Energy barriers}

In order to evaluate the height of the energy barrier separating the stable elastic solution from the broken state, it is necessary to keep all the elements equilibrated whereas only one is allowed to move from its equilibrium position to the transition thresholds. thus, it is necessary to evaluate the displacement required to break a generic $p$-th element and the resulting associated force. Here we begin with considering the case of the soft device, and we explore the energy landscape and rate-dependent response of the system under this condition. To do this, we introduce the non-dimensional shear force $\tau$, that, accordingly with the non-dimensional quantities introduced in Section~\ref{sec:ch5_model}, reads 
\begin{equation}
\tau=\frac{F_{\tau} L}{k_e u_d},
\label{eq:ch6_tau}
\end{equation}
where $F_{\tau}$ is the dimensional shear force. We apply this force to the generic element with index $p$, and we recall that it is always the element that is going to break first. We also recall that the mechanical energy of the system introduced in~\eqref{eq:ch5_phi1} is 
\begin{multline}
n\varphi=\frac{1}{4}\left[\boldsymbol{J}\boldsymbol{v}\cdot\boldsymbol{v}-\left(\boldsymbol{v}\cdot\boldsymbol{i}_{1}\right)^2-\left(\boldsymbol{v}\cdot\boldsymbol{i}_{n}\right)^2\right]+\frac{1}{4}\left[\boldsymbol{L}\boldsymbol{z}\cdot\boldsymbol{z}-\left(\boldsymbol{z}\cdot\boldsymbol{i}_{1}\right)^2-\left(\boldsymbol{z}\cdot\boldsymbol{i}_{n}\right)^2\right]\\
+\frac{1}{2}\nu^2(n-\boldsymbol{\chi}\cdot\boldsymbol{\chi}).
\label{eq:ch6_phi}
\end{multline}
and the vectors $\boldsymbol{v}=\boldsymbol{w}^{u}-\boldsymbol{w}^{l}$ and $\boldsymbol{z}=\boldsymbol{w}^{u}+\boldsymbol{w}^{l}$ represent the shears and the rigid displacement of the system, respectively. To evaluate the energy barriers in the case of soft devices we have to consider a mixed problem. Indeed, we assign the total force $f$ at the upper and lower free ends of the two elastic chains, whereas we fix the displacement on the $p$-th element of the system, resulting in a soft-hard device hypothesis. Thus, using~\eqref{eq:ch6_phi} and~\eqref{eq:ch6_tau}, the new energy function to minimize reads
\begin{equation}
h(f,\tau):=\varphi-\frac{f}{n}\Bigl(w_n^u-\delta\Bigr)+\frac{f}{n}\left(w_1^l+\delta\right)-\frac{\tau}{n}\Bigl(w_p^u-w_p^l-\delta_p\Bigr),
\end{equation}
namely
\begin{equation}
h(f,\tau)=\varphi-\frac{f}{n}\left(\frac{\boldsymbol{z}+\boldsymbol{v}}{2}\cdot\boldsymbol{i}_{n}-\delta\right)+\frac{f}{n}\left(\frac{\boldsymbol{z}-\boldsymbol{v}}{2}\cdot\boldsymbol{i}_{1}+\delta\right)-\frac{\tau}{n}\Bigl(\boldsymbol{v}\cdot\boldsymbol{i}_{p}-\delta_p\Bigr),
\end{equation}
where $\delta_p$ is the imposed displacement of the $p$-th element of the system. The variational problem is now 
\begin{equation}
\min_{\boldsymbol{w}^u,\,\,\boldsymbol{w}^l,\,\,w_p^u-w_p^l=\delta_{p}} \,\,h(f,\tau),
\end{equation}
and hence equilibrium requires  
\begin{equation}
\begin{split}
\frac{\partial h(f,\tau)}{\partial \boldsymbol{v}}&=\boldsymbol{J}\boldsymbol{v}-v_1\boldsymbol{i}_1-v_n\boldsymbol{i}_n-\frac{f}{n}\boldsymbol{i}_n-\frac{f}{n}\boldsymbol{i}_1-2\frac{\tau}{n}\boldsymbol{i}_p=\boldsymbol{0},\\
\frac{\partial h(f,\tau)}{\partial \boldsymbol{z}}&=\boldsymbol{L}\boldsymbol{z}-z_1\boldsymbol{i}_1-z_n\boldsymbol{i}_n-\frac{f}{n}\boldsymbol{i}_n+\frac{f}{n}\boldsymbol{i}_1=\boldsymbol{0}.
\end{split}
\label{eq:ch6_eq}
\end{equation}
%
%
By a direct inspection of~\eqref{eq:ch6_eq} it is clear that the displacement $\boldsymbol{z}$ does not dependent on $\tau$ but only on $f$, and gives the same solution obtained in~\eqref{eq:ch5_z1n} and~\eqref{eq:ch5_zz} that we report for clarity: 
\begin{equation}
\boldsymbol{z}(f)=\frac{\boldsymbol{L}^{-1}\boldsymbol{i}_{n}-\boldsymbol{L}^{-1}\boldsymbol{i}_{1}}{2\boldsymbol{L}^{-1}_{1,n}}\frac{f}{n}.
\label{eq:ch6_z}
\end{equation}
This feature can be addressed to the fact the $\tau$ is a full-fledged internal force of the system, that is equilibrated. On the other hand, from the first equation of~\eqref{eq:ch6_eq} we obtain the first and last displacement 
\begin{equation}
\begin{split}
v_1=&\Biggl[\left(\boldsymbol{J}^{-1}_{1,n}\right)^{2}+\boldsymbol{J}^{-1}_{1,n}+\boldsymbol{J}^{-1}_{1,1}\left(1-\boldsymbol{J}^{-1}_{n,n}\right)\Biggr]\frac{f}{n\mathcal{D}}+\Biggl[\boldsymbol{J}^{-1}_{p,n}\boldsymbol{J}^{-1}_{1,n}+\boldsymbol{J}^{-1}_{p,1}\left(1-\boldsymbol{J}^{-1}_{n,n}\right)\Biggr]\frac{2\tau}{n\mathcal{D}},\\
&\\
v_n=&\Biggl[\left(\boldsymbol{J}^{-1}_{1,n}\right)^{2}+\boldsymbol{J}^{-1}_{1,n}+\boldsymbol{J}^{-1}_{n,n}\left(1-\boldsymbol{J}^{-1}_{1,1}\right)\Biggr]\frac{f}{n\mathcal{D}}+\Biggl[\boldsymbol{J}^{-1}_{p,1}\boldsymbol{J}^{-1}_{1,n}+\boldsymbol{J}^{-1}_{p,n}\left(1-\boldsymbol{J}^{-1}_{1,1}\right)\Biggr]\frac{2\tau}{n\mathcal{D}},
\end{split}
\label{eq:ch6_v1vn}
\end{equation}
where to simplify the notation, we have indicated with $\mathcal{D}$ the denominator quantity
\begin{equation}
\mathcal{D}=\left(1-\boldsymbol{J}^{-1}_{1,1}\right)\left(1-\boldsymbol{J}^{-1}_{n,n}\right)-\left(\boldsymbol{J}^{-1}_{1,n}\right)^{2}.
\end{equation}
The shear vector thus reads 
\begin{multline}
\boldsymbol{v}(f,\tau):=\Biggl[\left(1+\boldsymbol{J}^{-1}_{1,n}-\boldsymbol{J}^{-1}_{n,n}\right)\boldsymbol{J}^{-1}\boldsymbol{i}_1+\left(1+\boldsymbol{J}^{-1}_{1,n}-\boldsymbol{J}^{-1}_{1,1}\right)\boldsymbol{J}^{-1}\boldsymbol{i}_n\Biggr]\frac{f}{n\mathcal{D}}\\
+\Biggr[\left(\boldsymbol{J}^{-1}_{p,n}\boldsymbol{J}^{-1}_{1,n}+\boldsymbol{J}^{-1}_{p,1}-\boldsymbol{J}^{-1}_{p,1}\boldsymbol{J}^{-1}_{n,n}\right)\boldsymbol{J}^{-1}\boldsymbol{i}_1\\
+\left(\boldsymbol{J}^{-1}_{p,1}\boldsymbol{J}^{-1}_{1,n}+\boldsymbol{J}^{-1}_{p,n}-\boldsymbol{J}^{-1}_{p,n}\boldsymbol{J}^{-1}_{1,1}\right)\boldsymbol{J}^{-1}\boldsymbol{i}_n+\mathcal{D}\boldsymbol{J}^{-1}\boldsymbol{i}_p\Biggr]\frac{2\tau}{n\mathcal{D}}.
\label{eq:ch6_v}
\end{multline}
It is worth noticing that in both~\eqref{eq:ch6_v} and consequently~\eqref{eq:ch6_v1vn}, the shears are expressed as a linear combination of the contribution of the force $f$ and the shear $\tau$ and are founded separately in the equations. Also, when $\tau=0$ we obtain the same results of the previous case, \textit{i.e.} Equation~\eqref{eq:ch5_vv}. We search for the solution that maximizes the shear force, that corresponds to the maximum value of the imposed displacement $\delta_p$. Thus, by recalling that the maximum displacement is reached when $\delta_p^{max}=1$, we impose 
\begin{equation}
\delta_p^{max}:=\boldsymbol{v}(f,\tau)\cdot\boldsymbol{i}_p=1
\end{equation}
to obtain the maximum value of $\tau$ with respect to the generic force $f$, obtaining the function
\begin{equation}
\tau(f)|_{\delta_p^{max}}.
\end{equation}
Thus, the  shear displacements in~\eqref{eq:ch6_v} can be rewritten as a function of only $f$ such that 
\begin{equation}
\boldsymbol{v}_{\tau}(f):=\boldsymbol{v}\bigl(f,\tau(f)|_{\delta_p^{max}}\bigr),
\label{eq:ch6_vtau}
\end{equation}
where, from now on, we indicate with the pedex $\tau$ the function where $\delta_p$ is maximized. 

Following, the resulting mechanical energy corresponding to the maximum attainable displacement, that corresponds to the energy in the `saddle' point, \textit{i.e} the energy associated to the maximum point before rupture, can be evaluated as
\begin{multline}
n\varphi_{\tau}(f):=\frac{1}{4}\left[\boldsymbol{J}\boldsymbol{v}_{\tau}(f)\cdot\boldsymbol{v}_{\tau}(f)-\left(\boldsymbol{v}_{\tau}(f)\cdot\boldsymbol{i}_{1}\right)^2-\left(\boldsymbol{v}_{\tau}(f)\cdot\boldsymbol{i}_{n}\right)^2\right]\\
+\frac{1}{4}\left[\boldsymbol{L}\boldsymbol{z}\cdot\boldsymbol{z}-\left(\boldsymbol{z}\cdot\boldsymbol{i}_{1}\right)^2-\left(\boldsymbol{z}\cdot\boldsymbol{i}_{n}\right)^2\right]+\frac{1}{2}\nu^2(n-\boldsymbol{\chi}\cdot\boldsymbol{\chi}),
\end{multline}
and the soft (Gibbs) energy reads 
\begin{equation}
g_{\tau}(f):=\varphi_{\tau}(f)-2\frac{f}{n}\delta_{\tau},
\end{equation}
where $\delta_{\tau}$, following Equation~\eqref{eq:ch5_deltadelta} and using~\eqref{eq:ch6_vtau} instead of~\eqref{eq:ch5_vv}, reads 
\begin{equation}
\delta_{\tau}=\frac{1}{4}\Bigl[\boldsymbol{z}(f)\cdot\boldsymbol{i}_n-\boldsymbol{z}(f)\cdot\boldsymbol{i}_1+\boldsymbol{v}_{\tau}(f)\cdot\boldsymbol{i}_n+\boldsymbol{v}_{\tau}(f)\cdot\boldsymbol{i}_1\Bigr]. 
\end{equation}

By using~\eqref{eq:ch6_phi} and~\eqref{eq:ch5_deltadelta}, the soft device energy, at equilibrium reads
\begin{equation}
g(f):=\varphi(f)-2\frac{f}{n}\delta,
\end{equation}
so that the height of the energy barriers in the soft device hypothesis is 
\begin{equation}
g_b(f)=g_{\tau}(f)-g(f).
\label{eq:ch6_softbarrier}
\end{equation}
%

\section{Rate Effects}

To introduce rate effects, we consider a time-dependent applied force of the type
\begin{equation}
f(t)=v_f t,
\label{eq:ch6_force}
\end{equation}
where $t$ is the time and $v_f$ is the assigned velocity of the load. Then, we the rate of loading, here indicated with the symbol $r(t)$, that reads 
\begin{equation}
r(t)=r_0 e^{-\beta g_b(f(t))},
\label{eq:ch6_kramer}
\end{equation}
where the barrier is given by equation~\eqref{eq:ch6_softbarrier} and $\beta=1/(k_B T)$. It is important to remark that here the temperature it is not to be intended as the actual temperature of the brain but more as the combination of many effects such as the actual temperature and thermal noise at the molecular scale. Moreover, it is worth remarking that with the Kramers' theory here adopted, the energy barrier is a pure mechanical quantity, computed at zero temperature and $\beta$ in~\eqref{eq:ch6_kramer} represents the internal noise corresponding to the energy fluctuations among the different configurations, and should be compared to the energy of the barrier. Other approaches such as the transition state theory discussed in Chapter~\ref{sec:ch2_rate} adopt different methods to compute the actual activation energy and will be the topic of further works, where the rate effects will be analyzed in more aspects and details (\cite{hanggi:1990}). Following the approaches of (\cite{hummer:2003,benichou:2016}) we define the probability that, at a certain time $t$, the system has still $p$ units attached, namely $\mathcal{P}_p(t)$, where $p=0,\dots,n$. In other words, it represents the probability that the $p$-th element identified is not yet broken. This probability satisfies the first-order equation
\begin{equation}
\frac{d\mathcal{P}_p(t)}{dt}=\dot{\mathcal{P}}_p(t)=-r(t)\mathcal{P}_p(t).
\label{eq:ch6_proba}
\end{equation}
%

%
\begin{figure}[t!]
\centering
  \includegraphics[width=0.95\textwidth]{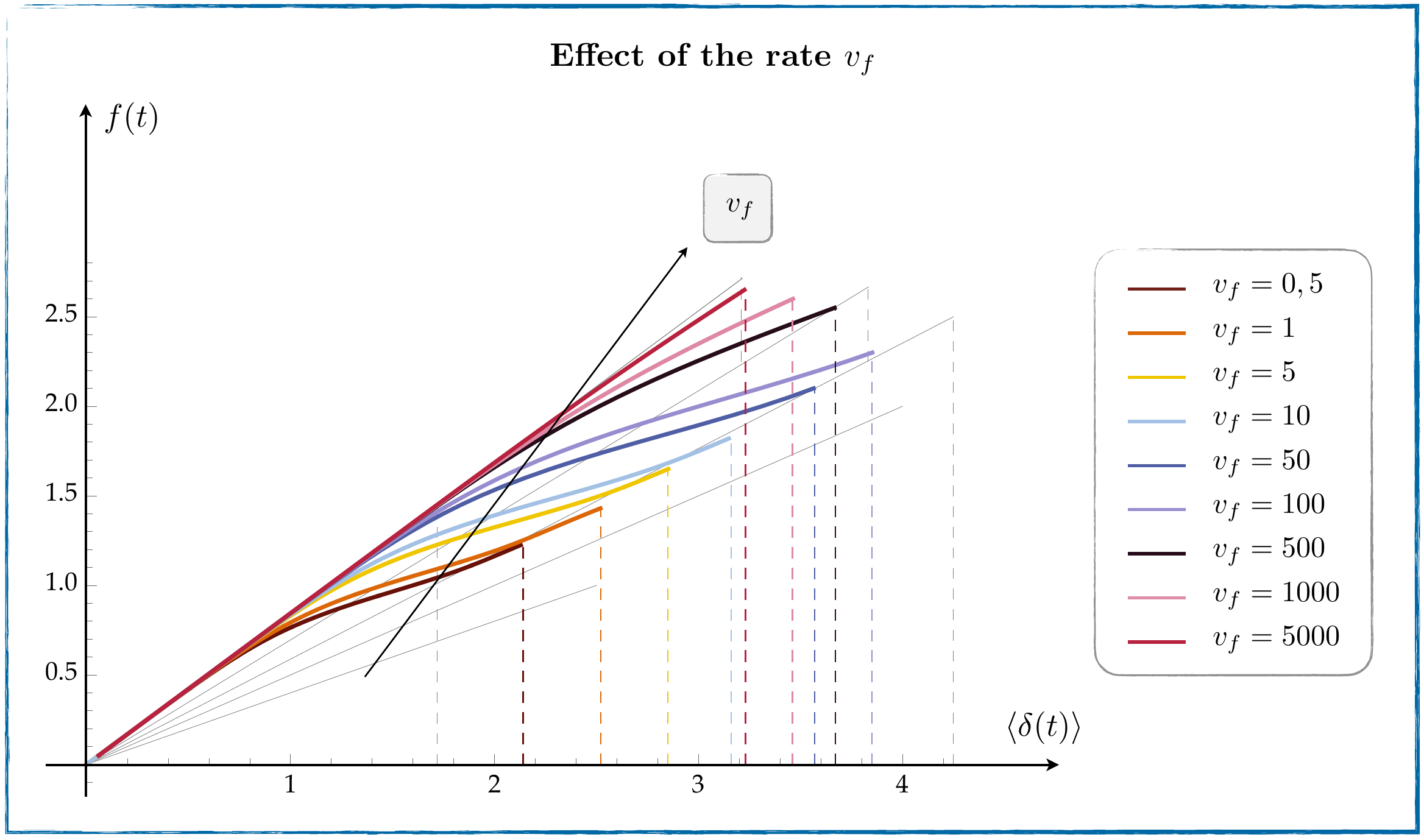}
  \caption[Effect of the rate of loading on microtubules and tau proteins]{Force-displacement time-dependent curves for varying rates of loading. It is possible to observe that as the rate $v_f$ increases, the rupture behaviour becomes more fragile and only the fully attached configuration is explored. For low rates, the system evolves within the different metastable configurations, the rupture becomes more ductile and it happens at lower force thresholds. The rate-dependent behaviour is compared with the equilibrium results of Section~\ref{sec:ch5_strategie}, represented with thin grey lines. Here $n=5$, $\beta=18$ and $r_0=10^3$.}
  \label{fig:ch6_rate}
\end{figure}

When we consider the evolution of the system for a certain time $t=0,\dots t_{max}$ and we apply the force~\eqref{eq:ch6_force}, the initial probability that the system is fully attached, \textit{i.e.} $p=n$ is $\mathcal{P}_n(t)=1$. Then the system starts to detach and this probability decreases with time. Simultaneously, as the probability of being in the configuration $p=n$ decreases one of the states $p=n-1$ increases of the same value. On the other hand, also the probability $\mathcal{P}_n-1(t)$ is affected by the time evolution and there is a `flux' from the next configuration, $p=n-2$. Thus, two rates have to be considered: $r_{p\to p-1}(t)$, from the state $p$ to the state $p-1$ that represents the probability of detaching, and $r_{p+1\to p}(t)$, which is the probability that the detached elements of the previous configuration $p+1$ can be found in $p$. In particular, the first configuration doesn't take probability from anyone as well as the last one doesn't give to anyone, thus
\begin{equation}
r_{n+1\to n}(t)=0, \qquad\qquad r_{0\to -1}(t)=0. 
\end{equation}
Then to perform the computation of the probability at each instant of time, we may proceed by expressing~\eqref{eq:ch6_proba} as a discretized function of the time, introducing the time discretization $\Delta t$ and the counter $j$ such that
\begin{multline}
\dot{\mathcal{P}}_p(t):=\frac{\mathcal{P}_p(j\Delta t)-\mathcal{P}_p((j-1)\Delta t)}{\Delta t}=\\
-r_{p\to p-1}(j\Delta t)\mathcal{P}_p((j-1)\Delta t)+r_{p+1\to p}(j\Delta t)\mathcal{P}_{p+1}((j-1)\Delta t).
\end{multline}

Eventually, we obtain the displacement of the system as a function of the time following the probability distribution in~\eqref{eq:ch6_proba}, which reads 
\begin{equation}
\langle\delta(t)\rangle=\sum_{p=0}^{n}\mathcal{P}_p(t)\delta(p,t),
\end{equation}
with $\delta(p,t)$ given by Equation~\eqref{eq:ch5_deltadelta}.

\section{Effect of rate of loading and temperature}

%
\begin{figure}[t!]
\centering
  \includegraphics[width=0.95\textwidth]{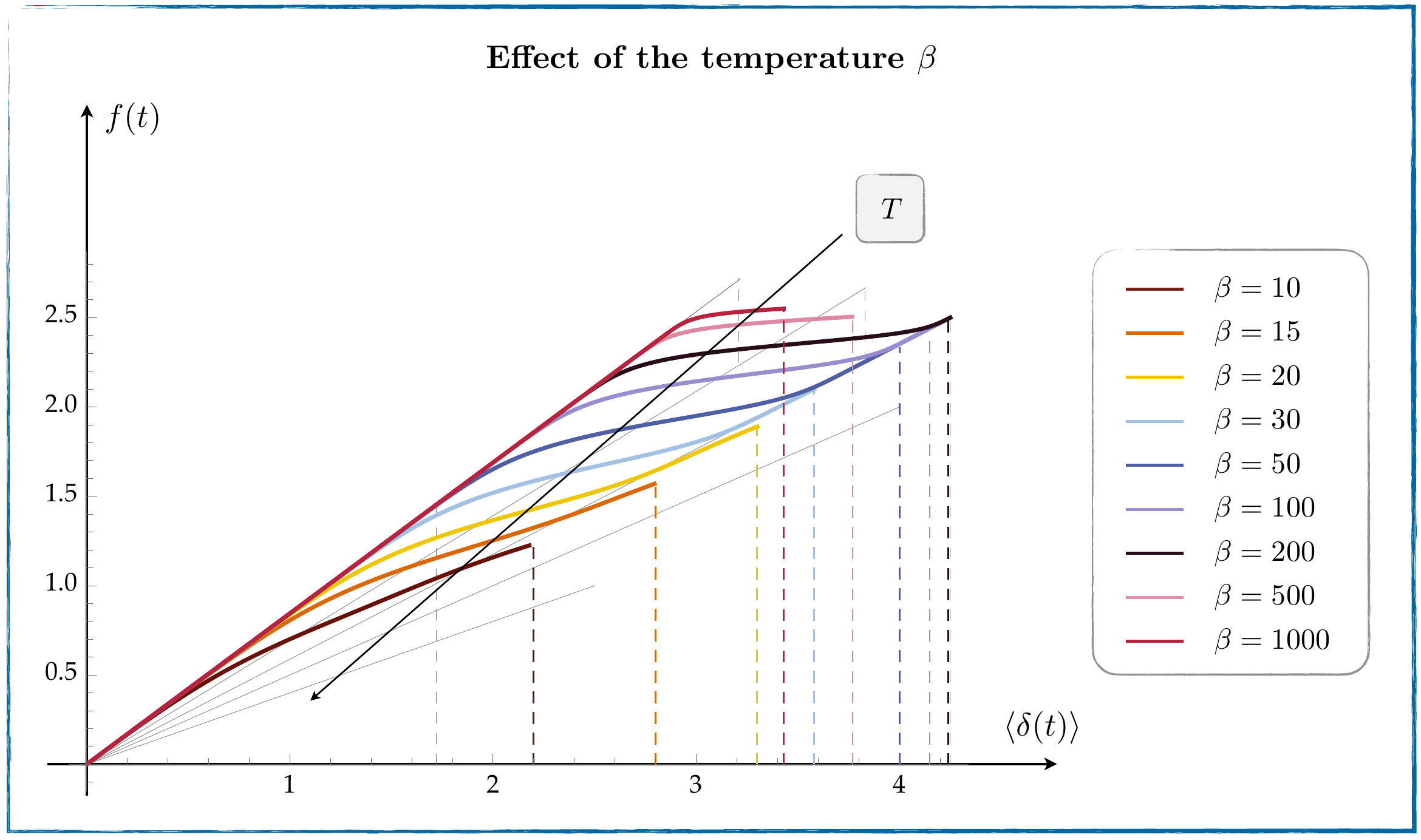}
  \caption[Effect of the temperature on microtubules and tau proteins]{The effect of temperature is represented. Specifically, as the temperature decreases, the maximum delay condition is approached and the system fails without exploring other metastable solutions. The rate dependent behaviour is compared with the equilibrium results of Section~\ref{sec:ch5_strategie}, represented with thin grey lines. Here $n=5$, $v=5$ and $r_0=10^3$.}
  \label{fig:ch6_rateTemp}
\end{figure}

In Figure~\ref{fig:ch6_rate} we observe the rate-dependent behaviour of the system made by microtubules and tau proteins, the main constituent of the neuronal axon. In particular, we show the force-displacement diagram for varying rates of the applied force. As we can observe, the force threshold at which the system breaks increases as the rate increases. Indeed, when a fast force is applied to a material, this usually stiffens and hence the resulting force is greater. On the other hand, when the rate is low, the force is better distributed along the length of the specimen. In this case, as we observe from the curves with low rates, the system evolves to other metastable configurations, and it changes phase continuously from the fully attached state to other solutions with some element detached. Indeed, as the velocity increases, the probability of changing phase decreases and thus the solution with the fully attached units is preferred (see red line in Figure~\ref{fig:ch6_rate}). When the energy barrier disappears and it is not possible to attain configurations with lower energy at a certain force threshold the system breaks and all the remaining units detach causing a drop in the force value to zero. Indeed, the strain curve is the average of the probability of being in a certain configuration but when the force associated with each equilibrium branch is exceeded, that configuration is no longer a feasible state achievable by the system. Moreover, when the rate increases, a `Maxwell' behaviour is followed, with the overall collapse of the system attained without exploring other metastable configurations. In low loading regimes, the system has the time to explore more metastable solutions.

This behaviour is the same as the one described in many papers on the topic of the rate-dependent response of tau proteins and microtubules (\cite{kuhl:2015,kuhl:2018,ahmadzadeh:2015}). In particular, here we derived this same overall response from a microscopic energetic description of the unit constituent of the system described, and based on a Kramers approach, the rate is introduced thanks to the possibility of evaluating the energy barriers separating the metastable configuration. 

In Figure~\ref{fig:ch6_rateTemp} we show the effect of temperature at a given rate of loading. In particular, it is possible to observe from the force-displacement curves that as the temperature increases, \textit{i.e.} $\beta$ decreases, the behaviour is more and more ductile even though the complete rupture of the system happen at a lower value of the force. Indeed, as the zero-temperature limit is approached (see the red curve in Figure~\ref{fig:ch6_rateTemp}), the system is frozen in the configuration with the element all attached because no thermal energy is provided. On the other hand, when the temperature grows and the thermal motion is activated and becomes predominant, the system has a given probability to escape a certain metastable local minimum even if the energy barrier is still not zero and other local solutions are explored. Thus, the more the temperature is high, the lowest is the force threshold at which the system breaks. 

Eventually, I remark the resulting behaviour obtained within this chapter is focused on the microscale level of the microtubules and tau proteins. Indeed, the model introduced at this level is of course not able to capture the complex problem of the viscoelastic behaviour of the brain tissue arising at the macroscopic scale but the aim here is to obtain simple analytical results to show the effect of rate, temperature and mechanical load at the level of the axons. As I will discuss in the conclusion, based on such microscopic results the aim is to develop a multiscale model able to describe the macroscopic response of the brain tissue considering also phenomena such as residual damages and the long time dependence on the acting forces.

	\clearpage


\renewcommand{\thefigure}
{\arabic{chapter}.\arabic{figure}}

\renewcommand{\theequation}
{7.\arabic{equation}}
\setcounter{equation}{0}

\chapter{Conclusions}
\label{ch_7}

%
Biological and bioinspired materials gained incredible interest in recent years due to their unique thermo-mechanical properties, and effort has been delivered to both understand the underlying mechanisms responsible for such incredible features and reproduce these materials. This topic has always fascinated me, and nowadays the process of understanding nature and developing ideas to be transferred into engineering solutions goes under the name of biomimetics, which has found applications in many fields such as medicine, biology, genetics, engineering, aerospace and material science. This fast development has been made possible by the advent of single molecule force spectroscopy experiments (\acs{SMFS}), which opened the possibility of both investigating the complex energy landscape characterizing biological materials and inspecting the hierarchical multiscale structure typical of these natural components. Indeed, one of the main open challenges in the field is the correct description of their structure at the molecular level to unveil the effects of mechanical and thermal fields at such a small scale.

%
In this thesis, I tackled the problem of studying biological systems with a bottom-up approach, describing the microscopic structure of a wide range of materials to derive the thermo-mechanical response of such systems under different boundaries conditions and evolution strategies. In particular, I entered the open challenge that nowadays tries to understand the role of temperature and forces in mechanisms such as unfolding, decohesion and fracture across different scales. To this matter, it is crucial to introduce models capable of correctly mimicking the main features of the system under investigation, which in most cases involves the description of the multiwells energy landscape that regulates the overall behaviour. Indeed, biological systems are characterized by non-convex energies, like the one pictured in Figure~\ref{fig:ch1_energy}, mainly due to the different scales involved and the microinstabilities observed at the molecular level. The resulting features derived from such types of energies unlock many interesting mechanisms that are widely studied by the scientific community. For instance, the conformational transition of proteins such as Titin immunoglobulin domains has been investigated by Matthias Rief and coworkers to unveil its typical folded $\to$ unfolded transition (\cite{rief:1997}) whereas the denaturation of a double-stranded DNA helix is a vital process for DNA and RNA transcription and replication (\cite{busta:2000}), where an open problem is the role of mechanical forces coupled with temperature effects at the level of the base pairs. Indeed, due to the action of mechanical and thermal fields, the detachment of the base pairs may happen abruptly or within a certain regime of sequential decohesion until a critical threshold of force is reached. Among many other interesting phenomena, we refer also to muscle mechanics, where the sarcomeres exhibit a conformational transition between two stable regimes (\cite{huxley:1971,darocha:2020}) and to the focal adhesion among cells and substrate, where the relative stiffness regulates the overall mechanical response (\cite{gao:2011,bell:1978}).

Microinstabilities at the small scale, where fundamental features such as non-local interaction or rate effects play a crucial role, have been historically approached with continuum mechanics models, also considering non-linear effects, history, the non-convexity of the energy wells and growth mechanisms (\cite{gurtin:2010,goriely:2017}). On the other hand, a common feature of all these examples is that it is almost always possible to identify elementary units at the microscopic levels that can be described with a specific energy regulating the conformational transition among different metastable states. Thus, in this thesis, we proposed a series of microscopic based models aiming at reproducing the simple behaviour of these units and, by coupling and increasing the level of complexity of these elements we have shown that is possible to describe a wide range of phenomena including mechanical, thermal and rate effects. 

\section{Discussion}

%
In the first part of the thesis, I showed how it is possible to approach the description of materials with microinstabilities by introducing a simple prototypical example of a chain undergoing a configurational folded/unfolded transition. To describe such a system, the idea of energy with multiple wells is introduced and two main cases are presented (see Figure~\ref{fig:ch1_energy}). When we consider the superimposition of two parabolas, in general having different properties, we approach the analysis of system undergoing conformational transition such as the folding $\to$ unfolding evolution of multistable proteins or the softening phenomena in focal adhesion. On the other hand, when the first elastic phase is followed by constant energy we mimic fracture or decohesion, which is typical of systems such as DNA or microtubules inside the neuronal axon. A remark about these assumptions is important. In general, real physical systems are characterized by non-convex energy of the type shown in Figure~\ref{fig:ch2_kramers}, where the two convex wells are linked by a concave --unstable-- region. This area represents the metastable configuration that separates two stable states and has to be crossed in order to attain the transition. In some existing examples, such as for Immunoglobulin domains, it can be very wide, thus quite affecting the overall response of the system in terms, for instance, of size of the energy barrier to be overcome or magnitude of the reference strain of the second configuration. With the aim of both pursuing the analytical treatment of the results of this thesis and catching the main physical features of the system under investigation, we consider two superimposed wells, thus neglecting this region. This approach was first introduced by (\cite{truskinovsky:1996}), and it has been shown that well suit a wide range of problem characterized by microinstabilities at the small scales. A possible approach to improve these models could be possible, for instance, to introduce the concave parabola by means of a spin variable with three values, \textit{e.g.} $\chi_i=-1$ for the folded region, $\chi_i=0$ for the unstable configuration and $\chi_i=1$ for the second phase but this would involve an enormous complication in the analytical treatment of the results, probably hiding the main features highlighted by the simple formulas we found. Similarly, the assumption of a piecewise quadratic wells for the stable configurations leads to linear stress-strain diagrams, while it is experimentally observed that systems such as the titin macromolecules undergoing unfolding are characterized by curve unfolding peaks (\cite{anderson:2013}). In (\cite{fp:2019}) has been demonstrated that our approach compared to other models capable of describing this feature such as the WLC does not lead to significant wrong estimation of the results whereas allows us to always obtain analytical results also when the temperature is considered (see Chapter~\ref{ch_3}). In any case, it is always possible to introduce all these mentioned features and numerically treat the problem, which is not the aim of this thesis.  

Thus, to introduce the effectiveness of our models and approaches, in Chapter~\ref{ch_2}, I considered a chain made by three units with the possibility of changing phase from a folded configuration to an unfolded one, (see Figure~\ref{fig:ch2_modelloesempio}). Specifically, both the `pure mechanical' case and the thermal effects are studied in both the hypothesis of the hard device (applied displacement) and soft device (applied force). The most relevant part of this Chapter concerns the analysis of the equilibrium path between the phases (see Section~\ref{sec:ch2_pt4}). Indeed, the evolution of a system with microinstabilities within the energy landscape is regulated by the possibility of overcoming energy barriers and by considering the simple example made by only three units, it is possible to visualize these paths in the configurational space, both for the hard and soft device hypotheses. Moreover, following (\cite{puglisi:2002,puglisi:2005}), we demonstrated that when a bi-parabolic type of energy is considered the minimum energy barrier path is the one across the spinodal point, \textit{i.e.} the point of intersection of the two parabolas (see Figure~\ref{fig:ch2_modelloesempio}), which is the equivalent of a saddle point for generic non-convex energy of the type in Figure~\ref{fig:ch2_kramers}. The knowledge of these paths is crucial to evaluate the energy barriers, fundamental for rate effects, evaluated for the case of the unfolding with three units (see Section~\ref{sec:ch2_rate}) and analyzed in Chapter~\ref{ch_6} for a more complex system. 

In Chapter~\ref{ch_3} I approached the problem of configurational transformation of proteins materials when the experimental set-up is considered (\cite{bustamante:2000}). When performing single molecule experiments, a known problem is that the effect of the pulling device (\textit{e.g.} the cantilever of the atomic force microscopy or the effect of the magnetic field in magnetic tweezers) regulates the overall response of the system under investigation (\cite{maitra:2010}). In the majority of cases, if these elements are not considered, they may lead to wrong estimates of the effective mechanical response up to a third of the true magnitude (\cite{biswas:2018}). To take into account the effect of the handling device we considered a microscale based model describing the titin protein molecule, constituted by unfolding domains pulled with an \acs{AFM} technique. Following (\cite{fp:2019}), I considered the protein and the device as a whole thermodynamical system and analysed both the regime of hard and soft devices also in the case of different temperatures. The key point of the analysis performed is that these two limit regimes are never attained in reality but the true overall response of the system lies in between them, and it is regulated by the stiffness of the loading device (\cite{luca2019}). In the thesis, I proved this feature by showing that the `ideal' hard and soft devices can be attained when the stiffness of the handling device goes to zero or infinity, whereas in real experiments there is a continuum transition between these two limiting cases. Thanks to the bi-parabolic energy hypothesis, we can deduce always fully analytical results describing the behaviour of the macromolecule undergoing phase transition in all the cases under investigation. In particular, we showed that when the temperature is considered, the constitutive equations are the same as in the case of pure mechanics (see Equations~\ref{eq:ch3_vfa}), and only the statistical averaged value of the unfolded phase fraction regulates the thermo-mechanical response of the system at varying temperatures. These results can be also extended when different mechanical properties such as stiffness or unfolding energy are considered for the two wells. Indeed, I studied the transition with a softening effect in Section~\ref{sec:ch3_aimeta}, and we discovered that both the relative stiffness between the wells and the transition energy may change completely the overall response. Eventually, we study the thermodynamical limit, \textit{i.e}, the case in which the number of units grows $n\to\infty$, and we demonstrate the important results that the Helmholtz and Gibbs ensemble are equivalent in this limit. 

These last result does not hold when non-local terms are introduced. Indeed, there are some materials in which the competition between interfacial and entropic energy terms is crucial. To address this feature, in Chapter~\ref{ch_4} I introduced next-to-nearest neighbour (\acf{NNN}) energy terms on the same model of Chapter~\ref{ch_3}, to reproduce the presence of interfacial energy terms penalizing the formation of phase interfaces. This model is applied to the case of memory shape nanowires, that exhibit an initial stress peak due to the nucleation of an initial phase before propagation. Indeed, when non-local terms are considered, the transition from a homogenous phase to a second configuration is delayed because of the energetically unfavourable event of nucleation of a new phase, generating phase interfaces (\cite{ball:1989,muller:1999,truskinovsky:2000, puglisi:2006,puglisi:2007}). In this analysis, I was able to describe this system by considering the competing effect of temperature, introduced within a statistical mechanics approach, and interfacial energy terms, regulating the macroscopic mechanical response. Specifically, I deduce fully analytical results describing the behaviour of the system under different conditions of temperature and interfacial energy (\cite{luca2020}). In particular, as a first result, we proved that both the `pure mechanical' case and the previous model without NNN units can be recovered as limit regimes of this newly updated framework. Moreover, as is observed in Figure~\ref{fig:ch4_phasen}, in the hard device hypothesis there is the presence of the nucleation stress peak that represents the energy required to nucleate and propagate the first phase. This peak strictly depends on the entity of the interfacial energy, as well as decreases for increasing values of temperature and discreteness of the system. In the last part of the Chapter, I compared the theoretical analytical results with MD simulation performed on SMA nanowires, where the effect of interfacial energy terms compared with entropic effects is crucial. I refer to two specific cases, one considering the effective energy of the phases (\cite{lao:2013}), whereas the second one is associated with the formation of a twin (\cite{zhang:2008}). I compared the results with these simulations remarking that all the parameters required to match the simulation results are derived from the data presented in the described papers or with an energetic scaling of the energy terms (see Section~\ref{sec:ch4_exp}), thus confirming the validity of the model and a very good agreement with the MD results. When the softening is introduced and the non-local terms are considered, it is possible to describe the fundamental dependence of the Maxwell stress on the temperature (see Figure~\ref{fig:ch4_maxT}). This feature has been recently addressed in a paper in which I am a co-author (\cite{luca2022:1}), but it is part of my thesis.

In the second part of the thesis, I tackled the problem of decohesion and fracture by introducing a model that can be schematized with a lattice of two elastic parallel units held together by breakable cross-links springs. In particular, this framework introduced in Chapter~\ref{ch_5} has been used to describe biological systems such as the double-stranded DNA helix or the microtubule and tau-proteins bundle equilibrium conditions. In this case, it is possible to analyze the mechanical response of the system and deduce the effect of the interaction among strands and ligands, which is typical of non-covalent bonds in biological systems. In particular, I found that the interaction strength of the bond is regulated by the relative stiffness of the two units and, depending on the specific properties of the system, the debonding or decohesion phenomena can happen abruptly or with a sub sequential detachment of the units (\cite{ferrer:2008, buehler:2008}). This is a well-known phenomenon observed in nature, and it is also regulated by the rate of loading. Indeed, the second goal is to introduce rate effects to mimic the rate-dependent behaviour of the brain tissue at the level of the axons, where the bond strength of tau proteins and microtubules depends on the external loads and the degradation within the time of such tissues can be responsible for diseases such as Parkins or Alzheimer (\acs{TBI}). Thus, as a first step, I introduced this new model and applied it to the prototypical example of the equilibrium mechanical behaviour of double-stranded DNA helices. I obtained also in this case fully analytical results. Specifically, I studied the system under the two hypotheses of the Maxwell and the maximum delay convention (see Figures~\ref{fig:ch5_global} and~\ref{fig:ch5_local}). In the first case, the system always follows the global minima solution, and the rupture is always abrupt or `fragile', independently on both the relative stiffness of the system (see Equation~\ref{eq:ch5_nu}) and its size. In the other case, the system is allowed to stay at a local minimum until it is locally stable, and a sequential rupture that spreads from the external region to the internal one is observed. Specifically, in this case, the `process' zone is characterized by a force plateau and it may be short or long depending on the specific properties of the system. Worth of importance is the continuum limit (Figure~\ref{fig:ch5_lc}), \textit{i.e} the limit in which $n\to\infty$ and $l\to0$, with the total overall length of the system $L=nl$ fixed. Simpler analytical formulas to handle are derived, and the mechanical properties such as the force plateau, the process zone and the rupture thresholds can be better understood. In particular, two limit regimes are observed for the stiffness parameter $\mu$ given by~\eqref{eq:ch5_mumu}: when $\mu\to\infty$ it means that the breakable units are much stiffer than the elastic backbones and the process of decohesion happens with a ductile type of rupture. In the opposite regime (\textit{i.e} when $\mu\to0$), the base pairs are much softer than the polyphosphate strands and the rupture is always of the fragile type. While developing our model, we found a paper from Pierre Gille De Gennes (\cite{degennes:2001}) where, with a simple ladder model, he described the behaviour of the `critical' force of a dsDNA helix with respect to his length, in terms of base pairs. I realize that our model is formally the same of De Gennes, but whereas he only studies the continuum limit to unveil how the rupture force depends on the length of the DNA helix, we studied the system both from the discrete and continuous point of view highlighting the fundamental transition from a fragile to a ductile type of rupture. Then, I compare his seminal results with our equations and we recover in terms of physical quantities the rupture force~\eqref{eq:ch5_fcnostra}, the critical force o a single pairs~\eqref{eq:ch5_chinostra} and the `persistence length'~\eqref{eq:ch5_fcnostra}, also known as `shear lag' in fracture mechanics, in terms of stiffnesses of the two elements, length and rupture displacement. Moreover, we use our theoretical results to fit the experimental data of Hatch and coworkers (\cite{hatch:2008,prakash:2011}), performed on double-stranded DNA helices with $3'3'$ and $5'5'$ ends, that the authors described with a modified version of the de Gennes' formula, and by simply fitting the data, without any link to the true physical quantities (see Equation~\eqref{eq:ch5_fcnostra}) and the experimental data (see Figure~\ref{fig:ch5_experiment}). 

Eventually, in Chapter~\ref{ch_6} of the thesis, I discussed about rate effects with the specific aim of modelling the bundle of tau proteins and microtubules, that can behave mainly within three loading regimes (\cite{kuhl:2015}). If both the strain and strain rate are low, the microtubules slide on each other and the tau proteins are able to detach and reattach in other positions, thus keeping the structural and biological functions intact. When the strain increases, it is possible that the tau proteins are no longer capable of reattaching and some functionality may be lost. From a clinical point of view, this regime is recognized with the presence of tau agglomerates within the axons, even though the mechanism underlined this feature is still poorly understood (\cite{smith:2012,shi:2021}). Finally, when both the strain and the strain rate are high, the microtubule fails abruptly and the force transmitted drops to zero, usually causing irreversible damage. To address these features, we consider a rate dependent model based on Kramer's rate formula, where the possibility of overcoming energy barriers is given by the interplay between temperature and the height of the barriers themselves. In particular, we are able to compute the value of the energy barriers assuming that the transition threshold separating the elastic and the broken phases is attained when the $p$-th single unit fails. Then, it is possible to deduce the probability of being in a certain attached or detached configuration and the resulting effect on the force-displacement curve is investigated. I remark that in this first approximation, the possibility that the cross-links reattach to existing vacant sites is neglected, thus the breaking is irreversible. We observe that as the rate of the applied force $v_f$ is increased, the system breaks at a high force threshold but with a fragile rupture and few or no elements are detached, as shown in Figure~\ref{fig:ch6_rate}. This behaviour is in agreement with the experimental observation of typical viscoelastic materials. On the other hand, when the rate is low a sequential detachment of the cross-links is observed, leaving the system evolving within a wider range of locally stable configurations resulting in a lower critical force and a ductile fracture. This is also in agreement with other results, where the analysis has been performed with continuum approaches (\cite{ahmadzadeh:2015}) or numerical FEM methods (\cite{kuhl:2018}). I also deduced the dependence of temperature at a fixed rate (see Equation~\ref{eq:ch6_kramer}). As temperature decreases, the zero-temperature limit, \textit{i.e.} the equilibrium mechanics described in Section~\ref{sec:ch5_strategie}, is approached and the system is frozen within the fully attached configuration. Indeed, in Figure~\ref{fig:ch6_rateTemp} one observes that for high values of $\beta$ the force threshold is reached before any other configuration with $p<n$ is explored, and a fragile rupture is attained. On the other hand, when the temperature decreases and more energy is provided, the ability to overcome energy barriers increases and different local configurations are explored. The rupture becomes then more ductile whereas the force threshold decreases.

%
\section{Future work and perspectives}

The possibility of describing systems with microinstabiliteis with simple mathematical models able to capture the main physical features allows the possibility of studying a wide range of biological and bio-inspired materials, and many challenges are still open. In this thesis, I have shown that phenomena such as the influence of the device stiffness, the interplay between entropic and interfacial energy terms or the decohesion and fracture in double-stranded polypeptide chains require a correct description at the microscopic level. Indeed, it is always possible to identify basic units characterized by a multiphase energy function that can be assembled to mimic the behaviour of a more complex system. Based on this idea and on the work described in this thesis, many developments of the models and systems introduced can be performed. 

The rate-dependent response of microtubules and tau protein bundles of Chapter~\ref{ch_6} has to be thorough considering also the hard device problem, by assigning the total displacement at the double-stranded chain. Moreover, due to the complexity of the rate theories and based on their specific application (see (\cite{hanggi:1990}) and reference therein), different considerations and approaches can be considered in this specific example. In particular, the choice of a Kramers' based theory relies on the idea of obtaining analytical results, which I am confident are possible to be obtained especially in some limit regime (see Section~\ref{sec:ch5_mulimit}), and it is the aim of a work in progress (\cite{luca2022:3}). Eventually, following for example the method introduced in (\cite{detommasi:2015}), the final aim is to deduce a multiscale model for the behaviour of the brain tissue considering phenomena arising from the microscopic description of such a system as residual damage, rate effects and the fragile-ductile rupture transition (\cite{garcia:2017,recho:2016,thompson:2020,kuhl:2016,kuhl:2018,kuhl:2018:2}). 

On the other hand, when rate effects are not considered, it is possible to address the behaviour of double-stranded polypeptide chains, which are a life-key component in the majority of living systems. On this topic, our aim is to study, based on the model introduced in Chapter~\ref{ch_5}, the cooperativity phenomenon typical of these molecules (\cite{luca2022:2}), which is strongly affected by thermal and mechanical loads. Moreover, in the case of the axons, the model can be also improved by considering phenomena such as the possibility that the breakable links may reattach to vacant sites, reproducing the `molecular motors' responsible for the overall stability of the axons (\cite{kuhl:2018}). Another area of exploration is the possibility of considering different loading directions of the applied force or displacement, that can be used to describe, for instance, the peeling phenomenon. 

The possibility of introducing the handling device to describe the effect of the experimental set-up unveiled in Chapter~\ref{ch_3} can be also easily extended to the other model I introduced in the thesis. In particular, an interesting area of application is the description of the effect of different pulling speeds on the mechanical response of molecules, recently made possible by the advent of \acf{HS-AFM} techniques. 

Also, the softening behaviour that is observed in the phase transition of many biological systems and materials (see Chapter~\ref{ch_4}) has been addressed in a recent work in which I am a co-author (\cite{luca2022:1}). There, the effect of the temperature when non-local terms and different stiffnesses for the two wells are considered has been studied and the main outcome of the work is the dependence of the Maxwell stress from the temperature, a phenomenon that is also experimentally observed~\ref{fig:ch4_maxT}. 

To conclude, in this thesis I proposed different models describing the response of a wide range of biological phenomena, introducing the microscopic description of the multi wells energy landscape characterizing these systems and considering the different effects of temperature, rate and mechanical loading. My aim is twofold. I focused on obtaining analytical results to unveil the main features and characteristics of such systems, analyzing different boundary conditions, evolution strategies and limit regimes. On the other hand, based on these results and on the hierarchical multiscale structure typical of biological systems, it can be possible to derive multiscale models considering arising phenomena from the microscopic thermo-mechanical behaviour.

	\clearpage
	



\nocite{*}
\printbibliography

@article{puglisi:2000,
  author = {Puglisi, G. and Truskinovsky, L.},
  title = {Mechanics of a discrete chain with bi-stable elements},
  journal = {Journal of the Mechanics and Physics of Solids},
  year = {2000},
  volume = {48},
  pages = {1-27},
  number = {1},
  doi = {10.1016/S0022-5096(99)00006-X},
}

@article{puglisi:2005,
  title={Thermodynamics of rate-independent plasticity},
  author={Puglisi, G. and Truskinovsky, L.},
  journal={Journal of the Mechanics and Physics of Solids},
  volume={53},
  number={3},
  pages={655--679},
  year={2005},
  publisher={Elsevier},
  doi={10.1016/j.jmps.2004.08.004}
}

@article{florio:2020,
  title={Role of temperature in the decohesion of an elastic chain tethered to a substrate by onsite breakable links},
  author={Florio, G. and Puglisi, G. and Giordano, S.},
  journal={Physical Review Research},
  volume={2},
  number={3},
  pages={033227},
  year={2020},
  publisher={APS},
  doi={10.1103/PhysRevResearch.2.033227}
}

@article{fp:2019,
  author = {Florio, G. and Puglisi, G.},
  title = {Unveiling the influence of device stiffness in single macromolecule unfolding},
  journal = {Scientific Reports},
  year = {2019},
  volume = {9},
  pages = {4997},
  doi = {10.1038/s41598-019-41330-x},
}

@article{puglisi:2013,
  title={Cohesion-decohesion asymmetry in geckos},
  author={Puglisi, G. and Truskinovsky, L.},
  journal={Physical Review E},
  volume={87},
  number={3},
  pages={032714},
  year={2013},
  publisher={APS},
  doi={10.1103/PhysRevE.87.032714}
}

@article{puglisi:2006,
   author = {Puglisi, G.},
   title = {Hysteresis in multi-stable lattices with non-local interactions},
   journal = {Journal of Mechanics and Physics of Solids},
   year = {2006},
   volume = {54},
   number = {10},
   pages = {2060--2088},
   publisher = {Elsevier},
   doi = {10.1016/j.jmps.2006.04.006}
}

@article{puglisi:2007,
   author = {Puglisi, G.},
   title = {Nucleation and phase propagation in a multistable lattice with weak nonlocal interactions},
   journal = {Continuum Mechanics and Thermodynamics},
   year = {2007},
   volume = {19},
   number = {5},
   pages = {299--319},
   publisher = {Springer},
   doi = {10.1007/s00161-007-0056-7}
}

@article{detom:2013,
  author = {De Tommasi, D. and Millardi, N. and Puglisi, G. and Saccomandi, G.},
  title = {An energetic model for macromolecules unfolding in stretching experiments},
  journal = {Journal of the Royal Society Interface},
  year = {2013},
  volume = {10},
  pages = {88},
  doi = {10.1098/rsif.2013.0651},
}

@article{linke:2002,
  title={PEVK domain of titin: an entropic spring with actin-binding properties},
  author={Linke, W. A. and Kulke, M. and Li, H. and Fujita-Becker, S. and Neagoe, C. and Manstein, D. J. and Gautel, M. and Fernandez, J. M.},
  journal={Journal of structural biology},
  volume={137},
  number={1-2},
  pages={194--205},
  year={2002},
  publisher={Elsevier},
  doi={10.1006/jsbi.2002.4468}
}

@article{detommasi:2015,
  title={Multiscale mechanics of macromolecular materials with unfolding domains},
  author={De Tommasi, D. and Puglisi, G. and Saccomandi, G.},
  journal={Journal of the Mechanics and Physics of Solids},
  volume={78},
  pages={154--172},
  year={2015},
  publisher={Elsevier},
doi={10.1016/j.jmps.2015.02.002}
}

@article{detommasi:2017,
  title={Micromechanical model for protein materials: From macromolecules to macroscopic fibers},
  author={Puglisi, G. and De Tommasi, D. and Pantano, M. F. and Pugno, N. M. and Saccomandi, G.},
  journal={Physical Review E},
  volume={96},
  number={4},
  pages={042407},
  year={2017},
  publisher={APS},
doi={10.1103/PhysRevE.96.042407}
}

@article{puglisi:2002,
  title={Rate independent hysteresis in a bi-stable chain},
  author={Puglisi, G. and Truskinovsky, L.},
  journal={Journal of the Mechanics and Physics of Solids},
  volume={50},
  number={2},
  pages={165--187},
  year={2002},
  publisher={Elsevier},
  doi={10.1016/S0022-5096(01)00055-2}
}

@book{aristotele,
   title={Historia animalium},
   author={Aristotele},
   year={343 B.C.},
   }

@article{autumn:2000,
  title={Adhesive force of a single gecko foot-hair},
  author={Autumn, K. and Liang, Y. A. and Hsieh, S. T. and Zesch, W. and Chan, W. P. and Kenny, T. W. and Fearing, R. and Full, R. J.},
  journal={Nature},
  volume={405},
  number={6787},
  pages={681--685},
  year={2000},
  publisher={Nature Publishing Group},
  doi={10.1038/35015073}
}

@article{feynman:1959,
   author={Feynman, R. P.},
   title={There's plenty of room at the bottom},
   journal={IEEE J. MEMS},
   volume={1},
   pages={60-66},
   year={1959},
   doi={10.1007/s12045-011-0109-x}
  }

@article{aliu:2018,
  title={Amino acid disorders},
  author={Aliu, E. and Kanungo, S. and Arnold, G. L.},
  journal={Annals of translational medicine},
  volume={6},
  number={24},
  year={2018},
  publisher={AME Publications},
  doi={10.21037/atm.2018.12.12}
}

@article{burnham:2014,
  title={Skewed brownian fluctuations in single-molecule magnetic tweezers},
  author={Burnham, D. R. and De Vlaminck, I. and Henighan, T. and Dekker, C.},
  journal={PloS one},
  volume={9},
  number={9},
  pages={e108271},
  year={2014},
  publisher={Public Library of Science San Francisco, USA},
  doi={10.1371/journal.pone.0108271}
}

@article{rees:2010,
  title={Sickle-cell disease},
  author={Rees, D. C. and Williams, T. N. and Gladwin, M. T.},
  journal={The Lancet},
  volume={376},
  number={9757},
  pages={2018--2031},
  year={2010},
  publisher={Elsevier},
  doi={10.1016/S0140-6736(10)61029-X}
}

@article{pauling:1951,
  title={The structure of proteins: two hydrogen-bonded helical configuration of the polypeptide chain},
  author={Pauling, L. and Corey, R. B. and Branson, H. T.},
  journal={Proceedings of the National Academy of Science PNAS},
  volume={37},
  number={4},
  pages={205--211},
  year={1951},
  doi={10.1073/pnas.37.4.205}
}

@article{edison:2001,
  title={Linus Pauling and the planar peptide bond},
  author={Edison, A. S.},
  journal={Nature structural biology},
  volume={8},
  number={3},
  pages={201--202},
  year={2001},
  publisher={Nature Publishing Group},
  doi={10.1038/84921}
}

@article{baldwin:2013,
  title={The new view of hydrophobic free energy},
  author={Baldwin, R. L.},
  journal={FEBS letters},
  volume={587},
  number={8},
  pages={1062--1066},
  year={2013},
  publisher={Elsevier},
  doi={10.1016/j.febslet.2013.01.006}
}

@article{kendrew:1958,
  title={A three-dimensional model of the myoglobin molecule obtained by x-ray analysis},
  author={Kendrew, J. C. and Bodo, G. and Dintzis, H. M. and Parrish, R. G. and Wyckoff, H. and Phillips, D. C.},
  journal={Nature},
  volume={181},
  number={4610},
  pages={662--666},
  year={1958},
  publisher={Nature Publishing Group},
  doi={10.1038/181662a0}
}

@article{baldock:2011,
  title={Shape of tropoelastin, the highly extensible protein that controls human tissue elasticity},
  author={Baldock, C. and Oberhauser, A. F. and Ma, L. and Lammie, D. and Siegler, V. and Mithieux, S. M. and Tu, Y. and Chow, J. Y. H. and Suleman, F. and Malfois, M. and others},
  journal={Proceedings of the National Academy of Sciences},
  volume={108},
  number={11},
  pages={4322--4327},
  year={2011},
  publisher={National Acad Sciences},
  doi={10.1073/pnas.1014280108}
}

@article{yeo:2016,
  title={Subtle balance of tropoelastin molecular shape and flexibility regulates dynamics and hierarchical assembly},
  author={Yeo, G. C. and Tarakanova, A. and Baldock, C. and Wise, S. G. and Buehler, M. J. and Weiss, A. S.},
  journal={Science advances},
  volume={2},
  number={2},
  pages={e1501145},
  year={2016},
  publisher={American Association for the Advancement of Science},
  doi={10.1126/sciadv.1501145}
}

@article{bustamante:2000,
  author = {Bustamante, C. and Macosko, J. C. and Wuite, G. J. L.},
  title = {Grabbing the cat by the tail: manipulating molecules one by one},
  journal = {{N}ature Reviews Molecular Cell Biology},
  year = {2000},
  volume = {1},
  pages = {130--136},
  doi = {10.1038/35040072 },
}

@article{neuman:2008,
  author = {Neuman, K. C.  and Nagy, A.},
  title = {Single-molecule force spectroscopy: optical tweezers, magnetic tweezers and Atomic Force Microscopy},
  journal = {Nature Methods},
  year = {2008},
  volume = {5},
  pages = {491--505},
  doi = {10.1038/nmeth.1218},
}

@article{dutta:2016,
  author = {Dutta, S. and Armitage, B. A. and Lyubchenko, Y. L.},
  title = {Probing of miniPEG$\gamma$-PNA-DNA hybrid duplex stability with AFM force spectroscopy},
  journal = {Biochemistry},
  year = {2016},
  volume = {55},
  pages = {1523--1528},
  doi = {10.1021/acs.biochem.5b01250},
}

@article{hughes:2016,
  author = {Hughes, M. L. and Dougan, L.},
  title = {The physics of pulling polyproteins: a review of single molecule force spectroscopy using the AFM to study protein unfolding},
  journal = {Reports on Progress in Physics},
  year = {2016},
  volume = {79},
  pages = {--076601},
  doi = {10.1088/0034-4885/79/7/076601},
}

@article{woodside:2006,
  author = {Woodside, M. T.  and Anthony, P. C. and Behnke-Parks, W. M.  and Larizadeh, K. and Herschlag, D. and Block, S. M.},
  title = {Direct measurement of the full, sequence-dependent folding landscape of a nucleic acid},
  journal = {Science},
  year = {2006},
  volume = {314},
  pages = {1001--1004},
  doi = {10.1126/science.1133601},
}

@article{binnig:1986,
  title={Atomic force microscope},
  author={Binnig, G. and Quate, C. F. and Gerber, C.},
  journal={Physical review letters},
  volume={56},
  number={9},
  pages={930},
  year={1986},
  publisher={APS},
  doi={10.1103/PhysRevLett.56.930}
}

@article{garcia:2002,
  title={Dynamic atomic force microscopy methods},
  author={Garc{\i}a, R. and Perez, R.},
  journal={Surface science reports},
  volume={47},
  number={6-8},
  pages={197--301},
  year={2002},
  publisher={Elsevier},
  doi={10.1016/S0167-5729(02)00077-8}
}

@article{francis:2010,
  title={Atomic force microscopy comes of age},
  author={Francis, L. W. and Lewis, P. D. and Wright, C. J. and Conlan, R. S.},
  journal={Biology of the Cell},
  volume={102},
  number={2},
  pages={133--143},
  year={2010},
  publisher={Wiley Online Library},
  doi={10.1042/BC20090127}
}

@article{ashkin:1986,
  title={Observation of a single-beam gradient force optical trap for dielectric particles},
  author={Ashkin, A. and Dziedzic, J. M. and Bjorkholm, J. E. and Chu, S.},
  journal={Optics letters},
  volume={11},
  number={5},
  pages={288--290},
  year={1986},
  publisher={Optical Society of America},
  doi={10.1364/OL.11.000288}
}

@article{strick:1996,
  title={The elasticity of a single supercoiled DNA molecule},
  author={Strick, T. R. and Allemand, J. F. and Bensimon, D. and Bensimon, A. and Croquette, V.},
  journal={Science},
  volume={271},
  number={5257},
  pages={1835--1837},
  year={1996},
  publisher={American Association for the Advancement of Science},
  doi={10.1126/science.271.5257.1835}
}

@article{szabo:2005,
  author = {Hummer, G. and Szabo, A.},
  title = {Free energy surfaces from Single Molecule Force Spectroscopy},
  journal = {Accounts of chemical research},
  year = {2005},
  volume = {38},
  pages = {504--516},
  doi = {10.1021/ar040148d},
}

@article{walder:2017,
  author = {Walder, R. and Van Patten, W. J. and Adhikari, A. and Perkins, T. T.},
  title = {Going vertical to improve the accuracy of Atomic Force Microscopy based Single-Molecule Force Spectroscopy.},
  journal = {ACS Nano},
  year = {2017},
  volume = {12},
  pages = {198--207},
  doi = {10.1021/acsnano.7b05721},
}

@article{benichou:2016,
  author = {Benichou, I. and Zhang, Y. and Dudko, O. K. and Givli, S.},
  title = {The rate dependent response of a bistable chain at finite temperature},
  journal = {Journal of the Mechanics and Physics of Solids},
  year = {2016},
  volume = {95},
  pages = {44--63},
  doi = {10.1016/j.jmps.2016.05.001},
}

@article{cossio:2015,
  author = {Cossio, P. and Hummer, G. and Szabo, A.},
  title = {On artifacts in Single-Molecule Force Spectroscopy},
  journal = {Proceedings of the National Academy of Science (PNAS)},
  year = {2015},
  volume = {112},
  pages = {14248--14253},
  doi = {10.1073/pnas.1519633112/-/DCSupplemental.},
}

@article{maitra:2010,
  author = {Maitra, A. and Arya, G.},
  title = {Model accounting for the effects of pulling-device stiffness in the analyses of Single-Molecule Force measurements},
  journal = {Physical Review Letters},
  year = {2010},
  volume = {104},
  pages = {108301},
  doi = {10.1103/PhysRevLett.104.108301},
}

@article{biswas:2018,
  author = {Biswas, S. and Leitao, S. and Theillaud, Q. and Erickson B. W. and Fantner, G. E.},
  title = {Reducing uncertainties in energy dissipation measurements in atomic force spectroscopy of molecular networks and cell-adhesion studies},
  journal = {Scientific Reports},
  year = {2018},
  volume = {8},
  pages = {9390},
  doi = {10.1038/s41598-018-26979-0},
}

@article{dudko:2008,
  author = {Dudko, O. K. and Hummer, G. and Szabo, A.},
  title = {Theory, analysis, and interpretation of single-molecule force spectroscopy experiments},
  journal = {Proceedings of the National Academy of Sciences (PNAS)},
  year = {2008},
  volume = {105},
  pages = {15755--15760},
  doi = {10.1073/pnas.0806085105},
}

@article{li:2014,
  author = {Dechang, L. and Baohua, J.},
  title = {Predicted rupture force of a single molecular bond becomes rate independent at ultralow loading rates},
  journal = {Physical Review Letters},
  year = {2014},
  volume = {112},
  pages = {078302},
  doi = {10.1103/PhysRevLett.112.078302},
}

@article{roberts:2002,
  title={Molecular biology of the cell},
  author={Roberts, K. and Alberts, B. and Johnson, A. and Walter, P. and Hunt, T.},
  journal={New York: Garland Science},
  volume={32},
  number={2},
  year={2002}
}

@article{herrmann:2004,
  title={Intermediate filaments: molecular structure, assembly mechanism, and integration into functionally distinct intracellular scaffolds},
  author={Herrmann, H. and Aebi, U.},
  journal={Annual review of biochemistry},
  volume={73},
  number={1},
  pages={749--789},
  year={2004},
  publisher={Annual Reviews 4139 El Camino Way, PO Box 10139, Palo Alto, CA 94303-0139, USA},
  doi={10.1146/annurev.biochem.73.011303.073823}
}

@article{kreplak:2007,
  title={Biomechanical properties of intermediate filaments: from tissues to single filaments and back},
  author={Kreplak, L. and Fudge, D.},
  journal={Bioessays},
  volume={29},
  number={1},
  pages={26--35},
  year={2007},
  publisher={Wiley Online Library},
  doi={10.1002/bies.20514}
}

@article{fratzl:2007,
  title={Nature’s hierarchical materials},
  author={Fratzl, P. and Weinkamer, R.},
  journal={Progress in materials Science},
  volume={52},
  number={8},
  pages={1263--1334},
  year={2007},
  publisher={Elsevier},
  doi={10.1016/j.pmatsci.2007.06.001}
}

@article{gelse:2003,
  title={Collagens—structure, function, and biosynthesis},
  author={Gelse, K. and P{\"o}schl, E. and Aigner, T.},
  journal={Advanced drug delivery reviews},
  volume={55},
  number={12},
  pages={1531--1546},
  year={2003},
  publisher={Elsevier},
  doi={10.1016/j.addr.2003.08.002}
}

@article{buehler:2009,
  title={Deformation and failure of protein materials in physiologically extreme conditions and disease},
  author={Buehler, M. J. and Yung, Y. C.},
  journal={Nature materials},
  volume={8},
  number={3},
  pages={175--188},
  year={2009},
  publisher={Nature Publishing Group},
  doi={10.1038/nmat2387}
}

@article{gao:2003,
  title={Materials become insensitive to flaws at nanoscale: lessons from nature},
  author={Gao, H. and Ji, B. and J{\"a}ger, I. L. and Arzt, E. and Fratzl, P.},
  journal={Proceedings of the national Academy of Sciences},
  volume={100},
  number={10},
  pages={5597--5600},
  year={2003},
  publisher={National Acad Sciences},
  doi={10.1073/pnas.0631609100}
}

@article{kitano:2002,
  title={Computational systems biology},
  author={Kitano, H.},
  journal={Nature},
  volume={420},
  number={6912},
  pages={206--210},
  year={2002},
  publisher={Nature Publishing Group},
  doi={10.1038/nature01254}
}

@article{taylor:2007,
  title={Living with cracks: damage and repair in human bone},
  author={Taylor, D. and Hazenberg, J. G. and Lee, T. C.},
  journal={Nature materials},
  volume={6},
  number={4},
  pages={263--268},
  year={2007},
  publisher={Nature Publishing Group},
  doi={10.1038/nmat1866}
}

@article{rammensee:2008,
  title={Assembly mechanism of recombinant spider silk proteins},
  author={Rammensee, S. and Slotta, U. and Scheibel, T. and Bausch, A. R.},
  journal={Proceedings of the National Academy of Sciences},
  volume={105},
  number={18},
  pages={6590--6595},
  year={2008},
  publisher={National Acad Sciences},
  doi={10.1073/pnas.0709246105}
}

@article{bini:2004,
  title={Mapping domain structures in silks from insects and spiders related to protein assembly},
  author={Bini, E. and Knight, D. P. and Kaplan, D. L.},
  journal={Journal of molecular biology},
  volume={335},
  number={1},
  pages={27--40},
  year={2004},
  publisher={Elsevier},
  doi={10.1016/j.jmb.2003.10.043}
}

@article{langer:2004,
  title={Designing materials for biology and medicine},
  author={Langer, R. and Tirrell, D. A.},
  journal={Nature},
  volume={428},
  number={6982},
  pages={487--492},
  year={2004},
  publisher={Nature Publishing Group},
  doi={10.1038/nature02388}
}

@article{hest:2001,
  title={Protein-based materials, toward a new level of structural control},
  author={van Hest, J. CM and Tirrell, D. A.},
  journal={Chemical communications},
  number={19},
  pages={1897--1904},
  year={2001},
  publisher={Royal Society of Chemistry},
  doi={10.1039/B105185G}
}

@article{mandal:2014,
  title={Self-assembly of peptides to nanostructures},
  author={Mandal, D. and Shirazi, A. N. and Parang, K.},
  journal={Organic \& biomolecular chemistry},
  volume={12},
  number={22},
  pages={3544--3561},
  year={2014},
  publisher={Royal Society of Chemistry},
  doi={10.1039/C4OB00447G}
}

@article{alon:2007,
  title={Simplicity in biology},
  author={Alon, U.},
  journal={Nature},
  volume={446},
  number={7135},
  pages={497--497},
  year={2007},
  publisher={Nature Publishing Group},
  doi={10.1038/446497a}
}

@article{buehler:2008,
  title={Theoretical and computational hierarchical nanomechanics of protein materials: Deformation and fracture},
  author={Buehler, M. J. and Keten, S. and Ackbarow, T.},
  journal={Progress in Materials Science},
  volume={53},
  number={8},
  pages={1101--1241},
  year={2008},
  publisher={Elsevier},
  doi={10.1016/j.pmatsci.2008.06.002}
}

@article{ferrer:2008,
  title={Measuring molecular rupture forces between single actin filaments and actin-binding proteins},
  author={Ferrer, J. M. and Lee, H. and Chen, J. and Pelz, B. and Nakamura, F. and Kamm, R. D. and Lang, M. J.},
  journal={Proceedings of the National Academy of Sciences},
  volume={105},
  number={27},
  pages={9221--9226},
  year={2008},
  publisher={National Acad Sciences},
  doi={10.1073/pnas.0706124105}
}

@article{ackbarow:2007,
  title={Hierarchies, multiple energy barriers, and robustness govern the fracture mechanics of $\alpha$-helical and $\beta$-sheet protein domains},
  author={Ackbarow, T. and Chen, X. and Keten, S. and Buehler, M. J.},
  journal={Proceedings of the National Academy of Sciences},
  volume={104},
  number={42},
  pages={16410--16415},
  year={2007},
  publisher={National Acad Sciences},
  doi={10.1073/pnas.0705759104}
}

@article{bell:1978,
  title={Models for the specific adhesion of cells to cells},
  author={Bell, G. I.},
  journal={Science},
  volume={200},
  number={4342},
  pages={618--627},
  year={1978},
  publisher={American Association for the Advancement of Science},
  doi={10.1126/science.347575}
}

@article{kramers:1940,
  title={Brownian motion in a field of force and the diffusion model of chemical reactions},
  author={Kramers, H. A.},
  journal={Physica},
  volume={7},
  number={4},
  pages={284--304},
  year={1940},
  publisher={Elsevier},
  doi={10.1016/S0031-8914(40)90098-2}
}

@article{hanggi:1990,
  title={Reaction-rate theory: fifty years after Kramers},
  author={H{\"a}nggi, P. and Talkner, P. and Borkovec, M.},
  journal={Reviews of modern physics},
  volume={62},
  number={2},
  pages={251},
  year={1990},
  publisher={APS},
  doi={10.1103/RevModPhys.62.251}
}

@article{zhurkov:1965,
  title={Kinetic concept of the strength of solids},
  author={Zhurkov, S. N.},
  journal={International Journal of Fracture Mechanics},
  volume={1},
  number={4},
  pages={311--323},
  year={1965},
  publisher={Springer},
  doi={10.1007/BF03545562}
}

@article{evans:1997,
  title={Dynamic strength of molecular adhesion bonds},
  author={Evans, E. and Ritchie, K.},
  journal={Biophysical journal},
  volume={72},
  number={4},
  pages={1541--1555},
  year={1997},
  publisher={Elsevier},
  doi={10.1016/S0006-3495(97)78802-7}
}

@article{evans:2001,
  title={Probing the relation between force—lifetime—and chemistry in single molecular bonds},
  author={Evans, E.},
  journal={Annual review of biophysics and biomolecular structure},
  volume={30},
  number={1},
  pages={105--128},
  year={2001},
  publisher={Annual Reviews 4139 El Camino Way, PO Box 10139, Palo Alto, CA 94303-0139, USA},
  doi={10.1146/annurev.biophys.30.1.105}
}

@article{hummer:2001,
  title={Free energy reconstruction from nonequilibrium single-molecule pulling experiments},
  author={Hummer, G. and Szabo, A.},
  journal={Proceedings of the National Academy of Sciences},
  volume={98},
  number={7},
  pages={3658--3661},
  year={2001},
  publisher={National Acad Sciences},
  doi={10.1073/pnas.071034098}
}

@article{hummer:2003,
  title={Kinetics from nonequilibrium single-molecule pulling experiments},
  author={Hummer, G. and Szabo, A.},
  journal={Biophysical journal},
  volume={85},
  number={1},
  pages={5--15},
  year={2003},
  publisher={Elsevier},
  doi={/10.1016/S0006-3495(03)74449-X}
}

@article{dudko:2006,
  title={Intrinsic rates and activation free energies from single-molecule pulling experiments},
  author={Dudko, O. K. and Hummer, G. and Szabo, A.},
  journal={Physical review letters},
  volume={96},
  number={10},
  pages={108101},
  year={2006},
  publisher={APS},
  doi={10.1103/PhysRevLett.96.108101}
}

@article{keten:2008,
  title={Strength limit of entropic elasticity in beta-sheet protein domains},
  author={Keten, S. and Buehler, M. J.},
  journal={Physical Review E},
  volume={78},
  number={6},
  pages={061913},
  year={2008},
  publisher={APS},
  doi={10.1103/PhysRevE.78.061913}
}

@article{rief:1998,
  title={Elastically coupled two-level systems as a model for biopolymer extensibility},
  author={Rief, M. and Fernandez, J. M. and Gaub, H. E.},
  journal={Physical Review Letters},
  volume={81},
  number={21},
  pages={4764},
  year={1998},
  publisher={APS},
  doi={10.1103/PhysRevLett.81.4764}
}

@book{courtney:1990,
  title={Mechanical behavior of materials},
  author={Courtney, T. H.},
  year={1990},
  publisher={Waveland Press},
  isbn={978-9332584785}
}

@book{timoshenko:1983,
  title={History of strength of materials: with a brief account of the history of theory of elasticity and theory of structures},
  author={Timoshenko, S.},
  year={1983},
  publisher={Courier Corporation},
  isbn={ 978-0486611877}
}

@book{lemaitre:1994,
  title={Mechanics of solid materials},
  author={Lemaitre, J. and Chaboche, J. L.},
  year={1994},
  publisher={Cambridge university press},
  isbn={978-0521328531}
}

@article{rief:1997,
  author = {Rief, M. and Gautel, M. and Oesterhelt, F. and Fernandez, J. M. and Gaub, H. E.},
  title = {Reversible unfolding of individual titin immunoglobulin domains by AFM},
  journal = {Science},
  year = {1997},
  volume = {276},
  pages = {1109--1112},
  doi = {10.1126/science.276.5315.1109},
}

@article{fantner:2005,
  title={Sacrificial bonds and hidden length dissipate energy as mineralized fibrils separate during bone fracture},
  author={Fantner, G. E. and Hassenkam, T. and Kindt, J. H. and Weaver, J. C. and Birkedal, H. and Pechenik, L. and Cutroni, J. A. and Cidade, G. and Stucky, G. D. and Morse, D. E. and others},
  journal={Nature materials},
  volume={4},
  number={8},
  pages={612--616},
  year={2005},
  publisher={Nature Publishing Group},
  doi={10.1038/nmat1428}
}

@book{lubliner:2008,
  title={Plasticity theory},
  author={Lubliner, J.},
  year={2008},
  publisher={Courier Corporation},
  isbn={978-0486462905}
}

@article{buehler:2007,
  title={Fracture mechanics of protein materials},
  author={Buehler, M. J. and Ackbarow, T.},
  journal={Materials Today},
  volume={10},
  number={9},
  pages={46--58},
  year={2007},
  publisher={Elsevier},
  doi={10.1016/S1369-7021(07)70208-0}
}

@book{weiner:1983,
   title={Statistical Mechanics of Elasticity},
   author={Weiner, J. H.},
   volume={},
   year={1983},
   publisher={Dover Publications},
   doi={10/0486422607}
}

@article{marko:1995,
  title={Stretching DNA},
  author={Marko, J. F. and Siggia, E. D.},
  journal={Macromolecules},
  volume={28},
  number={26},
  pages={8759--8770},
  year={1995},
  publisher={ACS Publications},
  doi={10.1021/ma00130a008}
}

@article{griffith:1921,
  title={VI. The phenomena of rupture and flow in solids},
  author={Griffith, A. A.},
  journal={Philosophical transactions of the royal society of london. Series A, containing papers of a mathematical or physical character},
  volume={221},
  number={582-593},
  pages={163--198},
  year={1921},
  publisher={The royal society London},
  doi={10.1098/rsta.1921.0006}
}

@article{irwin:1957,
  title={Analysis of stresses and strains near the end of a crack traversing a plate},
  author={Irwin, G. R.},
  year={1957},
  doi={10.1115/1.4011547}
}

@article{williams:1957,
  title={Some observations of Leonardo, Galileo, Mariotte and others relative to size effect},
  author={Williams, E.},
  journal={Annals of Science},
  volume={13},
  number={1},
  pages={23--29},
  year={1957},
  publisher={Taylor \& Francis},
  doi={10.1080/00033795700200031}
}

@article{fraser:2007,
  title={Nuclear organization of the genome and the potential for gene regulation},
  author={Fraser, P. and Bickmore, W.},
  journal={Nature},
  volume={447},
  number={7143},
  pages={413--417},
  year={2007},
  publisher={Nature Publishing Group},
  doi={10.1038/nature05916}
}

@article{rounsevell:2004,
  title={Atomic force microscopy: mechanical unfolding of proteins},
  author={Rounsevell, R. and Forman, J. R. and Clarke, J.},
  journal={Methods},
  volume={34},
  number={1},
  pages={100--111},
  year={2004},
  publisher={Elsevier},
  doi={10.1016/j.ymeth.2004.03.007}
}

@article{meyers:2013,
  title={Structural biological materials: critical mechanics-materials connections},
  author={Meyers, M. A. and McKittrick, J. and Chen, P. Y.},
  journal={science},
  volume={339},
  number={6121},
  pages={773--779},
  year={2013},
  publisher={American Association for the Advancement of Science},
  doi={10.1126/science.1220854}
}

@article{meyers:2008,
  title={Biological materials: structure and mechanical properties},
  author={Meyers, M. A. and Chen, P.-Y. and Lin, A. Y.-M. and Seki, Y.},
  journal={Progress in materials science},
  volume={53},
  number={1},
  pages={1--206},
  year={2008},
  publisher={Elsevier},
  doi={10.1016/j.pmatsci.2007.05.002}
}

@article{bechtle:2010,
  title={On the mechanical properties of hierarchically structured biological materials},
  author={Bechtle, S. and Ang, S. F. and Schneider, G. A.},
  journal={Biomaterials},
  volume={31},
  number={25},
  pages={6378--6385},
  year={2010},
  publisher={Elsevier},
  doi={10.1016/j.biomaterials.2010.05.044}
}

@article{pineau:2016,
  title={Failure of metals I: Brittle and ductile fracture},
  author={Pineau, A. and Benzerga, A. A. and Pardoen, T.},
  journal={Acta Materialia},
  volume={107},
  pages={424--483},
  year={2016},
  publisher={Elsevier},
  doi={10.1016/j.actamat.2015.12.034}
}

@article{monn:2020,
  title={Lamellar architectures in stiff biomaterials may not always be templates for enhancing toughness in composites},
  author={Monn, M. A. and Vijaykumar, K. and Kochiyama, S. and Kesari, H.},
  journal={Nature communications},
  volume={11},
  number={1},
  pages={1--12},
  year={2020},
  publisher={Nature Publishing Group},
  doi={10.1038/s41467-019-14128-8}
}

@article{gurtin:1989,
   title={Multiphase thermomechanics with interfacial structure 2. {E}volution of an isothermal interface},
   author={Angenent, S. and Gurtin, M. E.},
   journal={Archive for Rational Mechanics and Analysis},
   volume={108},
   number={3},
   pages={323--391},
   year={1989},
   doi={10.1007/BF01041068}
}

@book{khachaturyan:1983,
   title={Theory of structural transformations in solids},
   author={Khachaturyan, A. G.},
   year={1983},
   publisher={John Wiley \& Sons},
   isbn={9780471078739}
}

@article{ericksen:1975,
   title={Equilibrium of bars},
   author={Ericksen, J. L.},
   journal={Journal of elasticity},
   volume={5},
   number={3-4},
   pages={191--201},
   year={1975},
   publisher={Springer},
   doi={10.1007/BF00126984}
}

@incollection{ball:1989,
   title={Fine phase mixtures as minimizers of energy},
   author={Ball, J. M. and James, R. D.},
   booktitle={Analysis and Continuum Mechanics},
   pages={647--686},
   year={1989},
   publisher={Springer},
   doi={10.1007/BF00281246}
}

@incollection{muller:1999,
   title={Variational models for microstructure and phase transitions},
   author={M{\"u}ller, S.},
   booktitle={Calculus of variations and geometric evolution problems},
   pages={85--210},
   year={1999},
   publisher={Springer},
   doi={10.1007/BFb0092670}
}

@article{truskinovsky:2000,
   title={Finite scale microstructures in nonlocal elasticity},
   author={Ren, X. and Truskinovsky, L.},
   journal={Journal of elasticity and the physical science of solids},
   volume={59},
   number={1-3},
   pages={319--355},
   year={2000},
   publisher={Springer},
   doi={10.1023/A:1011003321453}
}

@article{villaggio:1977,
  title={A model for an elastic-plastic body},
  author={M{\"u}ller, I. and Villaggio, P.},
  journal={Archive for Rational Mechanics and Analysis},
  volume={65},
  number={1},
  pages={25--46},
  year={1977},
  publisher={Springer},
  doi={10.1007/BF00289355}
}

@article{zanzotto:1992,
  title={Hysteresis in discrete systems of possibly interacting elements with a double-well energy},
  author={Fedelich, B. and Zanzotto, G.},
  journal={Journal of Nonlinear Science},
  volume={2},
  number={3},
  pages={319--342},
  year={1992},
  publisher={Springer},
  doi={https://doi.org/10.1007/BF01208928}
}

@article{vainchtein:2004,
   author = {Truskinovsky, L. and Vainchtein, A.},
   title = {The origin of nucleation peak in transformation plasticity},
   journal = {Journal of the Mechanics and Physics of Solids},
   year = {2004},
   volume = {52},
   number = {6},
   pages = {1421--1446},
   publisher = {Elsevier},
   doi = {10.1016/j.jmps.2003.09.034}
}

@article{makarov:2009,
  title={A theoretical model for the mechanical unfolding of repeat proteins},
  author={Makarov, D. E.},
  journal={Biophysical journal},
  volume={96},
  number={6},
  pages={2160--2167},
  year={2009},
  publisher={Elsevier},
  doi={10.1016/j.bpj.2008.12.3899}
}

@article{giordano:2018,
  title={Helmholtz and \uppercase{G}ibbs ensembles, thermodynamic limit and bistability in polymer lattice models},
  author={Giordano, S.},
  journal={Continuum Mechanics and Thermodynamics},
  volume={30},
  number={},
  pages={459-483},
  year={2018},
  publisher={Springer},
  doi={10.1007/s00161-017-0615-5}
}

@article{giordano:2017,
 title={Spin variable approach for the statistical mechanics of folding and unfolding chains},
 author={Giordano, S.},
  journal={Soft matter},
  volume={13},
  number={38},
  pages={6877--6893},
  year={2017},
  publisher={Royal Society of Chemistry},
  doi={10.1039/C7SM00882A}
}

@article{benedito:2018:2,
  title={Isotensional and isometric force-extension response of chains with bistable units and \uppercase{I}sing interactions},
  author={Benedito, M. and Giordano, S.},
  journal={Physical Review E},
  volume={98},
  number={5},
  pages={052146},
  year={2018},
  publisher={APS},
  doi={10.1103/PhysRevE.98.052146}
}

@article{benedito:2020, 
   title={Unfolding pathway and its identifiability in heterogeneous chains of bistable units},  
   number={5}, 
   journal={Physics Letters A}, 
   publisher={Elsevier}, 
   author={Benedito, M. and Giordano, S.}, 
   year={2020}, 
   volume={384}, 
   pages={126124},
   doi={10.1016/j.physleta.2019.126124},
}

@article{manca:2013,
  title={Two-state theory of single-molecule stretching experiments},
  author={Manca, F. and Giordano, S. and  Palla, P. L. and  Cleri, F. and Colombo, L.},
  journal={Physical Review E},
  volume={88},
  number={3},
  pages={032705},
  year={2013},
  publisher={APS},
doi={10.1103/PhysRevE.87.032705}
}

@article{rosenberg:2007,
	title={Complementary dimerization of microtubule-associated tau protein: Implications for microtubule bundling and tau-mediated pathogenesis},
  	author={Rosenberg, K. J. and Ross, J. L. and Feinstein, H. E. and Feinstein, S. C. and Israelachvili, J.},
  	journal={Proceedings of the National Academy of Sciences},
  	volume={105},
 	number={21},
  	pages={7445--7450},
  	year={2008},
  	publisher={National Acad Sciences},
	doi={10.1073/pnas.0802036105}
}

@article{kuhl:2015,
  	title={Tau-ism: the Yin and Yang of microtubule sliding, detachment, and rupture},
  	author={Van den Bedem, H. and Kuhl, E.},
  	journal={Biophysical journal},
  	volume={109},
  	number={11},
  	pages={2215},
  	year={2015},
  	publisher={The Biophysical Society},
	doi={10.1016/j.bpj.2015.10.020}
}

@article{kuhl:2016,
  	title={Modeling molecular mechanisms in the axon},
  	author={De Rooij, R. and Miller, K. E. and Kuhl, E.},
  	journal={Computational mechanics},
  	volume={59},
  	number={3},
  	pages={523--537},
  	year={2017},
  	publisher={Springer},
	doi={10.1007/s00466-016-1359-y}
}

@article{smith:2000,
	title={Axonal damage in traumatic brain injury},
  	author={Smith, D. H. and Meaney, D. F.},
  	journal={The neuroscientist},
  	volume={6},
  	number={6},
  	pages={483--495},
  	year={2000},
  	publisher={Sage Publications Sage CA: Thousand Oaks, CA},
  	doi={10.1177/107385840000600611}
	}

@article{smith:2012,
	title={Partial interruption of axonal transport due to microtubule breakage accounts for the formation of periodic varicosities after traumatic axonal injury},
  	author={Tang-Schomer, M. D. and Johnson, V. E. and Baas, P. W. and Stewart, W. and Smith, D. H.},
  	journal={Experimental neurology},
  	volume={233},
  	number={1},
  	pages={364--372},
 	year={2012},
  	publisher={Elsevier},
	doi={10.1016/j.expneurol.2011.10.030}
}

@article{smith:2010,
  	title={Mechanical breaking of microtubules in axons during dynamic stretch injury underlies delayed elasticity, microtubule disassembly, and axon degeneration},
  	author={Tang-Schomer, M. D. and Patel, A. R. and Baas, P. W. and Smith, D. H.},
  	journal={The FASEB Journal},
  	volume={24},
  	number={5},
  	pages={1401--1410},
  	year={2010},
  	publisher={Wiley Online Library},
	doi={10.1096/fj.09-142844}
}

@article{degennes:2001,
  	title={Maximum pull out force on DNA hybrids},
  	author={de Gennes, PG.},
  	journal={Comptes Rendus de l'Acad{\'e}mie des Sciences-Series IV-Physics},
  	volume={2},
  	number={10},
  	pages={1505--1508},
  	year={2001},
  	publisher={Elsevier},
  	doi={10.1016/S1296-2147(01)01287-2}
	}

@article{hatch:2008,
  	title={Demonstration that the shear force required to separate short double-stranded DNA does not increase significantly with sequence length for sequences longer than 25 base pairs},
  	author={Hatch, K. and Danilowicz, C. and Coljee, V. and Prentiss, M.},
  	journal={Physical Review E},
  	volume={78},
  	number={1},
  	pages={011920},
  	year={2008},
  	publisher={APS},
  	doi={10.1103/PhysRevE.78.011920}
	}

@article{prakash:2011,
  	title={Shear unzipping of double-stranded DNA},
  	author={Prakash, S. and Singh, Y.},
  	journal={Physical Review E},
  	volume={84},
  	number={3},
  	pages={031905},
  	year={2011},
  	publisher={APS},
  	doi={10.1103/PhysRevE.84.031905}
	}

@article{cao:2015,
  	title={A chemomechanical model of matrix and nuclear rigidity regulation of focal adhesion size},
  	author={Cao, X. and Lin, Y. and Driscoll, T. P. and Franco-Barraza, J. and Cukierman, E. and Mauck, R. L. and Shenoy, V. B.},
  	journal={Biophysical journal},
  	volume={109},
  	number={9},
  	pages={1807--1817},
  	year={2015},
  	publisher={Elsevier},
  	doi={10.1016/j.bpj.2015.08.048}
	}

@book{goriely:2017,
  	title = {The mathematics and mechanics of biological growth},
 	author = {Goriely, A.},
  	publisher = {Springer},
  	year = {2017},
	isbn={978-0-387-87710-5}
}

@article{butler:2018,
  	title={Machine learning for molecular and materials science},
  	author={Butler, K. T. and Davies, D. W. and Cartwright, H. and Isayev, O. and Walsh, A.},
  	journal={Nature},
  	volume={559},
  	number={7715},
  	pages={547--555},
  	year={2018},
  	publisher={Nature Publishing Group},
  	doi={10.1038/s41586-018-0337-2}
	}

@article{karplus:2002,
  	title={Molecular dynamics simulations of biomolecules},
  	author={Karplus, M. and McCammon, J. A.},
  	journal={Nature structural biology},
  	volume={9},
  	number={9},
  	pages={646--652},
  	year={2002},
  	publisher={Nature Publishing Group},
  	doi={10.1038/nsb0902-646}
	}

@article{hsu:2020,
  	title={The order-disorder continuum: linking predictions of protein structure and disorder through molecular simulation},
  	author={Hsu, C. C. and Buehler, M. J. and Tarakanova, A.},
  	journal={Scientific reports},
  	volume={10},
  	number={1},
  	pages={1--14},
  	year={2020},
  	publisher={Nature Publishing Group},
  	doi={10.1038/s41598-020-58868-w}
	}

@article{busta:2000,
  	author = {Bustamante, C. and Smith, S. and Liphardt, L. and Smith, D.},
  	title = {Single-molecule studies of DNA mechanics},
  	journal = {Current Opinion in Structural Biology},
  	year = {2000},
  	volume = {10},
  	number = {3},
  	pages = {279-285},
  	doi = {10.1016/S0959-440X(00)00085-3},
	}

@article{gao:2011,
  	title={Probing mechanical principles of focal contacts in cell--matrix adhesion with a coupled stochastic--elastic modelling framework},
  	author={Gao, H. and Qian, J. and Chen, B.},
  	journal={Journal of the royal society Interface},
  	volume={8},
  	number={62},
  	pages={1217--1232},
  	year={2011},
  	publisher={The Royal Society},
  	doi={10.1098/rsif.2011.0157}
	}

@article{schwarz:2013,
  	title={Physics of adherent cells},
  	author={Schwarz, U. S. and Safran, S. A.},
  	journal={Reviews of Modern Physics},
  	volume={85},
  	number={3},
  	pages={1327},
  	year={2013},
  	publisher={APS},
  	doi={10.1103/RevModPhys.85.1327}
	}

@article{caruel:2013,
  	title={Muscle as a metamaterial operating near a critical point},
  	author={Caruel, M. and Allain, JM. and Truskinovsky, L.},
  	journal={Physical Review Letters},
  	volume={110},
  	number={24},
  	pages={248103},
  	year={2013},
  	publisher={APS},
  	doi={10.1103/PhysRevLett.110.248103}
	}

@article{huxley:1971,
  	title={Proposed mechanism of force generation in striated muscle},
  	author={Huxley, A. F. and Simmons, R. M.},
  	journal={Nature},
  	volume={233},
  	number={5321},
  	pages={533--538},
  	year={1971},
  	publisher={Nature Publishing Group},
  	doi={10.1038/233533a0}
	}

@article{darocha:2020,
  	title={Rigidity-controlled crossover: from spinodal to critical failure},
  	author={da Rocha, H. B. and Truskinovsky, L.},
  	journal={Physical review letters},
  	volume={124},
  	number={1},
  	pages={015501},
  	year={2020},
  	publisher={APS},
  	doi={10.1103/PhysRevLett.124.015501}
	}

@article{dudko:2007,
  title={Extracting kinetics from single-molecule force spectroscopy: nanopore unzipping of DNA hairpins},
  author={Dudko, O. K. and Math{\'e}, J. and Szabo, A. and Meller, A. and Hummer, G.},
  journal={Biophysical journal},
  volume={92},
  number={12},
  pages={4188--4195},
  year={2007},
  publisher={Elsevier},
  doi={10.1529/biophysj.106.102855}
}

@article{liphardt:2001,
  title={Reversible unfolding of single RNA molecules by mechanical force},
  author={Liphardt, J. and Onoa, B. and Smith, S. B. and Tinoco, I. and Bustamante, C.},
  journal={Science},
  volume={292},
  number={5517},
  pages={733--737},
  year={2001},
  publisher={American Association for the Advancement of Science},
  doi={10.1126/science.1058498}
}

@article{mathe:2004,
  title={Nanopore unzipping of individual DNA hairpin molecules},
  author={Math{\'e}, J. and Visram, H. and Viasnoff, V. and Rabin, Y. and Meller, A.},
  journal={Biophysical journal},
  volume={87},
  number={5},
  pages={3205--3212},
  year={2004},
  publisher={Elsevier},
 doi={10.1529/biophysj.104.047274}
}

@article{thompson:2020,
  title={Protein-protein interactions in neurodegenerative diseases: a conspiracy theory},
  author={Thompson, T. B. and Chaggar, P. and Kuhl, E. and Goriely, A. and Alzheimer’s Disease Neuroimaging Initiative},
  journal={PLoS computational biology},
  volume={16},
  number={10},
  pages={e1008267},
  year={2020},
  publisher={Public Library of Science San Francisco, CA USA},
  doi={10.1371/journal.pcbi.1008267}
}

@article{garcia:2017,
  title={Continuum mechanical modeling of axonal growth},
  author={Garcia-Grajales, J. A. and Jerusalem, A. and Goriely, A.},
  journal={Computer Methods in Applied Mechanics and Engineering},
  volume={314},
  pages={147--163},
  year={2017},
  publisher={Elsevier},
  doi={10.1016/j.cma.2016.07.032}
}

@book{gurtin:2010,
	author = {Gurtin, M. E. and Fried, E. and Anand, L.},
	publisher = {Cambridge University Press},
	title = {The Mechanics and Thermodynamics of Continua},
	year = {2010},
	isbn={978-0521405980}
	}

@book{stillinger:2015,
	author = {Stillinger, F. H.},
	publisher = {Princeton University Press Princeton and Oxford},
	title = {Energy Landscapes, Inherent Structures, and Condensed-Matter Phenomena},
	year = {2015},
	isbn={9780691166803},
	}

@techreport{carr:1984,
  title={Structured phase transitions on a finite interval},
  author={Carr, J. and Gurtin, M. E. and Slemrod, M.},
  year={1984},
  institution={Carnegie-Mellon Inst. Resear. Pittsburgh PA},
  doi={10.1007/BF00280031}
}

@article{friddle:2012,
  author = {Friddle, R.W. and Noy, A. and De Yoreo, J. J.},
  title = {Interpreting the widespread nonlinear force spectra of intermolecular bonds},
  journal = {PNAS},
  year = {2012},
  volume = {109},
  pages = {13573--13578},
  doi = {10.1073/pnas.1202946109},
}

@article{friddle:2008,
  author = {Friddle,R . W.  and Podsiadlo, P. and Artyukhin, A. B. and Noy, A.},
  title = {Near-Equilibrium Chemical Force Microscopy},
  journal = {J. Phys. Chem. C},
  year = {2008},
  volume = {112},
  pages = {4986--4990},
  doi = {10.1021/jp7095967},
}

@article{suzuki:2013,
  author = {Suzuki, Y. and Dudko, O. K.},
  title = {Single molecules in an extension clamp: Extracting rates and activation barriers},
  journal = {Phys. Rev. Lett.},
  year = {2013},
  volume = {110},
  pages = {158105},
  doi = {10.1103/PhysRevLett.110.158105},
}

@article{givli:2011,
  author = {Benichou, I. and Givli, S.},
  title = {The hidden ingenuity in titin structure},
  journal = {Applied Physics Letters},
  year = {2011},
  volume = {99},
  pages = {091904},
  doi = {10.1063/1.3558901},
}

@article{erdmann:2007,
  title={Impact of receptor-ligand distance on adhesion cluster stability},
  author={Erdmann, T. and Schwarz, U. S.},
  journal={The European Physical Journal E},
  volume={22},
  number={2},
  pages={123--137},
  year={2007},
  publisher={Springer},
  doi={10.1140/epje/e2007-00019-8}
}

@article{caruel:2018,
  title={Physics of muscle contraction},
  author={Caruel, M. and Truskinovsky, L.},
  journal={Reports on Progress in Physics},
  volume={81},
  number={3},
  pages={036602},
  year={2018},
  publisher={IOP Publishing},
  doi={10.1088/1361-6633/aa7b9e}
}

@book{goodstein:2014,
  title={States of matter},
  author={Goodstein, D. L.},
  year={2014},
  publisher={Courier Corporation},
  isbn={978-0486649276}
}

@book{callen:1960,
  title={Thermodynamics},
  author={Callen, H. B.},
  year={1960},
  publisher={Wiley, New York},
  isbn={978-0-4718-6256-7}
}

@book{arnold:2013,
  title={Mathematical methods of classical mechanics},
  author={Arnol'd, V. I.},
  volume={60},
  year={2013},
  publisher={Springer Science \& Business Media},
  isbn={978-1-4757-1693-1}
}

@book{sethna:2021,
  title={Statistical mechanics: entropy, order parameters, and complexity},
  author={Sethna, J.},
  volume={14},
  year={2021},
  publisher={Oxford University Press, USA},
  isbn={978-0198566779}
}

@article{brenner:1962,
  title={Mechanical behavior of sapphire whiskers at elevated temperatures},
  author={Brenner, S. S.},
  journal={Journal of Applied Physics},
  volume={33},
  number={1},
  pages={33--39},
  year={1962},
  publisher={American Institute of Physics},
  doi={10.1063/1.1728523}
}

@article{miracle:2017,
  title={A critical review of high entropy alloys and related concepts},
  author={Miracle, D. B. and Senkov, O. N.},
  journal={Acta Materialia},
  volume={122},
  pages={448--511},
  year={2017},
  publisher={Elsevier},
  doi={10.1016/j.actamat.2016.08.081}
}

@article{gali:2013,
  title={Tensile properties of high-and medium-entropy alloys},
  author={Gali, A. and George, E. P.},
  journal={Intermetallics},
  volume={39},
  pages={74--78},
  year={2013},
  publisher={Elsevier},
  doi={10.1016/j.intermet.2013.03.018}
}

@book{zinn:2004,
  title={Path integrals in quantum mechanics},
  author={Zinn-Justin, J.},
  year={2004},
  publisher={OUP Oxford},
  isbn={978-9814273565}
}

@book{vanthoff:1884,
  title={Etudes de dynamique chimique},
  author={Van't Hoff, J. H.},
  volume={1},
  year={1884},
  publisher={Muller},
  isbn={978-1160546317}
}

@article{arrhenius:1889,
  title={{\"U}ber die Reaktionsgeschwindigkeit bei der Inversion von Rohrzucker durch S{\"a}uren},
  author={Arrhenius, S.},
  journal={Zeitschrift f{\"u}r physikalische Chemie},
  volume={4},
  number={1},
  pages={226--248},
  year={1889},
  publisher={De Gruyter Oldenbourg},
  doi={10.1515/zpch-1889-0416}
}

@article{rayleigh:1891,
  title={Reprinted in \textit{Scientific Paper of Lord Rayleigh, 1902, VOL. III}},
  author={Rayleigh, L. (J. W. Strutt)},
  journal={Philos. Mag},
  volume={32},
  number={424},
  year={1891},
  publisher={Cambridge University Press},
  doi={10.1364/AO.3.001091}
}

@article{einstein:1905,
  title={On the motion of small particles suspended in liquids at rest required by the molecular-kinetic theory of heat},
  author={Einstein, A. and others},
  journal={Annalen der physik},
  volume={17},
  number={549-560},
  pages={208},
  year={1905},
  doi={10.1002/andp.19053220806}
}

@article{fokker:1914,
  title={Die mittlere Energie rotierender elektrischer Dipole im Strahlungsfeld},
  author={Fokker, A. D.},
  journal={Annalen der Physik},
  volume={348},
  number={5},
  pages={810--820},
  year={1914},
  publisher={WILEY-VCH Verlag Leipzig},
  doi={10.1002/andp.19143480507}
}

@article{planck:1917,
  title={Sitzungsber},
  author={Planck, M.},
  journal={Preuss. Akad. Wiss. Phys. Math. Kl},
  volume={325},
  number={3},
  year={1917},
  EAN={2560019150946}
}

@article{ornstein:917,
  title={Over de Brown’sche Beweging},
  author={Ornstein, L. S.},
  journal={Versl. Acad. Amst},
  volume={26},
  pages={1005--1017},
  year={1917},
  doi={88238/b1990005299590205131}
}

@article{pontryagin:1933,
  title={On statistical analysis of dynamical systems},
  author={Pontryagin, L. S. and Andronov, A. A. and Vitt, A. A.},
  journal={Zh. Eksp. Teor. Fiz},
  volume={3},
  number={3},
  pages={165--180},
  year={1933}
}

@article{farkas:1927,
  title={Keimbildungsgeschwindigkeit in {\"u}bers{\"a}ttigten D{\"a}mpfen},
  author={Farkas, L.},
  journal={Zeitschrift f{\"u}r physikalische Chemie},
  volume={125},
  number={1},
  pages={236--242},
  year={1927},
  publisher={De Gruyter Oldenbourg},
  doi={10.1515/zpch-1927-12513}
}

@article{lindemann:1922,
  title={Discussion on “the radiation theory of chemical action”},
  author={Lindemann, F. A. and Arrhenius, S. and Langmuir, I. and Dhar, N. R. and Perrin, J. and Lewis, W. C.},
  journal={Transactions of the Faraday Society},
  volume={17},
  pages={598--606},
  year={1922},
  publisher={Royal Society of Chemistry},
  doi={10.1039/TF9221700598}
}

@article{eyring:1935,
  title={The activated complex in chemical reactions},
  author={Eyring, H.},
  journal={The Journal of Chemical Physics},
  volume={3},
  number={2},
  pages={107--115},
  year={1935},
  publisher={American Institute of Physics},
  doi={10.1063/1.1749604}
}

@article{weiner:1977,
  author = {Weiner, J. H. and Pear, M. R.},
  title = {Computer simulation of conformational transitions in an idealized polymer model},
  journal = {Macromolecules},
  year = {1977},
  volume = {10},
  number = {2},
  pages = {317--325},
  doi = {10.1021/ma60056a017},
}

@article{kreuzer:2001,
  author = {Kreuzer, H. J. and Payne, S. H. and Livadaru, L.},
  title = {Stretching a macromolecule in an Atomic Force Microscope: Statistical Mechanical analysis},
  journal = {Biophysical Journal},
  year = {2001},
  volume = {80},
  pages = {2505--2514},
  doi = {10.1016/S0006-3495(01)76222-4},
}

@article{kim:2009,
  author = {Kim, K. and Saleh, O. A.},
  title = {A high-resolution magnetic tweezer for single-molecule measurements},
  journal = {Nucleic Acids Research},
  year = {2009},
  volume = {37},
  number = {20},
  pages = {e136},
  doi = {10.1093/nar/gkp725},
}

@article{capitano:2013,
  author = {Capitanio, M. and Pavone, F. S.},
  title = {Interrogating Biology with Force: Single Molecule High-Resolution
Measurements with Optical Tweezers},
  journal = {Biophysical Journal},
  year = {2013},
  volume = {105},
  pages = {129--1303},
  doi = {10.1016/j.bpj.2013.08.007},
}

@article{borgia:2008,
  author = {Borgia, A. and Williams, P. M. and Clarke, J.},
  title = {Single-Molecule Studies of Protein Folding},
  journal = {Annual Review Biochemistry},
  year = {2008},
  volume = {101},
  pages = {101},
  doi = {10.1146/annurev.biochem.77.060706.093102},
}

@article{mancagiordano:2014,
  author = {Manca, F. and Giordano, S. and Palla, P. L. and Cleri, F.},
  title = {On the equivalence of thermodynamics ensembles for flexible polymer chains},
  journal = {Physica A},
  year = {2014},
  volume = {395},
  pages = {154--170},
  doi = {10.1016/j.physa.2013.10.042},
}

@article{rief:2002,
  author = {Rief, M. and Grubm$\ddot{\mbox{u}}$ller, M.},
  title = {Force spectroscopy of single biomolecules},
  journal = {A European Journal of Chemical Physics and Physical Chemistry},
  year = {2002},
  volume = {3},
  number = {3},
  pages = {255-261},
  doi = {10.1002/1439-7641(20020315)3:3<255::AID-CPHC255>3.0.CO;2-M},
}

@article{chung:2013,
  author = {Chung, J. and Kushner, A. M. and Weisman, A. C. and Guan, Z.},
  title = {Direct correlation of single-molecule properties
with bulk mechanical performance for the
biomimetic design of polymers},
  journal = {Nature Mat.},
  year = {2014},
  volume = {13},
  pages = {1055-62},
  doi = {10.1038/NMAT4090},
}

@article{bustamante:2003,
  author = {Keller, D. and Swigon, D. and Bustamante, C.},
  title = {Relating single-molecule measurement to thermodynamic},
  journal = {Biophysical Journal},
  year = {2003},
  volume = {84},
  number = {2},
  pages = {733-738},
  doi = {10.1016/S0006-3495(03)74892-9},
}

@article{su:2009,
  author = {Su, T. and Purohit, P. K.},
  title = {Mechanics of forced unfolding of proteins},
  journal = {Acta Biomaterialia},
  year = {2009},
  volume = {5},
  number = {6},
  pages = {1855--1863},
  doi = {10.1016/j.actbio.2009.01.038},
}

@article{benedito:2018,
  author = {Benedito, M. and Giordano, S.},
  title = {Thermodynamics of small systems with conformational transitions: the case of two-state freely jointed chains with extensible units},
  journal = {Journal of Chemical Physics},
  year = {2018},
  volume = {149},
  Pages = {054901},
  doi = {10.1063/1.5026386},
}

@article{staple:2008,
  author = {Staple, D. B. and Payne, S. H. and Reddin, A. L. C. and Kreuzer, H. J.},
  title = {Model for stretching and unfolding the giant multidomain muscle protein
 using Single-Molecule Force Spectroscopy},
  journal = {Physical Review Letters},
  year = {2008},
  volume = {101},
  pages = {248301},
  doi = {10.1103/PhysRevLett.101.248301},
}

@article{froli:2000,
  author = {Froli, M. and Royer-Carfagni, G.},
  title = {A mechanical model for the elastic-plastic behavior of metallic bars},
  journal = {International Journal of Solids and Structures},
  year = {2000},
  volume = {37},
  number = {29},
  pages = {3901-3918},
  doi = {10.1016/S0020-7683(99)00069-4},
}

@article{hudson:2019,
  author = {Borja da Rocha, H. and Truskinovsky, L.},
  title = {Functionality of disorder in muscle mechanics},
  journal = {Physical Review Letters},
  year = {2019},
  volume = {122},
  pages = {088103},
  doi = {10.1103/PhysRevLett.122.088103},
}

@article{rief:1999,
  title={Single molecule force spectroscopy of spectrin repeats: low unfolding forces in helix bundles},
  author={Rief, M. and Pascual, J. and Saraste, M. and Gaub, H. E.},
  journal={Journal of molecular biology},
  volume={286},
  number={2},
  pages={553--561},
  year={1999},
  publisher={Elsevier},
  doi={10.1006/jmbi.1998.2466}
}

@article{rief:1999:2,
  title={Sequence-dependent mechanics of single DNA molecules},
  author={Rief, M. and Clausen-Schaumann, H. and Gaub, H. E.},
  journal={Nature structural biology},
  volume={6},
  number={4},
  pages={346--349},
  year={1999},
  publisher={Nature Publishing Group},
  doi={10.1038/7582}
}

@article{recho:2016,
  author = {Recho, P. and Jerusalem, J. and Goriely, A.},
  title = {Growth, collapse, and stalling in a mechanical model for neurite motility},
  journal = {Physical Review E},
  year = {2016},
  volume = {93},
  number = {},
  pages = {032410},
  doi = {10.1103/PhysRevE.93.032410},
}

@article{benamar:2011,
  author = {BenAmar, M. and Chatelain, C. and Ciarletta, P.},
  title = {Contour instabilities in early tumor growth models},
  journal = {Physical Review Letters},
  year = {2011},
  volume = {106},
  Pages = {148101},
  doi = {10.1103/PhysRevLett.106.148101},
}

@article{krasnoslobodtsev:2013,
  author = {Krasnoslobodtsev, A. V. and Volkov, I. L. and Asiago, J. M. and Hindupur, J. and Rochet, JC. and Lyubchenko, Y. L.},
  title = {$\alpha$ synuclein misfolding assessed with Single Molecule AFM Force Spectroscopy: effect of pathogenic mutations},
  journal = {Biochemistry},
  year = {2013},
  volume = {52},
  pages = {7377--7386},
  doi = {10.1021/bi401037z},
}

@article{gonzalez:2017,
  author = {Marin-Gonzalez, A. and Vilhena, J. G. and Perez, R. and Moreno-Herrero, F.},
  title = {Understanding the mechanical response of double-stranded DNA and RNA under constant stretching forces using all-atom molecular dynamics},
  journal = {Proceeding of the National Academy of Science (PNAS)},
  year = {2017},
  volume = {114},
  pages = {7049-7054},
  doi = {10.1073/pnas.1705642114/-/DCSupplemental},
}

@article{woodside:2008,
  author = {Woodside, M. T. and Garcia-Garcia, C. and Block, S. M.},
  title = {Folding and unfolding single RNA molecules under tension},
  journal = {Current Opinion in Chemical Biology},
  year = {2008},
  volume = {12},
  pages = {640--646},
  doi = {10.1016/j.cbpa.2008.08.011},
}

@article{li:2008,
  author = {Li, P. T. L. and Vieregg, J. and Tinoco, I.},
  title = {How RNA unfolds and refolds},
  journal = {Annual Review of Biochemistry},
  year = {2008},
  volume = {77},
  pages = {77--100},
  doi = {10.1146/annurev.biochem.77.061206.174353},
}

@article{rosa:2004,
  author = {Rosa, R. and Eckel, R. and Bartels, F. and Sischka, A. and Baumgarth, B. and Wilking, S. D. and P$\ddot{\mbox{u}}$hler, A. and Sewald, N. and Becker, A. and Anselmetti, D.},
  title = {Single molecule force spectroscopy on ligand-DNA complexes: from molecular binding mechanisms to biosensor applications},
  journal = {J. Biotechnology},
  year = {2004},
  volume = {112},
  pages = {5--12},
  doi = {10.1016/j.jbiotec.2004.04.029},
}

@article{winkler:2010,
  title={Equivalence of statistical ensembles in stretching single flexible polymers},
  author={Winkler, R. G.},
  journal={Soft Matter},
  volume={6},
  number={24},
  pages={6183--6191},
  year={2010},
  publisher={Royal Society of Chemistry},
  doi = {10.1039/C0SM00488J},
}

@article{bustamante:2010,
  author = {Shank, E. A. and Cecconi, C. and Dill, J. W.  and Marqusee, S. and Bustamante, C.},
  title = {The folding cooperativity of a protein is controlled by its chain topology},
  journal = {Nature},
  year = {2010},
  volume = {465},
  pages = {637--640},
  doi = {10.1038/nature09021},
}

@article{zhanget:2008,
  author = {Zhang, Y. and Sun, G. and L$\ddot{\mbox{u}}$, S. and Li, N. and Long, M.},
  title = {Low spring constant regulates P-Selectin-PSGL-1 bond rupture},
  journal = {Biophys. Jour},
  year = {2008},
  volume = {95},
  pages = {5439--5448},
   doi = {10.1529/biophysj.108.137141},
}

@article{allen:1979,
  title={A microscopic theory for antiphase boundary motion and its application to antiphase domain coarsening},
  author={Allen, S. M. and Cahn, J. W.},
  journal={Acta metallurgica},
  volume={27},
  number={6},
  pages={1085--1095},
  year={1979},
  publisher={Elsevier},
doi={10.1016/0001-6160(79)90196-2}
}

@article{bilal:2017,
  title={Bistable metamaterial for switching and cascading elastic vibrations},
  author={Bilal, O. R. and Foehr, A. and Daraio, C.},
  journal={Proc. Natl. Acad. Sci. U.S.A.},
  volume={114},
  number={18},
  pages={4603-4606},
  year={2017},
  publisher={National Academy of Sciences },
doi={10.1073/pnas.1618314114}
}

@article{shang:2018, 
   title={Durable bistable auxetics made of rigid solids}, 
   author={Shang, X. and Liu, L. and Rafsanjani, A. and Pasini, D.},
   number={3}, 
   journal={Journal of Materials Research}, 
   publisher={Cambridge University Press}, 
   volume={33}, 
   year={2018}, 
   pages={300-308},
   doi={10.1557/jmr.2017.417}
}

@article{shaw:1997,
   title={On the nucleation and propagation of phase transformation fronts in a {N}i{T}i alloy},
   author={Shaw, J. A. and Kyriakides, S.},
   journal={Acta materialia},
   volume={45},
   number={2},
   pages={683--700},
   year={1997},
   publisher={Elsevier},
   doi={10.1016/S1359-6454(96)00189-9}
}

@article{abeyaratne:1996,
   title={Kinetics of materials with wiggly energies: theory and application to the evolution of twinning microstructures in a Cu-Al-Ni shape memory alloy},
   author={Abeyaratne, R. and Chu, C. and James, R. D.},
   journal={Philosophical Magazine A},
   volume={73},
   number={2},
   pages={457--497},
   year={1996},
   publisher={Taylor \& Francis},
   doi={10.1080/01418619608244394}
}

@article{benichou:2013,
  title={Structures undergoing discrete phase transformation},
  author={Benichou, I. and Givli, S.},
  journal={Journal of the Mechanics and Physics of Solids},
  volume={61},
  number={1},
  pages={94--113},
 year={2013},
 publisher={Elsevier},
 doi={10.1016/j.jmps.2012.08.009}
}

@article{triantafyllidis:1993,
   author = {Triantafyllidis, N. and Bardenhagen, S.},
   title = {On higher order gradient continuum theories in 1-D nonlinear elasticity. {D}erivation from and comparison to the corresponding discrete models},
   journal = {Journal of Elasticity},
   year = {1993},
   volume = {33},
   number = {3},
   pages = {259--293},
   publisher = {Springer},
   doi = {10.1007/BF00043251}
}

@article{coleman:1983,
  title={Necking and drawing in polymeric fibers under tension},
  author={Coleman, B. D.},
  journal={Archive for Rational Mechanics and Analysis},
  volume={83},
  number={2},
  pages={115--137},
  year={1983},
  publisher={Springer},
  doi={10.1007/BF00282158}
}

@article{tanaka:1986,
  title={Thermomechanics of transformation pseudoelasticity and shape memory effect in alloys},
  author={Tanaka, K. and Kobayashi, S. and Sato, Y.},
  journal={International Journal of Plasticity},
  volume={2},
  number={1},
  pages={59--72},
  year={1986},
  publisher={Elsevier},
  doi={10.1016/0749-6419(86)90016-1}
}

@article{keller:1998,
  title={The role of metastability in polymer phase transitions},
  author={Keller, A. and Cheng, S. Z. D.},
  journal={Polymer},
  volume={39},
  number={19},
  pages={4461--4487},
  year={1998},
  publisher={Elsevier},
   doi={10.1016/S0032-3861(97)10320-2}
}

@article{muller:2001,
  title={Thermodynamic aspects of shape memory alloys},
  author={M{\"u}ller, I. and Seelecke, S.},
  journal={Mathematical and computer modelling},
  volume={34},
  number={12-13},
  pages={1307--1355},
  year={2001},
  publisher={Elsevier},
  doi={10.1016/S0895-7177(01)00134-0}
}

@article{peyrard:2004,
  title={Nonlinear dynamics and statistical physics of {DNA}},
  author={Peyrard, M.},
  journal={Nonlinearity},
  volume={17},
  number={2},
  pages={R1},
  year={2004},
  publisher={IOP Publishing},
  doi={10.1088/0951-7715/17/2/R01}
}

@article{denisov:2005,
   title={Domain statistics in a finite \uppercase{I}sing chain},
   author={Denisov, S. I. and H{\"a}nggi, P.},
   journal={Physical Review E},
   volume={71},
   number={4},
   pages={046137},
   year={2005},
   publisher={APS},
   doi={10.1103/PhysRevE.71.046137}
}

@article{ising:1925,
  title={Beitrag zur theorie des ferromagnetismus},
  author={Ising, E.},
  journal={Zeitschrift f{\"u}r Physik},
  volume={31},
  number={1},
  pages={253--258},
  year={1925},
  publisher={Springer},
  doi={10.1007/BF02980577}
}

@article{seth:2016,
  title={Combinatorial approach to exactly solve the 1{D} \uppercase{I}sing model},
  author={Seth, S.},
  journal={European Journal of Physics},
  volume={38},
  number={1},
  pages={015104},
  year={2016},
  publisher={IOP Publishing},
  doi={10.1088/1361-6404/38/1/015104}
}

@article{manca:2012,
   title={Elasticity of flexible and semiflexible polymers with extensible bonds in the \uppercase{G}ibbs and \uppercase{H}elmholtz ensembles},
   author={Manca, F. and Giordano, S. and  Palla, P. L. and  Zucca, R. and Cleri, F. and Colombo, L.},
   journal={Journal of Chemical Physics},
   volume={136},
   number={15},
   pages={154906},
   year={2012},
   publisher={AIP},
   doi={10.1063/1.4704607}
}

@article{li:2010,
  title={High-efficiency mechanical energy storage and retrieval using interfaces in nanowires},
  author={Li, S. and Ding, X. and Li, J. and Ren, X. and Sun, J. and Ma, E.},
  journal={Nano letters},
  volume={10},
  number={5},
  pages={1774--1779},
  year={2010},
  publisher={ACS Publications},
  doi={10.1021/nl100263p}
}

@article{seo:2013,
  title={Origin of size dependency in coherent-twin-propagation-mediated tensile deformation of noble metal nanowires},
  author={Seo, J. H. and Park, H. S. and Yoo, Y. and Seong, T.-Y. and Li, J. and Ahn, J.-P. and Kim, B. and Choi, I.-S.},
  journal={Nano letters},
  volume={13},
  number={11},
  pages={5112--5116},
  year={2013},
  publisher={ACS Publications},
  doi={10.1021/nl402282n}
}

@article{ma:2013,
  title={Surface-induced structural transformation in nanowires},
  author={Ma, F. and Xu, K.-W. and Chu, P. K.},
  journal={Materials Science and Engineering: R: Reports},
  volume={74},
  number={6},
  pages={173--209},
  year={2013},
  publisher={Elsevier},
doi={10.1016/j.mser.2013.05.001}
}

@article{liang:2007,
  title={A micromechanical continuum model for the tensile behavior of shape memory metal nanowires},
  author={Liang, W. and Srolovitz, D. J. and Zhou, M.},
  journal={Journal of the Mechanics and Physics of Solids},
  volume={55},
  number={8},
  pages={1729--1761},
  year={2007},
  publisher={Elsevier},
  doi={10.1016/j.jmps.2007.01.001}
}

@article{tadmor:2003,
  title={A {P}eierls criterion for the onset of deformation twinning at a crack tip},
  author={Tadmor, E. B. and Hai, S.},
  journal={Journal of the Mechanics and Physics of Solids},
  volume={51},
  number={5},
  pages={765--793},
  year={2003},
  publisher={Elsevier},
  doi={10.1007/978-3-642-18756-8_11}
}

@article{park:2006,
  title={Deformation of {FCC} nanowires by twinning and slip},
  author={Park, H. S. and Gall, K. and Zimmerman, J. A.},
  journal={Journal of the Mechanics and Physics of Solids},
  volume={54},
  number={9},
  pages={1862--1881},
  year={2006},
  publisher={Elsevier},
  doi={10.1016/j.jmps.2006.03.006}
}

@article{lu:2019,
  title={Tensile mechanical performance of {N}i--{C}o alloy nanowires by molecular dynamics simulation},
  author={Lu, X. and Yang, P. and Luo, J. and Ren, J. and Xue, H. and Ding, Y.},
  journal={RSC advances},
  volume={9},
  number={44},
  pages={25817--25828},
  year={2019},
  publisher={Royal Society of Chemistry},
  doi={10.1039/C9RA04294F}
}

@article{wu:2006,
  title={Molecular dynamics study on mechanics of metal nanowire},
  author={Wu, H. A.},
  journal={Mechanics Research Communications},
  volume={33},
  number={1},
  pages={9--16},
  year={2006},
  publisher={Elsevier},
  doi={10.1016/j.euromechsol.2005.11.008},
}

@article{lao:2013,
  title={Molecular dynamics simulation of {FCC} metallic nanowires: a review},
  author={Lao, J. and Tam, M. N.and Pinisetty, D. and Gupta, N.},
  journal={Jom},
  volume={65},
  number={2},
  pages={175--184},
  year={2013},
  publisher={Springer},
  doi={10.1007/s11837-012-0465-3}
}

@article{zhang:2008,
  title={Size dependence of twin formation energy in cubic {S}i{C} at the nanoscale},
  author={Zhang, Y. and Shim, H. W. and Huang, H.},
  journal={Applied Physics Letters},
  volume={92},
  number={26},
  pages={261908},
  year={2008},
  publisher={American Institute of Physics},
  doi={10.1063/1.2953976}
}

@article{seo:2011,
  title={Superplastic deformation of defect-free {A}u nanowires via coherent twin propagation},
  author={Seo, J.-H. and Yoo, Y. and Park, N.-Y. and Yoon, S.-W. and Lee, H. and Han, S. and Lee, S.-W. and Seong, T.-Y. and Lee, S.-C. and Lee, K.-B. and others},
  journal={Nano Letters},
  volume={11},
  number={8},
  pages={3499--3502},
  year={2011},
  publisher={ACS Publications},
  doi={10.1021/nl2022306}
}

@article{guo:2009,
  title={Mechanism for the pseudoelastic behavior of {FCC} shape memory nanowires},
  author={Guo, X. and Liang, W. and Zhou, M.},
  journal={Experimental mechanics},
  volume={49},
  number={2},
  pages={183--190},
  year={2009},
  publisher={Springer},
  doi={10.1007/s11340-008-9173-x}
}

@article{rezaei:2017,
  title={Pseudoelasticity and shape memory effects in cylindrical {FCC} metal nanowires},
  author={Rezaei, R. and Deng, C.},
  journal={Acta Materialia},
  volume={132},
  pages={49--56},
  year={2017},
  publisher={Elsevier},
  doi={10.1016/j.actamat.2017.04.039}
}

@article{law:2003,
  title={Cooperativity in forced unfolding of tandem spectrin repeats},
  author={Law, R. and Carl, P. and Harper, S. and Dalhaimer, P. and Speicher, D. W. and Discher, D. E.},
  journal={Biophysical journal},
  volume={84},
  number={1},
  pages={533--544},
  year={2003},
  publisher={Elsevier},
  doi={10.1016/S0006-3495(03)74872-3}
}

@article{bhaskara:2011,
  title={Stability of domain structures in multi-domain proteins},
  author={Bhaskara, R. M. and Srinivasan, N.},
  journal={Scientific reports},
  volume={1},
  pages={40},
  year={2011},
  publisher={Nature Publishing Group},
  doi={10.1038/srep00040}
}

@article{xu:2013,
  title={Domain-Domain Interactions in Filamin \uppercase{A} (16-23) Impose a Hierarchy of Unfolding Forces},
  author={Xu, T. and Lannon, H. and  Wolf, S. and Nakamura, F. and Brujic, J.},
  journal={Biophysical Journal},
  volume={104},
  number={9},
  pages={2022-2030},
  year={2013},
  publisher={Cell},
  doi={10.1016/j.bpj.2013.03.034}
}

@article{macdonald:2001,
  title={Free energies of urea and of thermal unfolding show that two tandem repeats of spectrin are thermodynamically more stable than a single repeat},
  author={MacDonald, R. I. and Pozharski, E. V.},
  journal={Biochemistry},
  volume={40},
  number={13},
  pages={3974--3984},
  year={2001},
  publisher={ACS Publications},
  doi={10.1021/bi0025159}
}

@article{daily:2009,
  title={Allosteric communication occurs via networks of tertiary and quaternary motions in proteins},
  author={Daily, M. D. and Gray, J. J.},
  journal={PLoS computational biology},
  volume={5},
  number={2},
  year={2009},
  publisher={Public Library of Science},
  doi={10.1371/journal.pcbi.1000293}
}

@article{batey:2005,
  title={Cooperative folding in a multi-domain protein},
  author={Batey, S. and Randles, L. G. and Steward, A. and Clarke, J.},
  journal={Journal of molecular biology},
  volume={349},
  number={5},
  pages={1045--1059},
  year={2005},
  publisher={Elsevier},
  doi={10.1016/j.jmb.2005.04.028}
}

@article{shoham:2020,
  title={Unfolding compactly folded molecular domains: Overall stiffness modifies the force-barrier relation},
  author={Shoham, A. and Givli, S.},
  journal={Chemical Physics Letters},
  volume={758},
  pages={137924},
  year={2020},
  publisher={Elsevier},
  doi={10.1016/j.cplett.2020.137924}
}

@book{saenger:2013,
  title={Principles of nucleic acid structure},
  author={Saenger, W.},
  year={2013},
  publisher={Springer Science \& Business Media},
  isbn={978-1-4612-5190-3}
}

@article{strunz:1999,
  title={Dynamic force spectroscopy of single DNA molecules},
  author={Strunz, T. and Oroszlan, K. and Sch{\"a}fer, R. and G{\"u}ntherodt, H.-J.},
  journal={Proceedings of the National Academy of Sciences},
  volume={96},
  number={20},
  pages={11277--11282},
  year={1999},
  publisher={National Acad Sciences},
  doi={10.1073/pnas.96.20.11277}
}

@article{lee:1994,
  title={Direct measurement of the forces between complementary strands of DNA},
  author={Lee, G. U. and Chrisey, L. A. and Colton, R. J.},
  journal={Science},
  volume={266},
  number={5186},
  pages={771--773},
  year={1994},
  publisher={American Association for the Advancement of Science},
  doi={10.1126/science.7973628}
}

@article{watson:1953,
  title={Molecular structure of nucleic acids: a structure for deoxyribose nucleic acid},
  author={Watson, J. D. and Crick, F. H. C.},
  journal={Nature},
  volume={171},
  number={4356},
  pages={737--738},
  year={1953},
  publisher={Nature Publishing Group},
  doi={10.1038/171737a0}
}

@article{schumakovitch:2002,
  title={Temperature dependence of unbinding forces between complementary DNA strands},
  author={Schumakovitch, I. and Grange, W. and Strunz, T. and Bertoncini, P. and G{\"u}ntherodt, H.-J. and Hegner, M.},
  journal={Biophysical journal},
  volume={82},
  number={1},
  pages={517--521},
  year={2002},
  publisher={Elsevier},
  doi={10.1016/S0006-3495(02)75416-7}
}

@article{pope:2001,
  title={Force-induced melting of a short DNA double helix},
  author={Pope, L. H. and Davies, M. C. and Laughton, C. A. and Roberts, C. J. and Tendler, S. J. B. and Williams, P. M.},
  journal={European Biophysics Journal},
  volume={30},
  number={1},
  pages={53--62},
  year={2001},
  publisher={Springer},
  doi={10.1007/s002490000107}
}

@article{smith:1996,
  title={Overstretching B-DNA: the elastic response of individual double-stranded and single-stranded DNA molecules},
  author={Smith, S. B. and Cui, Y. and Bustamante, C.},
  journal={Science},
  volume={271},
  number={5250},
  pages={795--799},
  year={1996},
  publisher={American Association for the Advancement of Science},
  doi={10.1126/science.271.5250.795}
}

@article{bustamante:2000:2,
  title={Single-molecule studies of DNA mechanics},
  author={Bustamante, C. and Smith, S. B. and Liphardt, J. and Smith, D.},
  journal={Current opinion in structural biology},
  volume={10},
  number={3},
  pages={279--285},
  year={2000},
  publisher={Elsevier},
  doi={10.1016/S0959-440X(00)00085-3}
}

@article{dobson:2003,
  title={Protein folding and misfolding},
  author={Dobson, C. M.},
  journal={Nature},
  volume={426},
  number={6968},
  pages={884--890},
  year={2003},
  publisher={Nature Publishing Group},
  doi={10.1038/nature02261}
}

@article{pugno:2021,
  title={A coarse-grained mechanical model for folding and unfolding of tropoelastin with possible mutations},
  author={Florio, G. and Pugno, N. M. and Buehler, M. J. and Puglisi, G.},
  journal={Acta Biomaterialia},
  volume={134},
  pages={477--489},
  year={2021},
  publisher={Elsevier},
  doi={10.1016/j.actbio.2021.07.032}
}

@article{salvo:2022,
  title={On the role of elasticity in focal adhesion within the passive regime},
  author={Di Stefano, S. and Florio, G. and Napoli, G. and Pugno, N. M. and Puglisi, G.},
  journal={International Journal of Nonlinear Mechanics},
  volume={},
  pages={},
  year={2022},
  publisher={Accepted.},
}

@article{lumry:1966,
  title={Validity of the “two-state” hypothesis for conformational transitions of proteins},
  author={Lumry, R. and Biltonen, R. and Brandts, J. F.},
  journal={Biopolymers: Original Research on Biomolecules},
  volume={4},
  number={8},
  pages={917--944},
  year={1966},
  publisher={Wiley Online Library},
  doi={10.1002/bip.1966.360040808}
}

@article{jha:2014,
  title={Kinetic evidence for a two-stage mechanism of protein denaturation by guanidinium chloride},
  author={Jha, S. K. and Marqusee, S.},
  journal={Proceedings of the National Academy of Sciences},
  volume={111},
  number={13},
  pages={4856--4861},
  year={2014},
  publisher={National Acad Sciences},
  doi={10.1073/pnas.1315453111}
}

@article{hyeon:2005,
  title={Mechanical unfolding of RNA hairpins},
  author={Hyeon, C. and Thirumalai, D.},
  journal={Proceedings of the National Academy of Sciences},
  volume={102},
  number={19},
  pages={6789--6794},
  year={2005},
  publisher={National Acad Sciences},
  doi={10.1073/pnas.0408314102}
}

@article{abkevich:1995,
  title={Impact of local and non-local interactions on thermodynamics and kinetics of protein folding},
  author={Abkevich, V. I. and Gutin, A. M. and Shakhnovich, E. I.},
  journal={Journal of molecular biology},
  volume={252},
  number={4},
  pages={460--471},
  year={1995},
  publisher={Elsevier},
  doi={10.1006/jmbi.1995.0511}
}

@article{antao:1991,
  title={A thermodynamic study of unusually stable RNA and DNA hairpins},
  author={Antao, V. P. and Lai, S. Y. and Tinoco Jr, I.},
  journal={Nucleic acids research},
  volume={19},
  number={21},
  pages={5901--5905},
  year={1991},
  publisher={Oxford University Press},
  doi={10.1093/nar/19.21.5901}
}

@article{dill:1993,
  title={Cooperativity in protein-folding kinetics.},
  author={Dill, K. A. and Fiebig, K. M. and Chan, H. S.},
  journal={Proceedings of the National Academy of Sciences},
  volume={90},
  number={5},
  pages={1942--1946},
  year={1993},
  publisher={National Acad Sciences},
  doi={10.1073/pnas.90.5.1942}
}

@article{kloss:2008,
  title={Repeat-protein folding: new insights into origins of cooperativity, stability, and topology},
  author={Kloss, E. and Courtemanche, N. and Barrick, D.},
  journal={Archives of biochemistry and biophysics},
  volume={469},
  number={1},
  pages={83--99},
  year={2008},
  publisher={Elsevier},
  doi={10.1016/j.abb.2007.08.034}
}

@article{searle:1992,
  title={The cost of conformational order: entropy changes in molecular associations},
  author={Searle, M. S. and Williams, D. H.},
  journal={Journal of the American Chemical Society},
  volume={114},
  number={27},
  pages={10690--10697},
  year={1992},
  publisher={ACS Publications},
  doi={10.1021/ja00053a002}
}

@article{baldwin:1999,
  title={Is protein folding hierarchic? I. Local structure and peptide folding},
  author={Baldwin, R. L. and Rose, G. D.},
  journal={Trends in biochemical sciences},
  volume={24},
  number={1},
  pages={26--33},
  year={1999},
  publisher={Elsevier},
  doi={10.1016/S0968-0004(98)01346-2}
}

@article{dauxois:1995,
  title={Entropy-driven transition in a one-dimensional system},
  author={Dauxois, T. and Peyrard, M.},
  journal={Physical Review E},
  volume={51},
  number={5},
  pages={4027},
  year={1995},
  publisher={APS},
  doi={10.1103/physreve.51.4027}
}

@article{puglisi:2022,
  title={On the role of temperature and elasticity in the mechanically-induced melting transition of DNA},
  author={Florio, G. and Puglisi, G.},
  journal={},
  volume={},
  number={},
  pages={},
  year={2022},
  publisher={Submitted}
}

@article{pugno:2003,
  title={Tubular adhesive joints under axial load},
  author={Pugno, N. and Carpinteri, A.},
  journal={J. Appl. Mech.},
  volume={70},
  number={6},
  pages={832--839},
  year={2003},
  doi={10.1115/1.1604835}
}

@article{balk:2001,
  title={Dynamics of chains with non-monotone stress--strain relations. I. Model and numerical experiments},
  author={Balk, A. M. and Cherkaev, A. V. and Slepyan, L. I.},
  journal={Journal of the Mechanics and Physics of Solids},
  volume={49},
  number={1},
  pages={131--148},
  year={2001},
  publisher={Elsevier},
  doi={10.1016/S0022-5096(00)00025-9}
}

@article{cherkaev:2005,
  title={Transition waves in bistable structures. I. Delocalization of damage},
  author={Cherkaev, A. and Cherkaev, E. and Slepyan, L.},
  journal={Journal of the Mechanics and Physics of Solids},
  volume={53},
  number={2},
  pages={383--405},
  year={2005},
  publisher={Elsevier},
  doi={10.1016/j.jmps.2004.08.002}
}

@article{vainchtein:2010,
  title={The role of spinodal region in the kinetics of lattice phase transitions},
  author={Vainchtein, A.},
  journal={Journal of the Mechanics and Physics of Solids},
  volume={58},
  number={2},
  pages={227--240},
  year={2010},
  publisher={Elsevier},
  doi={10.1016/j.jmps.2009.10.004}
}

@article{efendiev:2010,
  title={Thermalization of a driven bi-stable FPU chain},
  author={Efendiev, Y. R. and Truskinovsky, L.},
  journal={Continuum mechanics and thermodynamics},
  volume={22},
  number={6},
  pages={679--698},
  year={2010},
  publisher={Springer},
  doi = {10.1007/s00161-010-0166-5},
}

@article{cohen:2014,
  title={Dynamics of a discrete chain of bi-stable elements: A biomimetic shock absorbing mechanism},
  author={Cohen, T. and Givli, S.},
  journal={Journal of the Mechanics and Physics of Solids},
  volume={64},
  pages={426--439},
  year={2014},
  publisher={Elsevier},
  doi={10.1016/j.jmps.2013.12.010}
}

@article{givli:2009,
  title={A coarse-grained model of the myofibril: overall dynamics and the evolution of sarcomere non-uniformities},
  author={Givli, S. and Bhattacharya, K.},
  journal={Journal of the Mechanics and Physics of Solids},
  volume={57},
  number={2},
  pages={221--243},
  year={2009},
  publisher={Elsevier},
  doi={10.1016/j.jmps.2008.10.013}
}

@article{raj:2011,
  title={Phase boundaries as agents of structural change in macromolecules},
  author={Raj, R. and Purohit, P. K.},
  journal={Journal of the Mechanics and Physics of Solids},
  volume={59},
  number={10},
  pages={2044--2069},
  year={2011},
  publisher={Elsevier},
  doi={10.1016/j.jmps.2011.07.003}
}

@article{sinha:2005,
  title={Inequivalence of statistical ensembles in single molecule measurements},
  author={Sinha, S. and Samuel, J.},
  journal={Physical Review E},
  volume={71},
  number={2},
  pages={021104},
  year={2005},
  publisher={APS},
  doi={10.1103/PhysRevE.71.021104}
}

@article{manca:2014,
  title={On the equivalence of thermodynamics ensembles for flexible polymer chains},
  author={Manca, F. and Giordano, S. and Palla, P. L. and Cleri, F.},
  journal={Physica A: Statistical Mechanics and its Applications},
  volume={395},
  pages={154--170},
  year={2014},
  publisher={Elsevier},
  doi={10.1016/j.physa.2013.10.042}
}

@article{evans:1999,
  title={Strength of a weak bond connecting flexible polymer chains},
  author={Evans, E. and Ritchie, K.},
  journal={Biophysical Journal},
  volume={76},
  number={5},
  pages={2439--2447},
  year={1999},
  publisher={Elsevier},
  doi={10.1016/S0006-3495(99)77399-6}
}

@article{herrmann:2012,
  title={Kramers and non-Kramers phase transitions in many-particle systems with dynamical constraint},
  author={Herrmann, M. and Niethammer, B. and Vel{\'a}zquez, J. J. L.},
  journal={Multiscale Modeling \& Simulation},
  volume={10},
  number={3},
  pages={818--852},
  year={2012},
  publisher={SIAM},
  doi={10.1137/110851882}
}

@article{evans:1991,
  title={Detachment of agglutinin-bonded red blood cells. I. Forces to rupture molecular-point attachments},
  author={Evans, E. and Berk, D. and Leung, A.},
  journal={Biophysical journal},
  volume={59},
  number={4},
  pages={838--848},
  year={1991},
  publisher={Elsevier},
  doi={10.1016/S0006-3495(91)82296-2}
}

@article{dembo:1988,
  title={The reaction-limited kinetics of membrane-to-surface adhesion and detachment},
  author={Dembo, M. and Torney, D. C. and Saxman, K. and Hammer, D.},
  journal={Proceedings of the Royal Society of London. Series B. Biological Sciences},
  volume={234},
  number={1274},
  pages={55--83},
  year={1988},
  publisher={The Royal Society London},
  doi={10.1098/rspb.1988.0038}
}

@article{hyeon:2006,
  title={Forced-unfolding and force-quench refolding of RNA hairpins},
  author={Hyeon, C. and Thirumalai, D.},
  journal={Biophysical journal},
  volume={90},
  number={10},
  pages={3410--3427},
  year={2006},
  publisher={Elsevier},
  doi={10.1529/biophysj.105.078030}
}

@article{koculi:2006,
  title={Counterion charge density determines the position and plasticity of RNA folding transition states},
  author={Koculi, E. and Thirumalai, D. and Woodson, S. A.},
  journal={Journal of molecular biology},
  volume={359},
  number={2},
  pages={446--454},
  year={2006},
  publisher={Elsevier},
  doi={10.1016/j.jmb.2006.03.031}
}

@article{izrailev:1997,
  title={Molecular dynamics study of unbinding of the avidin-biotin complex},
  author={Izrailev, S. and Stepaniants, S. and Balsera, M. and Oono, Y. and Schulten, K.},
  journal={Biophysical journal},
  volume={72},
  number={4},
  pages={1568--1581},
  year={1997},
  publisher={Elsevier},
  doi={10.1016/S0006-3495(97)78804-0}
}

@article{shapiro:1997,
  title={A quantitative analysis of single protein-ligand complex separation with the atomic force microscope},
  author={Shapiro, B. E. and Qian, H.},
  journal={Biophysical chemistry},
  volume={67},
  number={1-3},
  pages={211--219},
  year={1997},
  publisher={Elsevier},
  doi={10.1016/S0301-4622(97)00045-8}
}

@article{goriely:2015,
  title={Mechanics of the brain: perspectives, challenges, and opportunities},
  author={Goriely, A. and Geers, M. G. D. and Holzapfel, G. A. and Jayamohan, J. and J{\'e}rusalem, A. and Sivaloganathan, S. and Squier, W. and van Dommelen, J. A. W. and Waters, S. and Kuhl, E.},
  journal={Biomechanics and modeling in mechanobiology},
  volume={14},
  number={5},
  pages={931--965},
  year={2015},
  publisher={Springer},
  doi={10.1007/s10237-015-0662-4}
}

@article{ahmed:2013,
  title={Neuromechanics: the role of tension in neuronal growth and memory},
  author={Ahmed, W. W. and Rajagopalan, J. and Tofangchi, A. and Saif, T. A.},
  journal={Nano and cell mechanics: fundamentals and frontiers},
  pages={35--61},
  year={2013},
  publisher={Wiley Online Library},
  doi={10.1002/9781118482568.ch3}
}

@article{otoole:2008,
  title={A physical model of axonal elongation: force, viscosity, and adhesions govern the mode of outgrowth},
  author={O’Toole, M. and Lamoureux, P. and Miller, K. E.},
  journal={Biophysical journal},
  volume={94},
  number={7},
  pages={2610--2620},
  year={2008},
  publisher={Elsevier},
  doi={10.1529/biophysj.107.117424}
}

@article{otoole:2015,
  title={Measurement of subcellular force generation in neurons},
  author={O’Toole, M. and Lamoureux, P. and Miller, K. E.},
  journal={Biophysical journal},
  volume={108},
  number={5},
  pages={1027--1037},
  year={2015},
  publisher={Elsevier},
  doi={10.1016/j.bpj.2015.01.021}
}

@article{soheilypour:2015,
  title={Buckling behavior of individual and bundled microtubules},
  author={Soheilypour, M. and Peyro, M. and Peter, S. J. and Mofrad, M. R. K.},
  journal={Biophysical journal},
  volume={108},
  number={7},
  pages={1718--1726},
  year={2015},
  publisher={Elsevier},
  doi={10.1016/j.bpj.2015.01.030}
}

@article{peter:2012,
  title={Computational modeling of axonal microtubule bundles under tension},
  author={Peter, S. J. and Mofrad, M. R. K.},
  journal={Biophysical journal},
  volume={102},
  number={4},
  pages={749--757},
  year={2012},
  publisher={Elsevier},
  doi={10.1016/j.bpj.2011.11.4024}
}

@article{kasas:2004,
  title={Oscillation modes of microtubules},
  author={Kasas, S. and Cibert, C. and Kis, A. and De Los Rios, P. and Riederer, B. M. and Forro, L. and Dietler, G. and Catsicas, S.},
  journal={Biology of the Cell},
  volume={96},
  number={9},
  pages={697--700},
  year={2004},
  publisher={Elsevier},
  doi={10.1016/j.biolcel.2004.09.002}
}

@article{ahmadzadeh:2014,
  title={Viscoelasticity of tau proteins leads to strain rate-dependent breaking of microtubules during axonal stretch injury: predictions from a mathematical model},
  author={Ahmadzadeh, H. and Smith, D. H. and Shenoy, V. B.},
  journal={Biophysical journal},
  volume={106},
  number={5},
  pages={1123--1133},
  year={2014},
  publisher={Elsevier},
  doi={10.1016/j.bpj.2014.01.024}
}

@article{jakobs:2015,
  title={Force generation by molecular-motor-powered microtubule bundles; implications for neuronal polarization and growth},
  author={Jakobs, M. and Franze, K. and Zemel, A.},
  journal={Frontiers in Cellular Neuroscience},
  volume={9},
  pages={441},
  year={2015},
  publisher={Frontiers},
  doi={10.3389/fncel.2015.00441}
}

@article{suter:2011,
  title={The emerging role of forces in axonal elongation},
  author={Suter, D. M. and Miller, K. E.},
  journal={Progress in neurobiology},
  volume={94},
  number={2},
  pages={91--101},
  year={2011},
  publisher={Elsevier},
  doi={10.1016/j.pneurobio.2011.04.002}
}

@incollection{howard:2002,
  title={Mechanics of motor proteins},
  author={Howard, J.},
  booktitle={Physics of bio-molecules and cells. Physique des biomol{\'e}cules et des cellules},
  pages={69--94},
  year={2002},
  publisher={Springer},
  doi={10.1007/3-540-45701-1_2}
}

@article{ballatore:2007,
  title={Tau-mediated neurodegeneration in Alzheimer's disease and related disorders},
  author={Ballatore, C. and Lee, V. M. Y. and Trojanowski, J. Q.},
  journal={Nature reviews neuroscience},
  volume={8},
  number={9},
  pages={663--672},
  year={2007},
  publisher={Nature Publishing Group},
  doi={10.1038/nrn2194}  
}

@article{anderson:2013,
  title={Single molecule force spectroscopy on titin implicates immunoglobulin domain stability as a cardiac disease mechanism},
  author={Anderson, B. R. and Bogomolovas, J. and Labeit, S. and Granzier, H.},
  journal={Journal of Biological Chemistry},
  volume={288},
  number={8},
  pages={5303--5315},
  year={2013},
  publisher={ASBMB},
  doi={10.1074/jbc.M112.401372}
}

@article{hu:1996,
  title={Analytical inversion of symmetric tridiagonal matrices},
  author={Hu, G. Y. and O'Connell, R. F.},
  journal={Journal of Physics A: Mathematical and General},
  volume={29},
  number={7},
  pages={1511},
  year={1996},
  publisher={IOP Publishing},
  doi={10.1088/0305-4470/29/7/020}
}

@article{nabben:1999,
  title={Decay rates of the inverse of nonsymmetric tridiagonal and band matrices},
  author={Nabben, R.},
  journal={SIAM Journal on Matrix Analysis and Applications},
  volume={20},
  number={3},
  pages={820--837},
  year={1999},
  publisher={SIAM},
  doi={10.1137/S0895479897317259}
}

@article{nabben:1999:2,
  title={Two-sided bounds on the inverses of diagonally dominant tridiagonal matrices},
  author={Nabben, R.},
  journal={Linear Algebra and its Applications},
  volume={287},
  number={1-3},
  pages={289--305},
  year={1999},
  publisher={Elsevier},
  doi={10.1016/S0024-3795(98)10146-5}
}

@article{ahmadzadeh:2015,
  title={Mechanical effects of dynamic binding between tau proteins on microtubules during axonal injury},
  author={Ahmadzadeh, H. and Smith, D. H. and Shenoy, V. B.},
  journal={Biophysical journal},
  volume={109},
  number={11},
  pages={2328--2337},
  year={2015},
  publisher={Elsevier},
  doi={10.1016/j.bpj.2015.09.010}
}

@article{lavery:2002,
  title={Structure and mechanics of single biomolecules: experiment and simulation},
  author={Lavery, R. and Lebrun, A. and Allemand, J.-F. and Bensimon, D. and Croquette, V.},
  journal={Journal of Physics: Condensed Matter},
  volume={14},
  number={14},
  pages={R383},
  year={2002},
  publisher={IOP Publishing},
  doi={10.1088/0953-8984/14/14/202}
}

@article{chakrabarti:2009,
  title={Shear unzipping of DNA},
  author={Chakrabarti, B. and Nelson, D. R.},
  journal={The Journal of Physical Chemistry B},
  volume={113},
  number={12},
  pages={3831--3836},
  year={2009},
  publisher={ACS Publications},
  doi={10.1021/jp808232p}
}

@article{kuhl:2018,
  title={Microtubule polymerization and cross-link dynamics explain axonal stiffness and damage},
  author={de Rooij, R. and Kuhl, E.},
  journal={Biophysical journal},
  volume={114},
  number={1},
  pages={201--212},
  year={2018},
  publisher={Elsevier},
  doi={10.1016/j.bpj.2017.11.010}
}

@article{kuhl:2018:2,
  title={Physical biology of axonal damage},
  author={De Rooij, R. and Kuhl, E.},
  journal={Frontiers in cellular neuroscience},
  volume={12},
  pages={144},
  year={2018},
  publisher={Frontiers},
  doi={10.3389/fncel.2018.00144}
}

@article{truskinovsky:1996,
  title={Ericksen's bar revisited: Energy wiggles},
  author={Truskinovsky, L. and Zanzotto, G.},
  journal={Journal of the Mechanics and Physics of Solids},
  volume={44},
  number={8},
  pages={1371--1408},
  year={1996},
  publisher={Elsevier},
  doi={10.1016/0022-5096(96)00020-8}
}

@article{fuina:2016,
  title={Thermo-mechanical response of rigid plastic laminates for greenhouse covering},
  author={Fuina, S. and Marano, G. C. and Puglisi, G. and De Tommasi, D. and Scarascia-Mugnozza, G.},
  journal={Journal of Agricultural Engineering},
  volume={47},
  number={3},
  pages={157--163},
  year={2016},
  doi={10.4081/jae.2016.549}
}

@article{fuina:2019,
  title={Polycarbonate laminates thermo-mechanical behaviour under different operating temperatures},
  author={Fuina, S. and Marano, G. C. and Scarascia-Mugnozza, G.},
  journal={Polymer Testing},
  volume={76},
  pages={344--349},
  year={2019},
  publisher={Elsevier},
  doi={10.1016/j.polymertesting.2019.03.031}
}

@article{shi:2021,
  title={Structure-based classification of tauopathies},
  author={Shi, Ya. and Zhang, W. and Yang, Y. and Murzin, A. G. and Falcon, B. and Kotecha, A. and van Beers, M. and Tarutani, A. and Kametani, F. and Garringer, H. J. and others},
  journal={Nature},
  volume={598},
  number={7880},
  pages={359--363},
  year={2021},
  publisher={Nature Publishing Group},
  doi={10.1038/s41586-021-03911-7}
}

@article{asken:2017,
  title={Research gaps and controversies in chronic traumatic encephalopathy: a review},
  author={Asken, B. M. and Sullan, M. J. and DeKosky, S. T. and Jaffee, M. S. and Bauer, R. M.},
  journal={JAMA neurology},
  volume={74},
  number={10},
  pages={1255--1262},
  year={2017},
  publisher={American Medical Association},
  doi={10.1001/jamaneurol.2017.2396}
}

@article{eisenberg:2017,
  title={Taming tangled tau},
  author={Eisenberg, D. S. and Sawaya, M. R.},
  journal={Nature},
  volume={547},
  number={7662},
  pages={170--171},
  year={2017},
  publisher={Nature Publishing Group},
  doi={10.1038/nature23094}
}

@article{ouyang:2013,
  title={Contribution of cytoskeletal elements to the axonal mechanical properties},
  author={Ouyang, H. and Nauman, E. and Shi, R.},
  journal={Journal of biological engineering},
  volume={7},
  number={1},
  pages={1--8},
  year={2013},
  publisher={BioMed Central},
  doi={10.1186/1754-1611-7-21}
}

@article{dubey:2020,
  title={The axonal actin-spectrin lattice acts as a tension buffering shock absorber},
  author={Dubey, S. and Bhembre, N. and Bodas, S. and Veer, S. and Ghose, A. and Callan-Jones, A. and Pullarkat, P.},
  journal={Elife},
  volume={9},
  pages={e51772},
  year={2020},
  publisher={eLife Sciences Publications Limited},
  doi={10.7554/eLife.51772}
}

@article{elbaum:2012,
  title={Identification of an aggregation-prone structure of tau},
  author={Elbaum-Garfinkle, S. and Rhoades, E.},
  journal={Journal of the American Chemical Society},
  volume={134},
  number={40},
  pages={16607--16613},
  year={2012},
  publisher={ACS Publications},
  doi={10.1021/ja305206m}
}

@article{cowan:2013,
  title={Are tau aggregates toxic or protective in tauopathies?},
  author={Cowan, C. M. and Mudher, A.},
  journal={Frontiers in neurology},
  volume={4},
  pages={114},
  year={2013},
  publisher={Frontiers},
  doi={10.3389/fneur.2013.00114}
}

@article{hawkins:2010,
  title={Mechanics of microtubules},
  author={Hawkins, T. and Mirigian, M. and Yasar, M. S. and Ross, J. L.},
  journal={Journal of biomechanics},
  volume={43},
  number={1},
  pages={23--30},
  year={2010},
  publisher={Elsevier},
  doi={10.1016/j.jbiomech.2009.09.005}
}

@article{li:2015,
  title={Tau binds to multiple tubulin dimers with helical structure},
  author={Li, X.-H. and Culver, J. A. and Rhoades, E.},
  journal={Journal of the American Chemical Society},
  volume={137},
  number={29},
  pages={9218--9221},
  year={2015},
  publisher={ACS Publications},
  doi={10.1021/jacs.5b04561}
}

@article{morris:2011,
  title={The many faces of tau},
  author={Morris, M. and Maeda, S. and Vossel, K. and Mucke, L.},
  journal={Neuron},
  volume={70},
  number={3},
  pages={410--426},
  year={2011},
  publisher={Elsevier},
  doi={10.1016/j.neuron.2011.04.009}
}

@book{zinn:1996,
  title = {Quantum field theory and critical phenomena},
  author = {Zinn-Justin, J.},
  publisher = {Oxford : Clarendon Press},
  year = {1996},
  isbn={978-0198509233}
}

@article{luca2019,
    author = {Bellino, L. and Florio, G. and Puglisi, G.},
    title = {The influence of device handles in single molecule experiments},
    journal = {Soft Matter},
    year = {2019},
    volume = {15},
    pages = {8680--8690},
    number = {43},
    doi = {10.1039/c9sm01376h}
}

@article{luca2020,
  	title={On the competition between interface energy and temperature in phase transition phenomena},
  	author={Bellino, L. and Florio, G. and Giordano, S. and Puglisi, G.},
  	journal={Applications in Engineering Science},
  	volume={2},
  	pages={100009},
  	year={2020},
  	publisher={Elsevier},
	doi = {10.1016/j.apples.2020.100009}
}

@article{luca2022:1,
  	title={Thermal control of nucleation and propagation transition stresses in discrete lattices with non-local interactions and non-convex energy},
  	author={Cannizzo, A. and Bellino, L. and Florio, G. and Puglisi, G. and Giordano, S.},
  	journal={Accepted on EPJ Plus},
  %	volume={2},
  %	pages={100009},
  	year={2022},
  %	publisher={Elsevier},
  %	doi = {10.1016/j.apples.2020.100009}
}

@article{luca2022:2,
  	title={On the role of local interaction in cooperativity of double stranded peptide chains},
  	author={Bellino, L. and Florio, G. and Goriely, A. and Puglisi, G.},
  	journal={Preprint},
  %	volume={2},
  %	pages={100009},
  	year={2022},
  %	publisher={Elsevier},
  %	doi = {10.1016/j.apples.2020.100009}
}

@article{luca2022:3,
  	title={Effects of rate and temperature on the mechanical response of microtubules and tau proteins},
  	author={Bellino, L. and Florio, G. and Goriely, A. and Puglisi, G.},
  	journal={Preprint},
  %	volume={2},
  %	pages={100009},
  	year={2022},
  %	publisher={Elsevier},
  %	doi = {10.1016/j.apples.2020.100009}
}



\appendix



\renewcommand{\thefigure}
{A.\arabic{figure}}
\setcounter{figure}{0}

\renewcommand{\theequation}
{A.\arabic{equation}}
\setcounter{equation}{0}

\chapter{Relations between Helmholtz and Gibbs ensembles}
\label{appA}

\vspace{0.5cm}
\noindent\textsc{Hamiltonians}\\

Let us consider the Hamiltonian energy of the case of applied displacement $\delta$, corresponding to the Helmholtz ensemble (subscript $\mathscr{H}$) such that $\varphi_{\mathscr{H}}=\varphi_{\mathscr{H}}(\delta)$. We obtain the force conjugated to the displacement as 
\begin{equation}
f=\frac{\partial \varphi_{\mathscr{H}}(\delta)}{\partial \delta},
\label{eq:appA_force}
\end{equation}
which provided the functional relationship 
\begin{equation}
f=f(\delta),
\end{equation}
with inverse
\begin{equation}
\delta=\delta(f).
\label{eq:appA_delta}
\end{equation}
Then, the Hamiltonian energy in the Gibbs framework (subscript $\mathscr{G}$) is the Legendre transform of $\varphi_{\mathscr{H}}$ and is the function defined as (\cite{callen:1960,arnold:2013})
\begin{equation}
\varphi_{\mathscr{G}}(f)=\varphi_{\mathscr{H}}(\delta)-\delta\,f,
\end{equation}
with $\delta$ given by~\eqref{eq:appA_delta}. Now consider an arbitrary variation $\text{d}f$, so that
\begin{equation}
\text{d}\varphi_{\mathscr{G}}=\frac{\partial \varphi_{\mathscr{G}}}{\partial f}\text{d}f=\frac{\partial \varphi_{\mathscr{H}}}{\partial \delta}\text{d}\delta-\left(f \text{d}\delta+\delta\text{d}f\right)=-\delta\text{d}f,
\end{equation}
and by using~\eqref{eq:appA_force} we obtain
\begin{equation}
\delta=-\frac{\partial \varphi_{\mathscr{G}}(f)}{\partial f},
\end{equation}
that provide an explicit interpretation of~\eqref{eq:appA_delta}.

\vspace{0.5cm}
\noindent\textsc{Partition functions}\\

Consider now the partition function of the Helmholtz ensemble, that reads
\begin{equation}
\mathcal{Z}_{\mathscr{H}}=\int\,e^{-\beta \varphi_{\mathscr{H}}(\boldsymbol{\varepsilon},\delta)}\text{d}\boldsymbol{\varepsilon}.
\end{equation}
The Gibbs partition function is given by
\begin{equation}
\mathcal{Z}_{\mathscr{G}}=\int\mathcal{Z}_{\mathscr{H}}\,e^{\beta \delta f}\text{d}\boldsymbol{\delta}.
\end{equation}
On the other hand, let us start from the Gibbs partition function. We may evaluate the inverse Laplace transform as a Fourier transform by performing the analytical continuation $f\rightarrow i\omega$ thus obtaining 
\begin{equation}
\mathcal{Z}_{\mathscr{H}}=\int\mathcal{Z}_{\mathscr{G}}e^{-\beta i \omega \delta}\text{d}\boldsymbol{\omega}.
\end{equation}
%



\renewcommand{\thefigure}
{B.\arabic{figure}}
\setcounter{figure}{0}

\renewcommand{\theequation}
{B.\arabic{equation}}
\setcounter{equation}{0}

\chapter{Gaussian integration}
\label{appB}

A Gaussian integral is defined as
\begin{equation}
G=\int_{-\infty}^{+\infty} e^{-x^{2}}\text{d}x.
\label{eq:appD_g}
\end{equation}
The integral in~\eqref{eq:appB_g}  can be solved with the following procedure (\cite{zinn:1996}). Consider the square of $G$ and let us compute 
\begin{equation}
	\begin{split}
	\left(\int_{-\infty}^{+\infty} e^{-x^{2}}\text{d}x\right)^{2}=&=\int_{-\infty}^{+\infty} e^{-x^{2}}\text{d}x\int_{-\infty}^{+\infty} e^{-y^{2}}\text{d}y= \int_{-\infty}^{+\infty} \int_{-\infty}^{+\infty} e^{-\left(x^{2}+y^{2}\right)}\text{d}x\,\text{d}y.
	\end{split}
\end{equation}
Let us now switch to polar coordinates, being $x^{2}+y^{2}=r^{2}$, then obtaining
\begin{equation}
	\begin{split}
	\int_{-\infty}^{+\infty} \int_{-\infty}^{+\infty} e^{-\left(x^{2}+y^{2}\right)}\text{d}x\,\text{d}y=&\int_{0}^{2\pi} \int_{0}^{+\infty} e^{-r^{2}}r\text{d}r\,\text{d}\theta=2\pi\int_{0}^{+\infty} e^{-r^{2}}r\text{d}r=\pi
	\end{split}
\end{equation}
Finally, being $G^{2}=\pi$ we get 
\begin{equation}
G=\int_{-\infty}^{+\infty} e^{-x^{2}}\text{d}x=\sqrt{\pi}.
\label{eq:appB_g}
\end{equation}
Let us now consider some possible generalization of the Gaussian integral that are used in the thesis. 

\vspace{0.5cm}
\noindent\textsc{Second order polynomial function}\\

Consider the function $-ax^2+bx+c$ with $a,b,c$ real constants and let $x$ be the variable. The integral has solution
\begin{equation}
\int e^{-ax^{2}+bx+c}\text{d}x=\sqrt{\frac{\pi}{a}}\,e^{\,\frac{b^2}{4a}+c}
\end{equation}

\vspace{0.5cm}
\noindent\textsc{$n$-dimensional function}\\

Consider $\boldsymbol{A}$ an $n$-dimensional  positive definite matrix, $\boldsymbol{b}\in \|\!R^n$ a vector with $n$ real elements and $\boldsymbol{x}$ the vector of the variables. Moreover, there can be a constant real term $c$. The integral has solution
\begin{equation}
\int_{\mathbb{R}^{n}} e^{-\frac{1}{2}\boldsymbol{A}\boldsymbol{x}\cdot\boldsymbol{x}+\boldsymbol{b}\cdot\boldsymbol{x}+c}\text{d}\boldsymbol{x}=\sqrt{\frac{(2\pi)^{n}}{\text{det}(\boldsymbol{A})}}\,e^{\,\frac{1}{2}\boldsymbol{A}^{-1}\boldsymbol{b}\cdot{\boldsymbol{b}}+c}
\end{equation}
%




\renewcommand{\thefigure}
{C.\arabic{figure}}
\setcounter{figure}{0}

\renewcommand{\theequation}
{C.\arabic{equation}}
\setcounter{equation}{0}

\chapter{Partition functions for the chain made by 3 units}
\label{appC}

In this appendix we report the calculation to evaluate the partition functions of the prototypical example in section~\ref{sec:ch2_pt3}, where temperature effects are introduce. Starting from the Gibbs ensemble we also show how it is possible t obtain the Helmholtz partition function with an inverse Laplace transform perfumed as Fourier transform in the complex plane.

%
\vspace{0.5cm}
\noindent\textsc{Gibbs ensemble}\\

Let us consider the energy of the system in the case of applied force (soft device hypothesis) introduced in~\eqref{eq:ch2_g}. By definition, the partition function for the Gibbs ensemble reads
\begin{equation}
\mathcal{Z}_{\mathscr{G}}(f)=\sum_{\boldsymbol{\chi}}\int_{\mathbb{R}^{3}}e^{\,-\beta g}\text{d}\boldsymbol{\varepsilon}=\sum_{\boldsymbol{\chi}}\int_{\mathbb{R}^{3}}e^{\,-\beta\left(\varphi-f\delta\right)}\text{d}\boldsymbol{\varepsilon},
\end{equation}
where $\varphi$ is the energy in~\ref{eq:ch2_matrixenergy}, $\beta=1/(k_B T)$, $\boldsymbol{\chi}=\{\chi_1,\chi_2,\chi_3\}$ is the vector denoting the phase configuration and $f$ is the assigned force. Let us now recall that the total displacement is $\delta=\boldsymbol{\varepsilon}\cdot\boldsymbol{1}$, thus we obtain 
\begin{equation}
\mathcal{Z}_{\mathscr{G}}(f)=\sum_{\boldsymbol{\chi}}\int_{\mathbb{R}^{3}}e^{\,-\frac{\beta}{2}\boldsymbol{\varepsilon}\cdot\boldsymbol{\varepsilon}+\beta\left(\varepsilon_u\boldsymbol{\chi}+f\boldsymbol{1}\right)\cdot\boldsymbol{\varepsilon}-\frac{\beta}{2}\varepsilon_u^2\boldsymbol{\chi}\cdot\boldsymbol{\chi}}\,\text{d}\boldsymbol{\varepsilon}.
\label{eq:appC_int2}
\end{equation}
By performing a direct evaluation of the Gaussian integral in~\eqref{eq:appC_int2} as shown in Appendix~\ref{appB} we obtain 
\begin{equation}
	\mathcal{Z}_{\mathscr{G}}(f)=(2\pi)^{\frac{3}{2}}\sum_{\boldsymbol{\chi}}e^{\,\frac{\beta}{2}\left(\varepsilon_u\boldsymbol{\chi}+f\boldsymbol{1}\right)\cdot\left(\varepsilon_u\boldsymbol{\chi}+f\boldsymbol{1}\right)-\frac{\beta}{2}\varepsilon_u^{2}\boldsymbol{\chi}\cdot\boldsymbol{\chi}}.
\label{eq:appC_int3}
\end{equation}
We may also observe that the energy of the solutions with the same unfolded fraction is invariant with respect to the permutation of the elements. Thus we may use the binomial function to compute all the possible configurations attained at varying $\boldsymbol{\chi}$, such that we get 
\begin{equation}
\mathcal{Z}_{\mathscr{G}}(f)=\left(2\pi\right)^{\frac{3}{2}}\sum_{p=0}^{n}\binom{n}{p}e^{\,\frac{n\beta}{2}f^{2}+\beta\varepsilon_u p f},
\end{equation}
the can be summed, giving the canonical partition function in the Gibbs ensemble
\begin{equation}
\mathcal{Z}_{\mathscr{G}}(f)=\left(2\pi\right)^{\frac{3}{2}}\left(1+e^{\,\beta\varepsilon_u f}\right)^{n}e^{\,\frac{n\beta}{2}f^{2}}.
\label{eq:appC_zg}
\end{equation}
%

%
\vspace{0.5cm}
\noindent\textsc{Helmholtz ensemble}\\

The partition function in the Helmholtz ensemble, that correctly suits the description of the hard device hypothesis with fixed displacement, can be obtained from the Gibbs one through an inverse Laplace transform. In particular, by performing the change of variable $f\to i\omega$, it is possible to compute a Fourier transform in the complex plane instead of the Laplace one. Thus, we may write
\begin{equation}
\mathcal{Z}_{\mathscr{H}}(\delta)=\int_{-\infty}^{+\infty}\mathcal{Z}_{\mathscr{G}}(f)\,e^{\,-\beta i \omega\delta}\text{d}\omega,
\end{equation}
where $\mathcal{Z}_{\mathscr{G}}(f)$ is given by~\eqref{eq:appC_int3} and $\delta$ is the fixed displacement. Following, we obtain 
%
\begin{equation}
\mathcal{Z}_{\mathscr{H}}(\delta)=(2\pi)^{\frac{3}{2}}\sum_{\boldsymbol{\chi}}\int_{-\infty}^{+\infty}\,e^{\,-\frac{\beta}{2}\omega^{2}\boldsymbol{1}\cdot\boldsymbol{1}+\beta\left(\varepsilon_u \boldsymbol{\chi}\cdot\boldsymbol{1}-\delta\right)i \omega}\text{d}\omega.
\end{equation}
The integral can be solved by performing the square completion and then the integration in the complex plane. By defining
\begin{equation}
a=\frac{\boldsymbol{1}\cdot\boldsymbol{1}}{n}, \qquad b=\varepsilon_u \boldsymbol{\chi}\cdot\boldsymbol{1}-\delta,
\end{equation}
the integral reads 
\begin{equation}
	\begin{split}
\mathcal{Z}_{\mathscr{H}}(\delta)&=(2\pi)^{\frac{3}{2}}\sum_{\boldsymbol{\chi}}\int_{-\infty}^{+\infty}\,e^{\,-\beta a \omega^2+\beta i b \omega}\text{d}\omega\\
	&=(2\pi)^{\frac{3}{2}}\sum_{\boldsymbol{\chi}}\int_{-\infty}^{+\infty}\,e^{\,-\beta a \left(\omega^2-\frac{i b}{a}\omega\right)}\text{d}\omega\\
	&=(2\pi)^{\frac{3}{2}}\sum_{\boldsymbol{\chi}}\int_{-\infty}^{+\infty}\,e^{\,-\beta a \left(\omega^2-\frac{i b}{a}\omega+\frac{b^2}{4a^2}-\frac{b^2}{4a^2}\right)}\text{d}\omega\\
	&=(2\pi)^{\frac{3}{2}}\sum_{\boldsymbol{\chi}}\int_{-\infty}^{+\infty}\,e^{\,-\beta a \left(\omega-\frac{i b}{2a}\right)^{2}-\frac{b^2}{4a}}\text{d}\omega\\
	&=(2\pi)^{\frac{3}{2}}\sum_{\boldsymbol{\chi}}e^{\,-\frac{b^2}{4a}}\int_{-\infty}^{+\infty}\,e^{\,-\beta a \left(\omega-\frac{i b}{2a}\right)^{2}}\text{d}\omega.
	\end{split}
\end{equation}
The integral can be solved with the change of variable $\omega-ib/(2a)=x$ in the complex plane, and by substituting the values of $a$ and $b$ previously defined we finally get 
\begin{equation}
\mathcal{Z}_{\mathscr{H}}(\delta)=(2\pi)^{\frac{3}{2}}\sqrt{\frac{\pi}{2\beta\boldsymbol{1}\cdot\boldsymbol{1}}}\sum_{\boldsymbol{\chi}}\,e^{\,-\beta\frac{\left(\delta-\varepsilon_u\boldsymbol{\chi}\cdot\boldsymbol{1}\right)^2}{2\boldsymbol{1}\cdot\boldsymbol{1}}}.
\end{equation}
Finally, counting all the possible configuration at varying phase we get the Helmholtz partition function: 
\begin{equation}
\mathcal{Z}_{\mathscr{H}}(\delta)=\frac{2\pi^2}{n\beta}\sum_{p=0}^{n=3}\binom{n}{p}\,e^{\,-\frac{\beta n}{2}\left(\delta-\varepsilon_u\frac{p}{n}\right)^2}.
\end{equation}
%




\renewcommand{\thefigure}
{D.\arabic{figure}}
\setcounter{figure}{0}

\renewcommand{\theequation}
{D.\arabic{equation}}
\setcounter{equation}{0}

\chapter{Numerical approximation of the bi-parabolic energy at high temperatures}
\label{appD}
\vspace{1.2cm}

\begin{figure}[h]
\begin{center}
\includegraphics[width=0.95\textwidth]{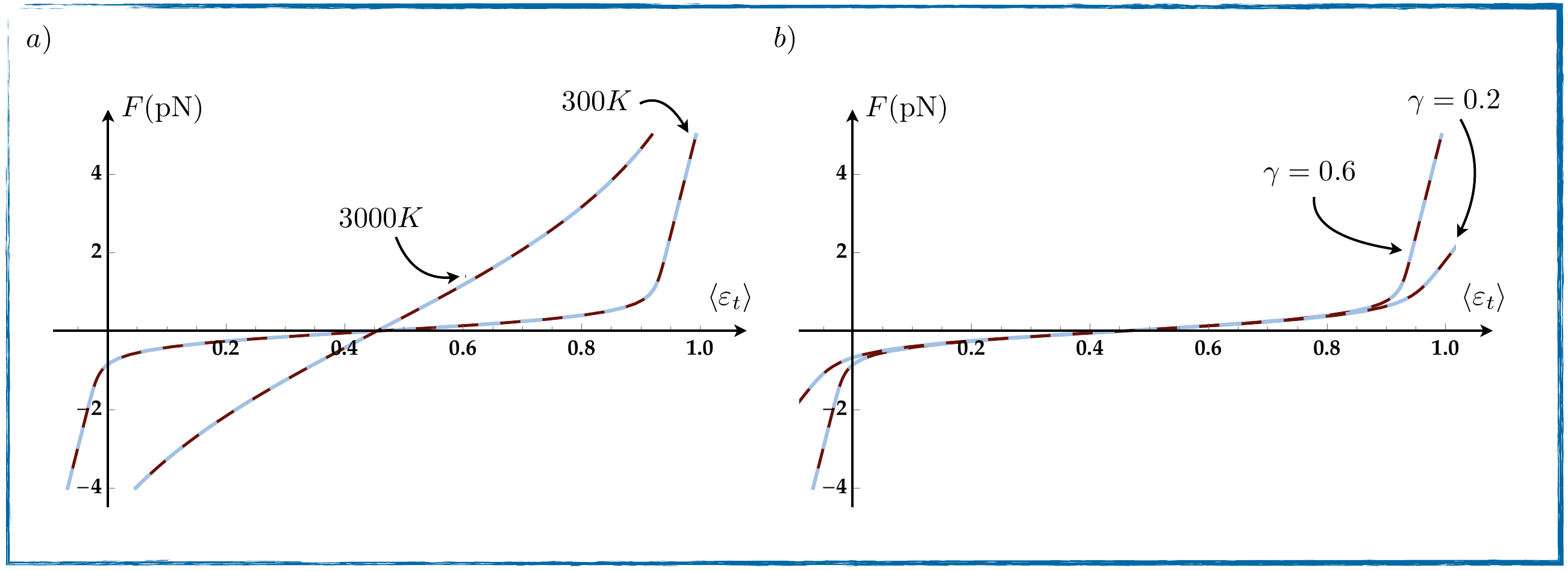}
\caption[Numerical approximation of the bi-parabolic energy at high temperatures]{Comparison of the force-strain curves obtained using the analytical formulas~\eqref{eq:ch3_gfet} and~\eqref{eq:ch3_gchi} and the numerical integration of the partition function without the approximation beyond the spinodal point described in the text. Panel a): We have considered two different temperatures i.e. $T=300\,\mbox{K}$, $T=3000\,\mbox{K}$ with $\gamma=0.6$. Panel b): We have considered two different spring constant of the measuring apparatus keeping $k_m$ fixed i.e. $\gamma=0.6$, $\gamma=0.2$ with $T=300\,\mbox{K}$. In both cases we have fixed $N=1$ with $l=20 \,\mbox{nm}, \varepsilon_u=1, k_m=90\,\mbox{pN}$, $\alpha=0.1$.}
\label{fig:appD_numeric1}
\end{center}
\end{figure}

In this Appendix, we numerically verify the approximation used in the paper, i.e. the extension of the energy (parabolic) function beyond the spinodal point. We focus on the case of the Gibbs ensemble whereas the results for the Helmholtz ensemble can be found in (\cite{fp:2019}).  In Figure \ref{fig:appD_numeric1}$_a$ we compare the force strain curves obtained via Equations~\eqref{eq:ch3_gfet} and~\eqref{eq:ch3_gchi} of the main paper and via the integration of the partition function without approximation depending on the temperature. In particular, we have considered two different temperatures, $T=300\mbox{K}$ and $T=3000\mbox{K}$. In both cases, the curves obtained analytically (continuous line) and numerically (dashed line) are perfectly coincident. We deduce that the approximation used is very robust. As a second check, we test the approximation for different values of the spring constant of the measurement apparatus, i.e. we choose different values of $\gamma$ keeping $k_m$ fixed. The results are shown in Fig.~\ref{fig:appD_numeric1}$_b$. Also in this case, by superposing the numerical and analytical results we observe that they are perfectly superimposed.

%
%



\renewcommand{\thefigure}
{E.\arabic{figure}}
\setcounter{figure}{0}

\renewcommand{\theequation}
{E.\arabic{equation}}
\setcounter{equation}{0}

\chapter{Explicit calculation of Chapter~\ref{ch_3}}
\label{appE}

In this Appendix, I report the analytical details of the model proposed in Chapter~\ref{ch_3}. Here I use the same notation introduced in the chapter to denote with the subscripts $m$, $d$, and $t$ the variables referred to the molecule, the device and the total (molecule plus device) system, respectively.

\section{Hamiltonian of the system and basic definitions}

The system is composed of $n$ mass points with mass $m$ connected by bistable springs with modulus $k_m$ and a loading device with mass $m$ represented as an $n+1$ spring with modulus $k_d$ (see Figure~\ref{fig:ch3_model}). The Hamiltonian function can be written as the sum of kinetic and elastic energy 
\begin{equation}
\mathcal{H} = E_{K} + \Phi_{t} = \sum_{i=1}^{n}\frac{1}{2 m}p_i^2+\frac{1}{2M}p^2_{n+1}+\sum_{i=1}^{n}\frac{1}{2}k_m l (\varepsilon_i-\varepsilon_u\chi_i)^2+\frac{1}{2}k_d \alpha L \varepsilon_d^2, 
\label{H}
\end{equation}
where $\varepsilon_u$ is the reference strain of the unfolded configuration, $l = L/n$ is the reference length of each element, $\alpha L \varepsilon_d$ is the elongation of the device and $p_i$ is the momentum of the $i$-th oscillator. Here $\chi_i$ is an internal variable that can assume values $0$ or $1$ if the $i$-th element is \textit{folded} or \textit{unfolded}, respectively. The total displacement can be expressed as
\begin{equation}
d = \sum_{i=1}^{n}l\varepsilon_i+\alpha L \varepsilon_d = L(\varepsilon_m +\alpha \varepsilon_d),
\label{eq:2}
\end{equation}
where $\varepsilon_m$ is the average strain of the molecule 
\begin{equation}
\varepsilon_m=\frac{1}{n}\sum_{i=1}^{n}\varepsilon_i.
\label{average}
\end{equation}
The relation between $\varepsilon_m$ and the total strain $\varepsilon_t$ is the following:
\begin{equation}
\varepsilon_t=\frac{d}{L(1+\alpha)}\Rightarrow (1+\alpha)\, \varepsilon_t=\varepsilon_m+\alpha\, \varepsilon_d.
\label{eq:3}
\end{equation}
By using \eqref{eq:3} we can express the device strain as
\begin{equation}
\alpha\, \varepsilon_d=(1+\alpha)\, \varepsilon_t-\varepsilon_m.
\end{equation}
The equilibrium condition requires a constant force $F$:
\begin{equation}
k_m(\varepsilon_i-\chi_i\varepsilon_u)=k_d\,\varepsilon_d=F.
\end{equation}
By averaging with respect to $i$ we then get the relation between the chain and the device strain
\begin{equation}
k_d \,\varepsilon_d = k_m(\varepsilon_m-\varepsilon_u \bar{\chi}),
\end{equation}
where $\bar{\chi}=\frac{\sum_{i=1}^n \chi_i}{n}$ is the fraction of unfolded domains. After introducing the non-dimensional parameter  
\begin{equation}
\gamma=\frac{k_d}{k_d+\alpha\, k_p}  \in [0,1],
\end{equation}
measuring the relative device vs total stiffness, by using \eqref{eq:3} we get
\begin{equation}
\varepsilon_m=(1+\alpha)\gamma\, \varepsilon_t+(1-\gamma)\bar{\chi}\,\varepsilon_u.
\label{mecstress}
\end{equation}
%

\section{Helmholtz Ensemble}

Consider first the case of a hard device when the total displacement $d$ is fixed. In this case, we have to consider the partition function in the Helmholtz ensemble  defined as 
\[
\mathcal{Z}_{\mathscr{H}}=\sum_{\boldsymbol{\chi}}\int_{\mathbb{R}^{2(n+1)}} e^{-\beta \mathcal{H}}\delta \Biggl(\sum_{i=1}^{n}l\varepsilon_i+\alpha L \varepsilon_d - d\Biggr) \prod_{i}^{n}dp_i \,dp_{d} \prod_{i}^{n} l d\varepsilon_{i} (\alpha L) d\varepsilon_d, 
\]
where $\beta = 1/k_BT$, $k_B$ being the Boltzmann constant, $T$ the absolute temperature and $\boldsymbol{\chi}= \{\chi_{1},\dots ,\chi_{n}\}\in \{0,1\}^{n}$ is the vector denoting the phase (folded of unfolded) configuration. For the sake of simplicity here and in what follows we drop the domain of the vector spin variable $\boldsymbol{\chi}$. We  used the Dirac function to enforce the displacement constraint (\ref{eq:2}).
We can separate the contributions to $\mathcal{Z}_{\mathscr{H}}$ of the kinetic energy and of the potential energy, so that we can split up the integral over the momenta and over the strains, respectively: 
\[
\mathcal{Z}_{\mathscr{H}}= \alpha L l^n \int_{\mathbb{R}^{(n+1)}} e^{-\beta E_k}\prod_{i=1}^{n} dp_i dp_{d}  \sum_{\boldsymbol{\chi}}\int_{\mathbb{R}^{(n+1)}} e^{-\beta \Phi_t}\delta \Biggl(\sum_{i=1}^{n}l\varepsilon_i+\alpha L \varepsilon_d - d\Biggr)\prod_{i=1}^{n} d\varepsilon_i d\varepsilon_d.
\]
We solve the Gaussian integral over the momenta to obtain 
\begin{multline}
\mathcal{Z}_{\mathscr{H}}=A \alpha L l^n  \sum_{\boldsymbol{\chi}} \int_{\mathbb{R}^{(n+1)}}e^{-\beta \Bigl(\sum_{i}\frac{1}{2}k_m l (\varepsilon_i-\varepsilon_u\chi_i)^2+\frac{1}{2}\frac{k_d}{\alpha L}(\alpha L \varepsilon_d)^2\Bigr)}\\
\times\delta \Biggl(\sum_{i}l\varepsilon_i+\alpha L \varepsilon_d - d\Biggr)\prod_{i} d\varepsilon_i\,d\varepsilon_d,
\end{multline}
where
\begin{equation}
A=(2\pi)^{\frac{n+1}{2}}\biggl(\frac{m}{\beta }\biggr)^{\frac{n}{2}} \biggl(\frac{M}{\beta}\biggr)^{\frac{1}{2}}.
\label{A}
\end{equation}
We can also integrate out the free variable $\varepsilon_d$ to obtain 
\[
\mathcal{Z}_{\mathscr{H}}=A\,l^n \sum_{\boldsymbol{\chi}} \int_{\mathbb{R}^{n}}e^{-\beta \Bigl[\sum_{i}\frac{1}{2}k_m l (\varepsilon_i-\varepsilon_u\chi_i)^2+\frac{1}{2}\frac{k_d}{\alpha L}\bigl(\sum_{i}l\varepsilon_i-d\bigr)^2\Bigr]}\prod_{i} d\varepsilon_i.
\]
By using (\ref{eq:3}) we get
\begin{equation}
\label{ZZ}
\mathcal{Z}_{\mathscr{H}}=A\,l^n\sum_{\boldsymbol{\chi}} \int_{\mathbb{R}^{n}}e^{-\frac{\beta l k_m}{2} \Bigl[\sum_{i} (\varepsilon_i-\varepsilon_u\chi_i)^2+\eta n\,\bigl(\frac{1}{n}\sum_{i}\varepsilon_i-\varepsilon_t(1+\alpha)\bigr)^2\Bigr]}\prod_{i} d\varepsilon_i,
\end{equation}
where 
\begin{equation}
\eta = \frac{k_d}{\alpha k_m}=\frac{\gamma}{1-\gamma} \quad\text{with}\quad \eta \in [0,+\infty[ \, .
\label{eq:16}
\end{equation}

In order to solve the Gaussian integrals, we rearrange the exponent in the partition function as follows:
\begin{multline}
-\frac{\beta l k_m}{2} \Biggl[\biggl(1+\frac{\eta}{n}\biggr)\sum_{i=1}^{n} \varepsilon_i^2+\frac{\eta}{n}\sum_{i,j=1, i\ne j}^{n} \varepsilon_i \varepsilon_j\Biggr.\\
\Biggl.-2\sum_{i=1}^{n}\Bigl(\varepsilon_u \chi_i + \eta \varepsilon_t(1+\alpha)\Bigr)\varepsilon_i+\eta n \varepsilon_t^2(1+\alpha)^2+\varepsilon_u^2\sum_{i=1}^{n}\chi_i^2\Biggr]
\end{multline}
that is equal to
\begin{equation}
-\frac{1}{2}\boldsymbol{A}\boldsymbol{\varepsilon} \cdot \boldsymbol{\varepsilon}+ \boldsymbol{b}\cdot\boldsymbol{\varepsilon}+C,
\end{equation}
where we have introduced 
\begin{equation}
\begin{split}
\boldsymbol{A} &= \,\beta k_m l\begin{pmatrix}
1+\frac{\eta}{n} & \frac{\eta}{n} & \dots & \frac{\eta}{n} \\
\frac{\eta}{n}  & \ddots & &\vdots \\
\vdots & & \ddots& \vdots\\
\frac{\eta}{n} & \dots &\dots & 1+\frac{\eta}{n}\\
\end{pmatrix},\\ 
&\\
\boldsymbol{b}&= \,\{\beta k_m l \bigl(\varepsilon_u \chi_1 + \eta \,\varepsilon_t(1+\alpha)\bigr),\dots, \beta k_m l \bigl(\varepsilon_u \chi_n + \eta \,\varepsilon_t(1+\alpha)\bigr) \}^T, \\ 
&\\
\boldsymbol{\varepsilon} &= \,\{\varepsilon_1,\dots,\varepsilon_n\}^T,\\
&\\
C&=-\frac{\beta k_m l}{2}\left(n\,\eta \varepsilon_t^2(1+\alpha)^2+\varepsilon_u^2\sum_{i=1}^{n}\chi_i^2\right).
\end{split}
\end{equation}
The  Gaussian integration of quadratic functions can be solved explicitly (\cite{zinn:1996}) giving
\begin{equation}
\int_{\mathbb{R}^{n}} e^{-\frac{1}{2}\boldsymbol{A}\boldsymbol{\varepsilon} \cdot \boldsymbol{\varepsilon}+ \boldsymbol{b}\cdot\boldsymbol{\varepsilon}+C}d\boldsymbol{\varepsilon} = \sqrt{\frac{(2\pi)^n}{\mbox{det}\boldsymbol{A}}}\,e^{\frac{1}{2}\boldsymbol{A}^{-1}\boldsymbol{b}\cdot\boldsymbol{b}+C}.
\end{equation}
Thus, we obtain 
\begin{multline}
\mathcal{Z}_{\mathscr{H}}=\mathcal{K}_{\mathscr{H}}\times\\
 \sum_{\boldsymbol{\chi}}  \,e^{\frac{\beta l k_m}{2}\biggl[\sum_{i}\bigl(\varepsilon_u \chi_i + \eta \varepsilon_t(1+\alpha)\bigr)-\frac{\gamma}{n}\Bigl(\sum_{i}\bigl(\varepsilon_u \chi_i + \eta \varepsilon_t(1+\alpha)\bigr)\Bigr)^2-\varepsilon_u^2\sum_i\chi_i^2-\eta n \varepsilon_t^2(1+\alpha)^2\biggr]}
\label{f}
\end{multline}
with
\[
\mathcal{K}_{\mathscr{H}} = A\,l^n\biggl(\frac{2\pi}{\beta k_m l}\biggr)^{\frac{n}{2}}(1-\gamma)^{\frac{1}{2}}.
\]
We observe that, due to the absence of non-local energy terms, all solutions with the same unfolded fraction $\bar{\chi}$ are characterized by the same energy. As a result, the partition function describing the chain and the apparatus as a whole is 
\begin{equation}
\mathcal{Z}_{\mathscr{H}}=\mathcal{K}_{\mathscr{H}}\sum_{p=0}^n\binom{n}{p}\,e^{-\frac{\beta k_m l n \gamma}{2}\Bigl[\frac{p}{n}\varepsilon_u-\varepsilon_t(1+\alpha)\Bigr]^2}.
\label{partitionhelm}
\end{equation}
Notice that the binomial coefficient gives the number of iso-energetic configurations for a fixed value of $p$. We then deduce that the Helmholtz free energy is given by
\[
\mathcal{F}=-\frac{1}{\beta}\ln \mathcal{Z}_{\mathscr{H}}
\]
and the expectation value of the force can be obtained as 
\begin{equation}
\label{FF}
\langle F \rangle = \frac{1}{L(1+\alpha)}\frac{\partial \mathcal{F}}{\partial \varepsilon_t}=-\frac{1}{\beta L (1+\alpha)}\frac{1}{\mathcal{Z}_{\mathscr{H}}}\frac{\partial \mathcal{Z}_{\mathscr{H}}}{\partial \varepsilon_t}.
\end{equation}
Observe that the force-strain relation can be written in the same form of Equation~\eqref{eq:ch3_fepsilont} of the main paper
\begin{equation}
\langle F \rangle = k_m \gamma \bigl[\varepsilon_t(1+\alpha)-\varepsilon_u \langle \bar{\chi} \rangle\bigr]
\label{etFhelm}
\end{equation}
after introducing the (temperature-dependent) expectation value of the unfolded fraction
\begin{equation}
\langle \bar{\chi} \rangle = \langle \bar{\chi} \rangle_{\mathscr{H}}(\beta, \varepsilon_t) = \frac{\displaystyle\sum_{p=0}^n\binom{n}{p}\frac{p}{n}\,e^{-\frac{\beta k_m l n \gamma}{2}\Bigl[\frac{p}{n}\varepsilon_u-\varepsilon_t(1+\alpha)\Bigr]^2}}{\displaystyle\sum_{p=0}^n\binom{n}{p}\,e^{-\frac{\beta k_m l n \gamma}{2}\Bigl[\frac{p}{n}\varepsilon_u-\varepsilon_t(1+\alpha)\Bigr]^2}}.
\label{chiH}
\end{equation}

In order to evaluate the expectation value of the molecule strain, it is convenient to start from the expression \eqref{ZZ}. We have
\begin{multline}
\langle \varepsilon_m \rangle = \frac{A}{Z_{\mathscr{H}}} \sum_{\boldsymbol{\chi}}\int_{\mathbb{R}^{n}}\Biggl(\frac{1}{n}\sum_{i=1}^n\varepsilon_i\Biggr)\,\times\\
e^{-\frac{\beta l k_m}{2} \Bigl[\sum_{i} (\varepsilon_i-\varepsilon_u\chi_i)^2+\eta n\bigl(\frac{1}{n}\sum_{i}\varepsilon_i-\varepsilon_t(1+\alpha)\bigr)^2\Bigr]}\prod_{i=1}^n l d\varepsilon_i,
\end{multline}
where $A$ is given by (\ref{A}). It is straightforward to show that
\begin{equation}
\frac{1}{L(1+\alpha)}\frac{1}{\mathcal{Z}_{\mathscr{H}}}\frac{\partial Z_{\mathscr{H}}}{\partial \varepsilon_t} = -\beta\, k_m \frac{\gamma}{1-\gamma}\Bigl[\varepsilon_t(1+\alpha)-\langle \varepsilon_m \rangle\Bigr],
\label{media}
\end{equation}
and, thus,
\begin{equation}
\langle \varepsilon_m \rangle=(1+\alpha) \varepsilon_t-\frac{1-\gamma}{k_m \gamma }\left(-\frac{1}{\beta L(1+\alpha)}\frac{1}{\mathcal{Z}_{\mathscr{H}}}\frac{\partial \mathcal{Z}_{\mathscr{H}}}{\partial \varepsilon_t} \right)=\varepsilon_u \langle \bar{\chi} \rangle + \gamma \Bigl[\varepsilon_t(1+\alpha)-\varepsilon_u\langle \bar{\chi} \rangle\Bigr].
\label{EPSM}
\end{equation}
with the same form of the mechanical limit in Equation~\eqref{eq:ch3_epsilontm}. Finally, we have
\begin{equation}
\langle F \rangle = k_m \Bigl[\langle \varepsilon_m\rangle- \varepsilon_u\langle \bar{\chi} \rangle\Bigr],
\end{equation}
again respecting the results in Equation~\eqref{eq:ch3_Fo} of the mechanical limit, with the variation due to the expectation value of $\bar \chi$ in \eqref{chiH}.

%
\section{Gibbs Ensemble}

Consider now the case of assigned force (soft device). The partition function for the Gibbs canonical ensemble is
\[
\mathcal{Z}_{\mathscr{G}}=\sum_{\boldsymbol{\chi}}\int_{\mathbb{R}^{2(n+1)}}e^{-\beta \Bigl[\mathcal{H}-F\bigl(\sum_{i=1}^{n}l\varepsilon_i+\alpha L \varepsilon_d\bigr)\Bigr]}\prod_{i}^n dp_{i}dp_d \prod_{i}^n  l\, d\varepsilon_i (\alpha L )d\varepsilon_d,
\]
where the Hamiltonian is defined in \eqref{H}. We obtain
\[
\begin{split}
\mathcal{Z}_{\mathscr{G}}&=\displaystyle A\alpha L\,l^n\sum_{\boldsymbol{\chi}}\int_{\mathbb{R}^{(n+1)}}e^{-\beta \Bigl[\frac{1}{2}\sum_{i}\bigl(k_m l (\varepsilon_i - \varepsilon_u \chi_i)^2 - F l\varepsilon_i \bigr) + \frac{1}{2}\frac{k_d}{\alpha L}(\alpha L \varepsilon_d)^2  -F\alpha L \varepsilon_d\Bigr]}\prod_{i=1}^{n}d\varepsilon_i d\varepsilon_d \\
&=\displaystyle A\alpha L\,l^n\sum_{\boldsymbol{\chi}}\int_{\mathbb{R}^{(n+1)}}e^{-\frac{\beta l k_m}{2} \Bigl[\sum_{i}\bigl((\varepsilon_i-\varepsilon_u \chi_i)^2-\frac{2F}{k_m}\varepsilon_i\bigr)+\frac{1-\gamma}{\gamma} n(\alpha \varepsilon_d)^2-\frac{2F}{k_m}n\alpha \varepsilon_d\Bigr]}\prod_{i}d\varepsilon_i d\varepsilon_d\\
&= A\alpha L\,l^n \, I_m \,I_d,
\end{split}
\]
where $A$ has the same value \eqref{A} obtained in the case of assigned displacement, we used Eq. (\ref{eq:16}), $I_m$ and $I_d$ correspond to the integration with respect to the $\varepsilon_i$ and $\varepsilon_d$, respectively. We easily obtain
\begin{equation}
I_d=C_{\mathscr{G}}\,e^{\frac{\beta l n}{2 k_m \gamma}(1-\gamma)\,F^2}
\label{eq:49}
\end{equation}
where we have defined the constant
\[
C_{\mathscr{G}}=\frac{1}{\alpha}\biggl[\frac{2\pi (1-\gamma)}{\beta k_m l \gamma n}\biggr]^{\frac{1}{2}}.
\]
On the other hand, the integral $I_m$ can be rewritten as
\begin{equation}
\begin{split}
I_m&=\sum_{\boldsymbol{\chi}}\prod_{i=1}^n\int_{\mathbb{R}}e^{-\frac{\beta l k_m}{2} \bigl[(\varepsilon_i-\varepsilon_u \chi_i)^2-\frac{2F}{k_m}\varepsilon_i \bigr]}d\varepsilon_i\\
 &=\biggl(\frac{2\pi}{\beta k_m l}\biggr)^{\frac{n}{2}}\,\sum_{\boldsymbol{\chi}}\prod_{i=1}^n e^{\frac{\beta l }{2 k_m}\,\bigl(F^2+F\,2 k_m \varepsilon_u \chi_i\bigr)}.
\end{split}
\end{equation}
Also, in this case, we may observe that, due to the absence of non-local energy terms, the energy of the solutions with the same unfolded fraction is invariant with respect to the permutation of the elements. Thus, we obtain the analytic expression
\begin{equation}
\begin{split}
I_m  =&\biggl(\frac{2\pi}{\beta k_m l}\biggr)^{\frac{n}{2}}\sum_{p=0}^n \binom{n}{p}\Bigl(e^{\frac{\beta l}{2 k_m}\,F^2}\Bigr)^{n-p}\Bigl(e^{\frac{\beta l}{2 k_m}(F^2+F\,2 k_m \varepsilon_u)}\Bigr)^{p}\\
=&\biggl(\frac{2\pi}{\beta k_m l}\biggr)^{\frac{n}{2}}\sum_{p=0}^n \binom{n}{p}\,e^{\frac{\beta l n}{2 k_m}\,\bigl(F^2+F\,2 k_m \varepsilon_u \frac{p}{n}\bigr)}\\
=&\biggl(\frac{2\pi}{\beta k_m l}\biggr)^{\frac{n}{2}}\,e^{\frac{\beta l n}{2k_m}F^2 }\left(1+e^{\beta l \varepsilon_u F}\right)^n.
\label{eq:55}
\end{split}
\end{equation}
Finally, we find the partition function in the Gibbs ensemble:
\begin{equation}
\mathcal{Z}_{\mathscr{G}}=\mathcal{K}_{\mathscr{G}}\,e^{\frac{\beta l n}{2 k_m \gamma}\,F^2}\left(1+e^{\beta l \varepsilon_u F}\right)^n,
\label{eq:partfgibbs}
\end{equation}
where 
\[
\mathcal{K}_{\mathscr{G}}=A\alpha Ll^n \biggl(\frac{2\pi}{\beta k_m l}\biggr)^{\frac{n}{2}}C_{\mathscr{G}}.
\]

Based on this result we can deduce the constitutive force-strain relation in the case of assigned force.  By using the definition of  average strain \eqref{average}, we get
\begin{equation}
\langle \varepsilon_m \rangle = \frac{A\,(\alpha L\,l^n )\,I_d}{\mathcal{Z}_{\mathscr{G}}}\,\sum_{\boldsymbol{\chi}}\int_{\mathbb{R}^{n}}\left(\frac{1}{n}\sum_{i=1}^{n}\varepsilon_i \right)\,e^{-\frac{\beta l k_m}{2} \sum_{i}\Bigl[(\varepsilon_i-\varepsilon_u \chi_i)^2-\frac{2F}{k_m}\varepsilon_i\Bigr]}\prod_{i=1}^{n}d\varepsilon_i.
\label{BBB}
\end{equation}
This can be rewritten as
\begin{multline}
\langle \varepsilon_m \rangle = \frac{A\alpha L l^n\,I_d}{\mathcal{Z}_{\mathscr{G}}}\sum_{\boldsymbol{\chi}}\frac{1}{n}\times\\
\sum_{i=1}^{n}\left(\prod_{j=1}^{i-1}\int_{\mathbb{R}^{i-1}}e^{-\tilde{\beta}h(\varepsilon_j,\chi_j)}d\varepsilon_j\,\int_{\mathbb{R}}  \varepsilon_i\,e^{-\tilde{\beta}h(\varepsilon_i,\chi_i)}d\varepsilon_i\prod_{k=i+1}^{n}\int_{\mathbb{R}^{n-i-1}}e^{-\tilde{\beta}h(\varepsilon_k,\chi_k)}d\varepsilon_k\right)
\end{multline}
where $\tilde{\beta}=\beta l k_m/2$ and 
\[
h(\varepsilon,\chi)=\left((\varepsilon-\varepsilon_u \chi)^2-\frac{2F}{k_m}\varepsilon\right).
\]
Thus, we have a product of simple Gaussian integrals (the integral over $\varepsilon_i$ requires integration by parts). The solution can be written as
\begin{equation}
\begin{split}
\langle \varepsilon_m \rangle =& \left(\frac{2\pi}{k_m \beta l}\right)^{\frac{n}{2}}\frac{A\alpha Ll^n \,I_d}{\mathcal{Z}_{\mathscr{G}}}\sum_{\boldsymbol{\chi}}\frac{1}{n}\sum_{i=1}^{n}\,
\Biggl(\frac{1}{k_m}\left(F+k_m \varepsilon_u \chi_i \right) e^{\frac{\beta l}{2 k_m}(F^2+2 F k_m \varepsilon_u \chi_i)}\times\\
\times&\prod_{j=1}^{i-1}e^{\frac{\beta l}{2 k_m}(F^2+2 F k_m  \varepsilon_u \chi_j)}\times \prod_{k=i+1}^{n}e^{\frac{\beta l}{2 k_m}(F^2+2 F k_m  \varepsilon_u \chi_k)}\Biggr)\\
=&\left(\frac{2\pi}{k_m \beta l}\right)^{\frac{n}{2}} \frac{A\alpha L l^n\,I_d}{\mathcal{Z}_{\mathscr{G}}}\sum_{\boldsymbol{\chi}}\frac{1}{n}\times\\
\times&\sum_{i=1}^{n}\left(\frac{1}{k_m}\left(F+k_m \varepsilon_u \chi_i \right)\prod_{j=1}^{n}e^{\frac{\beta l}{2 k_m}(F^2+2 F k_m \varepsilon_u \chi_j)}\right)
\label{integralegigante}
\end{split}
\end{equation}
By simplifying \eqref{integralegigante} we get
\begin{equation}
\begin{split}
\langle \varepsilon_m \rangle =&\left(\frac{2\pi}{k_m \beta l}\right)^{\frac{n}{2}} \frac{A\alpha Ll^n \,I_d}{\mathcal{Z}_{\mathscr{G}}}\sum_{\boldsymbol{\chi}}\frac{1}{n}\sum_{i=1}^{n}\left[\frac{1}{k_m}\left(F+k_m \varepsilon_u \chi_i \right)\prod_{j=1}^{n}e^{\frac{\beta l}{2 k_m}(F^2+2 F k_m \varepsilon_u \chi_j)} \right]\\
=&\left(\frac{2\pi}{k_m \beta l}\right)^{\frac{n}{2}}\frac{A\alpha Ll^n \,I_d}{\mathcal{Z}_{\mathscr{G}}}\sum_{\boldsymbol{\chi}}\left[\left(\frac{F}{k_m}+ \varepsilon_u \frac{1}{n}\sum_{i=1}^{n}\chi_i \right)\prod_{j=1}^{n}e^{\frac{\beta l}{2 k_m}(F^2+2 F k_m  \varepsilon_u \chi_j)} \right]\\
=&\left(\frac{2\pi}{k_m \beta l}\right)^{\frac{n}{2}}\frac{A\alpha Ll^n\,I_d}{\mathcal{Z}_{\mathscr{G}}}\left[\sum_{p=0}^n\binom{n}{p}\left(\frac{F}{k_m}+\varepsilon_u\frac{p}{n}\right)\,e^{\frac{\beta l n}{2 k_m}\,\bigl(F^2+F\,2 k_m \varepsilon_u \frac{p}{n}\bigr)}\right],
\end{split}
\end{equation}
where in the last equality we followed the same procedure used in \eqref{eq:55}. Finally, using the expression  \eqref{eq:partfgibbs} of the partition function  and the integrals $I_m$, $I_d$ we obtain 
\begin{equation}
\langle \varepsilon_m \rangle = \frac{F}{k_m}+\varepsilon_u \langle \bar{\chi} \rangle
\label{kmFgib}
\end{equation}
that again has the same form of the molecular response Equation~\eqref{eq:ch3_Fo} in the purely mechanical approximation, but  in this case, we consider the expectation value of the unfolded fraction $\langle \bar{\chi} \rangle$ in the Gibbs ensemble 
\begin{equation}
\langle \bar{\chi} \rangle = \langle \bar{\chi} \rangle_{\mathscr{G}}(\beta, F)=\frac{\displaystyle\sum_{p=0}^n\binom{n}{p}\frac{p}{n}e^{\frac{\beta l n}{2 k_m}\,\bigl(F^2+F\,2 k_m \varepsilon_u \frac{p}{n}\bigr)}}{\displaystyle\sum_{p=0}^n\binom{n}{p}e^{\frac{\beta l n}{2 k_m}\,\bigl(F^2+F\,2 k_m \varepsilon_u \frac{p}{n}\bigr)}}=\frac{e^{F l \beta \varepsilon_u}}{1+e^{F l \beta \varepsilon_u}}.
\label{chigibbs}
\end{equation}

By definition, the Gibbs free energy is
\[
\mathcal{G}=-\frac{1}{\beta}\ln \mathcal{Z}_{\mathscr{G}}
\]
and the expectation value of the total strain of the system, which is the variable conjugated to the force, can be obtained as 
\begin{equation}
\langle \varepsilon_t \rangle = \frac{1}{\beta L(1+\alpha)}\frac{1}{Z_{\mathscr{G}}}\frac{\partial}{\partial F}\mathcal{Z}_{\mathscr{G}}.
\label{defet}
\end{equation}
This leads to
\begin{equation}
\langle \varepsilon_t \rangle = \frac{1}{(1+\alpha)}\biggl(\frac{F}{k_m\gamma}+\varepsilon_u \langle \bar{\chi} \rangle\biggr),
\label{etF}
\end{equation}
where we have used \eqref{chigibbs}, that   has the same form of  Equation~\eqref{eq:ch3_fepsilont}. From \eqref{kmFgib} and \eqref{etF} we can obtain the relation between $\langle \varepsilon_t \rangle$ and $\langle \varepsilon_m \rangle$
\begin{equation}
\langle \varepsilon_m \rangle=\varepsilon_u \langle \bar{\chi} \rangle + \gamma \bigl[(1+\alpha)\langle\varepsilon_t\rangle-\langle \bar{\chi} \rangle\varepsilon_u\bigr].
\end{equation}
 consistent with Equation~\eqref{eq:ch3_fepsilont} of the mechanical limit.
%

\section{From Helmholtz to Gibbs ensembles: Laplace Transform}

As well known (\cite{weiner:1983}), the partition functions in the Gibbs and Helmholtz ensembles are connected by a Laplace transform with the force $F$ and the total displacement $d$ as conjugate variables. From \eqref{partitionhelm} we have
\begin{eqnarray}
\int_\mathbb{R} \mathcal{Z}_{\mathscr{H}}\,e^{\beta \, Fd} dd&=& (1+\alpha)L\int_\mathbb{R} \mathcal{Z}_{\mathscr{H}}\,e^{\beta \bigl(F\, \varepsilon_t L(1+\alpha)\bigr)}d\varepsilon_t\nonumber\\
&=&\mathcal{K}_{\mathscr{H}} (1+\alpha)L\sum_{p=0}^{n}\binom{n}{p}\int_\mathbb{R} e^{-\beta l n\Bigl(\frac{k_m \gamma}{2}\bigl(\frac{p}{n}\varepsilon_u-\varepsilon_t(1+\alpha)\bigr)^2-F\varepsilon_t (1+\alpha)\Bigr)}d\varepsilon_t\nonumber\\
&=&\mathcal{K}_{\mathscr{H}}L\left(\frac{2\pi}{k_m l n \beta\gamma}\right)^{1/2}\sum_{p=0}^{n}\binom{n}{p}\,e^{\frac{\beta l n}{2 k_m \gamma}\Bigl(F^2 + F\, 2 k_m \gamma \varepsilon_u \frac{p}{n}\Bigr)}\nonumber\\
&=&\mathcal{K}_{\mathscr{G}}\,e^{\frac{\beta l n}{2 k_m \gamma}\,F^2}\left(1+e^{\beta l \varepsilon_u F}\right)^n=\mathcal{Z}_{\mathscr{G}},
\label{partgibs}
\end{eqnarray}
which is exactly the result obtained in \eqref{eq:partfgibbs}. The other quantities $\langle \varepsilon_t \rangle$ and $\langle \varepsilon_m \rangle$ in the Gibbs ensemble can be obtained accordingly.

%

\section{Thermodynamic limit}

In this section, we show how to evaluate the expression of the phase fraction expression \eqref{chiH} in the thermodynamic limit by using the saddle point method (\cite{zinn:1996}). According to the previous discussion the dependence of the response from temperature, device stiffness and discreteness parameter $n$ is measured by the expectation value of the unfolded fraction, being the other expectation values of mechanical variable related by the same equations~\eqref{eq:ch3_fepsilont},~\eqref{eq:ch3_epsilontm} and~\eqref{eq:ch3_Fo}. Since in the Gibbs ensemble, the formula \eqref{chigibbs} does not depend on $n$ the thermodynamical limit behavior coincides with one of the systems with finite discreteness. We then need to study only the limit  the Helmholtz ensemble $\langle \bar{\chi} \rangle$ in \eqref{chiH}.  To this end, we start considering the function $f$ defined as
\begin{equation}
f(\varepsilon_t) = \sum_{p=0}^n\binom{n}{p}\,e^{-\frac{\beta k_m l n \gamma}{2}\bigl[\frac{p}{n}\varepsilon_u-\varepsilon_t\left(1+\alpha\right)\bigr]^2}.
\label{tlimit1}
\end{equation}
Using the Stirling approximation, $n!\sim \left (\frac{n}{e}\right )^n \sqrt{2\pi n}$ for $n\gg1$, \eqref{tlimit1} can be written as 
\[
f(\varepsilon_t) \simeq \frac{1}{\sqrt{2 \pi n}}\sum_{p=0}^n \, \sqrt{\frac{1}{\frac{p}{n}(1-\frac{p}{n})}}\,\displaystyle  e^{n\, \text{ln}\,n-p\, \text{ln}\, p-(n-p)\, \text{ln}(n-p)-\frac{\beta k_m l n \gamma}{2}\left(\frac{p}{n}\varepsilon_u-\left(1+\alpha\right)\varepsilon_t\right)^2},
\]
where we  considered both $n$ and $p$ large. To deduce the thermodynamic limit, let us introduce the variable  $x=p/n$. In the limit of large $n$ we obtain
\[
f(\varepsilon_t) \simeq \sqrt{\frac{n}{2 \pi}}\int_{0}^1\, \sqrt{\frac{1}{x(1-x)}}\,\displaystyle  e^{-n \left( S(x) +\frac{\beta k_m l  \gamma}{2}\left(x\,\varepsilon_u-\left(1+\alpha\right)\varepsilon_t\right)^2\right )}\text{d}x
\]
where we  defined the (entropy) function
\[
S(x) = x\, \text{ln}x + (1-x)\text{ln}\, (1-x).
\]
Finally, for large $n$ we can apply the saddle point approximation. We search for the minimum of the function
\[
 S(x)+\frac{\beta k_m l \gamma}{2}\Bigl[x\,\varepsilon_u - \varepsilon_t (1+\alpha)\Bigr]^2
 \]
which can be found in solving the equation
\begin{equation}
\text{ln}\left (\frac{x}{1-x}\right )+\varepsilon_u \beta k_m l \gamma \Bigl[x\,\varepsilon_u-(1+\alpha) \varepsilon_t\Bigr]= 0. 
\label{hlimitt}
\end{equation}
It is easy to see that there exists only one solution $\chi_{c}$ in the interval $]0,1[$. Thus, we can solve the integral with the saddle point method by considering the expansion around $\chi_c$ up to the second-order as it follows: 
\begin{multline}
f(\varepsilon_t) \simeq \sqrt{\frac{n}{2 \pi}}\int_{0}^1\, \sqrt{\frac{1}{\chi_{c}(1-\chi_{c})}}\times\\
e^{-n \left(S(\chi_{c}) -\frac{\beta k_m l n \gamma}{2}\left(\chi_{c}\,\varepsilon_u-\left(1+\alpha\right)\varepsilon_t\right)^2-\frac{1}{2}\left(S''(\chi_{c})+\beta l k_m \gamma \varepsilon_u^2\right)\left(x-\chi_{c}\right)^2\right)}\text{d}x.
\end{multline}
By substituting the variable $y=\sqrt{n}(x-\chi_{c})$  we get
\begin{multline}
f(\varepsilon_t) \simeq \sqrt{\frac{1}{2\pi \chi_{c}(1-\chi_{c})}}\,e^{-n \left(S(\chi_{c}) -\frac{\beta k_m l n \gamma}{2}\left(\chi_{c}\,\varepsilon_u-\varepsilon_t\left(1+\alpha\right)\right)^2\right)}\times\\
\times\int_{-\sqrt{n}\,\chi_{c}}^{\sqrt{n}\,\chi_{c}} e^{-\frac{1}{2}\left(S''(\chi_{c})+\beta l k_m \gamma \varepsilon_u^2\right)y^2}\text{d}y.
\end{multline}
In the limit $n\rightarrow\infty$, we obtain
\[
f(\varepsilon_t) \sim \frac{\displaystyle e^{-n\left(S(\chi_{c})+\frac{\beta k_m l \gamma}{2}\left(\chi_{c} \varepsilon_u-\varepsilon_t(1+\alpha)\right)^2\right)}}{\displaystyle \sqrt{1+\beta k_m l \gamma \varepsilon_u \chi_{c}(1-\chi_{c})}}.
\]
Similarly, we can show that
\[
g(\varepsilon_t)=\sum_{p=0}^n\binom{n}{p}\,\frac{p}{n}\,e^{-\frac{\beta k_m l n \gamma}{2}\left(\frac{p}{n}\varepsilon_u-\varepsilon_t\left(1+\alpha\right)\right)^2} \sim \frac{\displaystyle \chi_{c}\, e^{-n\left(S(\chi_{c})+\frac{\beta k_m l \gamma}{2}\left(\chi_{c} \varepsilon_u-\varepsilon_t(1+\alpha)\right)^2\right)}}{\displaystyle \sqrt{1+\beta k_m l \gamma \varepsilon_u \chi_{c}(1-\chi_{c})}}.
\]
Finally, we get
\begin{equation}
\langle \bar{\chi} \rangle  =\frac{g(\varepsilon_t)}{f(\varepsilon_t)} \sim \chi_{c}(\varepsilon_t).
\end{equation}

%
\section{Ideal Cases}

In this section we consider the \textit{ideal} cases, typically considered in the literature, when the device effect is neglected and the displacement ({\it ideal hard device}) or the force ({\it ideal soft device}) is directly applied to the unfolding molecule. In this case $ \varepsilon_m \equiv  \varepsilon_t$ and the Hamiltonian  is  
\begin{equation}
\mathcal{H}^{id} =\sum_{i=1}^{n}\frac{1}{2 m}p_i^2+ \frac{1}{2}k_m l\sum_{i=1}^{n}(\varepsilon_i-\varepsilon_u \chi_i)^2 .
\label{hideal}
\end{equation}
%

\subsection{Ideal Helmholtz ensemble}

Using \eqref{hideal}, the partition function in the Helmholtz ensemble for the ideal case is 
\begin{equation}
\mathcal{Z}_{\mathscr{H}}^{id} = \sum_{\boldsymbol{\chi}}l^n\int_{\mathbb{R}^{2n}}e^{-\beta \mathcal{H}^{id}}\delta\left(l\sum_{i=1}^n\varepsilon_i-d\right)\prod_{i=1}^{n}dp_i\prod_{i=1}^{n}d\varepsilon_i.
\end{equation}
The integrals over the momenta result in the constant 
\[
A_{\mathscr{H}}^{id}=l^n (2\pi)^{\frac{n}{2}}\left(\frac{m}{\beta}\right)^{\frac{n}{2}}.
\]
The constraint on the total displacement is imposed by the Dirac delta as follows: 
\begin{equation}
\begin{split}
\mathcal{Z}_{\mathscr{H}}^{id} &= A_{\mathscr{H}}^{id} \sum_{\boldsymbol{\chi}} \int_{\mathbb{R}^{n}} e^{\,-\beta\left(\frac{k_m l}{2}\sum_{i=1}^{n-1}\left(\varepsilon_i-\varepsilon_u \chi_i\right)^2+\frac{k_m l}{2}\left(\varepsilon_n-\varepsilon_u \chi_n\right)^2\right)}\times\\
&\times\delta\left(l\sum_{i=1}^{n-1}\varepsilon_i+l\varepsilon_n-d\right)\prod_{i=1}^{n}d\varepsilon_i\\
&=A_{\mathscr{H}}^{id} \sum_{\boldsymbol{\chi}} \int_{\mathbb{R}^{n-1}} e^{\,-\frac{\beta k_m l}{2}(\sum_{i=1}^{n-1}\left(\varepsilon_i-\varepsilon_u \chi_i)^2+\left(\sum_{i=1}^{n-1}\varepsilon_i-\varepsilon_u\chi_n-n\,\varepsilon_m\right)^2\right)}\prod_{i=1}^{n-1}d\varepsilon_i\\
&=A_{\mathscr{H}}^{id} \sum_{\boldsymbol{\chi}} \int_{\mathbb{R}^{n-1}} e^{-\frac{1}{2}\boldsymbol{A}\boldsymbol{\varepsilon} \cdot \boldsymbol{\varepsilon}+ \boldsymbol{b}\cdot\boldsymbol{\varepsilon}+C}\prod_{i=1}^{n-1}d\varepsilon_i \, ,
\end{split}
\end{equation}
where we have introduced 
\begin{equation}
\begin{split}
\boldsymbol{A} &= \beta k_m l\begin{pmatrix}
2 & 1 & \dots & 1 \\
1 & 2 & &\vdots \\
\vdots & & \ddots& \vdots\\
1 & \dots &\dots & 2\\
\end{pmatrix}, \\ 
&\\
\boldsymbol{b}&= \{\beta k_m l \left(\varepsilon_u \chi_1 +\varepsilon_u \chi_n + n\,\varepsilon_m\right),...,\beta k_m l \left(\varepsilon_u \chi_{n-1} +\varepsilon_u \chi_n + n\,\varepsilon_m\right) \}^T, \\ 
&\\
\boldsymbol{\varepsilon} &= \{\varepsilon_1,...,\varepsilon_{n-1}\}^T,\\
&\\
C&=\varepsilon_u^2\sum_{i=1}^{n-1}\chi_i^2+\varepsilon_u^2\chi_n^2+n^2\varepsilon_m^2+2\varepsilon_u\chi_n\varepsilon_m\,n.
\end{split}
\end{equation}
The Gaussian integration can be solved as before in the general case with the presence of the device (\cite{zinn:1996}). We obtain 
\begin{multline}
\mathcal{Z}_{\mathscr{H}}^{id} = \mathcal{K}_{\mathscr{H}}^{id} \sum_{\boldsymbol{\chi}} \,\text{exp}\Biggl[\frac{\beta k_m l }{2}\left(\sum_{i=1}^{n-1}\left(\varepsilon_u\chi_i+\varepsilon_u\chi_n+ n\,\varepsilon_m\right)^2 - \frac{1}{n}\left(\sum_{i=1}^{n-1}\left(\varepsilon_u\chi_i+\varepsilon_u\chi_n+ n\,\varepsilon_m\right)\right)^2\right.\\
\left.-\varepsilon_u^2\sum_{i=1}^{n-1}\chi_i^2-\varepsilon_u^2\chi_n^2-n^2\varepsilon_m^2-2\varepsilon_u\chi_n \varepsilon_m n \right)\Biggr]
\end{multline}
with
\[
\mathcal{K}_{\mathscr{H}}^{id} = A_{\mathscr{H}}^{id} \sqrt{\frac{(2\pi)^{n-1}}{(\beta k_m l )^{n-1}\,n}}.
\]
Finally, we obtain the  partition function for the ideal case in the Helmholtz ensemble: 
\[
\mathcal{K}_{\mathscr{H}}^{id} = K_{\mathscr{H}}\sum_{p=0}^n\binom{n}{p}\,e^{-\frac{\beta k_m l n}{2}\left(\frac{p}{n}\varepsilon_u-\varepsilon_t\right)^2}.
\]
Using a procedure analogous to the general case, we deduce the formula for the expectation value of the unfolded fraction in the ideal case:
\begin{equation}
\langle \bar{\chi} ^{id}\rangle = \langle \bar{\chi}^{id} \rangle_{\mathscr{H}}(\beta, \varepsilon_t) = \frac{\displaystyle\sum_{p=0}^n\binom{n}{p}\frac{p}{n}\,e^{-\frac{\beta k_m l n }{2}\left(\frac{p}{n}\varepsilon_u-\varepsilon_t\right)^2}}{\displaystyle\sum_{p=0}^n\binom{n}{p}\,e^{-\frac{\beta k_m l n }{2}\left(\frac{p}{n}\varepsilon_u-\varepsilon_t\right)^2}}.
\label{chiHideal}
\end{equation}
%

\subsection{Ideal Gibbs Ensemble}

If we apply a fixed force at the end of the chain of $n$ bistable elements without considering the measuring device we obtain the case of \textit{ideal soft device}. By using \eqref{hideal} we can write the partition function in the Gibbs ensemble as
\[
\mathcal{Z}_{\mathscr{G}}^{id} = \sum_{\boldsymbol{\chi}}\int_{\mathbb{R}^{2n}}e^{-\beta \left( \mathcal{H}^{id}- F l \sum_{i=1}^{n} \varepsilon_i \right)} \prod_{i=1}^{n}dp_i \prod_{i=1}^{n}l\, d\varepsilon_i.
\]
As in the Helmholtz ensemble the integral over the momenta gives the constant 
\[
A_{\mathscr{G}}^{id}=A_{\mathscr{H}}^{id}=l^n (2\pi)^{\frac{n}{2}}\left(\frac{m}{\beta}\right)^{\frac{n}{2}}.
\]
The integrals over the strains can be rewritten as 
\begin{eqnarray}
\mathcal{Z}_{\mathscr{G}}^{id} &=& A_{\mathscr{G}}^{id} \sum_{\boldsymbol{\chi}}\int_{\mathbb{R}^{n}}e^{-\frac{\beta k_m l }{2}\sum_{i=1}^{n}\left(\left(\varepsilon_i-\varepsilon_u\chi_i\right)^2-\frac{2F}{k_m}\varepsilon_i\right)}\prod_{i=1}^{n}d\varepsilon_i\nonumber\\
&=&A_{\mathscr{G}}^{id} \sum_{\boldsymbol{\chi}}\prod_{i=1}^{n}\int_{\mathbb{R}}e^{-\frac{\beta k_m l }{2}\left(\left(\varepsilon_i-\varepsilon_u\chi_i\right)^2-\frac{2F}{k_m}\varepsilon_i\right)}d\varepsilon_i.
\end{eqnarray}
The solution can be obtained exactly as in Section~\ref{sec:ch3_temp}. We have 
\begin{equation}
\mathcal{Z}_{\mathscr{G}}^{id} = \mathcal{K}_{\mathscr{G}}^{id}\sum_{p=0}^n \binom{n}{p}\,e^{\,\frac{\beta l n}{2 k_m}\,\bigl(F^2+2 k_m \varepsilon_u \frac{p}{n}F\bigr)} = \mathcal{K}_{\mathscr{G}}^{id} \,e^{\frac{\beta l n}{2 k_m \gamma}F^2}\Bigl(1+e^{l \beta \varepsilon_u F}\Bigr)^n.
\label{zidealg}
\end{equation}
From \eqref{zidealg} we can obtain, as in the previous cases, the expectation value of the strain of the molecule and the expectation value of the unfolded fraction in the ideal case 
\[
\langle \varepsilon_m \rangle = \frac{F}{k_m}+\varepsilon_u \langle \bar{\chi}^{id} \rangle,
\]
\[
\langle \bar{\chi}^{id} \rangle = \langle \bar{\chi}^{id} \rangle_{\mathscr{G}}(\beta, F) = \frac{e^{\, \beta l \varepsilon_u F}}{1+e^{\, \beta l \varepsilon_u F}}.
\]
%



\renewcommand{\thefigure}
{F.\arabic{figure}}
\setcounter{figure}{0}

\renewcommand{\theequation}
{F.\arabic{equation}}
\setcounter{equation}{0}

\chapter{Linearization of the tridiagonal matrix $\boldsymbol{J}$}
\label{appF}

In this appendix we derive the Taylor expansion of the inverse of a general tridiagonal matrix (\cite{puglisi:2006}). Let us consider the Hessian matrix
\begin{equation}
\boldsymbol{J}=\boldsymbol{L}+\alpha\boldsymbol{A},
\end{equation}
where $\alpha$ is a small parameter and $\boldsymbol{A}$ is generic matrix. We may write 
\begin{equation}
\boldsymbol{J}^{-1}=\sum_{i=0}^{n}\left(-\alpha\right)^{i}\boldsymbol{L}^{-1}\left(\boldsymbol{A}\boldsymbol{L}^{-1}\right)^{i}.
\end{equation}
Indeed since
\[
	\left(\frac{d}{d\alpha}\right)\left(\left(\boldsymbol{L}+\alpha\boldsymbol{A}\right)\left(\boldsymbol{L}+\alpha\boldsymbol{A}\right)^{-1}\right)=0
\]
we get
\[
	\left(\frac{d}{d\alpha}\right)\left(\boldsymbol{L}+\alpha\boldsymbol{A}\right)^{-1}=-\left(\boldsymbol{L}+\alpha\boldsymbol{A}\right)^{-1}\boldsymbol{A}\left(\boldsymbol{L}+\alpha\boldsymbol{A}\right)^{-1}
\]
Successive derivatives are
\[
	\begin{split}
	&\left(\frac{d}{d\alpha}\right)\left(i!\left(\boldsymbol{L}+\alpha\boldsymbol{A}\right)^{-1}\left(\boldsymbol{A}\left(\boldsymbol{L}+\alpha\boldsymbol{A}\right)^{-1}\right)^{i}\right)=\\
	=&-(i+1)!\left(\boldsymbol{L}+\alpha\boldsymbol{A}\right)^{-1}\left(\boldsymbol{A}\left(\boldsymbol{L}+\alpha\boldsymbol{A}\right)^{-1}\right)^{(i+1)}\\
	\end{split}
\]
and so we obtain
\[
	\begin{split}
	&\left(\frac{d^{i}}{d\alpha^{i}}\right)\left(\boldsymbol{L}+\alpha\boldsymbol{A}\right)^{-1}=i!(-1)^{i}\left(\boldsymbol{L}+\alpha\boldsymbol{A}\right)^{-1}\left(\boldsymbol{A}\left(\boldsymbol{L}+\alpha\boldsymbol{A}\right)^{-1}\right)^{(i)}\\
	\end{split}
\]
and
\[
	\left(\left(\frac{d^{i}}{d\alpha^{i}}\right)\left(\boldsymbol{L}+\alpha\boldsymbol{A}\right)^{-1}\right)\vline_{\alpha=0}=i!(-1)^{i}\boldsymbol{L}^{-1}\left(\boldsymbol{A}\boldsymbol{L}^{-1}\right)^{i}
\]
that gives the result.



\renewcommand{\thefigure}
{G.\arabic{figure}}
\setcounter{figure}{0}

\renewcommand{\theequation}
{G.\arabic{equation}}
\setcounter{equation}{0}

\chapter{Non-local behavior: relation between next-to-nearest neighbor (NNN) interactions and the Ising scheme}
\chaptermark{NNN and Ising models}
\label{appG}
\vspace{1.5cm}

In this Appendix, I show the relation between Hamiltonian energy in~\eqref{eq:ch4_toten} introduced in Chapter~\ref{ch_4} and the Ising model used, in a recent work in which I am coauthor, to describe the effects of temperature on the Maxwell stress when non-local interactions and a softening phenomenon are considered (\cite{luca2022:1}). Under specific hypotheses, here we prove that the model with \acs{NNN} interaction (\cite{luca2020, puglisi:2006}) can be reconnected to a typical Ising scheme, widely adopted to study the effect of temperature on microscopic systems (\cite{makarov:2009,benedito:2018:2}). 

%
\begin{figure}[b!]
\begin{center}
\includegraphics[width=0.95\textwidth]{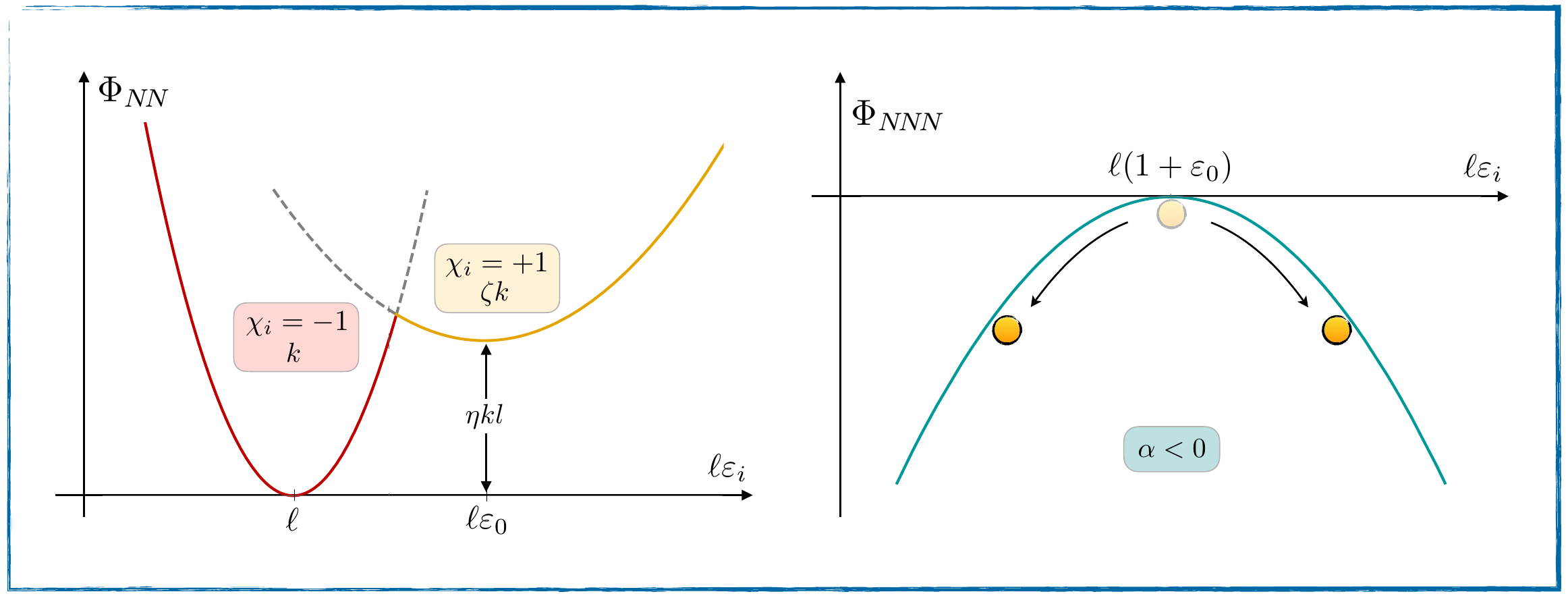}
\caption[Relation between NNN interactions and Ising scheme]{Energy of the bistable units (NN) and of the non local terms (NNN) in a generic configuration (see (\cite{luca2022:1}) for details).}
\label{fig:appG_energy}
\end{center}
\end{figure}
%

To begin with, let us consider the extended scheme with NNN harmonic springs reproduced in Figure~\ref{fig:appG_energy}, where more general equilibrium positions are considered with respect to the case presented in~\ref{fig:ch4_model}. The associated Hamiltonian energy for the unfolding units and for the non local interactions read, respectively, 
\begin{equation}
\begin{split}
\Phi_{N\!N}&=\frac{1}{2}k\ell\sum_{j=1}^{n}\Bigg[\Big(\varepsilon_j+1\Big)^2\frac{\left(1-\chi_j\right)}{2}+\bigg(\zeta \left(\varepsilon_j-\varepsilon_0\right)^2+2\eta\bigg)\frac{\left(1+\chi_j\right)}{2}\Bigg]\\
&\\
\Phi_{N\!N\!N}&=\frac{1}{2}\alpha k\ell \,\sum_{j=1}^{n-1}\Big(\varepsilon_j+\varepsilon_{j+1}-(1+\varepsilon_0)\Big)^2
\end{split}
\label{eq:appG_nnn}
\end{equation}
where $\ell$ and $\ell\varepsilon_0$ are the reference strains of the first and second well, respectively, $k/\ell$  is the stiffness of the first well ($k$ has the dimension of a force), $\zeta k/\ell$ is the stiffness of the second phase, $\alpha$ measures the relative stiffness of the non-local vs local springs and $\eta$ measures the transition energy with respect to the ground state (see Figure~\ref{fig:ch4_model}$_{b,c}$). In the ferromagnetic behaviour considered in Chapter~\ref{ch_4}, where the generation of phase interfaces is penalized, $\alpha<0$, even though this method is general and it can be extended also to the antiferromagnetic case ($\alpha>0$). 

Suppose now that the two stiffnesses of the folded and unfolded wells are sufficiently large, \textit{i.e.} as assumed in the main Chapter, that the effect of the non-local interactions is small, $\alpha\ll 1$. Then, we can assume that the strains of the units in the energy referred to the NNN terms can be approximated with their equilibrium strains of the explored wells such that
\begin{equation}
\begin{split}
\varepsilon_{j+1}+\varepsilon_{j}&\simeq\frac{1}{2}\Big[\big(1-\chi_{j+1}\big)+\varepsilon_0\big(1+\chi_{j+1}\big)\Big]+\frac{1}{2}\Big[\big(1-\chi_{j}\big)+\varepsilon_0\big(1+\chi_{j}\big)\Big]=\\
&\\
&=-\frac{1}{2}\chi_{j+1}\big(1-\varepsilon_0\big)-\frac{1}{2}\chi_{j}\big(1-\varepsilon_0\big)+\big(1+\varepsilon_0\big).
\end{split}
\label{eq:appG_ee}
\end{equation}
Then, by substituing~\eqref{eq:appG_ee} in~\eqref{eq:appG_nnn} we obtain 
\begin{equation}
\begin{split}
\Phi_{N\!N\!N}&=\frac{1}{2}\alpha k \ell \sum_{j=i}^{n-1}\Bigg[-\frac{1}{2}\big(1-\varepsilon_0\big)\big(\chi_{j+1}+\chi_j\big)\Bigg]^2\\
&=\frac{1}{4}\alpha k \ell\big(1-\varepsilon_0\big)^2\big(n-1\big)+\frac{1}{4}\alpha k \ell\big(1-\varepsilon_0\big)^2\sum_{j=1}^{n-1}\chi_{j+1}\chi_j.
\end{split}
\end{equation}

Here, by introducing the Ising coefficient $J$ such that
\begin{equation}
J=-\frac{1}{4}\alpha k \ell\big(1-\varepsilon_0\big)^2,
\end{equation}
we obtain that energy of the non-local terms in~\eqref{eq:appG_nnn} reads 
\begin{equation}
\Phi_{N\!N\!N}=+ \text{ const } -J\sum_{j=1}^{n-1}\chi_{j+1}\chi_j,
\end{equation}
where the constant can be neglected. This proves that the energy with NNN interactions in the hypothesis of small $\alpha$ can be approximated by a classical Ising chain.


\end{document}